\documentclass[11pt]{article} 

\usepackage{textcomp, amssymb}  
\usepackage{anysize}            
\usepackage{amsfonts}
\usepackage{amsmath}
\usepackage{dsfont}
\usepackage{MnSymbol}
\usepackage{mathtools}
\usepackage{graphicx}
\usepackage{epsfig}
\usepackage{epstopdf}
\usepackage{tikz}
\usepackage{amsthm}
\usepackage{ifpdf}

\usetikzlibrary{calc,decorations.pathmorphing,shapes,graphs}

\newcounter{sarrow}
\newcommand\xrsquigarrow[1]{%
\stepcounter{sarrow}%
\mathrel{\begin{tikzpicture}[baseline= {( $ (current bounding box.south) + (0,-0.5ex) $ )}]
\node[inner sep=.5ex] (\thesarrow) {$\scriptstyle #1$};
\path[draw,<-,decorate,
  decoration={zigzag,amplitude=0.7pt,segment length=1.2mm,pre=lineto,pre length=4pt}]
    (\thesarrow.south east) -- (\thesarrow.south west);
    \end{tikzpicture}}%
}

\newcounter{sarrow1}
\newcommand\xnrsquigarrow[1]{%
\stepcounter{sarrow1}%
\mathrel{\begin{tikzpicture}[baseline= {( $ (current bounding box.south) + (0,-0.5ex) $ )}]
\node[inner sep=.5ex] (\thesarrow) {$\scriptstyle #1$};
\path[draw,<-,decorate,
  decoration={zigzag,amplitude=0.7pt,segment length=1.2mm,pre=lineto,pre length=4pt}]
    (\thesarrow1.south east) -- (\thesarrow1.south west);
    $\slashedarrowfill@\relbar\relbar/$
    \end{tikzpicture}}%
}

\makeatletter
\def\slashedarrowfill@#1#2#3#4#5{%
  $\m@th\thickmuskip0mu\medmuskip\thickmuskip\thinmuskip\thickmuskip
   \relax#5#1\mkern-7mu%
   \cleaders\hbox{$#5\mkern-2mu#2\mkern-2mu$}\hfill
   \mathclap{#3}\mathclap{#2}%
   \cleaders\hbox{$#5\mkern-2mu#2\mkern-2mu$}\hfill
   \mkern-7mu#4$%
}
\def\rightslashedarrowfillb@{%
  \slashedarrowfill@\relbar\relbar/\rightarrow}
\newcommand\xnrightarrow[2][]{%
  \ext@arrow 0055{\rightslashedarrowfillb@}{#1}{#2}}

\def\rightslashedarrowfille@{%
  \slashedarrowfill@\relbar\relbar/\twoheadrightarrow}
\newcommand\xntworightarrow[2][]{%
  \ext@arrow 0055{\rightslashedarrowfille@}{#1}{#2}}

\def\rightslashedarrowfillg@{%
  \slashedarrowfill@\relbar\relbar{\raisebox{.12em}{}}\twoheadrightarrow}
\newcommand\xtworightarrow[2][]{%
  \ext@arrow 0055{\rightslashedarrowfillg@}{#1}{#2}}

\def\rightslashedarrowfillx@{%
  \slashedarrowfill@\Relbar\Relbar/\rightrightarrows}
\newcommand\xnTworightarrow[2][]{%
  \ext@arrow 0055{\rightslashedarrowfillx@}{#1}{#2}}

\def\rightslashedarrowfilly@{%
  \slashedarrowfill@\Relbar\Relbar{\raisebox{.12em}{}}\rightrightarrows}
\newcommand\xTworightarrow[2][]{%
  \ext@arrow 0055{\rightslashedarrowfilly@}{#1}{#2}}

\pgfdeclareshape{slash underlined}
{
  \inheritsavedanchors[from=rectangle] 
  \inheritanchorborder[from=rectangle]
  \inheritanchor[from=rectangle]{north}
  \inheritanchor[from=rectangle]{north west}
  \inheritanchor[from=rectangle]{north east}
  \inheritanchor[from=rectangle]{center}
  \inheritanchor[from=rectangle]{west}
  \inheritanchor[from=rectangle]{east}
  \inheritanchor[from=rectangle]{mid}
  \inheritanchor[from=rectangle]{mid west}
  \inheritanchor[from=rectangle]{mid east}
  \inheritanchor[from=rectangle]{base}
  \inheritanchor[from=rectangle]{base west}
  \inheritanchor[from=rectangle]{base east}
  \inheritanchor[from=rectangle]{south}
  \inheritanchor[from=rectangle]{south west}
  \inheritanchor[from=rectangle]{south east}
  \inheritanchorborder[from=rectangle]
  \foregroundpath{
    \southwest \pgf@xa=\pgf@x \pgf@ya=\pgf@y
    \northeast \pgf@xb=\pgf@x \pgf@yb=\pgf@y
    \pgf@xc=\pgf@xa
    \advance\pgf@xc by .5\pgf@xb
    \pgf@yc=\pgf@ya
    \advance\pgf@xc by -1.3pt
    \advance\pgf@yc by -1.8pt
    \pgfpathmoveto{\pgfqpoint{\pgf@xc}{\pgf@yc}}
    \advance\pgf@xc by  2.6pt
    \advance\pgf@yc by  3.6pt
    \pgfpathlineto{\pgfqpoint{\pgf@xc}{\pgf@yc}}
    \pgfpathmoveto{\pgfqpoint{\pgf@xa}{\pgf@ya}}
    \pgfpathlineto{\pgfqpoint{\pgf@xb}{\pgf@ya}}
 }
}
\tikzset{nomorepostaction/.code=\let\tikz@postactions\pgfutil@empty}

\newtheorem{theorem}{Theorem}[section]
\newtheorem{definition}[theorem]{Definition}
\newtheorem{proposition}[theorem]{Proposition}
\newtheorem{lemma}[theorem]{Lemma}

\begin{document}

\begin{titlepage}
\thispagestyle{empty}

\hrule
\begin{center}
{\bf\LARGE Truly Concurrent Calculi with Reversibility, Probabilism and Guards}

\vspace{0.7cm}
--- Yong Wang ---

\vspace{2cm}
\begin{figure}[!htbp]
 \centering
 \includegraphics[width=1.0\textwidth]{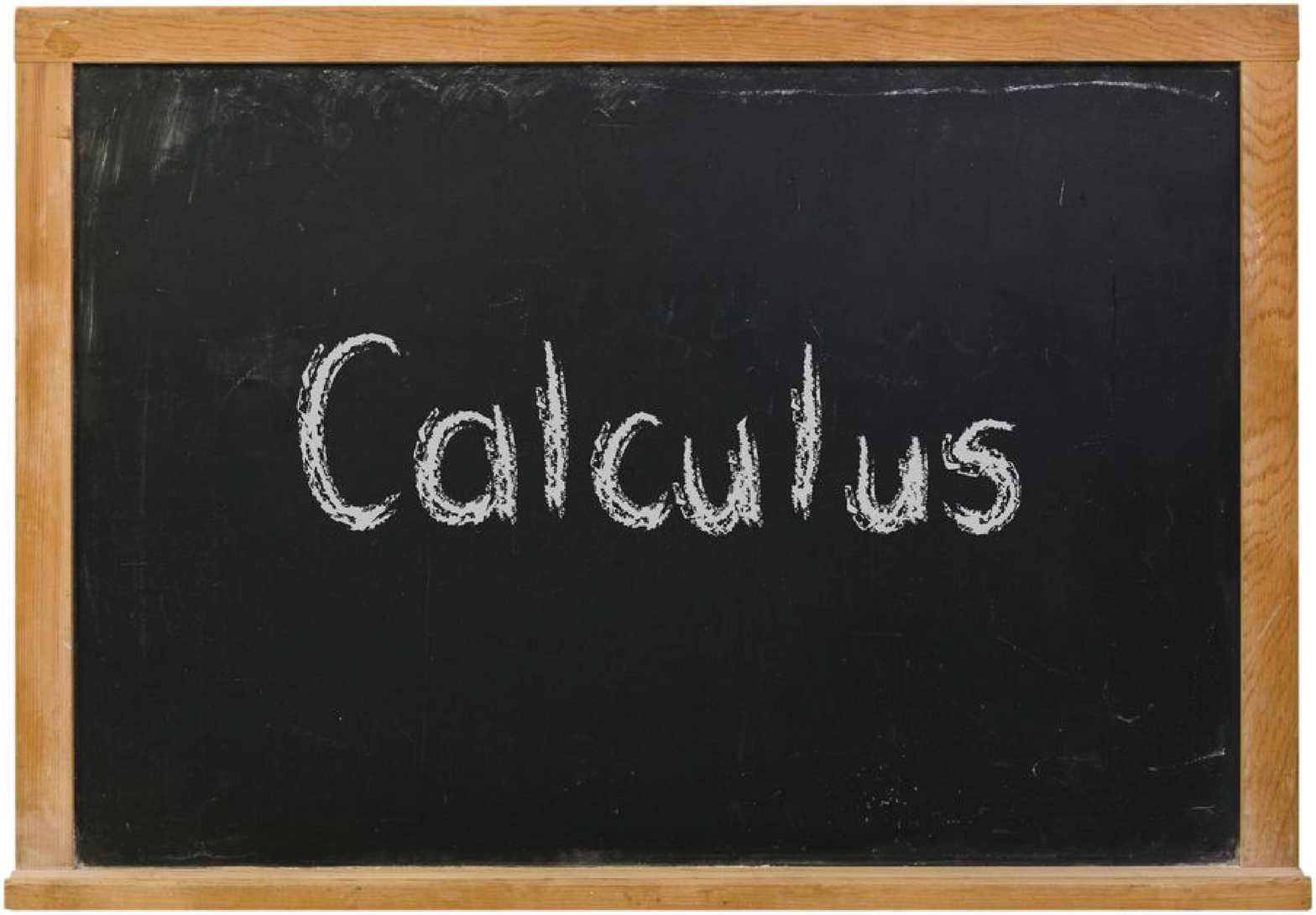}
\end{figure}

\end{center}
\end{titlepage}

\newpage 

\setcounter{page}{1}\pagenumbering{roman}

\tableofcontents

\newpage

\setcounter{page}{1}\pagenumbering{arabic}

        \section{Introduction}

The well-known process algebras, such as CCS \cite{CC} \cite{CCS}, ACP \cite{ACP} and $\pi$-calculus \cite{PI1} \cite{PI2}, capture the interleaving concurrency based on bisimilarity semantics.
We did some work on truly concurrent process algebras, such as CTC \cite{CTC}, APTC \cite{ATC} and $\pi_{tc}$ \cite{PITC}, capture the true concurrency based on truly concurrent bisimilarities, such as
pomset bisimilarity, step bisimilarity, history-preserving (hp-) bisimilarity and hereditary history-preserving (hhp-) bisimilarity. Truly concurrent process algebras are generalizations
of the corresponding traditional process algebras.

In this book, we introduce reversibility, probabilism, and guards into truly concurrent calculus CTC, based on the work on CTC \cite{CTC}, probabilistic process algebra \cite{PPA}
\cite{PPA2} \cite{PPA3}, reversible process algebra \cite{APRTC}, and process algebra with guards \cite{HLPA}. We introduce the
preliminaries in chapter \ref{bg}. CTC with guards in chapter \ref{ctcg}, CTC with probabilism and reversibility in chapter \ref{ctcpr}, CTC with probabilism and guards in chapter \ref{ctcpg},
CTC reversibility and guards in chapter \ref{ctcrg} and CTC with reversibility, probabilism and guards all together in chapter \ref{pa}. For CTC with reversibility, please refer to
\cite{APRTC}, and for CTC with probabilistic, please refer to \cite{APPTC}.

\newpage\section{Backgrounds}\label{bg}

To make this book self-satisfied, we introduce some preliminaries in this chapter, including some introductions on operational semantics, truly concurrent calculus CTC
 \cite{CTC}, which is based on truly concurrent operational semantics.

\subsection{Operational Semantics}\label{OS}

\begin{definition}[Bisimulation]
A bisimulation relation $R$ is a binary relation on processes such that: (1) if $p R q$ and $p\xrightarrow{a}p'$ then $q\xrightarrow{a}q'$ with $p' R q'$; (2) if $p R q$ and
$q\xrightarrow{a}q'$ then $p\xrightarrow{a}p'$ with $p' R q'$; (3) if $p R q$ and $pP$, then $qP$; (4) if $p R q$ and $qP$, then $pP$. Two processes $p$ and $q$ are bisimilar,
denoted by $p\sim_{HM} q$, if there is a bisimulation relation $R$ such that $p R q$.
\end{definition}

\begin{definition}[Congruence]
Let $\Sigma$ be a signature. An equivalence relation $R$ on $\mathcal{T}(\Sigma)$ is a congruence if for each $f\in\Sigma$, if $s_i R t_i$ for $i\in\{1,\cdots,ar(f)\}$, then
$f(s_1,\cdots,s_{ar(f)}) R f(t_1,\cdots,t_{ar(f)})$.
\end{definition}

\begin{definition}[Prime event structure with silent event]\label{PES}
Let $\Lambda$ be a fixed set of labels, ranged over $a,b,c,\cdots$ and $\tau$. A ($\Lambda$-labelled) prime event structure with silent event $\tau$ is a tuple
$\mathcal{E}=\langle \mathbb{E}, \leq, \sharp, \lambda\rangle$, where $\mathbb{E}$ is a denumerable set of events, including the silent event $\tau$. Let
$\hat{\mathbb{E}}=\mathbb{E}\backslash\{\tau\}$, exactly excluding $\tau$, it is obvious that $\hat{\tau^*}=\epsilon$, where $\epsilon$ is the empty event.
Let $\lambda:\mathbb{E}\rightarrow\Lambda$ be a labelling function and let $\lambda(\tau)=\tau$. And $\leq$, $\sharp$ are binary relations on $\mathbb{E}$,
called causality and conflict respectively, such that:

\begin{enumerate}
  \item $\leq$ is a partial order and $\lceil e \rceil = \{e'\in \mathbb{E}|e'\leq e\}$ is finite for all $e\in \mathbb{E}$. It is easy to see that
  $e\leq\tau^*\leq e'=e\leq\tau\leq\cdots\leq\tau\leq e'$, then $e\leq e'$.
  \item $\sharp$ is irreflexive, symmetric and hereditary with respect to $\leq$, that is, for all $e,e',e''\in \mathbb{E}$, if $e\sharp e'\leq e''$, then $e\sharp e''$.
\end{enumerate}

Then, the concepts of consistency and concurrency can be drawn from the above definition:

\begin{enumerate}
  \item $e,e'\in \mathbb{E}$ are consistent, denoted as $e\frown e'$, if $\neg(e\sharp e')$. A subset $X\subseteq \mathbb{E}$ is called consistent, if $e\frown e'$ for all
  $e,e'\in X$.
  \item $e,e'\in \mathbb{E}$ are concurrent, denoted as $e\parallel e'$, if $\neg(e\leq e')$, $\neg(e'\leq e)$, and $\neg(e\sharp e')$.
\end{enumerate}
\end{definition}

\begin{definition}[Configuration]
Let $\mathcal{E}$ be a PES. A (finite) configuration in $\mathcal{E}$ is a (finite) consistent subset of events $C\subseteq \mathcal{E}$, closed with respect to causality
(i.e. $\lceil C\rceil=C$). The set of finite configurations of $\mathcal{E}$ is denoted by $\mathcal{C}(\mathcal{E})$. We let $\hat{C}=C\backslash\{\tau\}$.
\end{definition}

A consistent subset of $X\subseteq \mathbb{E}$ of events can be seen as a pomset. Given $X, Y\subseteq \mathbb{E}$, $\hat{X}\sim \hat{Y}$ if $\hat{X}$ and $\hat{Y}$ are
isomorphic as pomsets. In the following of the paper, we say $C_1\sim C_2$, we mean $\hat{C_1}\sim\hat{C_2}$.

\begin{definition}[Pomset transitions and step]
Let $\mathcal{E}$ be a PES and let $C\in\mathcal{C}(\mathcal{E})$, and $\emptyset\neq X\subseteq \mathbb{E}$, if $C\cap X=\emptyset$ and $C'=C\cup X\in\mathcal{C}(\mathcal{E})$,
then $C\xrightarrow{X} C'$ is called a pomset transition from $C$ to $C'$. When the events in $X$ are pairwise concurrent, we say that $C\xrightarrow{X}C'$ is a step.
\end{definition}

\begin{definition}[Pomset, step bisimulation]\label{PSB}
Let $\mathcal{E}_1$, $\mathcal{E}_2$ be PESs. A pomset bisimulation is a relation $R\subseteq\mathcal{C}(\mathcal{E}_1)\times\mathcal{C}(\mathcal{E}_2)$, such that if
$(C_1,C_2)\in R$, and $C_1\xrightarrow{X_1}C_1'$ then $C_2\xrightarrow{X_2}C_2'$, with $X_1\subseteq \mathbb{E}_1$, $X_2\subseteq \mathbb{E}_2$, $X_1\sim X_2$ and $(C_1',C_2')\in R$,
and vice-versa. We say that $\mathcal{E}_1$, $\mathcal{E}_2$ are pomset bisimilar, written $\mathcal{E}_1\sim_p\mathcal{E}_2$, if there exists a pomset bisimulation $R$, such that
$(\emptyset,\emptyset)\in R$. By replacing pomset transitions with steps, we can get the definition of step bisimulation. When PESs $\mathcal{E}_1$ and $\mathcal{E}_2$ are step
bisimilar, we write $\mathcal{E}_1\sim_s\mathcal{E}_2$.
\end{definition}

\begin{definition}[Posetal product]
Given two PESs $\mathcal{E}_1$, $\mathcal{E}_2$, the posetal product of their configurations, denoted $\mathcal{C}(\mathcal{E}_1)\overline{\times}\mathcal{C}(\mathcal{E}_2)$,
is defined as

$$\{(C_1,f,C_2)|C_1\in\mathcal{C}(\mathcal{E}_1),C_2\in\mathcal{C}(\mathcal{E}_2),f:C_1\rightarrow C_2 \textrm{ isomorphism}\}.$$

A subset $R\subseteq\mathcal{C}(\mathcal{E}_1)\overline{\times}\mathcal{C}(\mathcal{E}_2)$ is called a posetal relation. We say that $R$ is downward closed when for any
$(C_1,f,C_2),(C_1',f',C_2')\in \mathcal{C}(\mathcal{E}_1)\overline{\times}\mathcal{C}(\mathcal{E}_2)$, if $(C_1,f,C_2)\subseteq (C_1',f',C_2')$ pointwise and $(C_1',f',C_2')\in R$,
then $(C_1,f,C_2)\in R$.

For $f:X_1\rightarrow X_2$, we define $f[x_1\mapsto x_2]:X_1\cup\{x_1\}\rightarrow X_2\cup\{x_2\}$, $z\in X_1\cup\{x_1\}$,(1)$f[x_1\mapsto x_2](z)=
x_2$,if $z=x_1$;(2)$f[x_1\mapsto x_2](z)=f(z)$, otherwise. Where $X_1\subseteq \mathbb{E}_1$, $X_2\subseteq \mathbb{E}_2$, $x_1\in \mathbb{E}_1$, $x_2\in \mathbb{E}_2$.
\end{definition}

\begin{definition}[(Hereditary) history-preserving bisimulation]\label{HHPB}
A history-preserving (hp-) bisimulation is a posetal relation $R\subseteq\mathcal{C}(\mathcal{E}_1)\overline{\times}\mathcal{C}(\mathcal{E}_2)$ such that if $(C_1,f,C_2)\in R$,
and $C_1\xrightarrow{e_1} C_1'$, then $C_2\xrightarrow{e_2} C_2'$, with $(C_1',f[e_1\mapsto e_2],C_2')\in R$, and vice-versa. $\mathcal{E}_1,\mathcal{E}_2$ are history-preserving
(hp-)bisimilar and are written $\mathcal{E}_1\sim_{hp}\mathcal{E}_2$ if there exists a hp-bisimulation $R$ such that $(\emptyset,\emptyset,\emptyset)\in R$.

A hereditary history-preserving (hhp-)bisimulation is a downward closed hp-bisimulation. $\mathcal{E}_1,\mathcal{E}_2$ are hereditary history-preserving (hhp-)bisimilar and are
written $\mathcal{E}_1\sim_{hhp}\mathcal{E}_2$.
\end{definition}

\begin{definition}[Weak pomset transitions and weak step]
Let $\mathcal{E}$ be a PES and let $C\in\mathcal{C}(\mathcal{E})$, and $\emptyset\neq X\subseteq \hat{\mathbb{E}}$, if $C\cap X=\emptyset$ and
$\hat{C'}=\hat{C}\cup X\in\mathcal{C}(\mathcal{E})$, then $C\xRightarrow{X} C'$ is called a weak pomset transition from $C$ to $C'$, where we define
$\xRightarrow{e}\triangleq\xrightarrow{\tau^*}\xrightarrow{e}\xrightarrow{\tau^*}$. And $\xRightarrow{X}\triangleq\xrightarrow{\tau^*}\xrightarrow{e}\xrightarrow{\tau^*}$, for every
$e\in X$. When the events in $X$ are pairwise concurrent, we say that $C\xRightarrow{X}C'$ is a weak step.
\end{definition}

We will also suppose that all the PESs in this paper are image finite, that is, for any PES $\mathcal{E}$ and $C\in \mathcal{C}(\mathcal{E})$ and
$a\in \Lambda$, $\{e\in \mathbb{E}|C\xrightarrow{e} C'\wedge \lambda(e)=a\}$ and $\{e\in\hat{\mathbb{E}}|C\xRightarrow{e} C'\wedge \lambda(e)=a\}$ is finite.

\begin{definition}[Weak pomset, step bisimulation]\label{WPSB}
Let $\mathcal{E}_1$, $\mathcal{E}_2$ be PESs. A weak pomset bisimulation is a relation $R\subseteq\mathcal{C}(\mathcal{E}_1)\times\mathcal{C}(\mathcal{E}_2)$, such that if
$(C_1,C_2)\in R$, and $C_1\xRightarrow{X_1}C_1'$ then $C_2\xRightarrow{X_2}C_2'$, with $X_1\subseteq \hat{\mathbb{E}_1}$, $X_2\subseteq \hat{\mathbb{E}_2}$, $X_1\sim X_2$ and
$(C_1',C_2')\in R$, and vice-versa. We say that $\mathcal{E}_1$, $\mathcal{E}_2$ are weak pomset bisimilar, written $\mathcal{E}_1\approx_p\mathcal{E}_2$, if there exists a weak pomset
bisimulation $R$, such that $(\emptyset,\emptyset)\in R$. By replacing weak pomset transitions with weak steps, we can get the definition of weak step bisimulation. When PESs
$\mathcal{E}_1$ and $\mathcal{E}_2$ are weak step bisimilar, we write $\mathcal{E}_1\approx_s\mathcal{E}_2$.
\end{definition}

\begin{definition}[Weakly posetal product]
Given two PESs $\mathcal{E}_1$, $\mathcal{E}_2$, the weakly posetal product of their configurations, denoted $\mathcal{C}(\mathcal{E}_1)\overline{\times}\mathcal{C}(\mathcal{E}_2)$, is
defined as

$$\{(C_1,f,C_2)|C_1\in\mathcal{C}(\mathcal{E}_1),C_2\in\mathcal{C}(\mathcal{E}_2),f:\hat{C_1}\rightarrow \hat{C_2} \textrm{ isomorphism}\}.$$

A subset $R\subseteq\mathcal{C}(\mathcal{E}_1)\overline{\times}\mathcal{C}(\mathcal{E}_2)$ is called a weakly posetal relation. We say that $R$ is downward closed when for any
$(C_1,f,C_2),(C_1',f,C_2')\in \mathcal{C}(\mathcal{E}_1)\overline{\times}\mathcal{C}(\mathcal{E}_2)$, if $(C_1,f,C_2)\subseteq (C_1',f',C_2')$ pointwise and $(C_1',f',C_2')\in R$, then
$(C_1,f,C_2)\in R$.

For $f:X_1\rightarrow X_2$, we define $f[x_1\mapsto x_2]:X_1\cup\{x_1\}\rightarrow X_2\cup\{x_2\}$, $z\in X_1\cup\{x_1\}$,(1)$f[x_1\mapsto x_2](z)=
x_2$,if $z=x_1$;(2)$f[x_1\mapsto x_2](z)=f(z)$, otherwise. Where $X_1\subseteq \hat{\mathbb{E}_1}$, $X_2\subseteq \hat{\mathbb{E}_2}$, $x_1\in \hat{\mathbb{E}}_1$,
$x_2\in \hat{\mathbb{E}}_2$. Also, we define $f(\tau^*)=f(\tau^*)$.
\end{definition}

\begin{definition}[Weak (hereditary) history-preserving bisimulation]\label{WHHPB}
A weak history-preserving (hp-) bisimulation is a weakly posetal relation $R\subseteq\mathcal{C}(\mathcal{E}_1)\overline{\times}\mathcal{C}(\mathcal{E}_2)$ such that if
$(C_1,f,C_2)\in R$, and $C_1\xRightarrow{e_1} C_1'$, then $C_2\xRightarrow{e_2} C_2'$, with $(C_1',f[e_1\mapsto e_2],C_2')\in R$, and vice-versa. $\mathcal{E}_1,\mathcal{E}_2$ are weak
history-preserving (hp-)bisimilar and are written $\mathcal{E}_1\approx_{hp}\mathcal{E}_2$ if there exists a weak hp-bisimulation $R$ such that $(\emptyset,\emptyset,\emptyset)\in R$.

A weakly hereditary history-preserving (hhp-)bisimulation is a downward closed weak hp-bisimulation. $\mathcal{E}_1,\mathcal{E}_2$ are weakly hereditary history-preserving
(hhp-)bisimilar and are written $\mathcal{E}_1\approx_{hhp}\mathcal{E}_2$.
\end{definition}

\begin{proposition}[Weakly concurrent behavioral equivalences]\label{WSCBE}
(Strongly) concurrent behavioral equivalences imply weakly concurrent behavioral equivalences. That is, $\sim_p$ implies $\approx_p$, $\sim_s$ implies $\approx_s$, $\sim_{hp}$ implies
$\approx_{hp}$, $\sim_{hhp}$ implies $\approx_{hhp}$.
\end{proposition}

\begin{proof}
From the definition of weak pomset transition, weak step transition, weakly posetal product and weakly concurrent behavioral equivalence, it is easy to see that
$\xrightarrow{e}=\xrightarrow{\epsilon}\xrightarrow{e}\xrightarrow{\epsilon}$ for $e\in \mathbb{E}$, where $\epsilon$ is the empty event.
\end{proof}

\subsection{CTC}

CTC \cite{CTC} is a calculus of truly concurrent systems. It includes syntax and semantics:

\begin{enumerate}
  \item Its syntax includes actions, process constant, and operators acting between actions, like Prefix, Summation, Composition, Restriction, Relabelling.
  \item Its semantics is based on labeled transition systems, Prefix, Summation, Composition, Restriction, Relabelling have their transition rules. CTC has good semantic properties
  based on the truly concurrent bisimulations. These properties include monoid laws, static laws, new expansion law for strongly truly concurrent bisimulations, $\tau$ laws for weakly
  truly concurrent bisimulations, and full congruences for strongly and weakly truly concurrent bisimulations, and also unique solution for recursion.
\end{enumerate}

CTC can be used widely in verification of computer systems with a truly concurrent flavor.

\newpage\section{CTC with Guards}\label{ctcg}

In this chapter, we design the calculus CTC with guards. This chapter is organized as follows. We introduce the operational semantics in section \ref{osctcg}, its syntax and operational
semantics in section \ref{sosctcg}, and its properties for strong bisimulations in section \ref{stcbctcg}, its properties for weak bisimulations in section \ref{wtcbctcg}.

\subsection{Operational Semantics}\label{osctcg}

\begin{definition}[Prime event structure with silent event and empty event]\label{PESG}
Let $\Lambda$ be a fixed set of labels, ranged over $a,b,c,\cdots$ and $\tau,\epsilon$. A ($\Lambda$-labelled) prime event structure with silent event $\tau$ and empty event $\epsilon$ 
is a tuple $\mathcal{E}=\langle \mathbb{E}, \leq, \sharp, \lambda\rangle$, where $\mathbb{E}$ is a denumerable set of events, including the silent event $\tau$ and empty event 
$\epsilon$. Let $\hat{\mathbb{E}}=\mathbb{E}\backslash\{\tau,\epsilon\}$, exactly excluding $\tau$ and $\epsilon$, it is obvious that $\hat{\tau^*}=\epsilon$. Let 
$\lambda:\mathbb{E}\rightarrow\Lambda$ be a labelling function and let $\lambda(\tau)=\tau$ and $\lambda(\epsilon)=\epsilon$. And $\leq$, $\sharp$ are binary relations on 
$\mathbb{E}$, called causality and conflict respectively, such that:

\begin{enumerate}
  \item $\leq$ is a partial order and $\lceil e \rceil = \{e'\in \mathbb{E}|e'\leq e\}$ is finite for all $e\in \mathbb{E}$. It is easy to see that 
  $e\leq\tau^*\leq e'=e\leq\tau\leq\cdots\leq\tau\leq e'$, then $e\leq e'$.
  \item $\sharp$ is irreflexive, symmetric and hereditary with respect to $\leq$, that is, for all $e,e',e''\in \mathbb{E}$, if $e\sharp e'\leq e''$, then $e\sharp e''$.
\end{enumerate}

Then, the concepts of consistency and concurrency can be drawn from the above definition:

\begin{enumerate}
  \item $e,e'\in \mathbb{E}$ are consistent, denoted as $e\frown e'$, if $\neg(e\sharp e')$. A subset $X\subseteq \mathbb{E}$ is called consistent, if $e\frown e'$ for all 
  $e,e'\in X$.
  \item $e,e'\in \mathbb{E}$ are concurrent, denoted as $e\parallel e'$, if $\neg(e\leq e')$, $\neg(e'\leq e)$, and $\neg(e\sharp e')$.
\end{enumerate}
\end{definition}

\begin{definition}[Configuration]
Let $\mathcal{E}$ be a PES. A (finite) configuration in $\mathcal{E}$ is a (finite) consistent subset of events $C\subseteq \mathcal{E}$, closed with respect to causality (i.e. 
$\lceil C\rceil=C$), and a data state $s\in S$ with $S$ the set of all data states, denoted $\langle C, s\rangle$. The set of finite configurations of $\mathcal{E}$ is denoted by 
$\langle\mathcal{C}(\mathcal{E}), S\rangle$. We let $\hat{C}=C\backslash\{\tau\}\cup\{\epsilon\}$.
\end{definition}

A consistent subset of $X\subseteq \mathbb{E}$ of events can be seen as a pomset. Given $X, Y\subseteq \mathbb{E}$, $\hat{X}\sim \hat{Y}$ if $\hat{X}$ and $\hat{Y}$ are isomorphic as 
pomsets. In the following of the paper, we say $C_1\sim C_2$, we mean $\hat{C_1}\sim\hat{C_2}$.

\begin{definition}[Pomset transitions and step]
Let $\mathcal{E}$ be a PES and let $C\in\mathcal{C}(\mathcal{E})$, and $\emptyset\neq X\subseteq \mathbb{E}$, if $C\cap X=\emptyset$ and $C'=C\cup X\in\mathcal{C}(\mathcal{E})$, then 
$\langle C,s\rangle\xrightarrow{X} \langle C',s'\rangle$ is called a pomset transition from $\langle C,s\rangle$ to $\langle C',s'\rangle$. When the events in $X$ are pairwise 
concurrent, we say that $\langle C,s\rangle\xrightarrow{X}\langle C',s'\rangle$ is a step. It is obvious that $\rightarrow^*\xrightarrow{X}\rightarrow^*=\xrightarrow{X}$ and 
$\rightarrow^*\xrightarrow{e}\rightarrow^*=\xrightarrow{e}$ for any $e\in\mathbb{E}$ and $X\subseteq\mathbb{E}$.
\end{definition}

\begin{definition}[Weak pomset transitions and weak step]
Let $\mathcal{E}$ be a PES and let $C\in\mathcal{C}(\mathcal{E})$, and $\emptyset\neq X\subseteq \hat{\mathbb{E}}$, if $C\cap X=\emptyset$ and 
$\hat{C'}=\hat{C}\cup X\in\mathcal{C}(\mathcal{E})$, then $\langle C,s\rangle\xRightarrow{X} \langle C',s'\rangle$ is called a weak pomset transition from 
$\langle C,s\rangle$ to $\langle C',s'\rangle$, where we define $\xRightarrow{e}\triangleq\xrightarrow{\tau^*}\xrightarrow{e}\xrightarrow{\tau^*}$. And 
$\xRightarrow{X}\triangleq\xrightarrow{\tau^*}\xrightarrow{e}\xrightarrow{\tau^*}$, for every $e\in X$. When the events in $X$ are pairwise concurrent, we say that 
$\langle C,s\rangle\xRightarrow{X}\langle C',s'\rangle$ is a weak step.
\end{definition}

We will also suppose that all the PESs in this paper are image finite, that is, for any PES $\mathcal{E}$ and $C\in \mathcal{C}(\mathcal{E})$ and 
$a\in \Lambda$, $\{e\in \mathbb{E}|\langle C,s\rangle\xrightarrow{e} \langle C',s'\rangle\wedge \lambda(e)=a\}$ and 
$\{e\in\hat{\mathbb{E}}|\langle C,s\rangle\xRightarrow{e} \langle C',s'\rangle\wedge \lambda(e)=a\}$ is finite.

\begin{definition}[Pomset, step bisimulation]\label{PSBG}
Let $\mathcal{E}_1$, $\mathcal{E}_2$ be PESs. A pomset bisimulation is a relation $R\subseteq\langle\mathcal{C}(\mathcal{E}_1),S\rangle\times\langle\mathcal{C}(\mathcal{E}_2),S\rangle$, 
such that if $(\langle C_1,s\rangle,\langle C_2,s\rangle)\in R$, and $\langle C_1,s\rangle\xrightarrow{X_1}\langle C_1',s'\rangle$ then 
$\langle C_2,s\rangle\xrightarrow{X_2}\langle C_2',s'\rangle$, with $X_1\subseteq \mathbb{E}_1$, $X_2\subseteq \mathbb{E}_2$, $X_1\sim X_2$ and 
$(\langle C_1',s'\rangle,\langle C_2',s'\rangle)\in R$ for all $s,s'\in S$, and vice-versa. We say that $\mathcal{E}_1$, $\mathcal{E}_2$ are pomset bisimilar, written 
$\mathcal{E}_1\sim_p\mathcal{E}_2$, if there exists a pomset bisimulation $R$, such that $(\langle\emptyset,\emptyset\rangle,\langle\emptyset,\emptyset\rangle)\in R$. By replacing 
pomset transitions with steps, we can get the definition of step bisimulation. When PESs $\mathcal{E}_1$ and $\mathcal{E}_2$ are step bisimilar, we write 
$\mathcal{E}_1\sim_s\mathcal{E}_2$.
\end{definition}

\begin{definition}[Weak pomset, step bisimulation]\label{WPSBG}
Let $\mathcal{E}_1$, $\mathcal{E}_2$ be PESs. A weak pomset bisimulation is a relation 
$R\subseteq\langle\mathcal{C}(\mathcal{E}_1),S\rangle\times\langle\mathcal{C}(\mathcal{E}_2),S\rangle$, such that if $(\langle C_1,s\rangle,\langle C_2,s\rangle)\in R$, and 
$\langle C_1,s\rangle\xRightarrow{X_1}\langle C_1',s'\rangle$ then $\langle C_2,s\rangle\xRightarrow{X_2}\langle C_2',s'\rangle$, with $X_1\subseteq \hat{\mathbb{E}_1}$, 
$X_2\subseteq \hat{\mathbb{E}_2}$, $X_1\sim X_2$ and $(\langle C_1',s'\rangle,\langle C_2',s'\rangle)\in R$ for all $s,s'\in S$, and vice-versa. We say that $\mathcal{E}_1$, 
$\mathcal{E}_2$ are weak pomset bisimilar, written $\mathcal{E}_1\approx_p\mathcal{E}_2$, if there exists a weak pomset bisimulation $R$, such that 
$(\langle\emptyset,\emptyset\rangle,\langle\emptyset,\emptyset\rangle)\in R$. By replacing weak pomset transitions with weak steps, we can get the definition of weak step bisimulation. 
When PESs $\mathcal{E}_1$ and $\mathcal{E}_2$ are weak step bisimilar, we write $\mathcal{E}_1\approx_s\mathcal{E}_2$.
\end{definition}

\begin{definition}[Posetal product]
Given two PESs $\mathcal{E}_1$, $\mathcal{E}_2$, the posetal product of their configurations, denoted 
$\langle\mathcal{C}(\mathcal{E}_1),S\rangle\overline{\times}\langle\mathcal{C}(\mathcal{E}_2),S\rangle$, is defined as

$$\{(\langle C_1,s\rangle,f,\langle C_2,s\rangle)|C_1\in\mathcal{C}(\mathcal{E}_1),C_2\in\mathcal{C}(\mathcal{E}_2),f:C_1\rightarrow C_2 \textrm{ isomorphism}\}.$$

A subset $R\subseteq\langle\mathcal{C}(\mathcal{E}_1),S\rangle\overline{\times}\langle\mathcal{C}(\mathcal{E}_2),S\rangle$ is called a posetal relation. We say that $R$ is downward 
closed when for any $(\langle C_1,s\rangle,f,\langle C_2,s\rangle),(\langle C_1',s'\rangle,f',\langle C_2',s'\rangle)\in \langle\mathcal{C}(\mathcal{E}_1),S\rangle\overline{\times}\langle\mathcal{C}(\mathcal{E}_2),S\rangle$, 
if $(\langle C_1,s\rangle,f,\langle C_2,s\rangle)\subseteq (\langle C_1',s'\rangle,f',\langle C_2',s'\rangle)$ pointwise and $(\langle C_1',s'\rangle,f',\langle C_2',s'\rangle)\in R$, 
then $(\langle C_1,s\rangle,f,\langle C_2,s\rangle)\in R$.

For $f:X_1\rightarrow X_2$, we define $f[x_1\mapsto x_2]:X_1\cup\{x_1\}\rightarrow X_2\cup\{x_2\}$, $z\in X_1\cup\{x_1\}$,(1)$f[x_1\mapsto x_2](z)=
x_2$,if $z=x_1$;(2)$f[x_1\mapsto x_2](z)=f(z)$, otherwise. Where $X_1\subseteq \mathbb{E}_1$, $X_2\subseteq \mathbb{E}_2$, $x_1\in \mathbb{E}_1$, $x_2\in \mathbb{E}_2$.
\end{definition}

\begin{definition}[Weakly posetal product]
Given two PESs $\mathcal{E}_1$, $\mathcal{E}_2$, the weakly posetal product of their configurations, denoted 
$\langle\mathcal{C}(\mathcal{E}_1),S\rangle\overline{\times}\langle\mathcal{C}(\mathcal{E}_2),S\rangle$, is defined as

$$\{(\langle C_1,s\rangle,f,\langle C_2,s\rangle)|C_1\in\mathcal{C}(\mathcal{E}_1),C_2\in\mathcal{C}(\mathcal{E}_2),f:\hat{C_1}\rightarrow \hat{C_2} \textrm{ isomorphism}\}.$$

A subset $R\subseteq\langle\mathcal{C}(\mathcal{E}_1),S\rangle\overline{\times}\langle\mathcal{C}(\mathcal{E}_2),S\rangle$ is called a weakly posetal relation. We say that $R$ is 
downward closed when for any $(\langle C_1,s\rangle,f,\langle C_2,s\rangle),(\langle C_1',s'\rangle,f,\langle C_2',s'\rangle)\in \langle\mathcal{C}(\mathcal{E}_1),S\rangle\overline{\times}\langle\mathcal{C}(\mathcal{E}_2),S\rangle$, 
if $(\langle C_1,s\rangle,f,\langle C_2,s\rangle)\subseteq (\langle C_1',s'\rangle,f',\langle C_2',s'\rangle)$ pointwise and $(\langle C_1',s'\rangle,f',\langle C_2',s'\rangle)\in R$, 
then $(\langle C_1,s\rangle,f,\langle C_2,s\rangle)\in R$.

For $f:X_1\rightarrow X_2$, we define $f[x_1\mapsto x_2]:X_1\cup\{x_1\}\rightarrow X_2\cup\{x_2\}$, $z\in X_1\cup\{x_1\}$,(1)$f[x_1\mapsto x_2](z)=
x_2$,if $z=x_1$;(2)$f[x_1\mapsto x_2](z)=f(z)$, otherwise. Where $X_1\subseteq \hat{\mathbb{E}_1}$, $X_2\subseteq \hat{\mathbb{E}_2}$, 
$x_1\in \hat{\mathbb{E}}_1$, $x_2\in \hat{\mathbb{E}}_2$. Also, we define $f(\tau^*)=f(\tau^*)$.
\end{definition}

\begin{definition}[(Hereditary) history-preserving bisimulation]\label{HHPBG}
A history-preserving (hp-) bisimulation is a posetal relation $R\subseteq\langle\mathcal{C}(\mathcal{E}_1),S\rangle\overline{\times}\langle\mathcal{C}(\mathcal{E}_2),S\rangle$ such 
that if $(\langle C_1,s\rangle,f,\langle C_2,s\rangle)\in R$, and $\langle C_1,s\rangle\xrightarrow{e_1} \langle C_1',s'\rangle$, then 
$\langle C_2,s\rangle\xrightarrow{e_2} \langle C_2',s'\rangle$, with $(\langle C_1',s'\rangle,f[e_1\mapsto e_2],\langle C_2',s'\rangle)\in R$ for all $s,s'\in S$, and vice-versa. 
$\mathcal{E}_1,\mathcal{E}_2$ are history-preserving (hp-)bisimilar and are written $\mathcal{E}_1\sim_{hp}\mathcal{E}_2$ if there exists a hp-bisimulation $R$ such that 
$(\langle\emptyset,\emptyset\rangle,\emptyset,\langle\emptyset,\emptyset\rangle)\in R$.

A hereditary history-preserving (hhp-)bisimulation is a downward closed hp-bisimulation. $\mathcal{E}_1,\mathcal{E}_2$ are hereditary history-preserving (hhp-)bisimilar and are written 
$\mathcal{E}_1\sim_{hhp}\mathcal{E}_2$.
\end{definition}

\begin{definition}[Weak (hereditary) history-preserving bisimulation]\label{WHHPBG}
A weak history-preserving (hp-) bisimulation is a weakly posetal relation 
$R\subseteq\langle\mathcal{C}(\mathcal{E}_1),S\rangle\overline{\times}\langle\mathcal{C}(\mathcal{E}_2),S\rangle$ such that if $(\langle C_1,s\rangle,f,\langle C_2,s\rangle)\in R$, and 
$\langle C_1,s\rangle\xRightarrow{e_1} \langle C_1',s'\rangle$, then $\langle C_2,s\rangle\xRightarrow{e_2} \langle C_2',s'\rangle$, with 
$(\langle C_1',s'\rangle,f[e_1\mapsto e_2],\langle C_2',s'\rangle)\in R$ for all $s,s'\in S$, and vice-versa. $\mathcal{E}_1,\mathcal{E}_2$ are weak history-preserving (hp-)bisimilar 
and are written $\mathcal{E}_1\approx_{hp}\mathcal{E}_2$ if there exists a weak hp-bisimulation $R$ such that 
$(\langle\emptyset,\emptyset\rangle,\emptyset,\langle\emptyset,\emptyset\rangle)\in R$.

A weakly hereditary history-preserving (hhp-)bisimulation is a downward closed weak hp-bisimulation. $\mathcal{E}_1,\mathcal{E}_2$ are weakly hereditary history-preserving 
(hhp-)bisimilar and are written $\mathcal{E}_1\approx_{hhp}\mathcal{E}_2$.
\end{definition}

\subsection{Syntax and Operational Semantics}\label{sosctcg}

We assume an infinite set $\mathcal{N}$ of (action or event) names, and use $a,b,c,\cdots$ to range over $\mathcal{N}$. We denote by $\overline{\mathcal{N}}$ the set of co-names and
let $\overline{a},\overline{b},\overline{c},\cdots$ range over $\overline{\mathcal{N}}$. Then we set $\mathcal{L}=\mathcal{N}\cup\overline{\mathcal{N}}$ as the set of labels, and use
$l,\overline{l}$ to range over $\mathcal{L}$. We extend complementation to $\mathcal{L}$ such that $\overline{\overline{a}}=a$. Let $\tau$ denote the silent step (internal action or
event) and define $Act=\mathcal{L}\cup\{\tau\}$ to be the set of actions, $\alpha,\beta$ range over $Act$. And $K,L$ are used to stand for subsets of $\mathcal{L}$ and $\overline{L}$
is used for the set of complements of labels in $L$. A relabelling function $f$ is a function from $\mathcal{L}$ to $\mathcal{L}$ such that $f(\overline{l})=\overline{f(l)}$. By
defining $f(\tau)=\tau$, we extend $f$ to $Act$.

Further, we introduce a set $\mathcal{X}$ of process variables, and a set $\mathcal{K}$ of process constants, and let $X,Y,\cdots$ range over $\mathcal{X}$, and $A,B,\cdots$ range over
$\mathcal{K}$, $\widetilde{X}$ is a tuple of distinct process variables, and also $E,F,\cdots$ range over the recursive expressions. We write $\mathcal{P}$ for the set of processes.
Sometimes, we use $I,J$ to stand for an indexing set, and we write $E_i:i\in I$ for a family of expressions indexed by $I$. $Id_D$ is the identity function or relation over set $D$.

For each process constant schema $A$, a defining equation of the form

$$A\overset{\text{def}}{=}P$$

is assumed, where $P$ is a process.

Let $G_{at}$ be the set of atomic guards, $\delta$ be the deadlock constant, and $\epsilon$ be the empty action, and extend $Act$ to $Act\cup\{\epsilon\}\cup\{\delta\}$. We extend
$G_{at}$ to the set of basic guards $G$ with element $\phi,\psi,\cdots$, which is generated by the following formation rules:

$$\phi::=\delta|\epsilon|\neg\phi|\psi\in G_{at}|\phi+\psi|\phi\cdot\psi$$

The predicate $test(\phi,s)$ represents that $\phi$ holds in the state $s$, and $test(\epsilon,s)$ holds and $test(\delta,s)$ does not hold. $effect(e,s)\in S$ denotes $s'$ in
$s\xrightarrow{e}s'$. The predicate weakest precondition $wp(e,\phi)$ denotes that $\forall s,s'\in S, test(\phi,effect(e,s))$ holds.

\subsubsection{Syntax}

We use the Prefix $.$ to model the causality relation $\leq$ in true concurrency, the Summation $+$ to model the conflict relation $\sharp$ in true concurrency, and the Composition
$\parallel$ to explicitly model concurrent relation in true concurrency. And we follow the conventions of process algebra.

\begin{definition}[Syntax]\label{syntax3}
Truly concurrent processes CTC with guards are defined inductively by the following formation rules:

\begin{enumerate}
  \item $A\in\mathcal{P}$;
  \item $\phi\in\mathcal{P}$;
  \item $\textbf{nil}\in\mathcal{P}$;
  \item if $P\in\mathcal{P}$, then the Prefix $\alpha.P\in\mathcal{P}$, for $\alpha\in Act$;
  \item if $P\in\mathcal{P}$, then the Prefix $\phi.P\in\mathcal{P}$, for $\phi\in G_{at}$;
  \item if $P,Q\in\mathcal{P}$, then the Summation $P+Q\in\mathcal{P}$;
  \item if $P,Q\in\mathcal{P}$, then the Composition $P\parallel Q\in\mathcal{P}$;
  \item if $P\in\mathcal{P}$, then the Prefix $(\alpha_1\parallel\cdots\parallel\alpha_n).P\in\mathcal{P}\quad(n\in I)$, for $\alpha_1,\cdots,\alpha_n\in Act$;
  \item if $P\in\mathcal{P}$, then the Restriction $P\setminus L\in\mathcal{P}$ with $L\in\mathcal{L}$;
  \item if $P\in\mathcal{P}$, then the Relabelling $P[f]\in\mathcal{P}$.
\end{enumerate}

The standard BNF grammar of syntax of CTC with guards can be summarized as follows:

$$P::=A|\textbf{nil}|\alpha.P| \phi.P| P+P | P\parallel P | (\alpha_1\parallel\cdots\parallel\alpha_n).P | P\setminus L | P[f].$$
\end{definition}

\subsubsection{Operational Semantics}

The operational semantics is defined by LTSs (labelled transition systems), and it is detailed by the following definition.

\begin{definition}[Semantics]\label{semantics3}
The operational semantics of CTC with guards corresponding to the syntax in Definition \ref{syntax3} is defined by a series of transition rules, named $\textbf{Act}$, $\textbf{Gur}$, $\textbf{Sum}$,
$\textbf{Com}$, $\textbf{Res}$, $\textbf{Rel}$ and $\textbf{Con}$ indicate that the rules are associated respectively with Prefix, Summation, Composition, Restriction, Relabelling and
Constants in Definition \ref{syntax3}. They are shown in Table \ref{TRForCTC3}.

\begin{center}
    \begin{table}
        \[\textbf{Act}_1\quad \frac{}{\langle\alpha.P,s\rangle\xrightarrow{\alpha}\langle P,s'\rangle}\]

        \[\textbf{Act}_2\quad \frac{}{\langle\epsilon,s\rangle\rightarrow\langle\surd,s\rangle}\]

        \[\textbf{Gur}\quad \frac{}{\langle\phi,s\rangle\rightarrow\langle\surd,s\rangle}\textrm{ if }test(\phi,s)\]

        \[\textbf{Sum}_1\quad \frac{\langle P,s\rangle\xrightarrow{\alpha}\langle P',s'\rangle}{\langle P+Q,s\rangle\xrightarrow{\alpha}\langle P',s'\rangle}\]

        \[\textbf{Com}_1\quad \frac{\langle P,s\rangle\xrightarrow{\alpha}\langle P',s'\rangle\quad Q\nrightarrow}{\langle P\parallel Q,s\rangle\xrightarrow{\alpha}\langle P'\parallel Q,s'\rangle}\]

        \[\textbf{Com}_2\quad \frac{\langle Q,s\rangle\xrightarrow{\alpha}\langle Q',s'\rangle\quad P\nrightarrow}{\langle P\parallel Q,s\rangle\xrightarrow{\alpha}\langle P\parallel Q',s'\rangle}\]

        \[\textbf{Com}_3\quad \frac{\langle P,s\rangle\xrightarrow{\alpha}\langle P',s'\rangle\quad \langle Q,s\rangle\xrightarrow{\beta}\langle Q',s''\rangle}{\langle P\parallel Q,s\rangle\xrightarrow{\{\alpha,\beta\}}\langle P'\parallel Q',s'\cup s''\rangle}\quad (\beta\neq\overline{\alpha})\]

        \[\textbf{Com}_4\quad \frac{\langle P,s\rangle\xrightarrow{l}\langle P',s'\rangle\quad \langle Q,s\rangle\xrightarrow{\overline{l}}\langle Q',s''\rangle}{\langle P\parallel Q,s\rangle\xrightarrow{\tau}\langle P'\parallel Q',s'\cup s''\rangle}\]

        \[\textbf{Act}_3\quad \frac{}{\langle (\alpha_1\parallel\cdots\parallel\alpha_n).P,s\rangle\xrightarrow{\{\alpha_1,\cdots,\alpha_n\}}\langle P,s'\rangle}\quad (\alpha_i\neq\overline{\alpha_j}\quad i,j\in\{1,\cdots,n\})\]

        \[\textbf{Sum}_2\quad \frac{\langle P,s\rangle\xrightarrow{\{\alpha_1,\cdots,\alpha_n\}}\langle P',s'\rangle}{\langle P+Q,s\rangle\xrightarrow{\{\alpha_1,\cdots,\alpha_n\}}\langle P',s'\rangle}\]

        \[\textbf{Res}_1\quad \frac{\langle P,s\rangle\xrightarrow{\alpha}\langle P',s'\rangle}{\langle P\setminus L,s\rangle\xrightarrow{\alpha}\langle P'\setminus L,s'\rangle}\quad (\alpha,\overline{\alpha}\notin L)\]

        \[\textbf{Res}_2\quad \frac{\langle P,s\rangle\xrightarrow{\{\alpha_1,\cdots,\alpha_n\}}\langle P',s'\rangle}{\langle P\setminus L,s\rangle\xrightarrow{\{\alpha_1,\cdots,\alpha_n\}}\langle P'\setminus L,s'\rangle}\quad (\alpha_1,\overline{\alpha_1},\cdots,\alpha_n,\overline{\alpha_n}\notin L)\]

        \[\textbf{Rel}_1\quad \frac{\langle P,s\rangle\xrightarrow{\alpha}\langle P',s'\rangle}{\langle P[f],s\rangle\xrightarrow{\langle f(\alpha)}P'[f],s'\rangle}\]

        \[\textbf{Rel}_2\quad \frac{\langle P,s\rangle\xrightarrow{\{\alpha_1,\cdots,\alpha_n\}}\langle P',s'\rangle}{\langle P[f],s\rangle\xrightarrow{\{f(\alpha_1),\cdots,f(\alpha_n)\}}\langle P'[f],s'\rangle}\]

        \[\textbf{Con}_1\quad\frac{\langle P,s\rangle\xrightarrow{\alpha}\langle P',s'\rangle}{\langle A,s\rangle\xrightarrow{\alpha}\langle P',s'\rangle}\quad (A\overset{\text{def}}{=}P)\]

        \[\textbf{Con}_2\quad\frac{\langle P,s\rangle\xrightarrow{\{\alpha_1,\cdots,\alpha_n\}}\langle P',s'\rangle}{\langle A,s\rangle\xrightarrow{\{\alpha_1,\cdots,\alpha_n\}}\langle P',s'\rangle}\quad (A\overset{\text{def}}{=}P)\]

        \caption{Transition rules of CTC with guards}
        \label{TRForCTC3}
    \end{table}
\end{center}
\end{definition}

\subsubsection{Properties of Transitions}

\begin{definition}[Sorts]\label{sorts3}
Given the sorts $\mathcal{L}(A)$ and $\mathcal{L}(X)$ of constants and variables, we define $\mathcal{L}(P)$ inductively as follows.

\begin{enumerate}
  \item $\mathcal{L}(l.P)=\{l\}\cup\mathcal{L}(P)$;
  \item $\mathcal{L}((l_1\parallel \cdots\parallel l_n).P)=\{l_1,\cdots,l_n\}\cup\mathcal{L}(P)$;
  \item $\mathcal{L}(\tau.P)=\mathcal{L}(P)$;
  \item $\mathcal{L}(\epsilon.P)=\mathcal{L}(P)$;
  \item $\mathcal{L}(\phi.P)=\mathcal{L}(P)$;
  \item $\mathcal{L}(P+Q)=\mathcal{L}(P)\cup\mathcal{L}(Q)$;
  \item $\mathcal{L}(P\parallel Q)=\mathcal{L}(P)\cup\mathcal{L}(Q)$;
  \item $\mathcal{L}(P\setminus L)=\mathcal{L}(P)-(L\cup\overline{L})$;
  \item $\mathcal{L}(P[f])=\{f(l):l\in\mathcal{L}(P)\}$;
  \item for $A\overset{\text{def}}{=}P$, $\mathcal{L}(P)\subseteq\mathcal{L}(A)$.
\end{enumerate}
\end{definition}

Now, we present some properties of the transition rules defined in Table \ref{TRForCTC3}.

\begin{proposition}
If $P\xrightarrow{\alpha}P'$, then
\begin{enumerate}
  \item $\alpha\in\mathcal{L}(P)\cup\{\tau\}\cup\{\epsilon\}$;
  \item $\mathcal{L}(P')\subseteq\mathcal{L}(P)$.
\end{enumerate}

If $P\xrightarrow{\{\alpha_1,\cdots,\alpha_n\}}P'$, then
\begin{enumerate}
  \item $\alpha_1,\cdots,\alpha_n\in\mathcal{L}(P)\cup\{\tau\}\cup\{\epsilon\}$;
  \item $\mathcal{L}(P')\subseteq\mathcal{L}(P)$.
\end{enumerate}
\end{proposition}

\begin{proof}
By induction on the inference of $P\xrightarrow{\alpha}P'$ and $P\xrightarrow{\{\alpha_1,\cdots,\alpha_n\}}P'$, there are several cases corresponding to the transition rules named
$\textbf{Act}_{1,2,3}$, $\textbf{Gur}$, $\textbf{Sum}_{1,2}$, $\textbf{Com}_{1,2,3,4}$, $\textbf{Res}_{1,2}$, $\textbf{Rel}_{1,2}$ and $\textbf{Con}_{1,2}$ in Table \ref{TRForCTC3},
we just prove the one case $\textbf{Act}_1$ and $\textbf{Act}_3$, and omit the others.

Case $\textbf{Act}_1$: by $\textbf{Act}_1$, with $P\equiv\alpha.P'$. Then by Definition \ref{sorts3}, we have (1) $\mathcal{L}(P)=\{\alpha\}\cup\mathcal{L}(P')$ if $\alpha\neq\tau$;
(2) $\mathcal{L}(P)=\mathcal{L}(P')$ if $\alpha=\tau$ or $\alpha=\epsilon$. So, $\alpha\in\mathcal{L}(P)\cup\{\tau\}\cup\{\epsilon\}$, and $\mathcal{L}(P')\subseteq\mathcal{L}(P)$, as
desired.

Case $\textbf{Act}_3$: by $\textbf{Act}_3$, with $P\equiv(\alpha_1\parallel\cdots\parallel\alpha_n).P'$. Then by Definition \ref{sorts3}, we have (1) $\mathcal{L}(P)=\{\alpha_1,\cdots,\alpha_n\}\cup\mathcal{L}(P')$ if
$\alpha_i\neq\tau$ for $i\leq n$; (2) $\mathcal{L}(P)=\mathcal{L}(P')$ if $\alpha_1,\cdots,\alpha_n=\tau$ or $\alpha_1,\cdots,\alpha_n=\epsilon$. So,
$\alpha_1,\cdots,\alpha_n\in\mathcal{L}(P)\cup\{\tau\}\cup\{\epsilon\}$, and $\mathcal{L}(P')\subseteq\mathcal{L}(P)$, as desired.
\end{proof}

\subsection{Strong Bisimulations}\label{stcbctcg}

\subsubsection{Laws and Congruence}

Based on the concepts of strongly truly concurrent bisimulation equivalences, we get the following laws.

\begin{proposition}[Monoid laws for strong pomset bisimulation] The monoid laws for strong pomset bisimulation are as follows.

\begin{enumerate}
  \item $P+Q\sim_p Q+P$;
  \item $P+(Q+R)\sim_p (P+Q)+R$;
  \item $P+P\sim_p P$;
  \item $P+\textbf{nil}\sim_p P$.
\end{enumerate}

\end{proposition}

\begin{proof}
\begin{enumerate}
  \item $P+Q\sim_p Q+P$. It is sufficient to prove the relation $R=\{(P+Q, Q+P)\}\cup \textbf{Id}$ is a strong pomset bisimulation. It can be proved similarly to the proof of 
  Monoid laws for strong pomset bisimulation in CTC, we omit it;
  \item $P+(Q+R)\sim_p (P+Q)+R$. It is sufficient to prove the relation $R=\{(P+(Q+R), (P+Q)+R)\}\cup \textbf{Id}$ is a strong pomset bisimulation. It can be proved similarly to the proof of
  Monoid laws for strong pomset bisimulation in CTC, we omit it;
  \item $P+P\sim_p P$. It is sufficient to prove the relation $R=\{(P+P, P)\}\cup \textbf{Id}$ is a strong pomset bisimulation. It can be proved similarly to the proof of
  Monoid laws for strong pomset bisimulation in CTC, we omit it;
  \item $P+\textbf{nil}\sim_p P$. It is sufficient to prove the relation $R=\{(P+\textbf{nil}, P)\}\cup \textbf{Id}$ is a strong pomset bisimulation. It can be proved similarly to the 
  proof of Monoid laws for strong pomset bisimulation in CTC, we omit it.
\end{enumerate}
\end{proof}

\begin{proposition}[Monoid laws for strong step bisimulation] The monoid laws for strong step bisimulation are as follows.
\begin{enumerate}
  \item $P+Q\sim_s Q+P$;
  \item $P+(Q+R)\sim_s (P+Q)+R$;
  \item $P+P\sim_s P$;
  \item $P+\textbf{nil}\sim_s P$.
\end{enumerate}
\end{proposition}

\begin{proof}
\begin{enumerate}
  \item $P+Q\sim_s Q+P$. It is sufficient to prove the relation $R=\{(P+Q, Q+P)\}\cup \textbf{Id}$ is a strong step bisimulation. It can be proved similarly to the proof of
  Monoid laws for strong step bisimulation in CTC, we omit it;
  \item $P+(Q+R)\sim_s (P+Q)+R$. It is sufficient to prove the relation $R=\{(P+(Q+R), (P+Q)+R)\}\cup \textbf{Id}$ is a strong step bisimulation. It can be proved similarly to the proof of
  Monoid laws for strong step bisimulation in CTC, we omit it;
  \item $P+P\sim_s P$. It is sufficient to prove the relation $R=\{(P+P, P)\}\cup \textbf{Id}$ is a strong step bisimulation. It can be proved similarly to the proof of
  Monoid laws for strong step bisimulation in CTC, we omit it;
  \item $P+\textbf{nil}\sim_s P$. It is sufficient to prove the relation $R=\{(P+\textbf{nil}, P)\}\cup \textbf{Id}$ is a strong step bisimulation. It can be proved similarly to the
  proof of Monoid laws for strong step bisimulation in CTC, we omit it.
\end{enumerate}
\end{proof}

\begin{proposition}[Monoid laws for strong hp-bisimulation] The monoid laws for strong hp-bisimulation are as follows.
\begin{enumerate}
  \item $P+Q\sim_{hp} Q+P$;
  \item $P+(Q+R)\sim_{hp} (P+Q)+R$;
  \item $P+P\sim_{hp} P$;
  \item $P+\textbf{nil}\sim_{hp} P$.
\end{enumerate}
\end{proposition}

\begin{proof}
\begin{enumerate}
  \item $P+Q\sim_{hp} Q+P$. It is sufficient to prove the relation $R=\{(P+Q, Q+P)\}\cup \textbf{Id}$ is a strong hp-bisimulation. It can be proved similarly to the proof of
  Monoid laws for strong hp-bisimulation in CTC, we omit it;
  \item $P+(Q+R)\sim_{hp} (P+Q)+R$. It is sufficient to prove the relation $R=\{(P+(Q+R), (P+Q)+R)\}\cup \textbf{Id}$ is a strong hp-bisimulation. It can be proved similarly to the proof of
  Monoid laws for strong hp-bisimulation in CTC, we omit it;
  \item $P+P\sim_{hp} P$. It is sufficient to prove the relation $R=\{(P+P, P)\}\cup \textbf{Id}$ is a strong hp-bisimulation. It can be proved similarly to the proof of
  Monoid laws for strong hp-bisimulation in CTC, we omit it;
  \item $P+\textbf{nil}\sim_{hp} P$. It is sufficient to prove the relation $R=\{(P+\textbf{nil}, P)\}\cup \textbf{Id}$ is a strong hp-bisimulation. It can be proved similarly to the
  proof of Monoid laws for strong hp-bisimulation in CTC, we omit it.
\end{enumerate}
\end{proof}

\begin{proposition}[Monoid laws for strongly hhp-bisimulation] The monoid laws for strongly hhp-bisimulation are as follows.
\begin{enumerate}
  \item $P+Q\sim_{hhp} Q+P$;
  \item $P+(Q+R)\sim_{hhp} (P+Q)+R$;
  \item $P+P\sim_{hhp} P$;
  \item $P+\textbf{nil}\sim_{hhp} P$.
\end{enumerate}
\end{proposition}

\begin{proof}
\begin{enumerate}
  \item $P+Q\sim_{hhp} Q+P$. It is sufficient to prove the relation $R=\{(P+Q, Q+P)\}\cup \textbf{Id}$ is a strong hhp-bisimulation. It can be proved similarly to the proof of
  Monoid laws for strong hhp-bisimulation in CTC, we omit it;
  \item $P+(Q+R)\sim_{hhp} (P+Q)+R$. It is sufficient to prove the relation $R=\{(P+(Q+R), (P+Q)+R)\}\cup \textbf{Id}$ is a strong hhp-bisimulation. It can be proved similarly to the proof of
  Monoid laws for strong hhp-bisimulation in CTC, we omit it;
  \item $P+P\sim_{hhp} P$. It is sufficient to prove the relation $R=\{(P+P, P)\}\cup \textbf{Id}$ is a strong hhp-bisimulation. It can be proved similarly to the proof of
  Monoid laws for strong hhp-bisimulation in CTC, we omit it;
  \item $P+\textbf{nil}\sim_{hhp} P$. It is sufficient to prove the relation $R=\{(P+\textbf{nil}, P)\}\cup \textbf{Id}$ is a strong hhp-bisimulation. It can be proved similarly to the
  proof of Monoid laws for strong hhp-bisimulation in CTC, we omit it.
\end{enumerate}
\end{proof}

\begin{proposition}[Static laws for strong pomset bisimulation]
The static laws for strong pomset bisimulation are as follows.
\begin{enumerate}
  \item $P\parallel Q\sim_p Q\parallel P$;
  \item $P\parallel(Q\parallel R)\sim_p (P\parallel Q)\parallel R$;
  \item $P\parallel \textbf{nil}\sim_p P$;
  \item $P\setminus L\sim_p P$, if $\mathcal{L}(P)\cap(L\cup\overline{L})=\emptyset$;
  \item $P\setminus K\setminus L\sim_p P\setminus(K\cup L)$;
  \item $P[f]\setminus L\sim_p P\setminus f^{-1}(L)[f]$;
  \item $(P\parallel Q)\setminus L\sim_p P\setminus L\parallel Q\setminus L$, if $\mathcal{L}(P)\cap\overline{\mathcal{L}(Q)}\cap(L\cup\overline{L})=\emptyset$;
  \item $P[Id]\sim_p P$;
  \item $P[f]\sim_p P[f']$, if $f\upharpoonright\mathcal{L}(P)=f'\upharpoonright\mathcal{L}(P)$;
  \item $P[f][f']\sim_p P[f'\circ f]$;
  \item $(P\parallel Q)[f]\sim_p P[f]\parallel Q[f]$, if $f\upharpoonright(L\cup\overline{L})$ is one-to-one, where $L=\mathcal{L}(P)\cup\mathcal{L}(Q)$.
\end{enumerate}
\end{proposition}

\begin{proof}
\begin{enumerate}
  \item $P\parallel Q\sim_p Q\parallel P$. It is sufficient to prove the relation $R=\{(P\parallel Q, Q\parallel P)\}\cup \textbf{Id}$ is a strong pomset bisimulation. It can be proved similarly to the proof of
  static laws for strong pomset bisimulation in CTC, we omit it;
  \item $P\parallel(Q\parallel R)\sim_p (P\parallel Q)\parallel R$. It is sufficient to prove the relation $R=\{(P\parallel(Q\parallel R), (P\parallel Q)\parallel R)\}\cup \textbf{Id}$ is a strong pomset bisimulation. It can be proved similarly to the proof of
  static laws for strong pomset bisimulation in CTC, we omit it;
  \item $P\parallel \textbf{nil}\sim_p P$. It is sufficient to prove the relation $R=\{(P\parallel \textbf{nil}, P)\}\cup \textbf{Id}$ is a strong pomset bisimulation. It can be proved similarly to the proof of
  static laws for strong pomset bisimulation in CTC, we omit it;
  \item $P\setminus L\sim_p P$, if $\mathcal{L}(P)\cap(L\cup\overline{L})=\emptyset$. It is sufficient to prove the relation $R=\{(P\setminus L, P)\}\cup \textbf{Id}$, if $\mathcal{L}(P)\cap(L\cup\overline{L})=\emptyset$, is a strong pomset bisimulation. It can be proved similarly to the proof of
  static laws for strong pomset bisimulation in CTC, we omit it;
  \item $P\setminus K\setminus L\sim_p P\setminus(K\cup L)$. It is sufficient to prove the relation $R=\{(P\setminus K\setminus L, P\setminus(K\cup L))\}\cup \textbf{Id}$ is a strong pomset bisimulation. It can be proved similarly to the proof of
  static laws for strong pomset bisimulation in CTC, we omit it;
  \item $P[f]\setminus L\sim_p P\setminus f^{-1}(L)[f]$. It is sufficient to prove the relation $R=\{(P[f]\setminus L, P\setminus f^{-1}(L)[f])\}\cup \textbf{Id}$ is a strong pomset bisimulation. It can be proved similarly to the proof of
  static laws for strong pomset bisimulation in CTC, we omit it;
  \item $(P\parallel Q)\setminus L\sim_p P\setminus L\parallel Q\setminus L$, if $\mathcal{L}(P)\cap\overline{\mathcal{L}(Q)}\cap(L\cup\overline{L})=\emptyset$. It is sufficient to prove the relation 
  $R=\{((P\parallel Q)\setminus L, P\setminus L\parallel Q\setminus L)\}\cup \textbf{Id}$, if $\mathcal{L}(P)\cap\overline{\mathcal{L}(Q)}\cap(L\cup\overline{L})=\emptyset$, is a strong pomset bisimulation. It can be proved similarly to the proof of
  static laws for strong pomset bisimulation in CTC, we omit it;
  \item $P[Id]\sim_p P$. It is sufficient to prove the relation $R=\{(P[Id], P)\}\cup \textbf{Id}$ is a strong pomset bisimulation. It can be proved similarly to the proof of
  static laws for strong pomset bisimulation in CTC, we omit it;
  \item $P[f]\sim_p P[f']$, if $f\upharpoonright\mathcal{L}(P)=f'\upharpoonright\mathcal{L}(P)$. It is sufficient to prove the relation $R=\{(P[f], P[f'])\}\cup \textbf{Id}$, if $f\upharpoonright\mathcal{L}(P)=f'\upharpoonright\mathcal{L}(P)$, is a strong pomset bisimulation. It can be proved similarly to the proof of
  static laws for strong pomset bisimulation in CTC, we omit it;
  \item $P[f][f']\sim_p P[f'\circ f]$. It is sufficient to prove the relation $R=\{(P[f][f'], P[f'\circ f])\}\cup \textbf{Id}$ is a strong pomset bisimulation. It can be proved similarly to the proof of
  static laws for strong pomset bisimulation in CTC, we omit it;
  \item $(P\parallel Q)[f]\sim_p P[f]\parallel Q[f]$, if $f\upharpoonright(L\cup\overline{L})$ is one-to-one, where $L=\mathcal{L}(P)\cup\mathcal{L}(Q)$. It is sufficient to prove the 
  relation $R=\{((P\parallel Q)[f], P[f]\parallel Q[f])\}\cup \textbf{Id}$, if $f\upharpoonright(L\cup\overline{L})$ is one-to-one, where $L=\mathcal{L}(P)\cup\mathcal{L}(Q)$, is a strong pomset bisimulation. It can be proved similarly to the proof of
  static laws for strong pomset bisimulation in CTC, we omit it.
\end{enumerate}
\end{proof}

\begin{proposition}[Static laws for strong step bisimulation]
The static laws for strong step bisimulation are as follows.
\begin{enumerate}
  \item $P\parallel Q\sim_s Q\parallel P$;
  \item $P\parallel(Q\parallel R)\sim_s (P\parallel Q)\parallel R$;
  \item $P\parallel \textbf{nil}\sim_s P$;
  \item $P\setminus L\sim_s P$, if $\mathcal{L}(P)\cap(L\cup\overline{L})=\emptyset$;
  \item $P\setminus K\setminus L\sim_s P\setminus(K\cup L)$;
  \item $P[f]\setminus L\sim_s P\setminus f^{-1}(L)[f]$;
  \item $(P\parallel Q)\setminus L\sim_s P\setminus L\parallel Q\setminus L$, if $\mathcal{L}(P)\cap\overline{\mathcal{L}(Q)}\cap(L\cup\overline{L})=\emptyset$;
  \item $P[Id]\sim_s P$;
  \item $P[f]\sim_s P[f']$, if $f\upharpoonright\mathcal{L}(P)=f'\upharpoonright\mathcal{L}(P)$;
  \item $P[f][f']\sim_s P[f'\circ f]$;
  \item $(P\parallel Q)[f]\sim_s P[f]\parallel Q[f]$, if $f\upharpoonright(L\cup\overline{L})$ is one-to-one, where $L=\mathcal{L}(P)\cup\mathcal{L}(Q)$.
\end{enumerate}
\end{proposition}

\begin{proof}
\begin{enumerate}
  \item $P\parallel Q\sim_s Q\parallel P$. It is sufficient to prove the relation $R=\{(P\parallel Q, Q\parallel P)\}\cup \textbf{Id}$ is a strong step bisimulation. It can be proved similarly to the proof of
  static laws for strong step bisimulation in CTC, we omit it;
  \item $P\parallel(Q\parallel R)\sim_s (P\parallel Q)\parallel R$. It is sufficient to prove the relation $R=\{(P\parallel(Q\parallel R), (P\parallel Q)\parallel R)\}\cup \textbf{Id}$ is a strong step bisimulation. It can be proved similarly to the proof of
  static laws for strong step bisimulation in CTC, we omit it;
  \item $P\parallel \textbf{nil}\sim_s P$. It is sufficient to prove the relation $R=\{(P\parallel \textbf{nil}, P)\}\cup \textbf{Id}$ is a strong step bisimulation. It can be proved similarly to the proof of
  static laws for strong step bisimulation in CTC, we omit it;
  \item $P\setminus L\sim_s P$, if $\mathcal{L}(P)\cap(L\cup\overline{L})=\emptyset$. It is sufficient to prove the relation $R=\{(P\setminus L, P)\}\cup \textbf{Id}$, if $\mathcal{L}(P)\cap(L\cup\overline{L})=\emptyset$, is a strong step bisimulation. It can be proved similarly to the proof of
  static laws for strong step bisimulation in CTC, we omit it;
  \item $P\setminus K\setminus L\sim_s P\setminus(K\cup L)$. It is sufficient to prove the relation $R=\{(P\setminus K\setminus L, P\setminus(K\cup L))\}\cup \textbf{Id}$ is a strong step bisimulation. It can be proved similarly to the proof of
  static laws for strong step bisimulation in CTC, we omit it;
  \item $P[f]\setminus L\sim_s P\setminus f^{-1}(L)[f]$. It is sufficient to prove the relation $R=\{(P[f]\setminus L, P\setminus f^{-1}(L)[f])\}\cup \textbf{Id}$ is a strong step bisimulation. It can be proved similarly to the proof of
  static laws for strong step bisimulation in CTC, we omit it;
  \item $(P\parallel Q)\setminus L\sim_s P\setminus L\parallel Q\setminus L$, if $\mathcal{L}(P)\cap\overline{\mathcal{L}(Q)}\cap(L\cup\overline{L})=\emptyset$. It is sufficient to prove the relation
  $R=\{((P\parallel Q)\setminus L, P\setminus L\parallel Q\setminus L)\}\cup \textbf{Id}$, if $\mathcal{L}(P)\cap\overline{\mathcal{L}(Q)}\cap(L\cup\overline{L})=\emptyset$, is a strong step bisimulation. It can be proved similarly to the proof of
  static laws for strong step bisimulation in CTC, we omit it;
  \item $P[Id]\sim_s P$. It is sufficient to prove the relation $R=\{(P[Id], P)\}\cup \textbf{Id}$ is a strong step bisimulation. It can be proved similarly to the proof of
  static laws for strong step bisimulation in CTC, we omit it;
  \item $P[f]\sim_s P[f']$, if $f\upharpoonright\mathcal{L}(P)=f'\upharpoonright\mathcal{L}(P)$. It is sufficient to prove the relation $R=\{(P[f], P[f'])\}\cup \textbf{Id}$, if $f\upharpoonright\mathcal{L}(P)=f'\upharpoonright\mathcal{L}(P)$, is a strong step bisimulation. It can be proved similarly to the proof of
  static laws for strong step bisimulation in CTC, we omit it;
  \item $P[f][f']\sim_s P[f'\circ f]$. It is sufficient to prove the relation $R=\{(P[f][f'], P[f'\circ f])\}\cup \textbf{Id}$ is a strong step bisimulation. It can be proved similarly to the proof of
  static laws for strong step bisimulation in CTC, we omit it;
  \item $(P\parallel Q)[f]\sim_s P[f]\parallel Q[f]$, if $f\upharpoonright(L\cup\overline{L})$ is one-to-one, where $L=\mathcal{L}(P)\cup\mathcal{L}(Q)$. It is sufficient to prove the
  relation $R=\{((P\parallel Q)[f], P[f]\parallel Q[f])\}\cup \textbf{Id}$, if $f\upharpoonright(L\cup\overline{L})$ is one-to-one, where $L=\mathcal{L}(P)\cup\mathcal{L}(Q)$, is a strong step bisimulation. It can be proved similarly to the proof of
  static laws for strong step bisimulation in CTC, we omit it.
\end{enumerate}
\end{proof}

\begin{proposition}[Static laws for strong hp-bisimulation]
The static laws for strong hp-bisimulation are as follows.
\begin{enumerate}
  \item $P\parallel Q\sim_{hp} Q\parallel P$;
  \item $P\parallel(Q\parallel R)\sim_{hp} (P\parallel Q)\parallel R$;
  \item $P\parallel \textbf{nil}\sim_{hp} P$;
  \item $P\setminus L\sim_{hp} P$, if $\mathcal{L}(P)\cap(L\cup\overline{L})=\emptyset$;
  \item $P\setminus K\setminus L\sim_{hp} P\setminus(K\cup L)$;
  \item $P[f]\setminus L\sim_{hp} P\setminus f^{-1}(L)[f]$;
  \item $(P\parallel Q)\setminus L\sim_{hp} P\setminus L\parallel Q\setminus L$, if $\mathcal{L}(P)\cap\overline{\mathcal{L}(Q)}\cap(L\cup\overline{L})=\emptyset$;
  \item $P[Id]\sim_{hp} P$;
  \item $P[f]\sim_{hp} P[f']$, if $f\upharpoonright\mathcal{L}(P)=f'\upharpoonright\mathcal{L}(P)$;
  \item $P[f][f']\sim_{hp} P[f'\circ f]$;
  \item $(P\parallel Q)[f]\sim_{hp} P[f]\parallel Q[f]$, if $f\upharpoonright(L\cup\overline{L})$ is one-to-one, where $L=\mathcal{L}(P)\cup\mathcal{L}(Q)$.
\end{enumerate}
\end{proposition}

\begin{proof}
\begin{enumerate}
  \item $P\parallel Q\sim_{hp} Q\parallel P$. It is sufficient to prove the relation $R=\{(P\parallel Q, Q\parallel P)\}\cup \textbf{Id}$ is a strong hp-bisimulation. It can be proved similarly to the proof of
  static laws for strong hp-bisimulation in CTC, we omit it;
  \item $P\parallel(Q\parallel R)\sim_{hp} (P\parallel Q)\parallel R$. It is sufficient to prove the relation $R=\{(P\parallel(Q\parallel R), (P\parallel Q)\parallel R)\}\cup \textbf{Id}$ is a strong hp-bisimulation. It can be proved similarly to the proof of
  static laws for strong hp-bisimulation in CTC, we omit it;
  \item $P\parallel \textbf{nil}\sim_{hp} P$. It is sufficient to prove the relation $R=\{(P\parallel \textbf{nil}, P)\}\cup \textbf{Id}$ is a strong hp-bisimulation. It can be proved similarly to the proof of
  static laws for strong hp-bisimulation in CTC, we omit it;
  \item $P\setminus L\sim_{hp} P$, if $\mathcal{L}(P)\cap(L\cup\overline{L})=\emptyset$. It is sufficient to prove the relation $R=\{(P\setminus L, P)\}\cup \textbf{Id}$, if $\mathcal{L}(P)\cap(L\cup\overline{L})=\emptyset$, is a strong hp-bisimulation. It can be proved similarly to the proof of
  static laws for strong hp-bisimulation in CTC, we omit it;
  \item $P\setminus K\setminus L\sim_{hp} P\setminus(K\cup L)$. It is sufficient to prove the relation $R=\{(P\setminus K\setminus L, P\setminus(K\cup L))\}\cup \textbf{Id}$ is a strong hp-bisimulation. It can be proved similarly to the proof of
  static laws for strong hp-bisimulation in CTC, we omit it;
  \item $P[f]\setminus L\sim_{hp} P\setminus f^{-1}(L)[f]$. It is sufficient to prove the relation $R=\{(P[f]\setminus L, P\setminus f^{-1}(L)[f])\}\cup \textbf{Id}$ is a strong hp-bisimulation. It can be proved similarly to the proof of
  static laws for strong hp-bisimulation in CTC, we omit it;
  \item $(P\parallel Q)\setminus L\sim_{hp} P\setminus L\parallel Q\setminus L$, if $\mathcal{L}(P)\cap\overline{\mathcal{L}(Q)}\cap(L\cup\overline{L})=\emptyset$. It is sufficient to prove the relation
  $R=\{((P\parallel Q)\setminus L, P\setminus L\parallel Q\setminus L)\}\cup \textbf{Id}$, if $\mathcal{L}(P)\cap\overline{\mathcal{L}(Q)}\cap(L\cup\overline{L})=\emptyset$, is a strong hp-bisimulation. It can be proved similarly to the proof of
  static laws for strong hp-bisimulation in CTC, we omit it;
  \item $P[Id]\sim_{hp} P$. It is sufficient to prove the relation $R=\{(P[Id], P)\}\cup \textbf{Id}$ is a strong hp-bisimulation. It can be proved similarly to the proof of
  static laws for strong hp-bisimulation in CTC, we omit it;
  \item $P[f]\sim_{hp} P[f']$, if $f\upharpoonright\mathcal{L}(P)=f'\upharpoonright\mathcal{L}(P)$. It is sufficient to prove the relation $R=\{(P[f], P[f'])\}\cup \textbf{Id}$, if $f\upharpoonright\mathcal{L}(P)=f'\upharpoonright\mathcal{L}(P)$, is a strong hp-bisimulation. It can be proved similarly to the proof of
  static laws for strong hp-bisimulation in CTC, we omit it;
  \item $P[f][f']\sim_{hp} P[f'\circ f]$. It is sufficient to prove the relation $R=\{(P[f][f'], P[f'\circ f])\}\cup \textbf{Id}$ is a strong hp-bisimulation. It can be proved similarly to the proof of
  static laws for strong hp-bisimulation in CTC, we omit it;
  \item $(P\parallel Q)[f]\sim_{hp} P[f]\parallel Q[f]$, if $f\upharpoonright(L\cup\overline{L})$ is one-to-one, where $L=\mathcal{L}(P)\cup\mathcal{L}(Q)$. It is sufficient to prove the
  relation $R=\{((P\parallel Q)[f], P[f]\parallel Q[f])\}\cup \textbf{Id}$, if $f\upharpoonright(L\cup\overline{L})$ is one-to-one, where $L=\mathcal{L}(P)\cup\mathcal{L}(Q)$, is a strong hp-bisimulation. It can be proved similarly to the proof of
  static laws for strong hp-bisimulation in CTC, we omit it.
\end{enumerate}
\end{proof}

\begin{proposition}[Static laws for strong hhp-bisimulation]
The static laws for strong hhp-bisimulation are as follows.
\begin{enumerate}
  \item $P\parallel Q\sim_{hhp} Q\parallel P$;
  \item $P\parallel(Q\parallel R)\sim_{hhp} (P\parallel Q)\parallel R$;
  \item $P\parallel \textbf{nil}\sim_{hhp} P$;
  \item $P\setminus L\sim_{hhp} P$, if $\mathcal{L}(P)\cap(L\cup\overline{L})=\emptyset$;
  \item $P\setminus K\setminus L\sim_{hhp} P\setminus(K\cup L)$;
  \item $P[f]\setminus L\sim_{hhp} P\setminus f^{-1}(L)[f]$;
  \item $(P\parallel Q)\setminus L\sim_{hhp} P\setminus L\parallel Q\setminus L$, if $\mathcal{L}(P)\cap\overline{\mathcal{L}(Q)}\cap(L\cup\overline{L})=\emptyset$;
  \item $P[Id]\sim_{hhp} P$;
  \item $P[f]\sim_{hhp} P[f']$, if $f\upharpoonright\mathcal{L}(P)=f'\upharpoonright\mathcal{L}(P)$;
  \item $P[f][f']\sim_{hhp} P[f'\circ f]$;
  \item $(P\parallel Q)[f]\sim_{hhp} P[f]\parallel Q[f]$, if $f\upharpoonright(L\cup\overline{L})$ is one-to-one, where $L=\mathcal{L}(P)\cup\mathcal{L}(Q)$.
\end{enumerate}
\end{proposition}

\begin{proof}
\begin{enumerate}
  \item $P\parallel Q\sim_{hhp} Q\parallel P$. It is sufficient to prove the relation $R=\{(P\parallel Q, Q\parallel P)\}\cup \textbf{Id}$ is a strong hhp-bisimulation. It can be proved similarly to the proof of
  static laws for strong hhp-bisimulation in CTC, we omit it;
  \item $P\parallel(Q\parallel R)\sim_{hhp} (P\parallel Q)\parallel R$. It is sufficient to prove the relation $R=\{(P\parallel(Q\parallel R), (P\parallel Q)\parallel R)\}\cup \textbf{Id}$ is a strong hhp-bisimulation. It can be proved similarly to the proof of
  static laws for strong hhp-bisimulation in CTC, we omit it;
  \item $P\parallel \textbf{nil}\sim_{hhp} P$. It is sufficient to prove the relation $R=\{(P\parallel \textbf{nil}, P)\}\cup \textbf{Id}$ is a strong hhp-bisimulation. It can be proved similarly to the proof of
  static laws for strong hhp-bisimulation in CTC, we omit it;
  \item $P\setminus L\sim_{hhp} P$, if $\mathcal{L}(P)\cap(L\cup\overline{L})=\emptyset$. It is sufficient to prove the relation $R=\{(P\setminus L, P)\}\cup \textbf{Id}$, if $\mathcal{L}(P)\cap(L\cup\overline{L})=\emptyset$, is a strong hhp-bisimulation. It can be proved similarly to the proof of
  static laws for strong hhp-bisimulation in CTC, we omit it;
  \item $P\setminus K\setminus L\sim_{hhp} P\setminus(K\cup L)$. It is sufficient to prove the relation $R=\{(P\setminus K\setminus L, P\setminus(K\cup L))\}\cup \textbf{Id}$ is a strong hhp-bisimulation. It can be proved similarly to the proof of
  static laws for strong hhp-bisimulation in CTC, we omit it;
  \item $P[f]\setminus L\sim_{hhp} P\setminus f^{-1}(L)[f]$. It is sufficient to prove the relation $R=\{(P[f]\setminus L, P\setminus f^{-1}(L)[f])\}\cup \textbf{Id}$ is a strong hhp-bisimulation. It can be proved similarly to the proof of
  static laws for strong hhp-bisimulation in CTC, we omit it;
  \item $(P\parallel Q)\setminus L\sim_{hhp} P\setminus L\parallel Q\setminus L$, if $\mathcal{L}(P)\cap\overline{\mathcal{L}(Q)}\cap(L\cup\overline{L})=\emptyset$. It is sufficient to prove the relation
  $R=\{((P\parallel Q)\setminus L, P\setminus L\parallel Q\setminus L)\}\cup \textbf{Id}$, if $\mathcal{L}(P)\cap\overline{\mathcal{L}(Q)}\cap(L\cup\overline{L})=\emptyset$, is a strong hhp-bisimulation. It can be proved similarly to the proof of
  static laws for strong hhp-bisimulation in CTC, we omit it;
  \item $P[Id]\sim_{hhp} P$. It is sufficient to prove the relation $R=\{(P[Id], P)\}\cup \textbf{Id}$ is a strong hhp-bisimulation. It can be proved similarly to the proof of
  static laws for strong hhp-bisimulation in CTC, we omit it;
  \item $P[f]\sim_{hhp} P[f']$, if $f\upharpoonright\mathcal{L}(P)=f'\upharpoonright\mathcal{L}(P)$. It is sufficient to prove the relation $R=\{(P[f], P[f'])\}\cup \textbf{Id}$, if $f\upharpoonright\mathcal{L}(P)=f'\upharpoonright\mathcal{L}(P)$, is a strong hhp-bisimulation. It can be proved similarly to the proof of
  static laws for strong hhp-bisimulation in CTC, we omit it;
  \item $P[f][f']\sim_{hhp} P[f'\circ f]$. It is sufficient to prove the relation $R=\{(P[f][f'], P[f'\circ f])\}\cup \textbf{Id}$ is a strong hhp-bisimulation. It can be proved similarly to the proof of
  static laws for strong hhp-bisimulation in CTC, we omit it;
  \item $(P\parallel Q)[f]\sim_{hhp} P[f]\parallel Q[f]$, if $f\upharpoonright(L\cup\overline{L})$ is one-to-one, where $L=\mathcal{L}(P)\cup\mathcal{L}(Q)$. It is sufficient to prove the
  relation $R=\{((P\parallel Q)[f], P[f]\parallel Q[f])\}\cup \textbf{Id}$, if $f\upharpoonright(L\cup\overline{L})$ is one-to-one, where $L=\mathcal{L}(P)\cup\mathcal{L}(Q)$, is a strong hhp-bisimulation. It can be proved similarly to the proof of
  static laws for strong hhp-bisimulation in CTC, we omit it.
\end{enumerate}
\end{proof}

\begin{proposition}[Guards laws for strong pomset bisimulation] The guards laws for strong pomset bisimulation are as follows.

\begin{enumerate}
  \item $P+\delta \sim_p P$;
  \item $\delta.P \sim_p \delta$;
  \item $\epsilon.P \sim_p P$;
  \item $P.\epsilon \sim_p P$;
  \item $\phi.\neg\phi \sim_p \delta$;
  \item $\phi+\neg\phi \sim_p \epsilon$;
  \item $\phi.\delta \sim_p \delta$;
  \item $\phi.(P+Q)\sim_p\phi.P+\phi.Q$;
  \item $\phi.(P.Q)\sim_p \phi.P.Q$;
  \item $(\phi+\psi).P \sim_p \phi.P + \psi.P$;
  \item $(\phi.\psi).P \sim_p \phi.(\psi.P)$;
  \item $\phi\sim_p\epsilon$ if $\forall s\in S.test(\phi,s)$;
  \item $\phi_0.\cdots.\phi_n \sim_p \delta$ if $\forall s\in S,\exists i\leq n.test(\neg\phi_i,s)$;
  \item $wp(\alpha,\phi).\alpha.\phi\sim_p wp(\alpha,\phi).\alpha$;
  \item $\neg wp(\alpha,\phi).\alpha.\neg\phi\sim_p\neg wp(\alpha,\phi).\alpha$;
  \item $\delta\parallel P \sim_p \delta$;
  \item $P\parallel \delta \sim_p \delta$;
  \item $\epsilon\parallel P \sim_p P$;
  \item $P\parallel \epsilon \sim_p P$;
  \item $\phi.(P\parallel Q) \sim_p\phi.P\parallel \phi.Q$;
  \item $\phi\parallel \delta \sim_p \delta$;
  \item $\delta\parallel \phi \sim_p \delta$;
  \item $\phi\parallel \epsilon \sim_p \phi$;
  \item $\epsilon\parallel \phi \sim_p \phi$;
  \item $\phi\parallel\neg\phi \sim_p \delta$;
  \item $\phi_0\parallel\cdots\parallel\phi_n \sim_p \delta$ if $\forall s_0,\cdots,s_n\in S,\exists i\leq n.test(\neg\phi_i,s_0\cup\cdots\cup s_n)$.
\end{enumerate}
\end{proposition}

\begin{proof}
\begin{enumerate}
  \item $P+\delta \sim_p P$. It is sufficient to prove the relation $R=\{(P+\delta, P)\}\cup \textbf{Id}$ is a strong pomset bisimulation, and we omit it;
  \item $\delta.P \sim_p \delta$. It is sufficient to prove the relation $R=\{(\delta.P, \delta)\}\cup \textbf{Id}$ is a strong pomset bisimulation, and we omit it;
  \item $\epsilon.P \sim_p P$. It is sufficient to prove the relation $R=\{(\epsilon.P, P)\}\cup \textbf{Id}$ is a strong pomset bisimulation, and we omit it;
  \item $P.\epsilon \sim_p P$. It is sufficient to prove the relation $R=\{(P.\epsilon, P)\}\cup \textbf{Id}$ is a strong pomset bisimulation, and we omit it;
  \item $\phi.\neg\phi \sim_p \delta$. It is sufficient to prove the relation $R=\{(\phi.\neg\phi, \delta)\}\cup \textbf{Id}$ is a strong pomset bisimulation, and we omit it;
  \item $\phi+\neg\phi \sim_p \epsilon$. It is sufficient to prove the relation $R=\{(\phi+\neg\phi, \epsilon)\}\cup \textbf{Id}$ is a strong pomset bisimulation, and we omit it;
  \item $\phi.\delta \sim_p \delta$. It is sufficient to prove the relation $R=\{(\phi.\delta, \delta)\}\cup \textbf{Id}$ is a strong pomset bisimulation, and we omit it;
  \item $\phi.(P+Q)\sim_p\phi.P+\phi.Q$. It is sufficient to prove the relation $R=\{(\phi.(P+Q), \phi.P+\phi.Q)\}\cup \textbf{Id}$ is a strong pomset bisimulation, and we omit it;
  \item $\phi.(P.Q)\sim_p \phi.P.Q$. It is sufficient to prove the relation $R=\{(\phi.(P.Q), \phi.P.Q)\}\cup \textbf{Id}$ is a strong pomset bisimulation, and we omit it;
  \item $(\phi+\psi).P \sim_p \phi.P + \psi.P$. It is sufficient to prove the relation $R=\{((\phi+\psi).P, \phi.P + \psi.P)\}\cup \textbf{Id}$ is a strong pomset bisimulation, and we omit it;
  \item $(\phi.\psi).P \sim_p \phi.(\psi.P)$. It is sufficient to prove the relation $R=\{((\phi.\psi).P, \phi.(\psi.P))\}\cup \textbf{Id}$ is a strong pomset bisimulation, and we omit it;
  \item $\phi\sim_p\epsilon$ if $\forall s\in S.test(\phi,s)$. It is sufficient to prove the relation $R=\{(\phi, \epsilon)\}\cup \textbf{Id}$, if $\forall s\in S.test(\phi,s)$, is a strong pomset bisimulation, and we omit it;
  \item $\phi_0.\cdots.\phi_n \sim_p \delta$ if $\forall s\in S,\exists i\leq n.test(\neg\phi_i,s)$. It is sufficient to prove the relation $R=\{(\phi_0.\cdots.\phi_n, \delta)\}\cup \textbf{Id}$, if $\forall s\in S,\exists i\leq n.test(\neg\phi_i,s)$, is a strong pomset bisimulation, and we omit it;
  \item $wp(\alpha,\phi).\alpha.\phi\sim_p wp(\alpha,\phi).\alpha$. It is sufficient to prove the relation $R=\{(wp(\alpha,\phi).\alpha.\phi, wp(\alpha,\phi).\alpha)\}\cup \textbf{Id}$ is a strong pomset bisimulation, and we omit it;
  \item $\neg wp(\alpha,\phi).\alpha.\neg\phi\sim_p\neg wp(\alpha,\phi).\alpha$. It is sufficient to prove the relation \\$R=\{(\neg wp(\alpha,\phi).\alpha.\neg\phi, \neg wp(\alpha,\phi).\alpha)\}\cup \textbf{Id}$ is a strong pomset bisimulation, and we omit it;
  \item $\delta\parallel P \sim_p \delta$. It is sufficient to prove the relation $R=\{(\delta\parallel P, \delta)\}\cup \textbf{Id}$ is a strong pomset bisimulation, and we omit it;
  \item $P\parallel \delta \sim_p \delta$. It is sufficient to prove the relation $R=\{(P\parallel \delta, \delta)\}\cup \textbf{Id}$ is a strong pomset bisimulation, and we omit it;
  \item $\epsilon\parallel P \sim_p P$. It is sufficient to prove the relation $R=\{(\epsilon\parallel P, P)\}\cup \textbf{Id}$ is a strong pomset bisimulation, and we omit it;
  \item $P\parallel \epsilon \sim_p P$. It is sufficient to prove the relation $R=\{(P\parallel \epsilon, P)\}\cup \textbf{Id}$ is a strong pomset bisimulation, and we omit it;
  \item $\phi.(P\parallel Q) \sim_p\phi.P\parallel \phi.Q$. It is sufficient to prove the relation $R=\{(\phi.(P\parallel Q), \phi.P\parallel \phi.Q)\}\cup \textbf{Id}$ is a strong pomset bisimulation, and we omit it;
  \item $\phi\parallel \delta \sim_p \delta$. It is sufficient to prove the relation $R=\{(\phi\parallel \delta, \delta)\}\cup \textbf{Id}$ is a strong pomset bisimulation, and we omit it;
  \item $\delta\parallel \phi \sim_p \delta$. It is sufficient to prove the relation $R=\{(\delta\parallel \phi, \delta)\}\cup \textbf{Id}$ is a strong pomset bisimulation, and we omit it;
  \item $\phi\parallel \epsilon \sim_p \phi$. It is sufficient to prove the relation $R=\{(\phi\parallel \epsilon, \phi)\}\cup \textbf{Id}$ is a strong pomset bisimulation, and we omit it;
  \item $\epsilon\parallel \phi \sim_p \phi$. It is sufficient to prove the relation $R=\{(\epsilon\parallel \phi, \phi)\}\cup \textbf{Id}$ is a strong pomset bisimulation, and we omit it;
  \item $\phi\parallel\neg\phi \sim_p \delta$. It is sufficient to prove the relation $R=\{(\phi\parallel\neg\phi, \delta)\}\cup \textbf{Id}$ is a strong pomset bisimulation, and we omit it;
  \item $\phi_0\parallel\cdots\parallel\phi_n \sim_p \delta$ if $\forall s_0,\cdots,s_n\in S,\exists i\leq n.test(\neg\phi_i,s_0\cup\cdots\cup s_n)$. It is sufficient to prove the relation $R=\{(\phi_0\parallel\cdots\parallel\phi_n, \delta)\}\cup \textbf{Id}$, if $\forall s_0,\cdots,s_n\in S,\exists i\leq n.test(\neg\phi_i,s_0\cup\cdots\cup s_n)$, is a strong pomset bisimulation, and we omit it.
\end{enumerate}
\end{proof}

\begin{proposition}[Guards laws for strong step bisimulation] The guards laws for strong step bisimulation are as follows.

\begin{enumerate}
  \item $P+\delta \sim_s P$;
  \item $\delta.P \sim_s \delta$;
  \item $\epsilon.P \sim_s P$;
  \item $P.\epsilon \sim_s P$;
  \item $\phi.\neg\phi \sim_s \delta$;
  \item $\phi+\neg\phi \sim_s \epsilon$;
  \item $\phi.\delta \sim_s \delta$;
  \item $\phi.(P+Q)\sim_s\phi.P+\phi.Q$;
  \item $\phi.(P.Q)\sim_s \phi.P.Q$;
  \item $(\phi+\psi).P \sim_s \phi.P + \psi.P$;
  \item $(\phi.\psi).P \sim_s \phi.(\psi.P)$;
  \item $\phi\sim_s\epsilon$ if $\forall s\in S.test(\phi,s)$;
  \item $\phi_0.\cdots.\phi_n \sim_s \delta$ if $\forall s\in S,\exists i\leq n.test(\neg\phi_i,s)$;
  \item $wp(\alpha,\phi).\alpha.\phi\sim_s wp(\alpha,\phi).\alpha$;
  \item $\neg wp(\alpha,\phi).\alpha.\neg\phi\sim_s\neg wp(\alpha,\phi).\alpha$;
  \item $\delta\parallel P \sim_s \delta$;
  \item $P\parallel \delta \sim_s \delta$;
  \item $\epsilon\parallel P \sim_s P$;
  \item $P\parallel \epsilon \sim_s P$;
  \item $\phi.(P\parallel Q) \sim_s\phi.P\parallel \phi.Q$;
  \item $\phi\parallel \delta \sim_s \delta$;
  \item $\delta\parallel \phi \sim_s \delta$;
  \item $\phi\parallel \epsilon \sim_s \phi$;
  \item $\epsilon\parallel \phi \sim_s \phi$;
  \item $\phi\parallel\neg\phi \sim_s \delta$;
  \item $\phi_0\parallel\cdots\parallel\phi_n \sim_s \delta$ if $\forall s_0,\cdots,s_n\in S,\exists i\leq n.test(\neg\phi_i,s_0\cup\cdots\cup s_n)$.
\end{enumerate}
\end{proposition}

\begin{proof}
\begin{enumerate}
  \item $P+\delta \sim_s P$. It is sufficient to prove the relation $R=\{(P+\delta, P)\}\cup \textbf{Id}$ is a strong step bisimulation, and we omit it;
  \item $\delta.P \sim_s \delta$. It is sufficient to prove the relation $R=\{(\delta.P, \delta)\}\cup \textbf{Id}$ is a strong step bisimulation, and we omit it;
  \item $\epsilon.P \sim_s P$. It is sufficient to prove the relation $R=\{(\epsilon.P, P)\}\cup \textbf{Id}$ is a strong step bisimulation, and we omit it;
  \item $P.\epsilon \sim_s P$. It is sufficient to prove the relation $R=\{(P.\epsilon, P)\}\cup \textbf{Id}$ is a strong step bisimulation, and we omit it;
  \item $\phi.\neg\phi \sim_s \delta$. It is sufficient to prove the relation $R=\{(\phi.\neg\phi, \delta)\}\cup \textbf{Id}$ is a strong step bisimulation, and we omit it;
  \item $\phi+\neg\phi \sim_s \epsilon$. It is sufficient to prove the relation $R=\{(\phi+\neg\phi, \epsilon)\}\cup \textbf{Id}$ is a strong step bisimulation, and we omit it;
  \item $\phi.\delta \sim_s \delta$. It is sufficient to prove the relation $R=\{(\phi.\delta, \delta)\}\cup \textbf{Id}$ is a strong step bisimulation, and we omit it;
  \item $\phi.(P+Q)\sim_s\phi.P+\phi.Q$. It is sufficient to prove the relation $R=\{(\phi.(P+Q), \phi.P+\phi.Q)\}\cup \textbf{Id}$ is a strong step bisimulation, and we omit it;
  \item $\phi.(P.Q)\sim_s \phi.P.Q$. It is sufficient to prove the relation $R=\{(\phi.(P.Q), \phi.P.Q)\}\cup \textbf{Id}$ is a strong step bisimulation, and we omit it;
  \item $(\phi+\psi).P \sim_s \phi.P + \psi.P$. It is sufficient to prove the relation $R=\{((\phi+\psi).P, \phi.P + \psi.P)\}\cup \textbf{Id}$ is a strong step bisimulation, and we omit it;
  \item $(\phi.\psi).P \sim_s \phi.(\psi.P)$. It is sufficient to prove the relation $R=\{((\phi.\psi).P, \phi.(\psi.P))\}\cup \textbf{Id}$ is a strong step bisimulation, and we omit it;
  \item $\phi\sim_s\epsilon$ if $\forall s\in S.test(\phi,s)$. It is sufficient to prove the relation $R=\{(\phi, \epsilon)\}\cup \textbf{Id}$, if $\forall s\in S.test(\phi,s)$, is a strong step bisimulation, and we omit it;
  \item $\phi_0.\cdots.\phi_n \sim_s \delta$ if $\forall s\in S,\exists i\leq n.test(\neg\phi_i,s)$. It is sufficient to prove the relation $R=\{(\phi_0.\cdots.\phi_n, \delta)\}\cup \textbf{Id}$, if $\forall s\in S,\exists i\leq n.test(\neg\phi_i,s)$, is a strong step bisimulation, and we omit it;
  \item $wp(\alpha,\phi).\alpha.\phi\sim_s wp(\alpha,\phi).\alpha$. It is sufficient to prove the relation $R=\{(wp(\alpha,\phi).\alpha.\phi, wp(\alpha,\phi).\alpha)\}\cup \textbf{Id}$ is a strong step bisimulation, and we omit it;
  \item $\neg wp(\alpha,\phi).\alpha.\neg\phi\sim_s\neg wp(\alpha,\phi).\alpha$. It is sufficient to prove the relation \\$R=\{(\neg wp(\alpha,\phi).\alpha.\neg\phi, \neg wp(\alpha,\phi).\alpha)\}\cup \textbf{Id}$ is a strong step bisimulation, and we omit it;
  \item $\delta\parallel P \sim_s \delta$. It is sufficient to prove the relation $R=\{(\delta\parallel P, \delta)\}\cup \textbf{Id}$ is a strong step bisimulation, and we omit it;
  \item $P\parallel \delta \sim_s \delta$. It is sufficient to prove the relation $R=\{(P\parallel \delta, \delta)\}\cup \textbf{Id}$ is a strong step bisimulation, and we omit it;
  \item $\epsilon\parallel P \sim_s P$. It is sufficient to prove the relation $R=\{(\epsilon\parallel P, P)\}\cup \textbf{Id}$ is a strong step bisimulation, and we omit it;
  \item $P\parallel \epsilon \sim_s P$. It is sufficient to prove the relation $R=\{(P\parallel \epsilon, P)\}\cup \textbf{Id}$ is a strong step bisimulation, and we omit it;
  \item $\phi.(P\parallel Q) \sim_s\phi.P\parallel \phi.Q$. It is sufficient to prove the relation $R=\{(\phi.(P\parallel Q), \phi.P\parallel \phi.Q)\}\cup \textbf{Id}$ is a strong step bisimulation, and we omit it;
  \item $\phi\parallel \delta \sim_s \delta$. It is sufficient to prove the relation $R=\{(\phi\parallel \delta, \delta)\}\cup \textbf{Id}$ is a strong step bisimulation, and we omit it;
  \item $\delta\parallel \phi \sim_s \delta$. It is sufficient to prove the relation $R=\{(\delta\parallel \phi, \delta)\}\cup \textbf{Id}$ is a strong step bisimulation, and we omit it;
  \item $\phi\parallel \epsilon \sim_s \phi$. It is sufficient to prove the relation $R=\{(\phi\parallel \epsilon, \phi)\}\cup \textbf{Id}$ is a strong step bisimulation, and we omit it;
  \item $\epsilon\parallel \phi \sim_s \phi$. It is sufficient to prove the relation $R=\{(\epsilon\parallel \phi, \phi)\}\cup \textbf{Id}$ is a strong step bisimulation, and we omit it;
  \item $\phi\parallel\neg\phi \sim_s \delta$. It is sufficient to prove the relation $R=\{(\phi\parallel\neg\phi, \delta)\}\cup \textbf{Id}$ is a strong step bisimulation, and we omit it;
  \item $\phi_0\parallel\cdots\parallel\phi_n \sim_s \delta$ if $\forall s_0,\cdots,s_n\in S,\exists i\leq n.test(\neg\phi_i,s_0\cup\cdots\cup s_n)$. It is sufficient to prove the relation $R=\{(\phi_0\parallel\cdots\parallel\phi_n, \delta)\}\cup \textbf{Id}$, if $\forall s_0,\cdots,s_n\in S,\exists i\leq n.test(\neg\phi_i,s_0\cup\cdots\cup s_n)$, is a strong step bisimulation, and we omit it.
\end{enumerate}
\end{proof}

\begin{proposition}[Guards laws for strong hp-bisimulation] The guards laws for strong hp-bisimulation are as follows.

\begin{enumerate}
  \item $P+\delta \sim_{hp} P$;
  \item $\delta.P \sim_{hp} \delta$;
  \item $\epsilon.P \sim_{hp} P$;
  \item $P.\epsilon \sim_{hp} P$;
  \item $\phi.\neg\phi \sim_{hp} \delta$;
  \item $\phi+\neg\phi \sim_{hp} \epsilon$;
  \item $\phi.\delta \sim_{hp} \delta$;
  \item $\phi.(P+Q)\sim_{hp}\phi.P+\phi.Q$;
  \item $\phi.(P.Q)\sim_{hp} \phi.P.Q$;
  \item $(\phi+\psi).P \sim_{hp} \phi.P + \psi.P$;
  \item $(\phi.\psi).P \sim_{hp} \phi.(\psi.P)$;
  \item $\phi\sim_{hp}\epsilon$ if $\forall s\in S.test(\phi,s)$;
  \item $\phi_0.\cdots.\phi_n \sim_{hp} \delta$ if $\forall s\in S,\exists i\leq n.test(\neg\phi_i,s)$;
  \item $wp(\alpha,\phi).\alpha.\phi\sim_{hp} wp(\alpha,\phi).\alpha$;
  \item $\neg wp(\alpha,\phi).\alpha.\neg\phi\sim_{hp}\neg wp(\alpha,\phi).\alpha$;
  \item $\delta\parallel P \sim_{hp} \delta$;
  \item $P\parallel \delta \sim_{hp} \delta$;
  \item $\epsilon\parallel P \sim_{hp} P$;
  \item $P\parallel \epsilon \sim_{hp} P$;
  \item $\phi.(P\parallel Q) \sim_{hp}\phi.P\parallel \phi.Q$;
  \item $\phi\parallel \delta \sim_{hp} \delta$;
  \item $\delta\parallel \phi \sim_{hp} \delta$;
  \item $\phi\parallel \epsilon \sim_{hp} \phi$;
  \item $\epsilon\parallel \phi \sim_{hp} \phi$;
  \item $\phi\parallel\neg\phi \sim_{hp} \delta$;
  \item $\phi_0\parallel\cdots\parallel\phi_n \sim_{hp} \delta$ if $\forall s_0,\cdots,s_n\in S,\exists i\leq n.test(\neg\phi_i,s_0\cup\cdots\cup s_n)$.
\end{enumerate}
\end{proposition}

\begin{proof}
\begin{enumerate}
  \item $P+\delta \sim_{hp} P$. It is sufficient to prove the relation $R=\{(P+\delta, P)\}\cup \textbf{Id}$ is a strong hp-bisimulation, and we omit it;
  \item $\delta.P \sim_{hp} \delta$. It is sufficient to prove the relation $R=\{(\delta.P, \delta)\}\cup \textbf{Id}$ is a strong hp-bisimulation, and we omit it;
  \item $\epsilon.P \sim_{hp} P$. It is sufficient to prove the relation $R=\{(\epsilon.P, P)\}\cup \textbf{Id}$ is a strong hp-bisimulation, and we omit it;
  \item $P.\epsilon \sim_{hp} P$. It is sufficient to prove the relation $R=\{(P.\epsilon, P)\}\cup \textbf{Id}$ is a strong hp-bisimulation, and we omit it;
  \item $\phi.\neg\phi \sim_{hp} \delta$. It is sufficient to prove the relation $R=\{(\phi.\neg\phi, \delta)\}\cup \textbf{Id}$ is a strong hp-bisimulation, and we omit it;
  \item $\phi+\neg\phi \sim_{hp} \epsilon$. It is sufficient to prove the relation $R=\{(\phi+\neg\phi, \epsilon)\}\cup \textbf{Id}$ is a strong hp-bisimulation, and we omit it;
  \item $\phi.\delta \sim_{hp} \delta$. It is sufficient to prove the relation $R=\{(\phi.\delta, \delta)\}\cup \textbf{Id}$ is a strong hp-bisimulation, and we omit it;
  \item $\phi.(P+Q)\sim_{hp}\phi.P+\phi.Q$. It is sufficient to prove the relation $R=\{(\phi.(P+Q), \phi.P+\phi.Q)\}\cup \textbf{Id}$ is a strong hp-bisimulation, and we omit it;
  \item $\phi.(P.Q)\sim_{hp} \phi.P.Q$. It is sufficient to prove the relation $R=\{(\phi.(P.Q), \phi.P.Q)\}\cup \textbf{Id}$ is a strong hp-bisimulation, and we omit it;
  \item $(\phi+\psi).P \sim_{hp} \phi.P + \psi.P$. It is sufficient to prove the relation $R=\{((\phi+\psi).P, \phi.P + \psi.P)\}\cup \textbf{Id}$ is a strong hp-bisimulation, and we omit it;
  \item $(\phi.\psi).P \sim_{hp} \phi.(\psi.P)$. It is sufficient to prove the relation $R=\{((\phi.\psi).P, \phi.(\psi.P))\}\cup \textbf{Id}$ is a strong hp-bisimulation, and we omit it;
  \item $\phi\sim_{hp}\epsilon$ if $\forall s\in S.test(\phi,s)$. It is sufficient to prove the relation $R=\{(\phi, \epsilon)\}\cup \textbf{Id}$, if $\forall s\in S.test(\phi,s)$, is a strong hp-bisimulation, and we omit it;
  \item $\phi_0.\cdots.\phi_n \sim_{hp} \delta$ if $\forall s\in S,\exists i\leq n.test(\neg\phi_i,s)$. It is sufficient to prove the relation $R=\{(\phi_0.\cdots.\phi_n, \delta)\}\cup \textbf{Id}$, if $\forall s\in S,\exists i\leq n.test(\neg\phi_i,s)$, is a strong hp-bisimulation, and we omit it;
  \item $wp(\alpha,\phi).\alpha.\phi\sim_{hp} wp(\alpha,\phi).\alpha$. It is sufficient to prove the relation $R=\{(wp(\alpha,\phi).\alpha.\phi, wp(\alpha,\phi).\alpha)\}\cup \textbf{Id}$ is a strong hp-bisimulation, and we omit it;
  \item $\neg wp(\alpha,\phi).\alpha.\neg\phi\sim_{hp}\neg wp(\alpha,\phi).\alpha$. It is sufficient to prove the relation \\$R=\{(\neg wp(\alpha,\phi).\alpha.\neg\phi, \neg wp(\alpha,\phi).\alpha)\}\cup \textbf{Id}$ is a strong hp-bisimulation, and we omit it;
  \item $\delta\parallel P \sim_{hp} \delta$. It is sufficient to prove the relation $R=\{(\delta\parallel P, \delta)\}\cup \textbf{Id}$ is a strong hp-bisimulation, and we omit it;
  \item $P\parallel \delta \sim_{hp} \delta$. It is sufficient to prove the relation $R=\{(P\parallel \delta, \delta)\}\cup \textbf{Id}$ is a strong hp-bisimulation, and we omit it;
  \item $\epsilon\parallel P \sim_{hp} P$. It is sufficient to prove the relation $R=\{(\epsilon\parallel P, P)\}\cup \textbf{Id}$ is a strong hp-bisimulation, and we omit it;
  \item $P\parallel \epsilon \sim_{hp} P$. It is sufficient to prove the relation $R=\{(P\parallel \epsilon, P)\}\cup \textbf{Id}$ is a strong hp-bisimulation, and we omit it;
  \item $\phi.(P\parallel Q) \sim_{hp}\phi.P\parallel \phi.Q$. It is sufficient to prove the relation $R=\{(\phi.(P\parallel Q), \phi.P\parallel \phi.Q)\}\cup \textbf{Id}$ is a strong hp-bisimulation, and we omit it;
  \item $\phi\parallel \delta \sim_{hp} \delta$. It is sufficient to prove the relation $R=\{(\phi\parallel \delta, \delta)\}\cup \textbf{Id}$ is a strong hp-bisimulation, and we omit it;
  \item $\delta\parallel \phi \sim_{hp} \delta$. It is sufficient to prove the relation $R=\{(\delta\parallel \phi, \delta)\}\cup \textbf{Id}$ is a strong hp-bisimulation, and we omit it;
  \item $\phi\parallel \epsilon \sim_{hp} \phi$. It is sufficient to prove the relation $R=\{(\phi\parallel \epsilon, \phi)\}\cup \textbf{Id}$ is a strong hp-bisimulation, and we omit it;
  \item $\epsilon\parallel \phi \sim_{hp} \phi$. It is sufficient to prove the relation $R=\{(\epsilon\parallel \phi, \phi)\}\cup \textbf{Id}$ is a strong hp-bisimulation, and we omit it;
  \item $\phi\parallel\neg\phi \sim_{hp} \delta$. It is sufficient to prove the relation $R=\{(\phi\parallel\neg\phi, \delta)\}\cup \textbf{Id}$ is a strong hp-bisimulation, and we omit it;
  \item $\phi_0\parallel\cdots\parallel\phi_n \sim_{hp} \delta$ if $\forall s_0,\cdots,s_n\in S,\exists i\leq n.test(\neg\phi_i,s_0\cup\cdots\cup s_n)$. It is sufficient to prove the relation $R=\{(\phi_0\parallel\cdots\parallel\phi_n, \delta)\}\cup \textbf{Id}$, if $\forall s_0,\cdots,s_n\in S,\exists i\leq n.test(\neg\phi_i,s_0\cup\cdots\cup s_n)$, is a strong hp-bisimulation, and we omit it.
\end{enumerate}
\end{proof}

\begin{proposition}[Guards laws for strong hhp-bisimulation] The guards laws for strong hhp-bisimulation are as follows.

\begin{enumerate}
  \item $P+\delta \sim_{hhp} P$;
  \item $\delta.P \sim_{hhp} \delta$;
  \item $\epsilon.P \sim_{hhp} P$;
  \item $P.\epsilon \sim_{hhp} P$;
  \item $\phi.\neg\phi \sim_{hhp} \delta$;
  \item $\phi+\neg\phi \sim_{hhp} \epsilon$;
  \item $\phi.\delta \sim_{hhp} \delta$;
  \item $\phi.(P+Q)\sim_{hhp}\phi.P+\phi.Q$;
  \item $\phi.(P.Q)\sim_{hhp} \phi.P.Q$;
  \item $(\phi+\psi).P \sim_{hhp} \phi.P + \psi.P$;
  \item $(\phi.\psi).P \sim_{hhp} \phi.(\psi.P)$;
  \item $\phi\sim_{hhp}\epsilon$ if $\forall s\in S.test(\phi,s)$;
  \item $\phi_0.\cdots.\phi_n \sim_{hhp} \delta$ if $\forall s\in S,\exists i\leq n.test(\neg\phi_i,s)$;
  \item $wp(\alpha,\phi).\alpha.\phi\sim_{hhp} wp(\alpha,\phi).\alpha$;
  \item $\neg wp(\alpha,\phi).\alpha.\neg\phi\sim_{hhp}\neg wp(\alpha,\phi).\alpha$;
  \item $\delta\parallel P \sim_{hhp} \delta$;
  \item $P\parallel \delta \sim_{hhp} \delta$;
  \item $\epsilon\parallel P \sim_{hhp} P$;
  \item $P\parallel \epsilon \sim_{hhp} P$;
  \item $\phi.(P\parallel Q) \sim_{hhp}\phi.P\parallel \phi.Q$;
  \item $\phi\parallel \delta \sim_{hhp} \delta$;
  \item $\delta\parallel \phi \sim_{hhp} \delta$;
  \item $\phi\parallel \epsilon \sim_{hhp} \phi$;
  \item $\epsilon\parallel \phi \sim_{hhp} \phi$;
  \item $\phi\parallel\neg\phi \sim_{hhp} \delta$;
  \item $\phi_0\parallel\cdots\parallel\phi_n \sim_{hhp} \delta$ if $\forall s_0,\cdots,s_n\in S,\exists i\leq n.test(\neg\phi_i,s_0\cup\cdots\cup s_n)$.
\end{enumerate}
\end{proposition}

\begin{proof}
\begin{enumerate}
  \item $P+\delta \sim_{hhp} P$. It is sufficient to prove the relation $R=\{(P+\delta, P)\}\cup \textbf{Id}$ is a strong hhp-bisimulation, and we omit it;
  \item $\delta.P \sim_{hhp} \delta$. It is sufficient to prove the relation $R=\{(\delta.P, \delta)\}\cup \textbf{Id}$ is a strong hhp-bisimulation, and we omit it;
  \item $\epsilon.P \sim_{hhp} P$. It is sufficient to prove the relation $R=\{(\epsilon.P, P)\}\cup \textbf{Id}$ is a strong hhp-bisimulation, and we omit it;
  \item $P.\epsilon \sim_{hhp} P$. It is sufficient to prove the relation $R=\{(P.\epsilon, P)\}\cup \textbf{Id}$ is a strong hhp-bisimulation, and we omit it;
  \item $\phi.\neg\phi \sim_{hhp} \delta$. It is sufficient to prove the relation $R=\{(\phi.\neg\phi, \delta)\}\cup \textbf{Id}$ is a strong hhp-bisimulation, and we omit it;
  \item $\phi+\neg\phi \sim_{hhp} \epsilon$. It is sufficient to prove the relation $R=\{(\phi+\neg\phi, \epsilon)\}\cup \textbf{Id}$ is a strong hhp-bisimulation, and we omit it;
  \item $\phi.\delta \sim_{hhp} \delta$. It is sufficient to prove the relation $R=\{(\phi.\delta, \delta)\}\cup \textbf{Id}$ is a strong hhp-bisimulation, and we omit it;
  \item $\phi.(P+Q)\sim_{hhp}\phi.P+\phi.Q$. It is sufficient to prove the relation $R=\{(\phi.(P+Q), \phi.P+\phi.Q)\}\cup \textbf{Id}$ is a strong hhp-bisimulation, and we omit it;
  \item $\phi.(P.Q)\sim_{hhp} \phi.P.Q$. It is sufficient to prove the relation $R=\{(\phi.(P.Q), \phi.P.Q)\}\cup \textbf{Id}$ is a strong hhp-bisimulation, and we omit it;
  \item $(\phi+\psi).P \sim_{hhp} \phi.P + \psi.P$. It is sufficient to prove the relation $R=\{((\phi+\psi).P, \phi.P + \psi.P)\}\cup \textbf{Id}$ is a strong hhp-bisimulation, and we omit it;
  \item $(\phi.\psi).P \sim_{hhp} \phi.(\psi.P)$. It is sufficient to prove the relation $R=\{((\phi.\psi).P, \phi.(\psi.P))\}\cup \textbf{Id}$ is a strong hhp-bisimulation, and we omit it;
  \item $\phi\sim_{hhp}\epsilon$ if $\forall s\in S.test(\phi,s)$. It is sufficient to prove the relation $R=\{(\phi, \epsilon)\}\cup \textbf{Id}$, if $\forall s\in S.test(\phi,s)$, is a strong hhp-bisimulation, and we omit it;
  \item $\phi_0.\cdots.\phi_n \sim_{hhp} \delta$ if $\forall s\in S,\exists i\leq n.test(\neg\phi_i,s)$. It is sufficient to prove the relation $R=\{(\phi_0.\cdots.\phi_n, \delta)\}\cup \textbf{Id}$, if $\forall s\in S,\exists i\leq n.test(\neg\phi_i,s)$, is a strong hhp-bisimulation, and we omit it;
  \item $wp(\alpha,\phi).\alpha.\phi\sim_{hhp} wp(\alpha,\phi).\alpha$. It is sufficient to prove the relation $R=\{(wp(\alpha,\phi).\alpha.\phi, wp(\alpha,\phi).\alpha)\}\cup \textbf{Id}$ is a strong hhp-bisimulation, and we omit it;
  \item $\neg wp(\alpha,\phi).\alpha.\neg\phi\sim_{hhp}\neg wp(\alpha,\phi).\alpha$. It is sufficient to prove the relation \\$R=\{(\neg wp(\alpha,\phi).\alpha.\neg\phi, \neg wp(\alpha,\phi).\alpha)\}\cup \textbf{Id}$ is a strong hhp-bisimulation, and we omit it;
  \item $\delta\parallel P \sim_{hhp} \delta$. It is sufficient to prove the relation $R=\{(\delta\parallel P, \delta)\}\cup \textbf{Id}$ is a strong hhp-bisimulation, and we omit it;
  \item $P\parallel \delta \sim_{hhp} \delta$. It is sufficient to prove the relation $R=\{(P\parallel \delta, \delta)\}\cup \textbf{Id}$ is a strong hhp-bisimulation, and we omit it;
  \item $\epsilon\parallel P \sim_{hhp} P$. It is sufficient to prove the relation $R=\{(\epsilon\parallel P, P)\}\cup \textbf{Id}$ is a strong hhp-bisimulation, and we omit it;
  \item $P\parallel \epsilon \sim_{hhp} P$. It is sufficient to prove the relation $R=\{(P\parallel \epsilon, P)\}\cup \textbf{Id}$ is a strong hhp-bisimulation, and we omit it;
  \item $\phi.(P\parallel Q) \sim_{hhp}\phi.P\parallel \phi.Q$. It is sufficient to prove the relation $R=\{(\phi.(P\parallel Q), \phi.P\parallel \phi.Q)\}\cup \textbf{Id}$ is a strong hhp-bisimulation, and we omit it;
  \item $\phi\parallel \delta \sim_{hhp} \delta$. It is sufficient to prove the relation $R=\{(\phi\parallel \delta, \delta)\}\cup \textbf{Id}$ is a strong hhp-bisimulation, and we omit it;
  \item $\delta\parallel \phi \sim_{hhp} \delta$. It is sufficient to prove the relation $R=\{(\delta\parallel \phi, \delta)\}\cup \textbf{Id}$ is a strong hhp-bisimulation, and we omit it;
  \item $\phi\parallel \epsilon \sim_{hhp} \phi$. It is sufficient to prove the relation $R=\{(\phi\parallel \epsilon, \phi)\}\cup \textbf{Id}$ is a strong hhp-bisimulation, and we omit it;
  \item $\epsilon\parallel \phi \sim_{hhp} \phi$. It is sufficient to prove the relation $R=\{(\epsilon\parallel \phi, \phi)\}\cup \textbf{Id}$ is a strong hhp-bisimulation, and we omit it;
  \item $\phi\parallel\neg\phi \sim_{hhp} \delta$. It is sufficient to prove the relation $R=\{(\phi\parallel\neg\phi, \delta)\}\cup \textbf{Id}$ is a strong hhp-bisimulation, and we omit it;
  \item $\phi_0\parallel\cdots\parallel\phi_n \sim_{hhp} \delta$ if $\forall s_0,\cdots,s_n\in S,\exists i\leq n.test(\neg\phi_i,s_0\cup\cdots\cup s_n)$. It is sufficient to prove the relation $R=\{(\phi_0\parallel\cdots\parallel\phi_n, \delta)\}\cup \textbf{Id}$, if $\forall s_0,\cdots,s_n\in S,\exists i\leq n.test(\neg\phi_i,s_0\cup\cdots\cup s_n)$, is a strong hhp-bisimulation, and we omit it.
\end{enumerate}
\end{proof}

\begin{proposition}[Expansion law for strong pomset bisimulation]
Let $P\equiv (P_1[f_1]\parallel\cdots\parallel P_n[f_n])\setminus L$, with $n\geq 1$. Then

\begin{eqnarray}
P\sim_p \{(f_1(\alpha_1)\parallel\cdots\parallel f_n(\alpha_n)).(P_1'[f_1]\parallel\cdots\parallel P_n'[f_n])\setminus L: \nonumber\\
\langle P_i,s_i\rangle\xrightarrow{\alpha_i}\langle P_i',s_i'\rangle,i\in\{1,\cdots,n\},f_i(\alpha_i)\notin L\cup\overline{L}\} \nonumber\\
+\sum\{\tau.(P_1[f_1]\parallel\cdots\parallel P_i'[f_i]\parallel\cdots\parallel P_j'[f_j]\parallel\cdots\parallel P_n[f_n])\setminus L: \nonumber\\
\langle P_i,s_i\rangle\xrightarrow{l_1}\langle P_i',s_i'\rangle,\langle P_j,s_j\rangle\xrightarrow{l_2}\langle P_j',s_j'\rangle,f_i(l_1)=\overline{f_j(l_2)},i<j\} \nonumber
\end{eqnarray}
\end{proposition}

\begin{proof}
Firstly, we consider the case without Restriction and Relabeling. That is, we suffice to prove the following case by induction on the size $n$.

For $P\equiv P_1\parallel\cdots\parallel P_n$, with $n\geq 1$, we need to prove

\begin{eqnarray}
P\sim_p \{(\alpha_1\parallel\cdots\parallel \alpha_n).(P_1'\parallel\cdots\parallel P_n'): \langle P_i,s_i\rangle\xrightarrow{\alpha_i}\langle P_i',s_i'\rangle,i\in\{1,\cdots,n\}\nonumber\\
+\sum\{\tau.(P_1\parallel\cdots\parallel P_i'\parallel\cdots\parallel P_j'\parallel\cdots\parallel P_n): \langle P_i,s_i\rangle\xrightarrow{l}\langle P_i',s_i'\rangle,\langle P_j,s_j\rangle\xrightarrow{\overline{l}}\langle P_j',s_j'\rangle,i<j\} \nonumber
\end{eqnarray}

For $n=1$, $P_1\sim_p \alpha_1.P_1':\langle P_1,s_1\rangle\xrightarrow{\alpha_1}\langle P_1',s_1'\rangle$ is obvious. Then with a hypothesis $n$, we consider $R\equiv P\parallel P_{n+1}$. By the transition rules $\textbf{Com}_{1,2,3,4}$, we can get

\begin{eqnarray}
R\sim_p \{(p\parallel \alpha_{n+1}).(P'\parallel P_{n+1}'): \langle P,s\rangle\xrightarrow{p}\langle P',s'\rangle,\langle P_{n+1},s\rangle\xrightarrow{\alpha_{n+1}}\langle P_{n+1}',s''\rangle,p\subseteq P\}\nonumber\\
+\sum\{\tau.(P'\parallel P_{n+1}'): \langle P,s\rangle\xrightarrow{l}\langle P',s'\rangle,\langle P_{n+1},s\rangle\xrightarrow{\overline{l}}\langle P_{n+1}',s''\rangle\} \nonumber
\end{eqnarray}

Now with the induction assumption $P\equiv P_1\parallel\cdots\parallel P_n$, the right-hand side can be reformulated as follows.

\begin{eqnarray}
\{(\alpha_1\parallel\cdots\parallel \alpha_n\parallel \alpha_{n+1}).(P_1'\parallel\cdots\parallel P_n'\parallel P_{n+1}'): \nonumber\\
\langle P_i,s_i\rangle\xrightarrow{\alpha_i}\langle P_i',s_i'\rangle,i\in\{1,\cdots,n+1\}\nonumber\\
+\sum\{\tau.(P_1\parallel\cdots\parallel P_i'\parallel\cdots\parallel P_j'\parallel\cdots\parallel P_n\parallel P_{n+1}): \nonumber\\
\langle P_i,s_i\rangle\xrightarrow{l}\langle P_i',s_i'\rangle,\langle P_j,s_j\rangle\xrightarrow{\overline{l}}\langle P_j',s_j'\rangle,i<j\} \nonumber\\
+\sum\{\tau.(P_1\parallel\cdots\parallel P_i'\parallel\cdots\parallel P_j\parallel\cdots\parallel P_n\parallel P_{n+1}'): \nonumber\\
\langle P_i,s_i\rangle\xrightarrow{l}\langle P_i',s_i'\rangle,\langle P_{n+1},s_{n+1}\rangle\xrightarrow{\overline{l}}\langle P_{n+1}',s_{n+1}'\rangle,i\in\{1,\cdots, n\}\} \nonumber
\end{eqnarray}

So,

\begin{eqnarray}
R\sim_p \{(\alpha_1\parallel\cdots\parallel \alpha_n\parallel \alpha_{n+1}).(P_1'\parallel\cdots\parallel P_n'\parallel P_{n+1}'): \nonumber\\
P_i\xrightarrow{\alpha_i}P_i',i\in\{1,\cdots,n+1\}\nonumber\\
+\sum\{\tau.(P_1\parallel\cdots\parallel P_i'\parallel\cdots\parallel P_j'\parallel\cdots\parallel P_n): \nonumber\\
P_i\xrightarrow{l}P_i',P_j\xrightarrow{\overline{l}}P_j',1 \leq i<j\geq n+1\} \nonumber
\end{eqnarray}

Then, we can easily add the full conditions with Restriction and Relabeling.
\end{proof}

\begin{proposition}[Expansion law for strong step bisimulation]
Let $P\equiv (P_1[f_1]\parallel\cdots\parallel P_n[f_n])\setminus L$, with $n\geq 1$. Then

\begin{eqnarray}
P\sim_p \{(f_1(\alpha_1)\parallel\cdots\parallel f_n(\alpha_n)).(P_1'[f_1]\parallel\cdots\parallel P_n'[f_n])\setminus L: \nonumber\\
\langle P_i,s_i\rangle\xrightarrow{\alpha_i}\langle P_i',s_i'\rangle,i\in\{1,\cdots,n\},f_i(\alpha_i)\notin L\cup\overline{L}\} \nonumber\\
+\sum\{\tau.(P_1[f_1]\parallel\cdots\parallel P_i'[f_i]\parallel\cdots\parallel P_j'[f_j]\parallel\cdots\parallel P_n[f_n])\setminus L: \nonumber\\
\langle P_i,s_i\rangle\xrightarrow{l_1}\langle P_i',s_i'\rangle,\langle P_j,s_j\rangle\xrightarrow{l_2}\langle P_j',s_j'\rangle,f_i(l_1)=\overline{f_j(l_2)},i<j\} \nonumber
\end{eqnarray}
\end{proposition}

\begin{proof}
Firstly, we consider the case without Restriction and Relabeling. That is, we suffice to prove the following case by induction on the size $n$.

For $P\equiv P_1\parallel\cdots\parallel P_n$, with $n\geq 1$, we need to prove

\begin{eqnarray}
P\sim_s \{(\alpha_1\parallel\cdots\parallel \alpha_n).(P_1'\parallel\cdots\parallel P_n'): \langle P_i,s_i\rangle\xrightarrow{\alpha_i}\langle P_i',s_i'\rangle,i\in\{1,\cdots,n\}\nonumber\\
+\sum\{\tau.(P_1\parallel\cdots\parallel P_i'\parallel\cdots\parallel P_j'\parallel\cdots\parallel P_n): \langle P_i,s_i\rangle\xrightarrow{l}\langle P_i',s_i'\rangle,\langle P_j,s_j\rangle\xrightarrow{\overline{l}}\langle P_j',s_j'\rangle,i<j\} \nonumber
\end{eqnarray}

For $n=1$, $P_1\sim_s \alpha_1.P_1':\langle P_1,s_1\rangle\xrightarrow{\alpha_1}\langle P_1',s_1'\rangle$ is obvious. Then with a hypothesis $n$, we consider $R\equiv P\parallel P_{n+1}$. By the transition rules $\textbf{Com}_{1,2,3,4}$, we can get

\begin{eqnarray}
R\sim_s \{(p\parallel \alpha_{n+1}).(P'\parallel P_{n+1}'): \langle P,s\rangle\xrightarrow{p}\langle P',s'\rangle,\langle P_{n+1},s\rangle\xrightarrow{\alpha_{n+1}}\langle P_{n+1}',s''\rangle,p\subseteq P\}\nonumber\\
+\sum\{\tau.(P'\parallel P_{n+1}'): \langle P,s\rangle\xrightarrow{l}\langle P',s'\rangle,\langle P_{n+1},s\rangle\xrightarrow{\overline{l}}\langle P_{n+1}',s''\rangle\} \nonumber
\end{eqnarray}

Now with the induction assumption $P\equiv P_1\parallel\cdots\parallel P_n$, the right-hand side can be reformulated as follows.

\begin{eqnarray}
\{(\alpha_1\parallel\cdots\parallel \alpha_n\parallel \alpha_{n+1}).(P_1'\parallel\cdots\parallel P_n'\parallel P_{n+1}'): \nonumber\\
\langle P_i,s_i\rangle\xrightarrow{\alpha_i}\langle P_i',s_i'\rangle,i\in\{1,\cdots,n+1\}\nonumber\\
+\sum\{\tau.(P_1\parallel\cdots\parallel P_i'\parallel\cdots\parallel P_j'\parallel\cdots\parallel P_n\parallel P_{n+1}): \nonumber\\
\langle P_i,s_i\rangle\xrightarrow{l}\langle P_i',s_i'\rangle,\langle P_j,s_j\rangle\xrightarrow{\overline{l}}\langle P_j',s_j'\rangle,i<j\} \nonumber\\
+\sum\{\tau.(P_1\parallel\cdots\parallel P_i'\parallel\cdots\parallel P_j\parallel\cdots\parallel P_n\parallel P_{n+1}'): \nonumber\\
\langle P_i,s_i\rangle\xrightarrow{l}\langle P_i',s_i'\rangle,\langle P_{n+1},s_{n+1}\rangle\xrightarrow{\overline{l}}\langle P_{n+1}',s_{n+1}'\rangle,i\in\{1,\cdots, n\}\} \nonumber
\end{eqnarray}

So,

\begin{eqnarray}
R\sim_s \{(\alpha_1\parallel\cdots\parallel \alpha_n\parallel \alpha_{n+1}).(P_1'\parallel\cdots\parallel P_n'\parallel P_{n+1}'): \nonumber\\
P_i\xrightarrow{\alpha_i}P_i',i\in\{1,\cdots,n+1\}\nonumber\\
+\sum\{\tau.(P_1\parallel\cdots\parallel P_i'\parallel\cdots\parallel P_j'\parallel\cdots\parallel P_n): \nonumber\\
P_i\xrightarrow{l}P_i',P_j\xrightarrow{\overline{l}}P_j',1 \leq i<j\geq n+1\} \nonumber
\end{eqnarray}

Then, we can easily add the full conditions with Restriction and Relabeling.
\end{proof}

\begin{proposition}[Expansion law for strong hp-bisimulation]
Let $P\equiv (P_1[f_1]\parallel\cdots\parallel P_n[f_n])\setminus L$, with $n\geq 1$. Then

\begin{eqnarray}
P\sim_{hp} \{(f_1(\alpha_1)\parallel\cdots\parallel f_n(\alpha_n)).(P_1'[f_1]\parallel\cdots\parallel P_n'[f_n])\setminus L: \nonumber\\
\langle P_i,s_i\rangle\xrightarrow{\alpha_i}\langle P_i',s_i'\rangle,i\in\{1,\cdots,n\},f_i(\alpha_i)\notin L\cup\overline{L}\} \nonumber\\
+\sum\{\tau.(P_1[f_1]\parallel\cdots\parallel P_i'[f_i]\parallel\cdots\parallel P_j'[f_j]\parallel\cdots\parallel P_n[f_n])\setminus L: \nonumber\\
\langle P_i,s_i\rangle\xrightarrow{l_1}\langle P_i',s_i'\rangle,\langle P_j,s_j\rangle\xrightarrow{l_2}\langle P_j',s_j'\rangle,f_i(l_1)=\overline{f_j(l_2)},i<j\} \nonumber
\end{eqnarray}
\end{proposition}

\begin{proof}
Firstly, we consider the case without Restriction and Relabeling. That is, we suffice to prove the following case by induction on the size $n$.

For $P\equiv P_1\parallel\cdots\parallel P_n$, with $n\geq 1$, we need to prove

\begin{eqnarray}
P\sim_{hp} \{(\alpha_1\parallel\cdots\parallel \alpha_n).(P_1'\parallel\cdots\parallel P_n'): \langle P_i,s_i\rangle\xrightarrow{\alpha_i}\langle P_i',s_i'\rangle,i\in\{1,\cdots,n\}\nonumber\\
+\sum\{\tau.(P_1\parallel\cdots\parallel P_i'\parallel\cdots\parallel P_j'\parallel\cdots\parallel P_n): \langle P_i,s_i\rangle\xrightarrow{l}\langle P_i',s_i'\rangle,\langle P_j,s_j\rangle\xrightarrow{\overline{l}}\langle P_j',s_j'\rangle,i<j\} \nonumber
\end{eqnarray}

For $n=1$, $P_1\sim_{hp} \alpha_1.P_1':\langle P_1,s_1\rangle\xrightarrow{\alpha_1}\langle P_1',s_1'\rangle$ is obvious. Then with a hypothesis $n$, we consider $R\equiv P\parallel P_{n+1}$. By the transition rules $\textbf{Com}_{1,2,3,4}$, we can get

\begin{eqnarray}
R\sim_{hp} \{(p\parallel \alpha_{n+1}).(P'\parallel P_{n+1}'): \langle P,s\rangle\xrightarrow{p}\langle P',s'\rangle,\langle P_{n+1},s\rangle\xrightarrow{\alpha_{n+1}}\langle P_{n+1}',s''\rangle,p\subseteq P\}\nonumber\\
+\sum\{\tau.(P'\parallel P_{n+1}'): \langle P,s\rangle\xrightarrow{l}\langle P',s'\rangle,\langle P_{n+1},s\rangle\xrightarrow{\overline{l}}\langle P_{n+1}',s''\rangle\} \nonumber
\end{eqnarray}

Now with the induction assumption $P\equiv P_1\parallel\cdots\parallel P_n$, the right-hand side can be reformulated as follows.

\begin{eqnarray}
\{(\alpha_1\parallel\cdots\parallel \alpha_n\parallel \alpha_{n+1}).(P_1'\parallel\cdots\parallel P_n'\parallel P_{n+1}'): \nonumber\\
\langle P_i,s_i\rangle\xrightarrow{\alpha_i}\langle P_i',s_i'\rangle,i\in\{1,\cdots,n+1\}\nonumber\\
+\sum\{\tau.(P_1\parallel\cdots\parallel P_i'\parallel\cdots\parallel P_j'\parallel\cdots\parallel P_n\parallel P_{n+1}): \nonumber\\
\langle P_i,s_i\rangle\xrightarrow{l}\langle P_i',s_i'\rangle,\langle P_j,s_j\rangle\xrightarrow{\overline{l}}\langle P_j',s_j'\rangle,i<j\} \nonumber\\
+\sum\{\tau.(P_1\parallel\cdots\parallel P_i'\parallel\cdots\parallel P_j\parallel\cdots\parallel P_n\parallel P_{n+1}'): \nonumber\\
\langle P_i,s_i\rangle\xrightarrow{l}\langle P_i',s_i'\rangle,\langle P_{n+1},s_{n+1}\rangle\xrightarrow{\overline{l}}\langle P_{n+1}',s_{n+1}'\rangle,i\in\{1,\cdots, n\}\} \nonumber
\end{eqnarray}

So,

\begin{eqnarray}
R\sim_{hp} \{(\alpha_1\parallel\cdots\parallel \alpha_n\parallel \alpha_{n+1}).(P_1'\parallel\cdots\parallel P_n'\parallel P_{n+1}'): \nonumber\\
P_i\xrightarrow{\alpha_i}P_i',i\in\{1,\cdots,n+1\}\nonumber\\
+\sum\{\tau.(P_1\parallel\cdots\parallel P_i'\parallel\cdots\parallel P_j'\parallel\cdots\parallel P_n): \nonumber\\
P_i\xrightarrow{l}P_i',P_j\xrightarrow{\overline{l}}P_j',1 \leq i<j\geq n+1\} \nonumber
\end{eqnarray}

Then, we can easily add the full conditions with Restriction and Relabeling.
\end{proof}

\begin{proposition}[Expansion law for strong hhp-bisimulation]
Let $P\equiv (P_1[f_1]\parallel\cdots\parallel P_n[f_n])\setminus L$, with $n\geq 1$. Then

\begin{eqnarray}
P\sim_{hhp} \{(f_1(\alpha_1)\parallel\cdots\parallel f_n(\alpha_n)).(P_1'[f_1]\parallel\cdots\parallel P_n'[f_n])\setminus L: \nonumber\\
\langle P_i,s_i\rangle\xrightarrow{\alpha_i}\langle P_i',s_i'\rangle,i\in\{1,\cdots,n\},f_i(\alpha_i)\notin L\cup\overline{L}\} \nonumber\\
+\sum\{\tau.(P_1[f_1]\parallel\cdots\parallel P_i'[f_i]\parallel\cdots\parallel P_j'[f_j]\parallel\cdots\parallel P_n[f_n])\setminus L: \nonumber\\
\langle P_i,s_i\rangle\xrightarrow{l_1}\langle P_i',s_i'\rangle,\langle P_j,s_j\rangle\xrightarrow{l_2}\langle P_j',s_j'\rangle,f_i(l_1)=\overline{f_j(l_2)},i<j\} \nonumber
\end{eqnarray}
\end{proposition}

\begin{proof}
Firstly, we consider the case without Restriction and Relabeling. That is, we suffice to prove the following case by induction on the size $n$.

For $P\equiv P_1\parallel\cdots\parallel P_n$, with $n\geq 1$, we need to prove

\begin{eqnarray}
P\sim_{hhp} \{(\alpha_1\parallel\cdots\parallel \alpha_n).(P_1'\parallel\cdots\parallel P_n'): \langle P_i,s_i\rangle\xrightarrow{\alpha_i}\langle P_i',s_i'\rangle,i\in\{1,\cdots,n\}\nonumber\\
+\sum\{\tau.(P_1\parallel\cdots\parallel P_i'\parallel\cdots\parallel P_j'\parallel\cdots\parallel P_n): \langle P_i,s_i\rangle\xrightarrow{l}\langle P_i',s_i'\rangle,\langle P_j,s_j\rangle\xrightarrow{\overline{l}}\langle P_j',s_j'\rangle,i<j\} \nonumber
\end{eqnarray}

For $n=1$, $P_1\sim_{hhp} \alpha_1.P_1':\langle P_1,s_1\rangle\xrightarrow{\alpha_1}\langle P_1',s_1'\rangle$ is obvious. Then with a hypothesis $n$, we consider $R\equiv P\parallel P_{n+1}$. By the transition rules $\textbf{Com}_{1,2,3,4}$, we can get

\begin{eqnarray}
R\sim_{hhp} \{(p\parallel \alpha_{n+1}).(P'\parallel P_{n+1}'): \langle P,s\rangle\xrightarrow{p}\langle P',s'\rangle,\langle P_{n+1},s\rangle\xrightarrow{\alpha_{n+1}}\langle P_{n+1}',s''\rangle,p\subseteq P\}\nonumber\\
+\sum\{\tau.(P'\parallel P_{n+1}'): \langle P,s\rangle\xrightarrow{l}\langle P',s'\rangle,\langle P_{n+1},s\rangle\xrightarrow{\overline{l}}\langle P_{n+1}',s''\rangle\} \nonumber
\end{eqnarray}

Now with the induction assumption $P\equiv P_1\parallel\cdots\parallel P_n$, the right-hand side can be reformulated as follows.

\begin{eqnarray}
\{(\alpha_1\parallel\cdots\parallel \alpha_n\parallel \alpha_{n+1}).(P_1'\parallel\cdots\parallel P_n'\parallel P_{n+1}'): \nonumber\\
\langle P_i,s_i\rangle\xrightarrow{\alpha_i}\langle P_i',s_i'\rangle,i\in\{1,\cdots,n+1\}\nonumber\\
+\sum\{\tau.(P_1\parallel\cdots\parallel P_i'\parallel\cdots\parallel P_j'\parallel\cdots\parallel P_n\parallel P_{n+1}): \nonumber\\
\langle P_i,s_i\rangle\xrightarrow{l}\langle P_i',s_i'\rangle,\langle P_j,s_j\rangle\xrightarrow{\overline{l}}\langle P_j',s_j'\rangle,i<j\} \nonumber\\
+\sum\{\tau.(P_1\parallel\cdots\parallel P_i'\parallel\cdots\parallel P_j\parallel\cdots\parallel P_n\parallel P_{n+1}'): \nonumber\\
\langle P_i,s_i\rangle\xrightarrow{l}\langle P_i',s_i'\rangle,\langle P_{n+1},s_{n+1}\rangle\xrightarrow{\overline{l}}\langle P_{n+1}',s_{n+1}'\rangle,i\in\{1,\cdots, n\}\} \nonumber
\end{eqnarray}

So,

\begin{eqnarray}
R\sim_{hhp} \{(\alpha_1\parallel\cdots\parallel \alpha_n\parallel \alpha_{n+1}).(P_1'\parallel\cdots\parallel P_n'\parallel P_{n+1}'): \nonumber\\
P_i\xrightarrow{\alpha_i}P_i',i\in\{1,\cdots,n+1\}\nonumber\\
+\sum\{\tau.(P_1\parallel\cdots\parallel P_i'\parallel\cdots\parallel P_j'\parallel\cdots\parallel P_n): \nonumber\\
P_i\xrightarrow{l}P_i',P_j\xrightarrow{\overline{l}}P_j',1 \leq i<j\geq n+1\} \nonumber
\end{eqnarray}

Then, we can easily add the full conditions with Restriction and Relabeling.
\end{proof}

\begin{theorem}[Congruence for strong pomset bisimulation]
We can enjoy the full congruence for strong pomset bisimulation as follows.
\begin{enumerate}
  \item If $A\overset{\text{def}}{=}P$, then $A\sim_{p} P$;
  \item Let $P_1\sim_{p} P_2$. Then
        \begin{enumerate}
           \item $\alpha.P_1\sim_{p} \alpha.P_2$;
           \item $\phi.P_1\sim_{p} \phi.P_2$;
           \item $(\alpha_1\parallel\cdots\parallel\alpha_n).P_1\sim_{p} (\alpha_1\parallel\cdots\parallel\alpha_n).P_2$;
           \item $P_1+Q\sim_{p} P_2 +Q$;
           \item $P_1\parallel Q\sim_{p} P_2\parallel Q$;
           \item $P_1\setminus L\sim_{p} P_2\setminus L$;
           \item $P_1[f]\sim_{p} P_2[f]$.
         \end{enumerate}
\end{enumerate}
\end{theorem}

\begin{proof}
\begin{enumerate}
  \item If $A\overset{\text{def}}{=}P$, then $A\sim_{p} P$. It is obvious.
  \item Let $P_1\sim_{p} P_2$. Then
        \begin{enumerate}
           \item $\alpha.P_1\sim_{p} \alpha.P_2$. It is sufficient to prove the relation $R=\{(\alpha.P_1, \alpha.P_2)\}\cup \textbf{Id}$ is a strong pomset bisimulation. It can be proved similarly to the proof of
            congruence for strong pomset bisimulation in CTC, we omit it;
           \item $\phi.P_1\sim_{p} \phi.P_2$. It is sufficient to prove the relation $R=\{(\phi.P_1, \phi.P_2)\}\cup \textbf{Id}$ is a strong pomset bisimulation. It can be proved similarly to the proof of
            congruence for strong pomset bisimulation in CTC, we omit it;
           \item $(\alpha_1\parallel\cdots\parallel\alpha_n).P_1\sim_{p} (\alpha_1\parallel\cdots\parallel\alpha_n).P_2$. It is sufficient to prove the relation $R=\{((\alpha_1\parallel\cdots\parallel\alpha_n).P_1, (\alpha_1\parallel\cdots\parallel\alpha_n).P_2)\}\cup \textbf{Id}$ is a strong pomset bisimulation. It can be proved similarly to the proof of
            congruence for strong pomset bisimulation in CTC, we omit it;
           \item $P_1+Q\sim_{p} P_2 +Q$. It is sufficient to prove the relation $R=\{(P_1+Q, P_2+Q)\}\cup \textbf{Id}$ is a strong pomset bisimulation. It can be proved similarly to the proof of
            congruence for strong pomset bisimulation in CTC, we omit it;
           \item $P_1\parallel Q\sim_{p} P_2\parallel Q$. It is sufficient to prove the relation $R=\{(P_1\parallel Q, P_2\parallel Q)\}\cup \textbf{Id}$ is a strong pomset bisimulation. It can be proved similarly to the proof of
            congruence for strong pomset bisimulation in CTC, we omit it;
           \item $P_1\setminus L\sim_{p} P_2\setminus L$. It is sufficient to prove the relation $R=\{(P_1\setminus L, P_2\setminus L)\}\cup \textbf{Id}$ is a strong pomset bisimulation. It can be proved similarly to the proof of
            congruence for strong pomset bisimulation in CTC, we omit it;
           \item $P_1[f]\sim_{p} P_2[f]$. It is sufficient to prove the relation $R=\{(P_1[f], P_2[f])\}\cup \textbf{Id}$ is a strong pomset bisimulation. It can be proved similarly to the proof of
            congruence for strong pomset bisimulation in CTC, we omit it.
         \end{enumerate}
\end{enumerate}
\end{proof}

\begin{theorem}[Congruence for strong step bisimulation]
We can enjoy the full congruence for strong step bisimulation as follows.
\begin{enumerate}
  \item If $A\overset{\text{def}}{=}P$, then $A\sim_s P$;
  \item Let $P_1\sim_s P_2$. Then
        \begin{enumerate}
           \item $\alpha.P_1\sim_s \alpha.P_2$;
           \item $\phi.P_1\sim_s \phi.P_2$;
           \item $(\alpha_1\parallel\cdots\parallel\alpha_n).P_1\sim_s (\alpha_1\parallel\cdots\parallel\alpha_n).P_2$;
           \item $P_1+Q\sim_s P_2 +Q$;
           \item $P_1\parallel Q\sim_s P_2\parallel Q$;
           \item $P_1\setminus L\sim_s P_2\setminus L$;
           \item $P_1[f]\sim_s P_2[f]$.
         \end{enumerate}
\end{enumerate}
\end{theorem}

\begin{proof}
\begin{enumerate}
  \item If $A\overset{\text{def}}{=}P$, then $A\sim_s P$. It is obvious.
  \item Let $P_1\sim_s P_2$. Then
        \begin{enumerate}
           \item $\alpha.P_1\sim_s \alpha.P_2$. It is sufficient to prove the relation $R=\{(\alpha.P_1, \alpha.P_2)\}\cup \textbf{Id}$ is a strong step bisimulation. It can be proved similarly to the proof of
            congruence for strong step bisimulation in CTC, we omit it;
           \item $\phi.P_1\sim_s \phi.P_2$. It is sufficient to prove the relation $R=\{(\phi.P_1, \phi.P_2)\}\cup \textbf{Id}$ is a strong step bisimulation. It can be proved similarly to the proof of
            congruence for strong step bisimulation in CTC, we omit it;
           \item $(\alpha_1\parallel\cdots\parallel\alpha_n).P_1\sim_s (\alpha_1\parallel\cdots\parallel\alpha_n).P_2$. It is sufficient to prove the relation $R=\{((\alpha_1\parallel\cdots\parallel\alpha_n).P_1, (\alpha_1\parallel\cdots\parallel\alpha_n).P_2)\}\cup \textbf{Id}$ is a strong step bisimulation. It can be proved similarly to the proof of
            congruence for strong step bisimulation in CTC, we omit it;
           \item $P_1+Q\sim_s P_2 +Q$. It is sufficient to prove the relation $R=\{(P_1+Q, P_2+Q)\}\cup \textbf{Id}$ is a strong step bisimulation. It can be proved similarly to the proof of
            congruence for strong step bisimulation in CTC, we omit it;
           \item $P_1\parallel Q\sim_s P_2\parallel Q$. It is sufficient to prove the relation $R=\{(P_1\parallel Q, P_2\parallel Q)\}\cup \textbf{Id}$ is a strong step bisimulation. It can be proved similarly to the proof of
            congruence for strong step bisimulation in CTC, we omit it;
           \item $P_1\setminus L\sim_s P_2\setminus L$. It is sufficient to prove the relation $R=\{(P_1\setminus L, P_2\setminus L)\}\cup \textbf{Id}$ is a strong step bisimulation. It can be proved similarly to the proof of
            congruence for strong step bisimulation in CTC, we omit it;
           \item $P_1[f]\sim_s P_2[f]$. It is sufficient to prove the relation $R=\{(P_1[f], P_2[f])\}\cup \textbf{Id}$ is a strong step bisimulation. It can be proved similarly to the proof of
            congruence for strong step bisimulation in CTC, we omit it.
         \end{enumerate}
\end{enumerate}
\end{proof}

\begin{theorem}[Congruence for strong hp-bisimulation]
We can enjoy the full congruence for strong hp-bisimulation as follows.
\begin{enumerate}
  \item If $A\overset{\text{def}}{=}P$, then $A\sim_{hp} P$;
  \item Let $P_1\sim_{hp} P_2$. Then
        \begin{enumerate}
           \item $\alpha.P_1\sim_{hp} \alpha.P_2$;
           \item $\phi.P_1\sim_{hp} \phi.P_2$;
           \item $(\alpha_1\parallel\cdots\parallel\alpha_n).P_1\sim_{hp} (\alpha_1\parallel\cdots\parallel\alpha_n).P_2$;
           \item $P_1+Q\sim_{hp} P_2 +Q$;
           \item $P_1\parallel Q\sim_{hp} P_2\parallel Q$;
           \item $P_1\setminus L\sim_{hp} P_2\setminus L$;
           \item $P_1[f]\sim_{hp} P_2[f]$.
         \end{enumerate}
\end{enumerate}
\end{theorem}

\begin{proof}
\begin{enumerate}
  \item If $A\overset{\text{def}}{=}P$, then $A\sim_{hp} P$. It is obvious.
  \item Let $P_1\sim_{hp} P_2$. Then
        \begin{enumerate}
           \item $\alpha.P_1\sim_{hp} \alpha.P_2$. It is sufficient to prove the relation $R=\{(\alpha.P_1, \alpha.P_2)\}\cup \textbf{Id}$ is a strong hp-bisimulation. It can be proved similarly to the proof of
            congruence for strong hp-bisimulation in CTC, we omit it;
           \item $\phi.P_1\sim_{hp} \phi.P_2$. It is sufficient to prove the relation $R=\{(\phi.P_1, \phi.P_2)\}\cup \textbf{Id}$ is a strong hp-bisimulation. It can be proved similarly to the proof of
            congruence for strong hp-bisimulation in CTC, we omit it;
           \item $(\alpha_1\parallel\cdots\parallel\alpha_n).P_1\sim_{hp} (\alpha_1\parallel\cdots\parallel\alpha_n).P_2$. It is sufficient to prove the relation $R=\{((\alpha_1\parallel\cdots\parallel\alpha_n).P_1, (\alpha_1\parallel\cdots\parallel\alpha_n).P_2)\}\cup \textbf{Id}$ is a strong hp-bisimulation. It can be proved similarly to the proof of
            congruence for strong hp-bisimulation in CTC, we omit it;
           \item $P_1+Q\sim_{hp} P_2 +Q$. It is sufficient to prove the relation $R=\{(P_1+Q, P_2+Q)\}\cup \textbf{Id}$ is a strong hp-bisimulation. It can be proved similarly to the proof of
            congruence for strong hp-bisimulation in CTC, we omit it;
           \item $P_1\parallel Q\sim_{hp} P_2\parallel Q$. It is sufficient to prove the relation $R=\{(P_1\parallel Q, P_2\parallel Q)\}\cup \textbf{Id}$ is a strong hp-bisimulation. It can be proved similarly to the proof of
            congruence for strong hp-bisimulation in CTC, we omit it;
           \item $P_1\setminus L\sim_{hp} P_2\setminus L$. It is sufficient to prove the relation $R=\{(P_1\setminus L, P_2\setminus L)\}\cup \textbf{Id}$ is a strong hp-bisimulation. It can be proved similarly to the proof of
            congruence for strong hp-bisimulation in CTC, we omit it;
           \item $P_1[f]\sim_{hp} P_2[f]$. It is sufficient to prove the relation $R=\{(P_1[f], P_2[f])\}\cup \textbf{Id}$ is a strong hp-bisimulation. It can be proved similarly to the proof of
            congruence for strong hp-bisimulation in CTC, we omit it.
         \end{enumerate}
\end{enumerate}
\end{proof}

\begin{theorem}[Congruence for strong hhp-bisimulation]
We can enjoy the full congruence for strong hhp-bisimulation as follows.
\begin{enumerate}
  \item If $A\overset{\text{def}}{=}P$, then $A\sim_{hhp} P$;
  \item Let $P_1\sim_{hhp} P_2$. Then
        \begin{enumerate}
           \item $\alpha.P_1\sim_{hhp} \alpha.P_2$;
           \item $\phi.P_1\sim_{hhp} \phi.P_2$;
           \item $(\alpha_1\parallel\cdots\parallel\alpha_n).P_1\sim_{hhp} (\alpha_1\parallel\cdots\parallel\alpha_n).P_2$;
           \item $P_1+Q\sim_{hhp} P_2 +Q$;
           \item $P_1\parallel Q\sim_{hhp} P_2\parallel Q$;
           \item $P_1\setminus L\sim_{hhp} P_2\setminus L$;
           \item $P_1[f]\sim_{hhp} P_2[f]$.
         \end{enumerate}
\end{enumerate}
\end{theorem}

\begin{proof}
\begin{enumerate}
  \item If $A\overset{\text{def}}{=}P$, then $A\sim_{hhp} P$. It is obvious.
  \item Let $P_1\sim_{hhp} P_2$. Then
        \begin{enumerate}
           \item $\alpha.P_1\sim_{hhp} \alpha.P_2$. It is sufficient to prove the relation $R=\{(\alpha.P_1, \alpha.P_2)\}\cup \textbf{Id}$ is a strong hhp-bisimulation. It can be proved similarly to the proof of
            congruence for strong hhp-bisimulation in CTC, we omit it;
           \item $\phi.P_1\sim_{hhp} \phi.P_2$. It is sufficient to prove the relation $R=\{(\phi.P_1, \phi.P_2)\}\cup \textbf{Id}$ is a strong hhp-bisimulation. It can be proved similarly to the proof of
            congruence for strong hhp-bisimulation in CTC, we omit it;
           \item $(\alpha_1\parallel\cdots\parallel\alpha_n).P_1\sim_{hhp} (\alpha_1\parallel\cdots\parallel\alpha_n).P_2$. It is sufficient to prove the relation $R=\{((\alpha_1\parallel\cdots\parallel\alpha_n).P_1, (\alpha_1\parallel\cdots\parallel\alpha_n).P_2)\}\cup \textbf{Id}$ is a strong hhp-bisimulation. It can be proved similarly to the proof of
            congruence for strong hhp-bisimulation in CTC, we omit it;
           \item $P_1+Q\sim_{hhp} P_2 +Q$. It is sufficient to prove the relation $R=\{(P_1+Q, P_2+Q)\}\cup \textbf{Id}$ is a strong hhp-bisimulation. It can be proved similarly to the proof of
            congruence for strong hhp-bisimulation in CTC, we omit it;
           \item $P_1\parallel Q\sim_{hhp} P_2\parallel Q$. It is sufficient to prove the relation $R=\{(P_1\parallel Q, P_2\parallel Q)\}\cup \textbf{Id}$ is a strong hhp-bisimulation. It can be proved similarly to the proof of
            congruence for strong hhp-bisimulation in CTC, we omit it;
           \item $P_1\setminus L\sim_{hhp} P_2\setminus L$. It is sufficient to prove the relation $R=\{(P_1\setminus L, P_2\setminus L)\}\cup \textbf{Id}$ is a strong hhp-bisimulation. It can be proved similarly to the proof of
            congruence for strong hhp-bisimulation in CTC, we omit it;
           \item $P_1[f]\sim_{hhp} P_2[f]$. It is sufficient to prove the relation $R=\{(P_1[f], P_2[f])\}\cup \textbf{Id}$ is a strong hhp-bisimulation. It can be proved similarly to the proof of
            congruence for strong hhp-bisimulation in CTC, we omit it.
         \end{enumerate}
\end{enumerate}
\end{proof}

\subsubsection{Recursion}

\begin{definition}[Weakly guarded recursive expression]
$X$ is weakly guarded in $E$ if each occurrence of $X$ is with some subexpression $\alpha.F$ or $(\alpha_1\parallel\cdots\parallel\alpha_n).F$ of $E$.
\end{definition}

\begin{lemma}\label{LUS3}
If the variables $\widetilde{X}$ are weakly guarded in $E$, and $\langle E\{\widetilde{P}/\widetilde{X}\},s\rangle\xrightarrow{\{\alpha_1,\cdots,\alpha_n\}}\langle P',s'\rangle$, then 
$P'$ takes the form $E'\{\widetilde{P}/\widetilde{X}\}$ for some expression $E'$, and moreover, for any $\widetilde{Q}$, 
$\langle E\{\widetilde{Q}/\widetilde{X}\},s\rangle\xrightarrow{\{\alpha_1,\cdots,\alpha_n\}}\langle E'\{\widetilde{Q}/\widetilde{X}\},s'\rangle$.
\end{lemma}

\begin{proof}
It needs to induct on the depth of the inference of $\langle E\{\widetilde{P}/\widetilde{X}\},s\rangle\xrightarrow{\{\alpha_1,\cdots,\alpha_n\}}\langle P',s'\rangle$.

\begin{enumerate}
  \item Case $E\equiv Y$, a variable. Then $Y\notin \widetilde{X}$. Since $\widetilde{X}$ are weakly guarded, $\langle Y\{\widetilde{P}/\widetilde{X}\equiv Y\},s\rangle\nrightarrow$, this case is 
  impossible.
  \item Case $E\equiv\beta.F$. Then we must have $\alpha=\beta$, and $P'\equiv F\{\widetilde{P}/\widetilde{X}\}$, and 
  $\langle E\{\widetilde{Q}/\widetilde{X}\},s\rangle\equiv \langle \beta.F\{\widetilde{Q}/\widetilde{X}\},s\rangle \xrightarrow{\beta}\langle F\{\widetilde{Q}/\widetilde{X}\},s'\rangle$, 
  then, let $E'$ be $F$, as desired.
  \item Case $E\equiv(\beta_1\parallel\cdots\parallel\beta_n).F$. Then we must have $\alpha_i=\beta_i$ for $1\leq i\leq n$, and $P'\equiv F\{\widetilde{P}/\widetilde{X}\}$, and 
  $\langle E\{\widetilde{Q}/\widetilde{X}\},s\rangle\equiv \langle(\beta_1\parallel\cdots\parallel\beta_n).F\{\widetilde{Q}/\widetilde{X}\},s\rangle \xrightarrow{\{\beta_1,\cdots,\beta_n\}}\langle F\{\widetilde{Q}/\widetilde{X}\},s'\rangle$, 
  then, let $E'$ be $F$, as desired.
  \item Case $E\equiv E_1+E_2$. Then either $\langle E_1\{\widetilde{P}/\widetilde{X}\},s\rangle \xrightarrow{\{\alpha_1,\cdots,\alpha_n\}}\langle P',s'\rangle$ or 
  $\langle E_2\{\widetilde{P}/\widetilde{X}\},s\rangle \xrightarrow{\{\alpha_1,\cdots,\alpha_n\}}\langle P',s'\rangle$, then, we can apply this lemma in either case, as desired.
  \item Case $E\equiv E_1\parallel E_2$. There are four possibilities.
  \begin{enumerate}
    \item We may have $\langle E_1\{\widetilde{P}/\widetilde{X}\},s\rangle \xrightarrow{\alpha}\langle P_1',s'\rangle$ and $\langle E_2\{\widetilde{P}/\widetilde{X}\},s\rangle\nrightarrow$ 
    with $P'\equiv P_1'\parallel (E_2\{\widetilde{P}/\widetilde{X}\})$, then by applying this lemma, $P_1'$ is of the form $E_1'\{\widetilde{P}/\widetilde{X}\}$, and for any $Q$, 
    $\langle E_1\{\widetilde{Q}/\widetilde{X}\},s\rangle\xrightarrow{\alpha} \langle E_1'\{\widetilde{Q}/\widetilde{X}\},s'\rangle$. So, $P'$ is of the form 
    $E_1'\parallel E_2\{\widetilde{P}/\widetilde{X}\}$, and for any $Q$, 
    $\langle E\{\widetilde{Q}/\widetilde{X}\}\equiv E_1\{\widetilde{Q}/\widetilde{X}\}\parallel E_2\{\widetilde{Q}/\widetilde{X}\},s\rangle\xrightarrow{\alpha} \langle(E_1'\parallel E_2)\{\widetilde{Q}/\widetilde{X}\},s'\rangle$, 
    then, let $E'$ be $E_1'\parallel E_2$, as desired.
    \item We may have $\langle E_2\{\widetilde{P}/\widetilde{X}\},s\rangle \xrightarrow{\alpha}\langle P_2',s'\rangle$ and $\langle E_1\{\widetilde{P}/\widetilde{X}\},s\rangle\nrightarrow$ 
    with $P'\equiv P_2'\parallel (E_1\{\widetilde{P}/\widetilde{X}\})$, this case can be prove similarly to the above subcase, as desired.
    \item We may have $\langle E_1\{\widetilde{P}/\widetilde{X}\},s\rangle \xrightarrow{\alpha}\langle P_1',s'\rangle$ and 
    $\langle E_2\{\widetilde{P}/\widetilde{X}\},s\rangle\xrightarrow{\beta}\langle P_2',s''\rangle$ with $\alpha\neq\overline{\beta}$ and $P'\equiv P_1'\parallel P_2'$, then by 
    applying this lemma, $P_1'$ is of the form $E_1'\{\widetilde{P}/\widetilde{X}\}$, and for any $Q$, 
    $\langle E_1\{\widetilde{Q}/\widetilde{X}\},s\rangle\xrightarrow{\alpha} \langle E_1'\{\widetilde{Q}/\widetilde{X}\},s'\rangle$; $P_2'$ is of the form 
    $E_2'\{\widetilde{P}/\widetilde{X}\}$, and for any $Q$, $\langle E_2\{\widetilde{Q}/\widetilde{X}\},s\rangle\xrightarrow{\alpha} \langle E_2'\{\widetilde{Q}/\widetilde{X}\},s''\rangle$. 
    So, $P'$ is of the form $E_1'\parallel E_2'\{\widetilde{P}/\widetilde{X}\}$, and for any $Q$, 
    $\langle E\{\widetilde{Q}/\widetilde{X}\}\equiv E_1\{\widetilde{Q}/\widetilde{X}\}\parallel E_2\{\widetilde{Q}/\widetilde{X}\},s\rangle\xrightarrow{\{\alpha,\beta\}} 
    \langle (E_1'\parallel E_2')\{\widetilde{Q}/\widetilde{X}\},s'\cup s''\rangle$, then, let $E'$ be $E_1'\parallel E_2'$, as desired.
    \item We may have $\langle E_1\{\widetilde{P}/\widetilde{X}\},s\rangle \xrightarrow{l}\langle P_1',s'\rangle$ and 
    $\langle E_2\{\widetilde{P}/\widetilde{X}\},s\rangle\xrightarrow{\overline{l}}\langle P_2',s''\rangle$ with $P'\equiv P_1'\parallel P_2'$, then by applying this lemma, 
    $P_1'$ is of the form $E_1'\{\widetilde{P}/\widetilde{X}\}$, and for any $Q$, $\langle E_1\{\widetilde{Q}/\widetilde{X}\},s\rangle\xrightarrow{l} \langle E_1'\{\widetilde{Q}/\widetilde{X}\},s'\rangle$; 
    $P_2'$ is of the form $E_2'\{\widetilde{P}/\widetilde{X}\}$, and for any $Q$, $\langle E_2\{\widetilde{Q}/\widetilde{X}\},s\rangle\xrightarrow{\overline{l}}\langle E_2'\{\widetilde{Q}/\widetilde{X}\},s''\rangle$. 
    So, $P'$ is of the form $E_1'\parallel E_2'\{\widetilde{P}/\widetilde{X}\}$, and for any $Q$, $\langle E\{\widetilde{Q}/\widetilde{X}\}\equiv E_1\{\widetilde{Q}/\widetilde{X}\}\parallel E_2\{\widetilde{Q}/\widetilde{X}\},s\rangle
    \xrightarrow{\tau} \langle (E_1'\parallel E_2')\{\widetilde{Q}/\widetilde{X}\},s'\cup s''\rangle$, then, let $E'$ be $E_1'\parallel E_2'$, as desired.
  \end{enumerate}
  \item Case $E\equiv F[R]$ and $E\equiv F\setminus L$. These cases can be prove similarly to the above case.
  \item Case $E\equiv C$, an agent constant defined by $C\overset{\text{def}}{=}R$. Then there is no $X\in\widetilde{X}$ occurring in $E$, so 
  $\langle C,s\rangle\xrightarrow{\{\alpha_1,\cdots,\alpha_n\}}\langle P',s'\rangle$, let $E'$ be $P'$, as desired.
\end{enumerate}
\end{proof}

\begin{theorem}[Unique solution of equations for strong pomset bisimulation]\label{USSSB3}
Let the recursive expressions $E_i(i\in I)$ contain at most the variables $X_i(i\in I)$, and let each $X_j(j\in I)$ be weakly guarded in each $E_i$. Then,

If $\widetilde{P}\sim_p \widetilde{E}\{\widetilde{P}/\widetilde{X}\}$ and $\widetilde{Q}\sim_p \widetilde{E}\{\widetilde{Q}/\widetilde{X}\}$, then $\widetilde{P}\sim_p \widetilde{Q}$.
\end{theorem}

\begin{proof}
It is sufficient to induct on the depth of the inference of $\langle E\{\widetilde{P}/\widetilde{X}\},s\rangle\xrightarrow{\{\alpha_1,\cdots,\alpha_n\}}\langle P',s'\rangle$.

\begin{enumerate}
  \item Case $E\equiv X_i$. Then we have $\langle E\{\widetilde{P}/\widetilde{X}\},s\rangle\equiv \langle P_i,s\rangle\xrightarrow{\{\alpha_1,\cdots,\alpha_n\}}\langle P',s'\rangle$, 
  since $P_i\sim_p E_i\{\widetilde{P}/\widetilde{X}\}$, we have $\langle E_i\{\widetilde{P}/\widetilde{X}\},s\rangle\xrightarrow{\{\alpha_1,\cdots,\alpha_n\}}\langle P'',s'\rangle\sim_p \langle P',s'\rangle$. 
  Since $\widetilde{X}$ are weakly guarded in $E_i$, by Lemma \ref{LUS3}, $P''\equiv E'\{\widetilde{P}/\widetilde{X}\}$ and $\langle E_i\{\widetilde{P}/\widetilde{X}\},s\rangle
  \xrightarrow{\{\alpha_1,\cdots,\alpha_n\}} \langle E'\{\widetilde{P}/\widetilde{X}\},s'\rangle$. Since 
  $E\{\widetilde{Q}/\widetilde{X}\}\equiv X_i\{\widetilde{Q}/\widetilde{X}\} \equiv Q_i\sim_p E_i\{\widetilde{Q}/\widetilde{X}\}$, $\langle E\{\widetilde{Q}/\widetilde{X}\},s\rangle\xrightarrow{\{\alpha_1,\cdots,\alpha_n\}}\langle Q',s'\rangle\sim_p \langle E'\{\widetilde{Q}/\widetilde{X}\},s'\rangle$. 
  So, $P'\sim_p Q'$, as desired.
  \item Case $E\equiv\alpha.F$. This case can be proven similarly.
  \item Case $E\equiv(\alpha_1\parallel\cdots\parallel\alpha_n).F$. This case can be proven similarly.
  \item Case $E\equiv E_1+E_2$. We have $\langle E_i\{\widetilde{P}/\widetilde{X}\},s\rangle \xrightarrow{\{\alpha_1,\cdots,\alpha_n\}}\langle P',s'\rangle$, 
  $\langle E_i\{\widetilde{Q}/\widetilde{X}\},s\rangle \xrightarrow{\{\alpha_1,\cdots,\alpha_n\}}\langle Q',s'\rangle$, then, $P'\sim_p Q'$, as desired.
  \item Case $E\equiv E_1\parallel E_2$, $E\equiv F[R]$ and $E\equiv F\setminus L$, $E\equiv C$. These cases can be prove similarly to the above case.
\end{enumerate}
\end{proof}

\begin{theorem}[Unique solution of equations for strong step bisimulation]\label{USSSB3}
Let the recursive expressions $E_i(i\in I)$ contain at most the variables $X_i(i\in I)$, and let each $X_j(j\in I)$ be weakly guarded in each $E_i$. Then,

If $\widetilde{P}\sim_s \widetilde{E}\{\widetilde{P}/\widetilde{X}\}$ and $\widetilde{Q}\sim_s \widetilde{E}\{\widetilde{Q}/\widetilde{X}\}$, then $\widetilde{P}\sim_s \widetilde{Q}$.
\end{theorem}

\begin{proof}
It is sufficient to induct on the depth of the inference of $\langle E\{\widetilde{P}/\widetilde{X}\},s\rangle\xrightarrow{\{\alpha_1,\cdots,\alpha_n\}}\langle P',s'\rangle$.

\begin{enumerate}
  \item Case $E\equiv X_i$. Then we have $\langle E\{\widetilde{P}/\widetilde{X}\},s\rangle\equiv \langle P_i,s\rangle\xrightarrow{\{\alpha_1,\cdots,\alpha_n\}}\langle P',s'\rangle$,
  since $P_i\sim_s E_i\{\widetilde{P}/\widetilde{X}\}$, we have $\langle E_i\{\widetilde{P}/\widetilde{X}\},s\rangle\xrightarrow{\{\alpha_1,\cdots,\alpha_n\}}\langle P'',s'\rangle\sim_s \langle P',s'\rangle$.
  Since $\widetilde{X}$ are weakly guarded in $E_i$, by Lemma \ref{LUS3}, $P''\equiv E'\{\widetilde{P}/\widetilde{X}\}$ and $\langle E_i\{\widetilde{P}/\widetilde{X}\},s\rangle
  \xrightarrow{\{\alpha_1,\cdots,\alpha_n\}} \langle E'\{\widetilde{P}/\widetilde{X}\},s'\rangle$. Since
  $E\{\widetilde{Q}/\widetilde{X}\}\equiv X_i\{\widetilde{Q}/\widetilde{X}\} \equiv Q_i\sim_s E_i\{\widetilde{Q}/\widetilde{X}\}$, $\langle E\{\widetilde{Q}/\widetilde{X}\},s\rangle\xrightarrow{\{\alpha_1,\cdots,\alpha_n\}}\langle Q',s'\rangle\sim_s \langle E'\{\widetilde{Q}/\widetilde{X}\},s'\rangle$.
  So, $P'\sim_s Q'$, as desired.
  \item Case $E\equiv\alpha.F$. This case can be proven similarly.
  \item Case $E\equiv(\alpha_1\parallel\cdots\parallel\alpha_n).F$. This case can be proven similarly.
  \item Case $E\equiv E_1+E_2$. We have $\langle E_i\{\widetilde{P}/\widetilde{X}\},s\rangle \xrightarrow{\{\alpha_1,\cdots,\alpha_n\}}\langle P',s'\rangle$,
  $\langle E_i\{\widetilde{Q}/\widetilde{X}\},s\rangle \xrightarrow{\{\alpha_1,\cdots,\alpha_n\}}\langle Q',s'\rangle$, then, $P'\sim_s Q'$, as desired.
  \item Case $E\equiv E_1\parallel E_2$, $E\equiv F[R]$ and $E\equiv F\setminus L$, $E\equiv C$. These cases can be prove similarly to the above case.
\end{enumerate}
\end{proof}

\begin{theorem}[Unique solution of equations for strong hp-bisimulation]\label{USSSB3}
Let the recursive expressions $E_i(i\in I)$ contain at most the variables $X_i(i\in I)$, and let each $X_j(j\in I)$ be weakly guarded in each $E_i$. Then,

If $\widetilde{P}\sim_{hp} \widetilde{E}\{\widetilde{P}/\widetilde{X}\}$ and $\widetilde{Q}\sim_{hp} \widetilde{E}\{\widetilde{Q}/\widetilde{X}\}$, then $\widetilde{P}\sim_{hp} \widetilde{Q}$.
\end{theorem}

\begin{proof}
It is sufficient to induct on the depth of the inference of $\langle E\{\widetilde{P}/\widetilde{X}\},s\rangle\xrightarrow{\{\alpha_1,\cdots,\alpha_n\}}\langle P',s'\rangle$.

\begin{enumerate}
  \item Case $E\equiv X_i$. Then we have $\langle E\{\widetilde{P}/\widetilde{X}\},s\rangle\equiv \langle P_i,s\rangle\xrightarrow{\{\alpha_1,\cdots,\alpha_n\}}\langle P',s'\rangle$,
  since $P_i\sim_{hp} E_i\{\widetilde{P}/\widetilde{X}\}$, we have $\langle E_i\{\widetilde{P}/\widetilde{X}\},s\rangle\xrightarrow{\{\alpha_1,\cdots,\alpha_n\}}\langle P'',s'\rangle\sim_{hp} \langle P',s'\rangle$.
  Since $\widetilde{X}$ are weakly guarded in $E_i$, by Lemma \ref{LUS3}, $P''\equiv E'\{\widetilde{P}/\widetilde{X}\}$ and $\langle E_i\{\widetilde{P}/\widetilde{X}\},s\rangle
  \xrightarrow{\{\alpha_1,\cdots,\alpha_n\}} \langle E'\{\widetilde{P}/\widetilde{X}\},s'\rangle$. Since
  $E\{\widetilde{Q}/\widetilde{X}\}\equiv X_i\{\widetilde{Q}/\widetilde{X}\} \equiv Q_i\sim_{hp} E_i\{\widetilde{Q}/\widetilde{X}\}$, $\langle E\{\widetilde{Q}/\widetilde{X}\},s\rangle\xrightarrow{\{\alpha_1,\cdots,\alpha_n\}}\langle Q',s'\rangle\sim_{hp} \langle E'\{\widetilde{Q}/\widetilde{X}\},s'\rangle$.
  So, $P'\sim_{hp} Q'$, as desired.
  \item Case $E\equiv\alpha.F$. This case can be proven similarly.
  \item Case $E\equiv(\alpha_1\parallel\cdots\parallel\alpha_n).F$. This case can be proven similarly.
  \item Case $E\equiv E_1+E_2$. We have $\langle E_i\{\widetilde{P}/\widetilde{X}\},s\rangle \xrightarrow{\{\alpha_1,\cdots,\alpha_n\}}\langle P',s'\rangle$,
  $\langle E_i\{\widetilde{Q}/\widetilde{X}\},s\rangle \xrightarrow{\{\alpha_1,\cdots,\alpha_n\}}\langle Q',s'\rangle$, then, $P'\sim_{hp} Q'$, as desired.
  \item Case $E\equiv E_1\parallel E_2$, $E\equiv F[R]$ and $E\equiv F\setminus L$, $E\equiv C$. These cases can be prove similarly to the above case.
\end{enumerate}
\end{proof}

\begin{theorem}[Unique solution of equations for strong hhp-bisimulation]\label{USSSB3}
Let the recursive expressions $E_i(i\in I)$ contain at most the variables $X_i(i\in I)$, and let each $X_j(j\in I)$ be weakly guarded in each $E_i$. Then,

If $\widetilde{P}\sim_{hhp} \widetilde{E}\{\widetilde{P}/\widetilde{X}\}$ and $\widetilde{Q}\sim_{hhp} \widetilde{E}\{\widetilde{Q}/\widetilde{X}\}$, then $\widetilde{P}\sim_{hhp} \widetilde{Q}$.
\end{theorem}

\begin{proof}
It is sufficient to induct on the depth of the inference of $\langle E\{\widetilde{P}/\widetilde{X}\},s\rangle\xrightarrow{\{\alpha_1,\cdots,\alpha_n\}}\langle P',s'\rangle$.

\begin{enumerate}
  \item Case $E\equiv X_i$. Then we have $\langle E\{\widetilde{P}/\widetilde{X}\},s\rangle\equiv \langle P_i,s\rangle\xrightarrow{\{\alpha_1,\cdots,\alpha_n\}}\langle P',s'\rangle$,
  since $P_i\sim_{hhp} E_i\{\widetilde{P}/\widetilde{X}\}$, we have $\langle E_i\{\widetilde{P}/\widetilde{X}\},s\rangle\xrightarrow{\{\alpha_1,\cdots,\alpha_n\}}\langle P'',s'\rangle\sim_{hhp} \langle P',s'\rangle$.
  Since $\widetilde{X}$ are weakly guarded in $E_i$, by Lemma \ref{LUS3}, $P''\equiv E'\{\widetilde{P}/\widetilde{X}\}$ and $\langle E_i\{\widetilde{P}/\widetilde{X}\},s\rangle
  \xrightarrow{\{\alpha_1,\cdots,\alpha_n\}} \langle E'\{\widetilde{P}/\widetilde{X}\},s'\rangle$. Since
  $E\{\widetilde{Q}/\widetilde{X}\}\equiv X_i\{\widetilde{Q}/\widetilde{X}\} \equiv Q_i\sim_{hhp} E_i\{\widetilde{Q}/\widetilde{X}\}$, $\langle E\{\widetilde{Q}/\widetilde{X}\},s\rangle\xrightarrow{\{\alpha_1,\cdots,\alpha_n\}}\langle Q',s'\rangle\sim_{hhp} \langle E'\{\widetilde{Q}/\widetilde{X}\},s'\rangle$.
  So, $P'\sim_{hhp} Q'$, as desired.
  \item Case $E\equiv\alpha.F$. This case can be proven similarly.
  \item Case $E\equiv(\alpha_1\parallel\cdots\parallel\alpha_n).F$. This case can be proven similarly.
  \item Case $E\equiv E_1+E_2$. We have $\langle E_i\{\widetilde{P}/\widetilde{X}\},s\rangle \xrightarrow{\{\alpha_1,\cdots,\alpha_n\}}\langle P',s'\rangle$,
  $\langle E_i\{\widetilde{Q}/\widetilde{X}\},s\rangle \xrightarrow{\{\alpha_1,\cdots,\alpha_n\}}\langle Q',s'\rangle$, then, $P'\sim_{hhp} Q'$, as desired.
  \item Case $E\equiv E_1\parallel E_2$, $E\equiv F[R]$ and $E\equiv F\setminus L$, $E\equiv C$. These cases can be prove similarly to the above case.
\end{enumerate}
\end{proof}

\subsection{Weak Bisimulations}\label{wtcbctcg}

The weak transition rules for CTC with guards are listed in Table \ref{WTRForCTC3}.

\begin{center}
    \begin{table}
        \[\textbf{Act}_1\quad \frac{}{\langle\alpha.P,s\rangle\xRightarrow{\alpha}\langle P,s'\rangle}\]

        \[\textbf{Act}_2\quad \frac{}{\langle\epsilon,s\rangle\Rightarrow\langle\surd,s\rangle}\]

        \[\textbf{Gur}\quad \frac{}{\langle\phi,s\rangle\Rightarrow\langle\surd,s\rangle}\textrm{ if }test(\phi,s)\]

        \[\textbf{Sum}_1\quad \frac{\langle P,s\rangle\xRightarrow{\alpha}\langle P',s'\rangle}{\langle P+Q,s\rangle\xRightarrow{\alpha}\langle P',s'\rangle}\]

        \[\textbf{Com}_1\quad \frac{\langle P,s\rangle\xRightarrow{\alpha}\langle P',s'\rangle\quad Q\nrightarrow}{\langle P\parallel Q,s\rangle\xRightarrow{\alpha}\langle P'\parallel Q,s'\rangle}\]

        \[\textbf{Com}_2\quad \frac{\langle Q,s\rangle\xRightarrow{\alpha}\langle Q',s'\rangle\quad P\nrightarrow}{\langle P\parallel Q,s\rangle\xRightarrow{\alpha}\langle P\parallel Q',s'\rangle}\]

        \[\textbf{Com}_3\quad \frac{\langle P,s\rangle\xRightarrow{\alpha}\langle P',s'\rangle\quad \langle Q,s\rangle\xRightarrow{\beta}\langle Q',s''\rangle}{\langle P\parallel Q,s\rangle\xRightarrow{\{\alpha,\beta\}}\langle P'\parallel Q',s'\cup s''\rangle}\quad (\beta\neq\overline{\alpha})\]

        \[\textbf{Com}_4\quad \frac{\langle P,s\rangle\xRightarrow{l}\langle P',s'\rangle\quad \langle Q,s\rangle\xRightarrow{\overline{l}}\langle Q',s''\rangle}{\langle P\parallel Q,s\rangle\xRightarrow{\tau}\langle P'\parallel Q',s'\cup s''\rangle}\]

        \[\textbf{Act}_3\quad \frac{}{\langle (\alpha_1\parallel\cdots\parallel\alpha_n).P,s\rangle\xRightarrow{\{\alpha_1,\cdots,\alpha_n\}}\langle P,s'\rangle}\quad (\alpha_i\neq\overline{\alpha_j}\quad i,j\in\{1,\cdots,n\})\]

        \[\textbf{Sum}_2\quad \frac{\langle P,s\rangle\xRightarrow{\{\alpha_1,\cdots,\alpha_n\}}\langle P',s'\rangle}{\langle P+Q,s\rangle\xRightarrow{\{\alpha_1,\cdots,\alpha_n\}}\langle P',s'\rangle}\]

        \[\textbf{Res}_1\quad \frac{\langle P,s\rangle\xRightarrow{\alpha}\langle P',s'\rangle}{\langle P\setminus L,s\rangle\xRightarrow{\alpha}\langle P'\setminus L,s'\rangle}\quad (\alpha,\overline{\alpha}\notin L)\]

        \[\textbf{Res}_2\quad \frac{\langle P,s\rangle\xRightarrow{\{\alpha_1,\cdots,\alpha_n\}}\langle P',s'\rangle}{\langle P\setminus L,s\rangle\xRightarrow{\{\alpha_1,\cdots,\alpha_n\}}\langle P'\setminus L,s'\rangle}\quad (\alpha_1,\overline{\alpha_1},\cdots,\alpha_n,\overline{\alpha_n}\notin L)\]

        \[\textbf{Rel}_1\quad \frac{\langle P,s\rangle\xRightarrow{\alpha}\langle P',s'\rangle}{\langle P[f],s\rangle\xRightarrow{\langle f(\alpha)}P'[f],s'\rangle}\]

        \[\textbf{Rel}_2\quad \frac{\langle P,s\rangle\xRightarrow{\{\alpha_1,\cdots,\alpha_n\}}\langle P',s'\rangle}{\langle P[f],s\rangle\xRightarrow{\{f(\alpha_1),\cdots,f(\alpha_n)\}}\langle P'[f],s'\rangle}\]

        \[\textbf{Con}_1\quad\frac{\langle P,s\rangle\xRightarrow{\alpha}\langle P',s'\rangle}{\langle A,s\rangle\xRightarrow{\alpha}\langle P',s'\rangle}\quad (A\overset{\text{def}}{=}P)\]

        \[\textbf{Con}_2\quad\frac{\langle P,s\rangle\xRightarrow{\{\alpha_1,\cdots,\alpha_n\}}\langle P',s'\rangle}{\langle A,s\rangle\xRightarrow{\{\alpha_1,\cdots,\alpha_n\}}\langle P',s'\rangle}\quad (A\overset{\text{def}}{=}P)\]
        \caption{Weak transition rules of CTC with guards}
        \label{WTRForCTC3}
    \end{table}
\end{center}

\subsubsection{Laws and Congruence}

Remembering that $\tau$ can neither be restricted nor relabeled, we know that the monoid laws, the static laws, the guards laws, and the expansion law in section \ref{stcbctcg} still 
hold with respect to the corresponding weakly truly concurrent bisimulations. And also, we can enjoy the full congruence of Prefix, Guards, Summation, Composition, Restriction, 
Relabelling and Constants with respect to corresponding weakly truly concurrent bisimulations. We will not retype these laws, and just give the $\tau$-specific laws.

\begin{proposition}[$\tau$ laws for weak pomset bisimulation]
The $\tau$ laws for weak pomset bisimulation is as follows.
\begin{enumerate}
  \item $P\approx_p \tau.P$;
  \item $\alpha.\tau.P\approx_p \alpha.P$;
  \item $(\alpha_1\parallel\cdots\parallel\alpha_n).\tau.P\approx_p (\alpha_1\parallel\cdots\parallel\alpha_n).P$;
  \item $P+\tau.P\approx_p \tau.P$;
  \item $\alpha.(P+\tau.Q)+\alpha.Q\approx_p\alpha.(P+\tau.Q)$;
  \item $\phi.(P+\tau.Q)+\phi.Q\approx_p\phi.(P+\tau.Q)$;
  \item $(\alpha_1\parallel\cdots\parallel\alpha_n).(P+\tau.Q)+ (\alpha_1\parallel\cdots\parallel\alpha_n).Q\approx_p (\alpha_1\parallel\cdots\parallel\alpha_n).(P+\tau.Q)$;
  \item $P\approx_p \tau\parallel P$.
\end{enumerate}
\end{proposition}

\begin{proof}
\begin{enumerate}
  \item $P\approx_p \tau.P$. It is sufficient to prove the relation $R=\{(P, \tau.P)\}\cup \textbf{Id}$ is a weak pomset bisimulation. It can be proved similarly to the proof of
  $\tau$ laws for weak pomset bisimulation in CTC, we omit it;
  \item $\alpha.\tau.P\approx_p \alpha.P$. It is sufficient to prove the relation $R=\{(\alpha.\tau.P, \alpha.P)\}\cup \textbf{Id}$ is a weak pomset bisimulation. It can be proved similarly to the proof of
  $\tau$ laws for weak pomset bisimulation in CTC, we omit it;
  \item $(\alpha_1\parallel\cdots\parallel\alpha_n).\tau.P\approx_p (\alpha_1\parallel\cdots\parallel\alpha_n).P$. It is sufficient to prove the relation $R=\{((\alpha_1\parallel\cdots\parallel\alpha_n).\tau.P, (\alpha_1\parallel\cdots\parallel\alpha_n).P)\}\cup \textbf{Id}$ is a weak pomset bisimulation. It can be proved similarly to the proof of
  $\tau$ laws for weak pomset bisimulation in CTC, we omit it;
  \item $P+\tau.P\approx_p \tau.P$. It is sufficient to prove the relation $R=\{(P+\tau.P, \tau.P)\}\cup \textbf{Id}$ is a weak pomset bisimulation. It can be proved similarly to the proof of
  $\tau$ laws for weak pomset bisimulation in CTC, we omit it;
  \item $\alpha.(P+\tau.Q)+\alpha.Q\approx_p\alpha.(P+\tau.Q)$. It is sufficient to prove the relation $R=\{(\alpha.(P+\tau.Q)+\alpha.Q, \alpha.(P+\tau.Q))\}\cup \textbf{Id}$ is a weak pomset bisimulation. It can be proved similarly to the proof of
  $\tau$ laws for weak pomset bisimulation in CTC, we omit it;
  \item $\phi.(P+\tau.Q)+\phi.Q\approx_p\phi.(P+\tau.Q)$. It is sufficient to prove the relation $R=\{(\phi.(P+\tau.Q)+\phi.Q, \phi.(P+\tau.Q))\}\cup \textbf{Id}$ is a weak pomset bisimulation, we omit it;
  \item $(\alpha_1\parallel\cdots\parallel\alpha_n).(P+\tau.Q)+ (\alpha_1\parallel\cdots\parallel\alpha_n).Q\approx_p (\alpha_1\parallel\cdots\parallel\alpha_n).(P+\tau.Q)$. It is sufficient to prove the relation $R=\{((\alpha_1\parallel\cdots\parallel\alpha_n).(P+\tau.Q)+ (\alpha_1\parallel\cdots\parallel\alpha_n).Q, (\alpha_1\parallel\cdots\parallel\alpha_n).(P+\tau.Q))\}\cup \textbf{Id}$ is a weak pomset bisimulation. It can be proved similarly to the proof of
  $\tau$ laws for weak pomset bisimulation in CTC, we omit it;
  \item $P\approx_p \tau\parallel P$. It is sufficient to prove the relation $R=\{(\tau\parallel P, P)\}\cup \textbf{Id}$ is a weak pomset bisimulation. It can be proved similarly to the proof of
  $\tau$ laws for weak pomset bisimulation in CTC, we omit it.
\end{enumerate}
\end{proof}

\begin{proposition}[$\tau$ laws for weak step bisimulation]
The $\tau$ laws for weak step bisimulation is as follows.
\begin{enumerate}
  \item $P\approx_s \tau.P$;
  \item $\alpha.\tau.P\approx_s \alpha.P$;
  \item $(\alpha_1\parallel\cdots\parallel\alpha_n).\tau.P\approx_s (\alpha_1\parallel\cdots\parallel\alpha_n).P$;
  \item $P+\tau.P\approx_s \tau.P$;
  \item $\alpha.(P+\tau.Q)+\alpha.Q\approx_s\alpha.(P+\tau.Q)$;
  \item $\phi.(P+\tau.Q)+\phi.Q\approx_s\phi.(P+\tau.Q)$;
  \item $(\alpha_1\parallel\cdots\parallel\alpha_n).(P+\tau.Q)+ (\alpha_1\parallel\cdots\parallel\alpha_n).Q\approx_s (\alpha_1\parallel\cdots\parallel\alpha_n).(P+\tau.Q)$;
  \item $P\approx_s \tau\parallel P$.
\end{enumerate}
\end{proposition}

\begin{proof}
\begin{enumerate}
  \item $P\approx_s \tau.P$. It is sufficient to prove the relation $R=\{(P, \tau.P)\}\cup \textbf{Id}$ is a weak step bisimulation. It can be proved similarly to the proof of
  $\tau$ laws for weak step bisimulation in CTC, we omit it;
  \item $\alpha.\tau.P\approx_s \alpha.P$. It is sufficient to prove the relation $R=\{(\alpha.\tau.P, \alpha.P)\}\cup \textbf{Id}$ is a weak step bisimulation. It can be proved similarly to the proof of
  $\tau$ laws for weak step bisimulation in CTC, we omit it;
  \item $(\alpha_1\parallel\cdots\parallel\alpha_n).\tau.P\approx_s (\alpha_1\parallel\cdots\parallel\alpha_n).P$. It is sufficient to prove the relation $R=\{((\alpha_1\parallel\cdots\parallel\alpha_n).\tau.P, (\alpha_1\parallel\cdots\parallel\alpha_n).P)\}\cup \textbf{Id}$ is a weak step bisimulation. It can be proved similarly to the proof of
  $\tau$ laws for weak step bisimulation in CTC, we omit it;
  \item $P+\tau.P\approx_s \tau.P$. It is sufficient to prove the relation $R=\{(P+\tau.P, \tau.P)\}\cup \textbf{Id}$ is a weak step bisimulation. It can be proved similarly to the proof of
  $\tau$ laws for weak step bisimulation in CTC, we omit it;
  \item $\alpha.(P+\tau.Q)+\alpha.Q\approx_s\alpha.(P+\tau.Q)$. It is sufficient to prove the relation $R=\{(\alpha.(P+\tau.Q)+\alpha.Q, \alpha.(P+\tau.Q))\}\cup \textbf{Id}$ is a weak step bisimulation. It can be proved similarly to the proof of
  $\tau$ laws for weak step bisimulation in CTC, we omit it;
  \item $\phi.(P+\tau.Q)+\phi.Q\approx_s\phi.(P+\tau.Q)$. It is sufficient to prove the relation $R=\{(\phi.(P+\tau.Q)+\phi.Q, \phi.(P+\tau.Q))\}\cup \textbf{Id}$ is a weak step bisimulation, we omit it;
  \item $(\alpha_1\parallel\cdots\parallel\alpha_n).(P+\tau.Q)+ (\alpha_1\parallel\cdots\parallel\alpha_n).Q\approx_s (\alpha_1\parallel\cdots\parallel\alpha_n).(P+\tau.Q)$. It is sufficient to prove the relation $R=\{((\alpha_1\parallel\cdots\parallel\alpha_n).(P+\tau.Q)+ (\alpha_1\parallel\cdots\parallel\alpha_n).Q, (\alpha_1\parallel\cdots\parallel\alpha_n).(P+\tau.Q))\}\cup \textbf{Id}$ is a weak step bisimulation. It can be proved similarly to the proof of
  $\tau$ laws for weak step bisimulation in CTC, we omit it;
  \item $P\approx_s \tau\parallel P$. It is sufficient to prove the relation $R=\{(\tau\parallel P, P)\}\cup \textbf{Id}$ is a weak step bisimulation. It can be proved similarly to the proof of
  $\tau$ laws for weak step bisimulation in CTC, we omit it.
\end{enumerate}
\end{proof}

\begin{proposition}[$\tau$ laws for weak hp-bisimulation]
The $\tau$ laws for weak hp-bisimulation is as follows.
\begin{enumerate}
  \item $P\approx_{hp} \tau.P$;
  \item $\alpha.\tau.P\approx_{hp} \alpha.P$;
  \item $(\alpha_1\parallel\cdots\parallel\alpha_n).\tau.P\approx_{hp} (\alpha_1\parallel\cdots\parallel\alpha_n).P$;
  \item $P+\tau.P\approx_{hp} \tau.P$;
  \item $\alpha.(P+\tau.Q)+\alpha.Q\approx_{hp}\alpha.(P+\tau.Q)$;
  \item $\phi.(P+\tau.Q)+\phi.Q\approx_{hp}\phi.(P+\tau.Q)$;
  \item $(\alpha_1\parallel\cdots\parallel\alpha_n).(P+\tau.Q)+ (\alpha_1\parallel\cdots\parallel\alpha_n).Q\approx_{hp} (\alpha_1\parallel\cdots\parallel\alpha_n).(P+\tau.Q)$;
  \item $P\approx_{hp} \tau\parallel P$.
\end{enumerate}
\end{proposition}

\begin{proof}
\begin{enumerate}
  \item $P\approx_{hp} \tau.P$. It is sufficient to prove the relation $R=\{(P, \tau.P)\}\cup \textbf{Id}$ is a weak hp-bisimulation. It can be proved similarly to the proof of
  $\tau$ laws for weak hp-bisimulation in CTC, we omit it;
  \item $\alpha.\tau.P\approx_{hp} \alpha.P$. It is sufficient to prove the relation $R=\{(\alpha.\tau.P, \alpha.P)\}\cup \textbf{Id}$ is a weak hp-bisimulation. It can be proved similarly to the proof of
  $\tau$ laws for weak hp-bisimulation in CTC, we omit it;
  \item $(\alpha_1\parallel\cdots\parallel\alpha_n).\tau.P\approx_{hp} (\alpha_1\parallel\cdots\parallel\alpha_n).P$. It is sufficient to prove the relation $R=\{((\alpha_1\parallel\cdots\parallel\alpha_n).\tau.P, (\alpha_1\parallel\cdots\parallel\alpha_n).P)\}\cup \textbf{Id}$ is a weak hp-bisimulation. It can be proved similarly to the proof of
  $\tau$ laws for weak hp-bisimulation in CTC, we omit it;
  \item $P+\tau.P\approx_{hp} \tau.P$. It is sufficient to prove the relation $R=\{(P+\tau.P, \tau.P)\}\cup \textbf{Id}$ is a weak hp-bisimulation. It can be proved similarly to the proof of
  $\tau$ laws for weak hp-bisimulation in CTC, we omit it;
  \item $\alpha.(P+\tau.Q)+\alpha.Q\approx_{hp}\alpha.(P+\tau.Q)$. It is sufficient to prove the relation $R=\{(\alpha.(P+\tau.Q)+\alpha.Q, \alpha.(P+\tau.Q))\}\cup \textbf{Id}$ is a weak hp-bisimulation. It can be proved similarly to the proof of
  $\tau$ laws for weak hp-bisimulation in CTC, we omit it;
  \item $\phi.(P+\tau.Q)+\phi.Q\approx_{hp}\phi.(P+\tau.Q)$. It is sufficient to prove the relation $R=\{(\phi.(P+\tau.Q)+\phi.Q, \phi.(P+\tau.Q))\}\cup \textbf{Id}$ is a weak hp-bisimulation, we omit it;
  \item $(\alpha_1\parallel\cdots\parallel\alpha_n).(P+\tau.Q)+ (\alpha_1\parallel\cdots\parallel\alpha_n).Q\approx_{hp} (\alpha_1\parallel\cdots\parallel\alpha_n).(P+\tau.Q)$. It is sufficient to prove the relation $R=\{((\alpha_1\parallel\cdots\parallel\alpha_n).(P+\tau.Q)+ (\alpha_1\parallel\cdots\parallel\alpha_n).Q, (\alpha_1\parallel\cdots\parallel\alpha_n).(P+\tau.Q))\}\cup \textbf{Id}$ is a weak hp-bisimulation. It can be proved similarly to the proof of
  $\tau$ laws for weak hp-bisimulation in CTC, we omit it;
  \item $P\approx_{hp} \tau\parallel P$. It is sufficient to prove the relation $R=\{(\tau\parallel P, P)\}\cup \textbf{Id}$ is a weak hp-bisimulation. It can be proved similarly to the proof of
  $\tau$ laws for weak hp-bisimulation in CTC, we omit it.
\end{enumerate}
\end{proof}

\begin{proposition}[$\tau$ laws for weak hhp-bisimulation]
The $\tau$ laws for weak hhp-bisimulation is as follows.
\begin{enumerate}
  \item $P\approx_{hhp} \tau.P$;
  \item $\alpha.\tau.P\approx_{hhp} \alpha.P$;
  \item $(\alpha_1\parallel\cdots\parallel\alpha_n).\tau.P\approx_{hhp} (\alpha_1\parallel\cdots\parallel\alpha_n).P$;
  \item $P+\tau.P\approx_{hhp} \tau.P$;
  \item $\alpha.(P+\tau.Q)+\alpha.Q\approx_{hhp}\alpha.(P+\tau.Q)$;
  \item $\phi.(P+\tau.Q)+\phi.Q\approx_{hhp}\phi.(P+\tau.Q)$;
  \item $(\alpha_1\parallel\cdots\parallel\alpha_n).(P+\tau.Q)+ (\alpha_1\parallel\cdots\parallel\alpha_n).Q\approx_{hhp} (\alpha_1\parallel\cdots\parallel\alpha_n).(P+\tau.Q)$;
  \item $P\approx_{hhp} \tau\parallel P$.
\end{enumerate}
\end{proposition}

\begin{proof}
\begin{enumerate}
  \item $P\approx_{hhp} \tau.P$. It is sufficient to prove the relation $R=\{(P, \tau.P)\}\cup \textbf{Id}$ is a weak hhp-bisimulation. It can be proved similarly to the proof of
  $\tau$ laws for weak hhp-bisimulation in CTC, we omit it;
  \item $\alpha.\tau.P\approx_{hhp} \alpha.P$. It is sufficient to prove the relation $R=\{(\alpha.\tau.P, \alpha.P)\}\cup \textbf{Id}$ is a weak hhp-bisimulation. It can be proved similarly to the proof of
  $\tau$ laws for weak hhp-bisimulation in CTC, we omit it;
  \item $(\alpha_1\parallel\cdots\parallel\alpha_n).\tau.P\approx_{hhp} (\alpha_1\parallel\cdots\parallel\alpha_n).P$. It is sufficient to prove the relation $R=\{((\alpha_1\parallel\cdots\parallel\alpha_n).\tau.P, (\alpha_1\parallel\cdots\parallel\alpha_n).P)\}\cup \textbf{Id}$ is a weak hhp-bisimulation. It can be proved similarly to the proof of
  $\tau$ laws for weak hhp-bisimulation in CTC, we omit it;
  \item $P+\tau.P\approx_{hhp} \tau.P$. It is sufficient to prove the relation $R=\{(P+\tau.P, \tau.P)\}\cup \textbf{Id}$ is a weak hhp-bisimulation. It can be proved similarly to the proof of
  $\tau$ laws for weak hhp-bisimulation in CTC, we omit it;
  \item $\alpha.(P+\tau.Q)+\alpha.Q\approx_{hhp}\alpha.(P+\tau.Q)$. It is sufficient to prove the relation $R=\{(\alpha.(P+\tau.Q)+\alpha.Q, \alpha.(P+\tau.Q))\}\cup \textbf{Id}$ is a weak hhp-bisimulation. It can be proved similarly to the proof of
  $\tau$ laws for weak hhp-bisimulation in CTC, we omit it;
  \item $\phi.(P+\tau.Q)+\phi.Q\approx_{hhp}\phi.(P+\tau.Q)$. It is sufficient to prove the relation $R=\{(\phi.(P+\tau.Q)+\phi.Q, \phi.(P+\tau.Q))\}\cup \textbf{Id}$ is a weak hhp-bisimulation, we omit it;
  \item $(\alpha_1\parallel\cdots\parallel\alpha_n).(P+\tau.Q)+ (\alpha_1\parallel\cdots\parallel\alpha_n).Q\approx_{hhp} (\alpha_1\parallel\cdots\parallel\alpha_n).(P+\tau.Q)$. It is sufficient to prove the relation $R=\{((\alpha_1\parallel\cdots\parallel\alpha_n).(P+\tau.Q)+ (\alpha_1\parallel\cdots\parallel\alpha_n).Q, (\alpha_1\parallel\cdots\parallel\alpha_n).(P+\tau.Q))\}\cup \textbf{Id}$ is a weak hhp-bisimulation. It can be proved similarly to the proof of
  $\tau$ laws for weak hhp-bisimulation in CTC, we omit it;
  \item $P\approx_{hhp} \tau\parallel P$. It is sufficient to prove the relation $R=\{(\tau\parallel P, P)\}\cup \textbf{Id}$ is a weak hhp-bisimulation. It can be proved similarly to the proof of
  $\tau$ laws for weak hhp-bisimulation in CTC, we omit it.
\end{enumerate}
\end{proof}

\subsubsection{Recursion}

\begin{definition}[Sequential]
$X$ is sequential in $E$ if every subexpression of $E$ which contains $X$, apart from $X$ itself, is of the form $\alpha.F$, or $(\alpha_1\parallel\cdots\parallel\alpha_n).F$, or 
$\sum\widetilde{F}$.
\end{definition}

\begin{definition}[Guarded recursive expression]
$X$ is guarded in $E$ if each occurrence of $X$ is with some subexpression $l.F$ or $(l_1\parallel\cdots\parallel l_n).F$ of $E$.
\end{definition}

\begin{lemma}\label{LUSWW3}
Let $G$ be guarded and sequential, $Vars(G)\subseteq\widetilde{X}$, and let $\langle G\{\widetilde{P}/\widetilde{X}\},s\rangle\xrightarrow{\{\alpha_1,\cdots,\alpha_n\}}\langle P',s'\rangle$. 
Then there is an expression $H$ such that $\langle G,s\rangle\xrightarrow{\{\alpha_1,\cdots,\alpha_n\}}\langle H,s'\rangle$, $P'\equiv H\{\widetilde{P}/\widetilde{X}\}$, and for any 
$\widetilde{Q}$, $\langle G\{\widetilde{Q}/\widetilde{X}\},s\rangle\xrightarrow{\{\alpha_1,\cdots,\alpha_n\}} \langle H\{\widetilde{Q}/\widetilde{X}\},s'\rangle$. Moreover $H$ is 
sequential, $Vars(H)\subseteq\widetilde{X}$, and if $\alpha_1=\cdots=\alpha_n=\tau$, then $H$ is also guarded.
\end{lemma}

\begin{proof}
We need to induct on the structure of $G$.

If $G$ is a Constant, a Composition, a Restriction or a Relabeling then it contains no variables, since $G$ is sequential and guarded, then 
$\langle G,s\rangle\xrightarrow{\{\alpha_1,\cdots,\alpha_n\}}\langle P',s'\rangle$, then let $H\equiv P'$, as desired.

$G$ cannot be a variable, since it is guarded.

If $G\equiv G_1+G_2$. Then either $\langle G_1\{\widetilde{P}/\widetilde{X}\},s\rangle \xrightarrow{\{\alpha_1,\cdots,\alpha_n\}}\langle P',s'\rangle$ or 
$\langle G_2\{\widetilde{P}/\widetilde{X}\},s\rangle \xrightarrow{\{\alpha_1,\cdots,\alpha_n\}}\langle P',s'\rangle$, then, we can apply this lemma in either case, as desired.

If $G\equiv\beta.H$. Then we must have $\alpha=\beta$, and $P'\equiv H\{\widetilde{P}/\widetilde{X}\}$, and 
$\langle G\{\widetilde{Q}/\widetilde{X}\},s\rangle\equiv \langle\beta.H\{\widetilde{Q}/\widetilde{X}\},s\rangle \xrightarrow{\beta}\langle H\{\widetilde{Q}/\widetilde{X}\},s'\rangle$, 
then, let $G'$ be $H$, as desired.

If $G\equiv(\beta_1\parallel\cdots\parallel\beta_n).H$. Then we must have $\alpha_i=\beta_i$ for $1\leq i\leq n$, and $P'\equiv H\{\widetilde{P}/\widetilde{X}\}$, and 
$\langle G\{\widetilde{Q}/\widetilde{X}\},s\rangle\equiv \langle(\beta_1\parallel\cdots\parallel\beta_n).H\{\widetilde{Q}/\widetilde{X}\},s\rangle \xrightarrow{\{\beta_1,\cdots,\beta_n\}}\langle H\{\widetilde{Q}/\widetilde{X}\},s'\rangle$, 
then, let $G'$ be $H$, as desired.

If $G\equiv\tau.H$. Then we must have $\tau=\tau$, and $P'\equiv H\{\widetilde{P}/\widetilde{X}\}$, and 
$\langle G\{\widetilde{Q}/\widetilde{X}\},s\rangle\equiv \langle\tau.H\{\widetilde{Q}/\widetilde{X}\},s\rangle \xrightarrow{\tau}\langle H\{\widetilde{Q}/\widetilde{X}\},s'\rangle$, 
then, let $G'$ be $H$, as desired.
\end{proof}

\begin{theorem}[Unique solution of equations for weak pomset bisimulation]
Let the guarded and sequential expressions $\widetilde{E}$ contain free variables $\subseteq \widetilde{X}$, then,

If $\widetilde{P}\approx_p \widetilde{E}\{\widetilde{P}/\widetilde{X}\}$ and $\widetilde{Q}\approx_p \widetilde{E}\{\widetilde{Q}/\widetilde{X}\}$, then 
$\widetilde{P}\approx_p \widetilde{Q}$.
\end{theorem}

\begin{proof}
Like the corresponding theorem in CCS, without loss of generality, we only consider a single equation $X=E$. So we assume $P\approx_p E(P)$, $Q\approx_p E(Q)$, then $P\approx_p Q$.

We will prove $\{(H(P),H(Q)): H\}$ sequential, if $\langle H(P),s\rangle\xrightarrow{\{\alpha_1,\cdots,\alpha_n\}}\langle P',s'\rangle$, then, for some $Q'$, 
$\langle H(Q),s\rangle\xRightarrow{\{\alpha_1.\cdots,\alpha_n\}}\langle Q',s'\rangle$ and $P'\approx_p Q'$.

Let $\langle H(P),s\rangle\xrightarrow{\{\alpha_1,\cdot,\alpha_n\}}\langle P',s'\rangle$, then $\langle H(E(P)),s\rangle\xRightarrow{\{\alpha_1,\cdots,\alpha_n\}}\langle P'',s''\rangle$ 
and $P'\approx_p P''$.

By Lemma \ref{LUSWW3}, we know there is a sequential $H'$ such that $\langle H(E(P)),s\rangle\xRightarrow{\{\alpha_1,\cdots,\alpha_n\}}\langle H'(P),s'\rangle\Rightarrow P''\approx_p P'$.

And, $\langle H(E(Q)),s\rangle\xRightarrow{\{\alpha_1,\cdots,\alpha_n\}}\langle H'(Q),s'\rangle\Rightarrow Q''$ and $P''\approx_p Q''$. And $\langle H(Q),s\rangle\xrightarrow{\{\alpha_1,\cdots,\alpha_n\}}\langle Q',s'\rangle\approx_p \Rightarrow Q'\approx_p Q''$. 
Hence, $P'\approx_p Q'$, as desired.
\end{proof}

\begin{theorem}[Unique solution of equations for weak step bisimulation]
Let the guarded and sequential expressions $\widetilde{E}$ contain free variables $\subseteq \widetilde{X}$, then,

If $\widetilde{P}\approx_s \widetilde{E}\{\widetilde{P}/\widetilde{X}\}$ and $\widetilde{Q}\approx_s \widetilde{E}\{\widetilde{Q}/\widetilde{X}\}$, then
$\widetilde{P}\approx_s \widetilde{Q}$.
\end{theorem}

\begin{proof}
Like the corresponding theorem in CCS, without loss of generality, we only consider a single equation $X=E$. So we assume $P\approx_s E(P)$, $Q\approx_s E(Q)$, then $P\approx_s Q$.

We will prove $\{(H(P),H(Q)): H\}$ sequential, if $\langle H(P),s\rangle\xrightarrow{\{\alpha_1,\cdots,\alpha_n\}}\langle P',s'\rangle$, then, for some $Q'$,
$\langle H(Q),s\rangle\xRightarrow{\{\alpha_1.\cdots,\alpha_n\}}\langle Q',s'\rangle$ and $P'\approx_s Q'$.

Let $\langle H(P),s\rangle\xrightarrow{\{\alpha_1,\cdot,\alpha_n\}}\langle P',s'\rangle$, then $\langle H(E(P)),s\rangle\xRightarrow{\{\alpha_1,\cdots,\alpha_n\}}\langle P'',s''\rangle$
and $P'\approx_s P''$.

By Lemma \ref{LUSWW3}, we know there is a sequential $H'$ such that $\langle H(E(P)),s\rangle\xRightarrow{\{\alpha_1,\cdots,\alpha_n\}}\langle H'(P),s'\rangle\Rightarrow P''\approx_s P'$.

And, $\langle H(E(Q)),s\rangle\xRightarrow{\{\alpha_1,\cdots,\alpha_n\}}\langle H'(Q),s'\rangle\Rightarrow Q''$ and $P''\approx_s Q''$. And $\langle H(Q),s\rangle\xrightarrow{\{\alpha_1,\cdots,\alpha_n\}}\langle Q',s'\rangle\approx_s \Rightarrow Q'\approx_s Q''$.
Hence, $P'\approx_s Q'$, as desired.
\end{proof}

\begin{theorem}[Unique solution of equations for weak hp-bisimulation]
Let the guarded and sequential expressions $\widetilde{E}$ contain free variables $\subseteq \widetilde{X}$, then,

If $\widetilde{P}\approx_{hp} \widetilde{E}\{\widetilde{P}/\widetilde{X}\}$ and $\widetilde{Q}\approx_{hp} \widetilde{E}\{\widetilde{Q}/\widetilde{X}\}$, then
$\widetilde{P}\approx_{hp} \widetilde{Q}$.
\end{theorem}

\begin{proof}
Like the corresponding theorem in CCS, without loss of generality, we only consider a single equation $X=E$. So we assume $P\approx_{hp} E(P)$, $Q\approx_{hp} E(Q)$, then $P\approx_{hp} Q$.

We will prove $\{(H(P),H(Q)): H\}$ sequential, if $\langle H(P),s\rangle\xrightarrow{\{\alpha_1,\cdots,\alpha_n\}}\langle P',s'\rangle$, then, for some $Q'$,
$\langle H(Q),s\rangle\xRightarrow{\{\alpha_1.\cdots,\alpha_n\}}\langle Q',s'\rangle$ and $P'\approx_{hp} Q'$.

Let $\langle H(P),s\rangle\xrightarrow{\{\alpha_1,\cdot,\alpha_n\}}\langle P',s'\rangle$, then $\langle H(E(P)),s\rangle\xRightarrow{\{\alpha_1,\cdots,\alpha_n\}}\langle P'',s''\rangle$
and $P'\approx_{hp} P''$.

By Lemma \ref{LUSWW3}, we know there is a sequential $H'$ such that $\langle H(E(P)),s\rangle\xRightarrow{\{\alpha_1,\cdots,\alpha_n\}}\langle H'(P),s'\rangle\Rightarrow P''\approx_{hp} P'$.

And, $\langle H(E(Q)),s\rangle\xRightarrow{\{\alpha_1,\cdots,\alpha_n\}}\langle H'(Q),s'\rangle\Rightarrow Q''$ and $P''\approx_{hp} Q''$. And $\langle H(Q),s\rangle\xrightarrow{\{\alpha_1,\cdots,\alpha_n\}}\langle Q',s'\rangle\approx_{hp} \Rightarrow Q'\approx_{hp} Q''$.
Hence, $P'\approx_{hp} Q'$, as desired.
\end{proof}

\begin{theorem}[Unique solution of equations for weak hhp-bisimulation]
Let the guarded and sequential expressions $\widetilde{E}$ contain free variables $\subseteq \widetilde{X}$, then,

If $\widetilde{P}\approx_{hhp} \widetilde{E}\{\widetilde{P}/\widetilde{X}\}$ and $\widetilde{Q}\approx_{hhp} \widetilde{E}\{\widetilde{Q}/\widetilde{X}\}$, then
$\widetilde{P}\approx_{hhp} \widetilde{Q}$.
\end{theorem}

\begin{proof}
Like the corresponding theorem in CCS, without loss of generality, we only consider a single equation $X=E$. So we assume $P\approx_{hhp} E(P)$, $Q\approx_{hhp} E(Q)$, then $P\approx_{hhp} Q$.

We will prove $\{(H(P),H(Q)): H\}$ sequential, if $\langle H(P),s\rangle\xrightarrow{\{\alpha_1,\cdots,\alpha_n\}}\langle P',s'\rangle$, then, for some $Q'$,
$\langle H(Q),s\rangle\xRightarrow{\{\alpha_1.\cdots,\alpha_n\}}\langle Q',s'\rangle$ and $P'\approx_{hhp} Q'$.

Let $\langle H(P),s\rangle\xrightarrow{\{\alpha_1,\cdot,\alpha_n\}}\langle P',s'\rangle$, then $\langle H(E(P)),s\rangle\xRightarrow{\{\alpha_1,\cdots,\alpha_n\}}\langle P'',s''\rangle$
and $P'\approx_{hhp} P''$.

By Lemma \ref{LUSWW3}, we know there is a sequential $H'$ such that $\langle H(E(P)),s\rangle\xRightarrow{\{\alpha_1,\cdots,\alpha_n\}}\langle H'(P),s'\rangle\Rightarrow P''\approx_{hhp} P'$.

And, $\langle H(E(Q)),s\rangle\xRightarrow{\{\alpha_1,\cdots,\alpha_n\}}\langle H'(Q),s'\rangle\Rightarrow Q''$ and $P''\approx_{hhp} Q''$. And $\langle H(Q),s\rangle\xrightarrow{\{\alpha_1,\cdots,\alpha_n\}}\langle Q',s'\rangle\approx_{hhp} \Rightarrow Q'\approx_{hhp} Q''$.
Hence, $P'\approx_{hhp} Q'$, as desired.
\end{proof}

\newpage\section{CTC with Probabilism and Reversibility}\label{ctcpr}

In this chapter, we design the calculus CTC with probabilism and reversibility. This chapter is organized as follows. We introduce the operational semantics in section \ref{osctcpr}, its syntax and operational
semantics in section \ref{sosctcpr}, and its properties for strong bisimulations in section \ref{stcbctcpr}, its properties for weak bisimulations in section \ref{wtcbctcpr}.

\subsection{Operational Semantics}\label{osctcpr}

\begin{definition}[Probabilistic transitions]
Let $\mathcal{E}$ be a PES and let $C\in\mathcal{C}(\mathcal{E})$, the transition $C\xrsquigarrow{\pi} C^{\pi}$ is called a probabilistic
transition
from $C$ to $C^{\pi}$.
\end{definition}

\begin{definition}[FR probabilistic pomset, step bisimulation]\label{PSBG}
Let $\mathcal{E}_1$, $\mathcal{E}_2$ be PESs. A FR probabilistic pomset bisimulation is a relation $R\subseteq\mathcal{C}(\mathcal{E}_1)\times\mathcal{C}(\mathcal{E}_2)$,
such that (1) if $(C_1,C_2)\in R$, and $ C_1\xrightarrow{X_1} C_1'$ then
$ C_2\xrightarrow{X_2} C_2'$, with $X_1\subseteq \mathbb{E}_1$, $X_2\subseteq \mathbb{E}_2$, $X_1\sim X_2$ and
$( C_1', C_2')\in R$, and vice-versa;
(2) if $( C_1, C_2)\in R$, and $ C_1\xtworightarrow{X_1[\mathcal{K}_1]} C_1'$ then
$ C_2\xrightarrow{X_2[\mathcal{K}_2]} C_2'$, with $X_1\subseteq \mathbb{E}_1$, $X_2\subseteq \mathbb{E}_2$, $X_1\sim X_2$ and
$( C_1', C_2')\in R$, and vice-versa;
(3) if $( C_1, C_2)\in R$, and $ C_1\xrsquigarrow{\pi} C_1^{\pi}$
then $ C_2\xrsquigarrow{\pi} C_2^{\pi}$ and $( C_1^{\pi}, C_2^{\pi})\in R$, and vice-versa; (4) if $( C_1, C_2)\in R$,
then $\mu(C_1,C)=\mu(C_2,C)$ for each $C\in\mathcal{C}(\mathcal{E})/R$; (5) $[\surd]_R=\{\surd\}$. We say that $\mathcal{E}_1$, $\mathcal{E}_2$ are FR probabilistic pomset bisimilar, written
$\mathcal{E}_1\sim_{pp}^{fr}\mathcal{E}_2$, if there exists a probabilistic pomset bisimulation $R$, such that $(\emptyset,\emptyset\rangle,\emptyset,\emptyset\rangle)\in R$.
By replacing FR probabilistic pomset transitions with FR probabilistic steps, we can get the definition of FR probabilistic step bisimulation. When PESs $\mathcal{E}_1$ and $\mathcal{E}_2$ are FR
probabilistic step bisimilar, we write $\mathcal{E}_1\sim_{ps}^{fr}\mathcal{E}_2$.
\end{definition}

\begin{definition}[FR weakly probabilistic pomset, step bisimulation]
Let $\mathcal{E}_1$, $\mathcal{E}_2$ be PESs. A FR weakly probabilistic pomset bisimulation is a relation $R\subseteq\mathcal{C}(\mathcal{E}_1)\times\mathcal{C}(\mathcal{E}_2)$,
such that (1) if $( C_1, C_2)\in R$, and $ C_1\xRightarrow{X_1} C_1'$ then
$ C_2\xRightarrow{X_2} C_2'$, with $X_1\subseteq \hat{\mathbb{E}_1}$, $X_2\subseteq \hat{\mathbb{E}_2}$, $X_1\sim X_2$ and
$( C_1', C_2')\in R$, and vice-versa;
(2) if $( C_1, C_2)\in R$, and $ C_1\xTworightarrow{X_1[\mathcal{K}_1]} C_1'$ then
$ C_2\xTworightarrow{X_2[\mathcal{K}_2]} C_2'$, with $X_1\subseteq \hat{\mathbb{E}_1}$, $X_2\subseteq \hat{\mathbb{E}_2}$, $X_1\sim X_2$ and
$( C_1', C_2')\in R$, and vice-versa;
(3) if $( C_1, C_2)\in R$, and $ C_1\xrsquigarrow{\pi} C_1^{\pi}$
then $ C_2\xrsquigarrow{\pi} C_2^{\pi}$ and $( C_1^{\pi}, C_2^{\pi})\in R$, and vice-versa; (4) if $( C_1, C_2)\in R$,
then $\mu(C_1,C)=\mu(C_2,C)$ for each $C\in\mathcal{C}(\mathcal{E})/R$; (5) $[\surd]_R=\{\surd\}$. We say that $\mathcal{E}_1$, $\mathcal{E}_2$ are FR weakly probabilistic pomset bisimilar,
written $\mathcal{E}_1\approx_{pp}^{fr}\mathcal{E}_2$, if there exists a FR weakly probabilistic pomset bisimulation $R$, such that
$(\emptyset,\emptyset\rangle,\emptyset,\emptyset\rangle)\in R$. By replacing FR weakly probabilistic pomset transitions with FR weakly probabilistic steps, we can get the
definition of FR weakly probabilistic step bisimulation. When PESs $\mathcal{E}_1$ and $\mathcal{E}_2$ are FR weakly probabilistic step bisimilar, we write
$\mathcal{E}_1\approx_{ps}^{FR}\mathcal{E}_2$.
\end{definition}

\begin{definition}[Posetal product]
Given two PESs $\mathcal{E}_1$, $\mathcal{E}_2$, the posetal product of their configurations, denoted
$\mathcal{C}(\mathcal{E}_1)\overline{\times}\mathcal{C}(\mathcal{E}_2)$, is defined as

$$\{( C_1,f, C_2)|C_1\in\mathcal{C}(\mathcal{E}_1),C_2\in\mathcal{C}(\mathcal{E}_2),f:C_1\rightarrow C_2 \textrm{ isomorphism}\}.$$

A subset $R\subseteq\mathcal{C}(\mathcal{E}_1)\overline{\times}\mathcal{C}(\mathcal{E}_2)$ is called a posetal relation. We say that $R$ is downward
closed when for any $( C_1,f, C_2),( C_1',f', C_2')\in \mathcal{C}(\mathcal{E}_1)\overline{\times}\mathcal{C}(\mathcal{E}_2)$,
if $( C_1,f, C_2)\subseteq ( C_1',f', C_2')$ pointwise and
$( C_1',f', C_2')\in R$, then $( C_1,f, C_2)\in R$.

For $f:X_1\rightarrow X_2$, we define $f[x_1\mapsto x_2]:X_1\cup\{x_1\}\rightarrow X_2\cup\{x_2\}$, $z\in X_1\cup\{x_1\}$,(1)$f[x_1\mapsto x_2](z)=
x_2$,if $z=x_1$;(2)$f[x_1\mapsto x_2](z)=f(z)$, otherwise. Where $X_1\subseteq \mathbb{E}_1$, $X_2\subseteq \mathbb{E}_2$, $x_1\in \mathbb{E}_1$, $x_2\in \mathbb{E}_2$.
\end{definition}

\begin{definition}[Weakly posetal product]
Given two PESs $\mathcal{E}_1$, $\mathcal{E}_2$, the weakly posetal product of their configurations, denoted
$\mathcal{C}(\mathcal{E}_1)\overline{\times}\mathcal{C}(\mathcal{E}_2)$, is defined as

$$\{( C_1,f, C_2)|C_1\in\mathcal{C}(\mathcal{E}_1),C_2\in\mathcal{C}(\mathcal{E}_2),f:\hat{C_1}\rightarrow \hat{C_2} \textrm{ isomorphism}\}.$$

A subset $R\subseteq\mathcal{C}(\mathcal{E}_1)\overline{\times}\mathcal{C}(\mathcal{E}_2)$ is called a weakly posetal relation. We say that $R$ is
downward closed when for any $( C_1,f, C_2),( C_1',f, C_2')\in \mathcal{C}(\mathcal{E}_1)\overline{\times}\mathcal{C}(\mathcal{E}_2)$,
if $( C_1,f, C_2)\subseteq ( C_1',f', C_2')$ pointwise and
$( C_1',f', C_2')\in R$, then $( C_1,f, C_2)\in R$.

For $f:X_1\rightarrow X_2$, we define $f[x_1\mapsto x_2]:X_1\cup\{x_1\}\rightarrow X_2\cup\{x_2\}$, $z\in X_1\cup\{x_1\}$,(1)$f[x_1\mapsto x_2](z)=
x_2$,if $z=x_1$;(2)$f[x_1\mapsto x_2](z)=f(z)$, otherwise. Where $X_1\subseteq \hat{\mathbb{E}_1}$, $X_2\subseteq \hat{\mathbb{E}_2}$, $x_1\in \hat{\mathbb{E}}_1$,
$x_2\in \hat{\mathbb{E}}_2$. Also, we define $f(\tau^*)=f(\tau^*)$.
\end{definition}

\begin{definition}[FR probabilistic (hereditary) history-preserving bisimulation]
A FR probabilistic history-preserving (hp-) bisimulation is a posetal relation
$R\subseteq\mathcal{C}(\mathcal{E}_1)\overline{\times}\mathcal{C}(\mathcal{E}_2)$ such that (1) if $( C_1,f, C_2)\in R$,
and $ C_1\xrightarrow{e_1}  C_1'$, then $ C_2\xrightarrow{e_2}  C_2'$, with
$( C_1',f[e_1\mapsto e_2], C_2')\in R$, and vice-versa;
(2) if $( C_1,f, C_2)\in R$,
and $ C_1\xtworightarrow{e_1[m]}  C_1'$, then $ C_2\xtworightarrow{e_2[n]}  C_2'$, with
$( C_1',f[e_1[m]\mapsto e_2[n]], C_2')\in R$, and vice-versa;
(3) if $( C_1,f, C_2)\in R$, and
$ C_1\xrsquigarrow{\pi} C_1^{\pi}$ then $ C_2\xrsquigarrow{\pi} C_2^{\pi}$ and $( C_1^{\pi},f, C_2^{\pi})\in R$,
and vice-versa; (4) if $(C_1,f,C_2)\in R$, then $\mu(C_1,C)=\mu(C_2,C)$ for each $C\in\mathcal{C}(\mathcal{E})/R$; (5) $[\surd]_R=\{\surd\}$. $\mathcal{E}_1,\mathcal{E}_2$ are
probabilistic history-preserving (hp-)bisimilar and are written $\mathcal{E}_1\sim_{php}\mathcal{E}_2$ if there exists a probabilistic hp-bisimulation $R$ such that
$(\emptyset,\emptyset\rangle,\emptyset,\emptyset,\emptyset\rangle)\in R$.

A FR probabilistic hereditary history-preserving (hhp-)bisimulation is a downward closed FR probabilistic hp-bisimulation. $\mathcal{E}_1,\mathcal{E}_2$ are FR probabilistic hereditary
history-preserving (hhp-)bisimilar and are written $\mathcal{E}_1\sim_{phhp}^{fr}\mathcal{E}_2$.
\end{definition}

\begin{definition}[FR weakly probabilistic (hereditary) history-preserving bisimulation]
A FR weakly probabilistic history-preserving (hp-) bisimulation is a weakly posetal relation
$R\subseteq\mathcal{C}(\mathcal{E}_1)\overline{\times}\mathcal{C}(\mathcal{E}_2)$ such that (1) if $( C_1,f, C_2)\in R$,
and $ C_1\xRightarrow{e_1}  C_1'$, then $ C_2\xRightarrow{e_2}  C_2'$, with
$( C_1',f[e_1\mapsto e_2], C_2')\in R$, and vice-versa;
(2) if $( C_1,f, C_2)\in R$,
and $ C_1\xTworightarrow{e_1[m]}  C_1'$, then $ C_2\xTworightarrow{e_2[n]}  C_2'$, with
$( C_1',f[e_1[m]\mapsto e_2[n]], C_2')\in R$, and vice-versa;
(3) if $( C_1,f, C_2)\in R$, and
$ C_1\xrsquigarrow{\pi} C_1^{\pi}$ then $ C_2\xrsquigarrow{\pi} C_2^{\pi}$ and
$( C_1^{\pi},f, C_2^{\pi})\in R$, and vice-versa; (4) if $(C_1,f,C_2)\in R$, then $\mu(C_1,C)=\mu(C_2,C)$ for each $C\in\mathcal{C}(\mathcal{E})/R$;
(5) $[\surd]_R=\{\surd\}$. $\mathcal{E}_1,\mathcal{E}_2$ are FR weakly probabilistic history-preserving (hp-)bisimilar and are written $\mathcal{E}_1\approx_{php}^{fr}\mathcal{E}_2$ if there
exists a FR weakly probabilistic hp-bisimulation $R$ such that $(\emptyset,\emptyset\rangle,\emptyset,\emptyset,\emptyset\rangle)\in R$.

A FR weakly probabilistic hereditary history-preserving (hhp-)bisimulation is a downward closed FR weakly probabilistic hp-bisimulation. $\mathcal{E}_1,\mathcal{E}_2$ are FR weakly
probabilistic hereditary history-preserving (hhp-)bisimilar and are written $\mathcal{E}_1\approx_{phhp}^{fr}\mathcal{E}_2$.
\end{definition}

\subsection{Syntax and Operational Semantics}\label{sosctcpr}

We assume an infinite set $\mathcal{N}$ of (action or event) names, and use $a,b,c,\cdots$ to range over $\mathcal{N}$. We denote by $\overline{\mathcal{N}}$ the set of co-names and
let $\overline{a},\overline{b},\overline{c},\cdots$ range over $\overline{\mathcal{N}}$. Then we set $\mathcal{L}=\mathcal{N}\cup\overline{\mathcal{N}}$ as the set of labels, and use
$l,\overline{l}$ to range over $\mathcal{L}$. We extend complementation to $\mathcal{L}$ such that $\overline{\overline{a}}=a$. Let $\tau$ denote the silent step (internal action or
event) and define $Act=\mathcal{L}\cup\{\tau\}\cup\mathcal{L}[\mathcal{K}]$ to be the set of actions, $\alpha,\beta$ range over $Act$. And $K,L$ are used to stand for subsets of
$\mathcal{L}$ and $\overline{L}$ is used for the set of complements of labels in $L$. A relabelling function $f$ is a function from $\mathcal{L}$ to $\mathcal{L}$ such that
$f(\overline{l})=\overline{f(l)}$. By defining $f(\tau)=\tau$, we extend $f$ to $Act$. We write $\mathcal{P}$ for the set of processes. Sometimes, we use $I,J$ to stand for an indexing
set, and we write $E_i:i\in I$ for a family of expressions indexed by $I$. $Id_D$ is the identity function or relation over set $D$.

For each process constant schema $A$, a defining equation of the form

$$A\overset{\text{def}}{=}P$$

is assumed, where $P$ is a process.

\subsubsection{Syntax}

We use the Prefix $.$ to model the causality relation $\leq$ in true concurrency, the Summation $+$ to model the conflict relation $\sharp$ in true concurrency, and the Composition
$\parallel$ to explicitly model concurrent relation in true concurrency. And we follow the conventions of process algebra.

\begin{definition}[Syntax]\label{syntax04}
Reversible truly concurrent processes CTC with probabilism and reversibility are defined inductively by the following formation rules:

\begin{enumerate}
  \item $A\in\mathcal{P}$;
  \item $\textbf{nil}\in\mathcal{P}$;
  \item if $P\in\mathcal{P}$, then the Prefix $\alpha.P\in\mathcal{P}$ and $P.\alpha[m]\in\mathcal{P}$, for $\alpha\in Act$ and $m\in\mathcal{K}$;
  \item if $P,Q\in\mathcal{P}$, then the Summation $P+Q\in\mathcal{P}$;
  \item if $P,Q\in\mathcal{P}$, then the Box-Summation $P\boxplus_{\pi}Q\in\mathcal{P}$;
  \item if $P,Q\in\mathcal{P}$, then the Composition $P\parallel Q\in\mathcal{P}$;
  \item if $P\in\mathcal{P}$, then the Prefix $(\alpha_1\parallel\cdots\parallel\alpha_n).P\in\mathcal{P}\quad(n\in I)$ and $P.(\alpha_1[m]\parallel\cdots\parallel\alpha_n[m])\in\mathcal{P}\quad(n\in I)$, for $\alpha_,\cdots,\alpha_n\in Act$ and $m\in\mathcal{K}$;
  \item if $P\in\mathcal{P}$, then the Restriction $P\setminus L\in\mathcal{P}$ with $L\in\mathcal{L}$;
  \item if $P\in\mathcal{P}$, then the Relabelling $P[f]\in\mathcal{P}$.
\end{enumerate}

The standard BNF grammar of syntax of CTC with probabilism and reversibility can be summarized as follows:

$P::=A|\textbf{nil}|\alpha.P| P.\alpha[m]| P+P |P\boxplus_{\pi} P| P\parallel P | (\alpha_1\parallel\cdots\parallel\alpha_n).P|  P.(\alpha_1[m]\parallel\cdots\parallel\alpha_n[m])  | P\setminus L | P[f].$
\end{definition}

\subsubsection{Operational Semantics}

The operational semantics is defined by LTSs (labelled transition systems), and it is detailed by the following definition.

\begin{definition}[Semantics]\label{semantics04}
The operational semantics of CTC with probabilism and reversibility corresponding to the syntax in Definition \ref{syntax04} is defined by a series of transition rules, they are shown in Table \ref{FTRForPS04},
\ref{RTRForPS04}, \ref{FTRForCom04}, \ref{RTRForCom04}, \ref{FTRForRRC04} and \ref{RTRForRRC04}. And the predicate
$\xrightarrow{\alpha}\alpha[m]$ represents successful forward termination after execution of the action $\alpha$, the predicate $\xtworightarrow{\alpha[m]}\alpha$ represents successful
reverse termination after execution of the event $\alpha[m]$, the the predicate $\textrm{Std(P)}$ represents that $p$ is a standard process containing no past events, the the predicate
$\textrm{NStd(P)}$ represents that $P$ is a process full of past events.
\end{definition}

\begin{center}
    \begin{table}
        $$\frac{}{\alpha.P\rightsquigarrow\breve{\alpha}.P}$$
        $$\frac{P\rightsquigarrow P'\quad Q\rightsquigarrow Q'}{P+Q\rightsquigarrow P'+Q'}$$
        $$\frac{P\rightsquigarrow P'}{P\boxplus_{\pi}Q\rightsquigarrow P'}\quad \frac{Q\rightsquigarrow Q'}{P\boxplus_{\pi}Q\rightsquigarrow Q'}$$
        $$\frac{P\rightsquigarrow P'\quad Q\rightsquigarrow Q'}{P\parallel Q\rightsquigarrow P'+Q'}$$
        $$\frac{}{(\alpha_1\parallel\cdots\parallel\alpha_n).P\rightsquigarrow(\breve{\alpha_1}\parallel\cdots\parallel\breve{\alpha_n}).P}$$
        $$\frac{P\rightsquigarrow P'}{P\setminus L\rightsquigarrow P'\setminus L}$$
        $$\frac{P\rightsquigarrow P'}{P[f]\rightsquigarrow P'[f]}$$
        $$\frac{P\rightsquigarrow P'}{A\rightsquigarrow P'}\quad (A\overset{\text{def}}{=}P)$$
        \caption{Probabilistic transition rules}
        \label{PTRForCTC04}
    \end{table}
\end{center}

The forward transition rules for Prefix and Summation are shown in Table \ref{FTRForPS04}.

\begin{center}
    \begin{table}
        $$\frac{}{\breve{\alpha}\xrightarrow{\alpha}\alpha[m]}$$

        $$\frac{ P\xrightarrow{\alpha}\alpha[m]}{ P+Q\xrightarrow{\alpha}\alpha[m]}
        \quad\frac{ P\xrightarrow{\alpha} P'}{ P+Q\xrightarrow{\alpha} P'}$$
        $$\frac{ Q\xrightarrow{\alpha} \alpha[m]}{ P+Q\xrightarrow{\alpha}\alpha[m]}
        \quad\frac{ Q\xrightarrow{\alpha} Q'}{ P+Q\xrightarrow{\alpha} Q'}$$

        $$\frac{ P\xrightarrow{\alpha} \alpha[m]\quad\textrm{Std}(Q)}{ P. Q\xrightarrow{\alpha} \alpha[m]. Q}
        \quad\frac{ P\xrightarrow{\alpha} P' \quad \textrm{Std}(Q)}{ P. Q\xrightarrow{\alpha} P'. Q}$$
        $$\frac{ Q\xrightarrow{\beta}\beta[n]\quad \textrm{NStd}(P)}{ P. Q\xrightarrow{\beta} P. \beta[n]}
        \quad\frac{ Q\xrightarrow{\beta} Q'\quad \textrm{NStd}(P)}{ P. Q\xrightarrow{\beta} P. Q'}$$
        \caption{Forward transition rules of Prefix and Summation}
        \label{FTRForPS04}
    \end{table}
\end{center}

The reverse transition rules for Prefix and Summation are shown in Table \ref{RTRForPS04}.

\begin{center}
    \begin{table}
        $$\frac{}{ \breve{\alpha[m]}\xtworightarrow{\alpha[m]}\alpha}$$

        $$\frac{ P\xtworightarrow{\alpha[m]} \alpha}{ P+Q\xtworightarrow{\alpha[m]}\alpha}
        \quad\frac{ P\xtworightarrow{\alpha[m]} P'}{ P+Q\xtworightarrow{\alpha[m]} P'}$$
        $$\frac{ Q\xtworightarrow{\alpha[m]}\alpha}{ P+Q\xtworightarrow{\alpha[m]}\alpha}
        \quad\frac{ Q\xtworightarrow{\alpha[m]} Q'}{ P+Q\xtworightarrow{\alpha[m]} Q'}$$

        $$\frac{ P\xtworightarrow{\alpha[m]}\alpha \quad \textrm{Std}(Q)}{ P. Q\xtworightarrow{\alpha[m]} \alpha. Q}
        \quad\frac{ P\xtworightarrow{\alpha[m]} P'\quad \textrm{Std}(Q)}{ P. Q\xtworightarrow{\alpha[m]} P'. Q}$$
        $$\frac{ Q\xtworightarrow{\beta[n]}\beta \quad \textrm{NStd}(P)}{ P. Q\xtworightarrow{\beta[n]} P. \beta}\quad
        \frac{ Q\xtworightarrow{\beta[n]} Q' \quad \textrm{NStd}(P)}{ P. Q\xtworightarrow{\beta[n]} P. Q'}$$
        \caption{Reverse transition rules of Prefix and Summation}
        \label{RTRForPS04}
    \end{table}
\end{center}

The forward transition rules for Composition are shown in Table \ref{FTRForCom04}.

\begin{center}
    \begin{table}
        $$\frac{ P\xrightarrow{\alpha} P'\quad  Q\nrightarrow}{ P\parallel Q\xrightarrow{\alpha} P'\parallel Q}$$
        $$\frac{ Q\xrightarrow{\alpha} Q'\quad  P\nrightarrow}{ P\parallel Q\xrightarrow{\alpha} P\parallel Q'}$$
        $$\frac{ P\xrightarrow{\alpha} P'\quad  Q\xrightarrow{\beta} Q'}{ P\parallel Q\xrightarrow{\{\alpha,\beta\}} P'\parallel Q'}\quad (\beta\neq\overline{\alpha})$$
        $$\frac{ P\xrightarrow{l} P'\quad  Q\xrightarrow{\overline{l}} Q'}{ P\parallel Q\xrightarrow{\tau} P'\parallel Q'}$$
        \caption{Forward transition rules of Composition}
        \label{FTRForCom04}
    \end{table}
\end{center}

The reverse transition rules for Composition are shown in Table \ref{RTRForCom04}.

\begin{center}
    \begin{table}
        $$\frac{ P\xtworightarrow{\alpha[m]} P'\quad  Q\xntworightarrow{}}{ P\parallel Q\xtworightarrow{\alpha[m]} P'\parallel Q}$$
        $$\frac{ Q\xtworightarrow{\alpha[m]} Q'\quad  P\xntworightarrow{}}{ P\parallel Q\xtworightarrow{\alpha[m]} P\parallel Q'}$$
        $$\frac{ P\xtworightarrow{\alpha[m]} P'\quad  Q\xtworightarrow{\beta[m]} Q'}{ P\parallel Q\xtworightarrow{\{\alpha[m],\beta[m]\}} P'\parallel Q'}\quad (\beta\neq\overline{\alpha})$$
        $$\frac{ P\xtworightarrow{l[m]} P'\quad  Q\xtworightarrow{\overline{l}[m]} Q'}{ P\parallel Q\xtworightarrow{\tau} P'\parallel Q'}$$
        \caption{Reverse transition rules of Composition}
        \label{RTRForCom04}
    \end{table}
\end{center}

The forward transition rules for Restriction, Relabelling and Constants are shown in Table \ref{FTRForRRC04}.

\begin{center}
    \begin{table}
        $$\frac{ P\xrightarrow{\alpha} P'}{ P\setminus L\xrightarrow{\alpha} P'\setminus L}\quad (\alpha,\overline{\alpha}\notin L)$$
        $$\frac{ P\xrightarrow{\{\alpha_1,\cdots,\alpha_n\}} P'}{ P\setminus L\xrightarrow{\{\alpha_1,\cdots,\alpha_n\}} P'\setminus L}\quad (\alpha_1,\overline{\alpha_1},\cdots,\alpha_n,\overline{\alpha_n}\notin L)$$
        $$\frac{ P\xrightarrow{\alpha} P'}{ P[f]\xrightarrow{f(\alpha)} P'[f]}$$
        $$\frac{ P\xrightarrow{\{\alpha_1,\cdots,\alpha_n\}} P'}{ P[f]\xrightarrow{\{f(\alpha_1),\cdots,f(\alpha_n)\}} P'[f]}$$
        $$\frac{ P\xrightarrow{\alpha} P'}{ A\xrightarrow{\alpha} P'}\quad (A\overset{\text{def}}{=}P)$$
        $$\frac{ P\xrightarrow{\{\alpha_1,\cdots,\alpha_n\}} P'}{ A\xrightarrow{\{\alpha_1,\cdots,\alpha_n\}} P'}\quad (A\overset{\text{def}}{=}P)$$
        \caption{Forward transition rules of Restriction, Relabelling and Constants}
        \label{FTRForRRC04}
    \end{table}
\end{center}

The reverse transition rules for Restriction, Relabelling and Constants are shown in Table \ref{RTRForRRC04}.

\begin{center}
    \begin{table}
        $$\frac{ P\xtworightarrow{\alpha[m]} P'}{ P\setminus L\xtworightarrow{\alpha[m]} P'\setminus L}\quad (\alpha,\overline{\alpha}\notin L)$$
        $$\frac{ P\xtworightarrow{\{\alpha_1[m],\cdots,\alpha_n[m]\}} P'}{ P\setminus L\xtworightarrow{\{\alpha_1[m],\cdots,\alpha_n[m]\}} P'\setminus L}\quad (\alpha_1,\overline{\alpha_1},\cdots,\alpha_n,\overline{\alpha_n}\notin L)$$
        $$\frac{ P\xtworightarrow{\alpha[m]} P'}{ P[f]\xtworightarrow{f(\alpha[m])} P'[f]}$$
        $$\frac{ P\xtworightarrow{\{\alpha_1[m],\cdots,\alpha_n[m]\}} P'}{ P[f]\xtworightarrow{\{f(\alpha_1)[m],\cdots,f(\alpha_n)[m]\}} P'[f]}$$
        $$\frac{ P\xtworightarrow{\alpha[m]} P'}{ A\xtworightarrow{\alpha[m]} P'}\quad (A\overset{\text{def}}{=}P)$$
        $$\frac{ P\xtworightarrow{\{\alpha_1[m],\cdots,\alpha_n[m]\}} P'}{ A\xtworightarrow{\{\alpha_1[m],\cdots,\alpha_n[m]\}} P'}\quad (A\overset{\text{def}}{=}P)$$
        \caption{Reverse transition rules of Restriction, Relabelling and Constants}
        \label{RTRForRRC04}
    \end{table}
\end{center}

\subsubsection{Properties of Transitions}

\begin{definition}[Sorts]\label{sorts04}
Given the sorts $\mathcal{L}(A)$ and $\mathcal{L}(X)$ of constants and variables, we define $\mathcal{L}(P)$ inductively as follows.

\begin{enumerate}
  \item $\mathcal{L}(l.P)=\{l\}\cup\mathcal{L}(P)$;
  \item $\mathcal{L}(P.l[m])=\{l\}\cup\mathcal{L}(P)$;
  \item $\mathcal{L}((l_1\parallel \cdots\parallel l_n).P)=\{l_1,\cdots,l_n\}\cup\mathcal{L}(P)$;
  \item $\mathcal{L}(P.(l_1[m]\parallel \cdots\parallel l_n[m]))=\{l_1,\cdots,l_n\}\cup\mathcal{L}(P)$;
  \item $\mathcal{L}(\tau.P)=\mathcal{L}(P)$;
  \item $\mathcal{L}(P+Q)=\mathcal{L}(P)\cup\mathcal{L}(Q)$;
  \item $\mathcal{L}(P\boxplus_{\pi}Q)=\mathcal{L}(P)\cup\mathcal{L}(Q)$;
  \item $\mathcal{L}(P\parallel Q)=\mathcal{L}(P)\cup\mathcal{L}(Q)$;
  \item $\mathcal{L}(P\setminus L)=\mathcal{L}(P)-(L\cup\overline{L})$;
  \item $\mathcal{L}(P[f])=\{f(l):l\in\mathcal{L}(P)\}$;
  \item for $A\overset{\text{def}}{=}P$, $\mathcal{L}(P)\subseteq\mathcal{L}(A)$.
\end{enumerate}
\end{definition}

Now, we present some properties of the transition rules defined in Definition \ref{semantics04}.

\begin{proposition}
If $P\xrightarrow{\alpha}P'$, then
\begin{enumerate}
  \item $\alpha\in\mathcal{L}(P)\cup\{\tau\}$;
  \item $\mathcal{L}(P')\subseteq\mathcal{L}(P)$.
\end{enumerate}

If $P\xrightarrow{\{\alpha_1,\cdots,\alpha_n\}}P'$, then
\begin{enumerate}
  \item $\alpha_1,\cdots,\alpha_n\in\mathcal{L}(P)\cup\{\tau\}$;
  \item $\mathcal{L}(P')\subseteq\mathcal{L}(P)$.
\end{enumerate}
\end{proposition}

\begin{proof}
By induction on the inference of $P\xrightarrow{\alpha}P'$ and $P\xrightarrow{\{\alpha_1,\cdots,\alpha_n\}}P'$, there are several cases corresponding to the forward transition rules in
Definition \ref{semantics04}, we omit them.
\end{proof}

\begin{proposition}
If $P\xtworightarrow{\alpha[m]}P'$, then
\begin{enumerate}
  \item $\alpha\in\mathcal{L}(P)\cup\{\tau\}$;
  \item $\mathcal{L}(P')\subseteq\mathcal{L}(P)$.
\end{enumerate}

If $P\xtworightarrow{\{\alpha_1[m],\cdots,\alpha_n[m]\}}P'$, then
\begin{enumerate}
  \item $\alpha_1,\cdots,\alpha_n\in\mathcal{L}(P)\cup\{\tau\}$;
  \item $\mathcal{L}(P')\subseteq\mathcal{L}(P)$.
\end{enumerate}
\end{proposition}

\begin{proof}
By induction on the inference of $P\xtworightarrow{\alpha}P'$ and $P\xtworightarrow{\{\alpha_1,\cdots,\alpha_n\}}P'$, there are several cases corresponding to the forward transition
rules in Definition \ref{semantics04}, we omit them.
\end{proof}

\subsection{Strong Bisimulations}\label{stcbctcpr}

\subsubsection{Laws and Congruence}

Based on the concepts of strongly FR truly concurrent bisimulation equivalences, we get the following laws.

\begin{proposition}[Monoid laws for FR strongly probabilistic pomset bisimulation] The monoid laws for FR strongly probabilistic pomset bisimulation are as follows.

\begin{enumerate}
  \item $P+Q\sim_{pp}^{fr} Q+P$;
  \item $P+(Q+R)\sim_{pp}^{fr} (P+Q)+R$;
  \item $P+P\sim_{pp}^{fr} P$;
  \item $P+\textbf{nil}\sim_{pp}^{fr} P$.
\end{enumerate}

\end{proposition}

\begin{proof}
\begin{enumerate}
  \item $P+Q\sim_{pp}^{fr} Q+P$. It is sufficient to prove the relation $R=\{(P+Q, Q+P)\}\cup \textbf{Id}$ is a FR strongly probabilistic pomset bisimulation, we omit it;
  \item $P+(Q+R)\sim_{pp}^{fr} (P+Q)+R$. It is sufficient to prove the relation $R=\{(P+(Q+R), (P+Q)+R)\}\cup \textbf{Id}$ is a FR strongly probabilistic pomset bisimulation, we omit it;
  \item $P+P\sim_{pp}^{fr} P$. It is sufficient to prove the relation $R=\{(P+P, P)\}\cup \textbf{Id}$ is a FR strongly probabilistic pomset bisimulation, we omit it;
  \item $P+\textbf{nil}\sim_{pp}^{fr} P$. It is sufficient to prove the relation $R=\{(P+\textbf{nil}, P)\}\cup \textbf{Id}$ is a FR strongly probabilistic pomset bisimulation, we omit it.
\end{enumerate}
\end{proof}

\begin{proposition}[Monoid laws for FR strongly probabilistic step bisimulation] The monoid laws for FR strongly probabilistic step bisimulation are as follows.
\begin{enumerate}
  \item $P+Q\sim_{ps}^{fr} Q+P$;
  \item $P+(Q+R)\sim_{ps}^{fr} (P+Q)+R$;
  \item $P+P\sim_{ps}^{fr} P$;
  \item $P+\textbf{nil}\sim_{ps}^{fr} P$.
\end{enumerate}
\end{proposition}

\begin{proof}
\begin{enumerate}
  \item $P+Q\sim_{ps}^{fr} Q+P$. It is sufficient to prove the relation $R=\{(P+Q, Q+P)\}\cup \textbf{Id}$ is a FR strongly probabilistic step bisimulation, we omit it;
  \item $P+(Q+R)\sim_{ps}^{fr} (P+Q)+R$. It is sufficient to prove the relation $R=\{(P+(Q+R), (P+Q)+R)\}\cup \textbf{Id}$ is a FR strongly probabilistic step bisimulation, we omit it;
  \item $P+P\sim_{ps}^{fr} P$. It is sufficient to prove the relation $R=\{(P+P, P)\}\cup \textbf{Id}$ is a FR strongly probabilistic step bisimulation, we omit it;
  \item $P+\textbf{nil}\sim_{ps}^{fr} P$. It is sufficient to prove the relation $R=\{(P+\textbf{nil}, P)\}\cup \textbf{Id}$ is a FR strongly probabilistic step bisimulation, we omit it.
\end{enumerate}
\end{proof}

\begin{proposition}[Monoid laws for FR strongly probabilistic hp-bisimulation] The monoid laws for FR strongly probabilistic hp-bisimulation are as follows.
\begin{enumerate}
  \item $P+Q\sim_{php}^{fr} Q+P$;
  \item $P+(Q+R)\sim_{php}^{fr} (P+Q)+R$;
  \item $P+P\sim_{php}^{fr} P$;
  \item $P+\textbf{nil}\sim_{php}^{fr} P$.
\end{enumerate}
\end{proposition}

\begin{proof}
\begin{enumerate}
  \item $P+Q\sim_{php}^{fr} Q+P$. It is sufficient to prove the relation $R=\{(P+Q, Q+P)\}\cup \textbf{Id}$ is a FR strongly probabilistic hp-bisimulation, we omit it;
  \item $P+(Q+R)\sim_{php}^{fr} (P+Q)+R$. It is sufficient to prove the relation $R=\{(P+(Q+R), (P+Q)+R)\}\cup \textbf{Id}$ is a FR strongly probabilistic hp-bisimulation, we omit it;
  \item $P+P\sim_{php}^{fr} P$. It is sufficient to prove the relation $R=\{(P+P, P)\}\cup \textbf{Id}$ is a FR strongly probabilistic hp-bisimulation, we omit it;
  \item $P+\textbf{nil}\sim_{php}^{fr} P$. It is sufficient to prove the relation $R=\{(P+\textbf{nil}, P)\}\cup \textbf{Id}$ is a FR strongly probabilistic hp-bisimulation, we omit it.
\end{enumerate}
\end{proof}

\begin{proposition}[Monoid laws for FR strongly probabilistic hhp-bisimulation] The monoid laws for FR strongly probabilistic hhp-bisimulation are as follows.
\begin{enumerate}
  \item $P+Q\sim_{phhp}^{fr} Q+P$;
  \item $P+(Q+R)\sim_{phhp}^{fr} (P+Q)+R$;
  \item $P+P\sim_{phhp}^{fr} P$;
  \item $P+\textbf{nil}\sim_{phhp}^{fr} P$.
\end{enumerate}
\end{proposition}

\begin{proof}
\begin{enumerate}
  \item $P+Q\sim_{phhp}^{fr} Q+P$. It is sufficient to prove the relation $R=\{(P+Q, Q+P)\}\cup \textbf{Id}$ is a FR strongly probabilistic hhp-bisimulation, we omit it;
  \item $P+(Q+R)\sim_{phhp}^{fr} (P+Q)+R$. It is sufficient to prove the relation $R=\{(P+(Q+R), (P+Q)+R)\}\cup \textbf{Id}$ is a FR strongly probabilistic hhp-bisimulation, we omit it;
  \item $P+P\sim_{phhp}^{fr} P$. It is sufficient to prove the relation $R=\{(P+P, P)\}\cup \textbf{Id}$ is a FR strongly probabilistic hhp-bisimulation, we omit it;
  \item $P+\textbf{nil}\sim_{phhp}^{fr} P$. It is sufficient to prove the relation $R=\{(P+\textbf{nil}, P)\}\cup \textbf{Id}$ is a FR strongly probabilistic hhp-bisimulation, we omit it.
\end{enumerate}
\end{proof}

\begin{proposition}[Monoid laws 2 for FR strongly probabilistic pomset bisimulation]
The monoid laws 2 for FR strongly probabilistic pomset bisimulation are as follows.

\begin{enumerate}
  \item $P\boxplus_{\pi} Q\sim_{pp}^{fr} Q\boxplus_{1-\pi} P$;
  \item $P\boxplus_{\pi}(Q\boxplus_{\rho} R)\sim_{pp}^{fr} (P\boxplus_{\frac{\pi}{\pi+\rho-\pi\rho}}Q)\boxplus_{\pi+\rho-\pi\rho} R$;
  \item $P\boxplus_{\pi}P\sim_{pp}^{fr} P$;
  \item $P\boxplus_{\pi}\textbf{nil}\sim_{pp}^{fr} P$.
\end{enumerate}
\end{proposition}

\begin{proof}
\begin{enumerate}
  \item $P\boxplus_{\pi} Q\sim_{pp}^{fr} Q\boxplus_{1-\pi} P$. It is sufficient to prove the relation $R=\{(P\boxplus_{\pi} Q, Q\boxplus_{1-\pi} P)\}\cup \textbf{Id}$ is a FR strongly probabilistic pomset bisimulation, we omit it;
  \item $P\boxplus_{\pi}(Q\boxplus_{\rho} R)\sim_{pp}^{fr} (P\boxplus_{\frac{\pi}{\pi+\rho-\pi\rho}}Q)\boxplus_{\pi+\rho-\pi\rho} R$. It is sufficient to prove the relation $R=\{(P\boxplus_{\pi}(Q\boxplus_{\rho} R), (P\boxplus_{\frac{\pi}{\pi+\rho-\pi\rho}}Q)\boxplus_{\pi+\rho-\pi\rho} R)\}\cup \textbf{Id}$ is a FR strongly probabilistic pomset bisimulation, we omit it;
  \item $P\boxplus_{\pi}P\sim_{pp}^{fr} P$. It is sufficient to prove the relation $R=\{(P\boxplus_{\pi}P, P)\}\cup \textbf{Id}$ is a FR strongly probabilistic pomset bisimulation, we omit it;
  \item $P\boxplus_{\pi}\textbf{nil}\sim_{pp}^{fr} P$. It is sufficient to prove the relation $R=\{(P\boxplus_{\pi}\textbf{nil}, P)\}\cup \textbf{Id}$ is a FR strongly probabilistic pomset bisimulation, we omit it.
\end{enumerate}
\end{proof}

\begin{proposition}[Monoid laws 2 for FR strongly probabilistic step bisimulation]
The monoid laws 2 for FR strongly probabilistic step bisimulation are as follows.

\begin{enumerate}
  \item $P\boxplus_{\pi} Q\sim_{ps}^{fr} Q\boxplus_{1-\pi} P$;
  \item $P\boxplus_{\pi}(Q\boxplus_{\rho} R)\sim_{ps}^{fr} (P\boxplus_{\frac{\pi}{\pi+\rho-\pi\rho}}Q)\boxplus_{\pi+\rho-\pi\rho} R$;
  \item $P\boxplus_{\pi}P\sim_{ps}^{fr} P$;
  \item $P\boxplus_{\pi}\textbf{nil}\sim_{ps}^{fr} P$.
\end{enumerate}
\end{proposition}

\begin{proof}
\begin{enumerate}
  \item $P\boxplus_{\pi} Q\sim_{ps}^{fr} Q\boxplus_{1-\pi} P$. It is sufficient to prove the relation $R=\{(P\boxplus_{\pi} Q, Q\boxplus_{1-\pi} P)\}\cup \textbf{Id}$ is a FR strongly probabilistic step bisimulation, we omit it;
  \item $P\boxplus_{\pi}(Q\boxplus_{\rho} R)\sim_{ps}^{fr} (P\boxplus_{\frac{\pi}{\pi+\rho-\pi\rho}}Q)\boxplus_{\pi+\rho-\pi\rho} R$. It is sufficient to prove the relation $R=\{(P\boxplus_{\pi}(Q\boxplus_{\rho} R), (P\boxplus_{\frac{\pi}{\pi+\rho-\pi\rho}}Q)\boxplus_{\pi+\rho-\pi\rho} R)\}\cup \textbf{Id}$ is a FR strongly probabilistic step bisimulation, we omit it;
  \item $P\boxplus_{\pi}P\sim_{ps}^{fr} P$. It is sufficient to prove the relation $R=\{(P\boxplus_{\pi}P, P)\}\cup \textbf{Id}$ is a FR strongly probabilistic step bisimulation, we omit it;
  \item $P\boxplus_{\pi}\textbf{nil}\sim_{ps}^{fr} P$. It is sufficient to prove the relation $R=\{(P\boxplus_{\pi}\textbf{nil}, P)\}\cup \textbf{Id}$ is a FR strongly probabilistic step bisimulation, we omit it.
\end{enumerate}
\end{proof}

\begin{proposition}[Monoid laws 2 for FR strongly probabilistic hp-bisimulation]
The monoid laws 2 for FR strongly probabilistic hp-bisimulation are as follows.

\begin{enumerate}
  \item $P\boxplus_{\pi} Q\sim_{php}^{fr} Q\boxplus_{1-\pi} P$;
  \item $P\boxplus_{\pi}(Q\boxplus_{\rho} R)\sim_{php}^{fr} (P\boxplus_{\frac{\pi}{\pi+\rho-\pi\rho}}Q)\boxplus_{\pi+\rho-\pi\rho} R$;
  \item $P\boxplus_{\pi}P\sim_{php}^{fr} P$;
  \item $P\boxplus_{\pi}\textbf{nil}\sim_{php}^{fr} P$.
\end{enumerate}
\end{proposition}

\begin{proof}
\begin{enumerate}
  \item $P\boxplus_{\pi} Q\sim_{php}^{fr} Q\boxplus_{1-\pi} P$. It is sufficient to prove the relation $R=\{(P\boxplus_{\pi} Q, Q\boxplus_{1-\pi} P)\}\cup \textbf{Id}$ is a FR strongly probabilistic hp-bisimulation, we omit it;
  \item $P\boxplus_{\pi}(Q\boxplus_{\rho} R)\sim_{php}^{fr} (P\boxplus_{\frac{\pi}{\pi+\rho-\pi\rho}}Q)\boxplus_{\pi+\rho-\pi\rho} R$. It is sufficient to prove the relation $R=\{(P\boxplus_{\pi}(Q\boxplus_{\rho} R), (P\boxplus_{\frac{\pi}{\pi+\rho-\pi\rho}}Q)\boxplus_{\pi+\rho-\pi\rho} R)\}\cup \textbf{Id}$ is a FR strongly probabilistic hp-bisimulation, we omit it;
  \item $P\boxplus_{\pi}P\sim_{php}^{fr} P$. It is sufficient to prove the relation $R=\{(P\boxplus_{\pi}P, P)\}\cup \textbf{Id}$ is a FR strongly probabilistic hp-bisimulation, we omit it;
  \item $P\boxplus_{\pi}\textbf{nil}\sim_{php}^{fr} P$. It is sufficient to prove the relation $R=\{(P\boxplus_{\pi}\textbf{nil}, P)\}\cup \textbf{Id}$ is a FR strongly probabilistic hp-bisimulation, we omit it.
\end{enumerate}
\end{proof}

\begin{proposition}[Monoid laws 2 for FR strongly probabilistic hhp-bisimulation]
The monoid laws 2 for FR strongly probabilistic hhp-bisimulation are as follows.

\begin{enumerate}
  \item $P\boxplus_{\pi} Q\sim_{phhp}^{fr} Q\boxplus_{1-\pi} P$;
  \item $P\boxplus_{\pi}(Q\boxplus_{\rho} R)\sim_{phhp}^{fr} (P\boxplus_{\frac{\pi}{\pi+\rho-\pi\rho}}Q)\boxplus_{\pi+\rho-\pi\rho} R$;
  \item $P\boxplus_{\pi}P\sim_{phhp}^{fr} P$;
  \item $P\boxplus_{\pi}\textbf{nil}\sim_{phhp}^{fr} P$.
\end{enumerate}
\end{proposition}

\begin{proof}
\begin{enumerate}
  \item $P\boxplus_{\pi} Q\sim_{phhp}^{fr} Q\boxplus_{1-\pi} P$. It is sufficient to prove the relation $R=\{(P\boxplus_{\pi} Q, Q\boxplus_{1-\pi} P)\}\cup \textbf{Id}$ is a FR strongly probabilistic hhp-bisimulation, we omit it;
  \item $P\boxplus_{\pi}(Q\boxplus_{\rho} R)\sim_{phhp}^{fr} (P\boxplus_{\frac{\pi}{\pi+\rho-\pi\rho}}Q)\boxplus_{\pi+\rho-\pi\rho} R$. It is sufficient to prove the relation $R=\{(P\boxplus_{\pi}(Q\boxplus_{\rho} R), (P\boxplus_{\frac{\pi}{\pi+\rho-\pi\rho}}Q)\boxplus_{\pi+\rho-\pi\rho} R)\}\cup \textbf{Id}$ is a FR strongly probabilistic hhp-bisimulation, we omit it;
  \item $P\boxplus_{\pi}P\sim_{phhp}^{fr} P$. It is sufficient to prove the relation $R=\{(P\boxplus_{\pi}P, P)\}\cup \textbf{Id}$ is a FR strongly probabilistic hhp-bisimulation, we omit it;
  \item $P\boxplus_{\pi}\textbf{nil}\sim_{phhp}^{fr} P$. It is sufficient to prove the relation $R=\{(P\boxplus_{\pi}\textbf{nil}, P)\}\cup \textbf{Id}$ is a FR strongly probabilistic hhp-bisimulation, we omit it.
\end{enumerate}
\end{proof}

\begin{proposition}[Static laws for FR strongly probabilistic pomset bisimulation]
The static laws for FR strongly probabilistic pomset bisimulation are as follows.
\begin{enumerate}
  \item $P\parallel Q\sim_{pp}^{fr} Q\parallel P$;
  \item $P\parallel(Q\parallel R)\sim_{pp}^{fr} (P\parallel Q)\parallel R$;
  \item $P\parallel \textbf{nil}\sim_{pp}^{fr} P$;
  \item $P\setminus L\sim_{pp}^{fr} P$, if $\mathcal{L}(P)\cap(L\cup\overline{L})=\emptyset$;
  \item $P\setminus K\setminus L\sim_{pp}^{fr} P\setminus(K\cup L)$;
  \item $P[f]\setminus L\sim_{pp}^{fr} P\setminus f^{-1}(L)[f]$;
  \item $(P\parallel Q)\setminus L\sim_{pp}^{fr} P\setminus L\parallel Q\setminus L$, if $\mathcal{L}(P)\cap\overline{\mathcal{L}(Q)}\cap(L\cup\overline{L})=\emptyset$;
  \item $P[Id]\sim_{pp}^{fr} P$;
  \item $P[f]\sim_{pp}^{fr} P[f']$, if $f\upharpoonright\mathcal{L}(P)=f'\upharpoonright\mathcal{L}(P)$;
  \item $P[f][f']\sim_{pp}^{fr} P[f'\circ f]$;
  \item $(P\parallel Q)[f]\sim_{pp}^{fr} P[f]\parallel Q[f]$, if $f\upharpoonright(L\cup\overline{L})$ is one-to-one, where $L=\mathcal{L}(P)\cup\mathcal{L}(Q)$.
\end{enumerate}
\end{proposition}

\begin{proof}
\begin{enumerate}
  \item $P\parallel Q\sim_{pp}^{fr} Q\parallel P$. It is sufficient to prove the relation $R=\{(P\parallel Q, Q\parallel P)\}\cup \textbf{Id}$ is a FR strongly probabilistic pomset bisimulation, we omit it;
  \item $P\parallel(Q\parallel R)\sim_{pp}^{fr} (P\parallel Q)\parallel R$. It is sufficient to prove the relation $R=\{(P\parallel(Q\parallel R), (P\parallel Q)\parallel R)\}\cup \textbf{Id}$ is a FR strongly probabilistic pomset bisimulation, we omit it;
  \item $P\parallel \textbf{nil}\sim_{pp}^{fr} P$. It is sufficient to prove the relation $R=\{(P\parallel \textbf{nil}, P)\}\cup \textbf{Id}$ is a FR strongly probabilistic pomset bisimulation, we omit it;
  \item $P\setminus L\sim_{pp}^{fr} P$, if $\mathcal{L}(P)\cap(L\cup\overline{L})=\emptyset$. It is sufficient to prove the relation $R=\{(P\setminus L, P)\}\cup \textbf{Id}$, if $\mathcal{L}(P)\cap(L\cup\overline{L})=\emptyset$, is a FR strongly probabilistic pomset bisimulation, we omit it;
  \item $P\setminus K\setminus L\sim_{pp}^{fr} P\setminus(K\cup L)$. It is sufficient to prove the relation $R=\{(P\setminus K\setminus L, P\setminus(K\cup L))\}\cup \textbf{Id}$ is a FR strongly probabilistic pomset bisimulation, we omit it;
  \item $P[f]\setminus L\sim_{pp}^{fr} P\setminus f^{-1}(L)[f]$. It is sufficient to prove the relation $R=\{(P[f]\setminus L, P\setminus f^{-1}(L)[f])\}\cup \textbf{Id}$ is a FR strongly probabilistic pomset bisimulation, we omit it;
  \item $(P\parallel Q)\setminus L\sim_{pp}^{fr} P\setminus L\parallel Q\setminus L$, if $\mathcal{L}(P)\cap\overline{\mathcal{L}(Q)}\cap(L\cup\overline{L})=\emptyset$. It is sufficient to prove the relation
  $R=\{((P\parallel Q)\setminus L, P\setminus L\parallel Q\setminus L)\}\cup \textbf{Id}$, if $\mathcal{L}(P)\cap\overline{\mathcal{L}(Q)}\cap(L\cup\overline{L})=\emptyset$, is a FR strongly probabilistic pomset bisimulation, we omit it;
  \item $P[Id]\sim_{pp}^{fr} P$. It is sufficient to prove the relation $R=\{(P[Id], P)\}\cup \textbf{Id}$ is a FR strongly probabilistic pomset bisimulation, we omit it;
  \item $P[f]\sim_{pp}^{fr} P[f']$, if $f\upharpoonright\mathcal{L}(P)=f'\upharpoonright\mathcal{L}(P)$. It is sufficient to prove the relation $R=\{(P[f], P[f'])\}\cup \textbf{Id}$, if $f\upharpoonright\mathcal{L}(P)=f'\upharpoonright\mathcal{L}(P)$, is a FR strongly probabilistic pomset bisimulation, we omit it;
  \item $P[f][f']\sim_{pp}^{fr} P[f'\circ f]$. It is sufficient to prove the relation $R=\{(P[f][f'], P[f'\circ f])\}\cup \textbf{Id}$ is a FR strongly probabilistic pomset bisimulation, we omit it;
  \item $(P\parallel Q)[f]\sim_{pp}^{fr} P[f]\parallel Q[f]$, if $f\upharpoonright(L\cup\overline{L})$ is one-to-one, where $L=\mathcal{L}(P)\cup\mathcal{L}(Q)$. It is sufficient to prove the
  relation $R=\{((P\parallel Q)[f], P[f]\parallel Q[f])\}\cup \textbf{Id}$, if $f\upharpoonright(L\cup\overline{L})$ is one-to-one, where $L=\mathcal{L}(P)\cup\mathcal{L}(Q)$, is a FR strongly probabilistic pomset bisimulation, we omit it.
\end{enumerate}
\end{proof}

\begin{proposition}[Static laws for FR strongly probabilistic step bisimulation]
The static laws for FR strongly probabilistic step bisimulation are as follows.
\begin{enumerate}
  \item $P\parallel Q\sim_{ps}^{fr} Q\parallel P$;
  \item $P\parallel(Q\parallel R)\sim_{ps}^{fr} (P\parallel Q)\parallel R$;
  \item $P\parallel \textbf{nil}\sim_{ps}^{fr} P$;
  \item $P\setminus L\sim_{ps}^{fr} P$, if $\mathcal{L}(P)\cap(L\cup\overline{L})=\emptyset$;
  \item $P\setminus K\setminus L\sim_{ps}^{fr} P\setminus(K\cup L)$;
  \item $P[f]\setminus L\sim_{ps}^{fr} P\setminus f^{-1}(L)[f]$;
  \item $(P\parallel Q)\setminus L\sim_{ps}^{fr} P\setminus L\parallel Q\setminus L$, if $\mathcal{L}(P)\cap\overline{\mathcal{L}(Q)}\cap(L\cup\overline{L})=\emptyset$;
  \item $P[Id]\sim_{ps}^{fr} P$;
  \item $P[f]\sim_{ps}^{fr} P[f']$, if $f\upharpoonright\mathcal{L}(P)=f'\upharpoonright\mathcal{L}(P)$;
  \item $P[f][f']\sim_{ps}^{fr} P[f'\circ f]$;
  \item $(P\parallel Q)[f]\sim_{ps}^{fr} P[f]\parallel Q[f]$, if $f\upharpoonright(L\cup\overline{L})$ is one-to-one, where $L=\mathcal{L}(P)\cup\mathcal{L}(Q)$.
\end{enumerate}
\end{proposition}

\begin{proof}
\begin{enumerate}
  \item $P\parallel Q\sim_{ps}^{fr} Q\parallel P$. It is sufficient to prove the relation $R=\{(P\parallel Q, Q\parallel P)\}\cup \textbf{Id}$ is a FR strongly probabilistic step bisimulation, we omit it;
  \item $P\parallel(Q\parallel R)\sim_{ps}^{fr} (P\parallel Q)\parallel R$. It is sufficient to prove the relation $R=\{(P\parallel(Q\parallel R), (P\parallel Q)\parallel R)\}\cup \textbf{Id}$ is a FR strongly probabilistic step bisimulation, we omit it;
  \item $P\parallel \textbf{nil}\sim_{ps}^{fr} P$. It is sufficient to prove the relation $R=\{(P\parallel \textbf{nil}, P)\}\cup \textbf{Id}$ is a FR strongly probabilistic step bisimulation, we omit it;
  \item $P\setminus L\sim_{ps}^{fr} P$, if $\mathcal{L}(P)\cap(L\cup\overline{L})=\emptyset$. It is sufficient to prove the relation $R=\{(P\setminus L, P)\}\cup \textbf{Id}$, if $\mathcal{L}(P)\cap(L\cup\overline{L})=\emptyset$, is a FR strongly probabilistic step bisimulation, we omit it;
  \item $P\setminus K\setminus L\sim_{ps}^{fr} P\setminus(K\cup L)$. It is sufficient to prove the relation $R=\{(P\setminus K\setminus L, P\setminus(K\cup L))\}\cup \textbf{Id}$ is a FR strongly probabilistic step bisimulation, we omit it;
  \item $P[f]\setminus L\sim_{ps}^{fr} P\setminus f^{-1}(L)[f]$. It is sufficient to prove the relation $R=\{(P[f]\setminus L, P\setminus f^{-1}(L)[f])\}\cup \textbf{Id}$ is a FR strongly probabilistic step bisimulation, we omit it;
  \item $(P\parallel Q)\setminus L\sim_{ps}^{fr} P\setminus L\parallel Q\setminus L$, if $\mathcal{L}(P)\cap\overline{\mathcal{L}(Q)}\cap(L\cup\overline{L})=\emptyset$. It is sufficient to prove the relation
  $R=\{((P\parallel Q)\setminus L, P\setminus L\parallel Q\setminus L)\}\cup \textbf{Id}$, if $\mathcal{L}(P)\cap\overline{\mathcal{L}(Q)}\cap(L\cup\overline{L})=\emptyset$, is a FR strongly probabilistic step bisimulation, we omit it;
  \item $P[Id]\sim_{ps}^{fr} P$. It is sufficient to prove the relation $R=\{(P[Id], P)\}\cup \textbf{Id}$ is a FR strongly probabilistic step bisimulation, we omit it;
  \item $P[f]\sim_{ps}^{fr} P[f']$, if $f\upharpoonright\mathcal{L}(P)=f'\upharpoonright\mathcal{L}(P)$. It is sufficient to prove the relation $R=\{(P[f], P[f'])\}\cup \textbf{Id}$, if $f\upharpoonright\mathcal{L}(P)=f'\upharpoonright\mathcal{L}(P)$, is a FR strongly probabilistic step bisimulation, we omit it;
  \item $P[f][f']\sim_{ps}^{fr} P[f'\circ f]$. It is sufficient to prove the relation $R=\{(P[f][f'], P[f'\circ f])\}\cup \textbf{Id}$ is a FR strongly probabilistic step bisimulation, we omit it;
  \item $(P\parallel Q)[f]\sim_{ps}^{fr} P[f]\parallel Q[f]$, if $f\upharpoonright(L\cup\overline{L})$ is one-to-one, where $L=\mathcal{L}(P)\cup\mathcal{L}(Q)$. It is sufficient to prove the
  relation $R=\{((P\parallel Q)[f], P[f]\parallel Q[f])\}\cup \textbf{Id}$, if $f\upharpoonright(L\cup\overline{L})$ is one-to-one, where $L=\mathcal{L}(P)\cup\mathcal{L}(Q)$, is a FR strongly probabilistic step bisimulation, we omit it.
\end{enumerate}
\end{proof}

\begin{proposition}[Static laws for FR strongly probabilistic hp-bisimulation]
The static laws for FR strongly probabilistic hp-bisimulation are as follows.
\begin{enumerate}
  \item $P\parallel Q\sim_{php}^{fr} Q\parallel P$;
  \item $P\parallel(Q\parallel R)\sim_{php}^{fr} (P\parallel Q)\parallel R$;
  \item $P\parallel \textbf{nil}\sim_{php}^{fr} P$;
  \item $P\setminus L\sim_{php}^{fr} P$, if $\mathcal{L}(P)\cap(L\cup\overline{L})=\emptyset$;
  \item $P\setminus K\setminus L\sim_{php}^{fr} P\setminus(K\cup L)$;
  \item $P[f]\setminus L\sim_{php}^{fr} P\setminus f^{-1}(L)[f]$;
  \item $(P\parallel Q)\setminus L\sim_{php}^{fr} P\setminus L\parallel Q\setminus L$, if $\mathcal{L}(P)\cap\overline{\mathcal{L}(Q)}\cap(L\cup\overline{L})=\emptyset$;
  \item $P[Id]\sim_{php}^{fr} P$;
  \item $P[f]\sim_{php}^{fr} P[f']$, if $f\upharpoonright\mathcal{L}(P)=f'\upharpoonright\mathcal{L}(P)$;
  \item $P[f][f']\sim_{php}^{fr} P[f'\circ f]$;
  \item $(P\parallel Q)[f]\sim_{php}^{fr} P[f]\parallel Q[f]$, if $f\upharpoonright(L\cup\overline{L})$ is one-to-one, where $L=\mathcal{L}(P)\cup\mathcal{L}(Q)$.
\end{enumerate}
\end{proposition}

\begin{proof}
\begin{enumerate}
  \item $P\parallel Q\sim_{php}^{fr} Q\parallel P$. It is sufficient to prove the relation $R=\{(P\parallel Q, Q\parallel P)\}\cup \textbf{Id}$ is a FR strongly probabilistic hp-bisimulation, we omit it;
  \item $P\parallel(Q\parallel R)\sim_{php}^{fr} (P\parallel Q)\parallel R$. It is sufficient to prove the relation $R=\{(P\parallel(Q\parallel R), (P\parallel Q)\parallel R)\}\cup \textbf{Id}$ is a FR strongly probabilistic hp-bisimulation, we omit it;
  \item $P\parallel \textbf{nil}\sim_{php}^{fr} P$. It is sufficient to prove the relation $R=\{(P\parallel \textbf{nil}, P)\}\cup \textbf{Id}$ is a FR strongly probabilistic hp-bisimulation, we omit it;
  \item $P\setminus L\sim_{php}^{fr} P$, if $\mathcal{L}(P)\cap(L\cup\overline{L})=\emptyset$. It is sufficient to prove the relation $R=\{(P\setminus L, P)\}\cup \textbf{Id}$, if $\mathcal{L}(P)\cap(L\cup\overline{L})=\emptyset$, is a FR strongly probabilistic hp-bisimulation, we omit it;
  \item $P\setminus K\setminus L\sim_{php}^{fr} P\setminus(K\cup L)$. It is sufficient to prove the relation $R=\{(P\setminus K\setminus L, P\setminus(K\cup L))\}\cup \textbf{Id}$ is a FR strongly probabilistic hp-bisimulation, we omit it;
  \item $P[f]\setminus L\sim_{php}^{fr} P\setminus f^{-1}(L)[f]$. It is sufficient to prove the relation $R=\{(P[f]\setminus L, P\setminus f^{-1}(L)[f])\}\cup \textbf{Id}$ is a FR strongly probabilistic hp-bisimulation, we omit it;
  \item $(P\parallel Q)\setminus L\sim_{php}^{fr} P\setminus L\parallel Q\setminus L$, if $\mathcal{L}(P)\cap\overline{\mathcal{L}(Q)}\cap(L\cup\overline{L})=\emptyset$. It is sufficient to prove the relation
  $R=\{((P\parallel Q)\setminus L, P\setminus L\parallel Q\setminus L)\}\cup \textbf{Id}$, if $\mathcal{L}(P)\cap\overline{\mathcal{L}(Q)}\cap(L\cup\overline{L})=\emptyset$, is a FR strongly probabilistic hp-bisimulation, we omit it;
  \item $P[Id]\sim_{php}^{fr} P$. It is sufficient to prove the relation $R=\{(P[Id], P)\}\cup \textbf{Id}$ is a FR strongly probabilistic hp-bisimulation, we omit it;
  \item $P[f]\sim_{php}^{fr} P[f']$, if $f\upharpoonright\mathcal{L}(P)=f'\upharpoonright\mathcal{L}(P)$. It is sufficient to prove the relation $R=\{(P[f], P[f'])\}\cup \textbf{Id}$, if $f\upharpoonright\mathcal{L}(P)=f'\upharpoonright\mathcal{L}(P)$, is a FR strongly probabilistic hp-bisimulation, we omit it;
  \item $P[f][f']\sim_{php}^{fr} P[f'\circ f]$. It is sufficient to prove the relation $R=\{(P[f][f'], P[f'\circ f])\}\cup \textbf{Id}$ is a FR strongly probabilistic hp-bisimulation, we omit it;
  \item $(P\parallel Q)[f]\sim_{php}^{fr} P[f]\parallel Q[f]$, if $f\upharpoonright(L\cup\overline{L})$ is one-to-one, where $L=\mathcal{L}(P)\cup\mathcal{L}(Q)$. It is sufficient to prove the
  relation $R=\{((P\parallel Q)[f], P[f]\parallel Q[f])\}\cup \textbf{Id}$, if $f\upharpoonright(L\cup\overline{L})$ is one-to-one, where $L=\mathcal{L}(P)\cup\mathcal{L}(Q)$, is a FR strongly probabilistic hp-bisimulation, we omit it.
\end{enumerate}
\end{proof}

\begin{proposition}[Static laws for FR strongly probabilistic hhp-bisimulation]
The static laws for FR strongly probabilistic hhp-bisimulation are as follows.
\begin{enumerate}
  \item $P\parallel Q\sim_{phhp}^{fr} Q\parallel P$;
  \item $P\parallel(Q\parallel R)\sim_{phhp}^{fr} (P\parallel Q)\parallel R$;
  \item $P\parallel \textbf{nil}\sim_{phhp}^{fr} P$;
  \item $P\setminus L\sim_{phhp}^{fr} P$, if $\mathcal{L}(P)\cap(L\cup\overline{L})=\emptyset$;
  \item $P\setminus K\setminus L\sim_{phhp}^{fr} P\setminus(K\cup L)$;
  \item $P[f]\setminus L\sim_{phhp}^{fr} P\setminus f^{-1}(L)[f]$;
  \item $(P\parallel Q)\setminus L\sim_{phhp}^{fr} P\setminus L\parallel Q\setminus L$, if $\mathcal{L}(P)\cap\overline{\mathcal{L}(Q)}\cap(L\cup\overline{L})=\emptyset$;
  \item $P[Id]\sim_{phhp}^{fr} P$;
  \item $P[f]\sim_{phhp}^{fr} P[f']$, if $f\upharpoonright\mathcal{L}(P)=f'\upharpoonright\mathcal{L}(P)$;
  \item $P[f][f']\sim_{phhp}^{fr} P[f'\circ f]$;
  \item $(P\parallel Q)[f]\sim_{phhp}^{fr} P[f]\parallel Q[f]$, if $f\upharpoonright(L\cup\overline{L})$ is one-to-one, where $L=\mathcal{L}(P)\cup\mathcal{L}(Q)$.
\end{enumerate}
\end{proposition}

\begin{proof}
\begin{enumerate}
  \item $P\parallel Q\sim_{phhp}^{fr} Q\parallel P$. It is sufficient to prove the relation $R=\{(P\parallel Q, Q\parallel P)\}\cup \textbf{Id}$ is a FR strongly probabilistic hhp-bisimulation, we omit it;
  \item $P\parallel(Q\parallel R)\sim_{phhp}^{fr} (P\parallel Q)\parallel R$. It is sufficient to prove the relation $R=\{(P\parallel(Q\parallel R), (P\parallel Q)\parallel R)\}\cup \textbf{Id}$ is a FR strongly probabilistic hhp-bisimulation, we omit it;
  \item $P\parallel \textbf{nil}\sim_{phhp}^{fr} P$. It is sufficient to prove the relation $R=\{(P\parallel \textbf{nil}, P)\}\cup \textbf{Id}$ is a FR strongly probabilistic hhp-bisimulation, we omit it;
  \item $P\setminus L\sim_{phhp}^{fr} P$, if $\mathcal{L}(P)\cap(L\cup\overline{L})=\emptyset$. It is sufficient to prove the relation $R=\{(P\setminus L, P)\}\cup \textbf{Id}$, if $\mathcal{L}(P)\cap(L\cup\overline{L})=\emptyset$, is a FR strongly probabilistic hhp-bisimulation, we omit it;
  \item $P\setminus K\setminus L\sim_{phhp}^{fr} P\setminus(K\cup L)$. It is sufficient to prove the relation $R=\{(P\setminus K\setminus L, P\setminus(K\cup L))\}\cup \textbf{Id}$ is a FR strongly probabilistic hhp-bisimulation, we omit it;
  \item $P[f]\setminus L\sim_{phhp}^{fr} P\setminus f^{-1}(L)[f]$. It is sufficient to prove the relation $R=\{(P[f]\setminus L, P\setminus f^{-1}(L)[f])\}\cup \textbf{Id}$ is a FR strongly probabilistic hhp-bisimulation, we omit it;
  \item $(P\parallel Q)\setminus L\sim_{phhp}^{fr} P\setminus L\parallel Q\setminus L$, if $\mathcal{L}(P)\cap\overline{\mathcal{L}(Q)}\cap(L\cup\overline{L})=\emptyset$. It is sufficient to prove the relation
  $R=\{((P\parallel Q)\setminus L, P\setminus L\parallel Q\setminus L)\}\cup \textbf{Id}$, if $\mathcal{L}(P)\cap\overline{\mathcal{L}(Q)}\cap(L\cup\overline{L})=\emptyset$, is a FR strongly probabilistic hhp-bisimulation, we omit it;
  \item $P[Id]\sim_{phhp}^{fr} P$. It is sufficient to prove the relation $R=\{(P[Id], P)\}\cup \textbf{Id}$ is a FR strongly probabilistic hhp-bisimulation, we omit it;
  \item $P[f]\sim_{phhp}^{fr} P[f']$, if $f\upharpoonright\mathcal{L}(P)=f'\upharpoonright\mathcal{L}(P)$. It is sufficient to prove the relation $R=\{(P[f], P[f'])\}\cup \textbf{Id}$, if $f\upharpoonright\mathcal{L}(P)=f'\upharpoonright\mathcal{L}(P)$, is a FR strongly probabilistic hhp-bisimulation, we omit it;
  \item $P[f][f']\sim_{phhp}^{fr} P[f'\circ f]$. It is sufficient to prove the relation $R=\{(P[f][f'], P[f'\circ f])\}\cup \textbf{Id}$ is a FR strongly probabilistic hhp-bisimulation, we omit it;
  \item $(P\parallel Q)[f]\sim_{phhp}^{fr} P[f]\parallel Q[f]$, if $f\upharpoonright(L\cup\overline{L})$ is one-to-one, where $L=\mathcal{L}(P)\cup\mathcal{L}(Q)$. It is sufficient to prove the
  relation $R=\{((P\parallel Q)[f], P[f]\parallel Q[f])\}\cup \textbf{Id}$, if $f\upharpoonright(L\cup\overline{L})$ is one-to-one, where $L=\mathcal{L}(P)\cup\mathcal{L}(Q)$, is a FR strongly probabilistic hhp-bisimulation, we omit it.
\end{enumerate}
\end{proof}

\begin{proposition}[Expansion law for FR strongly probabilistic pomset bisimulation]
Let $P\equiv (P_1[f_1]\parallel\cdots\parallel P_n[f_n])\setminus L$, with $n\geq 1$. Then

\begin{eqnarray}
P\sim_{pp}^{f} \{(f_1(\alpha_1)\parallel\cdots\parallel f_n(\alpha_n)).(P_1'[f_1]\parallel\cdots\parallel P_n'[f_n])\setminus L: \nonumber\\
 P_i\rightsquigarrow\xrightarrow{\alpha_i} P_i',i\in\{1,\cdots,n\},f_i(\alpha_i)\notin L\cup\overline{L}\} \nonumber\\
+\sum\{\tau.(P_1[f_1]\parallel\cdots\parallel P_i'[f_i]\parallel\cdots\parallel P_j'[f_j]\parallel\cdots\parallel P_n[f_n])\setminus L: \nonumber\\
 P_i\rightsquigarrow\xrightarrow{l_1} P_i', P_j\rightsquigarrow\xrightarrow{l_2} P_j',f_i(l_1)=\overline{f_j(l_2)},i<j\}\nonumber
\end{eqnarray}
\begin{eqnarray}
P\sim_{pp}^{r} \{(P_1'[f_1]\parallel\cdots\parallel P_n'[f_n]).(f_1(\alpha_1[m])\parallel\cdots\parallel f_n(\alpha_n)[m])\setminus L: \nonumber\\
 P_i\rightsquigarrow\xtworightarrow{\alpha_i[m]} P_i',i\in\{1,\cdots,n\},f_i(\alpha_i)\notin L\cup\overline{L}\} \nonumber\\
+\sum\{(P_1[f_1]\parallel\cdots\parallel P_i'[f_i]\parallel\cdots\parallel P_j'[f_j]\parallel\cdots\parallel P_n[f_n]).\tau\setminus L: \nonumber\\
 P_i\rightsquigarrow\xtworightarrow{l_1[m]} P_i', P_j\rightsquigarrow\xtworightarrow{l_2[m]} P_j',f_i(l_1)=\overline{f_j(l_2)},i<j\}\nonumber
\end{eqnarray}
\end{proposition}

\begin{proof}
(1) The case of forward strongly probabilistic pomset bisimulation.

Firstly, we consider the case without Restriction and Relabeling. That is, we suffice to prove the following case by induction on the size $n$.

For $P\equiv P_1\parallel\cdots\parallel P_n$, with $n\geq 1$, we need to prove

\begin{eqnarray}
P\sim_{pp} \{(\alpha_1\parallel\cdots\parallel \alpha_n).(P_1'\parallel\cdots\parallel P_n'):  P_i\rightsquigarrow\xrightarrow{\alpha_i} P_i',i\in\{1,\cdots,n\}\nonumber\\
+\sum\{\tau.(P_1\parallel\cdots\parallel P_i'\parallel\cdots\parallel P_j'\parallel\cdots\parallel P_n):  P_i\rightsquigarrow\xrightarrow{l} P_i', P_j\rightsquigarrow\xrightarrow{\overline{l}} P_j',i<j\} \nonumber
\end{eqnarray}

For $n=1$, $P_1\sim_{pp}^{f} \alpha_1.P_1': P_1\rightsquigarrow\xrightarrow{\alpha_1} P_1'$ is obvious. Then with a hypothesis $n$, we consider
$R\equiv P\parallel P_{n+1}$. By the forward transition rules of Composition, we can get

\begin{eqnarray}
R\sim_{pp}^{f} \{(p\parallel \alpha_{n+1}).(P'\parallel P_{n+1}'):  P\rightsquigarrow\xrightarrow{p} P', P_{n+1}\rightsquigarrow\xrightarrow{\alpha_{n+1}} P_{n+1}',p\subseteq P\}\nonumber\\
+\sum\{\tau.(P'\parallel P_{n+1}'):  P\rightsquigarrow\xrightarrow{l} P', P_{n+1}\rightsquigarrow\xrightarrow{\overline{l}} P_{n+1}'\} \nonumber
\end{eqnarray}

Now with the induction assumption $P\equiv P_1\parallel\cdots\parallel P_n$, the right-hand side can be reformulated as follows.

\begin{eqnarray}
\{(\alpha_1\parallel\cdots\parallel \alpha_n\parallel \alpha_{n+1}).(P_1'\parallel\cdots\parallel P_n'\parallel P_{n+1}'): \nonumber\\
 P_i\rightsquigarrow\xrightarrow{\alpha_i} P_i',i\in\{1,\cdots,n+1\}\nonumber\\
+\sum\{\tau.(P_1\parallel\cdots\parallel P_i'\parallel\cdots\parallel P_j'\parallel\cdots\parallel P_n\parallel P_{n+1}): \nonumber\\
 P_i\rightsquigarrow\xrightarrow{l} P_i', P_j\rightsquigarrow\xrightarrow{\overline{l}} P_j',i<j\} \nonumber\\
+\sum\{\tau.(P_1\parallel\cdots\parallel P_i'\parallel\cdots\parallel P_j\parallel\cdots\parallel P_n\parallel P_{n+1}'): \nonumber\\
 P_i\rightsquigarrow\xrightarrow{l} P_i', P_{n+1}\rightsquigarrow\xrightarrow{\overline{l}} P_{n+1}',i\in\{1,\cdots, n\}\} \nonumber
\end{eqnarray}

So,

\begin{eqnarray}
R\sim_{pp}^{f} \{(\alpha_1\parallel\cdots\parallel \alpha_n\parallel \alpha_{n+1}).(P_1'\parallel\cdots\parallel P_n'\parallel P_{n+1}'): \nonumber\\
 P_i\rightsquigarrow\xrightarrow{\alpha_i} P_i',i\in\{1,\cdots,n+1\}\nonumber\\
+\sum\{\tau.(P_1\parallel\cdots\parallel P_i'\parallel\cdots\parallel P_j'\parallel\cdots\parallel P_n): \nonumber\\
 P_i\rightsquigarrow\xrightarrow{l} P_i', P_j\rightsquigarrow\xrightarrow{\overline{l}} P_j',1 \leq i<j\geq n+1\} \nonumber
\end{eqnarray}

Then, we can easily add the full conditions with Restriction and Relabeling.

(2) The case of reverse strongly probabilistic pomset bisimulation.

Firstly, we consider the case without Restriction and Relabeling. That is, we suffice to prove the following case by induction on the size $n$.

For $P\equiv P_1\parallel\cdots\parallel P_n$, with $n\geq 1$, we need to prove

\begin{eqnarray}
P\sim_{pp}^{r} \{(P_1'\parallel\cdots\parallel P_n').(\alpha_1[m]\parallel\cdots\parallel \alpha_n[m]):  P_i\rightsquigarrow\xtworightarrow{\alpha_i[m]} P_i',i\in\{1,\cdots,n\}\nonumber\\
+\sum\{(P_1\parallel\cdots\parallel P_i'\parallel\cdots\parallel P_j'\parallel\cdots\parallel P_n).\tau:  P_i\rightsquigarrow\xtworightarrow{l[m]} P_i', P_j\rightsquigarrow\xtworightarrow{\overline{l}[m]} P_j',i<j\} \nonumber
\end{eqnarray}

For $n=1$, $P_1\sim_{pp}^{r} P_1'.\alpha_1[m]: P_1\rightsquigarrow\xtworightarrow{\alpha_1[m]} P_1'$ is obvious. Then with a hypothesis $n$, we consider
$R\equiv P\parallel P_{n+1}$. By the reverse transition rules of Composition, we can get

\begin{eqnarray}
R\sim_{pp}^{r} \{(P'\parallel P_{n+1}').(p[m]\parallel \alpha_{n+1}[m]):  P\rightsquigarrow\xtworightarrow{p[m]} P', P_{n+1}\rightsquigarrow\xtworightarrow{\alpha_{n+1}[m]} P_{n+1}',p\subseteq P\}\nonumber\\
+\sum\{(P'\parallel P_{n+1}').\tau:  P\rightsquigarrow\xtworightarrow{l[m]} P', P_{n+1}\rightsquigarrow\xtworightarrow{\overline{l}[m]} P_{n+1}'\} \nonumber
\end{eqnarray}

Now with the induction assumption $P\equiv P_1\parallel\cdots\parallel P_n$, the right-hand side can be reformulated as follows.

\begin{eqnarray}
\{(P_1'\parallel\cdots\parallel P_n'\parallel P_{n+1}).(\alpha_1[m]\parallel\cdots\parallel \alpha_n[m]\parallel \alpha_{n+1}[m]): \nonumber\\
 P_i\rightsquigarrow\xtworightarrow{\alpha_i[m]} P_i',i\in\{1,\cdots,n+1\}\nonumber\\
+\sum\{(P_1\parallel\cdots\parallel P_i'\parallel\cdots\parallel P_j'\parallel\cdots\parallel P_n\parallel P_{n+1}).\tau: \nonumber\\
 P_i\rightsquigarrow\xtworightarrow{l[m]} P_i', P_j\rightsquigarrow\xtworightarrow{\overline{l}[m]} P_j',i<j\} \nonumber\\
+\sum\{(P_1\parallel\cdots\parallel P_i'\parallel\cdots\parallel P_j'\parallel\cdots\parallel P_n\parallel P_{n+1}).\tau: \nonumber\\
 P_i\rightsquigarrow\xtworightarrow{l[m]} P_i', P_{n+1}\rightsquigarrow\xtworightarrow{\overline{l}[m]} P_{n+1}',i\in\{1,\cdots, n\}\} \nonumber
\end{eqnarray}

So,

\begin{eqnarray}
R\sim_{pp}^{r} \{(P_1'\parallel\cdots\parallel P_n'\parallel P_{n+1}').(\alpha_1[m]\parallel\cdots\parallel \alpha_n[m]\parallel \alpha_{n+1}[m]): \nonumber\\
 P_i\rightsquigarrow\xtworightarrow{\alpha_i[m]} P_i',i\in\{1,\cdots,n+1\}\nonumber\\
+\sum\{(P_1\parallel\cdots\parallel P_i'\parallel\cdots\parallel P_j'\parallel\cdots\parallel P_n).\tau: \nonumber\\
 P_i\rightsquigarrow\xtworightarrow{l[m]} P_i', P_j\rightsquigarrow\xtworightarrow{\overline{l}[m]} P_j',1 \leq i<j\geq n+1\} \nonumber
\end{eqnarray}

Then, we can easily add the full conditions with Restriction and Relabeling.
\end{proof}

\begin{proposition}[Expansion law for FR strongly probabilistic step bisimulation]
Let $P\equiv (P_1[f_1]\parallel\cdots\parallel P_n[f_n])\setminus L$, with $n\geq 1$. Then

\begin{eqnarray}
P\sim_{ps}^{f} \{(f_1(\alpha_1)\parallel\cdots\parallel f_n(\alpha_n)).(P_1'[f_1]\parallel\cdots\parallel P_n'[f_n])\setminus L: \nonumber\\
 P_i\rightsquigarrow\xrightarrow{\alpha_i} P_i',i\in\{1,\cdots,n\},f_i(\alpha_i)\notin L\cup\overline{L}\} \nonumber\\
+\sum\{\tau.(P_1[f_1]\parallel\cdots\parallel P_i'[f_i]\parallel\cdots\parallel P_j'[f_j]\parallel\cdots\parallel P_n[f_n])\setminus L: \nonumber\\
 P_i\rightsquigarrow\xrightarrow{l_1} P_i', P_j\rightsquigarrow\xrightarrow{l_2} P_j',f_i(l_1)=\overline{f_j(l_2)},i<j\}\nonumber
\end{eqnarray}
\begin{eqnarray}
P\sim_{ps}^{r} \{(P_1'[f_1]\parallel\cdots\parallel P_n'[f_n]).(f_1(\alpha_1[m])\parallel\cdots\parallel f_n(\alpha_n)[m])\setminus L: \nonumber\\
 P_i\rightsquigarrow\xtworightarrow{\alpha_i[m]} P_i',i\in\{1,\cdots,n\},f_i(\alpha_i)\notin L\cup\overline{L}\} \nonumber\\
+\sum\{(P_1[f_1]\parallel\cdots\parallel P_i'[f_i]\parallel\cdots\parallel P_j'[f_j]\parallel\cdots\parallel P_n[f_n]).\tau\setminus L: \nonumber\\
 P_i\rightsquigarrow\xtworightarrow{l_1[m]} P_i', P_j\rightsquigarrow\xtworightarrow{l_2[m]} P_j',f_i(l_1)=\overline{f_j(l_2)},i<j\}\nonumber
\end{eqnarray}
\end{proposition}

\begin{proof}
(1) The case of forward strongly probabilistic step bisimulation.

Firstly, we consider the case without Restriction and Relabeling. That is, we suffice to prove the following case by induction on the size $n$.

For $P\equiv P_1\parallel\cdots\parallel P_n$, with $n\geq 1$, we need to prove

\begin{eqnarray}
P\sim_{ps} \{(\alpha_1\parallel\cdots\parallel \alpha_n).(P_1'\parallel\cdots\parallel P_n'):  P_i\rightsquigarrow\xrightarrow{\alpha_i} P_i',i\in\{1,\cdots,n\}\nonumber\\
+\sum\{\tau.(P_1\parallel\cdots\parallel P_i'\parallel\cdots\parallel P_j'\parallel\cdots\parallel P_n):  P_i\rightsquigarrow\xrightarrow{l} P_i', P_j\rightsquigarrow\xrightarrow{\overline{l}} P_j',i<j\} \nonumber
\end{eqnarray}

For $n=1$, $P_1\sim_{ps}^{f} \alpha_1.P_1': P_1\rightsquigarrow\xrightarrow{\alpha_1} P_1'$ is obvious. Then with a hypothesis $n$, we consider
$R\equiv P\parallel P_{n+1}$. By the forward transition rules of Composition, we can get

\begin{eqnarray}
R\sim_{ps}^{f} \{(p\parallel \alpha_{n+1}).(P'\parallel P_{n+1}'):  P\rightsquigarrow\xrightarrow{p} P', P_{n+1}\rightsquigarrow\xrightarrow{\alpha_{n+1}} P_{n+1}',p\subseteq P\}\nonumber\\
+\sum\{\tau.(P'\parallel P_{n+1}'):  P\rightsquigarrow\xrightarrow{l} P', P_{n+1}\rightsquigarrow\xrightarrow{\overline{l}} P_{n+1}'\} \nonumber
\end{eqnarray}

Now with the induction assumption $P\equiv P_1\parallel\cdots\parallel P_n$, the right-hand side can be reformulated as follows.

\begin{eqnarray}
\{(\alpha_1\parallel\cdots\parallel \alpha_n\parallel \alpha_{n+1}).(P_1'\parallel\cdots\parallel P_n'\parallel P_{n+1}'): \nonumber\\
 P_i\rightsquigarrow\xrightarrow{\alpha_i} P_i',i\in\{1,\cdots,n+1\}\nonumber\\
+\sum\{\tau.(P_1\parallel\cdots\parallel P_i'\parallel\cdots\parallel P_j'\parallel\cdots\parallel P_n\parallel P_{n+1}): \nonumber\\
 P_i\rightsquigarrow\xrightarrow{l} P_i', P_j\rightsquigarrow\xrightarrow{\overline{l}} P_j',i<j\} \nonumber\\
+\sum\{\tau.(P_1\parallel\cdots\parallel P_i'\parallel\cdots\parallel P_j\parallel\cdots\parallel P_n\parallel P_{n+1}'): \nonumber\\
 P_i\rightsquigarrow\xrightarrow{l} P_i', P_{n+1}\rightsquigarrow\xrightarrow{\overline{l}} P_{n+1}',i\in\{1,\cdots, n\}\} \nonumber
\end{eqnarray}

So,

\begin{eqnarray}
R\sim_{ps}^{f} \{(\alpha_1\parallel\cdots\parallel \alpha_n\parallel \alpha_{n+1}).(P_1'\parallel\cdots\parallel P_n'\parallel P_{n+1}'): \nonumber\\
 P_i\rightsquigarrow\xrightarrow{\alpha_i} P_i',i\in\{1,\cdots,n+1\}\nonumber\\
+\sum\{\tau.(P_1\parallel\cdots\parallel P_i'\parallel\cdots\parallel P_j'\parallel\cdots\parallel P_n): \nonumber\\
 P_i\rightsquigarrow\xrightarrow{l} P_i', P_j\rightsquigarrow\xrightarrow{\overline{l}} P_j',1 \leq i<j\geq n+1\} \nonumber
\end{eqnarray}

Then, we can easily add the full conditions with Restriction and Relabeling.

(2) The case of reverse strongly probabilistic step bisimulation.

Firstly, we consider the case without Restriction and Relabeling. That is, we suffice to prove the following case by induction on the size $n$.

For $P\equiv P_1\parallel\cdots\parallel P_n$, with $n\geq 1$, we need to prove

\begin{eqnarray}
P\sim_{ps}^{r} \{(P_1'\parallel\cdots\parallel P_n').(\alpha_1[m]\parallel\cdots\parallel \alpha_n[m]):  P_i\rightsquigarrow\xtworightarrow{\alpha_i[m]} P_i',i\in\{1,\cdots,n\}\nonumber\\
+\sum\{(P_1\parallel\cdots\parallel P_i'\parallel\cdots\parallel P_j'\parallel\cdots\parallel P_n).\tau:  P_i\rightsquigarrow\xtworightarrow{l[m]} P_i', P_j\rightsquigarrow\xtworightarrow{\overline{l}[m]} P_j',i<j\} \nonumber
\end{eqnarray}

For $n=1$, $P_1\sim_{ps}^{r} P_1'.\alpha_1[m]: P_1\rightsquigarrow\xtworightarrow{\alpha_1[m]} P_1'$ is obvious. Then with a hypothesis $n$, we consider
$R\equiv P\parallel P_{n+1}$. By the reverse transition rules of Composition, we can get

\begin{eqnarray}
R\sim_{ps}^{r} \{(P'\parallel P_{n+1}').(p[m]\parallel \alpha_{n+1}[m]):  P\rightsquigarrow\xtworightarrow{p[m]} P', P_{n+1}\rightsquigarrow\xtworightarrow{\alpha_{n+1}[m]} P_{n+1}',p\subseteq P\}\nonumber\\
+\sum\{(P'\parallel P_{n+1}').\tau:  P\rightsquigarrow\xtworightarrow{l[m]} P', P_{n+1}\rightsquigarrow\xtworightarrow{\overline{l}[m]} P_{n+1}'\} \nonumber
\end{eqnarray}

Now with the induction assumption $P\equiv P_1\parallel\cdots\parallel P_n$, the right-hand side can be reformulated as follows.

\begin{eqnarray}
\{(P_1'\parallel\cdots\parallel P_n'\parallel P_{n+1}).(\alpha_1[m]\parallel\cdots\parallel \alpha_n[m]\parallel \alpha_{n+1}[m]): \nonumber\\
 P_i\rightsquigarrow\xtworightarrow{\alpha_i[m]} P_i',i\in\{1,\cdots,n+1\}\nonumber\\
+\sum\{(P_1\parallel\cdots\parallel P_i'\parallel\cdots\parallel P_j'\parallel\cdots\parallel P_n\parallel P_{n+1}).\tau: \nonumber\\
 P_i\rightsquigarrow\xtworightarrow{l[m]} P_i', P_j\rightsquigarrow\xtworightarrow{\overline{l}[m]} P_j',i<j\} \nonumber\\
+\sum\{(P_1\parallel\cdots\parallel P_i'\parallel\cdots\parallel P_j'\parallel\cdots\parallel P_n\parallel P_{n+1}).\tau: \nonumber\\
 P_i\rightsquigarrow\xtworightarrow{l[m]} P_i', P_{n+1}\rightsquigarrow\xtworightarrow{\overline{l}[m]} P_{n+1}',i\in\{1,\cdots, n\}\} \nonumber
\end{eqnarray}

So,

\begin{eqnarray}
R\sim_{ps}^{r} \{(P_1'\parallel\cdots\parallel P_n'\parallel P_{n+1}').(\alpha_1[m]\parallel\cdots\parallel \alpha_n[m]\parallel \alpha_{n+1}[m]): \nonumber\\
 P_i\rightsquigarrow\xtworightarrow{\alpha_i[m]} P_i',i\in\{1,\cdots,n+1\}\nonumber\\
+\sum\{(P_1\parallel\cdots\parallel P_i'\parallel\cdots\parallel P_j'\parallel\cdots\parallel P_n).\tau: \nonumber\\
 P_i\rightsquigarrow\xtworightarrow{l[m]} P_i', P_j\rightsquigarrow\xtworightarrow{\overline{l}[m]} P_j',1 \leq i<j\geq n+1\} \nonumber
\end{eqnarray}

Then, we can easily add the full conditions with Restriction and Relabeling.
\end{proof}

\begin{proposition}[Expansion law for FR strongly probabilistic hp-bisimulation]
Let $P\equiv (P_1[f_1]\parallel\cdots\parallel P_n[f_n])\setminus L$, with $n\geq 1$. Then

\begin{eqnarray}
P\sim_{php}^{f} \{(f_1(\alpha_1)\parallel\cdots\parallel f_n(\alpha_n)).(P_1'[f_1]\parallel\cdots\parallel P_n'[f_n])\setminus L: \nonumber\\
 P_i\rightsquigarrow\xrightarrow{\alpha_i} P_i',i\in\{1,\cdots,n\},f_i(\alpha_i)\notin L\cup\overline{L}\} \nonumber\\
+\sum\{\tau.(P_1[f_1]\parallel\cdots\parallel P_i'[f_i]\parallel\cdots\parallel P_j'[f_j]\parallel\cdots\parallel P_n[f_n])\setminus L: \nonumber\\
 P_i\rightsquigarrow\xrightarrow{l_1} P_i', P_j\rightsquigarrow\xrightarrow{l_2} P_j',f_i(l_1)=\overline{f_j(l_2)},i<j\}\nonumber
\end{eqnarray}
\begin{eqnarray}
P\sim_{php}^{r} \{(P_1'[f_1]\parallel\cdots\parallel P_n'[f_n]).(f_1(\alpha_1[m])\parallel\cdots\parallel f_n(\alpha_n)[m])\setminus L: \nonumber\\
 P_i\rightsquigarrow\xtworightarrow{\alpha_i[m]} P_i',i\in\{1,\cdots,n\},f_i(\alpha_i)\notin L\cup\overline{L}\} \nonumber\\
+\sum\{(P_1[f_1]\parallel\cdots\parallel P_i'[f_i]\parallel\cdots\parallel P_j'[f_j]\parallel\cdots\parallel P_n[f_n]).\tau\setminus L: \nonumber\\
 P_i\rightsquigarrow\xtworightarrow{l_1[m]} P_i', P_j\rightsquigarrow\xtworightarrow{l_2[m]} P_j',f_i(l_1)=\overline{f_j(l_2)},i<j\}\nonumber
\end{eqnarray}
\end{proposition}

\begin{proof}
(1) The case of forward strongly probabilistic hp-bisimulation.

Firstly, we consider the case without Restriction and Relabeling. That is, we suffice to prove the following case by induction on the size $n$.

For $P\equiv P_1\parallel\cdots\parallel P_n$, with $n\geq 1$, we need to prove

\begin{eqnarray}
P\sim_{php} \{(\alpha_1\parallel\cdots\parallel \alpha_n).(P_1'\parallel\cdots\parallel P_n'):  P_i\rightsquigarrow\xrightarrow{\alpha_i} P_i',i\in\{1,\cdots,n\}\nonumber\\
+\sum\{\tau.(P_1\parallel\cdots\parallel P_i'\parallel\cdots\parallel P_j'\parallel\cdots\parallel P_n):  P_i\rightsquigarrow\xrightarrow{l} P_i', P_j\rightsquigarrow\xrightarrow{\overline{l}} P_j',i<j\} \nonumber
\end{eqnarray}

For $n=1$, $P_1\sim_{php}^{f} \alpha_1.P_1': P_1\rightsquigarrow\xrightarrow{\alpha_1} P_1'$ is obvious. Then with a hypothesis $n$, we consider
$R\equiv P\parallel P_{n+1}$. By the forward transition rules of Composition, we can get

\begin{eqnarray}
R\sim_{php}^{f} \{(p\parallel \alpha_{n+1}).(P'\parallel P_{n+1}'):  P\rightsquigarrow\xrightarrow{p} P', P_{n+1}\rightsquigarrow\xrightarrow{\alpha_{n+1}} P_{n+1}',p\subseteq P\}\nonumber\\
+\sum\{\tau.(P'\parallel P_{n+1}'):  P\rightsquigarrow\xrightarrow{l} P', P_{n+1}\rightsquigarrow\xrightarrow{\overline{l}} P_{n+1}'\} \nonumber
\end{eqnarray}

Now with the induction assumption $P\equiv P_1\parallel\cdots\parallel P_n$, the right-hand side can be reformulated as follows.

\begin{eqnarray}
\{(\alpha_1\parallel\cdots\parallel \alpha_n\parallel \alpha_{n+1}).(P_1'\parallel\cdots\parallel P_n'\parallel P_{n+1}'): \nonumber\\
 P_i\rightsquigarrow\xrightarrow{\alpha_i} P_i',i\in\{1,\cdots,n+1\}\nonumber\\
+\sum\{\tau.(P_1\parallel\cdots\parallel P_i'\parallel\cdots\parallel P_j'\parallel\cdots\parallel P_n\parallel P_{n+1}): \nonumber\\
 P_i\rightsquigarrow\xrightarrow{l} P_i', P_j\rightsquigarrow\xrightarrow{\overline{l}} P_j',i<j\} \nonumber\\
+\sum\{\tau.(P_1\parallel\cdots\parallel P_i'\parallel\cdots\parallel P_j\parallel\cdots\parallel P_n\parallel P_{n+1}'): \nonumber\\
 P_i\rightsquigarrow\xrightarrow{l} P_i', P_{n+1}\rightsquigarrow\xrightarrow{\overline{l}} P_{n+1}',i\in\{1,\cdots, n\}\} \nonumber
\end{eqnarray}

So,

\begin{eqnarray}
R\sim_{php}^{f} \{(\alpha_1\parallel\cdots\parallel \alpha_n\parallel \alpha_{n+1}).(P_1'\parallel\cdots\parallel P_n'\parallel P_{n+1}'): \nonumber\\
 P_i\rightsquigarrow\xrightarrow{\alpha_i} P_i',i\in\{1,\cdots,n+1\}\nonumber\\
+\sum\{\tau.(P_1\parallel\cdots\parallel P_i'\parallel\cdots\parallel P_j'\parallel\cdots\parallel P_n): \nonumber\\
 P_i\rightsquigarrow\xrightarrow{l} P_i', P_j\rightsquigarrow\xrightarrow{\overline{l}} P_j',1 \leq i<j\geq n+1\} \nonumber
\end{eqnarray}

Then, we can easily add the full conditions with Restriction and Relabeling.

(2) The case of reverse strongly probabilistic hp-bisimulation.

Firstly, we consider the case without Restriction and Relabeling. That is, we suffice to prove the following case by induction on the size $n$.

For $P\equiv P_1\parallel\cdots\parallel P_n$, with $n\geq 1$, we need to prove

\begin{eqnarray}
P\sim_{php}^{r} \{(P_1'\parallel\cdots\parallel P_n').(\alpha_1[m]\parallel\cdots\parallel \alpha_n[m]):  P_i\rightsquigarrow\xtworightarrow{\alpha_i[m]} P_i',i\in\{1,\cdots,n\}\nonumber\\
+\sum\{(P_1\parallel\cdots\parallel P_i'\parallel\cdots\parallel P_j'\parallel\cdots\parallel P_n).\tau:  P_i\rightsquigarrow\xtworightarrow{l[m]} P_i', P_j\rightsquigarrow\xtworightarrow{\overline{l}[m]} P_j',i<j\} \nonumber
\end{eqnarray}

For $n=1$, $P_1\sim_{php}^{r} P_1'.\alpha_1[m]: P_1\rightsquigarrow\xtworightarrow{\alpha_1[m]} P_1'$ is obvious. Then with a hypothesis $n$, we consider
$R\equiv P\parallel P_{n+1}$. By the reverse transition rules of Composition, we can get

\begin{eqnarray}
R\sim_{php}^{r} \{(P'\parallel P_{n+1}').(p[m]\parallel \alpha_{n+1}[m]):  P\rightsquigarrow\xtworightarrow{p[m]} P', P_{n+1}\rightsquigarrow\xtworightarrow{\alpha_{n+1}[m]} P_{n+1}',p\subseteq P\}\nonumber\\
+\sum\{(P'\parallel P_{n+1}').\tau:  P\rightsquigarrow\xtworightarrow{l[m]} P', P_{n+1}\rightsquigarrow\xtworightarrow{\overline{l}[m]} P_{n+1}'\} \nonumber
\end{eqnarray}

Now with the induction assumption $P\equiv P_1\parallel\cdots\parallel P_n$, the right-hand side can be reformulated as follows.

\begin{eqnarray}
\{(P_1'\parallel\cdots\parallel P_n'\parallel P_{n+1}).(\alpha_1[m]\parallel\cdots\parallel \alpha_n[m]\parallel \alpha_{n+1}[m]): \nonumber\\
 P_i\rightsquigarrow\xtworightarrow{\alpha_i[m]} P_i',i\in\{1,\cdots,n+1\}\nonumber\\
+\sum\{(P_1\parallel\cdots\parallel P_i'\parallel\cdots\parallel P_j'\parallel\cdots\parallel P_n\parallel P_{n+1}).\tau: \nonumber\\
 P_i\rightsquigarrow\xtworightarrow{l[m]} P_i', P_j\rightsquigarrow\xtworightarrow{\overline{l}[m]} P_j',i<j\} \nonumber\\
+\sum\{(P_1\parallel\cdots\parallel P_i'\parallel\cdots\parallel P_j'\parallel\cdots\parallel P_n\parallel P_{n+1}).\tau: \nonumber\\
 P_i\rightsquigarrow\xtworightarrow{l[m]} P_i', P_{n+1}\rightsquigarrow\xtworightarrow{\overline{l}[m]} P_{n+1}',i\in\{1,\cdots, n\}\} \nonumber
\end{eqnarray}

So,

\begin{eqnarray}
R\sim_{php}^{r} \{(P_1'\parallel\cdots\parallel P_n'\parallel P_{n+1}').(\alpha_1[m]\parallel\cdots\parallel \alpha_n[m]\parallel \alpha_{n+1}[m]): \nonumber\\
 P_i\rightsquigarrow\xtworightarrow{\alpha_i[m]} P_i',i\in\{1,\cdots,n+1\}\nonumber\\
+\sum\{(P_1\parallel\cdots\parallel P_i'\parallel\cdots\parallel P_j'\parallel\cdots\parallel P_n).\tau: \nonumber\\
 P_i\rightsquigarrow\xtworightarrow{l[m]} P_i', P_j\rightsquigarrow\xtworightarrow{\overline{l}[m]} P_j',1 \leq i<j\geq n+1\} \nonumber
\end{eqnarray}

Then, we can easily add the full conditions with Restriction and Relabeling.
\end{proof}

\begin{proposition}[Expansion law for FR strongly probabilistic hhp-bisimulation]
Let $P\equiv (P_1[f_1]\parallel\cdots\parallel P_n[f_n])\setminus L$, with $n\geq 1$. Then

\begin{eqnarray}
P\sim_{phhp}^{f} \{(f_1(\alpha_1)\parallel\cdots\parallel f_n(\alpha_n)).(P_1'[f_1]\parallel\cdots\parallel P_n'[f_n])\setminus L: \nonumber\\
 P_i\rightsquigarrow\xrightarrow{\alpha_i} P_i',i\in\{1,\cdots,n\},f_i(\alpha_i)\notin L\cup\overline{L}\} \nonumber\\
+\sum\{\tau.(P_1[f_1]\parallel\cdots\parallel P_i'[f_i]\parallel\cdots\parallel P_j'[f_j]\parallel\cdots\parallel P_n[f_n])\setminus L: \nonumber\\
 P_i\rightsquigarrow\xrightarrow{l_1} P_i', P_j\rightsquigarrow\xrightarrow{l_2} P_j',f_i(l_1)=\overline{f_j(l_2)},i<j\}\nonumber
\end{eqnarray}
\begin{eqnarray}
P\sim_{phhp}^{r} \{(P_1'[f_1]\parallel\cdots\parallel P_n'[f_n]).(f_1(\alpha_1[m])\parallel\cdots\parallel f_n(\alpha_n)[m])\setminus L: \nonumber\\
 P_i\rightsquigarrow\xtworightarrow{\alpha_i[m]} P_i',i\in\{1,\cdots,n\},f_i(\alpha_i)\notin L\cup\overline{L}\} \nonumber\\
+\sum\{(P_1[f_1]\parallel\cdots\parallel P_i'[f_i]\parallel\cdots\parallel P_j'[f_j]\parallel\cdots\parallel P_n[f_n]).\tau\setminus L: \nonumber\\
 P_i\rightsquigarrow\xtworightarrow{l_1[m]} P_i', P_j\rightsquigarrow\xtworightarrow{l_2[m]} P_j',f_i(l_1)=\overline{f_j(l_2)},i<j\}\nonumber
\end{eqnarray}
\end{proposition}

\begin{proof}
(1) The case of forward strongly probabilistic hhp-bisimulation.

Firstly, we consider the case without Restriction and Relabeling. That is, we suffice to prove the following case by induction on the size $n$.

For $P\equiv P_1\parallel\cdots\parallel P_n$, with $n\geq 1$, we need to prove

\begin{eqnarray}
P\sim_{phhp} \{(\alpha_1\parallel\cdots\parallel \alpha_n).(P_1'\parallel\cdots\parallel P_n'):  P_i\rightsquigarrow\xrightarrow{\alpha_i} P_i',i\in\{1,\cdots,n\}\nonumber\\
+\sum\{\tau.(P_1\parallel\cdots\parallel P_i'\parallel\cdots\parallel P_j'\parallel\cdots\parallel P_n):  P_i\rightsquigarrow\xrightarrow{l} P_i', P_j\rightsquigarrow\xrightarrow{\overline{l}} P_j',i<j\} \nonumber
\end{eqnarray}

For $n=1$, $P_1\sim_{phhp}^{f} \alpha_1.P_1': P_1\rightsquigarrow\xrightarrow{\alpha_1} P_1'$ is obvious. Then with a hypothesis $n$, we consider
$R\equiv P\parallel P_{n+1}$. By the forward transition rules of Composition, we can get

\begin{eqnarray}
R\sim_{phhp}^{f} \{(p\parallel \alpha_{n+1}).(P'\parallel P_{n+1}'):  P\rightsquigarrow\xrightarrow{p} P', P_{n+1}\rightsquigarrow\xrightarrow{\alpha_{n+1}} P_{n+1}',p\subseteq P\}\nonumber\\
+\sum\{\tau.(P'\parallel P_{n+1}'):  P\rightsquigarrow\xrightarrow{l} P', P_{n+1}\rightsquigarrow\xrightarrow{\overline{l}} P_{n+1}'\} \nonumber
\end{eqnarray}

Now with the induction assumption $P\equiv P_1\parallel\cdots\parallel P_n$, the right-hand side can be reformulated as follows.

\begin{eqnarray}
\{(\alpha_1\parallel\cdots\parallel \alpha_n\parallel \alpha_{n+1}).(P_1'\parallel\cdots\parallel P_n'\parallel P_{n+1}'): \nonumber\\
 P_i\rightsquigarrow\xrightarrow{\alpha_i} P_i',i\in\{1,\cdots,n+1\}\nonumber\\
+\sum\{\tau.(P_1\parallel\cdots\parallel P_i'\parallel\cdots\parallel P_j'\parallel\cdots\parallel P_n\parallel P_{n+1}): \nonumber\\
 P_i\rightsquigarrow\xrightarrow{l} P_i', P_j\rightsquigarrow\xrightarrow{\overline{l}} P_j',i<j\} \nonumber\\
+\sum\{\tau.(P_1\parallel\cdots\parallel P_i'\parallel\cdots\parallel P_j\parallel\cdots\parallel P_n\parallel P_{n+1}'): \nonumber\\
 P_i\rightsquigarrow\xrightarrow{l} P_i', P_{n+1}\rightsquigarrow\xrightarrow{\overline{l}} P_{n+1}',i\in\{1,\cdots, n\}\} \nonumber
\end{eqnarray}

So,

\begin{eqnarray}
R\sim_{phhp}^{f} \{(\alpha_1\parallel\cdots\parallel \alpha_n\parallel \alpha_{n+1}).(P_1'\parallel\cdots\parallel P_n'\parallel P_{n+1}'): \nonumber\\
 P_i\rightsquigarrow\xrightarrow{\alpha_i} P_i',i\in\{1,\cdots,n+1\}\nonumber\\
+\sum\{\tau.(P_1\parallel\cdots\parallel P_i'\parallel\cdots\parallel P_j'\parallel\cdots\parallel P_n): \nonumber\\
 P_i\rightsquigarrow\xrightarrow{l} P_i', P_j\rightsquigarrow\xrightarrow{\overline{l}} P_j',1 \leq i<j\geq n+1\} \nonumber
\end{eqnarray}

Then, we can easily add the full conditions with Restriction and Relabeling.

(2) The case of reverse strongly probabilistic hhp-bisimulation.

Firstly, we consider the case without Restriction and Relabeling. That is, we suffice to prove the following case by induction on the size $n$.

For $P\equiv P_1\parallel\cdots\parallel P_n$, with $n\geq 1$, we need to prove

\begin{eqnarray}
P\sim_{phhp}^{r} \{(P_1'\parallel\cdots\parallel P_n').(\alpha_1[m]\parallel\cdots\parallel \alpha_n[m]):  P_i\rightsquigarrow\xtworightarrow{\alpha_i[m]} P_i',i\in\{1,\cdots,n\}\nonumber\\
+\sum\{(P_1\parallel\cdots\parallel P_i'\parallel\cdots\parallel P_j'\parallel\cdots\parallel P_n).\tau:  P_i\rightsquigarrow\xtworightarrow{l[m]} P_i', P_j\rightsquigarrow\xtworightarrow{\overline{l}[m]} P_j',i<j\} \nonumber
\end{eqnarray}

For $n=1$, $P_1\sim_{phhp}^{r} P_1'.\alpha_1[m]: P_1\rightsquigarrow\xtworightarrow{\alpha_1[m]} P_1'$ is obvious. Then with a hypothesis $n$, we consider
$R\equiv P\parallel P_{n+1}$. By the reverse transition rules of Composition, we can get

\begin{eqnarray}
R\sim_{phhp}^{r} \{(P'\parallel P_{n+1}').(p[m]\parallel \alpha_{n+1}[m]):  P\rightsquigarrow\xtworightarrow{p[m]} P', P_{n+1}\rightsquigarrow\xtworightarrow{\alpha_{n+1}[m]} P_{n+1}',p\subseteq P\}\nonumber\\
+\sum\{(P'\parallel P_{n+1}').\tau:  P\rightsquigarrow\xtworightarrow{l[m]} P', P_{n+1}\rightsquigarrow\xtworightarrow{\overline{l}[m]} P_{n+1}'\} \nonumber
\end{eqnarray}

Now with the induction assumption $P\equiv P_1\parallel\cdots\parallel P_n$, the right-hand side can be reformulated as follows.

\begin{eqnarray}
\{(P_1'\parallel\cdots\parallel P_n'\parallel P_{n+1}).(\alpha_1[m]\parallel\cdots\parallel \alpha_n[m]\parallel \alpha_{n+1}[m]): \nonumber\\
 P_i\rightsquigarrow\xtworightarrow{\alpha_i[m]} P_i',i\in\{1,\cdots,n+1\}\nonumber\\
+\sum\{(P_1\parallel\cdots\parallel P_i'\parallel\cdots\parallel P_j'\parallel\cdots\parallel P_n\parallel P_{n+1}).\tau: \nonumber\\
 P_i\rightsquigarrow\xtworightarrow{l[m]} P_i', P_j\rightsquigarrow\xtworightarrow{\overline{l}[m]} P_j',i<j\} \nonumber\\
+\sum\{(P_1\parallel\cdots\parallel P_i'\parallel\cdots\parallel P_j'\parallel\cdots\parallel P_n\parallel P_{n+1}).\tau: \nonumber\\
 P_i\rightsquigarrow\xtworightarrow{l[m]} P_i', P_{n+1}\rightsquigarrow\xtworightarrow{\overline{l}[m]} P_{n+1}',i\in\{1,\cdots, n\}\} \nonumber
\end{eqnarray}

So,

\begin{eqnarray}
R\sim_{phhp}^{r} \{(P_1'\parallel\cdots\parallel P_n'\parallel P_{n+1}').(\alpha_1[m]\parallel\cdots\parallel \alpha_n[m]\parallel \alpha_{n+1}[m]): \nonumber\\
 P_i\rightsquigarrow\xtworightarrow{\alpha_i[m]} P_i',i\in\{1,\cdots,n+1\}\nonumber\\
+\sum\{(P_1\parallel\cdots\parallel P_i'\parallel\cdots\parallel P_j'\parallel\cdots\parallel P_n).\tau: \nonumber\\
 P_i\rightsquigarrow\xtworightarrow{l[m]} P_i', P_j\rightsquigarrow\xtworightarrow{\overline{l}[m]} P_j',1 \leq i<j\geq n+1\} \nonumber
\end{eqnarray}

Then, we can easily add the full conditions with Restriction and Relabeling.
\end{proof}

\begin{theorem}[Congruence for FR strongly probabilistic pomset bisimulation] \label{CSSB05}
We can enjoy the congruence for FR strongly probabilistic pomset bisimulation as follows.
\begin{enumerate}
  \item If $A\overset{\text{def}}{=}P$, then $A\sim_{pp}^{fr} P$;
  \item Let $P_1\sim_{pp}^{fr} P_2$. Then
        \begin{enumerate}
           \item $\alpha.P_1\sim_{pp}^f \alpha.P_2$;
           \item $(\alpha_1\parallel\cdots\parallel\alpha_n).P_1\sim_{pp}^f (\alpha_1\parallel\cdots\parallel\alpha_n).P_2$;
           \item $P_1.\alpha[m]\sim_{pp}^r P_2.\alpha[m]$;
           \item $P_1.(\alpha_1[m]\parallel\cdots\parallel\alpha_n[m])\sim_{pp}^r P_2.(\alpha_1[m]\parallel\cdots\parallel\alpha_n[m])$;
           \item $P_1+Q\sim_{pp}^{fr} P_2 +Q$;
           \item $P_1\boxplus_{\pi}Q\sim_{pp}^{fr} P_2 \boxplus_{\pi}Q$;
           \item $P_1\parallel Q\sim_{pp}^{fr} P_2\parallel Q$;
           \item $P_1\setminus L\sim_{pp}^{fr} P_2\setminus L$;
           \item $P_1[f]\sim_{pp}^{fr} P_2[f]$.
         \end{enumerate}
\end{enumerate}
\end{theorem}

\begin{proof}
\begin{enumerate}
  \item If $A\overset{\text{def}}{=}P$, then $A\sim_{pp}^{fr} P$. It is obvious.
  \item Let $P_1\sim_{pp}^{fr} P_2$. Then
        \begin{enumerate}
           \item $\alpha.P_1\sim_{pp}^f \alpha.P_2$. It is sufficient to prove the relation $R=\{(\alpha.P_1, \alpha.P_2)\}\cup \textbf{Id}$ is a F strongly probabilistic pomset bisimulation, we omit it;
           \item $(\alpha_1\parallel\cdots\parallel\alpha_n).P_1\sim_{pp}^f (\alpha_1\parallel\cdots\parallel\alpha_n).P_2$. It is sufficient to prove the relation $R=\{((\alpha_1\parallel\cdots\parallel\alpha_n).P_1, (\alpha_1\parallel\cdots\parallel\alpha_n).P_2)\}\cup \textbf{Id}$ is a F strongly probabilistic pomset bisimulation, we omit it;
           \item $P_1.\alpha[m]\sim_{pp}^r P_2.\alpha[m]$. It is sufficient to prove the relation $R=\{(P_1.\alpha[m], P_2.\alpha[m])\}\cup \textbf{Id}$ is a R strongly probabilistic pomset bisimulation, we omit it;
           \item $P_1.(\alpha_1[m]\parallel\cdots\parallel\alpha_n[m])\sim_{pp}^r P_2.(\alpha_1[m]\parallel\cdots\parallel\alpha_n[m])$. It is sufficient to prove the relation $R=\{(P_1.(\alpha_1[m]\parallel\cdots\parallel\alpha_n[m]), P_2.(\alpha_1[m]\parallel\cdots\parallel\alpha_n[m]))\}\cup \textbf{Id}$ is a R strongly probabilistic pomset bisimulation, we omit it;
           \item $P_1+Q\sim_{pp}^{fr} P_2 +Q$. It is sufficient to prove the relation $R=\{(P_1+Q, P_2+Q)\}\cup \textbf{Id}$ is a FR strongly probabilistic pomset bisimulation, we omit it;
           \item $P_1\boxplus_{\pi}Q\sim_{pp}^{fr} P_2 \boxplus_{\pi}Q$. It is sufficient to prove the relation $R=\{(P_1\boxplus_{\pi}Q, P_2\boxplus_{\pi}Q)\}\cup \textbf{Id}$ is a FR strongly probabilistic pomset bisimulation, we omit it;
           \item $P_1\parallel Q\sim_{pp}^{fr} P_2\parallel Q$. It is sufficient to prove the relation $R=\{(P_1\parallel Q, P_2\parallel Q)\}\cup \textbf{Id}$ is a FR strongly probabilistic pomset bisimulation, we omit it;
           \item $P_1\setminus L\sim_{pp}^{fr} P_2\setminus L$. It is sufficient to prove the relation $R=\{(P_1\setminus L, P_2\setminus L)\}\cup \textbf{Id}$ is a FR strongly probabilistic pomset bisimulation, we omit it;
           \item $P_1[f]\sim_{pp}^{fr} P_2[f]$. It is sufficient to prove the relation $R=\{(P_1[f], P_2[f])\}\cup \textbf{Id}$ is a FR strongly probabilistic pomset bisimulation, we omit it.
         \end{enumerate}
\end{enumerate}
\end{proof}

\begin{theorem}[Congruence for FR strongly probabilistic step bisimulation] \label{CSSB05}
We can enjoy the congruence for FR strongly probabilistic step bisimulation as follows.
\begin{enumerate}
  \item If $A\overset{\text{def}}{=}P$, then $A\sim_{ps}^{fr} P$;
  \item Let $P_1\sim_{ps}^{fr} P_2$. Then
        \begin{enumerate}
           \item $\alpha.P_1\sim_{ps}^f \alpha.P_2$;
           \item $(\alpha_1\parallel\cdots\parallel\alpha_n).P_1\sim_{ps}^f (\alpha_1\parallel\cdots\parallel\alpha_n).P_2$;
           \item $P_1.\alpha[m]\sim_{ps}^r P_2.\alpha[m]$;
           \item $P_1.(\alpha_1[m]\parallel\cdots\parallel\alpha_n[m])\sim_{ps}^r P_2.(\alpha_1[m]\parallel\cdots\parallel\alpha_n[m])$;
           \item $P_1+Q\sim_{ps}^{fr} P_2 +Q$;
           \item $P_1\boxplus_{\pi}Q\sim_{ps}^{fr} P_2 \boxplus_{\pi}Q$;
           \item $P_1\parallel Q\sim_{ps}^{fr} P_2\parallel Q$;
           \item $P_1\setminus L\sim_{ps}^{fr} P_2\setminus L$;
           \item $P_1[f]\sim_{ps}^{fr} P_2[f]$.
         \end{enumerate}
\end{enumerate}
\end{theorem}

\begin{proof}
\begin{enumerate}
  \item If $A\overset{\text{def}}{=}P$, then $A\sim_{ps}^{fr} P$. It is obvious.
  \item Let $P_1\sim_{ps}^{fr} P_2$. Then
        \begin{enumerate}
           \item $\alpha.P_1\sim_{ps}^f \alpha.P_2$. It is sufficient to prove the relation $R=\{(\alpha.P_1, \alpha.P_2)\}\cup \textbf{Id}$ is a F strongly probabilistic step bisimulation, we omit it;
           \item $(\alpha_1\parallel\cdots\parallel\alpha_n).P_1\sim_{ps}^f (\alpha_1\parallel\cdots\parallel\alpha_n).P_2$. It is sufficient to prove the relation $R=\{((\alpha_1\parallel\cdots\parallel\alpha_n).P_1, (\alpha_1\parallel\cdots\parallel\alpha_n).P_2)\}\cup \textbf{Id}$ is a F strongly probabilistic step bisimulation, we omit it;
           \item $P_1.\alpha[m]\sim_{ps}^r P_2.\alpha[m]$. It is sufficient to prove the relation $R=\{(P_1.\alpha[m], P_2.\alpha[m])\}\cup \textbf{Id}$ is a R strongly probabilistic step bisimulation, we omit it;
           \item $P_1.(\alpha_1[m]\parallel\cdots\parallel\alpha_n[m])\sim_{ps}^r P_2.(\alpha_1[m]\parallel\cdots\parallel\alpha_n[m])$. It is sufficient to prove the relation $R=\{(P_1.(\alpha_1[m]\parallel\cdots\parallel\alpha_n[m]), P_2.(\alpha_1[m]\parallel\cdots\parallel\alpha_n[m]))\}\cup \textbf{Id}$ is a R strongly probabilistic step bisimulation, we omit it;
           \item $P_1+Q\sim_{ps}^{fr} P_2 +Q$. It is sufficient to prove the relation $R=\{(P_1+Q, P_2+Q)\}\cup \textbf{Id}$ is a FR strongly probabilistic step bisimulation, we omit it;
           \item $P_1\boxplus_{\pi}Q\sim_{ps}^{fr} P_2 \boxplus_{\pi}Q$. It is sufficient to prove the relation $R=\{(P_1\boxplus_{\pi}Q, P_2\boxplus_{\pi}Q)\}\cup \textbf{Id}$ is a FR strongly probabilistic step bisimulation, we omit it;
           \item $P_1\parallel Q\sim_{ps}^{fr} P_2\parallel Q$. It is sufficient to prove the relation $R=\{(P_1\parallel Q, P_2\parallel Q)\}\cup \textbf{Id}$ is a FR strongly probabilistic step bisimulation, we omit it;
           \item $P_1\setminus L\sim_{ps}^{fr} P_2\setminus L$. It is sufficient to prove the relation $R=\{(P_1\setminus L, P_2\setminus L)\}\cup \textbf{Id}$ is a FR strongly probabilistic step bisimulation, we omit it;
           \item $P_1[f]\sim_{ps}^{fr} P_2[f]$. It is sufficient to prove the relation $R=\{(P_1[f], P_2[f])\}\cup \textbf{Id}$ is a FR strongly probabilistic step bisimulation, we omit it.
         \end{enumerate}
\end{enumerate}
\end{proof}

\begin{theorem}[Congruence for FR strongly probabilistic hp-bisimulation] \label{CSSB05}
We can enjoy the congruence for FR strongly probabilistic hp-bisimulation as follows.
\begin{enumerate}
  \item If $A\overset{\text{def}}{=}P$, then $A\sim_{php}^{fr} P$;
  \item Let $P_1\sim_{php}^{fr} P_2$. Then
        \begin{enumerate}
           \item $\alpha.P_1\sim_{php}^f \alpha.P_2$;
           \item $(\alpha_1\parallel\cdots\parallel\alpha_n).P_1\sim_{php}^f (\alpha_1\parallel\cdots\parallel\alpha_n).P_2$;
           \item $P_1.\alpha[m]\sim_{php}^r P_2.\alpha[m]$;
           \item $P_1.(\alpha_1[m]\parallel\cdots\parallel\alpha_n[m])\sim_{php}^r P_2.(\alpha_1[m]\parallel\cdots\parallel\alpha_n[m])$;
           \item $P_1+Q\sim_{php}^{fr} P_2 +Q$;
           \item $P_1\boxplus_{\pi}Q\sim_{php}^{fr} P_2 \boxplus_{\pi}Q$;
           \item $P_1\parallel Q\sim_{php}^{fr} P_2\parallel Q$;
           \item $P_1\setminus L\sim_{php}^{fr} P_2\setminus L$;
           \item $P_1[f]\sim_{php}^{fr} P_2[f]$.
         \end{enumerate}
\end{enumerate}
\end{theorem}

\begin{proof}
\begin{enumerate}
  \item If $A\overset{\text{def}}{=}P$, then $A\sim_{php}^{fr} P$. It is obvious.
  \item Let $P_1\sim_{php}^{fr} P_2$. Then
        \begin{enumerate}
           \item $\alpha.P_1\sim_{php}^f \alpha.P_2$. It is sufficient to prove the relation $R=\{(\alpha.P_1, \alpha.P_2)\}\cup \textbf{Id}$ is a F strongly probabilistic hp-bisimulation, we omit it;
           \item $(\alpha_1\parallel\cdots\parallel\alpha_n).P_1\sim_{php}^f (\alpha_1\parallel\cdots\parallel\alpha_n).P_2$. It is sufficient to prove the relation $R=\{((\alpha_1\parallel\cdots\parallel\alpha_n).P_1, (\alpha_1\parallel\cdots\parallel\alpha_n).P_2)\}\cup \textbf{Id}$ is a F strongly probabilistic hp-bisimulation, we omit it;
           \item $P_1.\alpha[m]\sim_{php}^r P_2.\alpha[m]$. It is sufficient to prove the relation $R=\{(P_1.\alpha[m], P_2.\alpha[m])\}\cup \textbf{Id}$ is a R strongly probabilistic hp-bisimulation, we omit it;
           \item $P_1.(\alpha_1[m]\parallel\cdots\parallel\alpha_n[m])\sim_{php}^r P_2.(\alpha_1[m]\parallel\cdots\parallel\alpha_n[m])$. It is sufficient to prove the relation $R=\{(P_1.(\alpha_1[m]\parallel\cdots\parallel\alpha_n[m]), P_2.(\alpha_1[m]\parallel\cdots\parallel\alpha_n[m]))\}\cup \textbf{Id}$ is a R strongly probabilistic hp-bisimulation, we omit it;
           \item $P_1+Q\sim_{php}^{fr} P_2 +Q$. It is sufficient to prove the relation $R=\{(P_1+Q, P_2+Q)\}\cup \textbf{Id}$ is a FR strongly probabilistic hp-bisimulation, we omit it;
           \item $P_1\boxplus_{\pi}Q\sim_{php}^{fr} P_2 \boxplus_{\pi}Q$. It is sufficient to prove the relation $R=\{(P_1\boxplus_{\pi}Q, P_2\boxplus_{\pi}Q)\}\cup \textbf{Id}$ is a FR strongly probabilistic hp-bisimulation, we omit it;
           \item $P_1\parallel Q\sim_{php}^{fr} P_2\parallel Q$. It is sufficient to prove the relation $R=\{(P_1\parallel Q, P_2\parallel Q)\}\cup \textbf{Id}$ is a FR strongly probabilistic hp-bisimulation, we omit it;
           \item $P_1\setminus L\sim_{php}^{fr} P_2\setminus L$. It is sufficient to prove the relation $R=\{(P_1\setminus L, P_2\setminus L)\}\cup \textbf{Id}$ is a FR strongly probabilistic hp-bisimulation, we omit it;
           \item $P_1[f]\sim_{php}^{fr} P_2[f]$. It is sufficient to prove the relation $R=\{(P_1[f], P_2[f])\}\cup \textbf{Id}$ is a FR strongly probabilistic hp-bisimulation, we omit it.
         \end{enumerate}
\end{enumerate}
\end{proof}

\begin{theorem}[Congruence for FR strongly probabilistic hhp-bisimulation] \label{CSSB05}
We can enjoy the congruence for FR strongly probabilistic hhp-bisimulation as follows.
\begin{enumerate}
  \item If $A\overset{\text{def}}{=}P$, then $A\sim_{phhp}^{fr} P$;
  \item Let $P_1\sim_{phhp}^{fr} P_2$. Then
        \begin{enumerate}
           \item $\alpha.P_1\sim_{phhp}^f \alpha.P_2$;
           \item $(\alpha_1\parallel\cdots\parallel\alpha_n).P_1\sim_{phhp}^f (\alpha_1\parallel\cdots\parallel\alpha_n).P_2$;
           \item $P_1.\alpha[m]\sim_{phhp}^r P_2.\alpha[m]$;
           \item $P_1.(\alpha_1[m]\parallel\cdots\parallel\alpha_n[m])\sim_{phhp}^r P_2.(\alpha_1[m]\parallel\cdots\parallel\alpha_n[m])$;
           \item $P_1+Q\sim_{phhp}^{fr} P_2 +Q$;
           \item $P_1\boxplus_{\pi}Q\sim_{phhp}^{fr} P_2 \boxplus_{\pi}Q$;
           \item $P_1\parallel Q\sim_{phhp}^{fr} P_2\parallel Q$;
           \item $P_1\setminus L\sim_{phhp}^{fr} P_2\setminus L$;
           \item $P_1[f]\sim_{phhp}^{fr} P_2[f]$.
         \end{enumerate}
\end{enumerate}
\end{theorem}

\begin{proof}
\begin{enumerate}
  \item If $A\overset{\text{def}}{=}P$, then $A\sim_{phhp}^{fr} P$. It is obvious.
  \item Let $P_1\sim_{phhp}^{fr} P_2$. Then
        \begin{enumerate}
           \item $\alpha.P_1\sim_{phhp}^f \alpha.P_2$. It is sufficient to prove the relation $R=\{(\alpha.P_1, \alpha.P_2)\}\cup \textbf{Id}$ is a F strongly probabilistic hhp-bisimulation, we omit it;
           \item $(\alpha_1\parallel\cdots\parallel\alpha_n).P_1\sim_{phhp}^f (\alpha_1\parallel\cdots\parallel\alpha_n).P_2$. It is sufficient to prove the relation $R=\{((\alpha_1\parallel\cdots\parallel\alpha_n).P_1, (\alpha_1\parallel\cdots\parallel\alpha_n).P_2)\}\cup \textbf{Id}$ is a F strongly probabilistic hhp-bisimulation, we omit it;
           \item $P_1.\alpha[m]\sim_{phhp}^r P_2.\alpha[m]$. It is sufficient to prove the relation $R=\{(P_1.\alpha[m], P_2.\alpha[m])\}\cup \textbf{Id}$ is a R strongly probabilistic hhp-bisimulation, we omit it;
           \item $P_1.(\alpha_1[m]\parallel\cdots\parallel\alpha_n[m])\sim_{phhp}^r P_2.(\alpha_1[m]\parallel\cdots\parallel\alpha_n[m])$. It is sufficient to prove the relation $R=\{(P_1.(\alpha_1[m]\parallel\cdots\parallel\alpha_n[m]), P_2.(\alpha_1[m]\parallel\cdots\parallel\alpha_n[m]))\}\cup \textbf{Id}$ is a R strongly probabilistic hhp-bisimulation, we omit it;
           \item $P_1+Q\sim_{phhp}^{fr} P_2 +Q$. It is sufficient to prove the relation $R=\{(P_1+Q, P_2+Q)\}\cup \textbf{Id}$ is a FR strongly probabilistic hhp-bisimulation, we omit it;
           \item $P_1\boxplus_{\pi}Q\sim_{phhp}^{fr} P_2 \boxplus_{\pi}Q$. It is sufficient to prove the relation $R=\{(P_1\boxplus_{\pi}Q, P_2\boxplus_{\pi}Q)\}\cup \textbf{Id}$ is a FR strongly probabilistic hhp-bisimulation, we omit it;
           \item $P_1\parallel Q\sim_{phhp}^{fr} P_2\parallel Q$. It is sufficient to prove the relation $R=\{(P_1\parallel Q, P_2\parallel Q)\}\cup \textbf{Id}$ is a FR strongly probabilistic hhp-bisimulation, we omit it;
           \item $P_1\setminus L\sim_{phhp}^{fr} P_2\setminus L$. It is sufficient to prove the relation $R=\{(P_1\setminus L, P_2\setminus L)\}\cup \textbf{Id}$ is a FR strongly probabilistic hhp-bisimulation, we omit it;
           \item $P_1[f]\sim_{phhp}^{fr} P_2[f]$. It is sufficient to prove the relation $R=\{(P_1[f], P_2[f])\}\cup \textbf{Id}$ is a FR strongly probabilistic hhp-bisimulation, we omit it.
         \end{enumerate}
\end{enumerate}
\end{proof}

\subsubsection{Recursion}

\begin{definition}[Weakly guarded recursive expression]
$X$ is weakly guarded in $E$if each occurrence of $X$ is with some subexpression $\alpha.F$or $(\alpha_1\parallel\cdots\parallel\alpha_n).F$or $F.\alpha[m]$or
$F.(\alpha_1[m]\parallel\cdots\parallel\alpha_n[m])$of $E$.
\end{definition}

\begin{lemma}\label{LUS06}
If the variables $\widetilde{X}$are weakly guarded in $E$, and $E\{\widetilde{P}/\widetilde{X}\}\rightsquigarrow\xrightarrow{\{\alpha_1,\cdots,\alpha_n\}} P'$or
$E\{\widetilde{P}/\widetilde{X}\}\rightsquigarrow\xtworightarrow{\{\alpha_1[m],\cdots,\alpha_n[m]\}} P'$, then $P'$takes the form
$E'\{\widetilde{P}/\widetilde{X}\}$for some expression $E'$, and moreover, for any $\widetilde{Q}$,
$E\{\widetilde{Q}/\widetilde{X}\}\rightsquigarrow\xrightarrow{\{\alpha_1,\cdots,\alpha_n\}} E'\{\widetilde{Q}/\widetilde{X}\}$or
$E\{\widetilde{Q}/\widetilde{X}\}\rightsquigarrow\xtworightarrow{\{\alpha_1[m],\cdots,\alpha_n[m]\}} E'\{\widetilde{Q}/\widetilde{X}\}$.
\end{lemma}

\begin{proof}
We only prove the case of forward transition.

It needs to induct on the depth of the inference of $E\{\widetilde{P}/\widetilde{X}\}\rightsquigarrow\xrightarrow{\{\alpha_1,\cdots,\alpha_n\}} P'$.

\begin{enumerate}
  \item Case $E\equiv Y$, a variable. Then $Y\notin \widetilde{X}$. Since $\widetilde{X}$are weakly guarded, $Y\{\widetilde{P}/\widetilde{X}\equiv Y\}\nrightarrow$, this case is
  impossible.
  \item Case $E\equiv\beta.F$. Then we must have $\alpha=\beta$, and $P'\equiv F\{\widetilde{P}/\widetilde{X}\}$, and
  $E\{\widetilde{Q}/\widetilde{X}\}\equiv  \beta.F\{\widetilde{Q}/\widetilde{X}\} \rightsquigarrow\xrightarrow{\beta} F\{\widetilde{Q}/\widetilde{X}\}$,
  then, let $E'$be $F$, as desired.
  \item Case $E\equiv(\beta_1\parallel\cdots\parallel\beta_n).F$. Then we must have $\alpha_i=\beta_i$for $1\leq i\leq n$, and $P'\equiv F\{\widetilde{P}/\widetilde{X}\}$, and
  $E\{\widetilde{Q}/\widetilde{X}\}\equiv (\beta_1\parallel\cdots\parallel\beta_n).F\{\widetilde{Q}/\widetilde{X}\} \rightsquigarrow\xrightarrow{\{\beta_1,\cdots,\beta_n\}} F\{\widetilde{Q}/\widetilde{X}\}$,
  then, let $E'$be $F$, as desired.
  \item Case $E\equiv E_1+E_2$. Then either $E_1\{\widetilde{P}/\widetilde{X}\} \rightsquigarrow\xrightarrow{\{\alpha_1,\cdots,\alpha_n\}} P'$or
  $E_2\{\widetilde{P}/\widetilde{X}\} \rightsquigarrow\xrightarrow{\{\alpha_1,\cdots,\alpha_n\}} P'$, then, we can apply this lemma in either case, as desired.
  \item Case $E\equiv E_1\parallel E_2$. There are four possibilities.
  \begin{enumerate}
    \item We may have $E_1\{\widetilde{P}/\widetilde{X}\} \rightsquigarrow\xrightarrow{\alpha} P_1'$and $E_2\{\widetilde{P}/\widetilde{X}\}\nrightarrow$
    with $P'\equiv P_1'\parallel (E_2\{\widetilde{P}/\widetilde{X}\})$, then by applying this lemma, $P_1'$ is of the form $E_1'\{\widetilde{P}/\widetilde{X}\}$, and for any $Q$,
    $E_1\{\widetilde{Q}/\widetilde{X}\}\rightsquigarrow\xrightarrow{\alpha}  E_1'\{\widetilde{Q}/\widetilde{X}\}$. So, $P'$ is of the form
    $E_1'\parallel E_2\{\widetilde{P}/\widetilde{X}\}$, and for any $Q$,
    $E\{\widetilde{Q}/\widetilde{X}\}\equiv E_1\{\widetilde{Q}/\widetilde{X}\}\parallel E_2\{\widetilde{Q}/\widetilde{X}\}\rightsquigarrow\xrightarrow{\alpha} (E_1'\parallel E_2)\{\widetilde{Q}/\widetilde{X}\}$,
    then, let $E'$be $E_1'\parallel E_2$, as desired.
    \item We may have $E_2\{\widetilde{P}/\widetilde{X}\} \rightsquigarrow\xrightarrow{\alpha} P_2'$and $E_1\{\widetilde{P}/\widetilde{X}\}\nrightarrow$
    with $P'\equiv P_2'\parallel (E_1\{\widetilde{P}/\widetilde{X}\})$, this case can be prove similarly to the above subcase, as desired.
    \item We may have $E_1\{\widetilde{P}/\widetilde{X}\} \rightsquigarrow\xrightarrow{\alpha} P_1'$and
    $E_2\{\widetilde{P}/\widetilde{X}\}\rightsquigarrow\xrightarrow{\beta} P_2'$with $\alpha\neq\overline{\beta}$and $P'\equiv P_1'\parallel P_2'$, then by
    applying this lemma, $P_1'$ is of the form $E_1'\{\widetilde{P}/\widetilde{X}\}$, and for any $Q$,
    $E_1\{\widetilde{Q}/\widetilde{X}\}\rightsquigarrow\xrightarrow{\alpha}  E_1'\{\widetilde{Q}/\widetilde{X}\}$; $P_2'$ is of the form
    $E_2'\{\widetilde{P}/\widetilde{X}\}$, and for any $Q$, $E_2\{\widetilde{Q}/\widetilde{X}\}\rightsquigarrow\xrightarrow{\alpha}  E_2'\{\widetilde{Q}/\widetilde{X}\}$.
    So, $P'$ is of the form $E_1'\parallel E_2'\{\widetilde{P}/\widetilde{X}\}$, and for any $Q$,
    $E\{\widetilde{Q}/\widetilde{X}\}\equiv E_1\{\widetilde{Q}/\widetilde{X}\}\parallel E_2\{\widetilde{Q}/\widetilde{X}\}\rightsquigarrow\xrightarrow{\{\alpha,\beta\}}
     (E_1'\parallel E_2')\{\widetilde{Q}/\widetilde{X}\}$, then, let $E'$be $E_1'\parallel E_2'$, as desired.
    \item We may have $E_1\{\widetilde{P}/\widetilde{X}\} \rightsquigarrow\xrightarrow{l} P_1'$and
    $E_2\{\widetilde{P}/\widetilde{X}\}\rightsquigarrow\xrightarrow{\overline{l}} P_2'$with $P'\equiv P_1'\parallel P_2'$, then by applying this lemma,
    $P_1'$ is of the form $E_1'\{\widetilde{P}/\widetilde{X}\}$, and for any $Q$, $E_1\{\widetilde{Q}/\widetilde{X}\}\rightsquigarrow\xrightarrow{l}  E_1'\{\widetilde{Q}/\widetilde{X}\}$;
    $P_2'$ is of the form $E_2'\{\widetilde{P}/\widetilde{X}\}$, and for any $Q$, $E_2\{\widetilde{Q}/\widetilde{X}\}\rightsquigarrow\xrightarrow{\overline{l}} E_2'\{\widetilde{Q}/\widetilde{X}\}$.
    So, $P'$ is of the form $E_1'\parallel E_2'\{\widetilde{P}/\widetilde{X}\}$, and for any $Q$, $E\{\widetilde{Q}/\widetilde{X}\}\equiv E_1\{\widetilde{Q}/\widetilde{X}\}\parallel E_2\{\widetilde{Q}/\widetilde{X}\}
    \rightsquigarrow\xrightarrow{\tau}  (E_1'\parallel E_2')\{\widetilde{Q}/\widetilde{X}\}$, then, let $E'$be $E_1'\parallel E_2'$, as desired.
  \end{enumerate}
  \item Case $E\equiv F[R]$and $E\equiv F\setminus L$. These cases can be prove similarly to the above case.
  \item Case $E\equiv C$, an agent constant defined by $C\overset{\text{def}}{=}R$. Then there is no $X\in\widetilde{X}$occurring in $E$, so
  $C\rightsquigarrow\xrightarrow{\{\alpha_1,\cdots,\alpha_n\}} P'$, let $E'$be $P'$, as desired.
\end{enumerate}

For the case of reverse transition, it can be proven similarly, we omit it.
\end{proof}

\begin{theorem}[Unique solution of equations for FR strongly probabilistic pomset bisimulation]
Let the recursive expressions $E_i(i\in I)$contain at most the variables $X_i(i\in I)$, and let each $X_j(j\in I)$be weakly guarded in each $E_i$. Then,

If $\widetilde{P}\sim_{pp}^{fr} \widetilde{E}\{\widetilde{P}/\widetilde{X}\}$and $\widetilde{Q}\sim_{pp}^{fr} \widetilde{E}\{\widetilde{Q}/\widetilde{X}\}$, then
$\widetilde{P}\sim_{pp}^{fr} \widetilde{Q}$.
\end{theorem}

\begin{proof}
We only prove the case of forward transition.

It is sufficient to induct on the depth of the inference of $E\{\widetilde{P}/\widetilde{X}\}\rightsquigarrow\xrightarrow{\{\alpha_1,\cdots,\alpha_n\}} P'$.

\begin{enumerate}
  \item Case $E\equiv X_i$. Then we have $E\{\widetilde{P}/\widetilde{X}\}\equiv  P_i\rightsquigarrow\xrightarrow{\{\alpha_1,\cdots,\alpha_n\}} P'$,
  since $P_i\sim_{pp}^{fr} E_i\{\widetilde{P}/\widetilde{X}\}$, we have $E_i\{\widetilde{P}/\widetilde{X}\}\rightsquigarrow\xrightarrow{\{\alpha_1,\cdots,\alpha_n\}} P''\sim_{pp}^{fr}  P'$.
  Since $\widetilde{X}$are weakly guarded in $E_i$, by Lemma \ref{LUS06}, $P''\equiv E'\{\widetilde{P}/\widetilde{X}\}$and $E_i\{\widetilde{P}/\widetilde{X}\}
  \rightsquigarrow\xrightarrow{\{\alpha_1,\cdots,\alpha_n\}}  E'\{\widetilde{P}/\widetilde{X}\}$. Since
  $E\{\widetilde{Q}/\widetilde{X}\}\equiv X_i\{\widetilde{Q}/\widetilde{X}\} \equiv Q_i\sim_{pp}^{fr} E_i\{\widetilde{Q}/\widetilde{X}\}$, $E\{\widetilde{Q}/\widetilde{X}\}\rightsquigarrow\xrightarrow{\{\alpha_1,\cdots,\alpha_n\}} Q'\sim_{pp}^{fr}  E'\{\widetilde{Q}/\widetilde{X}\}$.
  So, $P'\sim_{pp}^{fr} Q'$, as desired.
  \item Case $E\equiv\alpha.F$. This case can be proven similarly.
  \item Case $E\equiv(\alpha_1\parallel\cdots\parallel\alpha_n).F$. This case can be proven similarly.
  \item Case $E\equiv E_1+E_2$. We have $E_i\{\widetilde{P}/\widetilde{X}\} \rightsquigarrow\xrightarrow{\{\alpha_1,\cdots,\alpha_n\}} P'$,
  $E_i\{\widetilde{Q}/\widetilde{X}\} \rightsquigarrow\xrightarrow{\{\alpha_1,\cdots,\alpha_n\}} Q'$, then, $P'\sim_{pp}^{fr} Q'$, as desired.
  \item Case $E\equiv E_1\parallel E_2$, $E\equiv F[R]$and $E\equiv F\setminus L$, $E\equiv C$. These cases can be prove similarly to the above case.
\end{enumerate}

For the case of reverse transition, it can be proven similarly, we omit it.
\end{proof}

\begin{theorem}[Unique solution of equations for FR strongly probabilistic step bisimulation]
Let the recursive expressions $E_i(i\in I)$contain at most the variables $X_i(i\in I)$, and let each $X_j(j\in I)$be weakly guarded in each $E_i$. Then,

If $\widetilde{P}\sim_{ps}^{fr} \widetilde{E}\{\widetilde{P}/\widetilde{X}\}$and $\widetilde{Q}\sim_{ps}^{fr} \widetilde{E}\{\widetilde{Q}/\widetilde{X}\}$, then
$\widetilde{P}\sim_{ps}^{fr} \widetilde{Q}$.
\end{theorem}

\begin{proof}
We only prove the case of forward transition.

It is sufficient to induct on the depth of the inference of $E\{\widetilde{P}/\widetilde{X}\}\rightsquigarrow\xrightarrow{\{\alpha_1,\cdots,\alpha_n\}} P'$.

\begin{enumerate}
  \item Case $E\equiv X_i$. Then we have $E\{\widetilde{P}/\widetilde{X}\}\equiv  P_i\rightsquigarrow\xrightarrow{\{\alpha_1,\cdots,\alpha_n\}} P'$,
  since $P_i\sim_{ps}^{fr} E_i\{\widetilde{P}/\widetilde{X}\}$, we have $E_i\{\widetilde{P}/\widetilde{X}\}\rightsquigarrow\xrightarrow{\{\alpha_1,\cdots,\alpha_n\}} P''\sim_{ps}^{fr}  P'$.
  Since $\widetilde{X}$are weakly guarded in $E_i$, by Lemma \ref{LUS06}, $P''\equiv E'\{\widetilde{P}/\widetilde{X}\}$and $E_i\{\widetilde{P}/\widetilde{X}\}
  \rightsquigarrow\xrightarrow{\{\alpha_1,\cdots,\alpha_n\}}  E'\{\widetilde{P}/\widetilde{X}\}$. Since
  $E\{\widetilde{Q}/\widetilde{X}\}\equiv X_i\{\widetilde{Q}/\widetilde{X}\} \equiv Q_i\sim_{ps}^{fr} E_i\{\widetilde{Q}/\widetilde{X}\}$, $E\{\widetilde{Q}/\widetilde{X}\}\rightsquigarrow\xrightarrow{\{\alpha_1,\cdots,\alpha_n\}} Q'\sim_{ps}^{fr}  E'\{\widetilde{Q}/\widetilde{X}\}$.
  So, $P'\sim_{ps}^{fr} Q'$, as desired.
  \item Case $E\equiv\alpha.F$. This case can be proven similarly.
  \item Case $E\equiv(\alpha_1\parallel\cdots\parallel\alpha_n).F$. This case can be proven similarly.
  \item Case $E\equiv E_1+E_2$. We have $E_i\{\widetilde{P}/\widetilde{X}\} \rightsquigarrow\xrightarrow{\{\alpha_1,\cdots,\alpha_n\}} P'$,
  $E_i\{\widetilde{Q}/\widetilde{X}\} \rightsquigarrow\xrightarrow{\{\alpha_1,\cdots,\alpha_n\}} Q'$, then, $P'\sim_{ps}^{fr} Q'$, as desired.
  \item Case $E\equiv E_1\parallel E_2$, $E\equiv F[R]$and $E\equiv F\setminus L$, $E\equiv C$. These cases can be prove similarly to the above case.
\end{enumerate}

For the case of reverse transition, it can be proven similarly, we omit it.
\end{proof}

\begin{theorem}[Unique solution of equations for FR strongly probabilistic hp-bisimulation]
Let the recursive expressions $E_i(i\in I)$contain at most the variables $X_i(i\in I)$, and let each $X_j(j\in I)$be weakly guarded in each $E_i$. Then,

If $\widetilde{P}\sim_{php}^{fr} \widetilde{E}\{\widetilde{P}/\widetilde{X}\}$and $\widetilde{Q}\sim_{php}^{fr} \widetilde{E}\{\widetilde{Q}/\widetilde{X}\}$, then
$\widetilde{P}\sim_{php}^{fr} \widetilde{Q}$.
\end{theorem}

\begin{proof}
We only prove the case of forward transition.

It is sufficient to induct on the depth of the inference of $E\{\widetilde{P}/\widetilde{X}\}\rightsquigarrow\xrightarrow{\{\alpha_1,\cdots,\alpha_n\}} P'$.

\begin{enumerate}
  \item Case $E\equiv X_i$. Then we have $E\{\widetilde{P}/\widetilde{X}\}\equiv  P_i\rightsquigarrow\xrightarrow{\{\alpha_1,\cdots,\alpha_n\}} P'$,
  since $P_i\sim_{php}^{fr} E_i\{\widetilde{P}/\widetilde{X}\}$, we have $E_i\{\widetilde{P}/\widetilde{X}\}\rightsquigarrow\xrightarrow{\{\alpha_1,\cdots,\alpha_n\}} P''\sim_{php}^{fr}  P'$.
  Since $\widetilde{X}$are weakly guarded in $E_i$, by Lemma \ref{LUS06}, $P''\equiv E'\{\widetilde{P}/\widetilde{X}\}$and $E_i\{\widetilde{P}/\widetilde{X}\}
  \rightsquigarrow\xrightarrow{\{\alpha_1,\cdots,\alpha_n\}}  E'\{\widetilde{P}/\widetilde{X}\}$. Since
  $E\{\widetilde{Q}/\widetilde{X}\}\equiv X_i\{\widetilde{Q}/\widetilde{X}\} \equiv Q_i\sim_{php}^{fr} E_i\{\widetilde{Q}/\widetilde{X}\}$, $E\{\widetilde{Q}/\widetilde{X}\}\rightsquigarrow\xrightarrow{\{\alpha_1,\cdots,\alpha_n\}} Q'\sim_{php}^{fr}  E'\{\widetilde{Q}/\widetilde{X}\}$.
  So, $P'\sim_{php}^{fr} Q'$, as desired.
  \item Case $E\equiv\alpha.F$. This case can be proven similarly.
  \item Case $E\equiv(\alpha_1\parallel\cdots\parallel\alpha_n).F$. This case can be proven similarly.
  \item Case $E\equiv E_1+E_2$. We have $E_i\{\widetilde{P}/\widetilde{X}\} \rightsquigarrow\xrightarrow{\{\alpha_1,\cdots,\alpha_n\}} P'$,
  $E_i\{\widetilde{Q}/\widetilde{X}\} \rightsquigarrow\xrightarrow{\{\alpha_1,\cdots,\alpha_n\}} Q'$, then, $P'\sim_{php}^{fr} Q'$, as desired.
  \item Case $E\equiv E_1\parallel E_2$, $E\equiv F[R]$and $E\equiv F\setminus L$, $E\equiv C$. These cases can be prove similarly to the above case.
\end{enumerate}

For the case of reverse transition, it can be proven similarly, we omit it.
\end{proof}

\begin{theorem}[Unique solution of equations for FR strongly probabilistic hhp-bisimulation]
Let the recursive expressions $E_i(i\in I)$contain at most the variables $X_i(i\in I)$, and let each $X_j(j\in I)$be weakly guarded in each $E_i$. Then,

If $\widetilde{P}\sim_{phhp}^{fr} \widetilde{E}\{\widetilde{P}/\widetilde{X}\}$and $\widetilde{Q}\sim_{phhp}^{fr} \widetilde{E}\{\widetilde{Q}/\widetilde{X}\}$, then
$\widetilde{P}\sim_{phhp}^{fr} \widetilde{Q}$.
\end{theorem}

\begin{proof}
We only prove the case of forward transition.

It is sufficient to induct on the depth of the inference of $E\{\widetilde{P}/\widetilde{X}\}\rightsquigarrow\xrightarrow{\{\alpha_1,\cdots,\alpha_n\}} P'$.

\begin{enumerate}
  \item Case $E\equiv X_i$. Then we have $E\{\widetilde{P}/\widetilde{X}\}\equiv  P_i\rightsquigarrow\xrightarrow{\{\alpha_1,\cdots,\alpha_n\}} P'$,
  since $P_i\sim_{phhp}^{fr} E_i\{\widetilde{P}/\widetilde{X}\}$, we have $E_i\{\widetilde{P}/\widetilde{X}\}\rightsquigarrow\xrightarrow{\{\alpha_1,\cdots,\alpha_n\}} P''\sim_{phhp}^{fr}  P'$.
  Since $\widetilde{X}$are weakly guarded in $E_i$, by Lemma \ref{LUS06}, $P''\equiv E'\{\widetilde{P}/\widetilde{X}\}$and $E_i\{\widetilde{P}/\widetilde{X}\}
  \rightsquigarrow\xrightarrow{\{\alpha_1,\cdots,\alpha_n\}}  E'\{\widetilde{P}/\widetilde{X}\}$. Since
  $E\{\widetilde{Q}/\widetilde{X}\}\equiv X_i\{\widetilde{Q}/\widetilde{X}\} \equiv Q_i\sim_{phhp}^{fr} E_i\{\widetilde{Q}/\widetilde{X}\}$, $E\{\widetilde{Q}/\widetilde{X}\}\rightsquigarrow\xrightarrow{\{\alpha_1,\cdots,\alpha_n\}} Q'\sim_{phhp}^{fr}  E'\{\widetilde{Q}/\widetilde{X}\}$.
  So, $P'\sim_{phhp}^{fr} Q'$, as desired.
  \item Case $E\equiv\alpha.F$. This case can be proven similarly.
  \item Case $E\equiv(\alpha_1\parallel\cdots\parallel\alpha_n).F$. This case can be proven similarly.
  \item Case $E\equiv E_1+E_2$. We have $E_i\{\widetilde{P}/\widetilde{X}\} \rightsquigarrow\xrightarrow{\{\alpha_1,\cdots,\alpha_n\}} P'$,
  $E_i\{\widetilde{Q}/\widetilde{X}\} \rightsquigarrow\xrightarrow{\{\alpha_1,\cdots,\alpha_n\}} Q'$, then, $P'\sim_{phhp}^{fr} Q'$, as desired.
  \item Case $E\equiv E_1\parallel E_2$, $E\equiv F[R]$and $E\equiv F\setminus L$, $E\equiv C$. These cases can be prove similarly to the above case.
\end{enumerate}

For the case of reverse transition, it can be proven similarly, we omit it.
\end{proof}

\subsection{Weak Bisimulations}\label{wtcbctcpr}

\subsubsection{Laws}

Remembering that $\tau$ can neither be restricted nor relabeled, we know that the monoid laws, the static laws, the guards laws, and the new expansion law
still hold with respect to the corresponding FR weakly probabilistic truly concurrent bisimulations. And also, we can enjoy the congruence of Prefix, Summation, Composition, Restriction, Relabelling
and Constants with respect to corresponding FR weakly probabilistic truly concurrent bisimulations. We will not retype these laws, and just give the $\tau$-specific laws. The forward and reverse
transition rules of $\tau$ are shown in Table \ref{TRForTAU04}, where $\rightsquigarrow\xrightarrow{\tau}\surd$ is a predicate which represents a successful termination after execution of the silent
step $\tau$.

\begin{center}
    \begin{table}
        $$\frac{}{\tau\rightsquigarrow\xrightarrow{\tau}\surd}$$
        $$\frac{}{\tau\rightsquigarrow\xtworightarrow{\tau}\surd}$$
        \caption{Forward and reverse transition rules of $\tau$}
        \label{TRForTAU04}
    \end{table}
\end{center}

\begin{proposition}[$\tau$ laws for FR weakly probabilistic pomset bisimulation]
The $\tau$ laws for FR weakly probabilistic pomset bisimulation is as follows.
\begin{enumerate}
  \item $P\approx_{pp}^f \tau.P$;
  \item $P\approx_{pp}^r P.\tau$;
  \item $\alpha.\tau.P\approx_{pp}^f \alpha.P$;
  \item $P.\tau.\alpha[m]\approx_{pp}^r P.\alpha[m]$;
  \item $(\alpha_1\parallel\cdots\parallel\alpha_n).\tau.P\approx_{pp}^f (\alpha_1\parallel\cdots\parallel\alpha_n).P$;
  \item $P.\tau.(\alpha_1[m]\parallel\cdots\parallel\alpha_n[m])\approx_{pp}^r P.(\alpha_1[m]\parallel\cdots\parallel\alpha_n[m])$;
  \item $P+\tau.P\approx_{pp}^f \tau.P$;
  \item $P+P.\tau\approx_{pp}^r P.\tau$;
  \item $P\cdot((Q+\tau\cdot(Q+R))\boxplus_{\pi}S)\approx_{pp}^{f}P\cdot((Q+R)\boxplus_{\pi}S)$;
  \item $((Q+(Q+R)\cdot\tau)\boxplus_{\pi}S)\cdot P\approx_{pp}^{r}((Q+R)\boxplus_{\pi}S)\cdot P$;
  \item $P\approx_{pp}^{fr} \tau\parallel P$.
\end{enumerate}
\end{proposition}

\begin{proof}
\begin{enumerate}
  \item $P\approx_{pp}^f \tau.P$. It is sufficient to prove the relation $R=\{(P, \tau.P)\}\cup \textbf{Id}$ is a F weakly probabilistic pomset bisimulation, we omit it;
  \item $P\approx_{pp}^r P.\tau$. It is sufficient to prove the relation $R=\{(P, P.\tau)\}\cup \textbf{Id}$ is a R weakly probabilistic pomset bisimulation, we omit it;
  \item $\alpha.\tau.P\approx_{pp}^f \alpha.P$. It is sufficient to prove the relation $R=\{(\alpha.\tau.P, \alpha.P)\}\cup \textbf{Id}$ is a F weakly probabilistic pomset bisimulation, we omit it;
  \item $P.\tau.\alpha[m]\approx_{pp}^r P.\alpha[m]$. It is sufficient to prove the relation $R=\{(P.\tau.\alpha[m], P.\alpha[m])\}\cup \textbf{Id}$ is a R weakly probabilistic pomset bisimulation, we omit it;
  \item $(\alpha_1\parallel\cdots\parallel\alpha_n).\tau.P\approx_{pp}^f (\alpha_1\parallel\cdots\parallel\alpha_n).P$. It is sufficient to prove the relation $R=\{((\alpha_1\parallel\cdots\parallel\alpha_n).\tau.P, (\alpha_1\parallel\cdots\parallel\alpha_n).P)\}\cup \textbf{Id}$ is a F weakly probabilistic pomset bisimulation, we omit it;
  \item $P.\tau.(\alpha_1[m]\parallel\cdots\parallel\alpha_n[m])\approx_{pp}^r P.(\alpha_1[m]\parallel\cdots\parallel\alpha_n[m])$. It is sufficient to prove the relation $R=\{(P.\tau.(\alpha_1[m]\parallel\cdots\parallel\alpha_n[m]), P.(\alpha_1[m]\parallel\cdots\parallel\alpha_n[m]))\}\cup \textbf{Id}$ is a R weakly probabilistic pomset bisimulation, we omit it;
  \item $P+\tau.P\approx_{pp}^f \tau.P$. It is sufficient to prove the relation $R=\{(P+\tau.P, \tau.P)\}\cup \textbf{Id}$ is a F weakly probabilistic pomset bisimulation, we omit it;
  \item $P+P.\tau\approx_{pp}^r P.\tau$. It is sufficient to prove the relation $R=\{(P+P.\tau, P.\tau)\}\cup \textbf{Id}$ is a R weakly probabilistic pomset bisimulation, we omit it;
  \item $P\cdot((Q+\tau\cdot(Q+R))\boxplus_{\pi}S)\approx_{pp}^{f}P\cdot((Q+R)\boxplus_{\pi}S)$. It is sufficient to prove the relation $R=\{(P\cdot((Q+\tau\cdot(Q+R))\boxplus_{\pi}S), P\cdot((Q+R)\boxplus_{\pi}S))\}\cup \textbf{Id}$ is a F weakly probabilistic pomset bisimulation, we omit it;
  \item $((Q+(Q+R)\cdot\tau)\boxplus_{\pi}S)\cdot P\approx_{pp}^{r}((Q+R)\boxplus_{\pi}S)\cdot P$. It is sufficient to prove the relation $R=\{(((Q+(Q+R)\cdot\tau)\boxplus_{\pi}S)\cdot P, ((Q+R)\boxplus_{\pi}S)\cdot P)\}\cup \textbf{Id}$ is a R weakly probabilistic pomset bisimulation, we omit it;
  \item $P\approx_{pp}^{fr} \tau\parallel P$. It is sufficient to prove the relation $R=\{(P, \tau\parallel P)\}\cup \textbf{Id}$ is a FR weakly probabilistic pomset bisimulation, we omit it.
\end{enumerate}
\end{proof}

\begin{proposition}[$\tau$ laws for FR weakly probabilistic step bisimulation]
The $\tau$ laws for FR weakly probabilistic step bisimulation is as follows.
\begin{enumerate}
  \item $P\approx_{ps}^f \tau.P$;
  \item $P\approx_{ps}^r P.\tau$;
  \item $\alpha.\tau.P\approx_{ps}^f \alpha.P$;
  \item $P.\tau.\alpha[m]\approx_{ps}^r P.\alpha[m]$;
  \item $(\alpha_1\parallel\cdots\parallel\alpha_n).\tau.P\approx_{ps}^f (\alpha_1\parallel\cdots\parallel\alpha_n).P$;
  \item $P.\tau.(\alpha_1[m]\parallel\cdots\parallel\alpha_n[m])\approx_{ps}^r P.(\alpha_1[m]\parallel\cdots\parallel\alpha_n[m])$;
  \item $P+\tau.P\approx_{ps}^f \tau.P$;
  \item $P+P.\tau\approx_{ps}^r P.\tau$;
  \item $P\cdot((Q+\tau\cdot(Q+R))\boxplus_{\pi}S)\approx_{ps}^{f}P\cdot((Q+R)\boxplus_{\pi}S)$;
  \item $((Q+(Q+R)\cdot\tau)\boxplus_{\pi}S)\cdot P\approx_{ps}^{r}((Q+R)\boxplus_{\pi}S)\cdot P$;
  \item $P\approx_{ps}^{fr} \tau\parallel P$.
\end{enumerate}
\end{proposition}

\begin{proof}
\begin{enumerate}
  \item $P\approx_{ps}^f \tau.P$. It is sufficient to prove the relation $R=\{(P, \tau.P)\}\cup \textbf{Id}$ is a F weakly probabilistic step bisimulation, we omit it;
  \item $P\approx_{ps}^r P.\tau$. It is sufficient to prove the relation $R=\{(P, P.\tau)\}\cup \textbf{Id}$ is a R weakly probabilistic step bisimulation, we omit it;
  \item $\alpha.\tau.P\approx_{ps}^f \alpha.P$. It is sufficient to prove the relation $R=\{(\alpha.\tau.P, \alpha.P)\}\cup \textbf{Id}$ is a F weakly probabilistic step bisimulation, we omit it;
  \item $P.\tau.\alpha[m]\approx_{ps}^r P.\alpha[m]$. It is sufficient to prove the relation $R=\{(P.\tau.\alpha[m], P.\alpha[m])\}\cup \textbf{Id}$ is a R weakly probabilistic step bisimulation, we omit it;
  \item $(\alpha_1\parallel\cdots\parallel\alpha_n).\tau.P\approx_{ps}^f (\alpha_1\parallel\cdots\parallel\alpha_n).P$. It is sufficient to prove the relation $R=\{((\alpha_1\parallel\cdots\parallel\alpha_n).\tau.P, (\alpha_1\parallel\cdots\parallel\alpha_n).P)\}\cup \textbf{Id}$ is a F weakly probabilistic step bisimulation, we omit it;
  \item $P.\tau.(\alpha_1[m]\parallel\cdots\parallel\alpha_n[m])\approx_{ps}^r P.(\alpha_1[m]\parallel\cdots\parallel\alpha_n[m])$. It is sufficient to prove the relation $R=\{(P.\tau.(\alpha_1[m]\parallel\cdots\parallel\alpha_n[m]), P.(\alpha_1[m]\parallel\cdots\parallel\alpha_n[m]))\}\cup \textbf{Id}$ is a R weakly probabilistic step bisimulation, we omit it;
  \item $P+\tau.P\approx_{ps}^f \tau.P$. It is sufficient to prove the relation $R=\{(P+\tau.P, \tau.P)\}\cup \textbf{Id}$ is a F weakly probabilistic step bisimulation, we omit it;
  \item $P+P.\tau\approx_{ps}^r P.\tau$. It is sufficient to prove the relation $R=\{(P+P.\tau, P.\tau)\}\cup \textbf{Id}$ is a R weakly probabilistic step bisimulation, we omit it;
  \item $P\cdot((Q+\tau\cdot(Q+R))\boxplus_{\pi}S)\approx_{ps}^{f}P\cdot((Q+R)\boxplus_{\pi}S)$. It is sufficient to prove the relation $R=\{(P\cdot((Q+\tau\cdot(Q+R))\boxplus_{\pi}S), P\cdot((Q+R)\boxplus_{\pi}S))\}\cup \textbf{Id}$ is a F weakly probabilistic step bisimulation, we omit it;
  \item $((Q+(Q+R)\cdot\tau)\boxplus_{\pi}S)\cdot P\approx_{ps}^{r}((Q+R)\boxplus_{\pi}S)\cdot P$. It is sufficient to prove the relation $R=\{(((Q+(Q+R)\cdot\tau)\boxplus_{\pi}S)\cdot P, ((Q+R)\boxplus_{\pi}S)\cdot P)\}\cup \textbf{Id}$ is a R weakly probabilistic step bisimulation, we omit it;
  \item $P\approx_{ps}^{fr} \tau\parallel P$. It is sufficient to prove the relation $R=\{(P, \tau\parallel P)\}\cup \textbf{Id}$ is a FR weakly probabilistic step bisimulation, we omit it.
\end{enumerate}
\end{proof}

\begin{proposition}[$\tau$ laws for FR weakly probabilistic hp-bisimulation]
The $\tau$ laws for FR weakly probabilistic hp-bisimulation is as follows.
\begin{enumerate}
  \item $P\approx_{php}^f \tau.P$;
  \item $P\approx_{php}^r P.\tau$;
  \item $\alpha.\tau.P\approx_{php}^f \alpha.P$;
  \item $P.\tau.\alpha[m]\approx_{php}^r P.\alpha[m]$;
  \item $(\alpha_1\parallel\cdots\parallel\alpha_n).\tau.P\approx_{php}^f (\alpha_1\parallel\cdots\parallel\alpha_n).P$;
  \item $P.\tau.(\alpha_1[m]\parallel\cdots\parallel\alpha_n[m])\approx_{php}^r P.(\alpha_1[m]\parallel\cdots\parallel\alpha_n[m])$;
  \item $P+\tau.P\approx_{php}^f \tau.P$;
  \item $P+P.\tau\approx_{php}^r P.\tau$;
  \item $P\cdot((Q+\tau\cdot(Q+R))\boxplus_{\pi}S)\approx_{php}^{f}P\cdot((Q+R)\boxplus_{\pi}S)$;
  \item $((Q+(Q+R)\cdot\tau)\boxplus_{\pi}S)\cdot P\approx_{php}^{r}((Q+R)\boxplus_{\pi}S)\cdot P$;
  \item $P\approx_{php}^{fr} \tau\parallel P$.
\end{enumerate}
\end{proposition}

\begin{proof}
\begin{enumerate}
  \item $P\approx_{php}^f \tau.P$. It is sufficient to prove the relation $R=\{(P, \tau.P)\}\cup \textbf{Id}$ is a F weakly probabilistic hp-bisimulation, we omit it;
  \item $P\approx_{php}^r P.\tau$. It is sufficient to prove the relation $R=\{(P, P.\tau)\}\cup \textbf{Id}$ is a R weakly probabilistic hp-bisimulation, we omit it;
  \item $\alpha.\tau.P\approx_{php}^f \alpha.P$. It is sufficient to prove the relation $R=\{(\alpha.\tau.P, \alpha.P)\}\cup \textbf{Id}$ is a F weakly probabilistic hp-bisimulation, we omit it;
  \item $P.\tau.\alpha[m]\approx_{php}^r P.\alpha[m]$. It is sufficient to prove the relation $R=\{(P.\tau.\alpha[m], P.\alpha[m])\}\cup \textbf{Id}$ is a R weakly probabilistic hp-bisimulation, we omit it;
  \item $(\alpha_1\parallel\cdots\parallel\alpha_n).\tau.P\approx_{php}^f (\alpha_1\parallel\cdots\parallel\alpha_n).P$. It is sufficient to prove the relation $R=\{((\alpha_1\parallel\cdots\parallel\alpha_n).\tau.P, (\alpha_1\parallel\cdots\parallel\alpha_n).P)\}\cup \textbf{Id}$ is a F weakly probabilistic hp-bisimulation, we omit it;
  \item $P.\tau.(\alpha_1[m]\parallel\cdots\parallel\alpha_n[m])\approx_{php}^r P.(\alpha_1[m]\parallel\cdots\parallel\alpha_n[m])$. It is sufficient to prove the relation $R=\{(P.\tau.(\alpha_1[m]\parallel\cdots\parallel\alpha_n[m]), P.(\alpha_1[m]\parallel\cdots\parallel\alpha_n[m]))\}\cup \textbf{Id}$ is a R weakly probabilistic hp-bisimulation, we omit it;
  \item $P+\tau.P\approx_{php}^f \tau.P$. It is sufficient to prove the relation $R=\{(P+\tau.P, \tau.P)\}\cup \textbf{Id}$ is a F weakly probabilistic hp-bisimulation, we omit it;
  \item $P+P.\tau\approx_{php}^r P.\tau$. It is sufficient to prove the relation $R=\{(P+P.\tau, P.\tau)\}\cup \textbf{Id}$ is a R weakly probabilistic hp-bisimulation, we omit it;
  \item $P\cdot((Q+\tau\cdot(Q+R))\boxplus_{\pi}S)\approx_{php}^{f}P\cdot((Q+R)\boxplus_{\pi}S)$. It is sufficient to prove the relation $R=\{(P\cdot((Q+\tau\cdot(Q+R))\boxplus_{\pi}S), P\cdot((Q+R)\boxplus_{\pi}S))\}\cup \textbf{Id}$ is a F weakly probabilistic hp-bisimulation, we omit it;
  \item $((Q+(Q+R)\cdot\tau)\boxplus_{\pi}S)\cdot P\approx_{php}^{r}((Q+R)\boxplus_{\pi}S)\cdot P$. It is sufficient to prove the relation $R=\{(((Q+(Q+R)\cdot\tau)\boxplus_{\pi}S)\cdot P, ((Q+R)\boxplus_{\pi}S)\cdot P)\}\cup \textbf{Id}$ is a R weakly probabilistic hp-bisimulation, we omit it;
  \item $P\approx_{php}^{fr} \tau\parallel P$. It is sufficient to prove the relation $R=\{(P, \tau\parallel P)\}\cup \textbf{Id}$ is a FR weakly probabilistic hp-bisimulation, we omit it.
\end{enumerate}
\end{proof}

\begin{proposition}[$\tau$ laws for FR weakly probabilistic hhp-bisimulation]
The $\tau$ laws for FR weakly probabilistic hhp-bisimulation is as follows.
\begin{enumerate}
  \item $P\approx_{phhp}^f \tau.P$;
  \item $P\approx_{phhp}^r P.\tau$;
  \item $\alpha.\tau.P\approx_{phhp}^f \alpha.P$;
  \item $P.\tau.\alpha[m]\approx_{phhp}^r P.\alpha[m]$;
  \item $(\alpha_1\parallel\cdots\parallel\alpha_n).\tau.P\approx_{phhp}^f (\alpha_1\parallel\cdots\parallel\alpha_n).P$;
  \item $P.\tau.(\alpha_1[m]\parallel\cdots\parallel\alpha_n[m])\approx_{phhp}^r P.(\alpha_1[m]\parallel\cdots\parallel\alpha_n[m])$;
  \item $P+\tau.P\approx_{phhp}^f \tau.P$;
  \item $P+P.\tau\approx_{phhp}^r P.\tau$;
  \item $P\cdot((Q+\tau\cdot(Q+R))\boxplus_{\pi}S)\approx_{phhp}^{f}P\cdot((Q+R)\boxplus_{\pi}S)$;
  \item $((Q+(Q+R)\cdot\tau)\boxplus_{\pi}S)\cdot P\approx_{phhp}^{r}((Q+R)\boxplus_{\pi}S)\cdot P$;
  \item $P\approx_{phhp}^{fr} \tau\parallel P$.
\end{enumerate}
\end{proposition}

\begin{proof}
\begin{enumerate}
  \item $P\approx_{phhp}^f \tau.P$. It is sufficient to prove the relation $R=\{(P, \tau.P)\}\cup \textbf{Id}$ is a F weakly probabilistic hhp-bisimulation, we omit it;
  \item $P\approx_{phhp}^r P.\tau$. It is sufficient to prove the relation $R=\{(P, P.\tau)\}\cup \textbf{Id}$ is a R weakly probabilistic hhp-bisimulation, we omit it;
  \item $\alpha.\tau.P\approx_{phhp}^f \alpha.P$. It is sufficient to prove the relation $R=\{(\alpha.\tau.P, \alpha.P)\}\cup \textbf{Id}$ is a F weakly probabilistic hhp-bisimulation, we omit it;
  \item $P.\tau.\alpha[m]\approx_{phhp}^r P.\alpha[m]$. It is sufficient to prove the relation $R=\{(P.\tau.\alpha[m], P.\alpha[m])\}\cup \textbf{Id}$ is a R weakly probabilistic hhp-bisimulation, we omit it;
  \item $(\alpha_1\parallel\cdots\parallel\alpha_n).\tau.P\approx_{phhp}^f (\alpha_1\parallel\cdots\parallel\alpha_n).P$. It is sufficient to prove the relation $R=\{((\alpha_1\parallel\cdots\parallel\alpha_n).\tau.P, (\alpha_1\parallel\cdots\parallel\alpha_n).P)\}\cup \textbf{Id}$ is a F weakly probabilistic hhp-bisimulation, we omit it;
  \item $P.\tau.(\alpha_1[m]\parallel\cdots\parallel\alpha_n[m])\approx_{phhp}^r P.(\alpha_1[m]\parallel\cdots\parallel\alpha_n[m])$. It is sufficient to prove the relation $R=\{(P.\tau.(\alpha_1[m]\parallel\cdots\parallel\alpha_n[m]), P.(\alpha_1[m]\parallel\cdots\parallel\alpha_n[m]))\}\cup \textbf{Id}$ is a R weakly probabilistic hhp-bisimulation, we omit it;
  \item $P+\tau.P\approx_{phhp}^f \tau.P$. It is sufficient to prove the relation $R=\{(P+\tau.P, \tau.P)\}\cup \textbf{Id}$ is a F weakly probabilistic hhp-bisimulation, we omit it;
  \item $P+P.\tau\approx_{phhp}^r P.\tau$. It is sufficient to prove the relation $R=\{(P+P.\tau, P.\tau)\}\cup \textbf{Id}$ is a R weakly probabilistic hhp-bisimulation, we omit it;
  \item $P\cdot((Q+\tau\cdot(Q+R))\boxplus_{\pi}S)\approx_{phhp}^{f}P\cdot((Q+R)\boxplus_{\pi}S)$. It is sufficient to prove the relation $R=\{(P\cdot((Q+\tau\cdot(Q+R))\boxplus_{\pi}S), P\cdot((Q+R)\boxplus_{\pi}S))\}\cup \textbf{Id}$ is a F weakly probabilistic hhp-bisimulation, we omit it;
  \item $((Q+(Q+R)\cdot\tau)\boxplus_{\pi}S)\cdot P\approx_{phhp}^{r}((Q+R)\boxplus_{\pi}S)\cdot P$. It is sufficient to prove the relation $R=\{(((Q+(Q+R)\cdot\tau)\boxplus_{\pi}S)\cdot P, ((Q+R)\boxplus_{\pi}S)\cdot P)\}\cup \textbf{Id}$ is a R weakly probabilistic hhp-bisimulation, we omit it;
  \item $P\approx_{phhp}^{fr} \tau\parallel P$. It is sufficient to prove the relation $R=\{(P, \tau\parallel P)\}\cup \textbf{Id}$ is a FR weakly probabilistic hhp-bisimulation, we omit it.
\end{enumerate}
\end{proof}

\subsubsection{Recursion}

\begin{definition}[Sequential]
$X$ is sequential in $E$ if every subexpression of $E$ which contains $X$, apart from $X$ itself, is of the form $\alpha.F$ or $F.\alpha[m]$, or
$(\alpha_1\parallel\cdots\parallel\alpha_n).F$ or $F.(\alpha_1[m]\parallel\cdots\parallel\alpha_n[m])$, or $\sum\widetilde{F}$.
\end{definition}

\begin{definition}[Guarded recursive expression]
$X$ is guarded in $E$ if each occurrence of $X$ is with some subexpression $l.F$ or $F.l[m]$, or $(l_1\parallel\cdots\parallel l_n).F$ or $F.(l_1[m]\parallel\cdots\parallel l_n[m])$ of
$E$.
\end{definition}

\begin{lemma}\label{LUSWW04}
Let $G$ be guarded and sequential, $Vars(G)\subseteq\widetilde{X}$, and let $ G\{\widetilde{P}/\widetilde{X}\}\rightsquigarrow\xrightarrow{\{\alpha_1,\cdots,\alpha_n\}} P'$
or $ G\{\widetilde{P}/\widetilde{X}\}\rightsquigarrow\xtworightarrow{\{\alpha_1[m],\cdots,\alpha_n[m]\}} P'$. Then there is an expression $H$ such that
$ G\rightsquigarrow\xrightarrow{\{\alpha_1,\cdots,\alpha_n\}} H$ or $ G\rightsquigarrow\xtworightarrow{\{\alpha_1[m],\cdots,\alpha_n[m]\}} H$,
$P'\equiv H\{\widetilde{P}/\widetilde{X}\}$, and for any $\widetilde{Q}$, $ G\{\widetilde{Q}/\widetilde{X}\}\rightsquigarrow\xrightarrow{\{\alpha_1,\cdots,\alpha_n\}}  H\{\widetilde{Q}/\widetilde{X}\}$
or $ G\{\widetilde{Q}/\widetilde{X}\}\rightsquigarrow\xtworightarrow{\{\alpha_1[m],\cdots,\alpha_n[m]\}}  H\{\widetilde{Q}/\widetilde{X}\}$. Moreover $H$ is sequential,
$Vars(H)\subseteq\widetilde{X}$, and if $\alpha_1=\cdots=\alpha_n=\alpha_1[m]=\cdots=\alpha_n[m]=\tau$, then $H$ is also guarded.
\end{lemma}

\begin{proof}
We only prove the case of forward transition.

We need to induct on the structure of $G$.

If $G$ is a Constant, a Composition, a Restriction or a Relabeling then it contains no variables, since $G$ is sequential and guarded, then
$ G\rightsquigarrow\xrightarrow{\{\alpha_1,\cdots,\alpha_n\}} P'$, then let $H\equiv P'$, as desired.

$G$ cannot be a variable, since it is guarded.

If $G\equiv G_1+G_2$. Then either $ G_1\{\widetilde{P}/\widetilde{X}\} \rightsquigarrow\xrightarrow{\{\alpha_1,\cdots,\alpha_n\}} P'$ or
$ G_2\{\widetilde{P}/\widetilde{X}\} \rightsquigarrow\xrightarrow{\{\alpha_1,\cdots,\alpha_n\}} P'$, then, we can apply this lemma in either case, as desired.

If $G\equiv\beta.H$. Then we must have $\alpha=\beta$, and $P'\equiv H\{\widetilde{P}/\widetilde{X}\}$, and
$ G\{\widetilde{Q}/\widetilde{X}\}\equiv \beta.H\{\widetilde{Q}/\widetilde{X}\} \rightsquigarrow\xrightarrow{\beta} H\{\widetilde{Q}/\widetilde{X}\}$,
then, let $G'$ be $H$, as desired.

If $G\equiv(\beta_1\parallel\cdots\parallel\beta_n).H$. Then we must have $\alpha_i=\beta_i$ for $1\leq i\leq n$, and $P'\equiv H\{\widetilde{P}/\widetilde{X}\}$, and
$ G\{\widetilde{Q}/\widetilde{X}\}\equiv (\beta_1\parallel\cdots\parallel\beta_n).H\{\widetilde{Q}/\widetilde{X}\} \rightsquigarrow\xrightarrow{\{\beta_1,\cdots,\beta_n\}} H\{\widetilde{Q}/\widetilde{X}\}$,
then, let $G'$ be $H$, as desired.

If $G\equiv\tau.H$. Then we must have $\tau=\tau$, and $P'\equiv H\{\widetilde{P}/\widetilde{X}\}$, and
$ G\{\widetilde{Q}/\widetilde{X}\}\equiv \tau.H\{\widetilde{Q}/\widetilde{X}\} \rightsquigarrow\xrightarrow{\tau} H\{\widetilde{Q}/\widetilde{X}\}$,
then, let $G'$ be $H$, as desired.

For the case of reverse transition, it can be proven similarly, we omit it.
\end{proof}

\begin{theorem}[Unique solution of equations for FR weakly probabilistic pomset bisimulation]
Let the guarded and sequential expressions $\widetilde{E}$ contain free variables $\subseteq \widetilde{X}$, then,

If $\widetilde{P}\approx_{pp}^{fr} \widetilde{E}\{\widetilde{P}/\widetilde{X}\}$ and $\widetilde{Q}\approx_{pp}^{fr} \widetilde{E}\{\widetilde{Q}/\widetilde{X}\}$, then
$\widetilde{P}\approx_{pp}^{fr} \widetilde{Q}$.
\end{theorem}

\begin{proof}
We only prove the case of forward transition.

Like the corresponding theorem in CCS, without loss of generality, we only consider a single equation $X=E$. So we assume $P\approx_{pp}^{fr} E(P)$, $Q\approx_{pp}^{fr} E(Q)$, then $P\approx_{pp}^{fr} Q$.

We will prove $\{(H(P),H(Q)): H\}$ sequential, if $ H(P)\rightsquigarrow\xrightarrow{\{\alpha_1,\cdots,\alpha_n\}} P'$, then, for some $Q'$,
$ H(Q)\rightsquigarrow\xRightarrow{\{\alpha_1.\cdots,\alpha_n\}} Q'$ and $P'\approx_{pp}^{fr} Q'$.

Let $ H(P)\rightsquigarrow\xrightarrow{\{\alpha_1,\cdot,\alpha_n\}} P'$, then $ H(E(P))\rightsquigarrow\xRightarrow{\{\alpha_1,\cdots,\alpha_n\}} P''$
and $P'\approx_{pp}^{fr} P''$.

By Lemma \ref{LUSWW04}, we know there is a sequential $H'$ such that $ H(E(P))\rightsquigarrow\xRightarrow{\{\alpha_1,\cdots,\alpha_n\}} H'(P)\Rightarrow P''\approx_{pp}^{fr} P'$.

And, $ H(E(Q))\rightsquigarrow\xRightarrow{\{\alpha_1,\cdots,\alpha_n\}} H'(Q)\Rightarrow Q''$ and $P''\approx_{pp}^{fr} Q''$. And $ H(Q)\rightsquigarrow\xrightarrow{\{\alpha_1,\cdots,\alpha_n\}} Q'\approx_{pp}^{fr} \Rightarrow Q'\approx_{pp}^{fr} Q''$.
Hence, $P'\approx_{pp}^{fr} Q'$, as desired.

For the case of reverse transition, it can be proven similarly, we omit it.
\end{proof}

\begin{theorem}[Unique solution of equations for FR weakly probabilistic step bisimulation]
Let the guarded and sequential expressions $\widetilde{E}$ contain free variables $\subseteq \widetilde{X}$, then,

If $\widetilde{P}\approx_{ps}^{fr} \widetilde{E}\{\widetilde{P}/\widetilde{X}\}$ and $\widetilde{Q}\approx_{ps}^{fr} \widetilde{E}\{\widetilde{Q}/\widetilde{X}\}$, then
$\widetilde{P}\approx_{ps}^{fr} \widetilde{Q}$.
\end{theorem}

\begin{proof}
We only prove the case of forward transition.

Like the corresponding theorem in CCS, without loss of generality, we only consider a single equation $X=E$. So we assume $P\approx_{ps}^{fr} E(P)$, $Q\approx_{ps}^{fr} E(Q)$, then $P\approx_{ps}^{fr} Q$.

We will prove $\{(H(P),H(Q)): H\}$ sequential, if $ H(P)\rightsquigarrow\xrightarrow{\{\alpha_1,\cdots,\alpha_n\}} P'$, then, for some $Q'$,
$ H(Q)\rightsquigarrow\xRightarrow{\{\alpha_1.\cdots,\alpha_n\}} Q'$ and $P'\approx_{ps}^{fr} Q'$.

Let $ H(P)\rightsquigarrow\xrightarrow{\{\alpha_1,\cdot,\alpha_n\}} P'$, then $ H(E(P))\rightsquigarrow\xRightarrow{\{\alpha_1,\cdots,\alpha_n\}} P''$
and $P'\approx_{ps}^{fr} P''$.

By Lemma \ref{LUSWW04}, we know there is a sequential $H'$ such that $ H(E(P))\rightsquigarrow\xRightarrow{\{\alpha_1,\cdots,\alpha_n\}} H'(P)\Rightarrow P''\approx_{ps}^{fr} P'$.

And, $ H(E(Q))\rightsquigarrow\xRightarrow{\{\alpha_1,\cdots,\alpha_n\}} H'(Q)\Rightarrow Q''$ and $P''\approx_{ps}^{fr} Q''$. And $ H(Q)\rightsquigarrow\xrightarrow{\{\alpha_1,\cdots,\alpha_n\}} Q'\approx_{ps}^{fr} \Rightarrow Q'\approx_{ps}^{fr} Q''$.
Hence, $P'\approx_{ps}^{fr} Q'$, as desired.

For the case of reverse transition, it can be proven similarly, we omit it.
\end{proof}

\begin{theorem}[Unique solution of equations for FR weakly probabilistic hp-bisimulation]
Let the guarded and sequential expressions $\widetilde{E}$ contain free variables $\subseteq \widetilde{X}$, then,

If $\widetilde{P}\approx_{php}^{fr} \widetilde{E}\{\widetilde{P}/\widetilde{X}\}$ and $\widetilde{Q}\approx_{php}^{fr} \widetilde{E}\{\widetilde{Q}/\widetilde{X}\}$, then
$\widetilde{P}\approx_{php}^{fr} \widetilde{Q}$.
\end{theorem}

\begin{proof}
We only prove the case of forward transition.

Like the corresponding theorem in CCS, without loss of generality, we only consider a single equation $X=E$. So we assume $P\approx_{php}^{fr} E(P)$, $Q\approx_{php}^{fr} E(Q)$, then $P\approx_{php}^{fr} Q$.

We will prove $\{(H(P),H(Q)): H\}$ sequential, if $ H(P)\rightsquigarrow\xrightarrow{\{\alpha_1,\cdots,\alpha_n\}} P'$, then, for some $Q'$,
$ H(Q)\rightsquigarrow\xRightarrow{\{\alpha_1.\cdots,\alpha_n\}} Q'$ and $P'\approx_{php}^{fr} Q'$.

Let $ H(P)\rightsquigarrow\xrightarrow{\{\alpha_1,\cdot,\alpha_n\}} P'$, then $ H(E(P))\rightsquigarrow\xRightarrow{\{\alpha_1,\cdots,\alpha_n\}} P''$
and $P'\approx_{php}^{fr} P''$.

By Lemma \ref{LUSWW04}, we know there is a sequential $H'$ such that $ H(E(P))\rightsquigarrow\xRightarrow{\{\alpha_1,\cdots,\alpha_n\}} H'(P)\Rightarrow P''\approx_{php}^{fr} P'$.

And, $ H(E(Q))\rightsquigarrow\xRightarrow{\{\alpha_1,\cdots,\alpha_n\}} H'(Q)\Rightarrow Q''$ and $P''\approx_{php}^{fr} Q''$. And $ H(Q)\rightsquigarrow\xrightarrow{\{\alpha_1,\cdots,\alpha_n\}} Q'\approx_{php}^{fr} \Rightarrow Q'\approx_{php}^{fr} Q''$.
Hence, $P'\approx_{php}^{fr} Q'$, as desired.

For the case of reverse transition, it can be proven similarly, we omit it.
\end{proof}

\begin{theorem}[Unique solution of equations for FR weakly probabilistic hhp-bisimulation]
Let the guarded and sequential expressions $\widetilde{E}$ contain free variables $\subseteq \widetilde{X}$, then,

If $\widetilde{P}\approx_{phhp}^{fr} \widetilde{E}\{\widetilde{P}/\widetilde{X}\}$ and $\widetilde{Q}\approx_{phhp}^{fr} \widetilde{E}\{\widetilde{Q}/\widetilde{X}\}$, then
$\widetilde{P}\approx_{phhp}^{fr} \widetilde{Q}$.
\end{theorem}

\begin{proof}
We only prove the case of forward transition.

Like the corresponding theorem in CCS, without loss of generality, we only consider a single equation $X=E$. So we assume $P\approx_{phhp}^{fr} E(P)$, $Q\approx_{phhp}^{fr} E(Q)$, then $P\approx_{phhp}^{fr} Q$.

We will prove $\{(H(P),H(Q)): H\}$ sequential, if $ H(P)\rightsquigarrow\xrightarrow{\{\alpha_1,\cdots,\alpha_n\}} P'$, then, for some $Q'$,
$ H(Q)\rightsquigarrow\xRightarrow{\{\alpha_1.\cdots,\alpha_n\}} Q'$ and $P'\approx_{phhp}^{fr} Q'$.

Let $ H(P)\rightsquigarrow\xrightarrow{\{\alpha_1,\cdot,\alpha_n\}} P'$, then $ H(E(P))\rightsquigarrow\xRightarrow{\{\alpha_1,\cdots,\alpha_n\}} P''$
and $P'\approx_{phhp}^{fr} P''$.

By Lemma \ref{LUSWW04}, we know there is a sequential $H'$ such that $ H(E(P))\rightsquigarrow\xRightarrow{\{\alpha_1,\cdots,\alpha_n\}} H'(P)\Rightarrow P''\approx_{phhp}^{fr} P'$.

And, $ H(E(Q))\rightsquigarrow\xRightarrow{\{\alpha_1,\cdots,\alpha_n\}} H'(Q)\Rightarrow Q''$ and $P''\approx_{phhp}^{fr} Q''$. And $ H(Q)\rightsquigarrow\xrightarrow{\{\alpha_1,\cdots,\alpha_n\}} Q'\approx_{phhp}^{fr} \Rightarrow Q'\approx_{phhp}^{fr} Q''$.
Hence, $P'\approx_{phhp}^{fr} Q'$, as desired.

For the case of reverse transition, it can be proven similarly, we omit it.
\end{proof}

\newpage\section{CTC with Probabilism and Guards}\label{ctcpg}

In this chapter, we design the calculus CTC with probabilism and guards. This chapter is organized as follows. We introduce the operational semantics in section \ref{osctcpg}, its syntax and operational
semantics in section \ref{sosctcpg}, and its properties for strong bisimulations in section \ref{stcbctcpg}, its properties for weak bisimulations in section \ref{wtcbctcpg}.

\subsection{Operational Semantics}\label{osctcpg}

\begin{definition}[Prime event structure with silent event and empty event]\label{PESG}
Let $\Lambda$ be a fixed set of labels, ranged over $a,b,c,\cdots$ and $\tau,\epsilon$. A ($\Lambda$-labelled) prime event structure with silent event $\tau$ and empty event $\epsilon$
is a tuple $\mathcal{E}=\langle \mathbb{E}, \leq, \sharp, \sharp_{\pi} \lambda\rangle$, where $\mathbb{E}$ is a denumerable set of events, including the silent event $\tau$ and
empty event $\epsilon$. Let $\hat{\mathbb{E}}=\mathbb{E}\backslash\{\tau,\epsilon\}$, exactly excluding $\tau$ and $\epsilon$, it is obvious that $\hat{\tau^*}=\epsilon$. Let
$\lambda:\mathbb{E}\rightarrow\Lambda$ be a labelling function and let $\lambda(\tau)=\tau$ and $\lambda(\epsilon)=\epsilon$. And $\leq$, $\sharp$, $\sharp_{\pi}$ are binary relations
on $\mathbb{E}$, called causality, conflict and probabilistic conflict respectively, such that:

\begin{enumerate}
  \item $\leq$ is a partial order and $\lceil e \rceil = \{e'\in \mathbb{E}|e'\leq e\}$ is finite for all $e\in \mathbb{E}$. It is easy to see that
  $e\leq\tau^*\leq e'=e\leq\tau\leq\cdots\leq\tau\leq e'$, then $e\leq e'$.
  \item $\sharp$ is irreflexive, symmetric and hereditary with respect to $\leq$, that is, for all $e,e',e''\in \mathbb{E}$, if $e\sharp e'\leq e''$, then $e\sharp e''$;
  \item $\sharp_{\pi}$ is irreflexive, symmetric and hereditary with respect to $\leq$, that is, for all $e,e',e''\in \mathbb{E}$, if $e\sharp_{\pi} e'\leq e''$, then $e\sharp_{\pi} e''$.
\end{enumerate}

Then, the concepts of consistency and concurrency can be drawn from the above definition:

\begin{enumerate}
  \item $e,e'\in \mathbb{E}$ are consistent, denoted as $e\frown e'$, if $\neg(e\sharp e')$ and $\neg(e\sharp_{\pi} e')$. A subset $X\subseteq \mathbb{E}$ is called consistent, if
  $e\frown e'$ for all $e,e'\in X$.
  \item $e,e'\in \mathbb{E}$ are concurrent, denoted as $e\parallel e'$, if $\neg(e\leq e')$, $\neg(e'\leq e)$, and $\neg(e\sharp e')$ and $\neg(e\sharp_{\pi} e')$.
\end{enumerate}
\end{definition}

\begin{definition}[Configuration]
Let $\mathcal{E}$ be a PES. A (finite) configuration in $\mathcal{E}$ is a (finite) consistent subset of events $C\subseteq \mathcal{E}$, closed with respect to causality
(i.e. $\lceil C\rceil=C$), and a data state $s\in S$ with $S$ the set of all data states, denoted $\langle C, s\rangle$. The set of finite configurations of $\mathcal{E}$ is denoted by
$\langle\mathcal{C}(\mathcal{E}), S\rangle$. We let $\hat{C}=C\backslash\{\tau\}\cup\{\epsilon\}$.
\end{definition}

A consistent subset of $X\subseteq \mathbb{E}$ of events can be seen as a pomset. Given $X, Y\subseteq \mathbb{E}$, $\hat{X}\sim \hat{Y}$ if $\hat{X}$ and $\hat{Y}$ are isomorphic as
pomsets. In the following of the paper, we say $C_1\sim C_2$, we mean $\hat{C_1}\sim\hat{C_2}$.

\begin{definition}[Pomset transitions and step]
Let $\mathcal{E}$ be a PES and let $C\in\mathcal{C}(\mathcal{E})$, and $\emptyset\neq X\subseteq \mathbb{E}$, if $C\cap X=\emptyset$ and $C'=C\cup X\in\mathcal{C}(\mathcal{E})$, then
$\langle C,s\rangle\xrightarrow{X} \langle C',s'\rangle$ is called a pomset transition from $\langle C,s\rangle$ to $\langle C',s'\rangle$. When the events in $X$ are pairwise
concurrent, we say that $\langle C,s\rangle\xrightarrow{X}\langle C',s'\rangle$ is a step. It is obvious that $\rightarrow^*\xrightarrow{X}\rightarrow^*=\xrightarrow{X}$ and
$\rightarrow^*\xrightarrow{e}\rightarrow^*=\xrightarrow{e}$ for any $e\in\mathbb{E}$ and $X\subseteq\mathbb{E}$.
\end{definition}

\begin{definition}[Probabilistic transitions]
Let $\mathcal{E}$ be a PES and let $C\in\mathcal{C}(\mathcal{E})$, the transition $\langle C,s\rangle\xrsquigarrow{\pi} \langle C^{\pi},s\rangle$ is called a probabilistic transition
from $\langle C,s\rangle$ to $\langle C^{\pi},s\rangle$.
\end{definition}

\begin{definition}[Weak pomset transitions and weak step]
Let $\mathcal{E}$ be a PES and let $C\in\mathcal{C}(\mathcal{E})$, and $\emptyset\neq X\subseteq \hat{\mathbb{E}}$, if $C\cap X=\emptyset$ and
$\hat{C'}=\hat{C}\cup X\in\mathcal{C}(\mathcal{E})$, then $\langle C,s\rangle\xRightarrow{X} \langle C',s'\rangle$ is called a weak pomset transition from $\langle C,s\rangle$ to
$\langle C',s'\rangle$, where we define $\xRightarrow{e}\triangleq\xrightarrow{\tau^*}\xrightarrow{e}\xrightarrow{\tau^*}$. And
$\xRightarrow{X}\triangleq\xrightarrow{\tau^*}\xrightarrow{e}\xrightarrow{\tau^*}$, for every $e\in X$. When the events in $X$ are pairwise concurrent, we say that
$\langle C,s\rangle\xRightarrow{X}\langle C',s'\rangle$ is a weak step.
\end{definition}

We will also suppose that all the PESs in this chapter are image finite, that is, for any PES $\mathcal{E}$ and $C\in \mathcal{C}(\mathcal{E})$ and $a\in \Lambda$,
$\{\langle C,s\rangle\xrsquigarrow{\pi} \langle C^{\pi},s\rangle\}$,
$\{e\in \mathbb{E}|\langle C,s\rangle\xrightarrow{e} \langle C',s'\rangle\wedge \lambda(e)=a\}$ and
$\{e\in\hat{\mathbb{E}}|\langle C,s\rangle\xRightarrow{e} \langle C',s'\rangle\wedge \lambda(e)=a\}$ is finite.

\begin{definition}[Probabilistic pomset, step bisimulation]\label{PSBG}
Let $\mathcal{E}_1$, $\mathcal{E}_2$ be PESs. A probabilistic pomset bisimulation is a relation $R\subseteq\langle\mathcal{C}(\mathcal{E}_1),S\rangle\times\langle\mathcal{C}(\mathcal{E}_2),S\rangle$,
such that (1) if $(\langle C_1,s\rangle,\langle C_2,s\rangle)\in R$, and $\langle C_1,s\rangle\xrightarrow{X_1}\langle C_1',s'\rangle$ then
$\langle C_2,s\rangle\xrightarrow{X_2}\langle C_2',s'\rangle$, with $X_1\subseteq \mathbb{E}_1$, $X_2\subseteq \mathbb{E}_2$, $X_1\sim X_2$ and
$(\langle C_1',s'\rangle,\langle C_2',s'\rangle)\in R$ for all $s,s'\in S$, and vice-versa; (2) if $(\langle C_1,s\rangle,\langle C_2,s\rangle)\in R$, and $\langle C_1,s\rangle\xrsquigarrow{\pi}\langle C_1^{\pi},s\rangle$
then $\langle C_2,s\rangle\xrsquigarrow{\pi}\langle C_2^{\pi},s\rangle$ and $(\langle C_1^{\pi},s\rangle,\langle C_2^{\pi},s\rangle)\in R$, and vice-versa; (3) if $(\langle C_1,s\rangle,\langle C_2,s\rangle)\in R$,
then $\mu(C_1,C)=\mu(C_2,C)$ for each $C\in\mathcal{C}(\mathcal{E})/R$; (4) $[\surd]_R=\{\surd\}$. We say that $\mathcal{E}_1$, $\mathcal{E}_2$ are probabilistic pomset bisimilar, written
$\mathcal{E}_1\sim_{pp}\mathcal{E}_2$, if there exists a probabilistic pomset bisimulation $R$, such that $(\langle\emptyset,\emptyset\rangle,\langle\emptyset,\emptyset\rangle)\in R$.
By replacing probabilistic pomset transitions with probabilistic steps, we can get the definition of probabilistic step bisimulation. When PESs $\mathcal{E}_1$ and $\mathcal{E}_2$ are
probabilistic step bisimilar, we write $\mathcal{E}_1\sim_{ps}\mathcal{E}_2$.
\end{definition}

\begin{definition}[Weakly probabilistic pomset, step bisimulation]\label{WPSBG}
Let $\mathcal{E}_1$, $\mathcal{E}_2$ be PESs. A weakly probabilistic pomset bisimulation is a relation $R\subseteq\langle\mathcal{C}(\mathcal{E}_1),S\rangle\times\langle\mathcal{C}(\mathcal{E}_2),S\rangle$,
such that (1) if $(\langle C_1,s\rangle,\langle C_2,s\rangle)\in R$, and $\langle C_1,s\rangle\xRightarrow{X_1}\langle C_1',s'\rangle$ then
$\langle C_2,s\rangle\xRightarrow{X_2}\langle C_2',s'\rangle$, with $X_1\subseteq \hat{\mathbb{E}_1}$, $X_2\subseteq \hat{\mathbb{E}_2}$, $X_1\sim X_2$ and
$(\langle C_1',s'\rangle,\langle C_2',s'\rangle)\in R$ for all $s,s'\in S$, and vice-versa; (2) if $(\langle C_1,s\rangle,\langle C_2,s\rangle)\in R$, and $\langle C_1,s\rangle\xrsquigarrow{\pi}\langle C_1^{\pi},s\rangle$
then $\langle C_2,s\rangle\xrsquigarrow{\pi}\langle C_2^{\pi},s\rangle$ and $(\langle C_1^{\pi},s\rangle,\langle C_2^{\pi},s\rangle)\in R$, and vice-versa; (3) if $(\langle C_1,s\rangle,\langle C_2,s\rangle)\in R$,
then $\mu(C_1,C)=\mu(C_2,C)$ for each $C\in\mathcal{C}(\mathcal{E})/R$; (4) $[\surd]_R=\{\surd\}$. We say that $\mathcal{E}_1$, $\mathcal{E}_2$ are weakly probabilistic pomset bisimilar,
written $\mathcal{E}_1\approx_{pp}\mathcal{E}_2$, if there exists a weakly probabilistic pomset bisimulation $R$, such that
$(\langle\emptyset,\emptyset\rangle,\langle\emptyset,\emptyset\rangle)\in R$. By replacing weakly probabilistic pomset transitions with weakly probabilistic steps, we can get the
definition of weakly probabilistic step bisimulation. When PESs $\mathcal{E}_1$ and $\mathcal{E}_2$ are weakly probabilistic step bisimilar, we write
$\mathcal{E}_1\approx_{ps}\mathcal{E}_2$.
\end{definition}

\begin{definition}[Posetal product]
Given two PESs $\mathcal{E}_1$, $\mathcal{E}_2$, the posetal product of their configurations, denoted
$\langle\mathcal{C}(\mathcal{E}_1),S\rangle\overline{\times}\langle\mathcal{C}(\mathcal{E}_2),S\rangle$, is defined as

$$\{(\langle C_1,s\rangle,f,\langle C_2,s\rangle)|C_1\in\mathcal{C}(\mathcal{E}_1),C_2\in\mathcal{C}(\mathcal{E}_2),f:C_1\rightarrow C_2 \textrm{ isomorphism}\}.$$

A subset $R\subseteq\langle\mathcal{C}(\mathcal{E}_1),S\rangle\overline{\times}\langle\mathcal{C}(\mathcal{E}_2),S\rangle$ is called a posetal relation. We say that $R$ is downward
closed when for any $(\langle C_1,s\rangle,f,\langle C_2,s\rangle),(\langle C_1',s'\rangle,f',\langle C_2',s'\rangle)\in \langle\mathcal{C}(\mathcal{E}_1),S\rangle\overline{\times}\langle\mathcal{C}(\mathcal{E}_2),S\rangle$,
if $(\langle C_1,s\rangle,f,\langle C_2,s\rangle)\subseteq (\langle C_1',s'\rangle,f',\langle C_2',s'\rangle)$ pointwise and
$(\langle C_1',s'\rangle,f',\langle C_2',s'\rangle)\in R$, then $(\langle C_1,s\rangle,f,\langle C_2,s\rangle)\in R$.

For $f:X_1\rightarrow X_2$, we define $f[x_1\mapsto x_2]:X_1\cup\{x_1\}\rightarrow X_2\cup\{x_2\}$, $z\in X_1\cup\{x_1\}$,(1)$f[x_1\mapsto x_2](z)=
x_2$,if $z=x_1$;(2)$f[x_1\mapsto x_2](z)=f(z)$, otherwise. Where $X_1\subseteq \mathbb{E}_1$, $X_2\subseteq \mathbb{E}_2$, $x_1\in \mathbb{E}_1$, $x_2\in \mathbb{E}_2$.
\end{definition}

\begin{definition}[Weakly posetal product]
Given two PESs $\mathcal{E}_1$, $\mathcal{E}_2$, the weakly posetal product of their configurations, denoted
$\langle\mathcal{C}(\mathcal{E}_1),S\rangle\overline{\times}\langle\mathcal{C}(\mathcal{E}_2),S\rangle$, is defined as

$$\{(\langle C_1,s\rangle,f,\langle C_2,s\rangle)|C_1\in\mathcal{C}(\mathcal{E}_1),C_2\in\mathcal{C}(\mathcal{E}_2),f:\hat{C_1}\rightarrow \hat{C_2} \textrm{ isomorphism}\}.$$

A subset $R\subseteq\langle\mathcal{C}(\mathcal{E}_1),S\rangle\overline{\times}\langle\mathcal{C}(\mathcal{E}_2),S\rangle$ is called a weakly posetal relation. We say that $R$ is
downward closed when for any $(\langle C_1,s\rangle,f,\langle C_2,s\rangle),(\langle C_1',s'\rangle,f,\langle C_2',s'\rangle)\in \langle\mathcal{C}(\mathcal{E}_1),S\rangle\overline{\times}\langle\mathcal{C}(\mathcal{E}_2),S\rangle$,
if $(\langle C_1,s\rangle,f,\langle C_2,s\rangle)\subseteq (\langle C_1',s'\rangle,f',\langle C_2',s'\rangle)$ pointwise and
$(\langle C_1',s'\rangle,f',\langle C_2',s'\rangle)\in R$, then $(\langle C_1,s\rangle,f,\langle C_2,s\rangle)\in R$.

For $f:X_1\rightarrow X_2$, we define $f[x_1\mapsto x_2]:X_1\cup\{x_1\}\rightarrow X_2\cup\{x_2\}$, $z\in X_1\cup\{x_1\}$,(1)$f[x_1\mapsto x_2](z)=
x_2$,if $z=x_1$;(2)$f[x_1\mapsto x_2](z)=f(z)$, otherwise. Where $X_1\subseteq \hat{\mathbb{E}_1}$, $X_2\subseteq \hat{\mathbb{E}_2}$, $x_1\in \hat{\mathbb{E}}_1$,
$x_2\in \hat{\mathbb{E}}_2$. Also, we define $f(\tau^*)=f(\tau^*)$.
\end{definition}

\begin{definition}[Probabilistic (hereditary) history-preserving bisimulation]\label{HHPBG}
A probabilistic history-preserving (hp-) bisimulation is a posetal relation
$R\subseteq\langle\mathcal{C}(\mathcal{E}_1),S\rangle\overline{\times}\langle\mathcal{C}(\mathcal{E}_2),S\rangle$ such that (1) if $(\langle C_1,s\rangle,f,\langle C_2,s\rangle)\in R$,
and $\langle C_1,s\rangle\xrightarrow{e_1} \langle C_1',s'\rangle$, then $\langle C_2,s\rangle\xrightarrow{e_2} \langle C_2',s'\rangle$, with
$(\langle C_1',s'\rangle,f[e_1\mapsto e_2],\langle C_2',s'\rangle)\in R$ for all $s,s'\in S$, and vice-versa; (2) if $(\langle C_1,s\rangle,f,\langle C_2,s\rangle)\in R$, and
$\langle C_1,s\rangle\xrsquigarrow{\pi}\langle C_1^{\pi},s\rangle$ then $\langle C_2,s\rangle\xrsquigarrow{\pi}\langle C_2^{\pi},s\rangle$ and $(\langle C_1^{\pi},s\rangle,f,\langle C_2^{\pi},s\rangle)\in R$,
and vice-versa; (3) if $(C_1,f,C_2)\in R$, then $\mu(C_1,C)=\mu(C_2,C)$ for each $C\in\mathcal{C}(\mathcal{E})/R$; (4) $[\surd]_R=\{\surd\}$. $\mathcal{E}_1,\mathcal{E}_2$ are
probabilistic history-preserving (hp-)bisimilar and are written $\mathcal{E}_1\sim_{php}\mathcal{E}_2$ if there exists a probabilistic hp-bisimulation $R$ such that
$(\langle\emptyset,\emptyset\rangle,\emptyset,\langle\emptyset,\emptyset\rangle)\in R$.

A probabilistic hereditary history-preserving (hhp-)bisimulation is a downward closed probabilistic hp-bisimulation. $\mathcal{E}_1,\mathcal{E}_2$ are probabilistic hereditary
history-preserving (hhp-)bisimilar and are written $\mathcal{E}_1\sim_{phhp}\mathcal{E}_2$.
\end{definition}

\begin{definition}[Weakly probabilistic (hereditary) history-preserving bisimulation]\label{WHHPBG}
A weakly probabilistic history-preserving (hp-) bisimulation is a weakly posetal relation
$R\subseteq\langle\mathcal{C}(\mathcal{E}_1),S\rangle\overline{\times}\langle\mathcal{C}(\mathcal{E}_2),S\rangle$ such that (1) if $(\langle C_1,s\rangle,f,\langle C_2,s\rangle)\in R$,
and $\langle C_1,s\rangle\xRightarrow{e_1} \langle C_1',s'\rangle$, then $\langle C_2,s\rangle\xRightarrow{e_2} \langle C_2',s'\rangle$, with
$(\langle C_1',s'\rangle,f[e_1\mapsto e_2],\langle C_2',s'\rangle)\in R$ for all $s,s'\in S$, and vice-versa; (2) if $(\langle C_1,s\rangle,f,\langle C_2,s\rangle)\in R$, and
$\langle C_1,s\rangle\xrsquigarrow{\pi}\langle C_1^{\pi},s\rangle$ then $\langle C_2,s\rangle\xrsquigarrow{\pi}\langle C_2^{\pi},s\rangle$ and
$(\langle C_1^{\pi},s\rangle,f,\langle C_2^{\pi},s\rangle)\in R$, and vice-versa; (3) if $(C_1,f,C_2)\in R$, then $\mu(C_1,C)=\mu(C_2,C)$ for each $C\in\mathcal{C}(\mathcal{E})/R$;
(4) $[\surd]_R=\{\surd\}$. $\mathcal{E}_1,\mathcal{E}_2$ are weakly probabilistic history-preserving (hp-)bisimilar and are written $\mathcal{E}_1\approx_{php}\mathcal{E}_2$ if there
exists a weakly probabilistic hp-bisimulation $R$ such that $(\langle\emptyset,\emptyset\rangle,\emptyset,\langle\emptyset,\emptyset\rangle)\in R$.

A weakly probabilistic hereditary history-preserving (hhp-)bisimulation is a downward closed weakly probabilistic hp-bisimulation. $\mathcal{E}_1,\mathcal{E}_2$ are weakly
probabilistic hereditary history-preserving (hhp-)bisimilar and are written $\mathcal{E}_1\approx_{phhp}\mathcal{E}_2$.
\end{definition}

\subsection{Syntax and Operational Semantics}\label{sosctcpg}

We assume an infinite set $\mathcal{N}$ of (action or event) names, and use $a,b,c,\cdots$ to range over $\mathcal{N}$. We denote by $\overline{\mathcal{N}}$ the set of co-names and
let $\overline{a},\overline{b},\overline{c},\cdots$ range over $\overline{\mathcal{N}}$. Then we set $\mathcal{L}=\mathcal{N}\cup\overline{\mathcal{N}}$ as the set of labels, and use
$l,\overline{l}$ to range over $\mathcal{L}$. We extend complementation to $\mathcal{L}$ such that $\overline{\overline{a}}=a$. Let $\tau$ denote the silent step (internal action or
event) and define $Act=\mathcal{L}\cup\{\tau\}$ to be the set of actions, $\alpha,\beta$ range over $Act$. And $K,L$ are used to stand for subsets of $\mathcal{L}$ and $\overline{L}$
is used for the set of complements of labels in $L$. A relabelling function $f$ is a function from $\mathcal{L}$ to $\mathcal{L}$ such that $f(\overline{l})=\overline{f(l)}$. By
defining $f(\tau)=\tau$, we extend $f$ to $Act$.

Further, we introduce a set $\mathcal{X}$ of process variables, and a set $\mathcal{K}$ of process constants, and let $X,Y,\cdots$ range over $\mathcal{X}$, and $A,B,\cdots$ range over
$\mathcal{K}$, $\widetilde{X}$ is a tuple of distinct process variables, and also $E,F,\cdots$ range over the recursive expressions. We write $\mathcal{P}$ for the set of processes.
Sometimes, we use $I,J$ to stand for an indexing set, and we write $E_i:i\in I$ for a family of expressions indexed by $I$. $Id_D$ is the identity function or relation over set $D$.

For each process constant schema $A$, a defining equation of the form

$$A\overset{\text{def}}{=}P$$

is assumed, where $P$ is a process.

Let $G_{at}$ be the set of atomic guards, $\delta$ be the deadlock constant, and $\epsilon$ be the empty action, and extend $Act$ to $Act\cup\{\epsilon\}\cup\{\delta\}$. We extend
$G_{at}$ to the set of basic guards $G$ with element $\phi,\psi,\cdots$, which is generated by the following formation rules:

$$\phi::=\delta|\epsilon|\neg\phi|\psi\in G_{at}|\phi+\psi|\phi\cdot\psi$$

The predicate $test(\phi,s)$ represents that $\phi$ holds in the state $s$, and $test(\epsilon,s)$ holds and $test(\delta,s)$ does not hold. $effect(e,s)\in S$ denotes $s'$ in
$s\xrightarrow{e}s'$. The predicate weakest precondition $wp(e,\phi)$ denotes that $\forall s,s'\in S, test(\phi,effect(e,s))$ holds.

\subsubsection{Syntax}

We use the Prefix $.$ to model the causality relation $\leq$ in true concurrency, the Summation $+$ to model the conflict relation $\sharp$, and the Box-Summation $\boxplus_{\pi}$ to
model the probabilistic conflict relation $\sharp_{\pi}$ in true concurrency, and the Composition $\parallel$ to explicitly model concurrent relation in true concurrency. And we
follow the conventions of process algebra.

\begin{definition}[Syntax]\label{syntax5}
Truly concurrent processes CTC with probabilism and guards are defined inductively by the following formation rules:

\begin{enumerate}
  \item $A\in\mathcal{P}$;
  \item $\phi\in\mathcal{P}$;
  \item $\textbf{nil}\in\mathcal{P}$;
  \item if $P\in\mathcal{P}$, then the Prefix $\alpha.P\in\mathcal{P}$, for $\alpha\in Act$;
  \item if $P\in\mathcal{P}$, then the Prefix $\phi.P\in\mathcal{P}$, for $\phi\in G_{at}$;
  \item if $P,Q\in\mathcal{P}$, then the Summation $P+Q\in\mathcal{P}$;
  \item if $P,Q\in\mathcal{P}$, then the Box-Summation $P\boxplus_{\pi}Q\in\mathcal{P}$;
  \item if $P,Q\in\mathcal{P}$, then the Composition $P\parallel Q\in\mathcal{P}$;
  \item if $P\in\mathcal{P}$, then the Prefix $(\alpha_1\parallel\cdots\parallel\alpha_n).P\in\mathcal{P}\quad(n\in I)$, for $\alpha_,\cdots,\alpha_n\in Act$;
  \item if $P\in\mathcal{P}$, then the Restriction $P\setminus L\in\mathcal{P}$ with $L\in\mathcal{L}$;
  \item if $P\in\mathcal{P}$, then the Relabelling $P[f]\in\mathcal{P}$.
\end{enumerate}

The standard BNF grammar of syntax of CTC with probabilism and guards can be summarized as follows:

$$P::=A|\textbf{nil}|\alpha.P|\phi.P| P+P | P\boxplus_{\pi} P | P\parallel P | (\alpha_1\parallel\cdots\parallel\alpha_n).P | P\setminus L | P[f].$$
\end{definition}

\subsubsection{Operational Semantics}

The operational semantics is defined by LTSs (labelled transition systems), and it is detailed by the following definition.

\begin{definition}[Semantics]\label{semantics5}
The operational semantics of CTC with probabilism and guards corresponding to the syntax in Definition \ref{syntax5} is defined by a series of transition rules, named $\textbf{PAct}$, $\textbf{PSum}$, $\textbf{PBox-Sum}$,
$\textbf{Com}$, $\textbf{Res}$, $\textbf{Rel}$ and $\textbf{Con}$ and named $\textbf{Act}$, $\textbf{Sum}$,
$\textbf{Com}$, $\textbf{Res}$, $\textbf{Rel}$ and $\textbf{Con}$ indicate that the rules are associated respectively with Prefix, Summation, Composition, Restriction, Relabelling and
Constants in Definition \ref{syntax5}. They are shown in Table \ref{PTRForCTC5} and \ref{TRForCTC5}.

\begin{center}
    \begin{table}
        \[\textbf{PAct}_1\quad \frac{}{\langle\alpha.P,s\rangle\rightsquigarrow\langle\breve{\alpha}.P,s\rangle}\]

        \[\textbf{PSum}\quad \frac{\langle P,s\rangle\rightsquigarrow \langle P',s\rangle\quad Q\rightsquigarrow Q'}{\langle P+Q,s\rangle\rightsquigarrow \langle P'+Q',s\rangle}\]

        \[\textbf{PBox-Sum}\quad \frac{\langle P,s\rangle\rightsquigarrow \langle P',s\rangle}{\langle P\boxplus_{\pi}Q,s\rangle\rightsquigarrow \langle P',s\rangle}\quad \frac{\langle Q,s\rangle\rightsquigarrow \langle Q',s\rangle}{\langle P\boxplus_{\pi}Q,s\rangle\rightsquigarrow \langle Q',s\rangle}\]

        \[\textbf{PCom}\quad \frac{\langle P,s\rangle\rightsquigarrow \langle P',s\rangle\quad \langle Q,s\rangle\rightsquigarrow \langle Q',s\rangle}{\langle P\parallel Q,s\rangle\rightsquigarrow \langle P'+Q',s\rangle}\]

        \[\textbf{PAct}_2\quad \frac{}{\langle (\alpha_1\parallel\cdots\parallel\alpha_n).P,s\rangle\rightsquigarrow\langle (\breve{\alpha_1}\parallel\cdots\parallel\breve{\alpha_n}).P,s\rangle}\]

        \[\textbf{PRes}\quad \frac{\langle P,s\rangle\rightsquigarrow \langle P',s\rangle}{\langle P\setminus L,s\rangle\rightsquigarrow \langle P'\setminus L,s\rangle}\]

        \[\textbf{PRel}\quad \frac{\langle P,s\rangle\rightsquigarrow \langle P',s\rangle}{\langle P[f],s\rangle\rightsquigarrow \langle P'[f],s\rangle}\]

        \[\textbf{PCon}\quad\frac{\langle P,s\rangle\rightsquigarrow \langle P',s\rangle}{\langle A,s\rangle\rightsquigarrow \langle P',s\rangle}\quad (A\overset{\text{def}}{=}P)\]

        \caption{Probabilistic transition rules of CTC with probabilism and guards}
        \label{PTRForCTC5}
    \end{table}
\end{center}

\begin{center}
    \begin{table}
        \[\textbf{Act}_1\quad \frac{}{\langle\breve{\alpha}.P,s\rangle\xrightarrow{\alpha}\langle P,s'\rangle}\]

        \[\textbf{Act}_2\quad \frac{}{\langle\epsilon,s\rangle\rightarrow\langle\surd,s\rangle}\]

        \[\textbf{Gur}\quad \frac{}{\langle\phi,s\rangle\rightarrow\langle\surd,s\rangle}\textrm{ if }test(\phi,s)\]

        \[\textbf{Sum}_1\quad \frac{\langle P,s\rangle\xrightarrow{\alpha}\langle P',s'\rangle}{\langle P+Q,s\rangle\xrightarrow{\alpha}\langle P',s'\rangle}\]

        \[\textbf{Com}_1\quad \frac{\langle P,s\rangle\xrightarrow{\alpha}\langle P',s'\rangle\quad Q\nrightarrow}{\langle P\parallel Q,s\rangle\xrightarrow{\alpha}\langle P'\parallel Q,s'\rangle}\]

        \[\textbf{Com}_2\quad \frac{\langle Q,s\rangle\xrightarrow{\alpha}\langle Q',s'\rangle\quad P\nrightarrow}{\langle P\parallel Q,s\rangle\xrightarrow{\alpha}\langle P\parallel Q',s'\rangle}\]

        \[\textbf{Com}_3\quad \frac{\langle P,s\rangle\xrightarrow{\alpha}\langle P',s'\rangle\quad \langle Q,s\rangle\xrightarrow{\beta}\langle Q',s''\rangle}{\langle P\parallel Q,s\rangle\xrightarrow{\{\alpha,\beta\}}\langle P'\parallel Q',s'\cup s''\rangle}\quad (\beta\neq\overline{\alpha})\]

        \[\textbf{Com}_4\quad \frac{\langle P,s\rangle\xrightarrow{l}\langle P',s'\rangle\quad \langle Q,s\rangle\xrightarrow{\overline{l}}\langle Q',s''\rangle}{\langle P\parallel Q,s\rangle\xrightarrow{\tau}\langle P'\parallel Q',s'\cup s''\rangle}\]

        \[\textbf{Act}_3\quad \frac{}{\langle(\breve{\alpha_1}\parallel\cdots\parallel\breve{\alpha_n}).P,s\rangle\xrightarrow{\{\alpha_1,\cdots,\alpha_n\}}\langle P,s'\rangle}\quad (\alpha_i\neq\overline{\alpha_j}\quad i,j\in\{1,\cdots,n\})\]

        \[\textbf{Sum}_2\quad \frac{\langle P,s\rangle\xrightarrow{\{\alpha_1,\cdots,\alpha_n\}}\langle P',s'\rangle}{\langle P+Q,s\rangle\xrightarrow{\{\alpha_1,\cdots,\alpha_n\}}\langle P',s'\rangle}\]

        \[\textbf{Res}_1\quad \frac{\langle P,s\rangle\xrightarrow{\alpha}\langle P',s'\rangle}{\langle P\setminus L,s\rangle\xrightarrow{\alpha}\langle P'\setminus L,s'\rangle}\quad (\alpha,\overline{\alpha}\notin L)\]

        \[\textbf{Res}_2\quad \frac{\langle P,s\rangle\xrightarrow{\{\alpha_1,\cdots,\alpha_n\}}\langle P',s'\rangle}{\langle P\setminus L,s\rangle\xrightarrow{\{\alpha_1,\cdots,\alpha_n\}}\langle P'\setminus L,s'\rangle}\quad (\alpha_1,\overline{\alpha_1},\cdots,\alpha_n,\overline{\alpha_n}\notin L)\]

        \[\textbf{Rel}_1\quad \frac{\langle P,s\rangle\xrightarrow{\alpha}\langle P',s'\rangle}{\langle P[f],s\rangle\xrightarrow{\langle f(\alpha)}P'[f],s'\rangle}\]

        \[\textbf{Rel}_2\quad \frac{\langle P,s\rangle\xrightarrow{\{\alpha_1,\cdots,\alpha_n\}}\langle P',s'\rangle}{\langle P[f],s\rangle\xrightarrow{\{f(\alpha_1),\cdots,f(\alpha_n)\}}\langle P'[f],s'\rangle}\]

        \[\textbf{Con}_1\quad\frac{\langle P,s\rangle\xrightarrow{\alpha}\langle P',s'\rangle}{\langle A,s\rangle\xrightarrow{\alpha}\langle P',s'\rangle}\quad (A\overset{\text{def}}{=}P)\]

        \[\textbf{Con}_2\quad\frac{\langle P,s\rangle\xrightarrow{\{\alpha_1,\cdots,\alpha_n\}}\langle P',s'\rangle}{\langle A,s\rangle\xrightarrow{\{\alpha_1,\cdots,\alpha_n\}}\langle P',s'\rangle}\quad (A\overset{\text{def}}{=}P)\]

        \caption{Action transition rules of CTC with probabilism and guards}
        \label{TRForCTC5}
    \end{table}
\end{center}
\end{definition}

\subsubsection{Properties of Transitions}

\begin{definition}[Sorts]\label{sorts3}
Given the sorts $\mathcal{L}(A)$ and $\mathcal{L}(X)$ of constants and variables, we define $\mathcal{L}(P)$ inductively as follows.

\begin{enumerate}
  \item $\mathcal{L}(l.P)=\{l\}\cup\mathcal{L}(P)$;
  \item $\mathcal{L}((l_1\parallel \cdots\parallel l_n).P)=\{l_1,\cdots,l_n\}\cup\mathcal{L}(P)$;
  \item $\mathcal{L}(\tau.P)=\mathcal{L}(P)$;
  \item $\mathcal{L}(\epsilon.P)=\mathcal{L}(P)$;
  \item $\mathcal{L}(\phi.P)=\mathcal{L}(P)$;
  \item $\mathcal{L}(P+Q)=\mathcal{L}(P)\cup\mathcal{L}(Q)$;
  \item $\mathcal{L}(P\boxplus_{\pi}Q)=\mathcal{L}(P)\cup\mathcal{L}(Q)$;
  \item $\mathcal{L}(P\parallel Q)=\mathcal{L}(P)\cup\mathcal{L}(Q)$;
  \item $\mathcal{L}(P\setminus L)=\mathcal{L}(P)-(L\cup\overline{L})$;
  \item $\mathcal{L}(P[f])=\{f(l):l\in\mathcal{L}(P)\}$;
  \item for $A\overset{\text{def}}{=}P$, $\mathcal{L}(P)\subseteq\mathcal{L}(A)$.
\end{enumerate}
\end{definition}

Now, we present some properties of the transition rules defined in Table \ref{TRForCTC3}.

\begin{proposition}
If $P\xrightarrow{\alpha}P'$, then
\begin{enumerate}
  \item $\alpha\in\mathcal{L}(P)\cup\{\tau\}\cup\{\epsilon\}$;
  \item $\mathcal{L}(P')\subseteq\mathcal{L}(P)$.
\end{enumerate}

If $P\xrightarrow{\{\alpha_1,\cdots,\alpha_n\}}P'$, then
\begin{enumerate}
  \item $\alpha_1,\cdots,\alpha_n\in\mathcal{L}(P)\cup\{\tau\}\cup\{\epsilon\}$;
  \item $\mathcal{L}(P')\subseteq\mathcal{L}(P)$.
\end{enumerate}
\end{proposition}

\begin{proof}
By induction on the inference of $P\xrightarrow{\alpha}P'$ and $P\xrightarrow{\{\alpha_1,\cdots,\alpha_n\}}P'$, there are several cases corresponding to the transition rules named
$\textbf{Act}_{1,2,3}$, $\textbf{Gur}$, $\textbf{Sum}_{1,2}$, $\textbf{Com}_{1,2,3,4}$, $\textbf{Res}_{1,2}$, $\textbf{Rel}_{1,2}$ and $\textbf{Con}_{1,2}$ in Table \ref{TRForCTC3},
we just prove the one case $\textbf{Act}_1$ and $\textbf{Act}_3$, and omit the others.

Case $\textbf{Act}_1$: by $\textbf{Act}_1$, with $P\equiv\alpha.P'$. Then by Definition \ref{sorts3}, we have (1) $\mathcal{L}(P)=\{\alpha\}\cup\mathcal{L}(P')$ if $\alpha\neq\tau$;
(2) $\mathcal{L}(P)=\mathcal{L}(P')$ if $\alpha=\tau$ or $\alpha=\epsilon$. So, $\alpha\in\mathcal{L}(P)\cup\{\tau\}\cup\{\epsilon\}$, and $\mathcal{L}(P')\subseteq\mathcal{L}(P)$, as
desired.

Case $\textbf{Act}_3$: by $\textbf{Act}_3$, with $P\equiv(\alpha_1\parallel\cdots\parallel\alpha_n).P'$. Then by Definition \ref{sorts3}, we have (1) $\mathcal{L}(P)=\{\alpha_1,\cdots,\alpha_n\}\cup\mathcal{L}(P')$ if
$\alpha_i\neq\tau$ for $i\leq n$; (2) $\mathcal{L}(P)=\mathcal{L}(P')$ if $\alpha_1,\cdots,\alpha_n=\tau$ or $\alpha_1,\cdots,\alpha_n=\epsilon$. So,
$\alpha_1,\cdots,\alpha_n\in\mathcal{L}(P)\cup\{\tau\}\cup\{\epsilon\}$, and $\mathcal{L}(P')\subseteq\mathcal{L}(P)$, as desired.
\end{proof}

\subsection{Strong Bisimulations}\label{stcbctcpg}

\subsubsection{Laws and Congruence}

Based on the concepts of strongly probabilistic truly concurrent bisimulation equivalences, we get the following laws.

\begin{proposition}[Monoid laws for strongly probabilistic pomset bisimulation] The monoid laws for strongly probabilistic pomset bisimulation are as follows.
\begin{enumerate}
  \item $P+Q\sim_{pp} Q+P$;
  \item $P+(Q+R)\sim_{pp} (P+Q)+R$;
  \item $P+P\sim_{pp} P$;
  \item $P+\textbf{nil}\sim_{pp} P$.
\end{enumerate}
\end{proposition}

\begin{proof}
\begin{enumerate}
  \item $P+Q\sim_{pp} Q+P$. It is sufficient to prove the relation $R=\{(P+Q, Q+P)\}\cup \textbf{Id}$ is a strongly probabilistic pomset bisimulation, we omit it;
  \item $P+(Q+R)\sim_{pp} (P+Q)+R$. It is sufficient to prove the relation $R=\{(P+(Q+R), (P+Q)+R)\}\cup \textbf{Id}$ is a strongly probabilistic pomset bisimulation, we omit it;
  \item $P+P\sim_{pp} P$. It is sufficient to prove the relation $R=\{(P+P, P)\}\cup \textbf{Id}$ is a strongly probabilistic pomset bisimulation, we omit it;
  \item $P+\textbf{nil}\sim_{pp} P$. It is sufficient to prove the relation $R=\{(P+\textbf{nil}, P)\}\cup \textbf{Id}$ is a strongly probabilistic pomset bisimulation, we omit it.
\end{enumerate}
\end{proof}

\begin{proposition}[Monoid laws for strongly probabilistic step bisimulation] The monoid laws for strongly probabilistic step bisimulation are as follows.
\begin{enumerate}
  \item $P+Q\sim_{ps} Q+P$;
  \item $P+(Q+R)\sim_{ps} (P+Q)+R$;
  \item $P+P\sim_{ps} P$;
  \item $P+\textbf{nil}\sim_{ps} P$.
\end{enumerate}
\end{proposition}

\begin{proof}
\begin{enumerate}
  \item $P+Q\sim_{ps} Q+P$. It is sufficient to prove the relation $R=\{(P+Q, Q+P)\}\cup \textbf{Id}$ is a strongly probabilistic step bisimulation, we omit it;
  \item $P+(Q+R)\sim_{ps} (P+Q)+R$. It is sufficient to prove the relation $R=\{(P+(Q+R), (P+Q)+R)\}\cup \textbf{Id}$ is a strongly probabilistic step bisimulation, we omit it;
  \item $P+P\sim_{ps} P$. It is sufficient to prove the relation $R=\{(P+P, P)\}\cup \textbf{Id}$ is a strongly probabilistic step bisimulation, we omit it;
  \item $P+\textbf{nil}\sim_{ps} P$. It is sufficient to prove the relation $R=\{(P+\textbf{nil}, P)\}\cup \textbf{Id}$ is a strongly probabilistic step bisimulation, we omit it.
\end{enumerate}
\end{proof}

\begin{proposition}[Monoid laws for strongly probabilistic hp-bisimulation] The monoid laws for strongly probabilistic hp-bisimulation are as follows.
\begin{enumerate}
  \item $P+Q\sim_{php} Q+P$;
  \item $P+(Q+R)\sim_{php} (P+Q)+R$;
  \item $P+P\sim_{php} P$;
  \item $P+\textbf{nil}\sim_{php} P$.
\end{enumerate}
\end{proposition}

\begin{proof}
\begin{enumerate}
  \item $P+Q\sim_{php} Q+P$. It is sufficient to prove the relation $R=\{(P+Q, Q+P)\}\cup \textbf{Id}$ is a strongly probabilistic hp-bisimulation, we omit it;
  \item $P+(Q+R)\sim_{php} (P+Q)+R$. It is sufficient to prove the relation $R=\{(P+(Q+R), (P+Q)+R)\}\cup \textbf{Id}$ is a strongly probabilistic hp-bisimulation, we omit it;
  \item $P+P\sim_{php} P$. It is sufficient to prove the relation $R=\{(P+P, P)\}\cup \textbf{Id}$ is a strongly probabilistic hp-bisimulation, we omit it;
  \item $P+\textbf{nil}\sim_{php} P$. It is sufficient to prove the relation $R=\{(P+\textbf{nil}, P)\}\cup \textbf{Id}$ is a strongly probabilistic hp-bisimulation, we omit it.
\end{enumerate}
\end{proof}

\begin{proposition}[Monoid laws for strongly probabilistic hhp-bisimulation] The monoid laws for strongly probabilistic hhp-bisimulation are as follows.
\begin{enumerate}
  \item $P+Q\sim_{phhp} Q+P$;
  \item $P+(Q+R)\sim_{phhp} (P+Q)+R$;
  \item $P+P\sim_{phhp} P$;
  \item $P+\textbf{nil}\sim_{phhp} P$.
\end{enumerate}
\end{proposition}

\begin{proof}
\begin{enumerate}
  \item $P+Q\sim_{phhp} Q+P$. It is sufficient to prove the relation $R=\{(P+Q, Q+P)\}\cup \textbf{Id}$ is a strongly probabilistic hhp-bisimulation, we omit it;
  \item $P+(Q+R)\sim_{phhp} (P+Q)+R$. It is sufficient to prove the relation $R=\{(P+(Q+R), (P+Q)+R)\}\cup \textbf{Id}$ is a strongly probabilistic hhp-bisimulation, we omit it;
  \item $P+P\sim_{phhp} P$. It is sufficient to prove the relation $R=\{(P+P, P)\}\cup \textbf{Id}$ is a strongly probabilistic hhp-bisimulation, we omit it;
  \item $P+\textbf{nil}\sim_{phhp} P$. It is sufficient to prove the relation $R=\{(P+\textbf{nil}, P)\}\cup \textbf{Id}$ is a strongly probabilistic hhp-bisimulation, we omit it.
\end{enumerate}
\end{proof}

\begin{proposition}[Monoid laws 2 for strongly probabilistic pomset bisimulation]
The monoid laws 2 for strongly probabilistic pomset bisimulation are as follows.

\begin{enumerate}
  \item $P\boxplus_{\pi} Q\sim_{pp} Q\boxplus_{1-\pi} P$;
  \item $P\boxplus_{\pi}(Q\boxplus_{\rho} R)\sim_{pp} (P\boxplus_{\frac{\pi}{\pi+\rho-\pi\rho}}Q)\boxplus_{\pi+\rho-\pi\rho} R$;
  \item $P\boxplus_{\pi}P\sim_{pp} P$;
  \item $P\boxplus_{\pi}\textbf{nil}\sim_{pp} P$.
\end{enumerate}
\end{proposition}

\begin{proof}
\begin{enumerate}
  \item $P\boxplus_{\pi} Q\sim_{pp} Q\boxplus_{1-\pi} P$. It is sufficient to prove the relation $R=\{(P\boxplus_{\pi} Q, Q\boxplus_{1-\pi} P)\}\cup \textbf{Id}$ is a strongly probabilistic pomset bisimulation, we omit it;
  \item $P\boxplus_{\pi}(Q\boxplus_{\rho} R)\sim_{pp} (P\boxplus_{\frac{\pi}{\pi+\rho-\pi\rho}}Q)\boxplus_{\pi+\rho-\pi\rho} R$. It is sufficient to prove the relation $R=\{(P\boxplus_{\pi}(Q\boxplus_{\rho} R), (P\boxplus_{\frac{\pi}{\pi+\rho-\pi\rho}}Q)\boxplus_{\pi+\rho-\pi\rho} R)\}\cup \textbf{Id}$ is a strongly probabilistic pomset bisimulation, we omit it;
  \item $P\boxplus_{\pi}P\sim_{pp} P$. It is sufficient to prove the relation $R=\{(P\boxplus_{\pi}P, P)\}\cup \textbf{Id}$ is a strongly probabilistic pomset bisimulation, we omit it;
  \item $P\boxplus_{\pi}\textbf{nil}\sim_{pp} P$. It is sufficient to prove the relation $R=\{(P\boxplus_{\pi}\textbf{nil}, P)\}\cup \textbf{Id}$ is a strongly probabilistic pomset bisimulation, we omit it.
\end{enumerate}
\end{proof}

\begin{proposition}[Monoid laws 2 for strongly probabilistic step bisimulation]
The monoid laws 2 for strongly probabilistic step bisimulation are as follows.

\begin{enumerate}
  \item $P\boxplus_{\pi} Q\sim_{ps} Q\boxplus_{1-\pi} P$;
  \item $P\boxplus_{\pi}(Q\boxplus_{\rho} R)\sim_{ps} (P\boxplus_{\frac{\pi}{\pi+\rho-\pi\rho}}Q)\boxplus_{\pi+\rho-\pi\rho} R$;
  \item $P\boxplus_{\pi}P\sim_{ps} P$;
  \item $P\boxplus_{\pi}\textbf{nil}\sim_{ps} P$.
\end{enumerate}
\end{proposition}

\begin{proof}
\begin{enumerate}
  \item $P\boxplus_{\pi} Q\sim_{ps} Q\boxplus_{1-\pi} P$. It is sufficient to prove the relation $R=\{(P\boxplus_{\pi} Q, Q\boxplus_{1-\pi} P)\}\cup \textbf{Id}$ is a strongly probabilistic step bisimulation, we omit it;
  \item $P\boxplus_{\pi}(Q\boxplus_{\rho} R)\sim_{ps} (P\boxplus_{\frac{\pi}{\pi+\rho-\pi\rho}}Q)\boxplus_{\pi+\rho-\pi\rho} R$. It is sufficient to prove the relation $R=\{(P\boxplus_{\pi}(Q\boxplus_{\rho} R), (P\boxplus_{\frac{\pi}{\pi+\rho-\pi\rho}}Q)\boxplus_{\pi+\rho-\pi\rho} R)\}\cup \textbf{Id}$ is a strongly probabilistic step bisimulation, we omit it;
  \item $P\boxplus_{\pi}P\sim_{ps} P$. It is sufficient to prove the relation $R=\{(P\boxplus_{\pi}P, P)\}\cup \textbf{Id}$ is a strongly probabilistic step bisimulation, we omit it;
  \item $P\boxplus_{\pi}\textbf{nil}\sim_{ps} P$. It is sufficient to prove the relation $R=\{(P\boxplus_{\pi}\textbf{nil}, P)\}\cup \textbf{Id}$ is a strongly probabilistic step bisimulation, we omit it.
\end{enumerate}
\end{proof}

\begin{proposition}[Monoid laws 2 for strongly probabilistic hp-bisimulation]
The monoid laws 2 for strongly probabilistic hp-bisimulation are as follows.

\begin{enumerate}
  \item $P\boxplus_{\pi} Q\sim_{php} Q\boxplus_{1-\pi} P$;
  \item $P\boxplus_{\pi}(Q\boxplus_{\rho} R)\sim_{php} (P\boxplus_{\frac{\pi}{\pi+\rho-\pi\rho}}Q)\boxplus_{\pi+\rho-\pi\rho} R$;
  \item $P\boxplus_{\pi}P\sim_{php} P$;
  \item $P\boxplus_{\pi}\textbf{nil}\sim_{php} P$.
\end{enumerate}
\end{proposition}

\begin{proof}
\begin{enumerate}
  \item $P\boxplus_{\pi} Q\sim_{php} Q\boxplus_{1-\pi} P$. It is sufficient to prove the relation $R=\{(P\boxplus_{\pi} Q, Q\boxplus_{1-\pi} P)\}\cup \textbf{Id}$ is a strongly probabilistic hp-bisimulation, we omit it;
  \item $P\boxplus_{\pi}(Q\boxplus_{\rho} R)\sim_{php} (P\boxplus_{\frac{\pi}{\pi+\rho-\pi\rho}}Q)\boxplus_{\pi+\rho-\pi\rho} R$. It is sufficient to prove the relation $R=\{(P\boxplus_{\pi}(Q\boxplus_{\rho} R), (P\boxplus_{\frac{\pi}{\pi+\rho-\pi\rho}}Q)\boxplus_{\pi+\rho-\pi\rho} R)\}\cup \textbf{Id}$ is a strongly probabilistic hp-bisimulation, we omit it;
  \item $P\boxplus_{\pi}P\sim_{php} P$. It is sufficient to prove the relation $R=\{(P\boxplus_{\pi}P, P)\}\cup \textbf{Id}$ is a strongly probabilistic hp-bisimulation, we omit it;
  \item $P\boxplus_{\pi}\textbf{nil}\sim_{php} P$. It is sufficient to prove the relation $R=\{(P\boxplus_{\pi}\textbf{nil}, P)\}\cup \textbf{Id}$ is a strongly probabilistic hp-bisimulation, we omit it.
\end{enumerate}
\end{proof}

\begin{proposition}[Monoid laws 2 for strongly probabilistic hhp-bisimulation]
The monoid laws 2 for strongly probabilistic hhp-bisimulation are as follows.

\begin{enumerate}
  \item $P\boxplus_{\pi} Q\sim_{phhp} Q\boxplus_{1-\pi} P$;
  \item $P\boxplus_{\pi}(Q\boxplus_{\rho} R)\sim_{phhp} (P\boxplus_{\frac{\pi}{\pi+\rho-\pi\rho}}Q)\boxplus_{\pi+\rho-\pi\rho} R$;
  \item $P\boxplus_{\pi}P\sim_{phhp} P$;
  \item $P\boxplus_{\pi}\textbf{nil}\sim_{phhp} P$.
\end{enumerate}
\end{proposition}

\begin{proof}
\begin{enumerate}
  \item $P\boxplus_{\pi} Q\sim_{phhp} Q\boxplus_{1-\pi} P$. It is sufficient to prove the relation $R=\{(P\boxplus_{\pi} Q, Q\boxplus_{1-\pi} P)\}\cup \textbf{Id}$ is a strongly probabilistic hhp-bisimulation, we omit it;
  \item $P\boxplus_{\pi}(Q\boxplus_{\rho} R)\sim_{phhp} (P\boxplus_{\frac{\pi}{\pi+\rho-\pi\rho}}Q)\boxplus_{\pi+\rho-\pi\rho} R$. It is sufficient to prove the relation $R=\{(P\boxplus_{\pi}(Q\boxplus_{\rho} R), (P\boxplus_{\frac{\pi}{\pi+\rho-\pi\rho}}Q)\boxplus_{\pi+\rho-\pi\rho} R)\}\cup \textbf{Id}$ is a strongly probabilistic hhp-bisimulation, we omit it;
  \item $P\boxplus_{\pi}P\sim_{phhp} P$. It is sufficient to prove the relation $R=\{(P\boxplus_{\pi}P, P)\}\cup \textbf{Id}$ is a strongly probabilistic hhp-bisimulation, we omit it;
  \item $P\boxplus_{\pi}\textbf{nil}\sim_{phhp} P$. It is sufficient to prove the relation $R=\{(P\boxplus_{\pi}\textbf{nil}, P)\}\cup \textbf{Id}$ is a strongly probabilistic hhp-bisimulation, we omit it.
\end{enumerate}
\end{proof}

\begin{proposition}[Static laws for strongly probabilistic pomset bisimulation]
The static laws for strongly probabilistic pomset bisimulation are as follows.

\begin{enumerate}
  \item $P\parallel Q\sim_{pp} Q\parallel P$;
  \item $P\parallel(Q\parallel R)\sim_{pp} (P\parallel Q)\parallel R$;
  \item $P\parallel \textbf{nil}\sim_{pp} P$;
  \item $P\setminus L\sim_{pp} P$, if $\mathcal{L}(P)\cap(L\cup\overline{L})=\emptyset$;
  \item $P\setminus K\setminus L\sim_{pp} P\setminus(K\cup L)$;
  \item $P[f]\setminus L\sim_{pp} P\setminus f^{-1}(L)[f]$;
  \item $(P\parallel Q)\setminus L\sim_{pp} P\setminus L\parallel Q\setminus L$, if $\mathcal{L}(P)\cap\overline{\mathcal{L}(Q)}\cap(L\cup\overline{L})=\emptyset$;
  \item $P[Id]\sim_{pp} P$;
  \item $P[f]\sim_{pp} P[f']$, if $f\upharpoonright\mathcal{L}(P)=f'\upharpoonright\mathcal{L}(P)$;
  \item $P[f][f']\sim_{pp} P[f'\circ f]$;
  \item $(P\parallel Q)[f]\sim_{pp} P[f]\parallel Q[f]$, if $f\upharpoonright(L\cup\overline{L})$ is one-to-one, where $L=\mathcal{L}(P)\cup\mathcal{L}(Q)$.
\end{enumerate}
\end{proposition}

\begin{proof}
\begin{enumerate}
  \item $P\parallel Q\sim_{pp} Q\parallel P$. It is sufficient to prove the relation $R=\{(P\parallel Q, Q\parallel P)\}\cup \textbf{Id}$ is a strongly probabilistic pomset bisimulation, we omit it;
  \item $P\parallel(Q\parallel R)\sim_{pp} (P\parallel Q)\parallel R$. It is sufficient to prove the relation $R=\{(P\parallel(Q\parallel R), (P\parallel Q)\parallel R)\}\cup \textbf{Id}$ is a strongly probabilistic pomset bisimulation, we omit it;
  \item $P\parallel \textbf{nil}\sim_{pp} P$. It is sufficient to prove the relation $R=\{(P\parallel \textbf{nil}, P)\}\cup \textbf{Id}$ is a strongly probabilistic pomset bisimulation, we omit it;
  \item $P\setminus L\sim_{pp} P$, if $\mathcal{L}(P)\cap(L\cup\overline{L})=\emptyset$. It is sufficient to prove the relation $R=\{(P\setminus L, P)\}\cup \textbf{Id}$, if $\mathcal{L}(P)\cap(L\cup\overline{L})=\emptyset$, is a strongly probabilistic pomset bisimulation, we omit it;
  \item $P\setminus K\setminus L\sim_{pp} P\setminus(K\cup L)$. It is sufficient to prove the relation $R=\{(P\setminus K\setminus L, P\setminus(K\cup L))\}\cup \textbf{Id}$ is a strongly probabilistic pomset bisimulation, we omit it;
  \item $P[f]\setminus L\sim_{pp} P\setminus f^{-1}(L)[f]$. It is sufficient to prove the relation $R=\{(P[f]\setminus L, P\setminus f^{-1}(L)[f])\}\cup \textbf{Id}$ is a strongly probabilistic pomset bisimulation, we omit it;
  \item $(P\parallel Q)\setminus L\sim_{pp} P\setminus L\parallel Q\setminus L$, if $\mathcal{L}(P)\cap\overline{\mathcal{L}(Q)}\cap(L\cup\overline{L})=\emptyset$. It is sufficient to prove the relation $R=\{(P+Q, Q+P)\}\cup \textbf{Id}$ is a strongly probabilistic pomset bisimulation, we omit it;
  \item $P[Id]\sim_{pp} P$. It is sufficient to prove the relation $R=\{(P[Id], P)\}\cup \textbf{Id}$ is a strongly probabilistic pomset bisimulation, we omit it;
  \item $P[f]\sim_{pp} P[f']$, if $f\upharpoonright\mathcal{L}(P)=f'\upharpoonright\mathcal{L}(P)$. It is sufficient to prove the relation $R=\{(P[f], P[f'])\}\cup \textbf{Id}$, if $f\upharpoonright\mathcal{L}(P)=f'\upharpoonright\mathcal{L}(P)$, is a strongly probabilistic pomset bisimulation, we omit it;
  \item $P[f][f']\sim_{pp} P[f'\circ f]$. It is sufficient to prove the relation $R=\{(P[f][f'], P[f'\circ f])\}\cup \textbf{Id}$ is a strongly probabilistic pomset bisimulation, we omit it;
  \item $(P\parallel Q)[f]\sim_{pp} P[f]\parallel Q[f]$, if $f\upharpoonright(L\cup\overline{L})$ is one-to-one, where $L=\mathcal{L}(P)\cup\mathcal{L}(Q)$. It is sufficient to prove the relation $R=\{((P\parallel Q)[f], P[f]\parallel Q[f])\}\cup \textbf{Id}$, if $f\upharpoonright(L\cup\overline{L})$ is one-to-one, where $L=\mathcal{L}(P)\cup\mathcal{L}(Q)$, is a strongly probabilistic pomset bisimulation, we omit it.
\end{enumerate}
\end{proof}

\begin{proposition}[Static laws for strongly probabilistic step bisimulation]
The static laws for strongly probabilistic step bisimulation are as follows.

\begin{enumerate}
  \item $P\parallel Q\sim_{ps} Q\parallel P$;
  \item $P\parallel(Q\parallel R)\sim_{ps} (P\parallel Q)\parallel R$;
  \item $P\parallel \textbf{nil}\sim_{ps} P$;
  \item $P\setminus L\sim_{ps} P$, if $\mathcal{L}(P)\cap(L\cup\overline{L})=\emptyset$;
  \item $P\setminus K\setminus L\sim_{ps} P\setminus(K\cup L)$;
  \item $P[f]\setminus L\sim_{ps} P\setminus f^{-1}(L)[f]$;
  \item $(P\parallel Q)\setminus L\sim_{ps} P\setminus L\parallel Q\setminus L$, if $\mathcal{L}(P)\cap\overline{\mathcal{L}(Q)}\cap(L\cup\overline{L})=\emptyset$;
  \item $P[Id]\sim_{ps} P$;
  \item $P[f]\sim_{ps} P[f']$, if $f\upharpoonright\mathcal{L}(P)=f'\upharpoonright\mathcal{L}(P)$;
  \item $P[f][f']\sim_{ps} P[f'\circ f]$;
  \item $(P\parallel Q)[f]\sim_{ps} P[f]\parallel Q[f]$, if $f\upharpoonright(L\cup\overline{L})$ is one-to-one, where $L=\mathcal{L}(P)\cup\mathcal{L}(Q)$.
\end{enumerate}
\end{proposition}

\begin{proof}
\begin{enumerate}
  \item $P\parallel Q\sim_{ps} Q\parallel P$. It is sufficient to prove the relation $R=\{(P\parallel Q, Q\parallel P)\}\cup \textbf{Id}$ is a strongly probabilistic step bisimulation, we omit it;
  \item $P\parallel(Q\parallel R)\sim_{ps} (P\parallel Q)\parallel R$. It is sufficient to prove the relation $R=\{(P\parallel(Q\parallel R), (P\parallel Q)\parallel R)\}\cup \textbf{Id}$ is a strongly probabilistic step bisimulation, we omit it;
  \item $P\parallel \textbf{nil}\sim_{ps} P$. It is sufficient to prove the relation $R=\{(P\parallel \textbf{nil}, P)\}\cup \textbf{Id}$ is a strongly probabilistic step bisimulation, we omit it;
  \item $P\setminus L\sim_{ps} P$, if $\mathcal{L}(P)\cap(L\cup\overline{L})=\emptyset$. It is sufficient to prove the relation $R=\{(P\setminus L, P)\}\cup \textbf{Id}$, if $\mathcal{L}(P)\cap(L\cup\overline{L})=\emptyset$, is a strongly probabilistic step bisimulation, we omit it;
  \item $P\setminus K\setminus L\sim_{ps} P\setminus(K\cup L)$. It is sufficient to prove the relation $R=\{(P\setminus K\setminus L, P\setminus(K\cup L))\}\cup \textbf{Id}$ is a strongly probabilistic step bisimulation, we omit it;
  \item $P[f]\setminus L\sim_{ps} P\setminus f^{-1}(L)[f]$. It is sufficient to prove the relation $R=\{(P[f]\setminus L, P\setminus f^{-1}(L)[f])\}\cup \textbf{Id}$ is a strongly probabilistic step bisimulation, we omit it;
  \item $(P\parallel Q)\setminus L\sim_{ps} P\setminus L\parallel Q\setminus L$, if $\mathcal{L}(P)\cap\overline{\mathcal{L}(Q)}\cap(L\cup\overline{L})=\emptyset$. It is sufficient to prove the relation $R=\{(P+Q, Q+P)\}\cup \textbf{Id}$ is a strongly probabilistic step bisimulation, we omit it;
  \item $P[Id]\sim_{ps} P$. It is sufficient to prove the relation $R=\{(P[Id], P)\}\cup \textbf{Id}$ is a strongly probabilistic step bisimulation, we omit it;
  \item $P[f]\sim_{ps} P[f']$, if $f\upharpoonright\mathcal{L}(P)=f'\upharpoonright\mathcal{L}(P)$. It is sufficient to prove the relation $R=\{(P[f], P[f'])\}\cup \textbf{Id}$, if $f\upharpoonright\mathcal{L}(P)=f'\upharpoonright\mathcal{L}(P)$, is a strongly probabilistic step bisimulation, we omit it;
  \item $P[f][f']\sim_{ps} P[f'\circ f]$. It is sufficient to prove the relation $R=\{(P[f][f'], P[f'\circ f])\}\cup \textbf{Id}$ is a strongly probabilistic step bisimulation, we omit it;
  \item $(P\parallel Q)[f]\sim_{ps} P[f]\parallel Q[f]$, if $f\upharpoonright(L\cup\overline{L})$ is one-to-one, where $L=\mathcal{L}(P)\cup\mathcal{L}(Q)$. It is sufficient to prove the relation $R=\{((P\parallel Q)[f], P[f]\parallel Q[f])\}\cup \textbf{Id}$, if $f\upharpoonright(L\cup\overline{L})$ is one-to-one, where $L=\mathcal{L}(P)\cup\mathcal{L}(Q)$, is a strongly probabilistic step bisimulation, we omit it.
\end{enumerate}
\end{proof}

\begin{proposition}[Static laws for strongly probabilistic hp-bisimulation]
The static laws for strongly probabilistic hp-bisimulation are as follows.

\begin{enumerate}
  \item $P\parallel Q\sim_{php} Q\parallel P$;
  \item $P\parallel(Q\parallel R)\sim_{php} (P\parallel Q)\parallel R$;
  \item $P\parallel \textbf{nil}\sim_{php} P$;
  \item $P\setminus L\sim_{php} P$, if $\mathcal{L}(P)\cap(L\cup\overline{L})=\emptyset$;
  \item $P\setminus K\setminus L\sim_{php} P\setminus(K\cup L)$;
  \item $P[f]\setminus L\sim_{php} P\setminus f^{-1}(L)[f]$;
  \item $(P\parallel Q)\setminus L\sim_{php} P\setminus L\parallel Q\setminus L$, if $\mathcal{L}(P)\cap\overline{\mathcal{L}(Q)}\cap(L\cup\overline{L})=\emptyset$;
  \item $P[Id]\sim_{php} P$;
  \item $P[f]\sim_{php} P[f']$, if $f\upharpoonright\mathcal{L}(P)=f'\upharpoonright\mathcal{L}(P)$;
  \item $P[f][f']\sim_{php} P[f'\circ f]$;
  \item $(P\parallel Q)[f]\sim_{php} P[f]\parallel Q[f]$, if $f\upharpoonright(L\cup\overline{L})$ is one-to-one, where $L=\mathcal{L}(P)\cup\mathcal{L}(Q)$.
\end{enumerate}
\end{proposition}

\begin{proof}
\begin{enumerate}
  \item $P\parallel Q\sim_{php} Q\parallel P$. It is sufficient to prove the relation $R=\{(P\parallel Q, Q\parallel P)\}\cup \textbf{Id}$ is a strongly probabilistic hp-bisimulation, we omit it;
  \item $P\parallel(Q\parallel R)\sim_{php} (P\parallel Q)\parallel R$. It is sufficient to prove the relation $R=\{(P\parallel(Q\parallel R), (P\parallel Q)\parallel R)\}\cup \textbf{Id}$ is a strongly probabilistic hp-bisimulation, we omit it;
  \item $P\parallel \textbf{nil}\sim_{php} P$. It is sufficient to prove the relation $R=\{(P\parallel \textbf{nil}, P)\}\cup \textbf{Id}$ is a strongly probabilistic hp-bisimulation, we omit it;
  \item $P\setminus L\sim_{php} P$, if $\mathcal{L}(P)\cap(L\cup\overline{L})=\emptyset$. It is sufficient to prove the relation $R=\{(P\setminus L, P)\}\cup \textbf{Id}$, if $\mathcal{L}(P)\cap(L\cup\overline{L})=\emptyset$, is a strongly probabilistic hp-bisimulation, we omit it;
  \item $P\setminus K\setminus L\sim_{php} P\setminus(K\cup L)$. It is sufficient to prove the relation $R=\{(P\setminus K\setminus L, P\setminus(K\cup L))\}\cup \textbf{Id}$ is a strongly probabilistic hp-bisimulation, we omit it;
  \item $P[f]\setminus L\sim_{php} P\setminus f^{-1}(L)[f]$. It is sufficient to prove the relation $R=\{(P[f]\setminus L, P\setminus f^{-1}(L)[f])\}\cup \textbf{Id}$ is a strongly probabilistic hp-bisimulation, we omit it;
  \item $(P\parallel Q)\setminus L\sim_{php} P\setminus L\parallel Q\setminus L$, if $\mathcal{L}(P)\cap\overline{\mathcal{L}(Q)}\cap(L\cup\overline{L})=\emptyset$. It is sufficient to prove the relation $R=\{(P+Q, Q+P)\}\cup \textbf{Id}$ is a strongly probabilistic hp-bisimulation, we omit it;
  \item $P[Id]\sim_{php} P$. It is sufficient to prove the relation $R=\{(P[Id], P)\}\cup \textbf{Id}$ is a strongly probabilistic hp-bisimulation, we omit it;
  \item $P[f]\sim_{php} P[f']$, if $f\upharpoonright\mathcal{L}(P)=f'\upharpoonright\mathcal{L}(P)$. It is sufficient to prove the relation $R=\{(P[f], P[f'])\}\cup \textbf{Id}$, if $f\upharpoonright\mathcal{L}(P)=f'\upharpoonright\mathcal{L}(P)$, is a strongly probabilistic hp-bisimulation, we omit it;
  \item $P[f][f']\sim_{php} P[f'\circ f]$. It is sufficient to prove the relation $R=\{(P[f][f'], P[f'\circ f])\}\cup \textbf{Id}$ is a strongly probabilistic hp-bisimulation, we omit it;
  \item $(P\parallel Q)[f]\sim_{php} P[f]\parallel Q[f]$, if $f\upharpoonright(L\cup\overline{L})$ is one-to-one, where $L=\mathcal{L}(P)\cup\mathcal{L}(Q)$. It is sufficient to prove the relation $R=\{((P\parallel Q)[f], P[f]\parallel Q[f])\}\cup \textbf{Id}$, if $f\upharpoonright(L\cup\overline{L})$ is one-to-one, where $L=\mathcal{L}(P)\cup\mathcal{L}(Q)$, is a strongly probabilistic hp-bisimulation, we omit it.
\end{enumerate}
\end{proof}

\begin{proposition}[Static laws for strongly probabilistic hhp-bisimulation]
The static laws for strongly probabilistic hhp-bisimulation are as follows.

\begin{enumerate}
  \item $P\parallel Q\sim_{phhp} Q\parallel P$;
  \item $P\parallel(Q\parallel R)\sim_{phhp} (P\parallel Q)\parallel R$;
  \item $P\parallel \textbf{nil}\sim_{phhp} P$;
  \item $P\setminus L\sim_{phhp} P$, if $\mathcal{L}(P)\cap(L\cup\overline{L})=\emptyset$;
  \item $P\setminus K\setminus L\sim_{phhp} P\setminus(K\cup L)$;
  \item $P[f]\setminus L\sim_{phhp} P\setminus f^{-1}(L)[f]$;
  \item $(P\parallel Q)\setminus L\sim_{phhp} P\setminus L\parallel Q\setminus L$, if $\mathcal{L}(P)\cap\overline{\mathcal{L}(Q)}\cap(L\cup\overline{L})=\emptyset$;
  \item $P[Id]\sim_{phhp} P$;
  \item $P[f]\sim_{phhp} P[f']$, if $f\upharpoonright\mathcal{L}(P)=f'\upharpoonright\mathcal{L}(P)$;
  \item $P[f][f']\sim_{phhp} P[f'\circ f]$;
  \item $(P\parallel Q)[f]\sim_{phhp} P[f]\parallel Q[f]$, if $f\upharpoonright(L\cup\overline{L})$ is one-to-one, where $L=\mathcal{L}(P)\cup\mathcal{L}(Q)$.
\end{enumerate}
\end{proposition}

\begin{proof}
\begin{enumerate}
  \item $P\parallel Q\sim_{phhp} Q\parallel P$. It is sufficient to prove the relation $R=\{(P\parallel Q, Q\parallel P)\}\cup \textbf{Id}$ is a strongly probabilistic hhp-bisimulation, we omit it;
  \item $P\parallel(Q\parallel R)\sim_{phhp} (P\parallel Q)\parallel R$. It is sufficient to prove the relation $R=\{(P\parallel(Q\parallel R), (P\parallel Q)\parallel R)\}\cup \textbf{Id}$ is a strongly probabilistic hhp-bisimulation, we omit it;
  \item $P\parallel \textbf{nil}\sim_{phhp} P$. It is sufficient to prove the relation $R=\{(P\parallel \textbf{nil}, P)\}\cup \textbf{Id}$ is a strongly probabilistic hhp-bisimulation, we omit it;
  \item $P\setminus L\sim_{phhp} P$, if $\mathcal{L}(P)\cap(L\cup\overline{L})=\emptyset$. It is sufficient to prove the relation $R=\{(P\setminus L, P)\}\cup \textbf{Id}$, if $\mathcal{L}(P)\cap(L\cup\overline{L})=\emptyset$, is a strongly probabilistic hhp-bisimulation, we omit it;
  \item $P\setminus K\setminus L\sim_{phhp} P\setminus(K\cup L)$. It is sufficient to prove the relation $R=\{(P\setminus K\setminus L, P\setminus(K\cup L))\}\cup \textbf{Id}$ is a strongly probabilistic hhp-bisimulation, we omit it;
  \item $P[f]\setminus L\sim_{phhp} P\setminus f^{-1}(L)[f]$. It is sufficient to prove the relation $R=\{(P[f]\setminus L, P\setminus f^{-1}(L)[f])\}\cup \textbf{Id}$ is a strongly probabilistic hhp-bisimulation, we omit it;
  \item $(P\parallel Q)\setminus L\sim_{phhp} P\setminus L\parallel Q\setminus L$, if $\mathcal{L}(P)\cap\overline{\mathcal{L}(Q)}\cap(L\cup\overline{L})=\emptyset$. It is sufficient to prove the relation $R=\{(P+Q, Q+P)\}\cup \textbf{Id}$ is a strongly probabilistic hhp-bisimulation, we omit it;
  \item $P[Id]\sim_{phhp} P$. It is sufficient to prove the relation $R=\{(P[Id], P)\}\cup \textbf{Id}$ is a strongly probabilistic hhp-bisimulation, we omit it;
  \item $P[f]\sim_{phhp} P[f']$, if $f\upharpoonright\mathcal{L}(P)=f'\upharpoonright\mathcal{L}(P)$. It is sufficient to prove the relation $R=\{(P[f], P[f'])\}\cup \textbf{Id}$, if $f\upharpoonright\mathcal{L}(P)=f'\upharpoonright\mathcal{L}(P)$, is a strongly probabilistic hhp-bisimulation, we omit it;
  \item $P[f][f']\sim_{phhp} P[f'\circ f]$. It is sufficient to prove the relation $R=\{(P[f][f'], P[f'\circ f])\}\cup \textbf{Id}$ is a strongly probabilistic hhp-bisimulation, we omit it;
  \item $(P\parallel Q)[f]\sim_{phhp} P[f]\parallel Q[f]$, if $f\upharpoonright(L\cup\overline{L})$ is one-to-one, where $L=\mathcal{L}(P)\cup\mathcal{L}(Q)$. It is sufficient to prove the relation $R=\{((P\parallel Q)[f], P[f]\parallel Q[f])\}\cup \textbf{Id}$, if $f\upharpoonright(L\cup\overline{L})$ is one-to-one, where $L=\mathcal{L}(P)\cup\mathcal{L}(Q)$, is a strongly probabilistic hhp-bisimulation, we omit it.
\end{enumerate}
\end{proof}

\begin{proposition}[Guards laws for strongly probabilistic pomset bisimulation] The guards laws for strongly probabilistic pomset bisimulation are as follows.

\begin{enumerate}
  \item $P+\delta \sim_{pp} P$;
  \item $\delta.P \sim_{pp} \delta$;
  \item $\epsilon.P \sim_{pp} P$;
  \item $P.\epsilon \sim_{pp} P$;
  \item $\phi.\neg\phi \sim_{pp} \delta$;
  \item $\phi+\neg\phi \sim_{pp} \epsilon$;
  \item $\phi.\delta \sim_{pp} \delta$;
  \item $\phi.(P+Q)\sim_{pp}\phi.P+\phi.Q$;
  \item $\phi.(P.Q)\sim_{pp} \phi.P.Q$;
  \item $(\phi+\psi).P \sim_{pp} \phi.P + \psi.P$;
  \item $(\phi.\psi).P \sim_{pp} \phi.(\psi.P)$;
  \item $\phi\sim_{pp}\epsilon$ if $\forall s\in S.test(\phi,s)$;
  \item $\phi_0.\cdots.\phi_n \sim_{pp} \delta$ if $\forall s\in S,\exists i\leq n.test(\neg\phi_i,s)$;
  \item $wp(\alpha,\phi).\alpha.\phi\sim_{pp} wp(\alpha,\phi).\alpha$;
  \item $\neg wp(\alpha,\phi).\alpha.\neg\phi\sim_{pp}\neg wp(\alpha,\phi).\alpha$;
  \item $\delta\parallel P \sim_{pp} \delta$;
  \item $P\parallel \delta \sim_{pp} \delta$;
  \item $\epsilon\parallel P \sim_{pp} P$;
  \item $P\parallel \epsilon \sim_{pp} P$;
  \item $\phi.(P\parallel Q) \sim_{pp}\phi.P\parallel \phi.Q$;
  \item $\phi\parallel \delta \sim_{pp} \delta$;
  \item $\delta\parallel \phi \sim_{pp} \delta$;
  \item $\phi\parallel \epsilon \sim_{pp} \phi$;
  \item $\epsilon\parallel \phi \sim_{pp} \phi$;
  \item $\phi\parallel\neg\phi \sim_{pp} \delta$;
  \item $\phi_0\parallel\cdots\parallel\phi_n \sim_{pp} \delta$ if $\forall s_0,\cdots,s_n\in S,\exists i\leq n.test(\neg\phi_i,s_0\cup\cdots\cup s_n)$.
\end{enumerate}

\end{proposition}

\begin{proof}
\begin{enumerate}
  \item $P+\delta \sim_{pp} P$. It is sufficient to prove the relation $R=\{(P+\delta, P)\}\cup \textbf{Id}$ is a strongly probabilistic pomset bisimulation, and we omit it;
  \item $\delta.P \sim_{pp} \delta$. It is sufficient to prove the relation $R=\{(\delta.P, \delta)\}\cup \textbf{Id}$ is a strongly probabilistic pomset bisimulation, and we omit it;
  \item $\epsilon.P \sim_{pp} P$. It is sufficient to prove the relation $R=\{(\epsilon.P, P)\}\cup \textbf{Id}$ is a strongly probabilistic pomset bisimulation, and we omit it;
  \item $P.\epsilon \sim_{pp} P$. It is sufficient to prove the relation $R=\{(P.\epsilon, P)\}\cup \textbf{Id}$ is a strongly probabilistic pomset bisimulation, and we omit it;
  \item $\phi.\neg\phi \sim_{pp} \delta$. It is sufficient to prove the relation $R=\{(\phi.\neg\phi, \delta)\}\cup \textbf{Id}$ is a strongly probabilistic pomset bisimulation, and we omit it;
  \item $\phi+\neg\phi \sim_{pp} \epsilon$. It is sufficient to prove the relation $R=\{(\phi+\neg\phi, \epsilon)\}\cup \textbf{Id}$ is a strongly probabilistic pomset bisimulation, and we omit it;
  \item $\phi.\delta \sim_{pp} \delta$. It is sufficient to prove the relation $R=\{(\phi.\delta, \delta)\}\cup \textbf{Id}$ is a strongly probabilistic pomset bisimulation, and we omit it;
  \item $\phi.(P+Q)\sim_{pp}\phi.P+\phi.Q$. It is sufficient to prove the relation $R=\{(\phi.(P+Q), \phi.P+\phi.Q)\}\cup \textbf{Id}$ is a strongly probabilistic pomset bisimulation, and we omit it;
  \item $\phi.(P.Q)\sim_{pp} \phi.P.Q$. It is sufficient to prove the relation $R=\{(\phi.(P.Q), \phi.P.Q)\}\cup \textbf{Id}$ is a strongly probabilistic pomset bisimulation, and we omit it;
  \item $(\phi+\psi).P \sim_{pp} \phi.P + \psi.P$. It is sufficient to prove the relation $R=\{((\phi+\psi).P, \phi.P + \psi.P)\}\cup \textbf{Id}$ is a strongly probabilistic pomset bisimulation, and we omit it;
  \item $(\phi.\psi).P \sim_{pp} \phi.(\psi.P)$. It is sufficient to prove the relation $R=\{((\phi.\psi).P, \phi.(\psi.P))\}\cup \textbf{Id}$ is a strongly probabilistic pomset bisimulation, and we omit it;
  \item $\phi\sim_{pp}\epsilon$ if $\forall s\in S.test(\phi,s)$. It is sufficient to prove the relation $R=\{(\phi, \epsilon)\}\cup \textbf{Id}$, if $\forall s\in S.test(\phi,s)$, is a strongly probabilistic pomset bisimulation, and we omit it;
  \item $\phi_0.\cdots.\phi_n \sim_{pp} \delta$ if $\forall s\in S,\exists i\leq n.test(\neg\phi_i,s)$. It is sufficient to prove the relation $R=\{(\phi_0.\cdots.\phi_n, \delta)\}\cup \textbf{Id}$, if $\forall s\in S,\exists i\leq n.test(\neg\phi_i,s)$, is a strongly probabilistic pomset bisimulation, and we omit it;
  \item $wp(\alpha,\phi).\alpha.\phi\sim_{pp} wp(\alpha,\phi).\alpha$. It is sufficient to prove the relation $R=\{(wp(\alpha,\phi).\alpha.\phi, wp(\alpha,\phi).\alpha)\}\cup \textbf{Id}$ is a strongly probabilistic pomset bisimulation, and we omit it;
  \item $\neg wp(\alpha,\phi).\alpha.\neg\phi\sim_{pp}\neg wp(\alpha,\phi).e$. It is sufficient to prove the relation \\$R=\{(\neg wp(\alpha,\phi).\alpha.\neg\phi, \neg wp(\alpha,\phi).\alpha)\}\cup \textbf{Id}$ is a strongly probabilistic pomset bisimulation, and we omit it;
  \item $\delta\parallel P \sim_{pp} \delta$. It is sufficient to prove the relation $R=\{(\delta\parallel P, \delta)\}\cup \textbf{Id}$ is a strongly probabilistic pomset bisimulation, and we omit it;
  \item $P\parallel \delta \sim_{pp} \delta$. It is sufficient to prove the relation $R=\{(P\parallel \delta, \delta)\}\cup \textbf{Id}$ is a strongly probabilistic pomset bisimulation, and we omit it;
  \item $\epsilon\parallel P \sim_{pp} P$. It is sufficient to prove the relation $R=\{(\epsilon\parallel P, P)\}\cup \textbf{Id}$ is a strongly probabilistic pomset bisimulation, and we omit it;
  \item $P\parallel \epsilon \sim_{pp} P$. It is sufficient to prove the relation $R=\{(P\parallel \epsilon, P)\}\cup \textbf{Id}$ is a strongly probabilistic pomset bisimulation, and we omit it;
  \item $\phi.(P\parallel Q) \sim_{pp}\phi.P\parallel \phi.Q$. It is sufficient to prove the relation $R=\{(\phi.(P\parallel Q), \phi.P\parallel \phi.Q)\}\cup \textbf{Id}$ is a strongly probabilistic pomset bisimulation, and we omit it;
  \item $\phi\parallel \delta \sim_{pp} \delta$. It is sufficient to prove the relation $R=\{(\phi\parallel \delta, \delta)\}\cup \textbf{Id}$ is a strongly probabilistic pomset bisimulation, and we omit it;
  \item $\delta\parallel \phi \sim_{pp} \delta$. It is sufficient to prove the relation $R=\{(\delta\parallel \phi, \delta)\}\cup \textbf{Id}$ is a strongly probabilistic pomset bisimulation, and we omit it;
  \item $\phi\parallel \epsilon \sim_{pp} \phi$. It is sufficient to prove the relation $R=\{(\phi\parallel \epsilon, \phi)\}\cup \textbf{Id}$ is a strongly probabilistic pomset bisimulation, and we omit it;
  \item $\epsilon\parallel \phi \sim_{pp} \phi$. It is sufficient to prove the relation $R=\{(\epsilon\parallel \phi, \phi)\}\cup \textbf{Id}$ is a strongly probabilistic pomset bisimulation, and we omit it;
  \item $\phi\parallel\neg\phi \sim_{pp} \delta$. It is sufficient to prove the relation $R=\{(\phi\parallel\neg\phi, \delta)\}\cup \textbf{Id}$ is a strongly probabilistic pomset bisimulation, and we omit it;
  \item $\phi_0\parallel\cdots\parallel\phi_n \sim_{pp} \delta$ if $\forall s_0,\cdots,s_n\in S,\exists i\leq n.test(\neg\phi_i,s_0\cup\cdots\cup s_n)$. It is sufficient to prove the relation $R=\{(\phi_0\parallel\cdots\parallel\phi_n, \delta)\}\cup \textbf{Id}$, if $\forall s_0,\cdots,s_n\in S,\exists i\leq n.test(\neg\phi_i,s_0\cup\cdots\cup s_n)$, is a strongly probabilistic pomset bisimulation, and we omit it.
\end{enumerate}
\end{proof}

\begin{proposition}[Guards laws for strongly probabilistic step bisimulation] The guards laws for strongly probabilistic step bisimulation are as follows.

\begin{enumerate}
  \item $P+\delta \sim_{ps} P$;
  \item $\delta.P \sim_{ps} \delta$;
  \item $\epsilon.P \sim_{ps} P$;
  \item $P.\epsilon \sim_{ps} P$;
  \item $\phi.\neg\phi \sim_{ps} \delta$;
  \item $\phi+\neg\phi \sim_{ps} \epsilon$;
  \item $\phi.\delta \sim_{ps} \delta$;
  \item $\phi.(P+Q)\sim_{ps}\phi.P+\phi.Q$;
  \item $\phi.(P.Q)\sim_{ps} \phi.P.Q$;
  \item $(\phi+\psi).P \sim_{ps} \phi.P + \psi.P$;
  \item $(\phi.\psi).P \sim_{ps} \phi.(\psi.P)$;
  \item $\phi\sim_{ps}\epsilon$ if $\forall s\in S.test(\phi,s)$;
  \item $\phi_0.\cdots.\phi_n \sim_{ps} \delta$ if $\forall s\in S,\exists i\leq n.test(\neg\phi_i,s)$;
  \item $wp(\alpha,\phi).\alpha.\phi\sim_{ps} wp(\alpha,\phi).\alpha$;
  \item $\neg wp(\alpha,\phi).\alpha.\neg\phi\sim_{ps}\neg wp(\alpha,\phi).\alpha$;
  \item $\delta\parallel P \sim_{ps} \delta$;
  \item $P\parallel \delta \sim_{ps} \delta$;
  \item $\epsilon\parallel P \sim_{ps} P$;
  \item $P\parallel \epsilon \sim_{ps} P$;
  \item $\phi.(P\parallel Q) \sim_{ps}\phi.P\parallel \phi.Q$;
  \item $\phi\parallel \delta \sim_{ps} \delta$;
  \item $\delta\parallel \phi \sim_{ps} \delta$;
  \item $\phi\parallel \epsilon \sim_{ps} \phi$;
  \item $\epsilon\parallel \phi \sim_{ps} \phi$;
  \item $\phi\parallel\neg\phi \sim_{ps} \delta$;
  \item $\phi_0\parallel\cdots\parallel\phi_n \sim_{ps} \delta$ if $\forall s_0,\cdots,s_n\in S,\exists i\leq n.test(\neg\phi_i,s_0\cup\cdots\cup s_n)$.
\end{enumerate}

\end{proposition}

\begin{proof}
\begin{enumerate}
  \item $P+\delta \sim_{ps} P$. It is sufficient to prove the relation $R=\{(P+\delta, P)\}\cup \textbf{Id}$ is a strongly probabilistic step bisimulation, and we omit it;
  \item $\delta.P \sim_{ps} \delta$. It is sufficient to prove the relation $R=\{(\delta.P, \delta)\}\cup \textbf{Id}$ is a strongly probabilistic step bisimulation, and we omit it;
  \item $\epsilon.P \sim_{ps} P$. It is sufficient to prove the relation $R=\{(\epsilon.P, P)\}\cup \textbf{Id}$ is a strongly probabilistic step bisimulation, and we omit it;
  \item $P.\epsilon \sim_{ps} P$. It is sufficient to prove the relation $R=\{(P.\epsilon, P)\}\cup \textbf{Id}$ is a strongly probabilistic step bisimulation, and we omit it;
  \item $\phi.\neg\phi \sim_{ps} \delta$. It is sufficient to prove the relation $R=\{(\phi.\neg\phi, \delta)\}\cup \textbf{Id}$ is a strongly probabilistic step bisimulation, and we omit it;
  \item $\phi+\neg\phi \sim_{ps} \epsilon$. It is sufficient to prove the relation $R=\{(\phi+\neg\phi, \epsilon)\}\cup \textbf{Id}$ is a strongly probabilistic step bisimulation, and we omit it;
  \item $\phi.\delta \sim_{ps} \delta$. It is sufficient to prove the relation $R=\{(\phi.\delta, \delta)\}\cup \textbf{Id}$ is a strongly probabilistic step bisimulation, and we omit it;
  \item $\phi.(P+Q)\sim_{ps}\phi.P+\phi.Q$. It is sufficient to prove the relation $R=\{(\phi.(P+Q), \phi.P+\phi.Q)\}\cup \textbf{Id}$ is a strongly probabilistic step bisimulation, and we omit it;
  \item $\phi.(P.Q)\sim_{ps} \phi.P.Q$. It is sufficient to prove the relation $R=\{(\phi.(P.Q), \phi.P.Q)\}\cup \textbf{Id}$ is a strongly probabilistic step bisimulation, and we omit it;
  \item $(\phi+\psi).P \sim_{ps} \phi.P + \psi.P$. It is sufficient to prove the relation $R=\{((\phi+\psi).P, \phi.P + \psi.P)\}\cup \textbf{Id}$ is a strongly probabilistic step bisimulation, and we omit it;
  \item $(\phi.\psi).P \sim_{ps} \phi.(\psi.P)$. It is sufficient to prove the relation $R=\{((\phi.\psi).P, \phi.(\psi.P))\}\cup \textbf{Id}$ is a strongly probabilistic step bisimulation, and we omit it;
  \item $\phi\sim_{ps}\epsilon$ if $\forall s\in S.test(\phi,s)$. It is sufficient to prove the relation $R=\{(\phi, \epsilon)\}\cup \textbf{Id}$, if $\forall s\in S.test(\phi,s)$, is a strongly probabilistic step bisimulation, and we omit it;
  \item $\phi_0.\cdots.\phi_n \sim_{ps} \delta$ if $\forall s\in S,\exists i\leq n.test(\neg\phi_i,s)$. It is sufficient to prove the relation $R=\{(\phi_0.\cdots.\phi_n, \delta)\}\cup \textbf{Id}$, if $\forall s\in S,\exists i\leq n.test(\neg\phi_i,s)$, is a strongly probabilistic step bisimulation, and we omit it;
  \item $wp(\alpha,\phi).\alpha.\phi\sim_{ps} wp(\alpha,\phi).\alpha$. It is sufficient to prove the relation $R=\{(wp(\alpha,\phi).\alpha.\phi, wp(\alpha,\phi).\alpha)\}\cup \textbf{Id}$ is a strongly probabilistic step bisimulation, and we omit it;
  \item $\neg wp(\alpha,\phi).\alpha.\neg\phi\sim_{ps}\neg wp(\alpha,\phi).\alpha$. It is sufficient to prove the relation \\$R=\{(\neg wp(\alpha,\phi).\alpha.\neg\phi, \neg wp(\alpha,\phi).\alpha)\}\cup \textbf{Id}$ is a strongly probabilistic step bisimulation, and we omit it;
  \item $\delta\parallel P \sim_{ps} \delta$. It is sufficient to prove the relation $R=\{(\delta\parallel P, \delta)\}\cup \textbf{Id}$ is a strongly probabilistic step bisimulation, and we omit it;
  \item $P\parallel \delta \sim_{ps} \delta$. It is sufficient to prove the relation $R=\{(P\parallel \delta, \delta)\}\cup \textbf{Id}$ is a strongly probabilistic step bisimulation, and we omit it;
  \item $\epsilon\parallel P \sim_{ps} P$. It is sufficient to prove the relation $R=\{(\epsilon\parallel P, P)\}\cup \textbf{Id}$ is a strongly probabilistic step bisimulation, and we omit it;
  \item $P\parallel \epsilon \sim_{ps} P$. It is sufficient to prove the relation $R=\{(P\parallel \epsilon, P)\}\cup \textbf{Id}$ is a strongly probabilistic step bisimulation, and we omit it;
  \item $\phi.(P\parallel Q) \sim_{ps}\phi.P\parallel \phi.Q$. It is sufficient to prove the relation $R=\{(\phi.(P\parallel Q), \phi.P\parallel \phi.Q)\}\cup \textbf{Id}$ is a strongly probabilistic step bisimulation, and we omit it;
  \item $\phi\parallel \delta \sim_{ps} \delta$. It is sufficient to prove the relation $R=\{(\phi\parallel \delta, \delta)\}\cup \textbf{Id}$ is a strongly probabilistic step bisimulation, and we omit it;
  \item $\delta\parallel \phi \sim_{ps} \delta$. It is sufficient to prove the relation $R=\{(\delta\parallel \phi, \delta)\}\cup \textbf{Id}$ is a strongly probabilistic step bisimulation, and we omit it;
  \item $\phi\parallel \epsilon \sim_{ps} \phi$. It is sufficient to prove the relation $R=\{(\phi\parallel \epsilon, \phi)\}\cup \textbf{Id}$ is a strongly probabilistic step bisimulation, and we omit it;
  \item $\epsilon\parallel \phi \sim_{ps} \phi$. It is sufficient to prove the relation $R=\{(\epsilon\parallel \phi, \phi)\}\cup \textbf{Id}$ is a strongly probabilistic step bisimulation, and we omit it;
  \item $\phi\parallel\neg\phi \sim_{ps} \delta$. It is sufficient to prove the relation $R=\{(\phi\parallel\neg\phi, \delta)\}\cup \textbf{Id}$ is a strongly probabilistic step bisimulation, and we omit it;
  \item $\phi_0\parallel\cdots\parallel\phi_n \sim_{ps} \delta$ if $\forall s_0,\cdots,s_n\in S,\exists i\leq n.test(\neg\phi_i,s_0\cup\cdots\cup s_n)$. It is sufficient to prove the relation $R=\{(\phi_0\parallel\cdots\parallel\phi_n, \delta)\}\cup \textbf{Id}$, if $\forall s_0,\cdots,s_n\in S,\exists i\leq n.test(\neg\phi_i,s_0\cup\cdots\cup s_n)$, is a strongly probabilistic step bisimulation, and we omit it.
\end{enumerate}
\end{proof}

\begin{proposition}[Guards laws for strongly probabilistic hp-bisimulation] The guards laws for strongly probabilistic hp-bisimulation are as follows.

\begin{enumerate}
  \item $P+\delta \sim_{php} P$;
  \item $\delta.P \sim_{php} \delta$;
  \item $\epsilon.P \sim_{php} P$;
  \item $P.\epsilon \sim_{php} P$;
  \item $\phi.\neg\phi \sim_{php} \delta$;
  \item $\phi+\neg\phi \sim_{php} \epsilon$;
  \item $\phi.\delta \sim_{php} \delta$;
  \item $\phi.(P+Q)\sim_{php}\phi.P+\phi.Q$;
  \item $\phi.(P.Q)\sim_{php} \phi.P.Q$;
  \item $(\phi+\psi).P \sim_{php} \phi.P + \psi.P$;
  \item $(\phi.\psi).P \sim_{php} \phi.(\psi.P)$;
  \item $\phi\sim_{php}\epsilon$ if $\forall s\in S.test(\phi,s)$;
  \item $\phi_0.\cdots.\phi_n \sim_{php} \delta$ if $\forall s\in S,\exists i\leq n.test(\neg\phi_i,s)$;
  \item $wp(\alpha,\phi).\alpha.\phi\sim_{php} wp(\alpha,\phi).\alpha$;
  \item $\neg wp(\alpha,\phi).\alpha.\neg\phi\sim_{php}\neg wp(\alpha,\phi).\alpha$;
  \item $\delta\parallel P \sim_{php} \delta$;
  \item $P\parallel \delta \sim_{php} \delta$;
  \item $\epsilon\parallel P \sim_{php} P$;
  \item $P\parallel \epsilon \sim_{php} P$;
  \item $\phi.(P\parallel Q) \sim_{php}\phi.P\parallel \phi.Q$;
  \item $\phi\parallel \delta \sim_{php} \delta$;
  \item $\delta\parallel \phi \sim_{php} \delta$;
  \item $\phi\parallel \epsilon \sim_{php} \phi$;
  \item $\epsilon\parallel \phi \sim_{php} \phi$;
  \item $\phi\parallel\neg\phi \sim_{php} \delta$;
  \item $\phi_0\parallel\cdots\parallel\phi_n \sim_{php} \delta$ if $\forall s_0,\cdots,s_n\in S,\exists i\leq n.test(\neg\phi_i,s_0\cup\cdots\cup s_n)$.
\end{enumerate}

\end{proposition}

\begin{proof}
\begin{enumerate}
  \item $P+\delta \sim_{php} P$. It is sufficient to prove the relation $R=\{(P+\delta, P)\}\cup \textbf{Id}$ is a strongly probabilistic hp-bisimulation, and we omit it;
  \item $\delta.P \sim_{php} \delta$. It is sufficient to prove the relation $R=\{(\delta.P, \delta)\}\cup \textbf{Id}$ is a strongly probabilistic hp-bisimulation, and we omit it;
  \item $\epsilon.P \sim_{php} P$. It is sufficient to prove the relation $R=\{(\epsilon.P, P)\}\cup \textbf{Id}$ is a strongly probabilistic hp-bisimulation, and we omit it;
  \item $P.\epsilon \sim_{php} P$. It is sufficient to prove the relation $R=\{(P.\epsilon, P)\}\cup \textbf{Id}$ is a strongly probabilistic hp-bisimulation, and we omit it;
  \item $\phi.\neg\phi \sim_{php} \delta$. It is sufficient to prove the relation $R=\{(\phi.\neg\phi, \delta)\}\cup \textbf{Id}$ is a strongly probabilistic hp-bisimulation, and we omit it;
  \item $\phi+\neg\phi \sim_{php} \epsilon$. It is sufficient to prove the relation $R=\{(\phi+\neg\phi, \epsilon)\}\cup \textbf{Id}$ is a strongly probabilistic hp-bisimulation, and we omit it;
  \item $\phi.\delta \sim_{php} \delta$. It is sufficient to prove the relation $R=\{(\phi.\delta, \delta)\}\cup \textbf{Id}$ is a strongly probabilistic hp-bisimulation, and we omit it;
  \item $\phi.(P+Q)\sim_{php}\phi.P+\phi.Q$. It is sufficient to prove the relation $R=\{(\phi.(P+Q), \phi.P+\phi.Q)\}\cup \textbf{Id}$ is a strongly probabilistic hp-bisimulation, and we omit it;
  \item $\phi.(P.Q)\sim_{php} \phi.P.Q$. It is sufficient to prove the relation $R=\{(\phi.(P.Q), \phi.P.Q)\}\cup \textbf{Id}$ is a strongly probabilistic hp-bisimulation, and we omit it;
  \item $(\phi+\psi).P \sim_{php} \phi.P + \psi.P$. It is sufficient to prove the relation $R=\{((\phi+\psi).P, \phi.P + \psi.P)\}\cup \textbf{Id}$ is a strongly probabilistic hp-bisimulation, and we omit it;
  \item $(\phi.\psi).P \sim_{php} \phi.(\psi.P)$. It is sufficient to prove the relation $R=\{((\phi.\psi).P, \phi.(\psi.P))\}\cup \textbf{Id}$ is a strongly probabilistic hp-bisimulation, and we omit it;
  \item $\phi\sim_{php}\epsilon$ if $\forall s\in S.test(\phi,s)$. It is sufficient to prove the relation $R=\{(\phi, \epsilon)\}\cup \textbf{Id}$, if $\forall s\in S.test(\phi,s)$, is a strongly probabilistic hp-bisimulation, and we omit it;
  \item $\phi_0.\cdots.\phi_n \sim_{php} \delta$ if $\forall s\in S,\exists i\leq n.test(\neg\phi_i,s)$. It is sufficient to prove the relation $R=\{(\phi_0.\cdots.\phi_n, \delta)\}\cup \textbf{Id}$, if $\forall s\in S,\exists i\leq n.test(\neg\phi_i,s)$, is a strongly probabilistic hp-bisimulation, and we omit it;
  \item $wp(\alpha,\phi).\alpha.\phi\sim_{php} wp(\alpha,\phi).\alpha$. It is sufficient to prove the relation $R=\{(wp(\alpha,\phi).\alpha.\phi, wp(\alpha,\phi).\alpha)\}\cup \textbf{Id}$ is a strongly probabilistic hp-bisimulation, and we omit it;
  \item $\neg wp(\alpha,\phi).\alpha.\neg\phi\sim_{php}\neg wp(\alpha,\phi).\alpha$. It is sufficient to prove the relation \\$R=\{(\neg wp(\alpha,\phi).\alpha.\neg\phi, \neg wp(\alpha,\phi).\alpha)\}\cup \textbf{Id}$ is a strongly probabilistic hp-bisimulation, and we omit it;
  \item $\delta\parallel P \sim_{php} \delta$. It is sufficient to prove the relation $R=\{(\delta\parallel P, \delta)\}\cup \textbf{Id}$ is a strongly probabilistic hp-bisimulation, and we omit it;
  \item $P\parallel \delta \sim_{php} \delta$. It is sufficient to prove the relation $R=\{(P\parallel \delta, \delta)\}\cup \textbf{Id}$ is a strongly probabilistic hp-bisimulation, and we omit it;
  \item $\epsilon\parallel P \sim_{php} P$. It is sufficient to prove the relation $R=\{(\epsilon\parallel P, P)\}\cup \textbf{Id}$ is a strongly probabilistic hp-bisimulation, and we omit it;
  \item $P\parallel \epsilon \sim_{php} P$. It is sufficient to prove the relation $R=\{(P\parallel \epsilon, P)\}\cup \textbf{Id}$ is a strongly probabilistic hp-bisimulation, and we omit it;
  \item $\phi.(P\parallel Q) \sim_{php}\phi.P\parallel \phi.Q$. It is sufficient to prove the relation $R=\{(\phi.(P\parallel Q), \phi.P\parallel \phi.Q)\}\cup \textbf{Id}$ is a strongly probabilistic hp-bisimulation, and we omit it;
  \item $\phi\parallel \delta \sim_{php} \delta$. It is sufficient to prove the relation $R=\{(\phi\parallel \delta, \delta)\}\cup \textbf{Id}$ is a strongly probabilistic hp-bisimulation, and we omit it;
  \item $\delta\parallel \phi \sim_{php} \delta$. It is sufficient to prove the relation $R=\{(\delta\parallel \phi, \delta)\}\cup \textbf{Id}$ is a strongly probabilistic hp-bisimulation, and we omit it;
  \item $\phi\parallel \epsilon \sim_{php} \phi$. It is sufficient to prove the relation $R=\{(\phi\parallel \epsilon, \phi)\}\cup \textbf{Id}$ is a strongly probabilistic hp-bisimulation, and we omit it;
  \item $\epsilon\parallel \phi \sim_{php} \phi$. It is sufficient to prove the relation $R=\{(\epsilon\parallel \phi, \phi)\}\cup \textbf{Id}$ is a strongly probabilistic hp-bisimulation, and we omit it;
  \item $\phi\parallel\neg\phi \sim_{php} \delta$. It is sufficient to prove the relation $R=\{(\phi\parallel\neg\phi, \delta)\}\cup \textbf{Id}$ is a strongly probabilistic hp-bisimulation, and we omit it;
  \item $\phi_0\parallel\cdots\parallel\phi_n \sim_{php} \delta$ if $\forall s_0,\cdots,s_n\in S,\exists i\leq n.test(\neg\phi_i,s_0\cup\cdots\cup s_n)$. It is sufficient to prove the relation $R=\{(\phi_0\parallel\cdots\parallel\phi_n, \delta)\}\cup \textbf{Id}$, if $\forall s_0,\cdots,s_n\in S,\exists i\leq n.test(\neg\phi_i,s_0\cup\cdots\cup s_n)$, is a strongly probabilistic hp-bisimulation, and we omit it.
\end{enumerate}
\end{proof}

\begin{proposition}[Guards laws for strongly probabilistic hhp-bisimulation] The guards laws for strongly probabilistic hhp-bisimulation are as follows.

\begin{enumerate}
  \item $P+\delta \sim_{phhp} P$;
  \item $\delta.P \sim_{phhp} \delta$;
  \item $\epsilon.P \sim_{phhp} P$;
  \item $P.\epsilon \sim_{phhp} P$;
  \item $\phi.\neg\phi \sim_{phhp} \delta$;
  \item $\phi+\neg\phi \sim_{phhp} \epsilon$;
  \item $\phi.\delta \sim_{phhp} \delta$;
  \item $\phi.(P+Q)\sim_{phhp}\phi.P+\phi.Q$;
  \item $\phi.(P.Q)\sim_{phhp} \phi.P.Q$;
  \item $(\phi+\psi).P \sim_{phhp} \phi.P + \psi.P$;
  \item $(\phi.\psi).P \sim_{phhp} \phi.(\psi.P)$;
  \item $\phi\sim_{phhp}\epsilon$ if $\forall s\in S.test(\phi,s)$;
  \item $\phi_0.\cdots.\phi_n \sim_{phhp} \delta$ if $\forall s\in S,\exists i\leq n.test(\neg\phi_i,s)$;
  \item $wp(\alpha,\phi).\alpha.\phi\sim_{phhp} wp(\alpha,\phi).\alpha$;
  \item $\neg wp(\alpha,\phi).\alpha.\neg\phi\sim_{phhp}\neg wp(\alpha,\phi).\alpha$;
  \item $\delta\parallel P \sim_{phhp} \delta$;
  \item $P\parallel \delta \sim_{phhp} \delta$;
  \item $\epsilon\parallel P \sim_{phhp} P$;
  \item $P\parallel \epsilon \sim_{phhp} P$;
  \item $\phi.(P\parallel Q) \sim_{phhp}\phi.P\parallel \phi.Q$;
  \item $\phi\parallel \delta \sim_{phhp} \delta$;
  \item $\delta\parallel \phi \sim_{phhp} \delta$;
  \item $\phi\parallel \epsilon \sim_{phhp} \phi$;
  \item $\epsilon\parallel \phi \sim_{phhp} \phi$;
  \item $\phi\parallel\neg\phi \sim_{phhp} \delta$;
  \item $\phi_0\parallel\cdots\parallel\phi_n \sim_{phhp} \delta$ if $\forall s_0,\cdots,s_n\in S,\exists i\leq n.test(\neg\phi_i,s_0\cup\cdots\cup s_n)$.
\end{enumerate}

\end{proposition}

\begin{proof}
\begin{enumerate}
  \item $P+\delta \sim_{phhp} P$. It is sufficient to prove the relation $R=\{(P+\delta, P)\}\cup \textbf{Id}$ is a strongly probabilistic hhp-bisimulation, and we omit it;
  \item $\delta.P \sim_{phhp} \delta$. It is sufficient to prove the relation $R=\{(\delta.P, \delta)\}\cup \textbf{Id}$ is a strongly probabilistic hhp-bisimulation, and we omit it;
  \item $\epsilon.P \sim_{phhp} P$. It is sufficient to prove the relation $R=\{(\epsilon.P, P)\}\cup \textbf{Id}$ is a strongly probabilistic hhp-bisimulation, and we omit it;
  \item $P.\epsilon \sim_{phhp} P$. It is sufficient to prove the relation $R=\{(P.\epsilon, P)\}\cup \textbf{Id}$ is a strongly probabilistic hhp-bisimulation, and we omit it;
  \item $\phi.\neg\phi \sim_{phhp} \delta$. It is sufficient to prove the relation $R=\{(\phi.\neg\phi, \delta)\}\cup \textbf{Id}$ is a strongly probabilistic hhp-bisimulation, and we omit it;
  \item $\phi+\neg\phi \sim_{phhp} \epsilon$. It is sufficient to prove the relation $R=\{(\phi+\neg\phi, \epsilon)\}\cup \textbf{Id}$ is a strongly probabilistic hhp-bisimulation, and we omit it;
  \item $\phi.\delta \sim_{phhp} \delta$. It is sufficient to prove the relation $R=\{(\phi.\delta, \delta)\}\cup \textbf{Id}$ is a strongly probabilistic hhp-bisimulation, and we omit it;
  \item $\phi.(P+Q)\sim_{phhp}\phi.P+\phi.Q$. It is sufficient to prove the relation $R=\{(\phi.(P+Q), \phi.P+\phi.Q)\}\cup \textbf{Id}$ is a strongly probabilistic hhp-bisimulation, and we omit it;
  \item $\phi.(P.Q)\sim_{phhp} \phi.P.Q$. It is sufficient to prove the relation $R=\{(\phi.(P.Q), \phi.P.Q)\}\cup \textbf{Id}$ is a strongly probabilistic hhp-bisimulation, and we omit it;
  \item $(\phi+\psi).P \sim_{phhp} \phi.P + \psi.P$. It is sufficient to prove the relation $R=\{((\phi+\psi).P, \phi.P + \psi.P)\}\cup \textbf{Id}$ is a strongly probabilistic hhp-bisimulation, and we omit it;
  \item $(\phi.\psi).P \sim_{phhp} \phi.(\psi.P)$. It is sufficient to prove the relation $R=\{((\phi.\psi).P, \phi.(\psi.P))\}\cup \textbf{Id}$ is a strongly probabilistic hhp-bisimulation, and we omit it;
  \item $\phi\sim_{phhp}\epsilon$ if $\forall s\in S.test(\phi,s)$. It is sufficient to prove the relation $R=\{(\phi, \epsilon)\}\cup \textbf{Id}$, if $\forall s\in S.test(\phi,s)$, is a strongly probabilistic hhp-bisimulation, and we omit it;
  \item $\phi_0.\cdots.\phi_n \sim_{phhp} \delta$ if $\forall s\in S,\exists i\leq n.test(\neg\phi_i,s)$. It is sufficient to prove the relation $R=\{(\phi_0.\cdots.\phi_n, \delta)\}\cup \textbf{Id}$, if $\forall s\in S,\exists i\leq n.test(\neg\phi_i,s)$, is a strongly probabilistic hhp-bisimulation, and we omit it;
  \item $wp(\alpha,\phi).\alpha.\phi\sim_{phhp} wp(\alpha,\phi).\alpha$. It is sufficient to prove the relation $R=\{(wp(\alpha,\phi).\alpha.\phi, wp(\alpha,\phi).\alpha)\}\cup \textbf{Id}$ is a strongly probabilistic hhp-bisimulation, and we omit it;
  \item $\neg wp(\alpha,\phi).\alpha.\neg\phi\sim_{phhp}\neg wp(\alpha,\phi).\alpha$. It is sufficient to prove the relation \\$R=\{(\neg wp(\alpha,\phi).\alpha.\neg\phi, \neg wp(\alpha,\phi).\alpha)\}\cup \textbf{Id}$ is a strongly probabilistic hhp-bisimulation, and we omit it;
  \item $\delta\parallel P \sim_{phhp} \delta$. It is sufficient to prove the relation $R=\{(\delta\parallel P, \delta)\}\cup \textbf{Id}$ is a strongly probabilistic hhp-bisimulation, and we omit it;
  \item $P\parallel \delta \sim_{phhp} \delta$. It is sufficient to prove the relation $R=\{(P\parallel \delta, \delta)\}\cup \textbf{Id}$ is a strongly probabilistic hhp-bisimulation, and we omit it;
  \item $\epsilon\parallel P \sim_{phhp} P$. It is sufficient to prove the relation $R=\{(\epsilon\parallel P, P)\}\cup \textbf{Id}$ is a strongly probabilistic hhp-bisimulation, and we omit it;
  \item $P\parallel \epsilon \sim_{phhp} P$. It is sufficient to prove the relation $R=\{(P\parallel \epsilon, P)\}\cup \textbf{Id}$ is a strongly probabilistic hhp-bisimulation, and we omit it;
  \item $\phi.(P\parallel Q) \sim_{phhp}\phi.P\parallel \phi.Q$. It is sufficient to prove the relation $R=\{(\phi.(P\parallel Q), \phi.P\parallel \phi.Q)\}\cup \textbf{Id}$ is a strongly probabilistic hhp-bisimulation, and we omit it;
  \item $\phi\parallel \delta \sim_{phhp} \delta$. It is sufficient to prove the relation $R=\{(\phi\parallel \delta, \delta)\}\cup \textbf{Id}$ is a strongly probabilistic hhp-bisimulation, and we omit it;
  \item $\delta\parallel \phi \sim_{phhp} \delta$. It is sufficient to prove the relation $R=\{(\delta\parallel \phi, \delta)\}\cup \textbf{Id}$ is a strongly probabilistic hhp-bisimulation, and we omit it;
  \item $\phi\parallel \epsilon \sim_{phhp} \phi$. It is sufficient to prove the relation $R=\{(\phi\parallel \epsilon, \phi)\}\cup \textbf{Id}$ is a strongly probabilistic hhp-bisimulation, and we omit it;
  \item $\epsilon\parallel \phi \sim_{phhp} \phi$. It is sufficient to prove the relation $R=\{(\epsilon\parallel \phi, \phi)\}\cup \textbf{Id}$ is a strongly probabilistic hhp-bisimulation, and we omit it;
  \item $\phi\parallel\neg\phi \sim_{phhp} \delta$. It is sufficient to prove the relation $R=\{(\phi\parallel\neg\phi, \delta)\}\cup \textbf{Id}$ is a strongly probabilistic hhp-bisimulation, and we omit it;
  \item $\phi_0\parallel\cdots\parallel\phi_n \sim_{phhp} \delta$ if $\forall s_0,\cdots,s_n\in S,\exists i\leq n.test(\neg\phi_i,s_0\cup\cdots\cup s_n)$. It is sufficient to prove the relation $R=\{(\phi_0\parallel\cdots\parallel\phi_n, \delta)\}\cup \textbf{Id}$, if $\forall s_0,\cdots,s_n\in S,\exists i\leq n.test(\neg\phi_i,s_0\cup\cdots\cup s_n)$, is a strongly probabilistic hhp-bisimulation, and we omit it.
\end{enumerate}
\end{proof}

\begin{proposition}[Expansion law for strongly probabilistic pomset bisimulation]
Let $P\equiv (P_1[f_1]\parallel\cdots\parallel P_n[f_n])\setminus L$, with $n\geq 1$. Then

\begin{eqnarray}
P\sim_{pp} \{(f_1(\alpha_1)\parallel\cdots\parallel f_n(\alpha_n)).(P_1'[f_1]\parallel\cdots\parallel P_n'[f_n])\setminus L: \nonumber\\
\langle P_i,s_i\rangle\rightsquigarrow\xrightarrow{\alpha_i}\langle P_i',s_i'\rangle,i\in\{1,\cdots,n\},f_i(\alpha_i)\notin L\cup\overline{L}\} \nonumber\\
+\sum\{\tau.(P_1[f_1]\parallel\cdots\parallel P_i'[f_i]\parallel\cdots\parallel P_j'[f_j]\parallel\cdots\parallel P_n[f_n])\setminus L: \nonumber\\
\langle P_i,s_i\rangle\rightsquigarrow\xrightarrow{l_1}\langle P_i',s_i'\rangle,\langle P_j,s_j\rangle\rightsquigarrow\xrightarrow{l_2}\langle P_j',s_j'\rangle,f_i(l_1)=\overline{f_j(l_2)},i<j\} \nonumber
\end{eqnarray}
\end{proposition}

\begin{proof}
Firstly, we consider the case without Restriction and Relabeling. That is, we suffice to prove the following case by induction on the size $n$.

For $P\equiv P_1\parallel\cdots\parallel P_n$, with $n\geq 1$, we need to prove

\begin{eqnarray}
P\sim_{pp} \{(\alpha_1\parallel\cdots\parallel \alpha_n).(P_1'\parallel\cdots\parallel P_n'): \langle P_i,s_i\rangle\rightsquigarrow\xrightarrow{\alpha_i}\langle P_i',s_i'\rangle,i\in\{1,\cdots,n\}\nonumber\\
+\sum\{\tau.(P_1\parallel\cdots\parallel P_i'\parallel\cdots\parallel P_j'\parallel\cdots\parallel P_n): \langle P_i,s_i\rangle\rightsquigarrow\xrightarrow{l}\langle P_i',s_i'\rangle,\langle P_j,s_j\rangle\rightsquigarrow\xrightarrow{\overline{l}}\langle P_j',s_j'\rangle,i<j\} \nonumber
\end{eqnarray}

For $n=1$, $P_1\sim_{pp} \alpha_1.P_1':\langle P_1,s_1\rangle\rightsquigarrow\xrightarrow{\alpha_1}\langle P_1',s_1'\rangle$ is obvious. Then with a hypothesis $n$, we consider $R\equiv P\parallel P_{n+1}$. By the transition rules $\textbf{Com}_{1,2,3,4}$, we can get

\begin{eqnarray}
R\sim_{pp} \{(p\parallel \alpha_{n+1}).(P'\parallel P_{n+1}'): \langle P,s\rangle\rightsquigarrow\xrightarrow{p}\langle P',s'\rangle,\langle P_{n+1},s\rangle\rightsquigarrow\xrightarrow{\alpha_{n+1}}\langle P_{n+1}',s''\rangle,p\subseteq P\}\nonumber\\
+\sum\{\tau.(P'\parallel P_{n+1}'): \langle P,s\rangle\rightsquigarrow\xrightarrow{l}\langle P',s'\rangle,\langle P_{n+1},s\rangle\rightsquigarrow\xrightarrow{\overline{l}}\langle P_{n+1}',s''\rangle\} \nonumber
\end{eqnarray}

Now with the induction assumption $P\equiv P_1\parallel\cdots\parallel P_n$, the right-hand side can be reformulated as follows.

\begin{eqnarray}
\{(\alpha_1\parallel\cdots\parallel \alpha_n\parallel \alpha_{n+1}).(P_1'\parallel\cdots\parallel P_n'\parallel P_{n+1}'): \nonumber\\
\langle P_i,s_i\rangle\rightsquigarrow\xrightarrow{\alpha_i}\langle P_i',s_i'\rangle,i\in\{1,\cdots,n+1\}\nonumber\\
+\sum\{\tau.(P_1\parallel\cdots\parallel P_i'\parallel\cdots\parallel P_j'\parallel\cdots\parallel P_n\parallel P_{n+1}): \nonumber\\
\langle P_i,s_i\rangle\rightsquigarrow\xrightarrow{l}\langle P_i',s_i'\rangle,\langle P_j,s_j\rangle\rightsquigarrow\xrightarrow{\overline{l}}\langle P_j',s_j'\rangle,i<j\} \nonumber\\
+\sum\{\tau.(P_1\parallel\cdots\parallel P_i'\parallel\cdots\parallel P_j\parallel\cdots\parallel P_n\parallel P_{n+1}'): \nonumber\\
\langle P_i,s_i\rangle\rightsquigarrow\xrightarrow{l}\langle P_i',s_i'\rangle,\langle P_{n+1},s_{n+1}\rangle\rightsquigarrow\xrightarrow{\overline{l}}\langle P_{n+1}',s_{n+1}'\rangle,i\in\{1,\cdots, n\}\} \nonumber
\end{eqnarray}

So,

\begin{eqnarray}
R\sim_{pp} \{(\alpha_1\parallel\cdots\parallel \alpha_n\parallel \alpha_{n+1}).(P_1'\parallel\cdots\parallel P_n'\parallel P_{n+1}'): \nonumber\\
P_i\rightsquigarrow\xrightarrow{\alpha_i}P_i',i\in\{1,\cdots,n+1\}\nonumber\\
+\sum\{\tau.(P_1\parallel\cdots\parallel P_i'\parallel\cdots\parallel P_j'\parallel\cdots\parallel P_n): \nonumber\\
P_i\rightsquigarrow\xrightarrow{l}P_i',P_j\rightsquigarrow\xrightarrow{\overline{l}}P_j',1 \leq i<j\geq n+1\} \nonumber
\end{eqnarray}

Then, we can easily add the full conditions with Restriction and Relabeling.
\end{proof}

\begin{proposition}[Expansion law for strongly probabilistic step bisimulation]
Let $P\equiv (P_1[f_1]\parallel\cdots\parallel P_n[f_n])\setminus L$, with $n\geq 1$. Then

\begin{eqnarray}
P\sim_{ps} \{(f_1(\alpha_1)\parallel\cdots\parallel f_n(\alpha_n)).(P_1'[f_1]\parallel\cdots\parallel P_n'[f_n])\setminus L: \nonumber\\
\langle P_i,s_i\rangle\rightsquigarrow\xrightarrow{\alpha_i}\langle P_i',s_i'\rangle,i\in\{1,\cdots,n\},f_i(\alpha_i)\notin L\cup\overline{L}\} \nonumber\\
+\sum\{\tau.(P_1[f_1]\parallel\cdots\parallel P_i'[f_i]\parallel\cdots\parallel P_j'[f_j]\parallel\cdots\parallel P_n[f_n])\setminus L: \nonumber\\
\langle P_i,s_i\rangle\rightsquigarrow\xrightarrow{l_1}\langle P_i',s_i'\rangle,\langle P_j,s_j\rangle\rightsquigarrow\xrightarrow{l_2}\langle P_j',s_j'\rangle,f_i(l_1)=\overline{f_j(l_2)},i<j\} \nonumber
\end{eqnarray}
\end{proposition}

\begin{proof}
Firstly, we consider the case without Restriction and Relabeling. That is, we suffice to prove the following case by induction on the size $n$.

For $P\equiv P_1\parallel\cdots\parallel P_n$, with $n\geq 1$, we need to prove

\begin{eqnarray}
P\sim_{ps} \{(\alpha_1\parallel\cdots\parallel \alpha_n).(P_1'\parallel\cdots\parallel P_n'): \langle P_i,s_i\rangle\rightsquigarrow\xrightarrow{\alpha_i}\langle P_i',s_i'\rangle,i\in\{1,\cdots,n\}\nonumber\\
+\sum\{\tau.(P_1\parallel\cdots\parallel P_i'\parallel\cdots\parallel P_j'\parallel\cdots\parallel P_n): \langle P_i,s_i\rangle\rightsquigarrow\xrightarrow{l}\langle P_i',s_i'\rangle,\langle P_j,s_j\rangle\rightsquigarrow\xrightarrow{\overline{l}}\langle P_j',s_j'\rangle,i<j\} \nonumber
\end{eqnarray}

For $n=1$, $P_1\sim_{ps} \alpha_1.P_1':\langle P_1,s_1\rangle\rightsquigarrow\xrightarrow{\alpha_1}\langle P_1',s_1'\rangle$ is obvious. Then with a hypothesis $n$, we consider $R\equiv P\parallel P_{n+1}$. By the transition rules $\textbf{Com}_{1,2,3,4}$, we can get

\begin{eqnarray}
R\sim_{ps} \{(p\parallel \alpha_{n+1}).(P'\parallel P_{n+1}'): \langle P,s\rangle\rightsquigarrow\xrightarrow{p}\langle P',s'\rangle,\langle P_{n+1},s\rangle\rightsquigarrow\xrightarrow{\alpha_{n+1}}\langle P_{n+1}',s''\rangle,p\subseteq P\}\nonumber\\
+\sum\{\tau.(P'\parallel P_{n+1}'): \langle P,s\rangle\rightsquigarrow\xrightarrow{l}\langle P',s'\rangle,\langle P_{n+1},s\rangle\rightsquigarrow\xrightarrow{\overline{l}}\langle P_{n+1}',s''\rangle\} \nonumber
\end{eqnarray}

Now with the induction assumption $P\equiv P_1\parallel\cdots\parallel P_n$, the right-hand side can be reformulated as follows.

\begin{eqnarray}
\{(\alpha_1\parallel\cdots\parallel \alpha_n\parallel \alpha_{n+1}).(P_1'\parallel\cdots\parallel P_n'\parallel P_{n+1}'): \nonumber\\
\langle P_i,s_i\rangle\rightsquigarrow\xrightarrow{\alpha_i}\langle P_i',s_i'\rangle,i\in\{1,\cdots,n+1\}\nonumber\\
+\sum\{\tau.(P_1\parallel\cdots\parallel P_i'\parallel\cdots\parallel P_j'\parallel\cdots\parallel P_n\parallel P_{n+1}): \nonumber\\
\langle P_i,s_i\rangle\rightsquigarrow\xrightarrow{l}\langle P_i',s_i'\rangle,\langle P_j,s_j\rangle\rightsquigarrow\xrightarrow{\overline{l}}\langle P_j',s_j'\rangle,i<j\} \nonumber\\
+\sum\{\tau.(P_1\parallel\cdots\parallel P_i'\parallel\cdots\parallel P_j\parallel\cdots\parallel P_n\parallel P_{n+1}'): \nonumber\\
\langle P_i,s_i\rangle\rightsquigarrow\xrightarrow{l}\langle P_i',s_i'\rangle,\langle P_{n+1},s_{n+1}\rangle\rightsquigarrow\xrightarrow{\overline{l}}\langle P_{n+1}',s_{n+1}'\rangle,i\in\{1,\cdots, n\}\} \nonumber
\end{eqnarray}

So,

\begin{eqnarray}
R\sim_{ps} \{(\alpha_1\parallel\cdots\parallel \alpha_n\parallel \alpha_{n+1}).(P_1'\parallel\cdots\parallel P_n'\parallel P_{n+1}'): \nonumber\\
P_i\rightsquigarrow\xrightarrow{\alpha_i}P_i',i\in\{1,\cdots,n+1\}\nonumber\\
+\sum\{\tau.(P_1\parallel\cdots\parallel P_i'\parallel\cdots\parallel P_j'\parallel\cdots\parallel P_n): \nonumber\\
P_i\rightsquigarrow\xrightarrow{l}P_i',P_j\rightsquigarrow\xrightarrow{\overline{l}}P_j',1 \leq i<j\geq n+1\} \nonumber
\end{eqnarray}

Then, we can easily add the full conditions with Restriction and Relabeling.
\end{proof}

\begin{proposition}[Expansion law for strongly probabilistic hp-bisimulation]
Let $P\equiv (P_1[f_1]\parallel\cdots\parallel P_n[f_n])\setminus L$, with $n\geq 1$. Then

\begin{eqnarray}
P\sim_{php} \{(f_1(\alpha_1)\parallel\cdots\parallel f_n(\alpha_n)).(P_1'[f_1]\parallel\cdots\parallel P_n'[f_n])\setminus L: \nonumber\\
\langle P_i,s_i\rangle\rightsquigarrow\xrightarrow{\alpha_i}\langle P_i',s_i'\rangle,i\in\{1,\cdots,n\},f_i(\alpha_i)\notin L\cup\overline{L}\} \nonumber\\
+\sum\{\tau.(P_1[f_1]\parallel\cdots\parallel P_i'[f_i]\parallel\cdots\parallel P_j'[f_j]\parallel\cdots\parallel P_n[f_n])\setminus L: \nonumber\\
\langle P_i,s_i\rangle\rightsquigarrow\xrightarrow{l_1}\langle P_i',s_i'\rangle,\langle P_j,s_j\rangle\rightsquigarrow\xrightarrow{l_2}\langle P_j',s_j'\rangle,f_i(l_1)=\overline{f_j(l_2)},i<j\} \nonumber
\end{eqnarray}
\end{proposition}

\begin{proof}
Firstly, we consider the case without Restriction and Relabeling. That is, we suffice to prove the following case by induction on the size $n$.

For $P\equiv P_1\parallel\cdots\parallel P_n$, with $n\geq 1$, we need to prove

\begin{eqnarray}
P\sim_{php} \{(\alpha_1\parallel\cdots\parallel \alpha_n).(P_1'\parallel\cdots\parallel P_n'): \langle P_i,s_i\rangle\rightsquigarrow\xrightarrow{\alpha_i}\langle P_i',s_i'\rangle,i\in\{1,\cdots,n\}\nonumber\\
+\sum\{\tau.(P_1\parallel\cdots\parallel P_i'\parallel\cdots\parallel P_j'\parallel\cdots\parallel P_n): \langle P_i,s_i\rangle\rightsquigarrow\xrightarrow{l}\langle P_i',s_i'\rangle,\langle P_j,s_j\rangle\rightsquigarrow\xrightarrow{\overline{l}}\langle P_j',s_j'\rangle,i<j\} \nonumber
\end{eqnarray}

For $n=1$, $P_1\sim_{php} \alpha_1.P_1':\langle P_1,s_1\rangle\rightsquigarrow\xrightarrow{\alpha_1}\langle P_1',s_1'\rangle$ is obvious. Then with a hypothesis $n$, we consider $R\equiv P\parallel P_{n+1}$. By the transition rules $\textbf{Com}_{1,2,3,4}$, we can get

\begin{eqnarray}
R\sim_{php} \{(p\parallel \alpha_{n+1}).(P'\parallel P_{n+1}'): \langle P,s\rangle\rightsquigarrow\xrightarrow{p}\langle P',s'\rangle,\langle P_{n+1},s\rangle\rightsquigarrow\xrightarrow{\alpha_{n+1}}\langle P_{n+1}',s''\rangle,p\subseteq P\}\nonumber\\
+\sum\{\tau.(P'\parallel P_{n+1}'): \langle P,s\rangle\rightsquigarrow\xrightarrow{l}\langle P',s'\rangle,\langle P_{n+1},s\rangle\rightsquigarrow\xrightarrow{\overline{l}}\langle P_{n+1}',s''\rangle\} \nonumber
\end{eqnarray}

Now with the induction assumption $P\equiv P_1\parallel\cdots\parallel P_n$, the right-hand side can be reformulated as follows.

\begin{eqnarray}
\{(\alpha_1\parallel\cdots\parallel \alpha_n\parallel \alpha_{n+1}).(P_1'\parallel\cdots\parallel P_n'\parallel P_{n+1}'): \nonumber\\
\langle P_i,s_i\rangle\rightsquigarrow\xrightarrow{\alpha_i}\langle P_i',s_i'\rangle,i\in\{1,\cdots,n+1\}\nonumber\\
+\sum\{\tau.(P_1\parallel\cdots\parallel P_i'\parallel\cdots\parallel P_j'\parallel\cdots\parallel P_n\parallel P_{n+1}): \nonumber\\
\langle P_i,s_i\rangle\rightsquigarrow\xrightarrow{l}\langle P_i',s_i'\rangle,\langle P_j,s_j\rangle\rightsquigarrow\xrightarrow{\overline{l}}\langle P_j',s_j'\rangle,i<j\} \nonumber\\
+\sum\{\tau.(P_1\parallel\cdots\parallel P_i'\parallel\cdots\parallel P_j\parallel\cdots\parallel P_n\parallel P_{n+1}'): \nonumber\\
\langle P_i,s_i\rangle\rightsquigarrow\xrightarrow{l}\langle P_i',s_i'\rangle,\langle P_{n+1},s_{n+1}\rangle\rightsquigarrow\xrightarrow{\overline{l}}\langle P_{n+1}',s_{n+1}'\rangle,i\in\{1,\cdots, n\}\} \nonumber
\end{eqnarray}

So,

\begin{eqnarray}
R\sim_{php} \{(\alpha_1\parallel\cdots\parallel \alpha_n\parallel \alpha_{n+1}).(P_1'\parallel\cdots\parallel P_n'\parallel P_{n+1}'): \nonumber\\
P_i\rightsquigarrow\xrightarrow{\alpha_i}P_i',i\in\{1,\cdots,n+1\}\nonumber\\
+\sum\{\tau.(P_1\parallel\cdots\parallel P_i'\parallel\cdots\parallel P_j'\parallel\cdots\parallel P_n): \nonumber\\
P_i\rightsquigarrow\xrightarrow{l}P_i',P_j\rightsquigarrow\xrightarrow{\overline{l}}P_j',1 \leq i<j\geq n+1\} \nonumber
\end{eqnarray}

Then, we can easily add the full conditions with Restriction and Relabeling.
\end{proof}

\begin{proposition}[Expansion law for strongly probabilistic hhp-bisimulation]
Let $P\equiv (P_1[f_1]\parallel\cdots\parallel P_n[f_n])\setminus L$, with $n\geq 1$. Then

\begin{eqnarray}
P\sim_{phhp} \{(f_1(\alpha_1)\parallel\cdots\parallel f_n(\alpha_n)).(P_1'[f_1]\parallel\cdots\parallel P_n'[f_n])\setminus L: \nonumber\\
\langle P_i,s_i\rangle\rightsquigarrow\xrightarrow{\alpha_i}\langle P_i',s_i'\rangle,i\in\{1,\cdots,n\},f_i(\alpha_i)\notin L\cup\overline{L}\} \nonumber\\
+\sum\{\tau.(P_1[f_1]\parallel\cdots\parallel P_i'[f_i]\parallel\cdots\parallel P_j'[f_j]\parallel\cdots\parallel P_n[f_n])\setminus L: \nonumber\\
\langle P_i,s_i\rangle\rightsquigarrow\xrightarrow{l_1}\langle P_i',s_i'\rangle,\langle P_j,s_j\rangle\rightsquigarrow\xrightarrow{l_2}\langle P_j',s_j'\rangle,f_i(l_1)=\overline{f_j(l_2)},i<j\} \nonumber
\end{eqnarray}
\end{proposition}

\begin{proof}
Firstly, we consider the case without Restriction and Relabeling. That is, we suffice to prove the following case by induction on the size $n$.

For $P\equiv P_1\parallel\cdots\parallel P_n$, with $n\geq 1$, we need to prove

\begin{eqnarray}
P\sim_{phhp} \{(\alpha_1\parallel\cdots\parallel \alpha_n).(P_1'\parallel\cdots\parallel P_n'): \langle P_i,s_i\rangle\rightsquigarrow\xrightarrow{\alpha_i}\langle P_i',s_i'\rangle,i\in\{1,\cdots,n\}\nonumber\\
+\sum\{\tau.(P_1\parallel\cdots\parallel P_i'\parallel\cdots\parallel P_j'\parallel\cdots\parallel P_n): \langle P_i,s_i\rangle\rightsquigarrow\xrightarrow{l}\langle P_i',s_i'\rangle,\langle P_j,s_j\rangle\rightsquigarrow\xrightarrow{\overline{l}}\langle P_j',s_j'\rangle,i<j\} \nonumber
\end{eqnarray}

For $n=1$, $P_1\sim_{phhp} \alpha_1.P_1':\langle P_1,s_1\rangle\rightsquigarrow\xrightarrow{\alpha_1}\langle P_1',s_1'\rangle$ is obvious. Then with a hypothesis $n$, we consider $R\equiv P\parallel P_{n+1}$. By the transition rules $\textbf{Com}_{1,2,3,4}$, we can get

\begin{eqnarray}
R\sim_{phhp} \{(p\parallel \alpha_{n+1}).(P'\parallel P_{n+1}'): \langle P,s\rangle\rightsquigarrow\xrightarrow{p}\langle P',s'\rangle,\langle P_{n+1},s\rangle\rightsquigarrow\xrightarrow{\alpha_{n+1}}\langle P_{n+1}',s''\rangle,p\subseteq P\}\nonumber\\
+\sum\{\tau.(P'\parallel P_{n+1}'): \langle P,s\rangle\rightsquigarrow\xrightarrow{l}\langle P',s'\rangle,\langle P_{n+1},s\rangle\rightsquigarrow\xrightarrow{\overline{l}}\langle P_{n+1}',s''\rangle\} \nonumber
\end{eqnarray}

Now with the induction assumption $P\equiv P_1\parallel\cdots\parallel P_n$, the right-hand side can be reformulated as follows.

\begin{eqnarray}
\{(\alpha_1\parallel\cdots\parallel \alpha_n\parallel \alpha_{n+1}).(P_1'\parallel\cdots\parallel P_n'\parallel P_{n+1}'): \nonumber\\
\langle P_i,s_i\rangle\rightsquigarrow\xrightarrow{\alpha_i}\langle P_i',s_i'\rangle,i\in\{1,\cdots,n+1\}\nonumber\\
+\sum\{\tau.(P_1\parallel\cdots\parallel P_i'\parallel\cdots\parallel P_j'\parallel\cdots\parallel P_n\parallel P_{n+1}): \nonumber\\
\langle P_i,s_i\rangle\rightsquigarrow\xrightarrow{l}\langle P_i',s_i'\rangle,\langle P_j,s_j\rangle\rightsquigarrow\xrightarrow{\overline{l}}\langle P_j',s_j'\rangle,i<j\} \nonumber\\
+\sum\{\tau.(P_1\parallel\cdots\parallel P_i'\parallel\cdots\parallel P_j\parallel\cdots\parallel P_n\parallel P_{n+1}'): \nonumber\\
\langle P_i,s_i\rangle\rightsquigarrow\xrightarrow{l}\langle P_i',s_i'\rangle,\langle P_{n+1},s_{n+1}\rangle\rightsquigarrow\xrightarrow{\overline{l}}\langle P_{n+1}',s_{n+1}'\rangle,i\in\{1,\cdots, n\}\} \nonumber
\end{eqnarray}

So,

\begin{eqnarray}
R\sim_{phhp} \{(\alpha_1\parallel\cdots\parallel \alpha_n\parallel \alpha_{n+1}).(P_1'\parallel\cdots\parallel P_n'\parallel P_{n+1}'): \nonumber\\
P_i\rightsquigarrow\xrightarrow{\alpha_i}P_i',i\in\{1,\cdots,n+1\}\nonumber\\
+\sum\{\tau.(P_1\parallel\cdots\parallel P_i'\parallel\cdots\parallel P_j'\parallel\cdots\parallel P_n): \nonumber\\
P_i\rightsquigarrow\xrightarrow{l}P_i',P_j\rightsquigarrow\xrightarrow{\overline{l}}P_j',1 \leq i<j\geq n+1\} \nonumber
\end{eqnarray}

Then, we can easily add the full conditions with Restriction and Relabeling.
\end{proof}

\begin{theorem}[Congruence for strongly probabilistic pomset bisimulation]
We can enjoy the full congruence for strongly probabilistic pomset bisimulation as follows.
\begin{enumerate}
  \item If $A\overset{\text{def}}{=}P$, then $A\sim_{pp} P$;
  \item Let $P_1\sim_{pp} P_2$. Then
        \begin{enumerate}
           \item $\alpha.P_1\sim_{pp} \alpha.P_2$;
           \item $\phi.P_1\sim_{pp} \phi.P_2$;
           \item $(\alpha_1\parallel\cdots\parallel\alpha_n).P_1\sim_{pp} (\alpha_1\parallel\cdots\parallel\alpha_n).P_2$;
           \item $P_1+Q\sim_{pp} P_2 +Q$;
           \item $P_1\boxplus_{pi}Q\sim_{pp} P_2 \boxplus_{\pi}Q$;
           \item $P_1\parallel Q\sim_{pp} P_2\parallel Q$;
           \item $P_1\setminus L\sim_{pp} P_2\setminus L$;
           \item $P_1[f]\sim_{pp} P_2[f]$.
         \end{enumerate}
\end{enumerate}
\end{theorem}

\begin{proof}
\begin{enumerate}
  \item If $A\overset{\text{def}}{=}P$, then $A\sim_{pp} P$. It is obvious.
  \item Let $P_1\sim_{pp} P_2$. Then
        \begin{enumerate}
           \item $\alpha.P_1\sim_{pp} \alpha.P_2$. It is sufficient to prove the relation $R=\{(\alpha.P_1, \alpha.P_2)\}\cup \textbf{Id}$ is a strongly probabilistic pomset bisimulation, we omit it;
           \item $\phi.P_1\sim_{pp} \phi.P_2$. It is sufficient to prove the relation $R=\{(\phi.P_1, \phi.P_2)\}\cup \textbf{Id}$ is a strongly probabilistic pomset bisimulation, we omit it;
           \item $(\alpha_1\parallel\cdots\parallel\alpha_n).P_1\sim_{pp} (\alpha_1\parallel\cdots\parallel\alpha_n).P_2$. It is sufficient to prove the relation $R=\{((\alpha_1\parallel\cdots\parallel\alpha_n).P_1, (\alpha_1\parallel\cdots\parallel\alpha_n).P_2)\}\cup \textbf{Id}$ is a strongly probabilistic pomset bisimulation, we omit it;
           \item $P_1+Q\sim_{pp} P_2 +Q$. It is sufficient to prove the relation $R=\{(P_1+Q, P_2+Q)\}\cup \textbf{Id}$ is a strongly probabilistic pomset bisimulation, we omit it;
           \item $P_1\boxplus_{pi}Q\sim_{pp} P_2 \boxplus_{pi}Q$. It is sufficient to prove the relation $R=\{(P_1\boxplus_{pi}Q, P_2\boxplus_{pi}Q)\}\cup \textbf{Id}$ is a strongly probabilistic pomset bisimulation, we omit it;
           \item $P_1\parallel Q\sim_{pp} P_2\parallel Q$. It is sufficient to prove the relation $R=\{(P_1\parallel Q, P_2\parallel Q)\}\cup \textbf{Id}$ is a strongly probabilistic pomset bisimulation, we omit it;
           \item $P_1\setminus L\sim_{pp} P_2\setminus L$. It is sufficient to prove the relation $R=\{(P_1\setminus L, P_2\setminus L)\}\cup \textbf{Id}$ is a strongly probabilistic pomset bisimulation, we omit it;
           \item $P_1[f]\sim_{pp} P_2[f]$. It is sufficient to prove the relation $R=\{(P_1[f], P_2[f])\}\cup \textbf{Id}$ is a strongly probabilistic pomset bisimulation, we omit it.
         \end{enumerate}
\end{enumerate}
\end{proof}

\begin{theorem}[Congruence for strongly probabilistic step bisimulation]
We can enjoy the full congruence for strongly probabilistic step bisimulation as follows.
\begin{enumerate}
  \item If $A\overset{\text{def}}{=}P$, then $A\sim_{ps} P$;
  \item Let $P_1\sim_{ps} P_2$. Then
        \begin{enumerate}
           \item $\alpha.P_1\sim_{ps} \alpha.P_2$;
           \item $\phi.P_1\sim_{ps} \phi.P_2$;
           \item $(\alpha_1\parallel\cdots\parallel\alpha_n).P_1\sim_{ps} (\alpha_1\parallel\cdots\parallel\alpha_n).P_2$;
           \item $P_1+Q\sim_{ps} P_2 +Q$;
           \item $P_1\boxplus_{pi}Q\sim_{ps} P_2 \boxplus_{\pi}Q$;
           \item $P_1\parallel Q\sim_{ps} P_2\parallel Q$;
           \item $P_1\setminus L\sim_{ps} P_2\setminus L$;
           \item $P_1[f]\sim_{ps} P_2[f]$.
         \end{enumerate}
\end{enumerate}
\end{theorem}

\begin{proof}
\begin{enumerate}
  \item If $A\overset{\text{def}}{=}P$, then $A\sim_{ps} P$. It is obvious.
  \item Let $P_1\sim_{ps} P_2$. Then
        \begin{enumerate}
           \item $\alpha.P_1\sim_{ps} \alpha.P_2$. It is sufficient to prove the relation $R=\{(\alpha.P_1, \alpha.P_2)\}\cup \textbf{Id}$ is a strongly probabilistic step bisimulation, we omit it;
           \item $\phi.P_1\sim_{ps} \phi.P_2$. It is sufficient to prove the relation $R=\{(\phi.P_1, \phi.P_2)\}\cup \textbf{Id}$ is a strongly probabilistic step bisimulation, we omit it;
           \item $(\alpha_1\parallel\cdots\parallel\alpha_n).P_1\sim_{ps} (\alpha_1\parallel\cdots\parallel\alpha_n).P_2$. It is sufficient to prove the relation $R=\{((\alpha_1\parallel\cdots\parallel\alpha_n).P_1, (\alpha_1\parallel\cdots\parallel\alpha_n).P_2)\}\cup \textbf{Id}$ is a strongly probabilistic step bisimulation, we omit it;
           \item $P_1+Q\sim_{ps} P_2 +Q$. It is sufficient to prove the relation $R=\{(P_1+Q, P_2+Q)\}\cup \textbf{Id}$ is a strongly probabilistic step bisimulation, we omit it;
           \item $P_1\boxplus_{pi}Q\sim_{ps} P_2 \boxplus_{pi}Q$. It is sufficient to prove the relation $R=\{(P_1\boxplus_{pi}Q, P_2\boxplus_{pi}Q)\}\cup \textbf{Id}$ is a strongly probabilistic step bisimulation, we omit it;
           \item $P_1\parallel Q\sim_{ps} P_2\parallel Q$. It is sufficient to prove the relation $R=\{(P_1\parallel Q, P_2\parallel Q)\}\cup \textbf{Id}$ is a strongly probabilistic step bisimulation, we omit it;
           \item $P_1\setminus L\sim_{ps} P_2\setminus L$. It is sufficient to prove the relation $R=\{(P_1\setminus L, P_2\setminus L)\}\cup \textbf{Id}$ is a strongly probabilistic step bisimulation, we omit it;
           \item $P_1[f]\sim_{ps} P_2[f]$. It is sufficient to prove the relation $R=\{(P_1[f], P_2[f])\}\cup \textbf{Id}$ is a strongly probabilistic step bisimulation, we omit it.
         \end{enumerate}
\end{enumerate}
\end{proof}

\begin{theorem}[Congruence for strongly probabilistic hp-bisimulation]
We can enjoy the full congruence for strongly probabilistic hp-bisimulation as follows.
\begin{enumerate}
  \item If $A\overset{\text{def}}{=}P$, then $A\sim_{php} P$;
  \item Let $P_1\sim_{php} P_2$. Then
        \begin{enumerate}
           \item $\alpha.P_1\sim_{php} \alpha.P_2$;
           \item $\phi.P_1\sim_{php} \phi.P_2$;
           \item $(\alpha_1\parallel\cdots\parallel\alpha_n).P_1\sim_{php} (\alpha_1\parallel\cdots\parallel\alpha_n).P_2$;
           \item $P_1+Q\sim_{php} P_2 +Q$;
           \item $P_1\boxplus_{pi}Q\sim_{php} P_2 \boxplus_{\pi}Q$;
           \item $P_1\parallel Q\sim_{php} P_2\parallel Q$;
           \item $P_1\setminus L\sim_{php} P_2\setminus L$;
           \item $P_1[f]\sim_{php} P_2[f]$.
         \end{enumerate}
\end{enumerate}
\end{theorem}

\begin{proof}
\begin{enumerate}
  \item If $A\overset{\text{def}}{=}P$, then $A\sim_{php} P$. It is obvious.
  \item Let $P_1\sim_{php} P_2$. Then
        \begin{enumerate}
           \item $\alpha.P_1\sim_{php} \alpha.P_2$. It is sufficient to prove the relation $R=\{(\alpha.P_1, \alpha.P_2)\}\cup \textbf{Id}$ is a strongly probabilistic hp-bisimulation, we omit it;
           \item $\phi.P_1\sim_{php} \phi.P_2$. It is sufficient to prove the relation $R=\{(\phi.P_1, \phi.P_2)\}\cup \textbf{Id}$ is a strongly probabilistic hp-bisimulation, we omit it;
           \item $(\alpha_1\parallel\cdots\parallel\alpha_n).P_1\sim_{php} (\alpha_1\parallel\cdots\parallel\alpha_n).P_2$. It is sufficient to prove the relation $R=\{((\alpha_1\parallel\cdots\parallel\alpha_n).P_1, (\alpha_1\parallel\cdots\parallel\alpha_n).P_2)\}\cup \textbf{Id}$ is a strongly probabilistic hp-bisimulation, we omit it;
           \item $P_1+Q\sim_{php} P_2 +Q$. It is sufficient to prove the relation $R=\{(P_1+Q, P_2+Q)\}\cup \textbf{Id}$ is a strongly probabilistic hp-bisimulation, we omit it;
           \item $P_1\boxplus_{pi}Q\sim_{php} P_2 \boxplus_{pi}Q$. It is sufficient to prove the relation $R=\{(P_1\boxplus_{pi}Q, P_2\boxplus_{pi}Q)\}\cup \textbf{Id}$ is a strongly probabilistic hp-bisimulation, we omit it;
           \item $P_1\parallel Q\sim_{php} P_2\parallel Q$. It is sufficient to prove the relation $R=\{(P_1\parallel Q, P_2\parallel Q)\}\cup \textbf{Id}$ is a strongly probabilistic hp-bisimulation, we omit it;
           \item $P_1\setminus L\sim_{php} P_2\setminus L$. It is sufficient to prove the relation $R=\{(P_1\setminus L, P_2\setminus L)\}\cup \textbf{Id}$ is a strongly probabilistic hp-bisimulation, we omit it;
           \item $P_1[f]\sim_{php} P_2[f]$. It is sufficient to prove the relation $R=\{(P_1[f], P_2[f])\}\cup \textbf{Id}$ is a strongly probabilistic hp-bisimulation, we omit it.
         \end{enumerate}
\end{enumerate}
\end{proof}

\begin{theorem}[Congruence for strongly probabilistic hhp-bisimulation]
We can enjoy the full congruence for strongly probabilistic hhp-bisimulation as follows.
\begin{enumerate}
  \item If $A\overset{\text{def}}{=}P$, then $A\sim_{phhp} P$;
  \item Let $P_1\sim_{phhp} P_2$. Then
        \begin{enumerate}
           \item $\alpha.P_1\sim_{phhp} \alpha.P_2$;
           \item $\phi.P_1\sim_{phhp} \phi.P_2$;
           \item $(\alpha_1\parallel\cdots\parallel\alpha_n).P_1\sim_{phhp} (\alpha_1\parallel\cdots\parallel\alpha_n).P_2$;
           \item $P_1+Q\sim_{phhp} P_2 +Q$;
           \item $P_1\boxplus_{pi}Q\sim_{phhp} P_2 \boxplus_{\pi}Q$;
           \item $P_1\parallel Q\sim_{phhp} P_2\parallel Q$;
           \item $P_1\setminus L\sim_{phhp} P_2\setminus L$;
           \item $P_1[f]\sim_{phhp} P_2[f]$.
         \end{enumerate}
\end{enumerate}
\end{theorem}

\begin{proof}
\begin{enumerate}
  \item If $A\overset{\text{def}}{=}P$, then $A\sim_{phhp} P$. It is obvious.
  \item Let $P_1\sim_{phhp} P_2$. Then
        \begin{enumerate}
           \item $\alpha.P_1\sim_{phhp} \alpha.P_2$. It is sufficient to prove the relation $R=\{(\alpha.P_1, \alpha.P_2)\}\cup \textbf{Id}$ is a strongly probabilistic hhp-bisimulation, we omit it;
           \item $\phi.P_1\sim_{phhp} \phi.P_2$. It is sufficient to prove the relation $R=\{(\phi.P_1, \phi.P_2)\}\cup \textbf{Id}$ is a strongly probabilistic hhp-bisimulation, we omit it;
           \item $(\alpha_1\parallel\cdots\parallel\alpha_n).P_1\sim_{phhp} (\alpha_1\parallel\cdots\parallel\alpha_n).P_2$. It is sufficient to prove the relation $R=\{((\alpha_1\parallel\cdots\parallel\alpha_n).P_1, (\alpha_1\parallel\cdots\parallel\alpha_n).P_2)\}\cup \textbf{Id}$ is a strongly probabilistic hhp-bisimulation, we omit it;
           \item $P_1+Q\sim_{phhp} P_2 +Q$. It is sufficient to prove the relation $R=\{(P_1+Q, P_2+Q)\}\cup \textbf{Id}$ is a strongly probabilistic hhp-bisimulation, we omit it;
           \item $P_1\boxplus_{pi}Q\sim_{phhp} P_2 \boxplus_{pi}Q$. It is sufficient to prove the relation $R=\{(P_1\boxplus_{pi}Q, P_2\boxplus_{pi}Q)\}\cup \textbf{Id}$ is a strongly probabilistic hhp-bisimulation, we omit it;
           \item $P_1\parallel Q\sim_{phhp} P_2\parallel Q$. It is sufficient to prove the relation $R=\{(P_1\parallel Q, P_2\parallel Q)\}\cup \textbf{Id}$ is a strongly probabilistic hhp-bisimulation, we omit it;
           \item $P_1\setminus L\sim_{phhp} P_2\setminus L$. It is sufficient to prove the relation $R=\{(P_1\setminus L, P_2\setminus L)\}\cup \textbf{Id}$ is a strongly probabilistic hhp-bisimulation, we omit it;
           \item $P_1[f]\sim_{phhp} P_2[f]$. It is sufficient to prove the relation $R=\{(P_1[f], P_2[f])\}\cup \textbf{Id}$ is a strongly probabilistic hhp-bisimulation, we omit it.
         \end{enumerate}
\end{enumerate}
\end{proof}

\subsubsection{Recursion}

\begin{definition}[Weakly guarded recursive expression]
$X$ is weakly guarded in $E$ if each occurrence of $X$ is with some subexpression $\alpha.F$ or $(\alpha_1\parallel\cdots\parallel\alpha_n).F$ of $E$.
\end{definition}

\begin{lemma}\label{LUS5}
If the variables $\widetilde{X}$ are weakly guarded in $E$, and $\langle E\{\widetilde{P}/\widetilde{X}\},s\rangle\rightsquigarrow\xrightarrow{\{\alpha_1,\cdots,\alpha_n\}}\langle P',s'\rangle$, then
$P'$ takes the form $E'\{\widetilde{P}/\widetilde{X}\}$ for some expression $E'$, and moreover, for any $\widetilde{Q}$,
$\langle E\{\widetilde{Q}/\widetilde{X}\},s\rangle\rightsquigarrow\xrightarrow{\{\alpha_1,\cdots,\alpha_n\}}\langle E'\{\widetilde{Q}/\widetilde{X}\},s'\rangle$.
\end{lemma}

\begin{proof}
It needs to induct on the depth of the inference of $\langle E\{\widetilde{P}/\widetilde{X}\},s\rangle\rightsquigarrow\xrightarrow{\{\alpha_1,\cdots,\alpha_n\}}\langle P',s'\rangle$.

\begin{enumerate}
  \item Case $E\equiv Y$, a variable. Then $Y\notin \widetilde{X}$. Since $\widetilde{X}$ are weakly guarded, $\langle Y\{\widetilde{P}/\widetilde{X}\equiv Y\},s\rangle\nrightarrow$, this case is
  impossible.
  \item Case $E\equiv\beta.F$. Then we must have $\alpha=\beta$, and $P'\equiv F\{\widetilde{P}/\widetilde{X}\}$, and
  $\langle E\{\widetilde{Q}/\widetilde{X}\},s\rangle\equiv \langle \beta.F\{\widetilde{Q}/\widetilde{X}\},s\rangle \rightsquigarrow\xrightarrow{\beta}\langle F\{\widetilde{Q}/\widetilde{X}\},s'\rangle$,
  then, let $E'$ be $F$, as desired.
  \item Case $E\equiv(\beta_1\parallel\cdots\parallel\beta_n).F$. Then we must have $\alpha_i=\beta_i$ for $1\leq i\leq n$, and $P'\equiv F\{\widetilde{P}/\widetilde{X}\}$, and
  $\langle E\{\widetilde{Q}/\widetilde{X}\},s\rangle\equiv \langle(\beta_1\parallel\cdots\parallel\beta_n).F\{\widetilde{Q}/\widetilde{X}\},s\rangle \rightsquigarrow\xrightarrow{\{\beta_1,\cdots,\beta_n\}}\langle F\{\widetilde{Q}/\widetilde{X}\},s'\rangle$,
  then, let $E'$ be $F$, as desired.
  \item Case $E\equiv E_1+E_2$. Then either $\langle E_1\{\widetilde{P}/\widetilde{X}\},s\rangle \rightsquigarrow\xrightarrow{\{\alpha_1,\cdots,\alpha_n\}}\langle P',s'\rangle$ or
  $\langle E_2\{\widetilde{P}/\widetilde{X}\},s\rangle \rightsquigarrow\xrightarrow{\{\alpha_1,\cdots,\alpha_n\}}\langle P',s'\rangle$, then, we can apply this lemma in either case, as desired.
  \item Case $E\equiv E_1\parallel E_2$. There are four possibilities.
  \begin{enumerate}
    \item We may have $\langle E_1\{\widetilde{P}/\widetilde{X}\},s\rangle \rightsquigarrow\xrightarrow{\alpha}\langle P_1',s'\rangle$ and $\langle E_2\{\widetilde{P}/\widetilde{X}\},s\rangle\nrightarrow$
    with $P'\equiv P_1'\parallel (E_2\{\widetilde{P}/\widetilde{X}\})$, then by applying this lemma, $P_1'$ is of the form $E_1'\{\widetilde{P}/\widetilde{X}\}$, and for any $Q$,
    $\langle E_1\{\widetilde{Q}/\widetilde{X}\},s\rangle\rightsquigarrow\xrightarrow{\alpha} \langle E_1'\{\widetilde{Q}/\widetilde{X}\},s'\rangle$. So, $P'$ is of the form
    $E_1'\parallel E_2\{\widetilde{P}/\widetilde{X}\}$, and for any $Q$,
    $\langle E\{\widetilde{Q}/\widetilde{X}\}\equiv E_1\{\widetilde{Q}/\widetilde{X}\}\parallel E_2\{\widetilde{Q}/\widetilde{X}\},s\rangle\rightsquigarrow\xrightarrow{\alpha} \langle(E_1'\parallel E_2)\{\widetilde{Q}/\widetilde{X}\},s'\rangle$,
    then, let $E'$ be $E_1'\parallel E_2$, as desired.
    \item We may have $\langle E_2\{\widetilde{P}/\widetilde{X}\},s\rangle \rightsquigarrow\xrightarrow{\alpha}\langle P_2',s'\rangle$ and $\langle E_1\{\widetilde{P}/\widetilde{X}\},s\rangle\nrightarrow$
    with $P'\equiv P_2'\parallel (E_1\{\widetilde{P}/\widetilde{X}\})$, this case can be prove similarly to the above subcase, as desired.
    \item We may have $\langle E_1\{\widetilde{P}/\widetilde{X}\},s\rangle \rightsquigarrow\xrightarrow{\alpha}\langle P_1',s'\rangle$ and
    $\langle E_2\{\widetilde{P}/\widetilde{X}\},s\rangle\rightsquigarrow\xrightarrow{\beta}\langle P_2',s''\rangle$ with $\alpha\neq\overline{\beta}$ and $P'\equiv P_1'\parallel P_2'$, then by
    applying this lemma, $P_1'$ is of the form $E_1'\{\widetilde{P}/\widetilde{X}\}$, and for any $Q$,
    $\langle E_1\{\widetilde{Q}/\widetilde{X}\},s\rangle\rightsquigarrow\xrightarrow{\alpha} \langle E_1'\{\widetilde{Q}/\widetilde{X}\},s'\rangle$; $P_2'$ is of the form
    $E_2'\{\widetilde{P}/\widetilde{X}\}$, and for any $Q$, $\langle E_2\{\widetilde{Q}/\widetilde{X}\},s\rangle\rightsquigarrow\xrightarrow{\alpha} \langle E_2'\{\widetilde{Q}/\widetilde{X}\},s''\rangle$.
    So, $P'$ is of the form $E_1'\parallel E_2'\{\widetilde{P}/\widetilde{X}\}$, and for any $Q$,
    $\langle E\{\widetilde{Q}/\widetilde{X}\}\equiv E_1\{\widetilde{Q}/\widetilde{X}\}\parallel E_2\{\widetilde{Q}/\widetilde{X}\},s\rangle\rightsquigarrow\xrightarrow{\{\alpha,\beta\}}
    \langle (E_1'\parallel E_2')\{\widetilde{Q}/\widetilde{X}\},s'\cup s''\rangle$, then, let $E'$ be $E_1'\parallel E_2'$, as desired.
    \item We may have $\langle E_1\{\widetilde{P}/\widetilde{X}\},s\rangle \rightsquigarrow\xrightarrow{l}\langle P_1',s'\rangle$ and
    $\langle E_2\{\widetilde{P}/\widetilde{X}\},s\rangle\rightsquigarrow\xrightarrow{\overline{l}}\langle P_2',s''\rangle$ with $P'\equiv P_1'\parallel P_2'$, then by applying this lemma,
    $P_1'$ is of the form $E_1'\{\widetilde{P}/\widetilde{X}\}$, and for any $Q$, $\langle E_1\{\widetilde{Q}/\widetilde{X}\},s\rangle\rightsquigarrow\xrightarrow{l} \langle E_1'\{\widetilde{Q}/\widetilde{X}\},s'\rangle$;
    $P_2'$ is of the form $E_2'\{\widetilde{P}/\widetilde{X}\}$, and for any $Q$, $\langle E_2\{\widetilde{Q}/\widetilde{X}\},s\rangle\rightsquigarrow\xrightarrow{\overline{l}}\langle E_2'\{\widetilde{Q}/\widetilde{X}\},s''\rangle$.
    So, $P'$ is of the form $E_1'\parallel E_2'\{\widetilde{P}/\widetilde{X}\}$, and for any $Q$, $\langle E\{\widetilde{Q}/\widetilde{X}\}\equiv E_1\{\widetilde{Q}/\widetilde{X}\}\parallel E_2\{\widetilde{Q}/\widetilde{X}\},s\rangle
    \rightsquigarrow\xrightarrow{\tau} \langle (E_1'\parallel E_2')\{\widetilde{Q}/\widetilde{X}\},s'\cup s''\rangle$, then, let $E'$ be $E_1'\parallel E_2'$, as desired.
  \end{enumerate}
  \item Case $E\equiv F[R]$ and $E\equiv F\setminus L$. These cases can be prove similarly to the above case.
  \item Case $E\equiv C$, an agent constant defined by $C\overset{\text{def}}{=}R$. Then there is no $X\in\widetilde{X}$ occurring in $E$, so
  $\langle C,s\rangle\rightsquigarrow\xrightarrow{\{\alpha_1,\cdots,\alpha_n\}}\langle P',s'\rangle$, let $E'$ be $P'$, as desired.
\end{enumerate}
\end{proof}

\begin{theorem}[Unique solution of equations for strongly probabilistic pomset bisimulation]
Let the recursive expressions $E_i(i\in I)$ contain at most the variables $X_i(i\in I)$, and let each $X_j(j\in I)$ be weakly guarded in each $E_i$. Then,

If $\widetilde{P}\sim_{pp} \widetilde{E}\{\widetilde{P}/\widetilde{X}\}$ and $\widetilde{Q}\sim_{pp} \widetilde{E}\{\widetilde{Q}/\widetilde{X}\}$, then $\widetilde{P}\sim_{pp} \widetilde{Q}$.
\end{theorem}

\begin{proof}
It is sufficient to induct on the depth of the inference of $\langle E\{\widetilde{P}/\widetilde{X}\},s\rangle\rightsquigarrow\xrightarrow{\{\alpha_1,\cdots,\alpha_n\}}\langle P',s'\rangle$.

\begin{enumerate}
  \item Case $E\equiv X_i$. Then we have $\langle E\{\widetilde{P}/\widetilde{X}\},s\rangle\equiv \langle P_i,s\rangle\rightsquigarrow\xrightarrow{\{\alpha_1,\cdots,\alpha_n\}}\langle P',s'\rangle$,
  since $P_i\sim_{pp} E_i\{\widetilde{P}/\widetilde{X}\}$, we have $\langle E_i\{\widetilde{P}/\widetilde{X}\},s\rangle\rightsquigarrow\xrightarrow{\{\alpha_1,\cdots,\alpha_n\}}\langle P'',s'\rangle\sim_{pp} \langle P',s'\rangle$.
  Since $\widetilde{X}$ are weakly guarded in $E_i$, by Lemma \ref{LUS5}, $P''\equiv E'\{\widetilde{P}/\widetilde{X}\}$ and $\langle E_i\{\widetilde{P}/\widetilde{X}\},s\rangle
  \rightsquigarrow\xrightarrow{\{\alpha_1,\cdots,\alpha_n\}} \langle E'\{\widetilde{P}/\widetilde{X}\},s'\rangle$. Since
  $E\{\widetilde{Q}/\widetilde{X}\}\equiv X_i\{\widetilde{Q}/\widetilde{X}\} \equiv Q_i\sim_{pp} E_i\{\widetilde{Q}/\widetilde{X}\}$, $\langle E\{\widetilde{Q}/\widetilde{X}\},s\rangle\rightsquigarrow\xrightarrow{\{\alpha_1,\cdots,\alpha_n\}}\langle Q',s'\rangle\sim_{pp} \langle E'\{\widetilde{Q}/\widetilde{X}\},s'\rangle$.
  So, $P'\sim_{pp} Q'$, as desired.
  \item Case $E\equiv\alpha.F$. This case can be proven similarly.
  \item Case $E\equiv(\alpha_1\parallel\cdots\parallel\alpha_n).F$. This case can be proven similarly.
  \item Case $E\equiv E_1+E_2$. We have $\langle E_i\{\widetilde{P}/\widetilde{X}\},s\rangle \rightsquigarrow\xrightarrow{\{\alpha_1,\cdots,\alpha_n\}}\langle P',s'\rangle$,
  $\langle E_i\{\widetilde{Q}/\widetilde{X}\},s\rangle \rightsquigarrow\xrightarrow{\{\alpha_1,\cdots,\alpha_n\}}\langle Q',s'\rangle$, then, $P'\sim_{pp} Q'$, as desired.
  \item Case $E\equiv E_1\parallel E_2$, $E\equiv F[R]$ and $E\equiv F\setminus L$, $E\equiv C$. These cases can be prove similarly to the above case.
\end{enumerate}
\end{proof}

\begin{theorem}[Unique solution of equations for strongly probabilistic step bisimulation]
Let the recursive expressions $E_i(i\in I)$ contain at most the variables $X_i(i\in I)$, and let each $X_j(j\in I)$ be weakly guarded in each $E_i$. Then,

If $\widetilde{P}\sim_{ps} \widetilde{E}\{\widetilde{P}/\widetilde{X}\}$ and $\widetilde{Q}\sim_{ps} \widetilde{E}\{\widetilde{Q}/\widetilde{X}\}$, then $\widetilde{P}\sim_{ps} \widetilde{Q}$.
\end{theorem}

\begin{proof}
It is sufficient to induct on the depth of the inference of $\langle E\{\widetilde{P}/\widetilde{X}\},s\rangle\rightsquigarrow\xrightarrow{\{\alpha_1,\cdots,\alpha_n\}}\langle P',s'\rangle$.

\begin{enumerate}
  \item Case $E\equiv X_i$. Then we have $\langle E\{\widetilde{P}/\widetilde{X}\},s\rangle\equiv \langle P_i,s\rangle\rightsquigarrow\xrightarrow{\{\alpha_1,\cdots,\alpha_n\}}\langle P',s'\rangle$,
  since $P_i\sim_{ps} E_i\{\widetilde{P}/\widetilde{X}\}$, we have $\langle E_i\{\widetilde{P}/\widetilde{X}\},s\rangle\rightsquigarrow\xrightarrow{\{\alpha_1,\cdots,\alpha_n\}}\langle P'',s'\rangle\sim_{ps} \langle P',s'\rangle$.
  Since $\widetilde{X}$ are weakly guarded in $E_i$, by Lemma \ref{LUS5}, $P''\equiv E'\{\widetilde{P}/\widetilde{X}\}$ and $\langle E_i\{\widetilde{P}/\widetilde{X}\},s\rangle
  \rightsquigarrow\xrightarrow{\{\alpha_1,\cdots,\alpha_n\}} \langle E'\{\widetilde{P}/\widetilde{X}\},s'\rangle$. Since
  $E\{\widetilde{Q}/\widetilde{X}\}\equiv X_i\{\widetilde{Q}/\widetilde{X}\} \equiv Q_i\sim_{ps} E_i\{\widetilde{Q}/\widetilde{X}\}$, $\langle E\{\widetilde{Q}/\widetilde{X}\},s\rangle\rightsquigarrow\xrightarrow{\{\alpha_1,\cdots,\alpha_n\}}\langle Q',s'\rangle\sim_{ps} \langle E'\{\widetilde{Q}/\widetilde{X}\},s'\rangle$.
  So, $P'\sim_{ps} Q'$, as desired.
  \item Case $E\equiv\alpha.F$. This case can be proven similarly.
  \item Case $E\equiv(\alpha_1\parallel\cdots\parallel\alpha_n).F$. This case can be proven similarly.
  \item Case $E\equiv E_1+E_2$. We have $\langle E_i\{\widetilde{P}/\widetilde{X}\},s\rangle \rightsquigarrow\xrightarrow{\{\alpha_1,\cdots,\alpha_n\}}\langle P',s'\rangle$,
  $\langle E_i\{\widetilde{Q}/\widetilde{X}\},s\rangle \rightsquigarrow\xrightarrow{\{\alpha_1,\cdots,\alpha_n\}}\langle Q',s'\rangle$, then, $P'\sim_{ps} Q'$, as desired.
  \item Case $E\equiv E_1\parallel E_2$, $E\equiv F[R]$ and $E\equiv F\setminus L$, $E\equiv C$. These cases can be prove similarly to the above case.
\end{enumerate}
\end{proof}

\begin{theorem}[Unique solution of equations for strongly probabilistic hp-bisimulation]
Let the recursive expressions $E_i(i\in I)$ contain at most the variables $X_i(i\in I)$, and let each $X_j(j\in I)$ be weakly guarded in each $E_i$. Then,

If $\widetilde{P}\sim_{php} \widetilde{E}\{\widetilde{P}/\widetilde{X}\}$ and $\widetilde{Q}\sim_{php} \widetilde{E}\{\widetilde{Q}/\widetilde{X}\}$, then $\widetilde{P}\sim_{php} \widetilde{Q}$.
\end{theorem}

\begin{proof}
It is sufficient to induct on the depth of the inference of $\langle E\{\widetilde{P}/\widetilde{X}\},s\rangle\rightsquigarrow\xrightarrow{\{\alpha_1,\cdots,\alpha_n\}}\langle P',s'\rangle$.

\begin{enumerate}
  \item Case $E\equiv X_i$. Then we have $\langle E\{\widetilde{P}/\widetilde{X}\},s\rangle\equiv \langle P_i,s\rangle\rightsquigarrow\xrightarrow{\{\alpha_1,\cdots,\alpha_n\}}\langle P',s'\rangle$,
  since $P_i\sim_{php} E_i\{\widetilde{P}/\widetilde{X}\}$, we have $\langle E_i\{\widetilde{P}/\widetilde{X}\},s\rangle\rightsquigarrow\xrightarrow{\{\alpha_1,\cdots,\alpha_n\}}\langle P'',s'\rangle\sim_{php} \langle P',s'\rangle$.
  Since $\widetilde{X}$ are weakly guarded in $E_i$, by Lemma \ref{LUS5}, $P''\equiv E'\{\widetilde{P}/\widetilde{X}\}$ and $\langle E_i\{\widetilde{P}/\widetilde{X}\},s\rangle
  \rightsquigarrow\xrightarrow{\{\alpha_1,\cdots,\alpha_n\}} \langle E'\{\widetilde{P}/\widetilde{X}\},s'\rangle$. Since
  $E\{\widetilde{Q}/\widetilde{X}\}\equiv X_i\{\widetilde{Q}/\widetilde{X}\} \equiv Q_i\sim_{php} E_i\{\widetilde{Q}/\widetilde{X}\}$, $\langle E\{\widetilde{Q}/\widetilde{X}\},s\rangle\rightsquigarrow\xrightarrow{\{\alpha_1,\cdots,\alpha_n\}}\langle Q',s'\rangle\sim_{php} \langle E'\{\widetilde{Q}/\widetilde{X}\},s'\rangle$.
  So, $P'\sim_{php} Q'$, as desired.
  \item Case $E\equiv\alpha.F$. This case can be proven similarly.
  \item Case $E\equiv(\alpha_1\parallel\cdots\parallel\alpha_n).F$. This case can be proven similarly.
  \item Case $E\equiv E_1+E_2$. We have $\langle E_i\{\widetilde{P}/\widetilde{X}\},s\rangle \rightsquigarrow\xrightarrow{\{\alpha_1,\cdots,\alpha_n\}}\langle P',s'\rangle$,
  $\langle E_i\{\widetilde{Q}/\widetilde{X}\},s\rangle \rightsquigarrow\xrightarrow{\{\alpha_1,\cdots,\alpha_n\}}\langle Q',s'\rangle$, then, $P'\sim_{php} Q'$, as desired.
  \item Case $E\equiv E_1\parallel E_2$, $E\equiv F[R]$ and $E\equiv F\setminus L$, $E\equiv C$. These cases can be prove similarly to the above case.
\end{enumerate}
\end{proof}

\begin{theorem}[Unique solution of equations for strongly probabilistic hhp-bisimulation]
Let the recursive expressions $E_i(i\in I)$ contain at most the variables $X_i(i\in I)$, and let each $X_j(j\in I)$ be weakly guarded in each $E_i$. Then,

If $\widetilde{P}\sim_{phhp} \widetilde{E}\{\widetilde{P}/\widetilde{X}\}$ and $\widetilde{Q}\sim_{phhp} \widetilde{E}\{\widetilde{Q}/\widetilde{X}\}$, then $\widetilde{P}\sim_{phhp} \widetilde{Q}$.
\end{theorem}

\begin{proof}
It is sufficient to induct on the depth of the inference of $\langle E\{\widetilde{P}/\widetilde{X}\},s\rangle\rightsquigarrow\xrightarrow{\{\alpha_1,\cdots,\alpha_n\}}\langle P',s'\rangle$.

\begin{enumerate}
  \item Case $E\equiv X_i$. Then we have $\langle E\{\widetilde{P}/\widetilde{X}\},s\rangle\equiv \langle P_i,s\rangle\rightsquigarrow\xrightarrow{\{\alpha_1,\cdots,\alpha_n\}}\langle P',s'\rangle$,
  since $P_i\sim_{phhp} E_i\{\widetilde{P}/\widetilde{X}\}$, we have $\langle E_i\{\widetilde{P}/\widetilde{X}\},s\rangle\rightsquigarrow\xrightarrow{\{\alpha_1,\cdots,\alpha_n\}}\langle P'',s'\rangle\sim_{phhp} \langle P',s'\rangle$.
  Since $\widetilde{X}$ are weakly guarded in $E_i$, by Lemma \ref{LUS5}, $P''\equiv E'\{\widetilde{P}/\widetilde{X}\}$ and $\langle E_i\{\widetilde{P}/\widetilde{X}\},s\rangle
  \rightsquigarrow\xrightarrow{\{\alpha_1,\cdots,\alpha_n\}} \langle E'\{\widetilde{P}/\widetilde{X}\},s'\rangle$. Since
  $E\{\widetilde{Q}/\widetilde{X}\}\equiv X_i\{\widetilde{Q}/\widetilde{X}\} \equiv Q_i\sim_{phhp} E_i\{\widetilde{Q}/\widetilde{X}\}$, $\langle E\{\widetilde{Q}/\widetilde{X}\},s\rangle\rightsquigarrow\xrightarrow{\{\alpha_1,\cdots,\alpha_n\}}\langle Q',s'\rangle\sim_{phhp} \langle E'\{\widetilde{Q}/\widetilde{X}\},s'\rangle$.
  So, $P'\sim_{phhp} Q'$, as desired.
  \item Case $E\equiv\alpha.F$. This case can be proven similarly.
  \item Case $E\equiv(\alpha_1\parallel\cdots\parallel\alpha_n).F$. This case can be proven similarly.
  \item Case $E\equiv E_1+E_2$. We have $\langle E_i\{\widetilde{P}/\widetilde{X}\},s\rangle \rightsquigarrow\xrightarrow{\{\alpha_1,\cdots,\alpha_n\}}\langle P',s'\rangle$,
  $\langle E_i\{\widetilde{Q}/\widetilde{X}\},s\rangle \rightsquigarrow\xrightarrow{\{\alpha_1,\cdots,\alpha_n\}}\langle Q',s'\rangle$, then, $P'\sim_{phhp} Q'$, as desired.
  \item Case $E\equiv E_1\parallel E_2$, $E\equiv F[R]$ and $E\equiv F\setminus L$, $E\equiv C$. These cases can be prove similarly to the above case.
\end{enumerate}
\end{proof}

\subsection{Weak Bisimulations}\label{wtcbctcpg}

The weak probabilistic transition rules of CTC with probabilism and guards are the same as the strong one in Table \ref{PTRForCTC5}. And the weak action transition rules of CTC with probabilism and guards are listed in Table
\ref{WTRForCTC5}.

\begin{center}
    \begin{table}
        \[\textbf{Act}_1\quad \frac{}{\langle\breve{\alpha}.P,s\rangle\xRightarrow{\alpha}\langle P,s'\rangle}\]

        \[\textbf{Act}_2\quad \frac{}{\langle\epsilon,s\rangle\Rightarrow\langle\surd,s\rangle}\]

        \[\textbf{Gur}\quad \frac{}{\langle\phi,s\rangle\Rightarrow\langle\surd,s\rangle}\textrm{ if }test(\phi,s)\]

        \[\textbf{Sum}_1\quad \frac{\langle P,s\rangle\xRightarrow{\alpha}\langle P',s'\rangle}{\langle P+Q,s\rangle\xRightarrow{\alpha}\langle P',s'\rangle}\]

        \[\textbf{Com}_1\quad \frac{\langle P,s\rangle\xRightarrow{\alpha}\langle P',s'\rangle\quad Q\nrightarrow}{\langle P\parallel Q,s\rangle\xRightarrow{\alpha}\langle P'\parallel Q,s'\rangle}\]

        \[\textbf{Com}_2\quad \frac{\langle Q,s\rangle\xRightarrow{\alpha}\langle Q',s'\rangle\quad P\nrightarrow}{\langle P\parallel Q,s\rangle\xRightarrow{\alpha}\langle P\parallel Q',s'\rangle}\]

        \[\textbf{Com}_3\quad \frac{\langle P,s\rangle\xRightarrow{\alpha}\langle P',s'\rangle\quad \langle Q,s\rangle\xRightarrow{\beta}\langle Q',s''\rangle}{\langle P\parallel Q,s\rangle\xRightarrow{\{\alpha,\beta\}}\langle P'\parallel Q',s'\cup s''\rangle}\quad (\beta\neq\overline{\alpha})\]

        \[\textbf{Com}_4\quad \frac{\langle P,s\rangle\xRightarrow{l}\langle P',s'\rangle\quad \langle Q,s\rangle\xRightarrow{\overline{l}}\langle Q',s''\rangle}{\langle P\parallel Q,s\rangle\xRightarrow{\tau}\langle P'\parallel Q',s'\cup s''\rangle}\]

        \[\textbf{Act}_3\quad \frac{}{\langle(\breve{\alpha_1}\parallel\cdots\parallel\breve{\alpha_n}).P,s\rangle\xRightarrow{\{\alpha_1,\cdots,\alpha_n\}}\langle P,s'\rangle}\quad (\alpha_i\neq\overline{\alpha_j}\quad i,j\in\{1,\cdots,n\})\]

        \[\textbf{Sum}_2\quad \frac{\langle P,s\rangle\xRightarrow{\{\alpha_1,\cdots,\alpha_n\}}\langle P',s'\rangle}{\langle P+Q,s\rangle\xRightarrow{\{\alpha_1,\cdots,\alpha_n\}}\langle P',s'\rangle}\]

        \[\textbf{Res}_1\quad \frac{\langle P,s\rangle\xRightarrow{\alpha}\langle P',s'\rangle}{\langle P\setminus L,s\rangle\xRightarrow{\alpha}\langle P'\setminus L,s'\rangle}\quad (\alpha,\overline{\alpha}\notin L)\]

        \[\textbf{Res}_2\quad \frac{\langle P,s\rangle\xRightarrow{\{\alpha_1,\cdots,\alpha_n\}}\langle P',s'\rangle}{\langle P\setminus L,s\rangle\xRightarrow{\{\alpha_1,\cdots,\alpha_n\}}\langle P'\setminus L,s'\rangle}\quad (\alpha_1,\overline{\alpha_1},\cdots,\alpha_n,\overline{\alpha_n}\notin L)\]

        \[\textbf{Rel}_1\quad \frac{\langle P,s\rangle\xRightarrow{\alpha}\langle P',s'\rangle}{\langle P[f],s\rangle\xRightarrow{\langle f(\alpha)}P'[f],s'\rangle}\]

        \[\textbf{Rel}_2\quad \frac{\langle P,s\rangle\xRightarrow{\{\alpha_1,\cdots,\alpha_n\}}\langle P',s'\rangle}{\langle P[f],s\rangle\xRightarrow{\{f(\alpha_1),\cdots,f(\alpha_n)\}}\langle P'[f],s'\rangle}\]

        \[\textbf{Con}_1\quad\frac{\langle P,s\rangle\xRightarrow{\alpha}\langle P',s'\rangle}{\langle A,s\rangle\xRightarrow{\alpha}\langle P',s'\rangle}\quad (A\overset{\text{def}}{=}P)\]

        \[\textbf{Con}_2\quad\frac{\langle P,s\rangle\xRightarrow{\{\alpha_1,\cdots,\alpha_n\}}\langle P',s'\rangle}{\langle A,s\rangle\xRightarrow{\{\alpha_1,\cdots,\alpha_n\}}\langle P',s'\rangle}\quad (A\overset{\text{def}}{=}P)\]

        \caption{Weak action transition rules of CTC with probabilism and guards}
        \label{WTRForCTC5}
    \end{table}
\end{center}

\subsubsection{Laws and Congruence}

Remembering that $\tau$ can neither be restricted nor relabeled, we know that the monoid laws, the monoid laws 2, the static laws and the expansion law in
section \ref{stcbctcpg} still hold with respect to the corresponding weakly probabilistic truly concurrent bisimulations. And also, we can enjoy the full congruence of Prefix, Summation, Composition,
Restriction, Relabelling and Constants with respect to corresponding weakly probabilistic truly concurrent bisimulations. We will not retype these laws, and just give the
$\tau$-specific laws.

\begin{proposition}[$\tau$ laws for weakly probabilistic pomset bisimulation]\label{TAUWPB3}
The $\tau$ laws for weakly probabilistic pomset bisimulation is as follows.

\begin{enumerate}
  \item $P\approx_{pp} \tau.P$;
  \item $\alpha.\tau.P\approx_{pp} \alpha.P$;
  \item $(\alpha_1\parallel\cdots\parallel\alpha_n).\tau.P\approx_{pp} (\alpha_1\parallel\cdots\parallel\alpha_n).P$;
  \item $P+\tau.P\approx_{pp} \tau.P$;
  \item $P\cdot((Q+\tau\cdot(Q+R))\boxplus_{\pi}S)\approx_{pp}P\cdot((Q+R)\boxplus_{\pi}S)$;
  \item $P\approx_{pp} \tau\parallel P$.
\end{enumerate}
\end{proposition}

\begin{proof}
\begin{enumerate}
  \item $P\approx_{pp} \tau.P$. It is sufficient to prove the relation $R=\{(P, \tau.P)\}\cup \textbf{Id}$ is a weakly probabilistic pomset bisimulation, we omit it;
  \item $\alpha.\tau.P\approx_{pp} \alpha.P$. It is sufficient to prove the relation $R=\{(\alpha.\tau.P, \alpha.P)\}\cup \textbf{Id}$ is a weakly probabilistic pomset bisimulation, we omit it;
  \item $(\alpha_1\parallel\cdots\parallel\alpha_n).\tau.P\approx_{pp} (\alpha_1\parallel\cdots\parallel\alpha_n).P$. It is sufficient to prove the relation $R=\{((\alpha_1\parallel\cdots\parallel\alpha_n).\tau.P, (\alpha_1\parallel\cdots\parallel\alpha_n).P)\}\cup \textbf{Id}$ is a weakly probabilistic pomset bisimulation, we omit it;
  \item $P+\tau.P\approx_{pp} \tau.P$. It is sufficient to prove the relation $R=\{(P+\tau.P, \tau.P)\}\cup \textbf{Id}$ is a weakly probabilistic pomset bisimulation, we omit it;
  \item $P\cdot((Q+\tau\cdot(Q+R))\boxplus_{\pi}S)\approx_{pp}P\cdot((Q+R)\boxplus_{\pi}S)$. It is sufficient to prove the relation $R=\{(P\cdot((Q+\tau\cdot(Q+R))\boxplus_{\pi}S), P\cdot((Q+R)\boxplus_{\pi}S))\}\cup \textbf{Id}$ is a weakly probabilistic pomset bisimulation, we omit it;
  \item $P\approx_{pp} \tau\parallel P$. It is sufficient to prove the relation $R=\{(P, \tau\parallel P)\}\cup \textbf{Id}$ is a weakly probabilistic pomset bisimulation, we omit it.
\end{enumerate}
\end{proof}

\begin{proposition}[$\tau$ laws for weakly probabilistic step bisimulation]\label{TAUWPB3}
The $\tau$ laws for weakly probabilistic step bisimulation is as follows.

\begin{enumerate}
  \item $P\approx_{ps} \tau.P$;
  \item $\alpha.\tau.P\approx_{ps} \alpha.P$;
  \item $(\alpha_1\parallel\cdots\parallel\alpha_n).\tau.P\approx_{ps} (\alpha_1\parallel\cdots\parallel\alpha_n).P$;
  \item $P+\tau.P\approx_{ps} \tau.P$;
  \item $P\cdot((Q+\tau\cdot(Q+R))\boxplus_{\pi}S)\approx_{ps}P\cdot((Q+R)\boxplus_{\pi}S)$;
  \item $P\approx_{ps} \tau\parallel P$.
\end{enumerate}
\end{proposition}

\begin{proof}
\begin{enumerate}
  \item $P\approx_{ps} \tau.P$. It is sufficient to prove the relation $R=\{(P, \tau.P)\}\cup \textbf{Id}$ is a weakly probabilistic step bisimulation, we omit it;
  \item $\alpha.\tau.P\approx_{ps} \alpha.P$. It is sufficient to prove the relation $R=\{(\alpha.\tau.P, \alpha.P)\}\cup \textbf{Id}$ is a weakly probabilistic step bisimulation, we omit it;
  \item $(\alpha_1\parallel\cdots\parallel\alpha_n).\tau.P\approx_{ps} (\alpha_1\parallel\cdots\parallel\alpha_n).P$. It is sufficient to prove the relation $R=\{((\alpha_1\parallel\cdots\parallel\alpha_n).\tau.P, (\alpha_1\parallel\cdots\parallel\alpha_n).P)\}\cup \textbf{Id}$ is a weakly probabilistic step bisimulation, we omit it;
  \item $P+\tau.P\approx_{ps} \tau.P$. It is sufficient to prove the relation $R=\{(P+\tau.P, \tau.P)\}\cup \textbf{Id}$ is a weakly probabilistic step bisimulation, we omit it;
  \item $P\cdot((Q+\tau\cdot(Q+R))\boxplus_{\pi}S)\approx_{ps}P\cdot((Q+R)\boxplus_{\pi}S)$. It is sufficient to prove the relation $R=\{(P\cdot((Q+\tau\cdot(Q+R))\boxplus_{\pi}S), P\cdot((Q+R)\boxplus_{\pi}S))\}\cup \textbf{Id}$ is a weakly probabilistic step bisimulation, we omit it;
  \item $P\approx_{ps} \tau\parallel P$. It is sufficient to prove the relation $R=\{(P, \tau\parallel P)\}\cup \textbf{Id}$ is a weakly probabilistic step bisimulation, we omit it.
\end{enumerate}
\end{proof}

\begin{proposition}[$\tau$ laws for weakly probabilistic hp-bisimulation]\label{TAUWPB3}
The $\tau$ laws for weakly probabilistic hp-bisimulation is as follows.

\begin{enumerate}
  \item $P\approx_{php} \tau.P$;
  \item $\alpha.\tau.P\approx_{php} \alpha.P$;
  \item $(\alpha_1\parallel\cdots\parallel\alpha_n).\tau.P\approx_{php} (\alpha_1\parallel\cdots\parallel\alpha_n).P$;
  \item $P+\tau.P\approx_{php} \tau.P$;
  \item $P\cdot((Q+\tau\cdot(Q+R))\boxplus_{\pi}S)\approx_{php}P\cdot((Q+R)\boxplus_{\pi}S)$;
  \item $P\approx_{php} \tau\parallel P$.
\end{enumerate}
\end{proposition}

\begin{proof}
\begin{enumerate}
  \item $P\approx_{php} \tau.P$. It is sufficient to prove the relation $R=\{(P, \tau.P)\}\cup \textbf{Id}$ is a weakly probabilistic hp-bisimulation, we omit it;
  \item $\alpha.\tau.P\approx_{php} \alpha.P$. It is sufficient to prove the relation $R=\{(\alpha.\tau.P, \alpha.P)\}\cup \textbf{Id}$ is a weakly probabilistic hp-bisimulation, we omit it;
  \item $(\alpha_1\parallel\cdots\parallel\alpha_n).\tau.P\approx_{php} (\alpha_1\parallel\cdots\parallel\alpha_n).P$. It is sufficient to prove the relation $R=\{((\alpha_1\parallel\cdots\parallel\alpha_n).\tau.P, (\alpha_1\parallel\cdots\parallel\alpha_n).P)\}\cup \textbf{Id}$ is a weakly probabilistic hp-bisimulation, we omit it;
  \item $P+\tau.P\approx_{php} \tau.P$. It is sufficient to prove the relation $R=\{(P+\tau.P, \tau.P)\}\cup \textbf{Id}$ is a weakly probabilistic hp-bisimulation, we omit it;
  \item $P\cdot((Q+\tau\cdot(Q+R))\boxplus_{\pi}S)\approx_{php}P\cdot((Q+R)\boxplus_{\pi}S)$. It is sufficient to prove the relation $R=\{(P\cdot((Q+\tau\cdot(Q+R))\boxplus_{\pi}S), P\cdot((Q+R)\boxplus_{\pi}S))\}\cup \textbf{Id}$ is a weakly probabilistic hp-bisimulation, we omit it;
  \item $P\approx_{php} \tau\parallel P$. It is sufficient to prove the relation $R=\{(P, \tau\parallel P)\}\cup \textbf{Id}$ is a weakly probabilistic hp-bisimulation, we omit it.
\end{enumerate}
\end{proof}

\begin{proposition}[$\tau$ laws for weakly probabilistic hhp-bisimulation]\label{TAUWPB3}
The $\tau$ laws for weakly probabilistic hhp-bisimulation is as follows.

\begin{enumerate}
  \item $P\approx_{phhp} \tau.P$;
  \item $\alpha.\tau.P\approx_{phhp} \alpha.P$;
  \item $(\alpha_1\parallel\cdots\parallel\alpha_n).\tau.P\approx_{phhp} (\alpha_1\parallel\cdots\parallel\alpha_n).P$;
  \item $P+\tau.P\approx_{phhp} \tau.P$;
  \item $P\cdot((Q+\tau\cdot(Q+R))\boxplus_{\pi}S)\approx_{phhp}P\cdot((Q+R)\boxplus_{\pi}S)$;
  \item $P\approx_{phhp} \tau\parallel P$.
\end{enumerate}
\end{proposition}

\begin{proof}
\begin{enumerate}
  \item $P\approx_{phhp} \tau.P$. It is sufficient to prove the relation $R=\{(P, \tau.P)\}\cup \textbf{Id}$ is a weakly probabilistic hhp-bisimulation, we omit it;
  \item $\alpha.\tau.P\approx_{phhp} \alpha.P$. It is sufficient to prove the relation $R=\{(\alpha.\tau.P, \alpha.P)\}\cup \textbf{Id}$ is a weakly probabilistic hhp-bisimulation, we omit it;
  \item $(\alpha_1\parallel\cdots\parallel\alpha_n).\tau.P\approx_{phhp} (\alpha_1\parallel\cdots\parallel\alpha_n).P$. It is sufficient to prove the relation $R=\{((\alpha_1\parallel\cdots\parallel\alpha_n).\tau.P, (\alpha_1\parallel\cdots\parallel\alpha_n).P)\}\cup \textbf{Id}$ is a weakly probabilistic hhp-bisimulation, we omit it;
  \item $P+\tau.P\approx_{phhp} \tau.P$. It is sufficient to prove the relation $R=\{(P+\tau.P, \tau.P)\}\cup \textbf{Id}$ is a weakly probabilistic hhp-bisimulation, we omit it;
  \item $P\cdot((Q+\tau\cdot(Q+R))\boxplus_{\pi}S)\approx_{phhp}P\cdot((Q+R)\boxplus_{\pi}S)$. It is sufficient to prove the relation $R=\{(P\cdot((Q+\tau\cdot(Q+R))\boxplus_{\pi}S), P\cdot((Q+R)\boxplus_{\pi}S))\}\cup \textbf{Id}$ is a weakly probabilistic hhp-bisimulation, we omit it;
  \item $P\approx_{phhp} \tau\parallel P$. It is sufficient to prove the relation $R=\{(P, \tau\parallel P)\}\cup \textbf{Id}$ is a weakly probabilistic hhp-bisimulation, we omit it.
\end{enumerate}
\end{proof}

\subsubsection{Recursion}

\begin{definition}[Sequential]
$X$ is sequential in $E$ if every subexpression of $E$ which contains $X$, apart from $X$ itself, is of the form $\alpha.F$, or $(\alpha_1\parallel\cdots\parallel\alpha_n).F$, or
$\sum\widetilde{F}$.
\end{definition}

\begin{definition}[Guarded recursive expression]
$X$ is guarded in $E$ if each occurrence of $X$ is with some subexpression $l.F$ or $(l_1\parallel\cdots\parallel l_n).F$ of $E$.
\end{definition}

\begin{lemma}\label{LUSWW5}
Let $G$ be guarded and sequential, $Vars(G)\subseteq\widetilde{X}$, and let $\langle G\{\widetilde{P}/\widetilde{X}\},s\rangle\rightsquigarrow\xrightarrow{\{\alpha_1,\cdots,\alpha_n\}}\langle P',s'\rangle$.
Then there is an expression $H$ such that $\langle G,s\rangle\rightsquigarrow\xrightarrow{\{\alpha_1,\cdots,\alpha_n\}}\langle H,s'\rangle$, $P'\equiv H\{\widetilde{P}/\widetilde{X}\}$, and for any
$\widetilde{Q}$, $\langle G\{\widetilde{Q}/\widetilde{X}\},s\rangle\rightsquigarrow\xrightarrow{\{\alpha_1,\cdots,\alpha_n\}} \langle H\{\widetilde{Q}/\widetilde{X}\},s'\rangle$. Moreover $H$ is
sequential, $Vars(H)\subseteq\widetilde{X}$, and if $\alpha_1=\cdots=\alpha_n=\tau$, then $H$ is also guarded.
\end{lemma}

\begin{proof}
We need to induct on the structure of $G$.

If $G$ is a Constant, a Composition, a Restriction or a Relabeling then it contains no variables, since $G$ is sequential and guarded, then
$\langle G,s\rangle\rightsquigarrow\xrightarrow{\{\alpha_1,\cdots,\alpha_n\}}\langle P',s'\rangle$, then let $H\equiv P'$, as desired.

$G$ cannot be a variable, since it is guarded.

If $G\equiv G_1+G_2$. Then either $\langle G_1\{\widetilde{P}/\widetilde{X}\},s\rangle \rightsquigarrow\xrightarrow{\{\alpha_1,\cdots,\alpha_n\}}\langle P',s'\rangle$ or
$\langle G_2\{\widetilde{P}/\widetilde{X}\},s\rangle \rightsquigarrow\xrightarrow{\{\alpha_1,\cdots,\alpha_n\}}\langle P',s'\rangle$, then, we can apply this lemma in either case, as desired.

If $G\equiv\beta.H$. Then we must have $\alpha=\beta$, and $P'\equiv H\{\widetilde{P}/\widetilde{X}\}$, and
$\langle G\{\widetilde{Q}/\widetilde{X}\},s\rangle\equiv \langle\beta.H\{\widetilde{Q}/\widetilde{X}\},s\rangle \rightsquigarrow\xrightarrow{\beta}\langle H\{\widetilde{Q}/\widetilde{X}\},s'\rangle$,
then, let $G'$ be $H$, as desired.

If $G\equiv(\beta_1\parallel\cdots\parallel\beta_n).H$. Then we must have $\alpha_i=\beta_i$ for $1\leq i\leq n$, and $P'\equiv H\{\widetilde{P}/\widetilde{X}\}$, and
$\langle G\{\widetilde{Q}/\widetilde{X}\},s\rangle\equiv \langle(\beta_1\parallel\cdots\parallel\beta_n).H\{\widetilde{Q}/\widetilde{X}\},s\rangle \rightsquigarrow\xrightarrow{\{\beta_1,\cdots,\beta_n\}}\langle H\{\widetilde{Q}/\widetilde{X}\},s'\rangle$,
then, let $G'$ be $H$, as desired.

If $G\equiv\tau.H$. Then we must have $\tau=\tau$, and $P'\equiv H\{\widetilde{P}/\widetilde{X}\}$, and
$\langle G\{\widetilde{Q}/\widetilde{X}\},s\rangle\equiv \langle\tau.H\{\widetilde{Q}/\widetilde{X}\},s\rangle \rightsquigarrow\xrightarrow{\tau}\langle H\{\widetilde{Q}/\widetilde{X}\},s'\rangle$,
then, let $G'$ be $H$, as desired.
\end{proof}

\begin{theorem}[Unique solution of equations for weakly probabilistic pomset bisimulation]
Let the guarded and sequential expressions $\widetilde{E}$ contain free variables $\subseteq \widetilde{X}$, then,

If $\widetilde{P}\approx_{pp} \widetilde{E}\{\widetilde{P}/\widetilde{X}\}$ and $\widetilde{Q}\approx_{pp} \widetilde{E}\{\widetilde{Q}/\widetilde{X}\}$, then
$\widetilde{P}\approx_{pp} \widetilde{Q}$.
\end{theorem}

\begin{proof}
Like the corresponding theorem in CCS, without loss of generality, we only consider a single equation $X=E$. So we assume $P\approx_{pp} E(P)$, $Q\approx_{pp} E(Q)$, then $P\approx_{pp} Q$.

We will prove $\{(H(P),H(Q)): H\}$ sequential, if $\langle H(P),s\rangle\rightsquigarrow\xrightarrow{\{\alpha_1,\cdots,\alpha_n\}}\langle P',s'\rangle$, then, for some $Q'$,
$\langle H(Q),s\rangle\rightsquigarrow\xRightarrow{\{\alpha_1.\cdots,\alpha_n\}}\langle Q',s'\rangle$ and $P'\approx_{pp} Q'$.

Let $\langle H(P),s\rangle\rightsquigarrow\xrightarrow{\{\alpha_1,\cdot,\alpha_n\}}\langle P',s'\rangle$, then $\langle H(E(P)),s\rangle\rightsquigarrow\xRightarrow{\{\alpha_1,\cdots,\alpha_n\}}\langle P'',s''\rangle$
and $P'\approx_{pp} P''$.

By Lemma \ref{LUSWW5}, we know there is a sequential $H'$ such that $\langle H(E(P)),s\rangle\rightsquigarrow\xRightarrow{\{\alpha_1,\cdots,\alpha_n\}}\langle H'(P),s'\rangle\Rightarrow P''\approx_{pp} P'$.

And, $\langle H(E(Q)),s\rangle\rightsquigarrow\xRightarrow{\{\alpha_1,\cdots,\alpha_n\}}\langle H'(Q),s'\rangle\Rightarrow Q''$ and $P''\approx_{pp} Q''$. And $\langle H(Q),s\rangle\rightsquigarrow\xrightarrow{\{\alpha_1,\cdots,\alpha_n\}}\langle Q',s'\rangle\approx_{pp} \Rightarrow Q'\approx_{pp} Q''$.
Hence, $P'\approx_{pp} Q'$, as desired.
\end{proof}

\begin{theorem}[Unique solution of equations for weakly probabilistic step bisimulation]
Let the guarded and sequential expressions $\widetilde{E}$ contain free variables $\subseteq \widetilde{X}$, then,

If $\widetilde{P}\approx_{ps} \widetilde{E}\{\widetilde{P}/\widetilde{X}\}$ and $\widetilde{Q}\approx_{ps} \widetilde{E}\{\widetilde{Q}/\widetilde{X}\}$, then
$\widetilde{P}\approx_{ps} \widetilde{Q}$.
\end{theorem}

\begin{proof}
Like the corresponding theorem in CCS, without loss of generality, we only consider a single equation $X=E$. So we assume $P\approx_{ps} E(P)$, $Q\approx_{ps} E(Q)$, then $P\approx_{ps} Q$.

We will prove $\{(H(P),H(Q)): H\}$ sequential, if $\langle H(P),s\rangle\rightsquigarrow\xrightarrow{\{\alpha_1,\cdots,\alpha_n\}}\langle P',s'\rangle$, then, for some $Q'$,
$\langle H(Q),s\rangle\rightsquigarrow\xRightarrow{\{\alpha_1.\cdots,\alpha_n\}}\langle Q',s'\rangle$ and $P'\approx_{ps} Q'$.

Let $\langle H(P),s\rangle\rightsquigarrow\xrightarrow{\{\alpha_1,\cdot,\alpha_n\}}\langle P',s'\rangle$, then $\langle H(E(P)),s\rangle\rightsquigarrow\xRightarrow{\{\alpha_1,\cdots,\alpha_n\}}\langle P'',s''\rangle$
and $P'\approx_{ps} P''$.

By Lemma \ref{LUSWW5}, we know there is a sequential $H'$ such that $\langle H(E(P)),s\rangle\rightsquigarrow\xRightarrow{\{\alpha_1,\cdots,\alpha_n\}}\langle H'(P),s'\rangle\Rightarrow P''\approx_{ps} P'$.

And, $\langle H(E(Q)),s\rangle\rightsquigarrow\xRightarrow{\{\alpha_1,\cdots,\alpha_n\}}\langle H'(Q),s'\rangle\Rightarrow Q''$ and $P''\approx_{ps} Q''$. And $\langle H(Q),s\rangle\rightsquigarrow\xrightarrow{\{\alpha_1,\cdots,\alpha_n\}}\langle Q',s'\rangle\approx_{ps} \Rightarrow Q'\approx_{ps} Q''$.
Hence, $P'\approx_{ps} Q'$, as desired.
\end{proof}

\begin{theorem}[Unique solution of equations for weakly probabilistic hp-bisimulation]
Let the guarded and sequential expressions $\widetilde{E}$ contain free variables $\subseteq \widetilde{X}$, then,

If $\widetilde{P}\approx_{php} \widetilde{E}\{\widetilde{P}/\widetilde{X}\}$ and $\widetilde{Q}\approx_{php} \widetilde{E}\{\widetilde{Q}/\widetilde{X}\}$, then
$\widetilde{P}\approx_{php} \widetilde{Q}$.
\end{theorem}

\begin{proof}
Like the corresponding theorem in CCS, without loss of generality, we only consider a single equation $X=E$. So we assume $P\approx_{php} E(P)$, $Q\approx_{php} E(Q)$, then $P\approx_{php} Q$.

We will prove $\{(H(P),H(Q)): H\}$ sequential, if $\langle H(P),s\rangle\rightsquigarrow\xrightarrow{\{\alpha_1,\cdots,\alpha_n\}}\langle P',s'\rangle$, then, for some $Q'$,
$\langle H(Q),s\rangle\rightsquigarrow\xRightarrow{\{\alpha_1.\cdots,\alpha_n\}}\langle Q',s'\rangle$ and $P'\approx_{php} Q'$.

Let $\langle H(P),s\rangle\rightsquigarrow\xrightarrow{\{\alpha_1,\cdot,\alpha_n\}}\langle P',s'\rangle$, then $\langle H(E(P)),s\rangle\rightsquigarrow\xRightarrow{\{\alpha_1,\cdots,\alpha_n\}}\langle P'',s''\rangle$
and $P'\approx_{php} P''$.

By Lemma \ref{LUSWW5}, we know there is a sequential $H'$ such that $\langle H(E(P)),s\rangle\rightsquigarrow\xRightarrow{\{\alpha_1,\cdots,\alpha_n\}}\langle H'(P),s'\rangle\Rightarrow P''\approx_{php} P'$.

And, $\langle H(E(Q)),s\rangle\rightsquigarrow\xRightarrow{\{\alpha_1,\cdots,\alpha_n\}}\langle H'(Q),s'\rangle\Rightarrow Q''$ and $P''\approx_{php} Q''$. And $\langle H(Q),s\rangle\rightsquigarrow\xrightarrow{\{\alpha_1,\cdots,\alpha_n\}}\langle Q',s'\rangle\approx_{php} \Rightarrow Q'\approx_{php} Q''$.
Hence, $P'\approx_{php} Q'$, as desired.
\end{proof}

\begin{theorem}[Unique solution of equations for weakly probabilistic hhp-bisimulation]
Let the guarded and sequential expressions $\widetilde{E}$ contain free variables $\subseteq \widetilde{X}$, then,

If $\widetilde{P}\approx_{phhp} \widetilde{E}\{\widetilde{P}/\widetilde{X}\}$ and $\widetilde{Q}\approx_{phhp} \widetilde{E}\{\widetilde{Q}/\widetilde{X}\}$, then
$\widetilde{P}\approx_{phhp} \widetilde{Q}$.
\end{theorem}

\begin{proof}
Like the corresponding theorem in CCS, without loss of generality, we only consider a single equation $X=E$. So we assume $P\approx_{phhp} E(P)$, $Q\approx_{phhp} E(Q)$, then $P\approx_{phhp} Q$.

We will prove $\{(H(P),H(Q)): H\}$ sequential, if $\langle H(P),s\rangle\rightsquigarrow\xrightarrow{\{\alpha_1,\cdots,\alpha_n\}}\langle P',s'\rangle$, then, for some $Q'$,
$\langle H(Q),s\rangle\rightsquigarrow\xRightarrow{\{\alpha_1.\cdots,\alpha_n\}}\langle Q',s'\rangle$ and $P'\approx_{phhp} Q'$.

Let $\langle H(P),s\rangle\rightsquigarrow\xrightarrow{\{\alpha_1,\cdot,\alpha_n\}}\langle P',s'\rangle$, then $\langle H(E(P)),s\rangle\rightsquigarrow\xRightarrow{\{\alpha_1,\cdots,\alpha_n\}}\langle P'',s''\rangle$
and $P'\approx_{phhp} P''$.

By Lemma \ref{LUSWW5}, we know there is a sequential $H'$ such that $\langle H(E(P)),s\rangle\rightsquigarrow\xRightarrow{\{\alpha_1,\cdots,\alpha_n\}}\langle H'(P),s'\rangle\Rightarrow P''\approx_{phhp} P'$.

And, $\langle H(E(Q)),s\rangle\rightsquigarrow\xRightarrow{\{\alpha_1,\cdots,\alpha_n\}}\langle H'(Q),s'\rangle\Rightarrow Q''$ and $P''\approx_{phhp} Q''$. And $\langle H(Q),s\rangle\rightsquigarrow\xrightarrow{\{\alpha_1,\cdots,\alpha_n\}}\langle Q',s'\rangle\approx_{phhp} \Rightarrow Q'\approx_{phhp} Q''$.
Hence, $P'\approx_{phhp} Q'$, as desired.
\end{proof}

\newpage\section{CTC with Reversibility and Guards}\label{ctcrg}

In this chapter, we design the calculus CTC with reversibility and guards. This chapter is organized as follows. We introduce the operational semantics in section \ref{osctcrg}, its syntax and operational
semantics in section \ref{sosctcrg}, and its properties for strong bisimulations in section \ref{sftcbctcrg}, its properties for weak bisimulations in section \ref{wftcbctcrg}.

\subsection{Operational Semantics}\label{osctcrg}

\begin{definition}[Prime event structure with silent event and empty event]
Let $\Lambda$ be a fixed set of labels, ranged over $a,b,c,\cdots$ and $\tau,\epsilon$. A ($\Lambda$-labelled) prime event structure with silent event $\tau$ and empty event
$\epsilon$ is a tuple $\mathcal{E}=\langle \mathbb{E}, \leq, \sharp, \lambda\rangle$, where $\mathbb{E}$ is a denumerable set of events, including the silent event $\tau$ and empty
event $\epsilon$. Let $\hat{\mathbb{E}}=\mathbb{E}\backslash\{\tau,\epsilon\}$, exactly excluding $\tau$ and $\epsilon$, it is obvious that $\hat{\tau^*}=\epsilon$. Let
$\lambda:\mathbb{E}\rightarrow\Lambda$ be a labelling function and let $\lambda(\tau)=\tau$ and $\lambda(\epsilon)=\epsilon$. And $\leq$, $\sharp$ are binary relations on $\mathbb{E}$,
called causality and conflict respectively, such that:

\begin{enumerate}
  \item $\leq$ is a partial order and $\lceil e \rceil = \{e'\in \mathbb{E}|e'\leq e\}$ is finite for all $e\in \mathbb{E}$. It is easy to see that
  $e\leq\tau^*\leq e'=e\leq\tau\leq\cdots\leq\tau\leq e'$, then $e\leq e'$.
  \item $\sharp$ is irreflexive, symmetric and hereditary with respect to $\leq$, that is, for all $e,e',e''\in \mathbb{E}$, if $e\sharp e'\leq e''$, then $e\sharp e''$.
\end{enumerate}

Then, the concepts of consistency and concurrency can be drawn from the above definition:

\begin{enumerate}
  \item $e,e'\in \mathbb{E}$ are consistent, denoted as $e\frown e'$, if $\neg(e\sharp e')$. A subset $X\subseteq \mathbb{E}$ is called consistent, if $e\frown e'$ for all
  $e,e'\in X$.
  \item $e,e'\in \mathbb{E}$ are concurrent, denoted as $e\parallel e'$, if $\neg(e\leq e')$, $\neg(e'\leq e)$, and $\neg(e\sharp e')$.
\end{enumerate}
\end{definition}

\begin{definition}[Configuration]
Let $\mathcal{E}$ be a PES. A (finite) configuration in $\mathcal{E}$ is a (finite) consistent subset of events $C\subseteq \mathcal{E}$, closed with respect to causality (i.e.
$\lceil C\rceil=C$), and a data state $s\in S$ with $S$ the set of all data states, denoted $\langle C, s\rangle$. The set of finite configurations of $\mathcal{E}$ is denoted by
$\langle\mathcal{C}(\mathcal{E}), S\rangle$. We let $\hat{C}=C\backslash\{\tau\}\cup\{\epsilon\}$.
\end{definition}

A consistent subset of $X\subseteq \mathbb{E}$ of events can be seen as a pomset. Given $X, Y\subseteq \mathbb{E}$, $\hat{X}\sim \hat{Y}$ if $\hat{X}$ and $\hat{Y}$ are isomorphic as
pomsets. In the following of the paper, we say $C_1\sim C_2$, we mean $\hat{C_1}\sim\hat{C_2}$.

\begin{definition}[FR pomset transitions and step]
Let $\mathcal{E}$ be a PES and let $C\in\mathcal{C}(\mathcal{E})$, and $\emptyset\neq X\subseteq \mathbb{E}$, if $C\cap X=\emptyset$ and $C'=C\cup X\in\mathcal{C}(\mathcal{E})$, then
$\langle C,s\rangle\xrightarrow{X} \langle C',s'\rangle$ is called a forward pomset transition from $\langle C,s\rangle$ to $\langle C',s'\rangle$ and
$\langle C',s'\rangle\xtworightarrow{X[\mathcal{K}]} \langle C,s\rangle$ is called a reverse pomset transition from $\langle C',s'\rangle$ to $\langle C,s\rangle$. When the events in
$X$ and $X[\mathcal{K}]$ are pairwise
concurrent, we say that $\langle C,s\rangle\xrightarrow{X}\langle C',s'\rangle$ is a forward step and $\langle C',s'\rangle\xrightarrow{X[\mathcal{K}]}\langle C,s\rangle$ is a reverse step.
It is obvious that $\rightarrow^*\xrightarrow{X}\rightarrow^*=\xrightarrow{X}$ and
$\rightarrow^*\xrightarrow{e}\rightarrow^*=\xrightarrow{e}$ for any $e\in\mathbb{E}$ and $X\subseteq\mathbb{E}$.
\end{definition}

\begin{definition}[FR weak pomset transitions and weak step]
Let $\mathcal{E}$ be a PES and let $C\in\mathcal{C}(\mathcal{E})$, and $\emptyset\neq X\subseteq \hat{\mathbb{E}}$, if $C\cap X=\emptyset$ and
$\hat{C'}=\hat{C}\cup X\in\mathcal{C}(\mathcal{E})$, then $\langle C,s\rangle\xRightarrow{X} \langle C',s'\rangle$ is called a forward weak pomset transition from $\langle C,s\rangle$ to
$\langle C',s'\rangle$, where we define $\xRightarrow{e}\triangleq\xrightarrow{\tau^*}\xrightarrow{e}\xrightarrow{\tau^*}$. And $\langle C',s'\rangle\xTworightarrow{X[\mathcal{K}]} \langle C,s\rangle$
is called a reverse weak pomset transition from $\langle C',s'\rangle$ to $\langle C,s\rangle$. When the events in $X$ are pairwise concurrent, we say that
$\langle C,s\rangle\xRightarrow{X}\langle C',s'\rangle$ is a forward weak step, when the events in $X[\mathcal{K}]$ are pairwise concurrent, we say that
$\langle C',s'\rangle\xTworightarrow{X}\langle C,s\rangle$ is a reverse weak step.
\end{definition}

We will also suppose that all the PESs are image finite, that is, for any PES $\mathcal{E}$ and $C\in \mathcal{C}(\mathcal{E})$ and $a\in \Lambda$,
$\{e\in \mathbb{E}|\langle C,s\rangle\xrightarrow{e} \langle C',s'\rangle\wedge \lambda(e)=a\}$ and
$\{e\in\hat{\mathbb{E}}|\langle C,s\rangle\xRightarrow{e} \langle C',s'\rangle\wedge \lambda(e)=a\}$ and
$\{e\in \mathbb{E}|\langle C',s'\rangle\xtworightarrow{e} \langle C,s\rangle\wedge \lambda(e)=a\}$ and
$\{e\in\hat{\mathbb{E}}|\langle C',s'\rangle\xTworightarrow{e} \langle C,s\rangle\wedge \lambda(e)=a\}$ are finite.

\begin{definition}[FR pomset, step bisimulation]
Let $\mathcal{E}_1$, $\mathcal{E}_2$ be PESs. A FR pomset bisimulation is a relation $R\subseteq\langle\mathcal{C}(\mathcal{E}_1),S\rangle\times\langle\mathcal{C}(\mathcal{E}_2),S\rangle$,
such that (1) if $(\langle C_1,s\rangle,\langle C_2,s\rangle)\in R$, and $\langle C_1,s\rangle\xrightarrow{X_1}\langle C_1',s'\rangle$ then
$\langle C_2,s\rangle\xrightarrow{X_2}\langle C_2',s'\rangle$, with $X_1\subseteq \mathbb{E}_1$, $X_2\subseteq \mathbb{E}_2$, $X_1\sim X_2$ and
$(\langle C_1',s'\rangle,\langle C_2',s'\rangle)\in R$ for all $s,s'\in S$, and vice-versa;
(2) if $(\langle C_1,s\rangle,\langle C_2,s\rangle)\in R$, and $\langle C_1,s\rangle\xtworightarrow{X_1[\mathcal{K}_1]}\langle C_1',s'\rangle$ then
$\langle C_2,s\rangle\xtworightarrow{X_2[\mathcal{K}_2]}\langle C_2',s'\rangle$, with $X_1\subseteq \mathbb{E}_1$, $X_2\subseteq \mathbb{E}_2$, $X_1\sim X_2$ and
$(\langle C_1',s'\rangle,\langle C_2',s'\rangle)\in R$ for all $s,s'\in S$, and vice-versa. We say that $\mathcal{E}_1$, $\mathcal{E}_2$ are FR pomset bisimilar, written
$\mathcal{E}_1\sim_p^{fr}\mathcal{E}_2$, if there exists a FR pomset bisimulation $R$, such that $(\langle\emptyset,\emptyset\rangle,\langle\emptyset,\emptyset\rangle)\in R$. By replacing
FR pomset transitions with FR steps, we can get the definition of FR step bisimulation. When PESs $\mathcal{E}_1$ and $\mathcal{E}_2$ are FR step bisimilar, we write
$\mathcal{E}_1\sim_s^{fr}\mathcal{E}_2$.
\end{definition}

\begin{definition}[FR weak pomset, step bisimulation]
Let $\mathcal{E}_1$, $\mathcal{E}_2$ be PESs. A FR weak pomset bisimulation is a relation
$R\subseteq\langle\mathcal{C}(\mathcal{E}_1),S\rangle\times\langle\mathcal{C}(\mathcal{E}_2),S\rangle$, such that (1) if $(\langle C_1,s\rangle,\langle C_2,s\rangle)\in R$, and
$\langle C_1,s\rangle\xRightarrow{X_1}\langle C_1',s'\rangle$ then $\langle C_2,s\rangle\xRightarrow{X_2}\langle C_2',s'\rangle$, with $X_1\subseteq \hat{\mathbb{E}_1}$,
$X_2\subseteq \hat{\mathbb{E}_2}$, $X_1\sim X_2$ and $(\langle C_1',s'\rangle,\langle C_2',s'\rangle)\in R$ for all $s,s'\in S$, and vice-versa;
(2) if $(\langle C_1,s\rangle,\langle C_2,s\rangle)\in R$, and
$\langle C_1,s\rangle\xTworightarrow{X_1[\mathcal{K}_1]}\langle C_1',s'\rangle$ then $\langle C_2,s\rangle\xTworightarrow{X_2[\mathcal{K}_2]}\langle C_2',s'\rangle$, with $X_1\subseteq \hat{\mathbb{E}_1}$,
$X_2\subseteq \hat{\mathbb{E}_2}$, $X_1\sim X_2$ and $(\langle C_1',s'\rangle,\langle C_2',s'\rangle)\in R$ for all $s,s'\in S$, and vice-versa. We say that $\mathcal{E}_1$,
$\mathcal{E}_2$ are FR weak pomset bisimilar, written $\mathcal{E}_1\approx_p^{fr}\mathcal{E}_2$, if there exists a FR weak pomset bisimulation $R$, such that
$(\langle\emptyset,\emptyset\rangle,\langle\emptyset,\emptyset\rangle)\in R$. By replacing FR weak pomset transitions with FR weak steps, we can get the definition of FR weak step bisimulation.
When PESs $\mathcal{E}_1$ and $\mathcal{E}_2$ are FR weak step bisimilar, we write $\mathcal{E}_1\approx_s^{fr}\mathcal{E}_2$.
\end{definition}

\begin{definition}[Posetal product]
Given two PESs $\mathcal{E}_1$, $\mathcal{E}_2$, the posetal product of their configurations, denoted
$\langle\mathcal{C}(\mathcal{E}_1),S\rangle\overline{\times}\langle\mathcal{C}(\mathcal{E}_2),S\rangle$, is defined as

$$\{(\langle C_1,s\rangle,f,\langle C_2,s\rangle)|C_1\in\mathcal{C}(\mathcal{E}_1),C_2\in\mathcal{C}(\mathcal{E}_2),f:C_1\rightarrow C_2 \textrm{ isomorphism}\}.$$

A subset $R\subseteq\langle\mathcal{C}(\mathcal{E}_1),S\rangle\overline{\times}\langle\mathcal{C}(\mathcal{E}_2),S\rangle$ is called a posetal relation. We say that $R$ is downward
closed when for any
$(\langle C_1,s\rangle,f,\langle C_2,s\rangle),(\langle C_1',s'\rangle,f',\langle C_2',s'\rangle)\in \langle\mathcal{C}(\mathcal{E}_1),S\rangle\overline{\times}\langle\mathcal{C}(\mathcal{E}_2),S\rangle$,
if $(\langle C_1,s\rangle,f,\langle C_2,s\rangle)\subseteq (\langle C_1',s'\rangle,f',\langle C_2',s'\rangle)$ pointwise and $(\langle C_1',s'\rangle,f',\langle C_2',s'\rangle)\in R$,
then $(\langle C_1,s\rangle,f,\langle C_2,s\rangle)\in R$.

For $f:X_1\rightarrow X_2$, we define $f[x_1\mapsto x_2]:X_1\cup\{x_1\}\rightarrow X_2\cup\{x_2\}$, $z\in X_1\cup\{x_1\}$,(1)$f[x_1\mapsto x_2](z)=
x_2$,if $z=x_1$;(2)$f[x_1\mapsto x_2](z)=f(z)$, otherwise. Where $X_1\subseteq \mathbb{E}_1$, $X_2\subseteq \mathbb{E}_2$, $x_1\in \mathbb{E}_1$, $x_2\in \mathbb{E}_2$.
\end{definition}

\begin{definition}[Weakly posetal product]
Given two PESs $\mathcal{E}_1$, $\mathcal{E}_2$, the weakly posetal product of their configurations, denoted
$\langle\mathcal{C}(\mathcal{E}_1),S\rangle\overline{\times}\langle\mathcal{C}(\mathcal{E}_2),S\rangle$, is defined as

$$\{(\langle C_1,s\rangle,f,\langle C_2,s\rangle)|C_1\in\mathcal{C}(\mathcal{E}_1),C_2\in\mathcal{C}(\mathcal{E}_2),f:\hat{C_1}\rightarrow \hat{C_2} \textrm{ isomorphism}\}.$$

A subset $R\subseteq\langle\mathcal{C}(\mathcal{E}_1),S\rangle\overline{\times}\langle\mathcal{C}(\mathcal{E}_2),S\rangle$ is called a weakly posetal relation. We say that $R$ is
downward closed when for any
$(\langle C_1,s\rangle,f,\langle C_2,s\rangle),(\langle C_1',s'\rangle,f,\langle C_2',s'\rangle)\in \langle\mathcal{C}(\mathcal{E}_1),S\rangle\overline{\times}\langle\mathcal{C}(\mathcal{E}_2),S\rangle$,
if $(\langle C_1,s\rangle,f,\langle C_2,s\rangle)\subseteq (\langle C_1',s'\rangle,f',\langle C_2',s'\rangle)$ pointwise and $(\langle C_1',s'\rangle,f',\langle C_2',s'\rangle)\in R$,
then $(\langle C_1,s\rangle,f,\langle C_2,s\rangle)\in R$.

For $f:X_1\rightarrow X_2$, we define $f[x_1\mapsto x_2]:X_1\cup\{x_1\}\rightarrow X_2\cup\{x_2\}$, $z\in X_1\cup\{x_1\}$,(1)$f[x_1\mapsto x_2](z)=
x_2$,if $z=x_1$;(2)$f[x_1\mapsto x_2](z)=f(z)$, otherwise. Where $X_1\subseteq \hat{\mathbb{E}_1}$, $X_2\subseteq \hat{\mathbb{E}_2}$, $x_1\in \hat{\mathbb{E}}_1$,
$x_2\in \hat{\mathbb{E}}_2$. Also, we define $f(\tau^*)=f(\tau^*)$.
\end{definition}

\begin{definition}[FR (hereditary) history-preserving bisimulation]
A FR history-preserving (hp-) bisimulation is a posetal relation $R\subseteq\langle\mathcal{C}(\mathcal{E}_1),S\rangle\overline{\times}\langle\mathcal{C}(\mathcal{E}_2),S\rangle$ such
that (1) if $(\langle C_1,s\rangle,f,\langle C_2,s\rangle)\in R$, and $\langle C_1,s\rangle\xrightarrow{e_1} \langle C_1',s'\rangle$, then
$\langle C_2,s\rangle\xrightarrow{e_2} \langle C_2',s'\rangle$, with $(\langle C_1',s'\rangle,f[e_1\mapsto e_2],\langle C_2',s'\rangle)\in R$ for all $s,s'\in S$, and vice-versa;
(2) if $(\langle C_1,s\rangle,f,\langle C_2,s\rangle)\in R$, and $\langle C_1,s\rangle\xtworightarrow{e_1[m]} \langle C_1',s'\rangle$, then
$\langle C_2,s\rangle\xtworightarrow{e_2[n]} \langle C_2',s'\rangle$, with $(\langle C_1',s'\rangle,f[e_1[m]\mapsto e_2[n],\langle C_2',s'\rangle)\in R$ for all $s,s'\in S$, and vice-versa.
$\mathcal{E}_1,\mathcal{E}_2$ are FR history-preserving (hp-)bisimilar and are written $\mathcal{E}_1\sim_{hp}^{fr}\mathcal{E}_2$ if there exists a FR hp-bisimulation $R$ such that
$(\langle\emptyset,\emptyset\rangle,\emptyset,\langle\emptyset,\emptyset\rangle)\in R$.

A FR hereditary history-preserving (hhp-)bisimulation is a downward closed FR hp-bisimulation. $\mathcal{E}_1,\mathcal{E}_2$ are FR hereditary history-preserving (hhp-)bisimilar and
are written $\mathcal{E}_1\sim_{hhp}^{fr}\mathcal{E}_2$.
\end{definition}

\begin{definition}[FR weak (hereditary) history-preserving bisimulation]
A FR weak history-preserving (hp-) bisimulation is a weakly posetal relation
$R\subseteq\langle\mathcal{C}(\mathcal{E}_1),S\rangle\overline{\times}\langle\mathcal{C}(\mathcal{E}_2),S\rangle$ such that (1) if $(\langle C_1,s\rangle,f,\langle C_2,s\rangle)\in R$, and
$\langle C_1,s\rangle\xRightarrow{e_1} \langle C_1',s'\rangle$, then $\langle C_2,s\rangle\xRightarrow{e_2} \langle C_2',s'\rangle$, with $(\langle C_1',s'\rangle,f[e_1\mapsto e_2],\langle C_2',s'\rangle)\in R$
for all $s,s'\in S$, and vice-versa;
(2) if $(\langle C_1,s\rangle,f,\langle C_2,s\rangle)\in R$, and
$\langle C_1,s\rangle\xTworightarrow{e_1[m]} \langle C_1',s'\rangle$, then $\langle C_2,s\rangle\xTworightarrow{e_2[n]} \langle C_2',s'\rangle$, with
$(\langle C_1',s'\rangle,f[e_1\mapsto e_2],\langle C_2',s'\rangle)\in R$
for all $s,s'\in S$, and vice-versa. $\mathcal{E}_1,\mathcal{E}_2$ are FR weak history-preserving (hp-)bisimilar and are written $\mathcal{E}_1\approx_{hp}^{fr}\mathcal{E}_2$ if there exists
a FR weak hp-bisimulation $R$ such that $(\langle\emptyset,\emptyset\rangle,\emptyset,\langle\emptyset,\emptyset\rangle)\in R$.

A FR weakly hereditary history-preserving (hhp-)bisimulation is a downward closed FR weak hp-bisimulation. $\mathcal{E}_1,\mathcal{E}_2$ are FR weakly hereditary history-preserving
(hhp-)bisimilar and are written $\mathcal{E}_1\approx_{hhp}^{fr}\mathcal{E}_2$.
\end{definition}

\subsection{Syntax and Operational Semantics}\label{sosctcrg}

We assume an infinite set $\mathcal{N}$ of (action or event) names, and use $a,b,c,\cdots$ to range over $\mathcal{N}$. We denote by $\overline{\mathcal{N}}$ the set of co-names and
let $\overline{a},\overline{b},\overline{c},\cdots$ range over $\overline{\mathcal{N}}$. Then we set $\mathcal{L}=\mathcal{N}\cup\overline{\mathcal{N}}$ as the set of labels, and use
$l,\overline{l}$ to range over $\mathcal{L}$. We extend complementation to $\mathcal{L}$ such that $\overline{\overline{a}}=a$. Let $\tau$ denote the silent step (internal action or
event) and define $Act=\mathcal{L}\cup\{\tau\}\cup\mathcal{L}[\mathcal{K}]$ to be the set of actions, $\alpha,\beta$ range over $Act$. And $K,L$ are used to stand for subsets of
$\mathcal{L}$ and $\overline{L}$ is used for the set of complements of labels in $L$. A relabelling function $f$ is a function from $\mathcal{L}$ to $\mathcal{L}$ such that
$f(\overline{l})=\overline{f(l)}$. By defining $f(\tau)=\tau$, we extend $f$ to $Act$. We write $\mathcal{P}$ for the set of processes. Sometimes, we use $I,J$ to stand for an indexing
set, and we write $E_i:i\in I$ for a family of expressions indexed by $I$. $Id_D$ is the identity function or relation over set $D$.

For each process constant schema $A$, a defining equation of the form

$$A\overset{\text{def}}{=}P$$

is assumed, where $P$ is a process.

Let $G_{at}$ be the set of atomic guards, $\delta$ be the deadlock constant, and $\epsilon$ be the empty action, and extend $Act$ to $Act\cup\{\epsilon\}\cup\{\delta\}$. We extend
$G_{at}$ to the set of basic guards $G$ with element $\phi,\psi,\cdots$, which is generated by the following formation rules:

$$\phi::=\delta|\epsilon|\neg\phi|\psi\in G_{at}|\phi+\psi|\phi\cdot\psi$$

The predicate $test(\phi,s)$ represents that $\phi$ holds in the state $s$, and $test(\epsilon,s)$ holds and $test(\delta,s)$ does not hold. $effect(e,s)\in S$ denotes $s'$ in
$s\xrightarrow{e}s'$. The predicate weakest precondition $wp(e,\phi)$ denotes that $\forall s,s'\in S, test(\phi,effect(e,s))$ holds.

\subsubsection{Syntax}

We use the Prefix $.$ to model the causality relation $\leq$ in true concurrency, the Summation $+$ to model the conflict relation $\sharp$ in true concurrency, and the Composition
$\parallel$ to explicitly model concurrent relation in true concurrency. And we follow the conventions of process algebra.

\begin{definition}[Syntax]\label{syntax06}
Reversible truly concurrent processes CTC with reversibility and guards are defined inductively by the following formation rules:

\begin{enumerate}
  \item $A\in\mathcal{P}$;
  \item $\phi\in\mathcal{P}$;
  \item $\textbf{nil}\in\mathcal{P}$;
  \item if $P\in\mathcal{P}$, then the Prefix $\alpha.P\in\mathcal{P}$ and $P.\alpha[m]\in\mathcal{P}$, for $\alpha\in Act$ and $m\in\mathcal{K}$;
  \item if $P\in\mathcal{P}$, then the Prefix $\phi.P\in\mathcal{P}$, for $\phi\in G_{at}$;
  \item if $P,Q\in\mathcal{P}$, then the Summation $P+Q\in\mathcal{P}$;
  \item if $P,Q\in\mathcal{P}$, then the Composition $P\parallel Q\in\mathcal{P}$;
  \item if $P\in\mathcal{P}$, then the Prefix $(\alpha_1\parallel\cdots\parallel\alpha_n).P\in\mathcal{P}\quad(n\in I)$ and $P.(\alpha_1[m]\parallel\cdots\parallel\alpha_n[m])\in\mathcal{P}\quad(n\in I)$, for $\alpha_,\cdots,\alpha_n\in Act$ and $m\in\mathcal{K}$;
  \item if $P\in\mathcal{P}$, then the Restriction $P\setminus L\in\mathcal{P}$ with $L\in\mathcal{L}$;
  \item if $P\in\mathcal{P}$, then the Relabelling $P[f]\in\mathcal{P}$.
\end{enumerate}

The standard BNF grammar of syntax of CTC with reversibility and guards can be summarized as follows:

$P::=A|\textbf{nil}|\alpha.P| P.\alpha[m]|\phi.P| P+P | P\parallel P | (\alpha_1\parallel\cdots\parallel\alpha_n).P|  P.(\alpha_1[m]\parallel\cdots\parallel\alpha_n[m])  | P\setminus L | P[f].$
\end{definition}

\subsubsection{Operational Semantics}

The operational semantics is defined by LTSs (labelled transition systems), and it is detailed by the following definition.

\begin{definition}[Semantics]\label{semantics06}
The operational semantics of CTC with reversibility and guards corresponding to the syntax in Definition \ref{syntax06} is defined by a series of transition rules, they are shown in Table \ref{FTRForPS06},
\ref{RTRForPS06}, \ref{FTRForCom06}, \ref{RTRForCom06}, \ref{FTRForRRC06} and \ref{RTRForRRC06}. And the predicate
$\xrightarrow{\alpha}\alpha[m]$ represents successful forward termination after execution of the action $\alpha$, the predicate $\xtworightarrow{\alpha[m]}\alpha$ represents successful
reverse termination after execution of the event $\alpha[m]$, the the predicate $\textrm{Std(P)}$ represents that $p$ is a standard process containing no past events, the the predicate
$\textrm{NStd(P)}$ represents that $P$ is a process full of past events.
\end{definition}

The forward transition rules for Prefix and Summation are shown in Table \ref{FTRForPS06}.

\begin{center}
    \begin{table}
        $$\frac{}{\langle\alpha,s\rangle\xrightarrow{\alpha}\langle\alpha[m],s'\rangle}$$
        $$\frac{}{\langle\epsilon,s\rangle\rightarrow\langle\surd,s\rangle}$$
        $$\frac{}{\langle\phi,s\rangle\rightarrow\langle\surd,s\rangle}\textrm{ if }test(\phi,s)$$

        $$\frac{\langle P,s\rangle\xrightarrow{\alpha}\langle\alpha[m],s'\rangle}{\langle P+Q,s\rangle\xrightarrow{\alpha}\langle\alpha[m],s'\rangle}
        \quad\frac{\langle P,s\rangle\xrightarrow{\alpha}\langle P',s'\rangle}{\langle P+Q,s\rangle\xrightarrow{\alpha}\langle P',s'\rangle}$$
        $$\frac{\langle Q,s\rangle\xrightarrow{\alpha}\langle \alpha[m],s'\rangle}{\langle P+Q,s\rangle\xrightarrow{\alpha}\langle\alpha[m],s'\rangle}
        \quad\frac{\langle Q,s\rangle\xrightarrow{\alpha}\langle Q',s'\rangle}{\langle P+Q,s\rangle\xrightarrow{\alpha}\langle Q',s'\rangle}$$


        $$\frac{\langle P,s\rangle\xrightarrow{\alpha}\langle \alpha[m],s'\rangle\quad\textrm{Std}(Q)}{\langle P. Q,s\rangle\xrightarrow{\alpha} \langle\alpha[m]. Q,s'\rangle}
        \quad\frac{\langle P,s\rangle\xrightarrow{\alpha}\langle P',s'\rangle \quad \textrm{Std}(Q)}{\langle P. Q,s\rangle\xrightarrow{\alpha}\langle P'. Q,s'\rangle}$$
        $$\frac{\langle Q,s\rangle\xrightarrow{\beta}\langle\beta[n],s'\rangle\quad \textrm{NStd}(P)}{\langle P. Q,s\rangle\xrightarrow{\beta}\langle P. \beta[n],s'\rangle}
        \quad\frac{\langle Q,s\rangle\xrightarrow{\beta}\langle Q',s'\rangle\quad \textrm{NStd}(P)}{\langle P. Q,s\rangle\xrightarrow{\beta}\langle P. Q',s'\rangle}$$
        \caption{Forward transition rules of Prefix and Summation}
        \label{FTRForPS06}
    \end{table}
\end{center}

The reverse transition rules for Prefix and Summation are shown in Table \ref{RTRForPS06}.

\begin{center}
    \begin{table}
        $$\frac{}{\langle \alpha[m],s\rangle\xtworightarrow{\alpha[m]}\langle\alpha,s'\rangle}$$
        $$\frac{}{\langle\epsilon,s\rangle\xtworightarrow{ }\langle\surd,s\rangle}$$
        $$\frac{}{\langle\phi,s\rangle\xtworightarrow{ }\langle\surd,s\rangle}\textrm{ if }test(\phi,s)$$

        $$\frac{\langle P,s\rangle\xtworightarrow{\alpha[m]}\langle \alpha,s'\rangle}{\langle P+Q,s\rangle\xtworightarrow{\alpha[m]}\langle\alpha,s'\rangle}
        \quad\frac{\langle P,s\rangle\xtworightarrow{\alpha[m]}\langle P',s'\rangle}{\langle P+Q,s\rangle\xtworightarrow{\alpha[m]}\langle P',s'\rangle}$$
        $$\frac{\langle Q,s\rangle\xtworightarrow{\alpha[m]}\langle\alpha,s'\rangle}{\langle P+Q,s\rangle\xtworightarrow{\alpha[m]}\langle\alpha,s'\rangle}
        \quad\frac{\langle Q,s\rangle\xtworightarrow{\alpha[m]}\langle Q',s'\rangle}{\langle P+Q,s\rangle\xtworightarrow{\alpha[m]}\langle Q',s'\rangle}$$


        $$\frac{\langle P,s\rangle\xtworightarrow{\alpha[m]}\langle\alpha,s'\rangle \quad \textrm{Std}(Q)}{\langle P. Q,s\rangle\xtworightarrow{\alpha[m]} \langle\alpha. Q,s'\rangle}
        \quad\frac{\langle P,s\rangle\xtworightarrow{\alpha[m]}\langle P',s'\rangle\quad \textrm{Std}(Q)}{\langle P. Q,s\rangle\xtworightarrow{\alpha[m]}\langle P'. Q,s'\rangle}$$
        $$\frac{\langle Q,s\rangle\xtworightarrow{\beta[n]}\langle\beta,s'\rangle \quad \textrm{NStd}(P)}{\langle P. Q,s\rangle\xtworightarrow{\beta[n]}\langle P. \beta,s'\rangle}\quad
        \frac{\langle Q,s\rangle\xtworightarrow{\beta[n]}\langle Q',s'\rangle \quad \textrm{NStd}(P)}{\langle P. Q,s\rangle\xtworightarrow{\beta[n]}\langle P. Q',s'\rangle}$$
        \caption{Reverse transition rules of Prefix and Summation}
        \label{RTRForPS06}
    \end{table}
\end{center}

The forward transition rules for Composition are shown in Table \ref{FTRForCom06}.

\begin{center}
    \begin{table}
        $$\frac{\langle P,s\rangle\xrightarrow{\alpha}\langle P',s'\rangle\quad \langle Q,s\rangle\nrightarrow}{\langle P\parallel Q,s\rangle\xrightarrow{\alpha}\langle P'\parallel Q,s'\rangle}$$
        $$\frac{\langle Q,s\rangle\xrightarrow{\alpha}\langle Q',s'\rangle\quad \langle P,s\rangle\nrightarrow}{\langle P\parallel Q,s\rangle\xrightarrow{\alpha}\langle P\parallel Q',s'\rangle}$$
        $$\frac{\langle P,s\rangle\xrightarrow{\alpha}\langle P',s'\rangle\quad \langle Q,s\rangle\xrightarrow{\beta}\langle Q',s''\rangle}{\langle P\parallel Q,s\rangle\xrightarrow{\{\alpha,\beta\}}\langle P'\parallel Q',s'\cup s''\rangle}\quad (\beta\neq\overline{\alpha})$$
        $$\frac{\langle P,s\rangle\xrightarrow{l}\langle P',s'\rangle\quad \langle Q,s\rangle\xrightarrow{\overline{l}}\langle Q',s''\rangle}{\langle P\parallel Q,s\rangle\xrightarrow{\tau}\langle P'\parallel Q',s'\cup s''\rangle}$$
        \caption{Forward transition rules of Composition}
        \label{FTRForCom06}
    \end{table}
\end{center}

The reverse transition rules for Composition are shown in Table \ref{RTRForCom06}.

\begin{center}
    \begin{table}
        $$\frac{\langle P,s\rangle\xtworightarrow{\alpha[m]}\langle P',s'\rangle\quad \langle Q,s\rangle\xntworightarrow{}}{\langle P\parallel Q,s\rangle\xtworightarrow{\alpha[m]}\langle P'\parallel Q,s'\rangle}$$
        $$\frac{\langle Q,s\rangle\xtworightarrow{\alpha[m]}\langle Q',s'\rangle\quad \langle P,s\rangle\xntworightarrow{}}{\langle P\parallel Q,s\rangle\xtworightarrow{\alpha[m]}\langle P\parallel Q',s'\rangle}$$
        $$\frac{\langle P,s\rangle\xtworightarrow{\alpha[m]}\langle P',s'\rangle\quad \langle Q,s\rangle\xtworightarrow{\beta[m]}\langle Q',s''\rangle}{\langle P\parallel Q,s\rangle\xtworightarrow{\{\alpha[m],\beta[m]\}}\langle P'\parallel Q',s'\cup s''\rangle}\quad (\beta\neq\overline{\alpha})$$
        $$\frac{\langle P,s\rangle\xtworightarrow{l[m]}\langle P',s'\rangle\quad \langle Q,s\rangle\xtworightarrow{\overline{l}[m]}\langle Q',s''\rangle}{\langle P\parallel Q,s\rangle\xtworightarrow{\tau}\langle P'\parallel Q',s'\cup s''\rangle}$$
        \caption{Reverse transition rules of Composition}
        \label{RTRForCom06}
    \end{table}
\end{center}

The forward transition rules for Restriction, Relabelling and Constants are shown in Table \ref{FTRForRRC06}.

\begin{center}
    \begin{table}
        $$\frac{\langle P,s\rangle\xrightarrow{\alpha}\langle P',s'\rangle}{\langle P\setminus L,s\rangle\xrightarrow{\alpha}\langle P'\setminus L,s'\rangle}\quad (\alpha,\overline{\alpha}\notin L)$$
        $$\frac{\langle P,s\rangle\xrightarrow{\{\alpha_1,\cdots,\alpha_n\}}\langle P',s'\rangle}{\langle P\setminus L,s\rangle\xrightarrow{\{\alpha_1,\cdots,\alpha_n\}}\langle P'\setminus L,s'\rangle}\quad (\alpha_1,\overline{\alpha_1},\cdots,\alpha_n,\overline{\alpha_n}\notin L)$$
        $$\frac{\langle P,s\rangle\xrightarrow{\alpha}\langle P',s'\rangle}{\langle P[f],s\rangle\xrightarrow{f(\alpha)}\langle P'[f],s'\rangle}$$
        $$\frac{\langle P,s\rangle\xrightarrow{\{\alpha_1,\cdots,\alpha_n\}}\langle P',s'\rangle}{\langle P[f],s\rangle\xrightarrow{\{f(\alpha_1),\cdots,f(\alpha_n)\}}\langle P'[f],s'\rangle}$$
        $$\frac{\langle P,s\rangle\xrightarrow{\alpha}\langle P',s'\rangle}{\langle A,s\rangle\xrightarrow{\alpha}\langle P',s'\rangle}\quad (A\overset{\text{def}}{=}P)$$
        $$\frac{\langle P,s\rangle\xrightarrow{\{\alpha_1,\cdots,\alpha_n\}}\langle P',s'\rangle}{\langle A,s\rangle\xrightarrow{\{\alpha_1,\cdots,\alpha_n\}}\langle P',s'\rangle}\quad (A\overset{\text{def}}{=}P)$$
        \caption{Forward transition rules of Restriction, Relabelling and Constants}
        \label{FTRForRRC06}
    \end{table}
\end{center}

The reverse transition rules for Restriction, Relabelling and Constants are shown in Table \ref{RTRForRRC06}.

\begin{center}
    \begin{table}
        $$\frac{\langle P,s\rangle\xtworightarrow{\alpha[m]}\langle P',s'\rangle}{\langle P\setminus L,s\rangle\xtworightarrow{\alpha[m]}\langle P'\setminus L,s'\rangle}\quad (\alpha,\overline{\alpha}\notin L)$$
        $$\frac{\langle P,s\rangle\xtworightarrow{\{\alpha_1[m],\cdots,\alpha_n[m]\}}\langle P',s'\rangle}{\langle P\setminus L,s\rangle\xtworightarrow{\{\alpha_1[m],\cdots,\alpha_n[m]\}}\langle P'\setminus L,s'\rangle}\quad (\alpha_1,\overline{\alpha_1},\cdots,\alpha_n,\overline{\alpha_n}\notin L)$$
        $$\frac{\langle P,s\rangle\xtworightarrow{\alpha[m]}\langle P',s'\rangle}{\langle P[f],s\rangle\xtworightarrow{f(\alpha[m])}\langle P'[f],s'\rangle}$$
        $$\frac{\langle P,s\rangle\xtworightarrow{\{\alpha_1[m],\cdots,\alpha_n[m]\}}\langle P',s'\rangle}{\langle P[f],s\rangle\xtworightarrow{\{f(\alpha_1)[m],\cdots,f(\alpha_n)[m]\}}\langle P'[f],s'\rangle}$$
        $$\frac{\langle P,s\rangle\xtworightarrow{\alpha[m]}\langle P',s'\rangle}{\langle A,s\rangle\xtworightarrow{\alpha[m]}\langle P',s'\rangle}\quad (A\overset{\text{def}}{=}P)$$
        $$\frac{\langle P,s\rangle\xtworightarrow{\{\alpha_1[m],\cdots,\alpha_n[m]\}}\langle P',s'\rangle}{\langle A,s\rangle\xtworightarrow{\{\alpha_1[m],\cdots,\alpha_n[m]\}}\langle P',s'\rangle}\quad (A\overset{\text{def}}{=}P)$$
        \caption{Reverse transition rules of Restriction, Relabelling and Constants}
        \label{RTRForRRC06}
    \end{table}
\end{center}

\subsubsection{Properties of Transitions}

\begin{definition}[Sorts]\label{sorts06}
Given the sorts $\mathcal{L}(A)$ and $\mathcal{L}(X)$ of constants and variables, we define $\mathcal{L}(P)$ inductively as follows.

\begin{enumerate}
  \item $\mathcal{L}(l.P)=\{l\}\cup\mathcal{L}(P)$;
  \item $\mathcal{L}(P.l[m])=\{l\}\cup\mathcal{L}(P)$;
  \item $\mathcal{L}((l_1\parallel \cdots\parallel l_n).P)=\{l_1,\cdots,l_n\}\cup\mathcal{L}(P)$;
  \item $\mathcal{L}(P.(l_1[m]\parallel \cdots\parallel l_n[m]))=\{l_1,\cdots,l_n\}\cup\mathcal{L}(P)$;
  \item $\mathcal{L}(\tau.P)=\mathcal{L}(P)$;
  \item $\mathcal{L}(\epsilon.P)=\mathcal{L}(P)$;
  \item $\mathcal{L}(\phi.P)=\mathcal{L}(P)$;
  \item $\mathcal{L}(P+Q)=\mathcal{L}(P)\cup\mathcal{L}(Q)$;
  \item $\mathcal{L}(P\parallel Q)=\mathcal{L}(P)\cup\mathcal{L}(Q)$;
  \item $\mathcal{L}(P\setminus L)=\mathcal{L}(P)-(L\cup\overline{L})$;
  \item $\mathcal{L}(P[f])=\{f(l):l\in\mathcal{L}(P)\}$;
  \item for $A\overset{\text{def}}{=}P$, $\mathcal{L}(P)\subseteq\mathcal{L}(A)$.
\end{enumerate}
\end{definition}

Now, we present some properties of the transition rules defined in Definition \ref{semantics06}.

\begin{proposition}
If $P\xrightarrow{\alpha}P'$, then
\begin{enumerate}
  \item $\alpha\in\mathcal{L}(P)\cup\{\tau\}\cup\{\epsilon\}$;
  \item $\mathcal{L}(P')\subseteq\mathcal{L}(P)$.
\end{enumerate}

If $P\xrightarrow{\{\alpha_1,\cdots,\alpha_n\}}P'$, then
\begin{enumerate}
  \item $\alpha_1,\cdots,\alpha_n\in\mathcal{L}(P)\cup\{\tau\}\cup\{\epsilon\}$;
  \item $\mathcal{L}(P')\subseteq\mathcal{L}(P)$.
\end{enumerate}
\end{proposition}

\begin{proof}
By induction on the inference of $P\xrightarrow{\alpha}P'$ and $P\xrightarrow{\{\alpha_1,\cdots,\alpha_n\}}P'$, there are several cases corresponding to the forward transition rules in
Definition \ref{semantics06}, we omit them.
\end{proof}

\begin{proposition}
If $P\xtworightarrow{\alpha[m]}P'$, then
\begin{enumerate}
  \item $\alpha\in\mathcal{L}(P)\cup\{\tau\}\cup\{\epsilon\}$;
  \item $\mathcal{L}(P')\subseteq\mathcal{L}(P)$.
\end{enumerate}

If $P\xtworightarrow{\{\alpha_1[m],\cdots,\alpha_n[m]\}}P'$, then
\begin{enumerate}
  \item $\alpha_1,\cdots,\alpha_n\in\mathcal{L}(P)\cup\{\tau\}\cup\{\epsilon\}$;
  \item $\mathcal{L}(P')\subseteq\mathcal{L}(P)$.
\end{enumerate}
\end{proposition}

\begin{proof}
By induction on the inference of $P\xtworightarrow{\alpha}P'$ and $P\xtworightarrow{\{\alpha_1,\cdots,\alpha_n\}}P'$, there are several cases corresponding to the forward transition
rules in Definition \ref{semantics06}, we omit them.
\end{proof}

\subsection{Strong Bisimulations}\label{sftcbctcrg}

\subsubsection{Laws and Congruence}

Based on the concepts of strongly FR truly concurrent bisimulation equivalences, we get the following laws.

\begin{proposition}[Monoid laws for FR strong pomset bisimulation] The monoid laws for FR strong pomset bisimulation are as follows.

\begin{enumerate}
  \item $P+Q\sim_p^{fr} Q+P$;
  \item $P+(Q+R)\sim_p^{fr} (P+Q)+R$;
  \item $P+P\sim_p^{fr} P$;
  \item $P+\textbf{nil}\sim_p^{fr} P$.
\end{enumerate}

\end{proposition}

\begin{proof}
\begin{enumerate}
  \item $P+Q\sim_p^{fr} Q+P$. It is sufficient to prove the relation $R=\{(P+Q, Q+P)\}\cup \textbf{Id}$ is a FR strong pomset bisimulation, we omit it;
  \item $P+(Q+R)\sim_p^{fr} (P+Q)+R$. It is sufficient to prove the relation $R=\{(P+(Q+R), (P+Q)+R)\}\cup \textbf{Id}$ is a FR strong pomset bisimulation, we omit it;
  \item $P+P\sim_p^{fr} P$. It is sufficient to prove the relation $R=\{(P+P, P)\}\cup \textbf{Id}$ is a FR strong pomset bisimulation, we omit it;
  \item $P+\textbf{nil}\sim_p^{fr} P$. It is sufficient to prove the relation $R=\{(P+\textbf{nil}, P)\}\cup \textbf{Id}$ is a FR strong pomset bisimulation, we omit it.
\end{enumerate}
\end{proof}

\begin{proposition}[Monoid laws for FR strong step bisimulation] The monoid laws for FR strong step bisimulation are as follows.
\begin{enumerate}
  \item $P+Q\sim_s^{fr} Q+P$;
  \item $P+(Q+R)\sim_s^{fr} (P+Q)+R$;
  \item $P+P\sim_s^{fr} P$;
  \item $P+\textbf{nil}\sim_s^{fr} P$.
\end{enumerate}
\end{proposition}

\begin{proof}
\begin{enumerate}
  \item $P+Q\sim_s^{fr} Q+P$. It is sufficient to prove the relation $R=\{(P+Q, Q+P)\}\cup \textbf{Id}$ is a FR strong step bisimulation, we omit it;
  \item $P+(Q+R)\sim_s^{fr} (P+Q)+R$. It is sufficient to prove the relation $R=\{(P+(Q+R), (P+Q)+R)\}\cup \textbf{Id}$ is a FR strong step bisimulation, we omit it;
  \item $P+P\sim_s^{fr} P$. It is sufficient to prove the relation $R=\{(P+P, P)\}\cup \textbf{Id}$ is a FR strong step bisimulation, we omit it;
  \item $P+\textbf{nil}\sim_s^{fr} P$. It is sufficient to prove the relation $R=\{(P+\textbf{nil}, P)\}\cup \textbf{Id}$ is a FR strong step bisimulation, we omit it.
\end{enumerate}
\end{proof}

\begin{proposition}[Monoid laws for FR strong hp-bisimulation] The monoid laws for FR strong hp-bisimulation are as follows.
\begin{enumerate}
  \item $P+Q\sim_{hp}^{fr} Q+P$;
  \item $P+(Q+R)\sim_{hp}^{fr} (P+Q)+R$;
  \item $P+P\sim_{hp}^{fr} P$;
  \item $P+\textbf{nil}\sim_{hp}^{fr} P$.
\end{enumerate}
\end{proposition}

\begin{proof}
\begin{enumerate}
  \item $P+Q\sim_{hp}^{fr} Q+P$. It is sufficient to prove the relation $R=\{(P+Q, Q+P)\}\cup \textbf{Id}$ is a FR strong hp-bisimulation, we omit it;
  \item $P+(Q+R)\sim_{hp}^{fr} (P+Q)+R$. It is sufficient to prove the relation $R=\{(P+(Q+R), (P+Q)+R)\}\cup \textbf{Id}$ is a FR strong hp-bisimulation, we omit it;
  \item $P+P\sim_{hp}^{fr} P$. It is sufficient to prove the relation $R=\{(P+P, P)\}\cup \textbf{Id}$ is a FR strong hp-bisimulation, we omit it;
  \item $P+\textbf{nil}\sim_{hp}^{fr} P$. It is sufficient to prove the relation $R=\{(P+\textbf{nil}, P)\}\cup \textbf{Id}$ is a FR strong hp-bisimulation, we omit it.
\end{enumerate}
\end{proof}

\begin{proposition}[Monoid laws for FR strong hhp-bisimulation] The monoid laws for FR strong hhp-bisimulation are as follows.
\begin{enumerate}
  \item $P+Q\sim_{hhp}^{fr} Q+P$;
  \item $P+(Q+R)\sim_{hhp}^{fr} (P+Q)+R$;
  \item $P+P\sim_{hhp}^{fr} P$;
  \item $P+\textbf{nil}\sim_{hhp}^{fr} P$.
\end{enumerate}
\end{proposition}

\begin{proof}
\begin{enumerate}
  \item $P+Q\sim_{hhp}^{fr} Q+P$. It is sufficient to prove the relation $R=\{(P+Q, Q+P)\}\cup \textbf{Id}$ is a FR strong hhp-bisimulation, we omit it;
  \item $P+(Q+R)\sim_{hhp}^{fr} (P+Q)+R$. It is sufficient to prove the relation $R=\{(P+(Q+R), (P+Q)+R)\}\cup \textbf{Id}$ is a FR strong hhp-bisimulation, we omit it;
  \item $P+P\sim_{hhp}^{fr} P$. It is sufficient to prove the relation $R=\{(P+P, P)\}\cup \textbf{Id}$ is a FR strong hhp-bisimulation, we omit it;
  \item $P+\textbf{nil}\sim_{hhp}^{fr} P$. It is sufficient to prove the relation $R=\{(P+\textbf{nil}, P)\}\cup \textbf{Id}$ is a FR strong hhp-bisimulation, we omit it.
\end{enumerate}
\end{proof}

\begin{proposition}[Static laws for FR strong pomset bisimulation]
The static laws for FR strong pomset bisimulation are as follows.
\begin{enumerate}
  \item $P\parallel Q\sim_p^{fr} Q\parallel P$;
  \item $P\parallel(Q\parallel R)\sim_p^{fr} (P\parallel Q)\parallel R$;
  \item $P\parallel \textbf{nil}\sim_p^{fr} P$;
  \item $P\setminus L\sim_p^{fr} P$, if $\mathcal{L}(P)\cap(L\cup\overline{L})=\emptyset$;
  \item $P\setminus K\setminus L\sim_p^{fr} P\setminus(K\cup L)$;
  \item $P[f]\setminus L\sim_p^{fr} P\setminus f^{-1}(L)[f]$;
  \item $(P\parallel Q)\setminus L\sim_p^{fr} P\setminus L\parallel Q\setminus L$, if $\mathcal{L}(P)\cap\overline{\mathcal{L}(Q)}\cap(L\cup\overline{L})=\emptyset$;
  \item $P[Id]\sim_p^{fr} P$;
  \item $P[f]\sim_p^{fr} P[f']$, if $f\upharpoonright\mathcal{L}(P)=f'\upharpoonright\mathcal{L}(P)$;
  \item $P[f][f']\sim_p^{fr} P[f'\circ f]$;
  \item $(P\parallel Q)[f]\sim_p^{fr} P[f]\parallel Q[f]$, if $f\upharpoonright(L\cup\overline{L})$ is one-to-one, where $L=\mathcal{L}(P)\cup\mathcal{L}(Q)$.
\end{enumerate}
\end{proposition}

\begin{proof}
\begin{enumerate}
  \item $P\parallel Q\sim_p^{fr} Q\parallel P$. It is sufficient to prove the relation $R=\{(P\parallel Q, Q\parallel P)\}\cup \textbf{Id}$ is a FR strong pomset bisimulation, we omit it;
  \item $P\parallel(Q\parallel R)\sim_p^{fr} (P\parallel Q)\parallel R$. It is sufficient to prove the relation $R=\{(P\parallel(Q\parallel R), (P\parallel Q)\parallel R)\}\cup \textbf{Id}$ is a FR strong pomset bisimulation, we omit it;
  \item $P\parallel \textbf{nil}\sim_p^{fr} P$. It is sufficient to prove the relation $R=\{(P\parallel \textbf{nil}, P)\}\cup \textbf{Id}$ is a FR strong pomset bisimulation, we omit it;
  \item $P\setminus L\sim_p^{fr} P$, if $\mathcal{L}(P)\cap(L\cup\overline{L})=\emptyset$. It is sufficient to prove the relation $R=\{(P\setminus L, P)\}\cup \textbf{Id}$, if $\mathcal{L}(P)\cap(L\cup\overline{L})=\emptyset$, is a FR strong pomset bisimulation, we omit it;
  \item $P\setminus K\setminus L\sim_p^{fr} P\setminus(K\cup L)$. It is sufficient to prove the relation $R=\{(P\setminus K\setminus L, P\setminus(K\cup L))\}\cup \textbf{Id}$ is a FR strong pomset bisimulation, we omit it;
  \item $P[f]\setminus L\sim_p^{fr} P\setminus f^{-1}(L)[f]$. It is sufficient to prove the relation $R=\{(P[f]\setminus L, P\setminus f^{-1}(L)[f])\}\cup \textbf{Id}$ is a FR strong pomset bisimulation, we omit it;
  \item $(P\parallel Q)\setminus L\sim_p^{fr} P\setminus L\parallel Q\setminus L$, if $\mathcal{L}(P)\cap\overline{\mathcal{L}(Q)}\cap(L\cup\overline{L})=\emptyset$. It is sufficient to prove the relation
  $R=\{((P\parallel Q)\setminus L, P\setminus L\parallel Q\setminus L)\}\cup \textbf{Id}$, if $\mathcal{L}(P)\cap\overline{\mathcal{L}(Q)}\cap(L\cup\overline{L})=\emptyset$, is a FR strong pomset bisimulation, we omit it;
  \item $P[Id]\sim_p^{fr} P$. It is sufficient to prove the relation $R=\{(P[Id], P)\}\cup \textbf{Id}$ is a FR strong pomset bisimulation, we omit it;
  \item $P[f]\sim_p^{fr} P[f']$, if $f\upharpoonright\mathcal{L}(P)=f'\upharpoonright\mathcal{L}(P)$. It is sufficient to prove the relation $R=\{(P[f], P[f'])\}\cup \textbf{Id}$, if $f\upharpoonright\mathcal{L}(P)=f'\upharpoonright\mathcal{L}(P)$, is a FR strong pomset bisimulation, we omit it;
  \item $P[f][f']\sim_p^{fr} P[f'\circ f]$. It is sufficient to prove the relation $R=\{(P[f][f'], P[f'\circ f])\}\cup \textbf{Id}$ is a FR strong pomset bisimulation, we omit it;
  \item $(P\parallel Q)[f]\sim_p^{fr} P[f]\parallel Q[f]$, if $f\upharpoonright(L\cup\overline{L})$ is one-to-one, where $L=\mathcal{L}(P)\cup\mathcal{L}(Q)$. It is sufficient to prove the
  relation $R=\{((P\parallel Q)[f], P[f]\parallel Q[f])\}\cup \textbf{Id}$, if $f\upharpoonright(L\cup\overline{L})$ is one-to-one, where $L=\mathcal{L}(P)\cup\mathcal{L}(Q)$, is a FR strong pomset bisimulation, we omit it.
\end{enumerate}
\end{proof}

\begin{proposition}[Static laws for FR strong step bisimulation]
The static laws for FR strong step bisimulation are as follows.
\begin{enumerate}
  \item $P\parallel Q\sim_s^{fr} Q\parallel P$;
  \item $P\parallel(Q\parallel R)\sim_s^{fr} (P\parallel Q)\parallel R$;
  \item $P\parallel \textbf{nil}\sim_s^{fr} P$;
  \item $P\setminus L\sim_s^{fr} P$, if $\mathcal{L}(P)\cap(L\cup\overline{L})=\emptyset$;
  \item $P\setminus K\setminus L\sim_s^{fr} P\setminus(K\cup L)$;
  \item $P[f]\setminus L\sim_s^{fr} P\setminus f^{-1}(L)[f]$;
  \item $(P\parallel Q)\setminus L\sim_s^{fr} P\setminus L\parallel Q\setminus L$, if $\mathcal{L}(P)\cap\overline{\mathcal{L}(Q)}\cap(L\cup\overline{L})=\emptyset$;
  \item $P[Id]\sim_s^{fr} P$;
  \item $P[f]\sim_s^{fr} P[f']$, if $f\upharpoonright\mathcal{L}(P)=f'\upharpoonright\mathcal{L}(P)$;
  \item $P[f][f']\sim_s^{fr} P[f'\circ f]$;
  \item $(P\parallel Q)[f]\sim_s^{fr} P[f]\parallel Q[f]$, if $f\upharpoonright(L\cup\overline{L})$ is one-to-one, where $L=\mathcal{L}(P)\cup\mathcal{L}(Q)$.
\end{enumerate}
\end{proposition}

\begin{proof}
\begin{enumerate}
  \item $P\parallel Q\sim_s^{fr} Q\parallel P$. It is sufficient to prove the relation $R=\{(P\parallel Q, Q\parallel P)\}\cup \textbf{Id}$ is a FR strong step bisimulation, we omit it;
  \item $P\parallel(Q\parallel R)\sim_s^{fr} (P\parallel Q)\parallel R$. It is sufficient to prove the relation $R=\{(P\parallel(Q\parallel R), (P\parallel Q)\parallel R)\}\cup \textbf{Id}$ is a FR strong step bisimulation, we omit it;
  \item $P\parallel \textbf{nil}\sim_s^{fr} P$. It is sufficient to prove the relation $R=\{(P\parallel \textbf{nil}, P)\}\cup \textbf{Id}$ is a FR strong step bisimulation, we omit it;
  \item $P\setminus L\sim_s^{fr} P$, if $\mathcal{L}(P)\cap(L\cup\overline{L})=\emptyset$. It is sufficient to prove the relation $R=\{(P\setminus L, P)\}\cup \textbf{Id}$, if $\mathcal{L}(P)\cap(L\cup\overline{L})=\emptyset$, is a FR strong step bisimulation, we omit it;
  \item $P\setminus K\setminus L\sim_s^{fr} P\setminus(K\cup L)$. It is sufficient to prove the relation $R=\{(P\setminus K\setminus L, P\setminus(K\cup L))\}\cup \textbf{Id}$ is a FR strong step bisimulation, we omit it;
  \item $P[f]\setminus L\sim_s^{fr} P\setminus f^{-1}(L)[f]$. It is sufficient to prove the relation $R=\{(P[f]\setminus L, P\setminus f^{-1}(L)[f])\}\cup \textbf{Id}$ is a FR strong step bisimulation, we omit it;
  \item $(P\parallel Q)\setminus L\sim_s^{fr} P\setminus L\parallel Q\setminus L$, if $\mathcal{L}(P)\cap\overline{\mathcal{L}(Q)}\cap(L\cup\overline{L})=\emptyset$. It is sufficient to prove the relation
  $R=\{((P\parallel Q)\setminus L, P\setminus L\parallel Q\setminus L)\}\cup \textbf{Id}$, if $\mathcal{L}(P)\cap\overline{\mathcal{L}(Q)}\cap(L\cup\overline{L})=\emptyset$, is a FR strong step bisimulation, we omit it;
  \item $P[Id]\sim_s^{fr} P$. It is sufficient to prove the relation $R=\{(P[Id], P)\}\cup \textbf{Id}$ is a FR strong step bisimulation, we omit it;
  \item $P[f]\sim_s^{fr} P[f']$, if $f\upharpoonright\mathcal{L}(P)=f'\upharpoonright\mathcal{L}(P)$. It is sufficient to prove the relation $R=\{(P[f], P[f'])\}\cup \textbf{Id}$, if $f\upharpoonright\mathcal{L}(P)=f'\upharpoonright\mathcal{L}(P)$, is a FR strong step bisimulation, we omit it;
  \item $P[f][f']\sim_s^{fr} P[f'\circ f]$. It is sufficient to prove the relation $R=\{(P[f][f'], P[f'\circ f])\}\cup \textbf{Id}$ is a FR strong step bisimulation, we omit it;
  \item $(P\parallel Q)[f]\sim_s^{fr} P[f]\parallel Q[f]$, if $f\upharpoonright(L\cup\overline{L})$ is one-to-one, where $L=\mathcal{L}(P)\cup\mathcal{L}(Q)$. It is sufficient to prove the
  relation $R=\{((P\parallel Q)[f], P[f]\parallel Q[f])\}\cup \textbf{Id}$, if $f\upharpoonright(L\cup\overline{L})$ is one-to-one, where $L=\mathcal{L}(P)\cup\mathcal{L}(Q)$, is a FR strong step bisimulation, we omit it.
\end{enumerate}
\end{proof}

\begin{proposition}[Static laws for FR strong hp-bisimulation]
The static laws for FR strong hp-bisimulation are as follows.
\begin{enumerate}
  \item $P\parallel Q\sim_{hp}^{fr} Q\parallel P$;
  \item $P\parallel(Q\parallel R)\sim_{hp}^{fr} (P\parallel Q)\parallel R$;
  \item $P\parallel \textbf{nil}\sim_{hp}^{fr} P$;
  \item $P\setminus L\sim_{hp}^{fr} P$, if $\mathcal{L}(P)\cap(L\cup\overline{L})=\emptyset$;
  \item $P\setminus K\setminus L\sim_{hp}^{fr} P\setminus(K\cup L)$;
  \item $P[f]\setminus L\sim_{hp}^{fr} P\setminus f^{-1}(L)[f]$;
  \item $(P\parallel Q)\setminus L\sim_{hp}^{fr} P\setminus L\parallel Q\setminus L$, if $\mathcal{L}(P)\cap\overline{\mathcal{L}(Q)}\cap(L\cup\overline{L})=\emptyset$;
  \item $P[Id]\sim_{hp}^{fr} P$;
  \item $P[f]\sim_{hp}^{fr} P[f']$, if $f\upharpoonright\mathcal{L}(P)=f'\upharpoonright\mathcal{L}(P)$;
  \item $P[f][f']\sim_{hp}^{fr} P[f'\circ f]$;
  \item $(P\parallel Q)[f]\sim_{hp}^{fr} P[f]\parallel Q[f]$, if $f\upharpoonright(L\cup\overline{L})$ is one-to-one, where $L=\mathcal{L}(P)\cup\mathcal{L}(Q)$.
\end{enumerate}
\end{proposition}

\begin{proof}
\begin{enumerate}
  \item $P\parallel Q\sim_{hp}^{fr} Q\parallel P$. It is sufficient to prove the relation $R=\{(P\parallel Q, Q\parallel P)\}\cup \textbf{Id}$ is a FR strong hp-bisimulation, we omit it;
  \item $P\parallel(Q\parallel R)\sim_{hp}^{fr} (P\parallel Q)\parallel R$. It is sufficient to prove the relation $R=\{(P\parallel(Q\parallel R), (P\parallel Q)\parallel R)\}\cup \textbf{Id}$ is a FR strong hp-bisimulation, we omit it;
  \item $P\parallel \textbf{nil}\sim_{hp}^{fr} P$. It is sufficient to prove the relation $R=\{(P\parallel \textbf{nil}, P)\}\cup \textbf{Id}$ is a FR strong hp-bisimulation, we omit it;
  \item $P\setminus L\sim_{hp}^{fr} P$, if $\mathcal{L}(P)\cap(L\cup\overline{L})=\emptyset$. It is sufficient to prove the relation $R=\{(P\setminus L, P)\}\cup \textbf{Id}$, if $\mathcal{L}(P)\cap(L\cup\overline{L})=\emptyset$, is a FR strong hp-bisimulation, we omit it;
  \item $P\setminus K\setminus L\sim_{hp}^{fr} P\setminus(K\cup L)$. It is sufficient to prove the relation $R=\{(P\setminus K\setminus L, P\setminus(K\cup L))\}\cup \textbf{Id}$ is a FR strong hp-bisimulation, we omit it;
  \item $P[f]\setminus L\sim_{hp}^{fr} P\setminus f^{-1}(L)[f]$. It is sufficient to prove the relation $R=\{(P[f]\setminus L, P\setminus f^{-1}(L)[f])\}\cup \textbf{Id}$ is a FR strong hp-bisimulation, we omit it;
  \item $(P\parallel Q)\setminus L\sim_{hp}^{fr} P\setminus L\parallel Q\setminus L$, if $\mathcal{L}(P)\cap\overline{\mathcal{L}(Q)}\cap(L\cup\overline{L})=\emptyset$. It is sufficient to prove the relation
  $R=\{((P\parallel Q)\setminus L, P\setminus L\parallel Q\setminus L)\}\cup \textbf{Id}$, if $\mathcal{L}(P)\cap\overline{\mathcal{L}(Q)}\cap(L\cup\overline{L})=\emptyset$, is a FR strong hp-bisimulation, we omit it;
  \item $P[Id]\sim_{hp}^{fr} P$. It is sufficient to prove the relation $R=\{(P[Id], P)\}\cup \textbf{Id}$ is a FR strong hp-bisimulation, we omit it;
  \item $P[f]\sim_{hp}^{fr} P[f']$, if $f\upharpoonright\mathcal{L}(P)=f'\upharpoonright\mathcal{L}(P)$. It is sufficient to prove the relation $R=\{(P[f], P[f'])\}\cup \textbf{Id}$, if $f\upharpoonright\mathcal{L}(P)=f'\upharpoonright\mathcal{L}(P)$, is a FR strong hp-bisimulation, we omit it;
  \item $P[f][f']\sim_{hp}^{fr} P[f'\circ f]$. It is sufficient to prove the relation $R=\{(P[f][f'], P[f'\circ f])\}\cup \textbf{Id}$ is a FR strong hp-bisimulation, we omit it;
  \item $(P\parallel Q)[f]\sim_{hp}^{fr} P[f]\parallel Q[f]$, if $f\upharpoonright(L\cup\overline{L})$ is one-to-one, where $L=\mathcal{L}(P)\cup\mathcal{L}(Q)$. It is sufficient to prove the
  relation $R=\{((P\parallel Q)[f], P[f]\parallel Q[f])\}\cup \textbf{Id}$, if $f\upharpoonright(L\cup\overline{L})$ is one-to-one, where $L=\mathcal{L}(P)\cup\mathcal{L}(Q)$, is a FR strong hp-bisimulation, we omit it.
\end{enumerate}
\end{proof}

\begin{proposition}[Static laws for FR strong hhp-bisimulation]
The static laws for FR strong hhp-bisimulation are as follows.
\begin{enumerate}
  \item $P\parallel Q\sim_{hhp}^{fr} Q\parallel P$;
  \item $P\parallel(Q\parallel R)\sim_{hhp}^{fr} (P\parallel Q)\parallel R$;
  \item $P\parallel \textbf{nil}\sim_{hhp}^{fr} P$;
  \item $P\setminus L\sim_{hhp}^{fr} P$, if $\mathcal{L}(P)\cap(L\cup\overline{L})=\emptyset$;
  \item $P\setminus K\setminus L\sim_{hhp}^{fr} P\setminus(K\cup L)$;
  \item $P[f]\setminus L\sim_{hhp}^{fr} P\setminus f^{-1}(L)[f]$;
  \item $(P\parallel Q)\setminus L\sim_{hhp}^{fr} P\setminus L\parallel Q\setminus L$, if $\mathcal{L}(P)\cap\overline{\mathcal{L}(Q)}\cap(L\cup\overline{L})=\emptyset$;
  \item $P[Id]\sim_{hhp}^{fr} P$;
  \item $P[f]\sim_{hhp}^{fr} P[f']$, if $f\upharpoonright\mathcal{L}(P)=f'\upharpoonright\mathcal{L}(P)$;
  \item $P[f][f']\sim_{hhp}^{fr} P[f'\circ f]$;
  \item $(P\parallel Q)[f]\sim_{hhp}^{fr} P[f]\parallel Q[f]$, if $f\upharpoonright(L\cup\overline{L})$ is one-to-one, where $L=\mathcal{L}(P)\cup\mathcal{L}(Q)$.
\end{enumerate}
\end{proposition}

\begin{proof}
\begin{enumerate}
  \item $P\parallel Q\sim_{hhp}^{fr} Q\parallel P$. It is sufficient to prove the relation $R=\{(P\parallel Q, Q\parallel P)\}\cup \textbf{Id}$ is a FR strong hhp-bisimulation, we omit it;
  \item $P\parallel(Q\parallel R)\sim_{hhp}^{fr} (P\parallel Q)\parallel R$. It is sufficient to prove the relation $R=\{(P\parallel(Q\parallel R), (P\parallel Q)\parallel R)\}\cup \textbf{Id}$ is a FR strong hhp-bisimulation, we omit it;
  \item $P\parallel \textbf{nil}\sim_{hhp}^{fr} P$. It is sufficient to prove the relation $R=\{(P\parallel \textbf{nil}, P)\}\cup \textbf{Id}$ is a FR strong hhp-bisimulation, we omit it;
  \item $P\setminus L\sim_{hhp}^{fr} P$, if $\mathcal{L}(P)\cap(L\cup\overline{L})=\emptyset$. It is sufficient to prove the relation $R=\{(P\setminus L, P)\}\cup \textbf{Id}$, if $\mathcal{L}(P)\cap(L\cup\overline{L})=\emptyset$, is a FR strong hhp-bisimulation, we omit it;
  \item $P\setminus K\setminus L\sim_{hhp}^{fr} P\setminus(K\cup L)$. It is sufficient to prove the relation $R=\{(P\setminus K\setminus L, P\setminus(K\cup L))\}\cup \textbf{Id}$ is a FR strong hhp-bisimulation, we omit it;
  \item $P[f]\setminus L\sim_{hhp}^{fr} P\setminus f^{-1}(L)[f]$. It is sufficient to prove the relation $R=\{(P[f]\setminus L, P\setminus f^{-1}(L)[f])\}\cup \textbf{Id}$ is a FR strong hhp-bisimulation, we omit it;
  \item $(P\parallel Q)\setminus L\sim_{hhp}^{fr} P\setminus L\parallel Q\setminus L$, if $\mathcal{L}(P)\cap\overline{\mathcal{L}(Q)}\cap(L\cup\overline{L})=\emptyset$. It is sufficient to prove the relation
  $R=\{((P\parallel Q)\setminus L, P\setminus L\parallel Q\setminus L)\}\cup \textbf{Id}$, if $\mathcal{L}(P)\cap\overline{\mathcal{L}(Q)}\cap(L\cup\overline{L})=\emptyset$, is a FR strong hhp-bisimulation, we omit it;
  \item $P[Id]\sim_{hhp}^{fr} P$. It is sufficient to prove the relation $R=\{(P[Id], P)\}\cup \textbf{Id}$ is a FR strong hhp-bisimulation, we omit it;
  \item $P[f]\sim_{hhp}^{fr} P[f']$, if $f\upharpoonright\mathcal{L}(P)=f'\upharpoonright\mathcal{L}(P)$. It is sufficient to prove the relation $R=\{(P[f], P[f'])\}\cup \textbf{Id}$, if $f\upharpoonright\mathcal{L}(P)=f'\upharpoonright\mathcal{L}(P)$, is a FR strong hhp-bisimulation, we omit it;
  \item $P[f][f']\sim_{hhp}^{fr} P[f'\circ f]$. It is sufficient to prove the relation $R=\{(P[f][f'], P[f'\circ f])\}\cup \textbf{Id}$ is a FR strong hhp-bisimulation, we omit it;
  \item $(P\parallel Q)[f]\sim_{hhp}^{fr} P[f]\parallel Q[f]$, if $f\upharpoonright(L\cup\overline{L})$ is one-to-one, where $L=\mathcal{L}(P)\cup\mathcal{L}(Q)$. It is sufficient to prove the
  relation $R=\{((P\parallel Q)[f], P[f]\parallel Q[f])\}\cup \textbf{Id}$, if $f\upharpoonright(L\cup\overline{L})$ is one-to-one, where $L=\mathcal{L}(P)\cup\mathcal{L}(Q)$, is a FR strong hhp-bisimulation, we omit it.
\end{enumerate}
\end{proof}

\begin{proposition}[Guards laws for FR strong pomset bisimulation] The guards laws for FR strong pomset bisimulation are as follows.

\begin{enumerate}
  \item $P+\delta \sim_p^{fr} P$;
  \item $\delta.P \sim_p^{fr} \delta$;
  \item $\epsilon.P \sim_p^{fr} P$;
  \item $P.\epsilon \sim_p^{fr} P$;
  \item $\phi.\neg\phi \sim_p^{fr} \delta$;
  \item $\phi+\neg\phi \sim_p^{fr} \epsilon$;
  \item $\phi.\delta \sim_p^{fr} \delta$;
  \item $\phi.(P+Q)\sim_p^{fr}\phi.P+\phi.Q\quad (Std(P),Std(Q))$;
  \item $(P+Q).\phi\sim_p^{fr} P.\phi+ Q.\phi\quad(NStd(P),NStd(Q))$;
  \item $\phi.(P.Q)\sim_p^{fr} \phi.P.Q\quad (Std(P),Std(Q))$;
  \item $(P.Q).\phi\sim_p^{fr} P.Q.\phi\quad(NStd(P),NStd(Q))$;
  \item $(\phi+\psi).P \sim_p^{fr} \phi.P + \psi.P\quad (Std(P))$;
  \item $P.(\phi+\psi) \sim_p^{fr} P.\phi + P.\psi\quad(NStd(P))$;
  \item $(\phi.\psi).P \sim_p^{fr} \phi.(\psi.P)\quad(Std(P))$;
  \item $P.(\phi.\psi) \sim_p^{fr}(P.\phi).\psi\quad(NStd(P))$;
  \item $\phi\sim_p^{fr}\epsilon$ if $\forall s\in S.test(\phi,s)$;
  \item $\phi_0\cdot\cdots\cdot\phi_n \sim_p^{fr} \delta$ if $\forall s\in S,\exists i\leq n.test(\neg\phi_i,s)$;
  \item $wp(\alpha,\phi).\alpha.\phi\sim_p^{fr} wp(\alpha,\phi).\alpha$;
  \item $\phi. \alpha[m]. wp(\alpha[m],\phi)\sim_p^{fr}\alpha[m].wp(\alpha[m],\phi)$;
  \item $\neg wp(\alpha,\phi).\alpha.\neg\phi\sim_p^{fr}\neg wp(\alpha,\phi).\alpha$;
  \item $\neg\phi .\alpha[m] .\neg wp(\alpha[m],\phi)\sim_p^{fr} \alpha[m]. \neg wp(\alpha[m],\phi)$; 
  \item $\delta\parallel P \sim_p^{fr} \delta$;
  \item $P\parallel \delta \sim_p^{fr} \delta$;
  \item $\epsilon\parallel P \sim_p^{fr} P$;
  \item $P\parallel \epsilon \sim_p^{fr} P$;
  \item $\phi.(P\parallel Q) \sim_p^{fr}\phi.P\parallel \phi.Q$;
  \item $\phi\parallel \delta \sim_p^{fr} \delta$;
  \item $\delta\parallel \phi \sim_p^{fr} \delta$;
  \item $\phi\parallel \epsilon \sim_p^{fr} \phi$;
  \item $\epsilon\parallel \phi \sim_p^{fr} \phi$;
  \item $\phi\parallel\neg\phi \sim_p^{fr} \delta$;
  \item $\phi_0\parallel\cdots\parallel\phi_n \sim_p^{fr} \delta$ if $\forall s_0,\cdots,s_n\in S,\exists i\leq n.test(\neg\phi_i,s_0\cup\cdots\cup s_n)$.
\end{enumerate}
\end{proposition}

\begin{proof}
\begin{enumerate}
  \item $P+\delta \sim_p^{fr} P$. It is sufficient to prove the relation $R=\{(P+\delta, P)\}\cup \textbf{Id}$ is a FR strong pomset bisimulation, and we omit it;
  \item $\delta.P \sim_p^{fr} \delta$. It is sufficient to prove the relation $R=\{(\delta.P, \delta)\}\cup \textbf{Id}$ is a FR strong pomset bisimulation, and we omit it;
  \item $\epsilon.P \sim_p^{fr} P$. It is sufficient to prove the relation $R=\{(\epsilon.P, P)\}\cup \textbf{Id}$ is a FR strong pomset bisimulation, and we omit it;
  \item $P.\epsilon \sim_p^{fr} P$. It is sufficient to prove the relation $R=\{(P.\epsilon, P)\}\cup \textbf{Id}$ is a FR strong pomset bisimulation, and we omit it;
  \item $\phi.\neg\phi \sim_p^{fr} \delta$. It is sufficient to prove the relation $R=\{(\phi.\neg\phi, \delta)\}\cup \textbf{Id}$ is a FR strong pomset bisimulation, and we omit it;
  \item $\phi+\neg\phi \sim_p^{fr} \epsilon$. It is sufficient to prove the relation $R=\{(\phi+\neg\phi, \epsilon)\}\cup \textbf{Id}$ is a FR strong pomset bisimulation, and we omit it;
  \item $\phi.\delta \sim_p^{fr} \delta$. It is sufficient to prove the relation $R=\{(\phi.\delta, \delta)\}\cup \textbf{Id}$ is a FR strong pomset bisimulation, and we omit it;
  \item $\phi.(P+Q)\sim_p^{fr}\phi.P+\phi.Q\quad (Std(P),Std(Q))$. It is sufficient to prove the relation $R=\{(\phi.(P+Q), \phi.P+\phi.Q)\}\cup \textbf{Id}\quad (Std(P),Std(Q))$ is a FR strong pomset bisimulation, and we omit it;
  \item $(P+Q).\phi\sim_p^{fr} P.\phi+ Q.\phi\quad(NStd(P),NStd(Q))$. It is sufficient to prove the relation $R=\{((P+Q).\phi, P.\phi+ Q.\phi)\}\cup \textbf{Id}\quad(NStd(P),NStd(Q))$ is a FR strong pomset bisimulation, and we omit it;
  \item $\phi.(P.Q)\sim_p^{fr} \phi.P.Q\quad (Std(P),Std(Q))$. It is sufficient to prove the relation $R=\{(\phi.(P.Q), \phi.P.Q)\}\cup \textbf{Id}\quad (Std(P),Std(Q))$ is a FR strong pomset bisimulation, and we omit it;
  \item $(P.Q).\phi\sim_p^{fr} P.Q.\phi\quad(NStd(P),NStd(Q))$. It is sufficient to prove the relation $R=\{((P.Q).\phi, P.Q.\phi)\}\cup \textbf{Id}\quad(NStd(P),NStd(Q))$ is a FR strong pomset bisimulation, and we omit it;
  \item $(\phi+\psi).P \sim_p^{fr} \phi.P + \psi.P\quad (Std(P))$. It is sufficient to prove the relation $R=\{((\phi+\psi).P, \phi.P + \psi.P)\}\cup \textbf{Id}\quad (Std(P))$, is a FR strong pomset bisimulation, and we omit it;
  \item $P.(\phi+\psi) \sim_p^{fr} P.\phi + P.\psi\quad(NStd(P))$. It is sufficient to prove the relation $R=\{(P.(\phi+\psi), P.\phi + P.\psi)\}\cup \textbf{Id}\quad(NStd(P))$, is a FR strong pomset bisimulation, and we omit it;
  \item $(\phi.\psi).P \sim_p^{fr} \phi.(\psi.P)\quad(Std(P))$. It is sufficient to prove the relation $R=\{((\phi.\psi).P, \phi.(\psi.P))\}\cup \textbf{Id}\quad(Std(P))$ is a FR strong pomset bisimulation, and we omit it;
  \item $P.(\phi.\psi) \sim_p^{fr}(P.\phi).\psi\quad(NStd(P))$. It is sufficient to prove the relation $R=\{(P.(\phi.\psi), (P.\phi).\psi)\}\cup \textbf{Id}\quad(NStd(P))$ is a FR strong pomset bisimulation, and we omit it;
  \item $\phi\sim_p^{fr}\epsilon$ if $\forall s\in S.test(\phi,s)$. It is sufficient to prove the relation $R=\{(\phi, \epsilon)\}\cup \textbf{Id}$, if $\forall s\in S.test(\phi,s)$, is a FR strong pomset bisimulation, and we omit it;
  \item $\phi_0\cdot\cdots\cdot\phi_n \sim_p^{fr} \delta$ if $\forall s\in S,\exists i\leq n.test(\neg\phi_i,s)$. It is sufficient to prove the relation $R=\{(\phi_0\cdot\cdots\cdot\phi_n, \delta)\}\cup \textbf{Id}$, if $\forall s\in S,\exists i\leq n.test(\neg\phi_i,s)$, is a FR strong pomset bisimulation, and we omit it;
  \item $wp(\alpha,\phi).\alpha.\phi\sim_p^{fr} wp(\alpha,\phi).\alpha$. It is sufficient to prove the relation $R=\{(wp(\alpha,\phi).\alpha.\phi, wp(\alpha,\phi).\alpha)\}\cup \textbf{Id}$ is a FR strong pomset bisimulation, and we omit it;
  \item $\phi. \alpha[m]. wp(\alpha[m],\phi)\sim_p^{fr}\alpha[m].wp(\alpha[m],\phi)$. It is sufficient to prove the relation $R=\{(\phi. \alpha[m]. wp(\alpha[m],\phi), \alpha[m].wp(\alpha[m],\phi))\}\cup \textbf{Id}$ is a FR strong pomset bisimulation, and we omit it;
  \item $\neg wp(\alpha,\phi).\alpha.\neg\phi\sim_p^{fr}\neg wp(\alpha,\phi).\alpha$. It is sufficient to prove the relation \\$R=\{(\neg wp(\alpha,\phi).\alpha.\neg\phi, \neg wp(\alpha,\phi).\alpha)\}\cup \textbf{Id}$, is a FR strong pomset bisimulation, and we omit it;
  \item $\neg\phi .\alpha[m] .\neg wp(\alpha[m],\phi)\sim_p^{fr} \alpha[m]. \neg wp(\alpha[m],\phi)$. It is sufficient to prove the relation $R=\{(\neg\phi .\alpha[m] .\neg wp(\alpha[m],\phi), \alpha[m]. \neg wp(\alpha[m],\phi))\}\cup \textbf{Id}$, is a FR strong pomset bisimulation, and we omit it;
  \item $\delta\parallel P \sim_p^{fr} \delta$. It is sufficient to prove the relation $R=\{(\delta\parallel P, \delta)\}\cup \textbf{Id}$ is a FR strong pomset bisimulation, and we omit it;
  \item $P\parallel \delta \sim_p^{fr} \delta$. It is sufficient to prove the relation $R=\{(P\parallel \delta, \delta)\}\cup \textbf{Id}$ is a FR strong pomset bisimulation, and we omit it;
  \item $\epsilon\parallel P \sim_p^{fr} P$. It is sufficient to prove the relation $R=\{(\epsilon\parallel P, P)\}\cup \textbf{Id}$ is a FR strong pomset bisimulation, and we omit it;
  \item $P\parallel \epsilon \sim_p^{fr} P$. It is sufficient to prove the relation $R=\{(P\parallel \epsilon, P)\}\cup \textbf{Id}$ is a FR strong pomset bisimulation, and we omit it;
  \item $\phi.(P\parallel Q) \sim_p^{fr}\phi.P\parallel \phi.Q$. It is sufficient to prove the relation $R=\{(\phi.(P\parallel Q), \phi.P\parallel \phi.Q)\}\cup \textbf{Id}$ is a FR strong pomset bisimulation, and we omit it;
  \item $\phi\parallel \delta \sim_p^{fr} \delta$. It is sufficient to prove the relation $R=\{(\phi\parallel \delta, \delta)\}\cup \textbf{Id}$ is a FR strong pomset bisimulation, and we omit it;
  \item $\delta\parallel \phi \sim_p^{fr} \delta$. It is sufficient to prove the relation $R=\{(\delta\parallel \phi, \delta)\}\cup \textbf{Id}$ is a FR strong pomset bisimulation, and we omit it;
  \item $\phi\parallel \epsilon \sim_p^{fr} \phi$. It is sufficient to prove the relation $R=\{(\phi\parallel \epsilon, \phi)\}\cup \textbf{Id}$ is a FR strong pomset bisimulation, and we omit it;
  \item $\epsilon\parallel \phi \sim_p^{fr} \phi$. It is sufficient to prove the relation $R=\{(\epsilon\parallel \phi, \phi)\}\cup \textbf{Id}$ is a FR strong pomset bisimulation, and we omit it;
  \item $\phi\parallel\neg\phi \sim_p^{fr} \delta$. It is sufficient to prove the relation $R=\{(\phi\parallel\neg\phi, \delta)\}\cup \textbf{Id}$ is a FR strong pomset bisimulation, and we omit it;
  \item $\phi_0\parallel\cdots\parallel\phi_n \sim_p^{fr} \delta$ if $\forall s_0,\cdots,s_n\in S,\exists i\leq n.test(\neg\phi_i,s_0\cup\cdots\cup s_n)$. It is sufficient to prove the relation $R=\{(\phi_0\parallel\cdots\parallel\phi_n, \delta)\}\cup \textbf{Id}$, if $\forall s_0,\cdots,s_n\in S,\exists i\leq n.test(\neg\phi_i,s_0\cup\cdots\cup s_n)$, is a FR strong pomset bisimulation, and we omit it.
\end{enumerate}
\end{proof}

\begin{proposition}[Guards laws for FR strong step bisimulation] The guards laws for FR strong step bisimulation are as follows.

\begin{enumerate}
  \item $P+\delta \sim_s^{fr} P$;
  \item $\delta.P \sim_s^{fr} \delta$;
  \item $\epsilon.P \sim_s^{fr} P$;
  \item $P.\epsilon \sim_s^{fr} P$;
  \item $\phi.\neg\phi \sim_s^{fr} \delta$;
  \item $\phi+\neg\phi \sim_s^{fr} \epsilon$;
  \item $\phi.\delta \sim_s^{fr} \delta$;
  \item $\phi.(P+Q)\sim_s^{fr}\phi.P+\phi.Q\quad (Std(P),Std(Q))$;
  \item $(P+Q).\phi\sim_s^{fr} P.\phi+ Q.\phi\quad(NStd(P),NStd(Q))$;
  \item $\phi.(P.Q)\sim_s^{fr} \phi.P.Q\quad (Std(P),Std(Q))$;
  \item $(P.Q).\phi\sim_s^{fr} P.Q.\phi\quad(NStd(P),NStd(Q))$;
  \item $(\phi+\psi).P \sim_s^{fr} \phi.P + \psi.P\quad (Std(P))$;
  \item $P.(\phi+\psi) \sim_s^{fr} P.\phi + P.\psi\quad(NStd(P))$;
  \item $(\phi.\psi).P \sim_s^{fr} \phi.(\psi.P)\quad(Std(P))$;
  \item $P.(\phi.\psi) \sim_s^{fr}(P.\phi).\psi\quad(NStd(P))$;
  \item $\phi\sim_s^{fr}\epsilon$ if $\forall s\in S.test(\phi,s)$;
  \item $\phi_0\cdot\cdots\cdot\phi_n \sim_s^{fr} \delta$ if $\forall s\in S,\exists i\leq n.test(\neg\phi_i,s)$;
  \item $wp(\alpha,\phi).\alpha.\phi\sim_s^{fr} wp(\alpha,\phi).\alpha$;
  \item $\phi. \alpha[m]. wp(\alpha[m],\phi)\sim_s^{fr}\alpha[m].wp(\alpha[m],\phi)$;
  \item $\neg wp(\alpha,\phi).\alpha.\neg\phi\sim_s^{fr}\neg wp(\alpha,\phi).\alpha$;
  \item $\neg\phi .\alpha[m] .\neg wp(\alpha[m],\phi)\sim_s^{fr} \alpha[m]. \neg wp(\alpha[m],\phi)$;
  \item $\delta\parallel P \sim_s^{fr} \delta$;
  \item $P\parallel \delta \sim_s^{fr} \delta$;
  \item $\epsilon\parallel P \sim_s^{fr} P$;
  \item $P\parallel \epsilon \sim_s^{fr} P$;
  \item $\phi.(P\parallel Q) \sim_s^{fr}\phi.P\parallel \phi.Q$;
  \item $\phi\parallel \delta \sim_s^{fr} \delta$;
  \item $\delta\parallel \phi \sim_s^{fr} \delta$;
  \item $\phi\parallel \epsilon \sim_s^{fr} \phi$;
  \item $\epsilon\parallel \phi \sim_s^{fr} \phi$;
  \item $\phi\parallel\neg\phi \sim_s^{fr} \delta$;
  \item $\phi_0\parallel\cdots\parallel\phi_n \sim_s^{fr} \delta$ if $\forall s_0,\cdots,s_n\in S,\exists i\leq n.test(\neg\phi_i,s_0\cup\cdots\cup s_n)$.
\end{enumerate}
\end{proposition}

\begin{proof}
\begin{enumerate}
  \item $P+\delta \sim_s^{fr} P$. It is sufficient to prove the relation $R=\{(P+\delta, P)\}\cup \textbf{Id}$ is a FR strong step bisimulation, and we omit it;
  \item $\delta.P \sim_s^{fr} \delta$. It is sufficient to prove the relation $R=\{(\delta.P, \delta)\}\cup \textbf{Id}$ is a FR strong step bisimulation, and we omit it;
  \item $\epsilon.P \sim_s^{fr} P$. It is sufficient to prove the relation $R=\{(\epsilon.P, P)\}\cup \textbf{Id}$ is a FR strong step bisimulation, and we omit it;
  \item $P.\epsilon \sim_s^{fr} P$. It is sufficient to prove the relation $R=\{(P.\epsilon, P)\}\cup \textbf{Id}$ is a FR strong step bisimulation, and we omit it;
  \item $\phi.\neg\phi \sim_s^{fr} \delta$. It is sufficient to prove the relation $R=\{(\phi.\neg\phi, \delta)\}\cup \textbf{Id}$ is a FR strong step bisimulation, and we omit it;
  \item $\phi+\neg\phi \sim_s^{fr} \epsilon$. It is sufficient to prove the relation $R=\{(\phi+\neg\phi, \epsilon)\}\cup \textbf{Id}$ is a FR strong step bisimulation, and we omit it;
  \item $\phi.\delta \sim_s^{fr} \delta$. It is sufficient to prove the relation $R=\{(\phi.\delta, \delta)\}\cup \textbf{Id}$ is a FR strong step bisimulation, and we omit it;
  \item $\phi.(P+Q)\sim_s^{fr}\phi.P+\phi.Q\quad (Std(P),Std(Q))$. It is sufficient to prove the relation $R=\{(\phi.(P+Q), \phi.P+\phi.Q)\}\cup \textbf{Id}\quad (Std(P),Std(Q))$ is a FR strong step bisimulation, and we omit it;
  \item $(P+Q).\phi\sim_s^{fr} P.\phi+ Q.\phi\quad(NStd(P),NStd(Q))$. It is sufficient to prove the relation $R=\{((P+Q).\phi, P.\phi+ Q.\phi)\}\cup \textbf{Id}\quad(NStd(P),NStd(Q))$ is a FR strong step bisimulation, and we omit it;
  \item $\phi.(P.Q)\sim_s^{fr} \phi.P.Q\quad (Std(P),Std(Q))$. It is sufficient to prove the relation $R=\{(\phi.(P.Q), \phi.P.Q)\}\cup \textbf{Id}\quad (Std(P),Std(Q))$ is a FR strong step bisimulation, and we omit it;
  \item $(P.Q).\phi\sim_s^{fr} P.Q.\phi\quad(NStd(P),NStd(Q))$. It is sufficient to prove the relation $R=\{((P.Q).\phi, P.Q.\phi)\}\cup \textbf{Id}\quad(NStd(P),NStd(Q))$ is a FR strong step bisimulation, and we omit it;
  \item $(\phi+\psi).P \sim_s^{fr} \phi.P + \psi.P\quad (Std(P))$. It is sufficient to prove the relation $R=\{((\phi+\psi).P, \phi.P + \psi.P)\}\cup \textbf{Id}\quad (Std(P))$, is a FR strong step bisimulation, and we omit it;
  \item $P.(\phi+\psi) \sim_s^{fr} P.\phi + P.\psi\quad(NStd(P))$. It is sufficient to prove the relation $R=\{(P.(\phi+\psi), P.\phi + P.\psi)\}\cup \textbf{Id}\quad(NStd(P))$, is a FR strong step bisimulation, and we omit it;
  \item $(\phi.\psi).P \sim_s^{fr} \phi.(\psi.P)\quad(Std(P))$. It is sufficient to prove the relation $R=\{((\phi.\psi).P, \phi.(\psi.P))\}\cup \textbf{Id}\quad(Std(P))$ is a FR strong step bisimulation, and we omit it;
  \item $P.(\phi.\psi) \sim_s^{fr}(P.\phi).\psi\quad(NStd(P))$. It is sufficient to prove the relation $R=\{(P.(\phi.\psi), (P.\phi).\psi)\}\cup \textbf{Id}\quad(NStd(P))$ is a FR strong step bisimulation, and we omit it;
  \item $\phi\sim_s^{fr}\epsilon$ if $\forall s\in S.test(\phi,s)$. It is sufficient to prove the relation $R=\{(\phi, \epsilon)\}\cup \textbf{Id}$, if $\forall s\in S.test(\phi,s)$, is a FR strong step bisimulation, and we omit it;
  \item $\phi_0\cdot\cdots\cdot\phi_n \sim_s^{fr} \delta$ if $\forall s\in S,\exists i\leq n.test(\neg\phi_i,s)$. It is sufficient to prove the relation $R=\{(\phi_0\cdot\cdots\cdot\phi_n, \delta)\}\cup \textbf{Id}$, if $\forall s\in S,\exists i\leq n.test(\neg\phi_i,s)$, is a FR strong step bisimulation, and we omit it;
  \item $wp(\alpha,\phi).\alpha.\phi\sim_s^{fr} wp(\alpha,\phi).\alpha$. It is sufficient to prove the relation $R=\{(wp(\alpha,\phi).\alpha.\phi, wp(\alpha,\phi).\alpha)\}\cup \textbf{Id}$ is a FR strong step bisimulation, and we omit it;
  \item $\phi. \alpha[m]. wp(\alpha[m],\phi)\sim_s^{fr}\alpha[m].wp(\alpha[m],\phi)$. It is sufficient to prove the relation $R=\{(\phi. \alpha[m]. wp(\alpha[m],\phi), \alpha[m].wp(\alpha[m],\phi))\}\cup \textbf{Id}$ is a FR strong step bisimulation, and we omit it;
  \item $\neg wp(\alpha,\phi).\alpha.\neg\phi\sim_s^{fr}\neg wp(\alpha,\phi).\alpha$. It is sufficient to prove the relation \\$R=\{(\neg wp(\alpha,\phi).\alpha.\neg\phi, \neg wp(\alpha,\phi).\alpha)\}\cup \textbf{Id}$, is a FR strong step bisimulation, and we omit it;
  \item $\neg\phi .\alpha[m] .\neg wp(\alpha[m],\phi)\sim_s^{fr} \alpha[m]. \neg wp(\alpha[m],\phi)$. It is sufficient to prove the relation $R=\{(\neg\phi .\alpha[m] .\neg wp(\alpha[m],\phi), \alpha[m]. \neg wp(\alpha[m],\phi))\}\cup \textbf{Id}$, is a FR strong step bisimulation, and we omit it;
  \item $\delta\parallel P \sim_s^{fr} \delta$. It is sufficient to prove the relation $R=\{(\delta\parallel P, \delta)\}\cup \textbf{Id}$ is a FR strong step bisimulation, and we omit it;
  \item $P\parallel \delta \sim_s^{fr} \delta$. It is sufficient to prove the relation $R=\{(P\parallel \delta, \delta)\}\cup \textbf{Id}$ is a FR strong step bisimulation, and we omit it;
  \item $\epsilon\parallel P \sim_s^{fr} P$. It is sufficient to prove the relation $R=\{(\epsilon\parallel P, P)\}\cup \textbf{Id}$ is a FR strong step bisimulation, and we omit it;
  \item $P\parallel \epsilon \sim_s^{fr} P$. It is sufficient to prove the relation $R=\{(P\parallel \epsilon, P)\}\cup \textbf{Id}$ is a FR strong step bisimulation, and we omit it;
  \item $\phi.(P\parallel Q) \sim_s^{fr}\phi.P\parallel \phi.Q$. It is sufficient to prove the relation $R=\{(\phi.(P\parallel Q), \phi.P\parallel \phi.Q)\}\cup \textbf{Id}$ is a FR strong step bisimulation, and we omit it;
  \item $\phi\parallel \delta \sim_s^{fr} \delta$. It is sufficient to prove the relation $R=\{(\phi\parallel \delta, \delta)\}\cup \textbf{Id}$ is a FR strong step bisimulation, and we omit it;
  \item $\delta\parallel \phi \sim_s^{fr} \delta$. It is sufficient to prove the relation $R=\{(\delta\parallel \phi, \delta)\}\cup \textbf{Id}$ is a FR strong step bisimulation, and we omit it;
  \item $\phi\parallel \epsilon \sim_s^{fr} \phi$. It is sufficient to prove the relation $R=\{(\phi\parallel \epsilon, \phi)\}\cup \textbf{Id}$ is a FR strong step bisimulation, and we omit it;
  \item $\epsilon\parallel \phi \sim_s^{fr} \phi$. It is sufficient to prove the relation $R=\{(\epsilon\parallel \phi, \phi)\}\cup \textbf{Id}$ is a FR strong step bisimulation, and we omit it;
  \item $\phi\parallel\neg\phi \sim_s^{fr} \delta$. It is sufficient to prove the relation $R=\{(\phi\parallel\neg\phi, \delta)\}\cup \textbf{Id}$ is a FR strong step bisimulation, and we omit it;
  \item $\phi_0\parallel\cdots\parallel\phi_n \sim_s^{fr} \delta$ if $\forall s_0,\cdots,s_n\in S,\exists i\leq n.test(\neg\phi_i,s_0\cup\cdots\cup s_n)$. It is sufficient to prove the relation $R=\{(\phi_0\parallel\cdots\parallel\phi_n, \delta)\}\cup \textbf{Id}$, if $\forall s_0,\cdots,s_n\in S,\exists i\leq n.test(\neg\phi_i,s_0\cup\cdots\cup s_n)$, is a FR strong step bisimulation, and we omit it.
\end{enumerate}
\end{proof}

\begin{proposition}[Guards laws for FR strong hp-bisimulation] The guards laws for FR strong hp-bisimulation are as follows.

\begin{enumerate}
  \item $P+\delta \sim_{hp}^{fr} P$;
  \item $\delta.P \sim_{hp}^{fr} \delta$;
  \item $\epsilon.P \sim_{hp}^{fr} P$;
  \item $P.\epsilon \sim_{hp}^{fr} P$;
  \item $\phi.\neg\phi \sim_{hp}^{fr} \delta$;
  \item $\phi+\neg\phi \sim_{hp}^{fr} \epsilon$;
  \item $\phi.\delta \sim_{hp}^{fr} \delta$;
  \item $\phi.(P+Q)\sim_{hp}^{fr}\phi.P+\phi.Q\quad (Std(P),Std(Q))$;
  \item $(P+Q).\phi\sim_{hp}^{fr} P.\phi+ Q.\phi\quad(NStd(P),NStd(Q))$;
  \item $\phi.(P.Q)\sim_{hp}^{fr} \phi.P.Q\quad (Std(P),Std(Q))$;
  \item $(P.Q).\phi\sim_{hp}^{fr} P.Q.\phi\quad(NStd(P),NStd(Q))$;
  \item $(\phi+\psi).P \sim_{hp}^{fr} \phi.P + \psi.P\quad (Std(P))$;
  \item $P.(\phi+\psi) \sim_{hp}^{fr} P.\phi + P.\psi\quad(NStd(P))$;
  \item $(\phi.\psi).P \sim_{hp}^{fr} \phi.(\psi.P)\quad(Std(P))$;
  \item $P.(\phi.\psi) \sim_{hp}^{fr}(P.\phi).\psi\quad(NStd(P))$;
  \item $\phi\sim_{hp}^{fr}\epsilon$ if $\forall s\in S.test(\phi,s)$;
  \item $\phi_0\cdot\cdots\cdot\phi_n \sim_{hp}^{fr} \delta$ if $\forall s\in S,\exists i\leq n.test(\neg\phi_i,s)$;
  \item $wp(\alpha,\phi).\alpha.\phi\sim_{hp}^{fr} wp(\alpha,\phi).\alpha$;
  \item $\phi. \alpha[m]. wp(\alpha[m],\phi)\sim_{hp}^{fr}\alpha[m].wp(\alpha[m],\phi)$;
  \item $\neg wp(\alpha,\phi).\alpha.\neg\phi\sim_{hp}^{fr}\neg wp(\alpha,\phi).\alpha$;
  \item $\neg\phi .\alpha[m] .\neg wp(\alpha[m],\phi)\sim_{hp}^{fr} \alpha[m]. \neg wp(\alpha[m],\phi)$;
  \item $\delta\parallel P \sim_{hp}^{fr} \delta$;
  \item $P\parallel \delta \sim_{hp}^{fr} \delta$;
  \item $\epsilon\parallel P \sim_{hp}^{fr} P$;
  \item $P\parallel \epsilon \sim_{hp}^{fr} P$;
  \item $\phi.(P\parallel Q) \sim_{hp}^{fr}\phi.P\parallel \phi.Q$;
  \item $\phi\parallel \delta \sim_{hp}^{fr} \delta$;
  \item $\delta\parallel \phi \sim_{hp}^{fr} \delta$;
  \item $\phi\parallel \epsilon \sim_{hp}^{fr} \phi$;
  \item $\epsilon\parallel \phi \sim_{hp}^{fr} \phi$;
  \item $\phi\parallel\neg\phi \sim_{hp}^{fr} \delta$;
  \item $\phi_0\parallel\cdots\parallel\phi_n \sim_{hp}^{fr} \delta$ if $\forall s_0,\cdots,s_n\in S,\exists i\leq n.test(\neg\phi_i,s_0\cup\cdots\cup s_n)$.
\end{enumerate}
\end{proposition}

\begin{proof}
\begin{enumerate}
  \item $P+\delta \sim_{hp}^{fr} P$. It is sufficient to prove the relation $R=\{(P+\delta, P)\}\cup \textbf{Id}$ is a FR strong hp-bisimulation, and we omit it;
  \item $\delta.P \sim_{hp}^{fr} \delta$. It is sufficient to prove the relation $R=\{(\delta.P, \delta)\}\cup \textbf{Id}$ is a FR strong hp-bisimulation, and we omit it;
  \item $\epsilon.P \sim_{hp}^{fr} P$. It is sufficient to prove the relation $R=\{(\epsilon.P, P)\}\cup \textbf{Id}$ is a FR strong hp-bisimulation, and we omit it;
  \item $P.\epsilon \sim_{hp}^{fr} P$. It is sufficient to prove the relation $R=\{(P.\epsilon, P)\}\cup \textbf{Id}$ is a FR strong hp-bisimulation, and we omit it;
  \item $\phi.\neg\phi \sim_{hp}^{fr} \delta$. It is sufficient to prove the relation $R=\{(\phi.\neg\phi, \delta)\}\cup \textbf{Id}$ is a FR strong hp-bisimulation, and we omit it;
  \item $\phi+\neg\phi \sim_{hp}^{fr} \epsilon$. It is sufficient to prove the relation $R=\{(\phi+\neg\phi, \epsilon)\}\cup \textbf{Id}$ is a FR strong hp-bisimulation, and we omit it;
  \item $\phi.\delta \sim_{hp}^{fr} \delta$. It is sufficient to prove the relation $R=\{(\phi.\delta, \delta)\}\cup \textbf{Id}$ is a FR strong hp-bisimulation, and we omit it;
  \item $\phi.(P+Q)\sim_{hp}^{fr}\phi.P+\phi.Q\quad (Std(P),Std(Q))$. It is sufficient to prove the relation $R=\{(\phi.(P+Q), \phi.P+\phi.Q)\}\cup \textbf{Id}\quad (Std(P),Std(Q))$ is a FR strong hp-bisimulation, and we omit it;
  \item $(P+Q).\phi\sim_{hp}^{fr} P.\phi+ Q.\phi\quad(NStd(P),NStd(Q))$. It is sufficient to prove the relation $R=\{((P+Q).\phi, P.\phi+ Q.\phi)\}\cup \textbf{Id}\quad(NStd(P),NStd(Q))$ is a FR strong hp-bisimulation, and we omit it;
  \item $\phi.(P.Q)\sim_{hp}^{fr} \phi.P.Q\quad (Std(P),Std(Q))$. It is sufficient to prove the relation $R=\{(\phi.(P.Q), \phi.P.Q)\}\cup \textbf{Id}\quad (Std(P),Std(Q))$ is a FR strong hp-bisimulation, and we omit it;
  \item $(P.Q).\phi\sim_{hp}^{fr} P.Q.\phi\quad(NStd(P),NStd(Q))$. It is sufficient to prove the relation $R=\{((P.Q).\phi, P.Q.\phi)\}\cup \textbf{Id}\quad(NStd(P),NStd(Q))$ is a FR strong hp-bisimulation, and we omit it;
  \item $(\phi+\psi).P \sim_{hp}^{fr} \phi.P + \psi.P\quad (Std(P))$. It is sufficient to prove the relation $R=\{((\phi+\psi).P, \phi.P + \psi.P)\}\cup \textbf{Id}\quad (Std(P))$, is a FR strong hp-bisimulation, and we omit it;
  \item $P.(\phi+\psi) \sim_{hp}^{fr} P.\phi + P.\psi\quad(NStd(P))$. It is sufficient to prove the relation $R=\{(P.(\phi+\psi), P.\phi + P.\psi)\}\cup \textbf{Id}\quad(NStd(P))$, is a FR strong hp-bisimulation, and we omit it;
  \item $(\phi.\psi).P \sim_{hp}^{fr} \phi.(\psi.P)\quad(Std(P))$. It is sufficient to prove the relation $R=\{((\phi.\psi).P, \phi.(\psi.P))\}\cup \textbf{Id}\quad(Std(P))$ is a FR strong hp-bisimulation, and we omit it;
  \item $P.(\phi.\psi) \sim_{hp}^{fr}(P.\phi).\psi\quad(NStd(P))$. It is sufficient to prove the relation $R=\{(P.(\phi.\psi), (P.\phi).\psi)\}\cup \textbf{Id}\quad(NStd(P))$ is a FR strong hp-bisimulation, and we omit it;
  \item $\phi\sim_{hp}^{fr}\epsilon$ if $\forall s\in S.test(\phi,s)$. It is sufficient to prove the relation $R=\{(\phi, \epsilon)\}\cup \textbf{Id}$, if $\forall s\in S.test(\phi,s)$, is a FR strong hp-bisimulation, and we omit it;
  \item $\phi_0\cdot\cdots\cdot\phi_n \sim_{hp}^{fr} \delta$ if $\forall s\in S,\exists i\leq n.test(\neg\phi_i,s)$. It is sufficient to prove the relation $R=\{(\phi_0\cdot\cdots\cdot\phi_n, \delta)\}\cup \textbf{Id}$, if $\forall s\in S,\exists i\leq n.test(\neg\phi_i,s)$, is a FR strong hp-bisimulation, and we omit it;
  \item $wp(\alpha,\phi).\alpha.\phi\sim_{hp}^{fr} wp(\alpha,\phi).\alpha$. It is sufficient to prove the relation $R=\{(wp(\alpha,\phi).\alpha.\phi, wp(\alpha,\phi).\alpha)\}\cup \textbf{Id}$ is a FR strong hp-bisimulation, and we omit it;
  \item $\phi. \alpha[m]. wp(\alpha[m],\phi)\sim_{hp}^{fr}\alpha[m].wp(\alpha[m],\phi)$. It is sufficient to prove the relation $R=\{(\phi. \alpha[m]. wp(\alpha[m],\phi), \alpha[m].wp(\alpha[m],\phi))\}\cup \textbf{Id}$ is a FR strong hp-bisimulation, and we omit it;
  \item $\neg wp(\alpha,\phi).\alpha.\neg\phi\sim_{hp}^{fr}\neg wp(\alpha,\phi).\alpha$. It is sufficient to prove the relation \\$R=\{(\neg wp(\alpha,\phi).\alpha.\neg\phi, \neg wp(\alpha,\phi).\alpha)\}\cup \textbf{Id}$, is a FR strong hp-bisimulation, and we omit it;
  \item $\neg\phi .\alpha[m] .\neg wp(\alpha[m],\phi)\sim_{hp}^{fr} \alpha[m]. \neg wp(\alpha[m],\phi)$. It is sufficient to prove the relation $R=\{(\neg\phi .\alpha[m] .\neg wp(\alpha[m],\phi), \alpha[m]. \neg wp(\alpha[m],\phi))\}\cup \textbf{Id}$, is a FR strong hp-bisimulation, and we omit it;
  \item $\delta\parallel P \sim_{hp}^{fr} \delta$. It is sufficient to prove the relation $R=\{(\delta\parallel P, \delta)\}\cup \textbf{Id}$ is a FR strong hp-bisimulation, and we omit it;
  \item $P\parallel \delta \sim_{hp}^{fr} \delta$. It is sufficient to prove the relation $R=\{(P\parallel \delta, \delta)\}\cup \textbf{Id}$ is a FR strong hp-bisimulation, and we omit it;
  \item $\epsilon\parallel P \sim_{hp}^{fr} P$. It is sufficient to prove the relation $R=\{(\epsilon\parallel P, P)\}\cup \textbf{Id}$ is a FR strong hp-bisimulation, and we omit it;
  \item $P\parallel \epsilon \sim_{hp}^{fr} P$. It is sufficient to prove the relation $R=\{(P\parallel \epsilon, P)\}\cup \textbf{Id}$ is a FR strong hp-bisimulation, and we omit it;
  \item $\phi.(P\parallel Q) \sim_{hp}^{fr}\phi.P\parallel \phi.Q$. It is sufficient to prove the relation $R=\{(\phi.(P\parallel Q), \phi.P\parallel \phi.Q)\}\cup \textbf{Id}$ is a FR strong hp-bisimulation, and we omit it;
  \item $\phi\parallel \delta \sim_{hp}^{fr} \delta$. It is sufficient to prove the relation $R=\{(\phi\parallel \delta, \delta)\}\cup \textbf{Id}$ is a FR strong hp-bisimulation, and we omit it;
  \item $\delta\parallel \phi \sim_{hp}^{fr} \delta$. It is sufficient to prove the relation $R=\{(\delta\parallel \phi, \delta)\}\cup \textbf{Id}$ is a FR strong hp-bisimulation, and we omit it;
  \item $\phi\parallel \epsilon \sim_{hp}^{fr} \phi$. It is sufficient to prove the relation $R=\{(\phi\parallel \epsilon, \phi)\}\cup \textbf{Id}$ is a FR strong hp-bisimulation, and we omit it;
  \item $\epsilon\parallel \phi \sim_{hp}^{fr} \phi$. It is sufficient to prove the relation $R=\{(\epsilon\parallel \phi, \phi)\}\cup \textbf{Id}$ is a FR strong hp-bisimulation, and we omit it;
  \item $\phi\parallel\neg\phi \sim_{hp}^{fr} \delta$. It is sufficient to prove the relation $R=\{(\phi\parallel\neg\phi, \delta)\}\cup \textbf{Id}$ is a FR strong hp-bisimulation, and we omit it;
  \item $\phi_0\parallel\cdots\parallel\phi_n \sim_{hp}^{fr} \delta$ if $\forall s_0,\cdots,s_n\in S,\exists i\leq n.test(\neg\phi_i,s_0\cup\cdots\cup s_n)$. It is sufficient to prove the relation $R=\{(\phi_0\parallel\cdots\parallel\phi_n, \delta)\}\cup \textbf{Id}$, if $\forall s_0,\cdots,s_n\in S,\exists i\leq n.test(\neg\phi_i,s_0\cup\cdots\cup s_n)$, is a FR strong hp-bisimulation, and we omit it.
\end{enumerate}
\end{proof}

\begin{proposition}[Guards laws for FR strong hhp-bisimulation] The guards laws for FR strong hhp-bisimulation are as follows.

\begin{enumerate}
  \item $P+\delta \sim_{hhp}^{fr} P$;
  \item $\delta.P \sim_{hhp}^{fr} \delta$;
  \item $\epsilon.P \sim_{hhp}^{fr} P$;
  \item $P.\epsilon \sim_{hhp}^{fr} P$;
  \item $\phi.\neg\phi \sim_{hhp}^{fr} \delta$;
  \item $\phi+\neg\phi \sim_{hhp}^{fr} \epsilon$;
  \item $\phi.\delta \sim_{hhp}^{fr} \delta$;
  \item $\phi.(P+Q)\sim_{hhp}^{fr}\phi.P+\phi.Q\quad (Std(P),Std(Q))$;
  \item $(P+Q).\phi\sim_{hhp}^{fr} P.\phi+ Q.\phi\quad(NStd(P),NStd(Q))$;
  \item $\phi.(P.Q)\sim_{hhp}^{fr} \phi.P.Q\quad (Std(P),Std(Q))$;
  \item $(P.Q).\phi\sim_{hhp}^{fr} P.Q.\phi\quad(NStd(P),NStd(Q))$;
  \item $(\phi+\psi).P \sim_{hhp}^{fr} \phi.P + \psi.P\quad (Std(P))$;
  \item $P.(\phi+\psi) \sim_{hhp}^{fr} P.\phi + P.\psi\quad(NStd(P))$;
  \item $(\phi.\psi).P \sim_{hhp}^{fr} \phi.(\psi.P)\quad(Std(P))$;
  \item $P.(\phi.\psi) \sim_{hhp}^{fr}(P.\phi).\psi\quad(NStd(P))$;
  \item $\phi\sim_{hhp}^{fr}\epsilon$ if $\forall s\in S.test(\phi,s)$;
  \item $\phi_0\cdot\cdots\cdot\phi_n \sim_{hhp}^{fr} \delta$ if $\forall s\in S,\exists i\leq n.test(\neg\phi_i,s)$;
  \item $wp(\alpha,\phi).\alpha.\phi\sim_{hhp}^{fr} wp(\alpha,\phi).\alpha$;
  \item $\phi. \alpha[m]. wp(\alpha[m],\phi)\sim_{hhp}^{fr}\alpha[m].wp(\alpha[m],\phi)$;
  \item $\neg wp(\alpha,\phi).\alpha.\neg\phi\sim_{hhp}^{fr}\neg wp(\alpha,\phi).\alpha$;
  \item $\neg\phi .\alpha[m] .\neg wp(\alpha[m],\phi)\sim_{hhp}^{fr} \alpha[m]. \neg wp(\alpha[m],\phi)$;
  \item $\delta\parallel P \sim_{hhp}^{fr} \delta$;
  \item $P\parallel \delta \sim_{hhp}^{fr} \delta$;
  \item $\epsilon\parallel P \sim_{hhp}^{fr} P$;
  \item $P\parallel \epsilon \sim_{hhp}^{fr} P$;
  \item $\phi.(P\parallel Q) \sim_{hhp}^{fr}\phi.P\parallel \phi.Q$;
  \item $\phi\parallel \delta \sim_{hhp}^{fr} \delta$;
  \item $\delta\parallel \phi \sim_{hhp}^{fr} \delta$;
  \item $\phi\parallel \epsilon \sim_{hhp}^{fr} \phi$;
  \item $\epsilon\parallel \phi \sim_{hhp}^{fr} \phi$;
  \item $\phi\parallel\neg\phi \sim_{hhp}^{fr} \delta$;
  \item $\phi_0\parallel\cdots\parallel\phi_n \sim_{hhp}^{fr} \delta$ if $\forall s_0,\cdots,s_n\in S,\exists i\leq n.test(\neg\phi_i,s_0\cup\cdots\cup s_n)$.
\end{enumerate}
\end{proposition}

\begin{proof}
\begin{enumerate}
  \item $P+\delta \sim_{hhp}^{fr} P$. It is sufficient to prove the relation $R=\{(P+\delta, P)\}\cup \textbf{Id}$ is a FR strong hhp-bisimulation, and we omit it;
  \item $\delta.P \sim_{hhp}^{fr} \delta$. It is sufficient to prove the relation $R=\{(\delta.P, \delta)\}\cup \textbf{Id}$ is a FR strong hhp-bisimulation, and we omit it;
  \item $\epsilon.P \sim_{hhp}^{fr} P$. It is sufficient to prove the relation $R=\{(\epsilon.P, P)\}\cup \textbf{Id}$ is a FR strong hhp-bisimulation, and we omit it;
  \item $P.\epsilon \sim_{hhp}^{fr} P$. It is sufficient to prove the relation $R=\{(P.\epsilon, P)\}\cup \textbf{Id}$ is a FR strong hhp-bisimulation, and we omit it;
  \item $\phi.\neg\phi \sim_{hhp}^{fr} \delta$. It is sufficient to prove the relation $R=\{(\phi.\neg\phi, \delta)\}\cup \textbf{Id}$ is a FR strong hhp-bisimulation, and we omit it;
  \item $\phi+\neg\phi \sim_{hhp}^{fr} \epsilon$. It is sufficient to prove the relation $R=\{(\phi+\neg\phi, \epsilon)\}\cup \textbf{Id}$ is a FR strong hhp-bisimulation, and we omit it;
  \item $\phi.\delta \sim_{hhp}^{fr} \delta$. It is sufficient to prove the relation $R=\{(\phi.\delta, \delta)\}\cup \textbf{Id}$ is a FR strong hhp-bisimulation, and we omit it;
  \item $\phi.(P+Q)\sim_{hhp}^{fr}\phi.P+\phi.Q\quad (Std(P),Std(Q))$. It is sufficient to prove the relation $R=\{(\phi.(P+Q), \phi.P+\phi.Q)\}\cup \textbf{Id}\quad (Std(P),Std(Q))$ is a FR strong hhp-bisimulation, and we omit it;
  \item $(P+Q).\phi\sim_{hhp}^{fr} P.\phi+ Q.\phi\quad(NStd(P),NStd(Q))$. It is sufficient to prove the relation $R=\{((P+Q).\phi, P.\phi+ Q.\phi)\}\cup \textbf{Id}\quad(NStd(P),NStd(Q))$ is a FR strong hhp-bisimulation, and we omit it;
  \item $\phi.(P.Q)\sim_{hhp}^{fr} \phi.P.Q\quad (Std(P),Std(Q))$. It is sufficient to prove the relation $R=\{(\phi.(P.Q), \phi.P.Q)\}\cup \textbf{Id}\quad (Std(P),Std(Q))$ is a FR strong hhp-bisimulation, and we omit it;
  \item $(P.Q).\phi\sim_{hhp}^{fr} P.Q.\phi\quad(NStd(P),NStd(Q))$. It is sufficient to prove the relation $R=\{((P.Q).\phi, P.Q.\phi)\}\cup \textbf{Id}\quad(NStd(P),NStd(Q))$ is a FR strong hhp-bisimulation, and we omit it;
  \item $(\phi+\psi).P \sim_{hhp}^{fr} \phi.P + \psi.P\quad (Std(P))$. It is sufficient to prove the relation $R=\{((\phi+\psi).P, \phi.P + \psi.P)\}\cup \textbf{Id}\quad (Std(P))$, is a FR strong hhp-bisimulation, and we omit it;
  \item $P.(\phi+\psi) \sim_{hhp}^{fr} P.\phi + P.\psi\quad(NStd(P))$. It is sufficient to prove the relation $R=\{(P.(\phi+\psi), P.\phi + P.\psi)\}\cup \textbf{Id}\quad(NStd(P))$, is a FR strong hhp-bisimulation, and we omit it;
  \item $(\phi.\psi).P \sim_{hhp}^{fr} \phi.(\psi.P)\quad(Std(P))$. It is sufficient to prove the relation $R=\{((\phi.\psi).P, \phi.(\psi.P))\}\cup \textbf{Id}\quad(Std(P))$ is a FR strong hhp-bisimulation, and we omit it;
  \item $P.(\phi.\psi) \sim_{hhp}^{fr}(P.\phi).\psi\quad(NStd(P))$. It is sufficient to prove the relation $R=\{(P.(\phi.\psi), (P.\phi).\psi)\}\cup \textbf{Id}\quad(NStd(P))$ is a FR strong hhp-bisimulation, and we omit it;
  \item $\phi\sim_{hhp}^{fr}\epsilon$ if $\forall s\in S.test(\phi,s)$. It is sufficient to prove the relation $R=\{(\phi, \epsilon)\}\cup \textbf{Id}$, if $\forall s\in S.test(\phi,s)$, is a FR strong hhp-bisimulation, and we omit it;
  \item $\phi_0\cdot\cdots\cdot\phi_n \sim_{hhp}^{fr} \delta$ if $\forall s\in S,\exists i\leq n.test(\neg\phi_i,s)$. It is sufficient to prove the relation $R=\{(\phi_0\cdot\cdots\cdot\phi_n, \delta)\}\cup \textbf{Id}$, if $\forall s\in S,\exists i\leq n.test(\neg\phi_i,s)$, is a FR strong hhp-bisimulation, and we omit it;
  \item $wp(\alpha,\phi).\alpha.\phi\sim_{hhp}^{fr} wp(\alpha,\phi).\alpha$. It is sufficient to prove the relation $R=\{(wp(\alpha,\phi).\alpha.\phi, wp(\alpha,\phi).\alpha)\}\cup \textbf{Id}$ is a FR strong hhp-bisimulation, and we omit it;
  \item $\phi. \alpha[m]. wp(\alpha[m],\phi)\sim_{hhp}^{fr}\alpha[m].wp(\alpha[m],\phi)$. It is sufficient to prove the relation $R=\{(\phi. \alpha[m]. wp(\alpha[m],\phi), \alpha[m].wp(\alpha[m],\phi))\}\cup \textbf{Id}$ is a FR strong hhp-bisimulation, and we omit it;
  \item $\neg wp(\alpha,\phi).\alpha.\neg\phi\sim_{hhp}^{fr}\neg wp(\alpha,\phi).\alpha$. It is sufficient to prove the relation \\$R=\{(\neg wp(\alpha,\phi).\alpha.\neg\phi, \neg wp(\alpha,\phi).\alpha)\}\cup \textbf{Id}$, is a FR strong hhp-bisimulation, and we omit it;
  \item $\neg\phi .\alpha[m] .\neg wp(\alpha[m],\phi)\sim_{hhp}^{fr} \alpha[m]. \neg wp(\alpha[m],\phi)$. It is sufficient to prove the relation $R=\{(\neg\phi .\alpha[m] .\neg wp(\alpha[m],\phi), \alpha[m]. \neg wp(\alpha[m],\phi))\}\cup \textbf{Id}$, is a FR strong hhp-bisimulation, and we omit it;
  \item $\delta\parallel P \sim_{hhp}^{fr} \delta$. It is sufficient to prove the relation $R=\{(\delta\parallel P, \delta)\}\cup \textbf{Id}$ is a FR strong hhp-bisimulation, and we omit it;
  \item $P\parallel \delta \sim_{hhp}^{fr} \delta$. It is sufficient to prove the relation $R=\{(P\parallel \delta, \delta)\}\cup \textbf{Id}$ is a FR strong hhp-bisimulation, and we omit it;
  \item $\epsilon\parallel P \sim_{hhp}^{fr} P$. It is sufficient to prove the relation $R=\{(\epsilon\parallel P, P)\}\cup \textbf{Id}$ is a FR strong hhp-bisimulation, and we omit it;
  \item $P\parallel \epsilon \sim_{hhp}^{fr} P$. It is sufficient to prove the relation $R=\{(P\parallel \epsilon, P)\}\cup \textbf{Id}$ is a FR strong hhp-bisimulation, and we omit it;
  \item $\phi.(P\parallel Q) \sim_{hhp}^{fr}\phi.P\parallel \phi.Q$. It is sufficient to prove the relation $R=\{(\phi.(P\parallel Q), \phi.P\parallel \phi.Q)\}\cup \textbf{Id}$ is a FR strong hhp-bisimulation, and we omit it;
  \item $\phi\parallel \delta \sim_{hhp}^{fr} \delta$. It is sufficient to prove the relation $R=\{(\phi\parallel \delta, \delta)\}\cup \textbf{Id}$ is a FR strong hhp-bisimulation, and we omit it;
  \item $\delta\parallel \phi \sim_{hhp}^{fr} \delta$. It is sufficient to prove the relation $R=\{(\delta\parallel \phi, \delta)\}\cup \textbf{Id}$ is a FR strong hhp-bisimulation, and we omit it;
  \item $\phi\parallel \epsilon \sim_{hhp}^{fr} \phi$. It is sufficient to prove the relation $R=\{(\phi\parallel \epsilon, \phi)\}\cup \textbf{Id}$ is a FR strong hhp-bisimulation, and we omit it;
  \item $\epsilon\parallel \phi \sim_{hhp}^{fr} \phi$. It is sufficient to prove the relation $R=\{(\epsilon\parallel \phi, \phi)\}\cup \textbf{Id}$ is a FR strong hhp-bisimulation, and we omit it;
  \item $\phi\parallel\neg\phi \sim_{hhp}^{fr} \delta$. It is sufficient to prove the relation $R=\{(\phi\parallel\neg\phi, \delta)\}\cup \textbf{Id}$ is a FR strong hhp-bisimulation, and we omit it;
  \item $\phi_0\parallel\cdots\parallel\phi_n \sim_{hhp}^{fr} \delta$ if $\forall s_0,\cdots,s_n\in S,\exists i\leq n.test(\neg\phi_i,s_0\cup\cdots\cup s_n)$. It is sufficient to prove the relation $R=\{(\phi_0\parallel\cdots\parallel\phi_n, \delta)\}\cup \textbf{Id}$, if $\forall s_0,\cdots,s_n\in S,\exists i\leq n.test(\neg\phi_i,s_0\cup\cdots\cup s_n)$, is a FR strong hhp-bisimulation, and we omit it.
\end{enumerate}
\end{proof}

\begin{proposition}[Expansion law for FR strong pomset bisimulation]
Let $P\equiv (P_1[f_1]\parallel\cdots\parallel P_n[f_n])\setminus L$, with $n\geq 1$. Then

\begin{eqnarray}
P\sim_p^{f} \{(f_1(\alpha_1)\parallel\cdots\parallel f_n(\alpha_n)).(P_1'[f_1]\parallel\cdots\parallel P_n'[f_n])\setminus L: \nonumber\\
\langle P_i,s_i\rangle\xrightarrow{\alpha_i}\langle P_i',s_i'\rangle,i\in\{1,\cdots,n\},f_i(\alpha_i)\notin L\cup\overline{L}\} \nonumber\\
+\sum\{\tau.(P_1[f_1]\parallel\cdots\parallel P_i'[f_i]\parallel\cdots\parallel P_j'[f_j]\parallel\cdots\parallel P_n[f_n])\setminus L: \nonumber\\
\langle P_i,s_i\rangle\xrightarrow{l_1}\langle P_i',s_i'\rangle,\langle P_j,s_j\rangle\xrightarrow{l_2}\langle P_j',s_j'\rangle,f_i(l_1)=\overline{f_j(l_2)},i<j\}\nonumber
\end{eqnarray}
\begin{eqnarray}
P\sim_p^{r} \{(P_1'[f_1]\parallel\cdots\parallel P_n'[f_n]).(f_1(\alpha_1[m])\parallel\cdots\parallel f_n(\alpha_n)[m])\setminus L: \nonumber\\
\langle P_i,s_i\rangle\xtworightarrow{\alpha_i[m]}\langle P_i',s_i'\rangle,i\in\{1,\cdots,n\},f_i(\alpha_i)\notin L\cup\overline{L}\} \nonumber\\
+\sum\{(P_1[f_1]\parallel\cdots\parallel P_i'[f_i]\parallel\cdots\parallel P_j'[f_j]\parallel\cdots\parallel P_n[f_n]).\tau\setminus L: \nonumber\\
\langle P_i,s_i\rangle\xtworightarrow{l_1[m]}\langle P_i',s_i'\rangle,\langle P_j,s_j\rangle\xtworightarrow{l_2[m]}\langle P_j',s_j'\rangle,f_i(l_1)=\overline{f_j(l_2)},i<j\}\nonumber
\end{eqnarray}
\end{proposition}

\begin{proof}
(1) The case of forward strong pomset bisimulation.

Firstly, we consider the case without Restriction and Relabeling. That is, we suffice to prove the following case by induction on the size $n$.

For $P\equiv P_1\parallel\cdots\parallel P_n$, with $n\geq 1$, we need to prove

\begin{eqnarray}
P\sim_p \{(\alpha_1\parallel\cdots\parallel \alpha_n).(P_1'\parallel\cdots\parallel P_n'): \langle P_i,s_i\rangle\xrightarrow{\alpha_i}\langle P_i',s_i'\rangle,i\in\{1,\cdots,n\}\nonumber\\
+\sum\{\tau.(P_1\parallel\cdots\parallel P_i'\parallel\cdots\parallel P_j'\parallel\cdots\parallel P_n): \langle P_i,s_i\rangle\xrightarrow{l}\langle P_i',s_i'\rangle,\langle P_j,s_j\rangle\xrightarrow{\overline{l}}\langle P_j',s_j'\rangle,i<j\} \nonumber
\end{eqnarray}

For $n=1$, $P_1\sim_p^{f} \alpha_1.P_1':\langle P_1,s_1\rangle\xrightarrow{\alpha_1}\langle P_1',s_1'\rangle$ is obvious. Then with a hypothesis $n$, we consider 
$R\equiv P\parallel P_{n+1}$. By the forward transition rules of Composition, we can get

\begin{eqnarray}
R\sim_p^{f} \{(p\parallel \alpha_{n+1}).(P'\parallel P_{n+1}'): \langle P,s\rangle\xrightarrow{p}\langle P',s'\rangle,\langle P_{n+1},s_{n+1}\rangle\xrightarrow{\alpha_{n+1}}\langle P_{n+1}',s_{n+1}'\rangle,p\subseteq P\}\nonumber\\
+\sum\{\tau.(P'\parallel P_{n+1}'): \langle P,s\rangle\xrightarrow{l}\langle P',s'\rangle,\langle P_{n+1},s_{n+1}\rangle\xrightarrow{\overline{l}}\langle P_{n+1}',s_{n+1}'\rangle\} \nonumber
\end{eqnarray}

Now with the induction assumption $P\equiv P_1\parallel\cdots\parallel P_n$, the right-hand side can be reformulated as follows.

\begin{eqnarray}
\{(\alpha_1\parallel\cdots\parallel \alpha_n\parallel \alpha_{n+1}).(P_1'\parallel\cdots\parallel P_n'\parallel P_{n+1}'): \nonumber\\
\langle P_i,s_i\rangle\xrightarrow{\alpha_i}\langle P_i',s_i'\rangle,i\in\{1,\cdots,n+1\}\nonumber\\
+\sum\{\tau.(P_1\parallel\cdots\parallel P_i'\parallel\cdots\parallel P_j'\parallel\cdots\parallel P_n\parallel P_{n+1}): \nonumber\\
\langle P_i,s_i\rangle\xrightarrow{l}\langle P_i',s_i'\rangle,\langle P_j,s_j\rangle\xrightarrow{\overline{l}}\langle P_j',s_j'\rangle,i<j\} \nonumber\\
+\sum\{\tau.(P_1\parallel\cdots\parallel P_i'\parallel\cdots\parallel P_j\parallel\cdots\parallel P_n\parallel P_{n+1}'): \nonumber\\
\langle P_i,s_i\rangle\xrightarrow{l}\langle P_i',s_i'\rangle,\langle P_{n+1},s_{n+1}\rangle\xrightarrow{\overline{l}}\langle P_{n+1}',s_{n+1}'\rangle,i\in\{1,\cdots, n\}\} \nonumber
\end{eqnarray}

So,

\begin{eqnarray}
R\sim_p^{f} \{(\alpha_1\parallel\cdots\parallel \alpha_n\parallel \alpha_{n+1}).(P_1'\parallel\cdots\parallel P_n'\parallel P_{n+1}'): \nonumber\\
\langle P_i,s_i\rangle\xrightarrow{\alpha_i}\langle P_i',s_i'\rangle,i\in\{1,\cdots,n+1\}\nonumber\\
+\sum\{\tau.(P_1\parallel\cdots\parallel P_i'\parallel\cdots\parallel P_j'\parallel\cdots\parallel P_n): \nonumber\\
\langle P_i,s_i\rangle\xrightarrow{l}\langle P_i',s_i'\rangle,\langle P_j,s_j\rangle\xrightarrow{\overline{l}}\langle P_j',s_j'\rangle,1 \leq i<j\geq n+1\} \nonumber
\end{eqnarray}

Then, we can easily add the full conditions with Restriction and Relabeling.

(2) The case of reverse strong pomset bisimulation.

Firstly, we consider the case without Restriction and Relabeling. That is, we suffice to prove the following case by induction on the size $n$.

For $P\equiv P_1\parallel\cdots\parallel P_n$, with $n\geq 1$, we need to prove

\begin{eqnarray}
P\sim_p^{r} \{(P_1'\parallel\cdots\parallel P_n').(\alpha_1[m]\parallel\cdots\parallel \alpha_n[m]): \langle P_i,s_i\rangle\xtworightarrow{\alpha_i[m]}\langle P_i',s_i'\rangle,i\in\{1,\cdots,n\}\nonumber\\
+\sum\{(P_1\parallel\cdots\parallel P_i'\parallel\cdots\parallel P_j'\parallel\cdots\parallel P_n).\tau: \langle P_i,s_i\rangle\xtworightarrow{l[m]}\langle P_i',s_i'\rangle,\langle P_j,s_j\rangle\xtworightarrow{\overline{l}[m]}\langle P_j',s_j'\rangle,i<j\} \nonumber
\end{eqnarray}

For $n=1$, $P_1\sim_p^{r} P_1'.\alpha_1[m]:\langle P_1,s_1\rangle\xtworightarrow{\alpha_1[m]}\langle P_1',s_1'\rangle$ is obvious. Then with a hypothesis $n$, we consider 
$R\equiv P\parallel P_{n+1}$. By the reverse transition rules of Composition, we can get

\begin{eqnarray}
R\sim_p^{r} \{(P'\parallel P_{n+1}').(p[m]\parallel \alpha_{n+1}[m]): \langle P,s\rangle\xtworightarrow{p[m]}\langle P',s'\rangle,\langle P_{n+1},s_{n+1}\rangle\xtworightarrow{\alpha_{n+1}[m]}\langle P_{n+1}',s_{n+1}'\rangle,p\subseteq P\}\nonumber\\
+\sum\{(P'\parallel P_{n+1}').\tau: \langle P,s\rangle\xtworightarrow{l[m]}\langle P',s'\rangle,\langle P_{n+1},s_{n+1}\rangle\xtworightarrow{\overline{l}[m]}\langle P_{n+1}',s_{n+1}'\rangle\} \nonumber
\end{eqnarray}

Now with the induction assumption $P\equiv P_1\parallel\cdots\parallel P_n$, the right-hand side can be reformulated as follows.

\begin{eqnarray}
\{(P_1'\parallel\cdots\parallel P_n'\parallel P_{n+1}).(\alpha_1[m]\parallel\cdots\parallel \alpha_n[m]\parallel \alpha_{n+1}[m]): \nonumber\\
\langle P_i,s_i\rangle\xtworightarrow{\alpha_i[m]}\langle P_i',s_i'\rangle,i\in\{1,\cdots,n+1\}\nonumber\\
+\sum\{(P_1\parallel\cdots\parallel P_i'\parallel\cdots\parallel P_j'\parallel\cdots\parallel P_n\parallel P_{n+1}).\tau: \nonumber\\
\langle P_i,s_i\rangle\xtworightarrow{l[m]}\langle P_i',s_i'\rangle,\langle P_j,s_j\rangle\xtworightarrow{\overline{l}[m]}\langle P_j',s_j'\rangle,i<j\} \nonumber\\
+\sum\{(P_1\parallel\cdots\parallel P_i'\parallel\cdots\parallel P_j'\parallel\cdots\parallel P_n\parallel P_{n+1}).\tau: \nonumber\\
\langle P_i,s_i\rangle\xtworightarrow{l[m]}\langle P_i',s_i'\rangle,\langle P_{n+1},s_{n+1}\rangle\xtworightarrow{\overline{l}[m]}\langle P_{n+1}',s_{n+1}'\rangle,i\in\{1,\cdots, n\}\} \nonumber
\end{eqnarray}

So,

\begin{eqnarray}
R\sim_p^{r} \{(P_1'\parallel\cdots\parallel P_n'\parallel P_{n+1}').(\alpha_1[m]\parallel\cdots\parallel \alpha_n[m]\parallel \alpha_{n+1}[m]): \nonumber\\
\langle P_i,s_i\rangle\xtworightarrow{\alpha_i[m]}\langle P_i',s_i'\rangle,i\in\{1,\cdots,n+1\}\nonumber\\
+\sum\{(P_1\parallel\cdots\parallel P_i'\parallel\cdots\parallel P_j'\parallel\cdots\parallel P_n).\tau: \nonumber\\
\langle P_i,s_i\rangle\xtworightarrow{l[m]}\langle P_i',s_i'\rangle,\langle P_j,s_j\rangle\xtworightarrow{\overline{l}[m]}\langle P_j',s_j'\rangle,1 \leq i<j\geq n+1\} \nonumber
\end{eqnarray}

Then, we can easily add the full conditions with Restriction and Relabeling.
\end{proof}

\begin{proposition}[Expansion law for FR strong step bisimulation]
Let $P\equiv (P_1[f_1]\parallel\cdots\parallel P_n[f_n])\setminus L$, with $n\geq 1$. Then

\begin{eqnarray}
P\sim_s^{f} \{(f_1(\alpha_1)\parallel\cdots\parallel f_n(\alpha_n)).(P_1'[f_1]\parallel\cdots\parallel P_n'[f_n])\setminus L: \nonumber\\
\langle P_i,s_i\rangle\xrightarrow{\alpha_i}\langle P_i',s_i'\rangle,i\in\{1,\cdots,n\},f_i(\alpha_i)\notin L\cup\overline{L}\} \nonumber\\
+\sum\{\tau.(P_1[f_1]\parallel\cdots\parallel P_i'[f_i]\parallel\cdots\parallel P_j'[f_j]\parallel\cdots\parallel P_n[f_n])\setminus L: \nonumber\\
\langle P_i,s_i\rangle\xrightarrow{l_1}\langle P_i',s_i'\rangle,\langle P_j,s_j\rangle\xrightarrow{l_2}\langle P_j',s_j'\rangle,f_i(l_1)=\overline{f_j(l_2)},i<j\}\nonumber
\end{eqnarray}
\begin{eqnarray}
P\sim_s^{r} \{(P_1'[f_1]\parallel\cdots\parallel P_n'[f_n]).(f_1(\alpha_1[m])\parallel\cdots\parallel f_n(\alpha_n)[m])\setminus L: \nonumber\\
\langle P_i,s_i\rangle\xtworightarrow{\alpha_i[m]}\langle P_i',s_i'\rangle,i\in\{1,\cdots,n\},f_i(\alpha_i)\notin L\cup\overline{L}\} \nonumber\\
+\sum\{(P_1[f_1]\parallel\cdots\parallel P_i'[f_i]\parallel\cdots\parallel P_j'[f_j]\parallel\cdots\parallel P_n[f_n]).\tau\setminus L: \nonumber\\
\langle P_i,s_i\rangle\xtworightarrow{l_1[m]}\langle P_i',s_i'\rangle,\langle P_j,s_j\rangle\xtworightarrow{l_2[m]}\langle P_j',s_j'\rangle,f_i(l_1)=\overline{f_j(l_2)},i<j\}\nonumber
\end{eqnarray}
\end{proposition}

\begin{proof}
(1) The case of forward strong step bisimulation.

Firstly, we consider the case without Restriction and Relabeling. That is, we suffice to prove the following case by induction on the size $n$.

For $P\equiv P_1\parallel\cdots\parallel P_n$, with $n\geq 1$, we need to prove

\begin{eqnarray}
P\sim_s \{(\alpha_1\parallel\cdots\parallel \alpha_n).(P_1'\parallel\cdots\parallel P_n'): \langle P_i,s_i\rangle\xrightarrow{\alpha_i}\langle P_i',s_i'\rangle,i\in\{1,\cdots,n\}\nonumber\\
+\sum\{\tau.(P_1\parallel\cdots\parallel P_i'\parallel\cdots\parallel P_j'\parallel\cdots\parallel P_n): \langle P_i,s_i\rangle\xrightarrow{l}\langle P_i',s_i'\rangle,\langle P_j,s_j\rangle\xrightarrow{\overline{l}}\langle P_j',s_j'\rangle,i<j\} \nonumber
\end{eqnarray}

For $n=1$, $P_1\sim_s^{f} \alpha_1.P_1':\langle P_1,s_1\rangle\xrightarrow{\alpha_1}\langle P_1',s_1'\rangle$ is obvious. Then with a hypothesis $n$, we consider
$R\equiv P\parallel P_{n+1}$. By the forward transition rules of Composition, we can get

\begin{eqnarray}
R\sim_s^{f} \{(p\parallel \alpha_{n+1}).(P'\parallel P_{n+1}'): \langle P,s\rangle\xrightarrow{p}\langle P',s'\rangle,\langle P_{n+1},s_{n+1}\rangle\xrightarrow{\alpha_{n+1}}\langle P_{n+1}',s_{n+1}'\rangle,p\subseteq P\}\nonumber\\
+\sum\{\tau.(P'\parallel P_{n+1}'): \langle P,s\rangle\xrightarrow{l}\langle P',s'\rangle,\langle P_{n+1},s_{n+1}\rangle\xrightarrow{\overline{l}}\langle P_{n+1}',s_{n+1}'\rangle\} \nonumber
\end{eqnarray}

Now with the induction assumption $P\equiv P_1\parallel\cdots\parallel P_n$, the right-hand side can be reformulated as follows.

\begin{eqnarray}
\{(\alpha_1\parallel\cdots\parallel \alpha_n\parallel \alpha_{n+1}).(P_1'\parallel\cdots\parallel P_n'\parallel P_{n+1}'): \nonumber\\
\langle P_i,s_i\rangle\xrightarrow{\alpha_i}\langle P_i',s_i'\rangle,i\in\{1,\cdots,n+1\}\nonumber\\
+\sum\{\tau.(P_1\parallel\cdots\parallel P_i'\parallel\cdots\parallel P_j'\parallel\cdots\parallel P_n\parallel P_{n+1}): \nonumber\\
\langle P_i,s_i\rangle\xrightarrow{l}\langle P_i',s_i'\rangle,\langle P_j,s_j\rangle\xrightarrow{\overline{l}}\langle P_j',s_j'\rangle,i<j\} \nonumber\\
+\sum\{\tau.(P_1\parallel\cdots\parallel P_i'\parallel\cdots\parallel P_j\parallel\cdots\parallel P_n\parallel P_{n+1}'): \nonumber\\
\langle P_i,s_i\rangle\xrightarrow{l}\langle P_i',s_i'\rangle,\langle P_{n+1},s_{n+1}\rangle\xrightarrow{\overline{l}}\langle P_{n+1}',s_{n+1}'\rangle,i\in\{1,\cdots, n\}\} \nonumber
\end{eqnarray}

So,

\begin{eqnarray}
R\sim_s^{f} \{(\alpha_1\parallel\cdots\parallel \alpha_n\parallel \alpha_{n+1}).(P_1'\parallel\cdots\parallel P_n'\parallel P_{n+1}'): \nonumber\\
\langle P_i,s_i\rangle\xrightarrow{\alpha_i}\langle P_i',s_i'\rangle,i\in\{1,\cdots,n+1\}\nonumber\\
+\sum\{\tau.(P_1\parallel\cdots\parallel P_i'\parallel\cdots\parallel P_j'\parallel\cdots\parallel P_n): \nonumber\\
\langle P_i,s_i\rangle\xrightarrow{l}\langle P_i',s_i'\rangle,\langle P_j,s_j\rangle\xrightarrow{\overline{l}}\langle P_j',s_j'\rangle,1 \leq i<j\geq n+1\} \nonumber
\end{eqnarray}

Then, we can easily add the full conditions with Restriction and Relabeling.

(2) The case of reverse strong step bisimulation.

Firstly, we consider the case without Restriction and Relabeling. That is, we suffice to prove the following case by induction on the size $n$.

For $P\equiv P_1\parallel\cdots\parallel P_n$, with $n\geq 1$, we need to prove

\begin{eqnarray}
P\sim_s^{r} \{(P_1'\parallel\cdots\parallel P_n').(\alpha_1[m]\parallel\cdots\parallel \alpha_n[m]): \langle P_i,s_i\rangle\xtworightarrow{\alpha_i[m]}\langle P_i',s_i'\rangle,i\in\{1,\cdots,n\}\nonumber\\
+\sum\{(P_1\parallel\cdots\parallel P_i'\parallel\cdots\parallel P_j'\parallel\cdots\parallel P_n).\tau: \langle P_i,s_i\rangle\xtworightarrow{l[m]}\langle P_i',s_i'\rangle,\langle P_j,s_j\rangle\xtworightarrow{\overline{l}[m]}\langle P_j',s_j'\rangle,i<j\} \nonumber
\end{eqnarray}

For $n=1$, $P_1\sim_s^{r} P_1'.\alpha_1[m]:\langle P_1,s_1\rangle\xtworightarrow{\alpha_1[m]}\langle P_1',s_1'\rangle$ is obvious. Then with a hypothesis $n$, we consider
$R\equiv P\parallel P_{n+1}$. By the reverse transition rules of Composition, we can get

\begin{eqnarray}
R\sim_s^{r} \{(P'\parallel P_{n+1}').(p[m]\parallel \alpha_{n+1}[m]): \langle P,s\rangle\xtworightarrow{p[m]}\langle P',s'\rangle,\langle P_{n+1},s_{n+1}\rangle\xtworightarrow{\alpha_{n+1}[m]}\langle P_{n+1}',s_{n+1}'\rangle,p\subseteq P\}\nonumber\\
+\sum\{(P'\parallel P_{n+1}').\tau: \langle P,s\rangle\xtworightarrow{l[m]}\langle P',s'\rangle,\langle P_{n+1},s_{n+1}\rangle\xtworightarrow{\overline{l}[m]}\langle P_{n+1}',s_{n+1}'\rangle\} \nonumber
\end{eqnarray}

Now with the induction assumption $P\equiv P_1\parallel\cdots\parallel P_n$, the right-hand side can be reformulated as follows.

\begin{eqnarray}
\{(P_1'\parallel\cdots\parallel P_n'\parallel P_{n+1}).(\alpha_1[m]\parallel\cdots\parallel \alpha_n[m]\parallel \alpha_{n+1}[m]): \nonumber\\
\langle P_i,s_i\rangle\xtworightarrow{\alpha_i[m]}\langle P_i',s_i'\rangle,i\in\{1,\cdots,n+1\}\nonumber\\
+\sum\{(P_1\parallel\cdots\parallel P_i'\parallel\cdots\parallel P_j'\parallel\cdots\parallel P_n\parallel P_{n+1}).\tau: \nonumber\\
\langle P_i,s_i\rangle\xtworightarrow{l[m]}\langle P_i',s_i'\rangle,\langle P_j,s_j\rangle\xtworightarrow{\overline{l}[m]}\langle P_j',s_j'\rangle,i<j\} \nonumber\\
+\sum\{(P_1\parallel\cdots\parallel P_i'\parallel\cdots\parallel P_j'\parallel\cdots\parallel P_n\parallel P_{n+1}).\tau: \nonumber\\
\langle P_i,s_i\rangle\xtworightarrow{l[m]}\langle P_i',s_i'\rangle,\langle P_{n+1},s_{n+1}\rangle\xtworightarrow{\overline{l}[m]}\langle P_{n+1}',s_{n+1}'\rangle,i\in\{1,\cdots, n\}\} \nonumber
\end{eqnarray}

So,

\begin{eqnarray}
R\sim_s^{r} \{(P_1'\parallel\cdots\parallel P_n'\parallel P_{n+1}').(\alpha_1[m]\parallel\cdots\parallel \alpha_n[m]\parallel \alpha_{n+1}[m]): \nonumber\\
\langle P_i,s_i\rangle\xtworightarrow{\alpha_i[m]}\langle P_i',s_i'\rangle,i\in\{1,\cdots,n+1\}\nonumber\\
+\sum\{(P_1\parallel\cdots\parallel P_i'\parallel\cdots\parallel P_j'\parallel\cdots\parallel P_n).\tau: \nonumber\\
\langle P_i,s_i\rangle\xtworightarrow{l[m]}\langle P_i',s_i'\rangle,\langle P_j,s_j\rangle\xtworightarrow{\overline{l}[m]}\langle P_j',s_j'\rangle,1 \leq i<j\geq n+1\} \nonumber
\end{eqnarray}

Then, we can easily add the full conditions with Restriction and Relabeling.
\end{proof}

\begin{proposition}[Expansion law for FR strong hp-bisimulation]
Let $P\equiv (P_1[f_1]\parallel\cdots\parallel P_n[f_n])\setminus L$, with $n\geq 1$. Then

\begin{eqnarray}
P\sim_{hp}^{f} \{(f_1(\alpha_1)\parallel\cdots\parallel f_n(\alpha_n)).(P_1'[f_1]\parallel\cdots\parallel P_n'[f_n])\setminus L: \nonumber\\
\langle P_i,s_i\rangle\xrightarrow{\alpha_i}\langle P_i',s_i'\rangle,i\in\{1,\cdots,n\},f_i(\alpha_i)\notin L\cup\overline{L}\} \nonumber\\
+\sum\{\tau.(P_1[f_1]\parallel\cdots\parallel P_i'[f_i]\parallel\cdots\parallel P_j'[f_j]\parallel\cdots\parallel P_n[f_n])\setminus L: \nonumber\\
\langle P_i,s_i\rangle\xrightarrow{l_1}\langle P_i',s_i'\rangle,\langle P_j,s_j\rangle\xrightarrow{l_2}\langle P_j',s_j'\rangle,f_i(l_1)=\overline{f_j(l_2)},i<j\}\nonumber
\end{eqnarray}
\begin{eqnarray}
P\sim_{hp}^{r} \{(P_1'[f_1]\parallel\cdots\parallel P_n'[f_n]).(f_1(\alpha_1[m])\parallel\cdots\parallel f_n(\alpha_n)[m])\setminus L: \nonumber\\
\langle P_i,s_i\rangle\xtworightarrow{\alpha_i[m]}\langle P_i',s_i'\rangle,i\in\{1,\cdots,n\},f_i(\alpha_i)\notin L\cup\overline{L}\} \nonumber\\
+\sum\{(P_1[f_1]\parallel\cdots\parallel P_i'[f_i]\parallel\cdots\parallel P_j'[f_j]\parallel\cdots\parallel P_n[f_n]).\tau\setminus L: \nonumber\\
\langle P_i,s_i\rangle\xtworightarrow{l_1[m]}\langle P_i',s_i'\rangle,\langle P_j,s_j\rangle\xtworightarrow{l_2[m]}\langle P_j',s_j'\rangle,f_i(l_1)=\overline{f_j(l_2)},i<j\}\nonumber
\end{eqnarray}
\end{proposition}

\begin{proof}
(1) The case of forward strong hp-bisimulation.

Firstly, we consider the case without Restriction and Relabeling. That is, we suffice to prove the following case by induction on the size $n$.

For $P\equiv P_1\parallel\cdots\parallel P_n$, with $n\geq 1$, we need to prove

\begin{eqnarray}
P\sim_{hp} \{(\alpha_1\parallel\cdots\parallel \alpha_n).(P_1'\parallel\cdots\parallel P_n'): \langle P_i,s_i\rangle\xrightarrow{\alpha_i}\langle P_i',s_i'\rangle,i\in\{1,\cdots,n\}\nonumber\\
+\sum\{\tau.(P_1\parallel\cdots\parallel P_i'\parallel\cdots\parallel P_j'\parallel\cdots\parallel P_n): \langle P_i,s_i\rangle\xrightarrow{l}\langle P_i',s_i'\rangle,\langle P_j,s_j\rangle\xrightarrow{\overline{l}}\langle P_j',s_j'\rangle,i<j\} \nonumber
\end{eqnarray}

For $n=1$, $P_1\sim_{hp}^{f} \alpha_1.P_1':\langle P_1,s_1\rangle\xrightarrow{\alpha_1}\langle P_1',s_1'\rangle$ is obvious. Then with a hypothesis $n$, we consider
$R\equiv P\parallel P_{n+1}$. By the forward transition rules of Composition, we can get

\begin{eqnarray}
R\sim_{hp}^{f} \{(p\parallel \alpha_{n+1}).(P'\parallel P_{n+1}'): \langle P,s\rangle\xrightarrow{p}\langle P',s'\rangle,\langle P_{n+1},s_{n+1}\rangle\xrightarrow{\alpha_{n+1}}\langle P_{n+1}',s_{n+1}'\rangle,p\subseteq P\}\nonumber\\
+\sum\{\tau.(P'\parallel P_{n+1}'): \langle P,s\rangle\xrightarrow{l}\langle P',s'\rangle,\langle P_{n+1},s_{n+1}\rangle\xrightarrow{\overline{l}}\langle P_{n+1}',s_{n+1}'\rangle\} \nonumber
\end{eqnarray}

Now with the induction assumption $P\equiv P_1\parallel\cdots\parallel P_n$, the right-hand side can be reformulated as follows.

\begin{eqnarray}
\{(\alpha_1\parallel\cdots\parallel \alpha_n\parallel \alpha_{n+1}).(P_1'\parallel\cdots\parallel P_n'\parallel P_{n+1}'): \nonumber\\
\langle P_i,s_i\rangle\xrightarrow{\alpha_i}\langle P_i',s_i'\rangle,i\in\{1,\cdots,n+1\}\nonumber\\
+\sum\{\tau.(P_1\parallel\cdots\parallel P_i'\parallel\cdots\parallel P_j'\parallel\cdots\parallel P_n\parallel P_{n+1}): \nonumber\\
\langle P_i,s_i\rangle\xrightarrow{l}\langle P_i',s_i'\rangle,\langle P_j,s_j\rangle\xrightarrow{\overline{l}}\langle P_j',s_j'\rangle,i<j\} \nonumber\\
+\sum\{\tau.(P_1\parallel\cdots\parallel P_i'\parallel\cdots\parallel P_j\parallel\cdots\parallel P_n\parallel P_{n+1}'): \nonumber\\
\langle P_i,s_i\rangle\xrightarrow{l}\langle P_i',s_i'\rangle,\langle P_{n+1},s_{n+1}\rangle\xrightarrow{\overline{l}}\langle P_{n+1}',s_{n+1}'\rangle,i\in\{1,\cdots, n\}\} \nonumber
\end{eqnarray}

So,

\begin{eqnarray}
R\sim_{hp}^{f} \{(\alpha_1\parallel\cdots\parallel \alpha_n\parallel \alpha_{n+1}).(P_1'\parallel\cdots\parallel P_n'\parallel P_{n+1}'): \nonumber\\
\langle P_i,s_i\rangle\xrightarrow{\alpha_i}\langle P_i',s_i'\rangle,i\in\{1,\cdots,n+1\}\nonumber\\
+\sum\{\tau.(P_1\parallel\cdots\parallel P_i'\parallel\cdots\parallel P_j'\parallel\cdots\parallel P_n): \nonumber\\
\langle P_i,s_i\rangle\xrightarrow{l}\langle P_i',s_i'\rangle,\langle P_j,s_j\rangle\xrightarrow{\overline{l}}\langle P_j',s_j'\rangle,1 \leq i<j\geq n+1\} \nonumber
\end{eqnarray}

Then, we can easily add the full conditions with Restriction and Relabeling.

(2) The case of reverse strong hp-bisimulation.

Firstly, we consider the case without Restriction and Relabeling. That is, we suffice to prove the following case by induction on the size $n$.

For $P\equiv P_1\parallel\cdots\parallel P_n$, with $n\geq 1$, we need to prove

\begin{eqnarray}
P\sim_{hp}^{r} \{(P_1'\parallel\cdots\parallel P_n').(\alpha_1[m]\parallel\cdots\parallel \alpha_n[m]): \langle P_i,s_i\rangle\xtworightarrow{\alpha_i[m]}\langle P_i',s_i'\rangle,i\in\{1,\cdots,n\}\nonumber\\
+\sum\{(P_1\parallel\cdots\parallel P_i'\parallel\cdots\parallel P_j'\parallel\cdots\parallel P_n).\tau: \langle P_i,s_i\rangle\xtworightarrow{l[m]}\langle P_i',s_i'\rangle,\langle P_j,s_j\rangle\xtworightarrow{\overline{l}[m]}\langle P_j',s_j'\rangle,i<j\} \nonumber
\end{eqnarray}

For $n=1$, $P_1\sim_{hp}^{r} P_1'.\alpha_1[m]:\langle P_1,s_1\rangle\xtworightarrow{\alpha_1[m]}\langle P_1',s_1'\rangle$ is obvious. Then with a hypothesis $n$, we consider
$R\equiv P\parallel P_{n+1}$. By the reverse transition rules of Composition, we can get

\begin{eqnarray}
R\sim_{hp}^{r} \{(P'\parallel P_{n+1}').(p[m]\parallel \alpha_{n+1}[m]): \langle P,s\rangle\xtworightarrow{p[m]}\langle P',s'\rangle,\langle P_{n+1},s_{n+1}\rangle\xtworightarrow{\alpha_{n+1}[m]}\langle P_{n+1}',s_{n+1}'\rangle,p\subseteq P\}\nonumber\\
+\sum\{(P'\parallel P_{n+1}').\tau: \langle P,s\rangle\xtworightarrow{l[m]}\langle P',s'\rangle,\langle P_{n+1},s_{n+1}\rangle\xtworightarrow{\overline{l}[m]}\langle P_{n+1}',s_{n+1}'\rangle\} \nonumber
\end{eqnarray}

Now with the induction assumption $P\equiv P_1\parallel\cdots\parallel P_n$, the right-hand side can be reformulated as follows.

\begin{eqnarray}
\{(P_1'\parallel\cdots\parallel P_n'\parallel P_{n+1}).(\alpha_1[m]\parallel\cdots\parallel \alpha_n[m]\parallel \alpha_{n+1}[m]): \nonumber\\
\langle P_i,s_i\rangle\xtworightarrow{\alpha_i[m]}\langle P_i',s_i'\rangle,i\in\{1,\cdots,n+1\}\nonumber\\
+\sum\{(P_1\parallel\cdots\parallel P_i'\parallel\cdots\parallel P_j'\parallel\cdots\parallel P_n\parallel P_{n+1}).\tau: \nonumber\\
\langle P_i,s_i\rangle\xtworightarrow{l[m]}\langle P_i',s_i'\rangle,\langle P_j,s_j\rangle\xtworightarrow{\overline{l}[m]}\langle P_j',s_j'\rangle,i<j\} \nonumber\\
+\sum\{(P_1\parallel\cdots\parallel P_i'\parallel\cdots\parallel P_j'\parallel\cdots\parallel P_n\parallel P_{n+1}).\tau: \nonumber\\
\langle P_i,s_i\rangle\xtworightarrow{l[m]}\langle P_i',s_i'\rangle,\langle P_{n+1},s_{n+1}\rangle\xtworightarrow{\overline{l}[m]}\langle P_{n+1}',s_{n+1}'\rangle,i\in\{1,\cdots, n\}\} \nonumber
\end{eqnarray}

So,

\begin{eqnarray}
R\sim_{hp}^{r} \{(P_1'\parallel\cdots\parallel P_n'\parallel P_{n+1}').(\alpha_1[m]\parallel\cdots\parallel \alpha_n[m]\parallel \alpha_{n+1}[m]): \nonumber\\
\langle P_i,s_i\rangle\xtworightarrow{\alpha_i[m]}\langle P_i',s_i'\rangle,i\in\{1,\cdots,n+1\}\nonumber\\
+\sum\{(P_1\parallel\cdots\parallel P_i'\parallel\cdots\parallel P_j'\parallel\cdots\parallel P_n).\tau: \nonumber\\
\langle P_i,s_i\rangle\xtworightarrow{l[m]}\langle P_i',s_i'\rangle,\langle P_j,s_j\rangle\xtworightarrow{\overline{l}[m]}\langle P_j',s_j'\rangle,1 \leq i<j\geq n+1\} \nonumber
\end{eqnarray}

Then, we can easily add the full conditions with Restriction and Relabeling.
\end{proof}

\begin{proposition}[Expansion law for FR strong hhp-bisimulation]
Let $P\equiv (P_1[f_1]\parallel\cdots\parallel P_n[f_n])\setminus L$, with $n\geq 1$. Then

\begin{eqnarray}
P\sim_{hhp}^{f} \{(f_1(\alpha_1)\parallel\cdots\parallel f_n(\alpha_n)).(P_1'[f_1]\parallel\cdots\parallel P_n'[f_n])\setminus L: \nonumber\\
\langle P_i,s_i\rangle\xrightarrow{\alpha_i}\langle P_i',s_i'\rangle,i\in\{1,\cdots,n\},f_i(\alpha_i)\notin L\cup\overline{L}\} \nonumber\\
+\sum\{\tau.(P_1[f_1]\parallel\cdots\parallel P_i'[f_i]\parallel\cdots\parallel P_j'[f_j]\parallel\cdots\parallel P_n[f_n])\setminus L: \nonumber\\
\langle P_i,s_i\rangle\xrightarrow{l_1}\langle P_i',s_i'\rangle,\langle P_j,s_j\rangle\xrightarrow{l_2}\langle P_j',s_j'\rangle,f_i(l_1)=\overline{f_j(l_2)},i<j\}\nonumber
\end{eqnarray}
\begin{eqnarray}
P\sim_{hhp}^{r} \{(P_1'[f_1]\parallel\cdots\parallel P_n'[f_n]).(f_1(\alpha_1[m])\parallel\cdots\parallel f_n(\alpha_n)[m])\setminus L: \nonumber\\
\langle P_i,s_i\rangle\xtworightarrow{\alpha_i[m]}\langle P_i',s_i'\rangle,i\in\{1,\cdots,n\},f_i(\alpha_i)\notin L\cup\overline{L}\} \nonumber\\
+\sum\{(P_1[f_1]\parallel\cdots\parallel P_i'[f_i]\parallel\cdots\parallel P_j'[f_j]\parallel\cdots\parallel P_n[f_n]).\tau\setminus L: \nonumber\\
\langle P_i,s_i\rangle\xtworightarrow{l_1[m]}\langle P_i',s_i'\rangle,\langle P_j,s_j\rangle\xtworightarrow{l_2[m]}\langle P_j',s_j'\rangle,f_i(l_1)=\overline{f_j(l_2)},i<j\}\nonumber
\end{eqnarray}
\end{proposition}

\begin{proof}
(1) The case of forward strong hhp-bisimulation.

Firstly, we consider the case without Restriction and Relabeling. That is, we suffice to prove the following case by induction on the size $n$.

For $P\equiv P_1\parallel\cdots\parallel P_n$, with $n\geq 1$, we need to prove

\begin{eqnarray}
P\sim_{hhp} \{(\alpha_1\parallel\cdots\parallel \alpha_n).(P_1'\parallel\cdots\parallel P_n'): \langle P_i,s_i\rangle\xrightarrow{\alpha_i}\langle P_i',s_i'\rangle,i\in\{1,\cdots,n\}\nonumber\\
+\sum\{\tau.(P_1\parallel\cdots\parallel P_i'\parallel\cdots\parallel P_j'\parallel\cdots\parallel P_n): \langle P_i,s_i\rangle\xrightarrow{l}\langle P_i',s_i'\rangle,\langle P_j,s_j\rangle\xrightarrow{\overline{l}}\langle P_j',s_j'\rangle,i<j\} \nonumber
\end{eqnarray}

For $n=1$, $P_1\sim_{hhp}^{f} \alpha_1.P_1':\langle P_1,s_1\rangle\xrightarrow{\alpha_1}\langle P_1',s_1'\rangle$ is obvious. Then with a hypothesis $n$, we consider
$R\equiv P\parallel P_{n+1}$. By the forward transition rules of Composition, we can get

\begin{eqnarray}
R\sim_{hhp}^{f} \{(p\parallel \alpha_{n+1}).(P'\parallel P_{n+1}'): \langle P,s\rangle\xrightarrow{p}\langle P',s'\rangle,\langle P_{n+1},s_{n+1}\rangle\xrightarrow{\alpha_{n+1}}\langle P_{n+1}',s_{n+1}'\rangle,p\subseteq P\}\nonumber\\
+\sum\{\tau.(P'\parallel P_{n+1}'): \langle P,s\rangle\xrightarrow{l}\langle P',s'\rangle,\langle P_{n+1},s_{n+1}\rangle\xrightarrow{\overline{l}}\langle P_{n+1}',s_{n+1}'\rangle\} \nonumber
\end{eqnarray}

Now with the induction assumption $P\equiv P_1\parallel\cdots\parallel P_n$, the right-hand side can be reformulated as follows.

\begin{eqnarray}
\{(\alpha_1\parallel\cdots\parallel \alpha_n\parallel \alpha_{n+1}).(P_1'\parallel\cdots\parallel P_n'\parallel P_{n+1}'): \nonumber\\
\langle P_i,s_i\rangle\xrightarrow{\alpha_i}\langle P_i',s_i'\rangle,i\in\{1,\cdots,n+1\}\nonumber\\
+\sum\{\tau.(P_1\parallel\cdots\parallel P_i'\parallel\cdots\parallel P_j'\parallel\cdots\parallel P_n\parallel P_{n+1}): \nonumber\\
\langle P_i,s_i\rangle\xrightarrow{l}\langle P_i',s_i'\rangle,\langle P_j,s_j\rangle\xrightarrow{\overline{l}}\langle P_j',s_j'\rangle,i<j\} \nonumber\\
+\sum\{\tau.(P_1\parallel\cdots\parallel P_i'\parallel\cdots\parallel P_j\parallel\cdots\parallel P_n\parallel P_{n+1}'): \nonumber\\
\langle P_i,s_i\rangle\xrightarrow{l}\langle P_i',s_i'\rangle,\langle P_{n+1},s_{n+1}\rangle\xrightarrow{\overline{l}}\langle P_{n+1}',s_{n+1}'\rangle,i\in\{1,\cdots, n\}\} \nonumber
\end{eqnarray}

So,

\begin{eqnarray}
R\sim_{hhp}^{f} \{(\alpha_1\parallel\cdots\parallel \alpha_n\parallel \alpha_{n+1}).(P_1'\parallel\cdots\parallel P_n'\parallel P_{n+1}'): \nonumber\\
\langle P_i,s_i\rangle\xrightarrow{\alpha_i}\langle P_i',s_i'\rangle,i\in\{1,\cdots,n+1\}\nonumber\\
+\sum\{\tau.(P_1\parallel\cdots\parallel P_i'\parallel\cdots\parallel P_j'\parallel\cdots\parallel P_n): \nonumber\\
\langle P_i,s_i\rangle\xrightarrow{l}\langle P_i',s_i'\rangle,\langle P_j,s_j\rangle\xrightarrow{\overline{l}}\langle P_j',s_j'\rangle,1 \leq i<j\geq n+1\} \nonumber
\end{eqnarray}

Then, we can easily add the full conditions with Restriction and Relabeling.

(2) The case of reverse strong hhp-bisimulation.

Firstly, we consider the case without Restriction and Relabeling. That is, we suffice to prove the following case by induction on the size $n$.

For $P\equiv P_1\parallel\cdots\parallel P_n$, with $n\geq 1$, we need to prove

\begin{eqnarray}
P\sim_{hhp}^{r} \{(P_1'\parallel\cdots\parallel P_n').(\alpha_1[m]\parallel\cdots\parallel \alpha_n[m]): \langle P_i,s_i\rangle\xtworightarrow{\alpha_i[m]}\langle P_i',s_i'\rangle,i\in\{1,\cdots,n\}\nonumber\\
+\sum\{(P_1\parallel\cdots\parallel P_i'\parallel\cdots\parallel P_j'\parallel\cdots\parallel P_n).\tau: \langle P_i,s_i\rangle\xtworightarrow{l[m]}\langle P_i',s_i'\rangle,\langle P_j,s_j\rangle\xtworightarrow{\overline{l}[m]}\langle P_j',s_j'\rangle,i<j\} \nonumber
\end{eqnarray}

For $n=1$, $P_1\sim_{hhp}^{r} P_1'.\alpha_1[m]:\langle P_1,s_1\rangle\xtworightarrow{\alpha_1[m]}\langle P_1',s_1'\rangle$ is obvious. Then with a hypothesis $n$, we consider
$R\equiv P\parallel P_{n+1}$. By the reverse transition rules of Composition, we can get

\begin{eqnarray}
R\sim_{hhp}^{r} \{(P'\parallel P_{n+1}').(p[m]\parallel \alpha_{n+1}[m]): \langle P,s\rangle\xtworightarrow{p[m]}\langle P',s'\rangle,\langle P_{n+1},s_{n+1}\rangle\xtworightarrow{\alpha_{n+1}[m]}\langle P_{n+1}',s_{n+1}'\rangle,p\subseteq P\}\nonumber\\
+\sum\{(P'\parallel P_{n+1}').\tau: \langle P,s\rangle\xtworightarrow{l[m]}\langle P',s'\rangle,\langle P_{n+1},s_{n+1}\rangle\xtworightarrow{\overline{l}[m]}\langle P_{n+1}',s_{n+1}'\rangle\} \nonumber
\end{eqnarray}

Now with the induction assumption $P\equiv P_1\parallel\cdots\parallel P_n$, the right-hand side can be reformulated as follows.

\begin{eqnarray}
\{(P_1'\parallel\cdots\parallel P_n'\parallel P_{n+1}).(\alpha_1[m]\parallel\cdots\parallel \alpha_n[m]\parallel \alpha_{n+1}[m]): \nonumber\\
\langle P_i,s_i\rangle\xtworightarrow{\alpha_i[m]}\langle P_i',s_i'\rangle,i\in\{1,\cdots,n+1\}\nonumber\\
+\sum\{(P_1\parallel\cdots\parallel P_i'\parallel\cdots\parallel P_j'\parallel\cdots\parallel P_n\parallel P_{n+1}).\tau: \nonumber\\
\langle P_i,s_i\rangle\xtworightarrow{l[m]}\langle P_i',s_i'\rangle,\langle P_j,s_j\rangle\xtworightarrow{\overline{l}[m]}\langle P_j',s_j'\rangle,i<j\} \nonumber\\
+\sum\{(P_1\parallel\cdots\parallel P_i'\parallel\cdots\parallel P_j'\parallel\cdots\parallel P_n\parallel P_{n+1}).\tau: \nonumber\\
\langle P_i,s_i\rangle\xtworightarrow{l[m]}\langle P_i',s_i'\rangle,\langle P_{n+1},s_{n+1}\rangle\xtworightarrow{\overline{l}[m]}\langle P_{n+1}',s_{n+1}'\rangle,i\in\{1,\cdots, n\}\} \nonumber
\end{eqnarray}

So,

\begin{eqnarray}
R\sim_{hhp}^{r} \{(P_1'\parallel\cdots\parallel P_n'\parallel P_{n+1}').(\alpha_1[m]\parallel\cdots\parallel \alpha_n[m]\parallel \alpha_{n+1}[m]): \nonumber\\
\langle P_i,s_i\rangle\xtworightarrow{\alpha_i[m]}\langle P_i',s_i'\rangle,i\in\{1,\cdots,n+1\}\nonumber\\
+\sum\{(P_1\parallel\cdots\parallel P_i'\parallel\cdots\parallel P_j'\parallel\cdots\parallel P_n).\tau: \nonumber\\
\langle P_i,s_i\rangle\xtworightarrow{l[m]}\langle P_i',s_i'\rangle,\langle P_j,s_j\rangle\xtworightarrow{\overline{l}[m]}\langle P_j',s_j'\rangle,1 \leq i<j\geq n+1\} \nonumber
\end{eqnarray}

Then, we can easily add the full conditions with Restriction and Relabeling.
\end{proof}

\begin{theorem}[Congruence for FR strong pomset bisimulation] \label{CSSB05}
We can enjoy the congruence for FR strong pomset bisimulation as follows.
\begin{enumerate}
  \item If $A\overset{\text{def}}{=}P$, then $A\sim_p^{fr} P$;
  \item Let $P_1\sim_p^{fr} P_2$. Then
        \begin{enumerate}
           \item $\alpha.P_1\sim_p^f \alpha.P_2$;
           \item $\phi.P_1\sim_p^f \phi.P_2$;
           \item $(\alpha_1\parallel\cdots\parallel\alpha_n).P_1\sim_p^f (\alpha_1\parallel\cdots\parallel\alpha_n).P_2$;
           \item $P_1.\alpha[m]\sim_p^r P_2.\alpha[m]$;
           \item $P_1.\phi\sim_p^r P_2.\phi$;
           \item $P_1.(\alpha_1[m]\parallel\cdots\parallel\alpha_n[m])\sim_p^r P_2.(\alpha_1[m]\parallel\cdots\parallel\alpha_n[m])$;
           \item $P_1+Q\sim_p^{fr} P_2 +Q$;
           \item $P_1\parallel Q\sim_p^{fr} P_2\parallel Q$;
           \item $P_1\setminus L\sim_p^{fr} P_2\setminus L$;
           \item $P_1[f]\sim_p^{fr} P_2[f]$.
         \end{enumerate}
\end{enumerate}
\end{theorem}

\begin{proof}
\begin{enumerate}
  \item If $A\overset{\text{def}}{=}P$, then $A\sim_p^{fr} P$. It is obvious.
  \item Let $P_1\sim_p^{fr} P_2$. Then
        \begin{enumerate}
           \item $\alpha.P_1\sim_p^f \alpha.P_2$. It is sufficient to prove the relation $R=\{(\alpha.P_1, \alpha.P_2)\}\cup \textbf{Id}$ is a F strong pomset bisimulation, we omit it;
           \item $\phi.P_1\sim_p^f \phi.P_2$. It is sufficient to prove the relation $R=\{(\phi.P_1, \phi.P_2)\}\cup \textbf{Id}$ is a F strong pomset bisimulation, we omit it;
           \item $(\alpha_1\parallel\cdots\parallel\alpha_n).P_1\sim_p^f (\alpha_1\parallel\cdots\parallel\alpha_n).P_2$. It is sufficient to prove the relation $R=\{((\alpha_1\parallel\cdots\parallel\alpha_n).P_1, (\alpha_1\parallel\cdots\parallel\alpha_n).P_2)\}\cup \textbf{Id}$ is a F strong pomset bisimulation, we omit it;
           \item $P_1.\alpha[m]\sim_p^r P_2.\alpha[m]$. It is sufficient to prove the relation $R=\{(P_1.\alpha[m], P_2.\alpha[m])\}\cup \textbf{Id}$ is a R strong pomset bisimulation, we omit it;
           \item $P_1.\phi\sim_p^r P_2.\phi$. It is sufficient to prove the relation $R=\{(P_1.\phi, P_2.\phi)\}\cup \textbf{Id}$ is a R strong pomset bisimulation, we omit it;
           \item $P_1.(\alpha_1[m]\parallel\cdots\parallel\alpha_n[m])\sim_p^r P_2.(\alpha_1[m]\parallel\cdots\parallel\alpha_n[m])$. It is sufficient to prove the relation $R=\{(P_1.(\alpha_1[m]\parallel\cdots\parallel\alpha_n[m]), P_2.(\alpha_1[m]\parallel\cdots\parallel\alpha_n[m]))\}\cup \textbf{Id}$ is a R strong pomset bisimulation, we omit it;
           \item $P_1+Q\sim_p^{fr} P_2 +Q$. It is sufficient to prove the relation $R=\{(P_1+Q, P_2+Q)\}\cup \textbf{Id}$ is a FR strong pomset bisimulation, we omit it;
           \item $P_1\parallel Q\sim_p^{fr} P_2\parallel Q$. It is sufficient to prove the relation $R=\{(P_1\parallel Q, P_2\parallel Q)\}\cup \textbf{Id}$ is a FR strong pomset bisimulation, we omit it;
           \item $P_1\setminus L\sim_p^{fr} P_2\setminus L$. It is sufficient to prove the relation $R=\{(P_1\setminus L, P_2\setminus L)\}\cup \textbf{Id}$ is a FR strong pomset bisimulation, we omit it;
           \item $P_1[f]\sim_p^{fr} P_2[f]$. It is sufficient to prove the relation $R=\{(P_1[f], P_2[f])\}\cup \textbf{Id}$ is a FR strong pomset bisimulation, we omit it.
         \end{enumerate}
\end{enumerate}
\end{proof}

\begin{theorem}[Congruence for FR strong step bisimulation] \label{CSSB05}
We can enjoy the congruence for FR strong step bisimulation as follows.
\begin{enumerate}
  \item If $A\overset{\text{def}}{=}P$, then $A\sim_s^{fr} P$;
  \item Let $P_1\sim_s^{fr} P_2$. Then
        \begin{enumerate}
           \item $\alpha.P_1\sim_s^f \alpha.P_2$;
           \item $\phi.P_1\sim_s^f \phi.P_2$;
           \item $(\alpha_1\parallel\cdots\parallel\alpha_n).P_1\sim_s^f (\alpha_1\parallel\cdots\parallel\alpha_n).P_2$;
           \item $P_1.\alpha[m]\sim_s^r P_2.\alpha[m]$;
           \item $P_1.\phi\sim_s^r P_2.\phi$;
           \item $P_1.(\alpha_1[m]\parallel\cdots\parallel\alpha_n[m])\sim_s^r P_2.(\alpha_1[m]\parallel\cdots\parallel\alpha_n[m])$;
           \item $P_1+Q\sim_s^{fr} P_2 +Q$;
           \item $P_1\parallel Q\sim_s^{fr} P_2\parallel Q$;
           \item $P_1\setminus L\sim_s^{fr} P_2\setminus L$;
           \item $P_1[f]\sim_s^{fr} P_2[f]$.
         \end{enumerate}
\end{enumerate}
\end{theorem}

\begin{proof}
\begin{enumerate}
  \item If $A\overset{\text{def}}{=}P$, then $A\sim_s^{fr} P$. It is obvious.
  \item Let $P_1\sim_s^{fr} P_2$. Then
        \begin{enumerate}
           \item $\alpha.P_1\sim_s^f \alpha.P_2$. It is sufficient to prove the relation $R=\{(\alpha.P_1, \alpha.P_2)\}\cup \textbf{Id}$ is a F strong step bisimulation, we omit it;
           \item $\phi.P_1\sim_s^f \phi.P_2$. It is sufficient to prove the relation $R=\{(\phi.P_1, \phi.P_2)\}\cup \textbf{Id}$ is a F strong step bisimulation, we omit it;
           \item $(\alpha_1\parallel\cdots\parallel\alpha_n).P_1\sim_s^f (\alpha_1\parallel\cdots\parallel\alpha_n).P_2$. It is sufficient to prove the relation $R=\{((\alpha_1\parallel\cdots\parallel\alpha_n).P_1, (\alpha_1\parallel\cdots\parallel\alpha_n).P_2)\}\cup \textbf{Id}$ is a F strong step bisimulation, we omit it;
           \item $P_1.\alpha[m]\sim_s^r P_2.\alpha[m]$. It is sufficient to prove the relation $R=\{(P_1.\alpha[m], P_2.\alpha[m])\}\cup \textbf{Id}$ is a R strong step bisimulation, we omit it;
           \item $P_1.\phi\sim_s^r P_2.\phi$. It is sufficient to prove the relation $R=\{(P_1.\phi, P_2.\phi)\}\cup \textbf{Id}$ is a R strong step bisimulation, we omit it;
           \item $P_1.(\alpha_1[m]\parallel\cdots\parallel\alpha_n[m])\sim_s^r P_2.(\alpha_1[m]\parallel\cdots\parallel\alpha_n[m])$. It is sufficient to prove the relation $R=\{(P_1.(\alpha_1[m]\parallel\cdots\parallel\alpha_n[m]), P_2.(\alpha_1[m]\parallel\cdots\parallel\alpha_n[m]))\}\cup \textbf{Id}$ is a R strong step bisimulation, we omit it;
           \item $P_1+Q\sim_s^{fr} P_2 +Q$. It is sufficient to prove the relation $R=\{(P_1+Q, P_2+Q)\}\cup \textbf{Id}$ is a FR strong step bisimulation, we omit it;
           \item $P_1\parallel Q\sim_s^{fr} P_2\parallel Q$. It is sufficient to prove the relation $R=\{(P_1\parallel Q, P_2\parallel Q)\}\cup \textbf{Id}$ is a FR strong step bisimulation, we omit it;
           \item $P_1\setminus L\sim_s^{fr} P_2\setminus L$. It is sufficient to prove the relation $R=\{(P_1\setminus L, P_2\setminus L)\}\cup \textbf{Id}$ is a FR strong step bisimulation, we omit it;
           \item $P_1[f]\sim_s^{fr} P_2[f]$. It is sufficient to prove the relation $R=\{(P_1[f], P_2[f])\}\cup \textbf{Id}$ is a FR strong step bisimulation, we omit it.
         \end{enumerate}
\end{enumerate}
\end{proof}

\begin{theorem}[Congruence for FR strong hp-bisimulation] \label{CSSB05}
We can enjoy the congruence for FR strong hp-bisimulation as follows.
\begin{enumerate}
  \item If $A\overset{\text{def}}{=}P$, then $A\sim_{hp}^{fr} P$;
  \item Let $P_1\sim_{hp}^{fr} P_2$. Then
        \begin{enumerate}
           \item $\alpha.P_1\sim_{hp}^f \alpha.P_2$;
           \item $\phi.P_1\sim_{hp}^f \phi.P_2$;
           \item $(\alpha_1\parallel\cdots\parallel\alpha_n).P_1\sim_{hp}^f (\alpha_1\parallel\cdots\parallel\alpha_n).P_2$;
           \item $P_1.\alpha[m]\sim_{hp}^r P_2.\alpha[m]$;
           \item $P_1.\phi\sim_{hp}^r P_2.\phi$;
           \item $P_1.(\alpha_1[m]\parallel\cdots\parallel\alpha_n[m])\sim_{hp}^r P_2.(\alpha_1[m]\parallel\cdots\parallel\alpha_n[m])$;
           \item $P_1+Q\sim_{hp}^{fr} P_2 +Q$;
           \item $P_1\parallel Q\sim_{hp}^{fr} P_2\parallel Q$;
           \item $P_1\setminus L\sim_{hp}^{fr} P_2\setminus L$;
           \item $P_1[f]\sim_{hp}^{fr} P_2[f]$.
         \end{enumerate}
\end{enumerate}
\end{theorem}

\begin{proof}
\begin{enumerate}
  \item If $A\overset{\text{def}}{=}P$, then $A\sim_{hp}^{fr} P$. It is obvious.
  \item Let $P_1\sim_{hp}^{fr} P_2$. Then
        \begin{enumerate}
           \item $\alpha.P_1\sim_{hp}^f \alpha.P_2$. It is sufficient to prove the relation $R=\{(\alpha.P_1, \alpha.P_2)\}\cup \textbf{Id}$ is a F strong hp-bisimulation, we omit it;
           \item $\phi.P_1\sim_{hp}^f \phi.P_2$. It is sufficient to prove the relation $R=\{(\phi.P_1, \phi.P_2)\}\cup \textbf{Id}$ is a F strong hp-bisimulation, we omit it;
           \item $(\alpha_1\parallel\cdots\parallel\alpha_n).P_1\sim_{hp}^f (\alpha_1\parallel\cdots\parallel\alpha_n).P_2$. It is sufficient to prove the relation $R=\{((\alpha_1\parallel\cdots\parallel\alpha_n).P_1, (\alpha_1\parallel\cdots\parallel\alpha_n).P_2)\}\cup \textbf{Id}$ is a F strong hp-bisimulation, we omit it;
           \item $P_1.\alpha[m]\sim_{hp}^r P_2.\alpha[m]$. It is sufficient to prove the relation $R=\{(P_1.\alpha[m], P_2.\alpha[m])\}\cup \textbf{Id}$ is a R strong hp-bisimulation, we omit it;
           \item $P_1.\phi\sim_{hp}^r P_2.\phi$. It is sufficient to prove the relation $R=\{(P_1.\phi, P_2.\phi)\}\cup \textbf{Id}$ is a R strong hp-bisimulation, we omit it;
           \item $P_1.(\alpha_1[m]\parallel\cdots\parallel\alpha_n[m])\sim_{hp}^r P_2.(\alpha_1[m]\parallel\cdots\parallel\alpha_n[m])$. It is sufficient to prove the relation $R=\{(P_1.(\alpha_1[m]\parallel\cdots\parallel\alpha_n[m]), P_2.(\alpha_1[m]\parallel\cdots\parallel\alpha_n[m]))\}\cup \textbf{Id}$ is a R strong hp-bisimulation, we omit it;
           \item $P_1+Q\sim_{hp}^{fr} P_2 +Q$. It is sufficient to prove the relation $R=\{(P_1+Q, P_2+Q)\}\cup \textbf{Id}$ is a FR strong hp-bisimulation, we omit it;
           \item $P_1\parallel Q\sim_{hp}^{fr} P_2\parallel Q$. It is sufficient to prove the relation $R=\{(P_1\parallel Q, P_2\parallel Q)\}\cup \textbf{Id}$ is a FR strong hp-bisimulation, we omit it;
           \item $P_1\setminus L\sim_{hp}^{fr} P_2\setminus L$. It is sufficient to prove the relation $R=\{(P_1\setminus L, P_2\setminus L)\}\cup \textbf{Id}$ is a FR strong hp-bisimulation, we omit it;
           \item $P_1[f]\sim_{hp}^{fr} P_2[f]$. It is sufficient to prove the relation $R=\{(P_1[f], P_2[f])\}\cup \textbf{Id}$ is a FR strong hp-bisimulation, we omit it.
         \end{enumerate}
\end{enumerate}
\end{proof}

\begin{theorem}[Congruence for FR strong hhp-bisimulation] \label{CSSB05}
We can enjoy the congruence for FR strong hhp-bisimulation as follows.
\begin{enumerate}
  \item If $A\overset{\text{def}}{=}P$, then $A\sim_{hhp}^{fr} P$;
  \item Let $P_1\sim_{hhp}^{fr} P_2$. Then
        \begin{enumerate}
           \item $\alpha.P_1\sim_{hhp}^f \alpha.P_2$;
           \item $\phi.P_1\sim_{hhp}^f \phi.P_2$;
           \item $(\alpha_1\parallel\cdots\parallel\alpha_n).P_1\sim_{hhp}^f (\alpha_1\parallel\cdots\parallel\alpha_n).P_2$;
           \item $P_1.\alpha[m]\sim_{hhp}^r P_2.\alpha[m]$;
           \item $P_1.\phi\sim_{hhp}^r P_2.\phi$;
           \item $P_1.(\alpha_1[m]\parallel\cdots\parallel\alpha_n[m])\sim_{hhp}^r P_2.(\alpha_1[m]\parallel\cdots\parallel\alpha_n[m])$;
           \item $P_1+Q\sim_{hhp}^{fr} P_2 +Q$;
           \item $P_1\parallel Q\sim_{hhp}^{fr} P_2\parallel Q$;
           \item $P_1\setminus L\sim_{hhp}^{fr} P_2\setminus L$;
           \item $P_1[f]\sim_{hhp}^{fr} P_2[f]$.
         \end{enumerate}
\end{enumerate}
\end{theorem}

\begin{proof}
\begin{enumerate}
  \item If $A\overset{\text{def}}{=}P$, then $A\sim_{hhp}^{fr} P$. It is obvious.
  \item Let $P_1\sim_{hhp}^{fr} P_2$. Then
        \begin{enumerate}
           \item $\alpha.P_1\sim_{hhp}^f \alpha.P_2$. It is sufficient to prove the relation $R=\{(\alpha.P_1, \alpha.P_2)\}\cup \textbf{Id}$ is a F strong hhp-bisimulation, we omit it;
           \item $\phi.P_1\sim_{hhp}^f \phi.P_2$. It is sufficient to prove the relation $R=\{(\phi.P_1, \phi.P_2)\}\cup \textbf{Id}$ is a F strong hhp-bisimulation, we omit it;
           \item $(\alpha_1\parallel\cdots\parallel\alpha_n).P_1\sim_{hhp}^f (\alpha_1\parallel\cdots\parallel\alpha_n).P_2$. It is sufficient to prove the relation $R=\{((\alpha_1\parallel\cdots\parallel\alpha_n).P_1, (\alpha_1\parallel\cdots\parallel\alpha_n).P_2)\}\cup \textbf{Id}$ is a F strong hhp-bisimulation, we omit it;
           \item $P_1.\alpha[m]\sim_{hhp}^r P_2.\alpha[m]$. It is sufficient to prove the relation $R=\{(P_1.\alpha[m], P_2.\alpha[m])\}\cup \textbf{Id}$ is a R strong hhp-bisimulation, we omit it;
           \item $P_1.\phi\sim_{hhp}^r P_2.\phi$. It is sufficient to prove the relation $R=\{(P_1.\phi, P_2.\phi)\}\cup \textbf{Id}$ is a R strong hhp-bisimulation, we omit it;
           \item $P_1.(\alpha_1[m]\parallel\cdots\parallel\alpha_n[m])\sim_{hhp}^r P_2.(\alpha_1[m]\parallel\cdots\parallel\alpha_n[m])$. It is sufficient to prove the relation $R=\{(P_1.(\alpha_1[m]\parallel\cdots\parallel\alpha_n[m]), P_2.(\alpha_1[m]\parallel\cdots\parallel\alpha_n[m]))\}\cup \textbf{Id}$ is a R strong hhp-bisimulation, we omit it;
           \item $P_1+Q\sim_{hhp}^{fr} P_2 +Q$. It is sufficient to prove the relation $R=\{(P_1+Q, P_2+Q)\}\cup \textbf{Id}$ is a FR strong hhp-bisimulation, we omit it;
           \item $P_1\parallel Q\sim_{hhp}^{fr} P_2\parallel Q$. It is sufficient to prove the relation $R=\{(P_1\parallel Q, P_2\parallel Q)\}\cup \textbf{Id}$ is a FR strong hhp-bisimulation, we omit it;
           \item $P_1\setminus L\sim_{hhp}^{fr} P_2\setminus L$. It is sufficient to prove the relation $R=\{(P_1\setminus L, P_2\setminus L)\}\cup \textbf{Id}$ is a FR strong hhp-bisimulation, we omit it;
           \item $P_1[f]\sim_{hhp}^{fr} P_2[f]$. It is sufficient to prove the relation $R=\{(P_1[f], P_2[f])\}\cup \textbf{Id}$ is a FR strong hhp-bisimulation, we omit it.
         \end{enumerate}
\end{enumerate}
\end{proof}

\subsubsection{Recursion}

\begin{definition}[Weakly guarded recursive expression]
$X$ is weakly guarded in $E$ if each occurrence of $X$ is with some subexpression $\alpha.F$ or $(\alpha_1\parallel\cdots\parallel\alpha_n).F$ or $F.\alpha[m]$ or 
$F.(\alpha_1[m]\parallel\cdots\parallel\alpha_n[m])$ of $E$.
\end{definition}

\begin{lemma}\label{LUS06}
If the variables $\widetilde{X}$ are weakly guarded in $E$, and $\langle E\{\widetilde{P}/\widetilde{X}\},s\rangle\xrightarrow{\{\alpha_1,\cdots,\alpha_n\}}\langle P',s'\rangle$ or 
$\langle E\{\widetilde{P}/\widetilde{X}\},s\rangle\xtworightarrow{\{\alpha_1[m],\cdots,\alpha_n[m]\}}\langle P',s'\rangle$, then $P'$ takes the form 
$E'\{\widetilde{P}/\widetilde{X}\}$ for some expression $E'$, and moreover, for any $\widetilde{Q}$, 
$\langle E\{\widetilde{Q}/\widetilde{X}\},s\rangle\xrightarrow{\{\alpha_1,\cdots,\alpha_n\}}\langle E'\{\widetilde{Q}/\widetilde{X}\},s'\rangle$ or 
$\langle E\{\widetilde{Q}/\widetilde{X}\},s\rangle\xtworightarrow{\{\alpha_1[m],\cdots,\alpha_n[m]\}}\langle E'\{\widetilde{Q}/\widetilde{X}\},s'\rangle$.
\end{lemma}

\begin{proof}
We only prove the case of forward transition. 

It needs to induct on the depth of the inference of $\langle E\{\widetilde{P}/\widetilde{X}\},s\rangle\xrightarrow{\{\alpha_1,\cdots,\alpha_n\}}\langle P',s'\rangle$.

\begin{enumerate}
  \item Case $E\equiv Y$, a variable. Then $Y\notin \widetilde{X}$. Since $\widetilde{X}$ are weakly guarded, $\langle Y\{\widetilde{P}/\widetilde{X}\equiv Y\},s\rangle\nrightarrow$, this case is
  impossible.
  \item Case $E\equiv\beta.F$. Then we must have $\alpha=\beta$, and $P'\equiv F\{\widetilde{P}/\widetilde{X}\}$, and
  $\langle E\{\widetilde{Q}/\widetilde{X}\},s\rangle\equiv \langle \beta.F\{\widetilde{Q}/\widetilde{X}\},s\rangle \xrightarrow{\beta}\langle F\{\widetilde{Q}/\widetilde{X}\},s'\rangle$,
  then, let $E'$ be $F$, as desired.
  \item Case $E\equiv(\beta_1\parallel\cdots\parallel\beta_n).F$. Then we must have $\alpha_i=\beta_i$ for $1\leq i\leq n$, and $P'\equiv F\{\widetilde{P}/\widetilde{X}\}$, and
  $\langle E\{\widetilde{Q}/\widetilde{X}\},s\rangle\equiv \langle(\beta_1\parallel\cdots\parallel\beta_n).F\{\widetilde{Q}/\widetilde{X}\},s\rangle \xrightarrow{\{\beta_1,\cdots,\beta_n\}}\langle F\{\widetilde{Q}/\widetilde{X}\},s'\rangle$,
  then, let $E'$ be $F$, as desired.
  \item Case $E\equiv E_1+E_2$. Then either $\langle E_1\{\widetilde{P}/\widetilde{X}\},s\rangle \xrightarrow{\{\alpha_1,\cdots,\alpha_n\}}\langle P',s'\rangle$ or
  $\langle E_2\{\widetilde{P}/\widetilde{X}\},s\rangle \xrightarrow{\{\alpha_1,\cdots,\alpha_n\}}\langle P',s'\rangle$, then, we can apply this lemma in either case, as desired.
  \item Case $E\equiv E_1\parallel E_2$. There are four possibilities.
  \begin{enumerate}
    \item We may have $\langle E_1\{\widetilde{P}/\widetilde{X}\},s\rangle \xrightarrow{\alpha}\langle P_1',s'\rangle$ and $\langle E_2\{\widetilde{P}/\widetilde{X}\},s\rangle\nrightarrow$
    with $P'\equiv P_1'\parallel (E_2\{\widetilde{P}/\widetilde{X}\})$, then by applying this lemma, $P_1'$ is of the form $E_1'\{\widetilde{P}/\widetilde{X}\}$, and for any $Q$,
    $\langle E_1\{\widetilde{Q}/\widetilde{X}\},s\rangle\xrightarrow{\alpha} \langle E_1'\{\widetilde{Q}/\widetilde{X}\},s'\rangle$. So, $P'$ is of the form
    $E_1'\parallel E_2\{\widetilde{P}/\widetilde{X}\}$, and for any $Q$,
    $\langle E\{\widetilde{Q}/\widetilde{X}\}\equiv E_1\{\widetilde{Q}/\widetilde{X}\}\parallel E_2\{\widetilde{Q}/\widetilde{X}\},s\rangle\xrightarrow{\alpha} \langle(E_1'\parallel E_2)\{\widetilde{Q}/\widetilde{X}\},s'\rangle$,
    then, let $E'$ be $E_1'\parallel E_2$, as desired.
    \item We may have $\langle E_2\{\widetilde{P}/\widetilde{X}\},s\rangle \xrightarrow{\alpha}\langle P_2',s'\rangle$ and $\langle E_1\{\widetilde{P}/\widetilde{X}\},s\rangle\nrightarrow$
    with $P'\equiv P_2'\parallel (E_1\{\widetilde{P}/\widetilde{X}\})$, this case can be prove similarly to the above subcase, as desired.
    \item We may have $\langle E_1\{\widetilde{P}/\widetilde{X}\},s\rangle \xrightarrow{\alpha}\langle P_1',s'\rangle$ and
    $\langle E_2\{\widetilde{P}/\widetilde{X}\},s\rangle\xrightarrow{\beta}\langle P_2',s''\rangle$ with $\alpha\neq\overline{\beta}$ and $P'\equiv P_1'\parallel P_2'$, then by
    applying this lemma, $P_1'$ is of the form $E_1'\{\widetilde{P}/\widetilde{X}\}$, and for any $Q$,
    $\langle E_1\{\widetilde{Q}/\widetilde{X}\},s\rangle\xrightarrow{\alpha} \langle E_1'\{\widetilde{Q}/\widetilde{X}\},s'\rangle$; $P_2'$ is of the form
    $E_2'\{\widetilde{P}/\widetilde{X}\}$, and for any $Q$, $\langle E_2\{\widetilde{Q}/\widetilde{X}\},s\rangle\xrightarrow{\alpha} \langle E_2'\{\widetilde{Q}/\widetilde{X}\},s''\rangle$.
    So, $P'$ is of the form $E_1'\parallel E_2'\{\widetilde{P}/\widetilde{X}\}$, and for any $Q$,
    $\langle E\{\widetilde{Q}/\widetilde{X}\}\equiv E_1\{\widetilde{Q}/\widetilde{X}\}\parallel E_2\{\widetilde{Q}/\widetilde{X}\},s\rangle\xrightarrow{\{\alpha,\beta\}}
    \langle (E_1'\parallel E_2')\{\widetilde{Q}/\widetilde{X}\},s'\cup s''\rangle$, then, let $E'$ be $E_1'\parallel E_2'$, as desired.
    \item We may have $\langle E_1\{\widetilde{P}/\widetilde{X}\},s\rangle \xrightarrow{l}\langle P_1',s'\rangle$ and
    $\langle E_2\{\widetilde{P}/\widetilde{X}\},s\rangle\xrightarrow{\overline{l}}\langle P_2',s''\rangle$ with $P'\equiv P_1'\parallel P_2'$, then by applying this lemma,
    $P_1'$ is of the form $E_1'\{\widetilde{P}/\widetilde{X}\}$, and for any $Q$, $\langle E_1\{\widetilde{Q}/\widetilde{X}\},s\rangle\xrightarrow{l} \langle E_1'\{\widetilde{Q}/\widetilde{X}\},s'\rangle$;
    $P_2'$ is of the form $E_2'\{\widetilde{P}/\widetilde{X}\}$, and for any $Q$, $\langle E_2\{\widetilde{Q}/\widetilde{X}\},s\rangle\xrightarrow{\overline{l}}\langle E_2'\{\widetilde{Q}/\widetilde{X}\},s''\rangle$.
    So, $P'$ is of the form $E_1'\parallel E_2'\{\widetilde{P}/\widetilde{X}\}$, and for any $Q$, $\langle E\{\widetilde{Q}/\widetilde{X}\}\equiv E_1\{\widetilde{Q}/\widetilde{X}\}\parallel E_2\{\widetilde{Q}/\widetilde{X}\},s\rangle
    \xrightarrow{\tau} \langle (E_1'\parallel E_2')\{\widetilde{Q}/\widetilde{X}\},s'\cup s''\rangle$, then, let $E'$ be $E_1'\parallel E_2'$, as desired.
  \end{enumerate}
  \item Case $E\equiv F[R]$ and $E\equiv F\setminus L$. These cases can be prove similarly to the above case.
  \item Case $E\equiv C$, an agent constant defined by $C\overset{\text{def}}{=}R$. Then there is no $X\in\widetilde{X}$ occurring in $E$, so
  $\langle C,s\rangle\xrightarrow{\{\alpha_1,\cdots,\alpha_n\}}\langle P',s'\rangle$, let $E'$ be $P'$, as desired.
\end{enumerate}

For the case of reverse transition, it can be proven similarly, we omit it.
\end{proof}

\begin{theorem}[Unique solution of equations for FR strong pomset bisimulation]
Let the recursive expressions $E_i(i\in I)$ contain at most the variables $X_i(i\in I)$, and let each $X_j(j\in I)$ be weakly guarded in each $E_i$. Then,

If $\widetilde{P}\sim_p^{fr} \widetilde{E}\{\widetilde{P}/\widetilde{X}\}$ and $\widetilde{Q}\sim_p^{fr} \widetilde{E}\{\widetilde{Q}/\widetilde{X}\}$, then 
$\widetilde{P}\sim_p^{fr} \widetilde{Q}$.
\end{theorem}

\begin{proof}
We only prove the case of forward transition. 

It is sufficient to induct on the depth of the inference of $\langle E\{\widetilde{P}/\widetilde{X}\},s\rangle\xrightarrow{\{\alpha_1,\cdots,\alpha_n\}}\langle P',s'\rangle$.

\begin{enumerate}
  \item Case $E\equiv X_i$. Then we have $\langle E\{\widetilde{P}/\widetilde{X}\},s\rangle\equiv \langle P_i,s\rangle\xrightarrow{\{\alpha_1,\cdots,\alpha_n\}}\langle P',s'\rangle$,
  since $P_i\sim_p^{fr} E_i\{\widetilde{P}/\widetilde{X}\}$, we have $\langle E_i\{\widetilde{P}/\widetilde{X}\},s\rangle\xrightarrow{\{\alpha_1,\cdots,\alpha_n\}}\langle P'',s'\rangle\sim_p^{fr} \langle P',s'\rangle$.
  Since $\widetilde{X}$ are weakly guarded in $E_i$, by Lemma \ref{LUS06}, $P''\equiv E'\{\widetilde{P}/\widetilde{X}\}$ and $\langle E_i\{\widetilde{P}/\widetilde{X}\},s\rangle
  \xrightarrow{\{\alpha_1,\cdots,\alpha_n\}} \langle E'\{\widetilde{P}/\widetilde{X}\},s'\rangle$. Since
  $E\{\widetilde{Q}/\widetilde{X}\}\equiv X_i\{\widetilde{Q}/\widetilde{X}\} \equiv Q_i\sim_p^{fr} E_i\{\widetilde{Q}/\widetilde{X}\}$, $\langle E\{\widetilde{Q}/\widetilde{X}\},s\rangle\xrightarrow{\{\alpha_1,\cdots,\alpha_n\}}\langle Q',s'\rangle\sim_p^{fr} \langle E'\{\widetilde{Q}/\widetilde{X}\},s'\rangle$.
  So, $P'\sim_p^{fr} Q'$, as desired.
  \item Case $E\equiv\alpha.F$. This case can be proven similarly.
  \item Case $E\equiv(\alpha_1\parallel\cdots\parallel\alpha_n).F$. This case can be proven similarly.
  \item Case $E\equiv E_1+E_2$. We have $\langle E_i\{\widetilde{P}/\widetilde{X}\},s\rangle \xrightarrow{\{\alpha_1,\cdots,\alpha_n\}}\langle P',s'\rangle$,
  $\langle E_i\{\widetilde{Q}/\widetilde{X}\},s\rangle \xrightarrow{\{\alpha_1,\cdots,\alpha_n\}}\langle Q',s'\rangle$, then, $P'\sim_p^{fr} Q'$, as desired.
  \item Case $E\equiv E_1\parallel E_2$, $E\equiv F[R]$ and $E\equiv F\setminus L$, $E\equiv C$. These cases can be prove similarly to the above case.
\end{enumerate}

For the case of reverse transition, it can be proven similarly, we omit it.
\end{proof}

\begin{theorem}[Unique solution of equations for FR strong step bisimulation]
Let the recursive expressions $E_i(i\in I)$ contain at most the variables $X_i(i\in I)$, and let each $X_j(j\in I)$ be weakly guarded in each $E_i$. Then,

If $\widetilde{P}\sim_s^{fr} \widetilde{E}\{\widetilde{P}/\widetilde{X}\}$ and $\widetilde{Q}\sim_s^{fr} \widetilde{E}\{\widetilde{Q}/\widetilde{X}\}$, then
$\widetilde{P}\sim_s^{fr} \widetilde{Q}$.
\end{theorem}

\begin{proof}
We only prove the case of forward transition.

It is sufficient to induct on the depth of the inference of $\langle E\{\widetilde{P}/\widetilde{X}\},s\rangle\xrightarrow{\{\alpha_1,\cdots,\alpha_n\}}\langle P',s'\rangle$.

\begin{enumerate}
  \item Case $E\equiv X_i$. Then we have $\langle E\{\widetilde{P}/\widetilde{X}\},s\rangle\equiv \langle P_i,s\rangle\xrightarrow{\{\alpha_1,\cdots,\alpha_n\}}\langle P',s'\rangle$,
  since $P_i\sim_s^{fr} E_i\{\widetilde{P}/\widetilde{X}\}$, we have $\langle E_i\{\widetilde{P}/\widetilde{X}\},s\rangle\xrightarrow{\{\alpha_1,\cdots,\alpha_n\}}\langle P'',s'\rangle\sim_s^{fr} \langle P',s'\rangle$.
  Since $\widetilde{X}$ are weakly guarded in $E_i$, by Lemma \ref{LUS06}, $P''\equiv E'\{\widetilde{P}/\widetilde{X}\}$ and $\langle E_i\{\widetilde{P}/\widetilde{X}\},s\rangle
  \xrightarrow{\{\alpha_1,\cdots,\alpha_n\}} \langle E'\{\widetilde{P}/\widetilde{X}\},s'\rangle$. Since
  $E\{\widetilde{Q}/\widetilde{X}\}\equiv X_i\{\widetilde{Q}/\widetilde{X}\} \equiv Q_i\sim_s^{fr} E_i\{\widetilde{Q}/\widetilde{X}\}$, $\langle E\{\widetilde{Q}/\widetilde{X}\},s\rangle\xrightarrow{\{\alpha_1,\cdots,\alpha_n\}}\langle Q',s'\rangle\sim_s^{fr} \langle E'\{\widetilde{Q}/\widetilde{X}\},s'\rangle$.
  So, $P'\sim_s^{fr} Q'$, as desired.
  \item Case $E\equiv\alpha.F$. This case can be proven similarly.
  \item Case $E\equiv(\alpha_1\parallel\cdots\parallel\alpha_n).F$. This case can be proven similarly.
  \item Case $E\equiv E_1+E_2$. We have $\langle E_i\{\widetilde{P}/\widetilde{X}\},s\rangle \xrightarrow{\{\alpha_1,\cdots,\alpha_n\}}\langle P',s'\rangle$,
  $\langle E_i\{\widetilde{Q}/\widetilde{X}\},s\rangle \xrightarrow{\{\alpha_1,\cdots,\alpha_n\}}\langle Q',s'\rangle$, then, $P'\sim_s^{fr} Q'$, as desired.
  \item Case $E\equiv E_1\parallel E_2$, $E\equiv F[R]$ and $E\equiv F\setminus L$, $E\equiv C$. These cases can be prove similarly to the above case.
\end{enumerate}

For the case of reverse transition, it can be proven similarly, we omit it.
\end{proof}

\begin{theorem}[Unique solution of equations for FR strong hp-bisimulation]
Let the recursive expressions $E_i(i\in I)$ contain at most the variables $X_i(i\in I)$, and let each $X_j(j\in I)$ be weakly guarded in each $E_i$. Then,

If $\widetilde{P}\sim_{hp}^{fr} \widetilde{E}\{\widetilde{P}/\widetilde{X}\}$ and $\widetilde{Q}\sim_{hp}^{fr} \widetilde{E}\{\widetilde{Q}/\widetilde{X}\}$, then
$\widetilde{P}\sim_{hp}^{fr} \widetilde{Q}$.
\end{theorem}

\begin{proof}
We only prove the case of forward transition.

It is sufficient to induct on the depth of the inference of $\langle E\{\widetilde{P}/\widetilde{X}\},s\rangle\xrightarrow{\{\alpha_1,\cdots,\alpha_n\}}\langle P',s'\rangle$.

\begin{enumerate}
  \item Case $E\equiv X_i$. Then we have $\langle E\{\widetilde{P}/\widetilde{X}\},s\rangle\equiv \langle P_i,s\rangle\xrightarrow{\{\alpha_1,\cdots,\alpha_n\}}\langle P',s'\rangle$,
  since $P_i\sim_{hp}^{fr} E_i\{\widetilde{P}/\widetilde{X}\}$, we have $\langle E_i\{\widetilde{P}/\widetilde{X}\},s\rangle\xrightarrow{\{\alpha_1,\cdots,\alpha_n\}}\langle P'',s'\rangle\sim_{hp}^{fr} \langle P',s'\rangle$.
  Since $\widetilde{X}$ are weakly guarded in $E_i$, by Lemma \ref{LUS06}, $P''\equiv E'\{\widetilde{P}/\widetilde{X}\}$ and $\langle E_i\{\widetilde{P}/\widetilde{X}\},s\rangle
  \xrightarrow{\{\alpha_1,\cdots,\alpha_n\}} \langle E'\{\widetilde{P}/\widetilde{X}\},s'\rangle$. Since
  $E\{\widetilde{Q}/\widetilde{X}\}\equiv X_i\{\widetilde{Q}/\widetilde{X}\} \equiv Q_i\sim_{hp}^{fr} E_i\{\widetilde{Q}/\widetilde{X}\}$, $\langle E\{\widetilde{Q}/\widetilde{X}\},s\rangle\xrightarrow{\{\alpha_1,\cdots,\alpha_n\}}\langle Q',s'\rangle\sim_{hp}^{fr} \langle E'\{\widetilde{Q}/\widetilde{X}\},s'\rangle$.
  So, $P'\sim_{hp}^{fr} Q'$, as desired.
  \item Case $E\equiv\alpha.F$. This case can be proven similarly.
  \item Case $E\equiv(\alpha_1\parallel\cdots\parallel\alpha_n).F$. This case can be proven similarly.
  \item Case $E\equiv E_1+E_2$. We have $\langle E_i\{\widetilde{P}/\widetilde{X}\},s\rangle \xrightarrow{\{\alpha_1,\cdots,\alpha_n\}}\langle P',s'\rangle$,
  $\langle E_i\{\widetilde{Q}/\widetilde{X}\},s\rangle \xrightarrow{\{\alpha_1,\cdots,\alpha_n\}}\langle Q',s'\rangle$, then, $P'\sim_{hp}^{fr} Q'$, as desired.
  \item Case $E\equiv E_1\parallel E_2$, $E\equiv F[R]$ and $E\equiv F\setminus L$, $E\equiv C$. These cases can be prove similarly to the above case.
\end{enumerate}

For the case of reverse transition, it can be proven similarly, we omit it.
\end{proof}

\begin{theorem}[Unique solution of equations for FR strong hhp-bisimulation]
Let the recursive expressions $E_i(i\in I)$ contain at most the variables $X_i(i\in I)$, and let each $X_j(j\in I)$ be weakly guarded in each $E_i$. Then,

If $\widetilde{P}\sim_{hhp}^{fr} \widetilde{E}\{\widetilde{P}/\widetilde{X}\}$ and $\widetilde{Q}\sim_{hhp}^{fr} \widetilde{E}\{\widetilde{Q}/\widetilde{X}\}$, then
$\widetilde{P}\sim_{hhp}^{fr} \widetilde{Q}$.
\end{theorem}

\begin{proof}
We only prove the case of forward transition.

It is sufficient to induct on the depth of the inference of $\langle E\{\widetilde{P}/\widetilde{X}\},s\rangle\xrightarrow{\{\alpha_1,\cdots,\alpha_n\}}\langle P',s'\rangle$.

\begin{enumerate}
  \item Case $E\equiv X_i$. Then we have $\langle E\{\widetilde{P}/\widetilde{X}\},s\rangle\equiv \langle P_i,s\rangle\xrightarrow{\{\alpha_1,\cdots,\alpha_n\}}\langle P',s'\rangle$,
  since $P_i\sim_{hhp}^{fr} E_i\{\widetilde{P}/\widetilde{X}\}$, we have $\langle E_i\{\widetilde{P}/\widetilde{X}\},s\rangle\xrightarrow{\{\alpha_1,\cdots,\alpha_n\}}\langle P'',s'\rangle\sim_{hhp}^{fr} \langle P',s'\rangle$.
  Since $\widetilde{X}$ are weakly guarded in $E_i$, by Lemma \ref{LUS06}, $P''\equiv E'\{\widetilde{P}/\widetilde{X}\}$ and $\langle E_i\{\widetilde{P}/\widetilde{X}\},s\rangle
  \xrightarrow{\{\alpha_1,\cdots,\alpha_n\}} \langle E'\{\widetilde{P}/\widetilde{X}\},s'\rangle$. Since
  $E\{\widetilde{Q}/\widetilde{X}\}\equiv X_i\{\widetilde{Q}/\widetilde{X}\} \equiv Q_i\sim_{hhp}^{fr} E_i\{\widetilde{Q}/\widetilde{X}\}$, $\langle E\{\widetilde{Q}/\widetilde{X}\},s\rangle\xrightarrow{\{\alpha_1,\cdots,\alpha_n\}}\langle Q',s'\rangle\sim_{hhp}^{fr} \langle E'\{\widetilde{Q}/\widetilde{X}\},s'\rangle$.
  So, $P'\sim_{hhp}^{fr} Q'$, as desired.
  \item Case $E\equiv\alpha.F$. This case can be proven similarly.
  \item Case $E\equiv(\alpha_1\parallel\cdots\parallel\alpha_n).F$. This case can be proven similarly.
  \item Case $E\equiv E_1+E_2$. We have $\langle E_i\{\widetilde{P}/\widetilde{X}\},s\rangle \xrightarrow{\{\alpha_1,\cdots,\alpha_n\}}\langle P',s'\rangle$,
  $\langle E_i\{\widetilde{Q}/\widetilde{X}\},s\rangle \xrightarrow{\{\alpha_1,\cdots,\alpha_n\}}\langle Q',s'\rangle$, then, $P'\sim_{hhp}^{fr} Q'$, as desired.
  \item Case $E\equiv E_1\parallel E_2$, $E\equiv F[R]$ and $E\equiv F\setminus L$, $E\equiv C$. These cases can be prove similarly to the above case.
\end{enumerate}

For the case of reverse transition, it can be proven similarly, we omit it.
\end{proof}

\subsection{Weak Bisimulations}\label{wftcbctcrg}

\subsubsection{Laws}

Remembering that $\tau$ can neither be restricted nor relabeled, we know that the monoid laws, the static laws, the guards laws, and the new expansion law
still hold with respect to the corresponding FR weak truly concurrent bisimulations. And also, we can enjoy the congruence of Prefix, Summation, Composition, Restriction, Relabelling 
and Constants with respect to corresponding FR weak truly concurrent bisimulations. We will not retype these laws, and just give the $\tau$-specific laws. The forward and reverse 
transition rules of $\tau$ are shown in Table \ref{TRForTAU06}, where $\xrightarrow{\tau}\surd$ is a predicate which represents a successful termination after execution of the silent 
step $\tau$.

\begin{center}
    \begin{table}
        $$\frac{}{\langle \tau,s\rangle\xrightarrow{\tau}\langle\surd,\tau(s)\rangle}$$
        $$\frac{}{\langle \tau,s\rangle\xtworightarrow{\tau}\langle\surd,\tau(s)\rangle}$$
        \caption{Forward and reverse transition rules of $\tau$}
        \label{TRForTAU06}
    \end{table}
\end{center}

\begin{proposition}[$\tau$ laws for FR weak pomset bisimulation]
The $\tau$ laws for FR weak pomset bisimulation is as follows.
\begin{enumerate}
  \item $P\approx_p^f \tau.P$;
  \item $P\approx_p^r P.\tau$;
  \item $\alpha.\tau.P\approx_p^f \alpha.P$;
  \item $P.\tau.\alpha[m]\approx_p^r P.\alpha[m]$;
  \item $(\alpha_1\parallel\cdots\parallel\alpha_n).\tau.P\approx_p^f (\alpha_1\parallel\cdots\parallel\alpha_n).P$;
  \item $P.\tau.(\alpha_1[m]\parallel\cdots\parallel\alpha_n[m])\approx_p^r P.(\alpha_1[m]\parallel\cdots\parallel\alpha_n[m])$;
  \item $P+\tau.P\approx_p^f \tau.P$;
  \item $P+P.\tau\approx_p^r P.\tau$;
  \item $\alpha.(\tau.(P+Q)+P)\approx_p^f\alpha.(P+Q)$;
  \item $((P+Q).\tau+P).\alpha[m]\approx_p^r(P+Q).\alpha[m]$;
  \item $\phi.(\tau.(P+Q)+P)\approx_p^f\phi.(P+Q)$;
  \item $((P+Q).\tau+P).\phi\approx_p^r(P+Q).\phi$;
  \item $(\alpha_1\parallel\cdots\parallel\alpha_n).(\tau.(P+Q)+P)\approx_p^f(\alpha_1\parallel\cdots\parallel\alpha_n).(P+Q)$;
  \item $((P+Q).\tau+P).(\alpha_1[m]\parallel\cdots\parallel\alpha_n[m])\approx_p^r(P+Q).(\alpha_1[m]\parallel\cdots\parallel\alpha_n[m])$;
  \item $P\approx_p^{fr} \tau\parallel P$.
\end{enumerate}
\end{proposition}

\begin{proof}
\begin{enumerate}
  \item $P\approx_p^f \tau.P$. It is sufficient to prove the relation $R=\{(P, \tau.P)\}\cup \textbf{Id}$ is a F weak pomset bisimulation, we omit it;
  \item $P\approx_p^r P.\tau$. It is sufficient to prove the relation $R=\{(P, P.\tau)\}\cup \textbf{Id}$ is a R weak pomset bisimulation, we omit it;
  \item $\alpha.\tau.P\approx_p^f \alpha.P$. It is sufficient to prove the relation $R=\{(\alpha.\tau.P, \alpha.P)\}\cup \textbf{Id}$ is a F weak pomset bisimulation, we omit it;
  \item $P.\tau.\alpha[m]\approx_p^r P.\alpha[m]$. It is sufficient to prove the relation $R=\{(P.\tau.\alpha[m], P.\alpha[m])\}\cup \textbf{Id}$ is a R weak pomset bisimulation, we omit it;
  \item $(\alpha_1\parallel\cdots\parallel\alpha_n).\tau.P\approx_p^f (\alpha_1\parallel\cdots\parallel\alpha_n).P$. It is sufficient to prove the relation $R=\{((\alpha_1\parallel\cdots\parallel\alpha_n).\tau.P, (\alpha_1\parallel\cdots\parallel\alpha_n).P)\}\cup \textbf{Id}$ is a F weak pomset bisimulation, we omit it;
  \item $P.\tau.(\alpha_1[m]\parallel\cdots\parallel\alpha_n[m])\approx_p^r P.(\alpha_1[m]\parallel\cdots\parallel\alpha_n[m])$. It is sufficient to prove the relation $R=\{(P.\tau.(\alpha_1[m]\parallel\cdots\parallel\alpha_n[m]), P.(\alpha_1[m]\parallel\cdots\parallel\alpha_n[m]))\}\cup \textbf{Id}$ is a R weak pomset bisimulation, we omit it;
  \item $P+\tau.P\approx_p^f \tau.P$. It is sufficient to prove the relation $R=\{(P+\tau.P, \tau.P)\}\cup \textbf{Id}$ is a F weak pomset bisimulation, we omit it;
  \item $P+P.\tau\approx_p^r P.\tau$. It is sufficient to prove the relation $R=\{(P+P.\tau, P.\tau)\}\cup \textbf{Id}$ is a R weak pomset bisimulation, we omit it;
  \item $\alpha.(\tau.(P+Q)+P)\approx_p^f\alpha.(P+Q)$. It is sufficient to prove the relation $R=\{(\alpha.(\tau.(P+Q)+P), \alpha.(P+Q))\}\cup \textbf{Id}$ is a F weak pomset bisimulation, we omit it;
  \item $((P+Q).\tau+P).\alpha[m]\approx_p^r(P+Q).\alpha[m]$. It is sufficient to prove the relation $R=\{(((P+Q).\tau+P).\alpha[m], (P+Q).\alpha[m])\}\cup \textbf{Id}$ is a R weak pomset bisimulation, we omit it;
  \item $\phi.(\tau.(P+Q)+P)\approx_p^f\phi.(P+Q)$. It is sufficient to prove the relation $R=\{(\phi.(\tau.(P+Q)+P), \phi.(P+Q))\}\cup \textbf{Id}$ is a F weak pomset bisimulation, we omit it;
  \item $((P+Q).\tau+P).\phi\approx_p^r(P+Q).\phi$. It is sufficient to prove the relation $R=\{(((P+Q).\tau+P).\phi, (P+Q).\phi)\}\cup \textbf{Id}$ is a R weak pomset bisimulation, we omit it;
  \item $(\alpha_1\parallel\cdots\parallel\alpha_n).(\tau.(P+Q)+P)\approx_p^f(\alpha_1\parallel\cdots\parallel\alpha_n).(P+Q)$. It is sufficient to prove the relation $R=\{((\alpha_1\parallel\cdots\parallel\alpha_n).(\tau.(P+Q)+P), (\alpha_1\parallel\cdots\parallel\alpha_n).(P+Q))\}\cup \textbf{Id}$ is a F weak pomset bisimulation, we omit it;
  \item $((P+Q).\tau+P).(\alpha_1[m]\parallel\cdots\parallel\alpha_n[m])\approx_p^r(P+Q).(\alpha_1[m]\parallel\cdots\parallel\alpha_n[m])$. It is sufficient to prove the relation $R=\{(((P+Q).\tau+P).(\alpha_1[m]\parallel\cdots\parallel\alpha_n[m]), (P+Q).(\alpha_1[m]\parallel\cdots\parallel\alpha_n[m]))\}\cup \textbf{Id}$ is a R weak pomset bisimulation, we omit it;
  \item $P\approx_p^{fr} \tau\parallel P$. It is sufficient to prove the relation $R=\{(P, \tau\parallel P)\}\cup \textbf{Id}$ is a FR weak pomset bisimulation, we omit it.
\end{enumerate}
\end{proof}

\begin{proposition}[$\tau$ laws for FR weak step bisimulation]
The $\tau$ laws for FR weak step bisimulation is as follows.
\begin{enumerate}
  \item $P\approx_s^f \tau.P$;
  \item $P\approx_s^r P.\tau$;
  \item $\alpha.\tau.P\approx_s^f \alpha.P$;
  \item $P.\tau.\alpha[m]\approx_s^r P.\alpha[m]$;
  \item $(\alpha_1\parallel\cdots\parallel\alpha_n).\tau.P\approx_s^f (\alpha_1\parallel\cdots\parallel\alpha_n).P$;
  \item $P.\tau.(\alpha_1[m]\parallel\cdots\parallel\alpha_n[m])\approx_s^r P.(\alpha_1[m]\parallel\cdots\parallel\alpha_n[m])$;
  \item $P+\tau.P\approx_s^f \tau.P$;
  \item $P+P.\tau\approx_s^r P.\tau$;
  \item $\alpha.(\tau.(P+Q)+P)\approx_s^f\alpha.(P+Q)$;
  \item $((P+Q).\tau+P).\alpha[m]\approx_s^r(P+Q).\alpha[m]$;
  \item $\phi.(\tau.(P+Q)+P)\approx_s^f\phi.(P+Q)$;
  \item $((P+Q).\tau+P).\phi\approx_s^r(P+Q).\phi$;
  \item $(\alpha_1\parallel\cdots\parallel\alpha_n).(\tau.(P+Q)+P)\approx_s^f(\alpha_1\parallel\cdots\parallel\alpha_n).(P+Q)$;
  \item $((P+Q).\tau+P).(\alpha_1[m]\parallel\cdots\parallel\alpha_n[m])\approx_s^r(P+Q).(\alpha_1[m]\parallel\cdots\parallel\alpha_n[m])$;
  \item $P\approx_s^{fr} \tau\parallel P$.
\end{enumerate}
\end{proposition}

\begin{proof}
\begin{enumerate}
  \item $P\approx_s^f \tau.P$. It is sufficient to prove the relation $R=\{(P, \tau.P)\}\cup \textbf{Id}$ is a F weak step bisimulation, we omit it;
  \item $P\approx_s^r P.\tau$. It is sufficient to prove the relation $R=\{(P, P.\tau)\}\cup \textbf{Id}$ is a R weak step bisimulation, we omit it;
  \item $\alpha.\tau.P\approx_s^f \alpha.P$. It is sufficient to prove the relation $R=\{(\alpha.\tau.P, \alpha.P)\}\cup \textbf{Id}$ is a F weak step bisimulation, we omit it;
  \item $P.\tau.\alpha[m]\approx_s^r P.\alpha[m]$. It is sufficient to prove the relation $R=\{(P.\tau.\alpha[m], P.\alpha[m])\}\cup \textbf{Id}$ is a R weak step bisimulation, we omit it;
  \item $(\alpha_1\parallel\cdots\parallel\alpha_n).\tau.P\approx_s^f (\alpha_1\parallel\cdots\parallel\alpha_n).P$. It is sufficient to prove the relation $R=\{((\alpha_1\parallel\cdots\parallel\alpha_n).\tau.P, (\alpha_1\parallel\cdots\parallel\alpha_n).P)\}\cup \textbf{Id}$ is a F weak step bisimulation, we omit it;
  \item $P.\tau.(\alpha_1[m]\parallel\cdots\parallel\alpha_n[m])\approx_s^r P.(\alpha_1[m]\parallel\cdots\parallel\alpha_n[m])$. It is sufficient to prove the relation $R=\{(P.\tau.(\alpha_1[m]\parallel\cdots\parallel\alpha_n[m]), P.(\alpha_1[m]\parallel\cdots\parallel\alpha_n[m]))\}\cup \textbf{Id}$ is a R weak step bisimulation, we omit it;
  \item $P+\tau.P\approx_s^f \tau.P$. It is sufficient to prove the relation $R=\{(P+\tau.P, \tau.P)\}\cup \textbf{Id}$ is a F weak step bisimulation, we omit it;
  \item $P+P.\tau\approx_s^r P.\tau$. It is sufficient to prove the relation $R=\{(P+P.\tau, P.\tau)\}\cup \textbf{Id}$ is a R weak step bisimulation, we omit it;
  \item $\alpha.(\tau.(P+Q)+P)\approx_s^f\alpha.(P+Q)$. It is sufficient to prove the relation $R=\{(\alpha.(\tau.(P+Q)+P), \alpha.(P+Q))\}\cup \textbf{Id}$ is a F weak step bisimulation, we omit it;
  \item $((P+Q).\tau+P).\alpha[m]\approx_s^r(P+Q).\alpha[m]$. It is sufficient to prove the relation $R=\{(((P+Q).\tau+P).\alpha[m], (P+Q).\alpha[m])\}\cup \textbf{Id}$ is a R weak step bisimulation, we omit it;
  \item $\phi.(\tau.(P+Q)+P)\approx_s^f\phi.(P+Q)$. It is sufficient to prove the relation $R=\{(\phi.(\tau.(P+Q)+P), \phi.(P+Q))\}\cup \textbf{Id}$ is a F weak step bisimulation, we omit it;
  \item $((P+Q).\tau+P).\phi\approx_s^r(P+Q).\phi$. It is sufficient to prove the relation $R=\{(((P+Q).\tau+P).\phi, (P+Q).\phi)\}\cup \textbf{Id}$ is a R weak step bisimulation, we omit it;
  \item $(\alpha_1\parallel\cdots\parallel\alpha_n).(\tau.(P+Q)+P)\approx_s^f(\alpha_1\parallel\cdots\parallel\alpha_n).(P+Q)$. It is sufficient to prove the relation $R=\{((\alpha_1\parallel\cdots\parallel\alpha_n).(\tau.(P+Q)+P), (\alpha_1\parallel\cdots\parallel\alpha_n).(P+Q))\}\cup \textbf{Id}$ is a F weak step bisimulation, we omit it;
  \item $((P+Q).\tau+P).(\alpha_1[m]\parallel\cdots\parallel\alpha_n[m])\approx_s^r(P+Q).(\alpha_1[m]\parallel\cdots\parallel\alpha_n[m])$. It is sufficient to prove the relation $R=\{(((P+Q).\tau+P).(\alpha_1[m]\parallel\cdots\parallel\alpha_n[m]), (P+Q).(\alpha_1[m]\parallel\cdots\parallel\alpha_n[m]))\}\cup \textbf{Id}$ is a R weak step bisimulation, we omit it;
  \item $P\approx_s^{fr} \tau\parallel P$. It is sufficient to prove the relation $R=\{(P, \tau\parallel P)\}\cup \textbf{Id}$ is a FR weak step bisimulation, we omit it.
\end{enumerate}
\end{proof}

\begin{proposition}[$\tau$ laws for FR weak hp-bisimulation]
The $\tau$ laws for FR weak hp-bisimulation is as follows.
\begin{enumerate}
  \item $P\approx_{hp}^f \tau.P$;
  \item $P\approx_{hp}^r P.\tau$;
  \item $\alpha.\tau.P\approx_{hp}^f \alpha.P$;
  \item $P.\tau.\alpha[m]\approx_{hp}^r P.\alpha[m]$;
  \item $(\alpha_1\parallel\cdots\parallel\alpha_n).\tau.P\approx_{hp}^f (\alpha_1\parallel\cdots\parallel\alpha_n).P$;
  \item $P.\tau.(\alpha_1[m]\parallel\cdots\parallel\alpha_n[m])\approx_{hp}^r P.(\alpha_1[m]\parallel\cdots\parallel\alpha_n[m])$;
  \item $P+\tau.P\approx_{hp}^f \tau.P$;
  \item $P+P.\tau\approx_{hp}^r P.\tau$;
  \item $\alpha.(\tau.(P+Q)+P)\approx_{hp}^f\alpha.(P+Q)$;
  \item $((P+Q).\tau+P).\alpha[m]\approx_{hp}^r(P+Q).\alpha[m]$;
  \item $\phi.(\tau.(P+Q)+P)\approx_{hp}^f\phi.(P+Q)$;
  \item $((P+Q).\tau+P).\phi\approx_{hp}^r(P+Q).\phi$;
  \item $(\alpha_1\parallel\cdots\parallel\alpha_n).(\tau.(P+Q)+P)\approx_{hp}^f(\alpha_1\parallel\cdots\parallel\alpha_n).(P+Q)$;
  \item $((P+Q).\tau+P).(\alpha_1[m]\parallel\cdots\parallel\alpha_n[m])\approx_{hp}^r(P+Q).(\alpha_1[m]\parallel\cdots\parallel\alpha_n[m])$;
  \item $P\approx_{hp}^{fr} \tau\parallel P$.
\end{enumerate}
\end{proposition}

\begin{proof}
\begin{enumerate}
  \item $P\approx_{hp}^f \tau.P$. It is sufficient to prove the relation $R=\{(P, \tau.P)\}\cup \textbf{Id}$ is a F weak hp-bisimulation, we omit it;
  \item $P\approx_{hp}^r P.\tau$. It is sufficient to prove the relation $R=\{(P, P.\tau)\}\cup \textbf{Id}$ is a R weak hp-bisimulation, we omit it;
  \item $\alpha.\tau.P\approx_{hp}^f \alpha.P$. It is sufficient to prove the relation $R=\{(\alpha.\tau.P, \alpha.P)\}\cup \textbf{Id}$ is a F weak hp-bisimulation, we omit it;
  \item $P.\tau.\alpha[m]\approx_{hp}^r P.\alpha[m]$. It is sufficient to prove the relation $R=\{(P.\tau.\alpha[m], P.\alpha[m])\}\cup \textbf{Id}$ is a R weak hp-bisimulation, we omit it;
  \item $(\alpha_1\parallel\cdots\parallel\alpha_n).\tau.P\approx_{hp}^f (\alpha_1\parallel\cdots\parallel\alpha_n).P$. It is sufficient to prove the relation $R=\{((\alpha_1\parallel\cdots\parallel\alpha_n).\tau.P, (\alpha_1\parallel\cdots\parallel\alpha_n).P)\}\cup \textbf{Id}$ is a F weak hp-bisimulation, we omit it;
  \item $P.\tau.(\alpha_1[m]\parallel\cdots\parallel\alpha_n[m])\approx_{hp}^r P.(\alpha_1[m]\parallel\cdots\parallel\alpha_n[m])$. It is sufficient to prove the relation $R=\{(P.\tau.(\alpha_1[m]\parallel\cdots\parallel\alpha_n[m]), P.(\alpha_1[m]\parallel\cdots\parallel\alpha_n[m]))\}\cup \textbf{Id}$ is a R weak hp-bisimulation, we omit it;
  \item $P+\tau.P\approx_{hp}^f \tau.P$. It is sufficient to prove the relation $R=\{(P+\tau.P, \tau.P)\}\cup \textbf{Id}$ is a F weak hp-bisimulation, we omit it;
  \item $P+P.\tau\approx_{hp}^r P.\tau$. It is sufficient to prove the relation $R=\{(P+P.\tau, P.\tau)\}\cup \textbf{Id}$ is a R weak hp-bisimulation, we omit it;
  \item $\alpha.(\tau.(P+Q)+P)\approx_{hp}^f\alpha.(P+Q)$. It is sufficient to prove the relation $R=\{(\alpha.(\tau.(P+Q)+P), \alpha.(P+Q))\}\cup \textbf{Id}$ is a F weak hp-bisimulation, we omit it;
  \item $((P+Q).\tau+P).\alpha[m]\approx_{hp}^r(P+Q).\alpha[m]$. It is sufficient to prove the relation $R=\{(((P+Q).\tau+P).\alpha[m], (P+Q).\alpha[m])\}\cup \textbf{Id}$ is a R weak hp-bisimulation, we omit it;
  \item $\phi.(\tau.(P+Q)+P)\approx_{hp}^f\phi.(P+Q)$. It is sufficient to prove the relation $R=\{(\phi.(\tau.(P+Q)+P), \phi.(P+Q))\}\cup \textbf{Id}$ is a F weak hp-bisimulation, we omit it;
  \item $((P+Q).\tau+P).\phi\approx_{hp}^r(P+Q).\phi$. It is sufficient to prove the relation $R=\{(((P+Q).\tau+P).\phi, (P+Q).\phi)\}\cup \textbf{Id}$ is a R weak hp-bisimulation, we omit it;
  \item $(\alpha_1\parallel\cdots\parallel\alpha_n).(\tau.(P+Q)+P)\approx_{hp}^f(\alpha_1\parallel\cdots\parallel\alpha_n).(P+Q)$. It is sufficient to prove the relation $R=\{((\alpha_1\parallel\cdots\parallel\alpha_n).(\tau.(P+Q)+P), (\alpha_1\parallel\cdots\parallel\alpha_n).(P+Q))\}\cup \textbf{Id}$ is a F weak hp-bisimulation, we omit it;
  \item $((P+Q).\tau+P).(\alpha_1[m]\parallel\cdots\parallel\alpha_n[m])\approx_{hp}^r(P+Q).(\alpha_1[m]\parallel\cdots\parallel\alpha_n[m])$. It is sufficient to prove the relation $R=\{(((P+Q).\tau+P).(\alpha_1[m]\parallel\cdots\parallel\alpha_n[m]), (P+Q).(\alpha_1[m]\parallel\cdots\parallel\alpha_n[m]))\}\cup \textbf{Id}$ is a R weak hp-bisimulation, we omit it;
  \item $P\approx_{hp}^{fr} \tau\parallel P$. It is sufficient to prove the relation $R=\{(P, \tau\parallel P)\}\cup \textbf{Id}$ is a FR weak hp-bisimulation, we omit it.
\end{enumerate}
\end{proof}

\begin{proposition}[$\tau$ laws for FR weak hhp-bisimulation]
The $\tau$ laws for FR weak hhp-bisimulation is as follows.
\begin{enumerate}
  \item $P\approx_{hhp}^f \tau.P$;
  \item $P\approx_{hhp}^r P.\tau$;
  \item $\alpha.\tau.P\approx_{hhp}^f \alpha.P$;
  \item $P.\tau.\alpha[m]\approx_{hhp}^r P.\alpha[m]$;
  \item $(\alpha_1\parallel\cdots\parallel\alpha_n).\tau.P\approx_{hhp}^f (\alpha_1\parallel\cdots\parallel\alpha_n).P$;
  \item $P.\tau.(\alpha_1[m]\parallel\cdots\parallel\alpha_n[m])\approx_{hhp}^r P.(\alpha_1[m]\parallel\cdots\parallel\alpha_n[m])$;
  \item $P+\tau.P\approx_{hhp}^f \tau.P$;
  \item $P+P.\tau\approx_{hhp}^r P.\tau$;
  \item $\alpha.(\tau.(P+Q)+P)\approx_{hhp}^f\alpha.(P+Q)$;
  \item $((P+Q).\tau+P).\alpha[m]\approx_{hhp}^r(P+Q).\alpha[m]$;
  \item $\phi.(\tau.(P+Q)+P)\approx_{hhp}^f\phi.(P+Q)$;
  \item $((P+Q).\tau+P).\phi\approx_{hhp}^r(P+Q).\phi$;
  \item $(\alpha_1\parallel\cdots\parallel\alpha_n).(\tau.(P+Q)+P)\approx_{hhp}^f(\alpha_1\parallel\cdots\parallel\alpha_n).(P+Q)$;
  \item $((P+Q).\tau+P).(\alpha_1[m]\parallel\cdots\parallel\alpha_n[m])\approx_{hhp}^r(P+Q).(\alpha_1[m]\parallel\cdots\parallel\alpha_n[m])$;
  \item $P\approx_{hhp}^{fr} \tau\parallel P$.
\end{enumerate}
\end{proposition}

\begin{proof}
\begin{enumerate}
  \item $P\approx_{hhp}^f \tau.P$. It is sufficient to prove the relation $R=\{(P, \tau.P)\}\cup \textbf{Id}$ is a F weak hhp-bisimulation, we omit it;
  \item $P\approx_{hhp}^r P.\tau$. It is sufficient to prove the relation $R=\{(P, P.\tau)\}\cup \textbf{Id}$ is a R weak hhp-bisimulation, we omit it;
  \item $\alpha.\tau.P\approx_{hhp}^f \alpha.P$. It is sufficient to prove the relation $R=\{(\alpha.\tau.P, \alpha.P)\}\cup \textbf{Id}$ is a F weak hhp-bisimulation, we omit it;
  \item $P.\tau.\alpha[m]\approx_{hhp}^r P.\alpha[m]$. It is sufficient to prove the relation $R=\{(P.\tau.\alpha[m], P.\alpha[m])\}\cup \textbf{Id}$ is a R weak hhp-bisimulation, we omit it;
  \item $(\alpha_1\parallel\cdots\parallel\alpha_n).\tau.P\approx_{hhp}^f (\alpha_1\parallel\cdots\parallel\alpha_n).P$. It is sufficient to prove the relation $R=\{((\alpha_1\parallel\cdots\parallel\alpha_n).\tau.P, (\alpha_1\parallel\cdots\parallel\alpha_n).P)\}\cup \textbf{Id}$ is a F weak hhp-bisimulation, we omit it;
  \item $P.\tau.(\alpha_1[m]\parallel\cdots\parallel\alpha_n[m])\approx_{hhp}^r P.(\alpha_1[m]\parallel\cdots\parallel\alpha_n[m])$. It is sufficient to prove the relation $R=\{(P.\tau.(\alpha_1[m]\parallel\cdots\parallel\alpha_n[m]), P.(\alpha_1[m]\parallel\cdots\parallel\alpha_n[m]))\}\cup \textbf{Id}$ is a R weak hhp-bisimulation, we omit it;
  \item $P+\tau.P\approx_{hhp}^f \tau.P$. It is sufficient to prove the relation $R=\{(P+\tau.P, \tau.P)\}\cup \textbf{Id}$ is a F weak hhp-bisimulation, we omit it;
  \item $P+P.\tau\approx_{hhp}^r P.\tau$. It is sufficient to prove the relation $R=\{(P+P.\tau, P.\tau)\}\cup \textbf{Id}$ is a R weak hhp-bisimulation, we omit it;
  \item $\alpha.(\tau.(P+Q)+P)\approx_{hhp}^f\alpha.(P+Q)$. It is sufficient to prove the relation $R=\{(\alpha.(\tau.(P+Q)+P), \alpha.(P+Q))\}\cup \textbf{Id}$ is a F weak hhp-bisimulation, we omit it;
  \item $((P+Q).\tau+P).\alpha[m]\approx_{hhp}^r(P+Q).\alpha[m]$. It is sufficient to prove the relation $R=\{(((P+Q).\tau+P).\alpha[m], (P+Q).\alpha[m])\}\cup \textbf{Id}$ is a R weak hhp-bisimulation, we omit it;
  \item $\phi.(\tau.(P+Q)+P)\approx_{hhp}^f\phi.(P+Q)$. It is sufficient to prove the relation $R=\{(\phi.(\tau.(P+Q)+P), \phi.(P+Q))\}\cup \textbf{Id}$ is a F weak hhp-bisimulation, we omit it;
  \item $((P+Q).\tau+P).\phi\approx_{hhp}^r(P+Q).\phi$. It is sufficient to prove the relation $R=\{(((P+Q).\tau+P).\phi, (P+Q).\phi)\}\cup \textbf{Id}$ is a R weak hhp-bisimulation, we omit it;
  \item $(\alpha_1\parallel\cdots\parallel\alpha_n).(\tau.(P+Q)+P)\approx_{hhp}^f(\alpha_1\parallel\cdots\parallel\alpha_n).(P+Q)$. It is sufficient to prove the relation $R=\{((\alpha_1\parallel\cdots\parallel\alpha_n).(\tau.(P+Q)+P), (\alpha_1\parallel\cdots\parallel\alpha_n).(P+Q))\}\cup \textbf{Id}$ is a F weak hhp-bisimulation, we omit it;
  \item $((P+Q).\tau+P).(\alpha_1[m]\parallel\cdots\parallel\alpha_n[m])\approx_{hhp}^r(P+Q).(\alpha_1[m]\parallel\cdots\parallel\alpha_n[m])$. It is sufficient to prove the relation $R=\{(((P+Q).\tau+P).(\alpha_1[m]\parallel\cdots\parallel\alpha_n[m]), (P+Q).(\alpha_1[m]\parallel\cdots\parallel\alpha_n[m]))\}\cup \textbf{Id}$ is a R weak hhp-bisimulation, we omit it;
  \item $P\approx_{hhp}^{fr} \tau\parallel P$. It is sufficient to prove the relation $R=\{(P, \tau\parallel P)\}\cup \textbf{Id}$ is a FR weak hhp-bisimulation, we omit it.
\end{enumerate}
\end{proof}

\subsubsection{Recursion}

\begin{definition}[Sequential]
$X$ is sequential in $E$ if every subexpression of $E$ which contains $X$, apart from $X$ itself, is of the form $\alpha.F$ or $F.\alpha[m]$, or 
$(\alpha_1\parallel\cdots\parallel\alpha_n).F$ or $F.(\alpha_1[m]\parallel\cdots\parallel\alpha_n[m])$, or $\sum\widetilde{F}$.
\end{definition}

\begin{definition}[Guarded recursive expression]
$X$ is guarded in $E$ if each occurrence of $X$ is with some subexpression $l.F$ or $F.l[m]$, or $(l_1\parallel\cdots\parallel l_n).F$ or $F.(l_1[m]\parallel\cdots\parallel l_n[m])$ of 
$E$.
\end{definition}

\begin{lemma}\label{LUSWW06}
Let $G$ be guarded and sequential, $Vars(G)\subseteq\widetilde{X}$, and let $\langle G\{\widetilde{P}/\widetilde{X}\},s\rangle\xrightarrow{\{\alpha_1,\cdots,\alpha_n\}}\langle P',s'\rangle$ 
or $\langle G\{\widetilde{P}/\widetilde{X}\},s\rangle\xtworightarrow{\{\alpha_1[m],\cdots,\alpha_n[m]\}}\langle P',s'\rangle$. Then there is an expression $H$ such that 
$\langle G,s\rangle\xrightarrow{\{\alpha_1,\cdots,\alpha_n\}}\langle H,s'\rangle$ or $\langle G,s\rangle\xtworightarrow{\{\alpha_1[m],\cdots,\alpha_n[m]\}}\langle H,s'\rangle$, 
$P'\equiv H\{\widetilde{P}/\widetilde{X}\}$, and for any $\widetilde{Q}$, \\$\langle G\{\widetilde{Q}/\widetilde{X}\},s\rangle\xrightarrow{\{\alpha_1,\cdots,\alpha_n\}} \langle H\{\widetilde{Q}/\widetilde{X}\},s'\rangle$ 
or $\langle G\{\widetilde{Q}/\widetilde{X}\},s\rangle\xtworightarrow{\{\alpha_1[m],\cdots,\alpha_n[m]\}} \langle H\{\widetilde{Q}/\widetilde{X}\},s'\rangle$. Moreover $H$ is sequential, 
$Vars(H)\subseteq\widetilde{X}$, and if $\alpha_1=\cdots=\alpha_n=\alpha_1[m]=\cdots=\alpha_n[m]=\tau$, then $H$ is also guarded.
\end{lemma}

\begin{proof}
We only prove the case of forward transition. 

We need to induct on the structure of $G$.

If $G$ is a Constant, a Composition, a Restriction or a Relabeling then it contains no variables, since $G$ is sequential and guarded, then
$\langle G,s\rangle\xrightarrow{\{\alpha_1,\cdots,\alpha_n\}}\langle P',s'\rangle$, then let $H\equiv P'$, as desired.

$G$ cannot be a variable, since it is guarded.

If $G\equiv G_1+G_2$. Then either $\langle G_1\{\widetilde{P}/\widetilde{X}\},s\rangle \xrightarrow{\{\alpha_1,\cdots,\alpha_n\}}\langle P',s'\rangle$ or
$\langle G_2\{\widetilde{P}/\widetilde{X}\},s\rangle \xrightarrow{\{\alpha_1,\cdots,\alpha_n\}}\langle P',s'\rangle$, then, we can apply this lemma in either case, as desired.

If $G\equiv\beta.H$. Then we must have $\alpha=\beta$, and $P'\equiv H\{\widetilde{P}/\widetilde{X}\}$, and
$\langle G\{\widetilde{Q}/\widetilde{X}\},s\rangle\equiv \langle\beta.H\{\widetilde{Q}/\widetilde{X}\},s\rangle \xrightarrow{\beta}\langle H\{\widetilde{Q}/\widetilde{X}\},s'\rangle$,
then, let $G'$ be $H$, as desired.

If $G\equiv(\beta_1\parallel\cdots\parallel\beta_n).H$. Then we must have $\alpha_i=\beta_i$ for $1\leq i\leq n$, and $P'\equiv H\{\widetilde{P}/\widetilde{X}\}$, and
$\langle G\{\widetilde{Q}/\widetilde{X}\},s\rangle\equiv \langle(\beta_1\parallel\cdots\parallel\beta_n).H\{\widetilde{Q}/\widetilde{X}\},s\rangle \xrightarrow{\{\beta_1,\cdots,\beta_n\}}\langle H\{\widetilde{Q}/\widetilde{X}\},s'\rangle$,
then, let $G'$ be $H$, as desired.

If $G\equiv\tau.H$. Then we must have $\tau=\tau$, and $P'\equiv H\{\widetilde{P}/\widetilde{X}\}$, and
$\langle G\{\widetilde{Q}/\widetilde{X}\},s\rangle\equiv \langle\tau.H\{\widetilde{Q}/\widetilde{X}\},s\rangle \xrightarrow{\tau}\langle H\{\widetilde{Q}/\widetilde{X}\},s'\rangle$,
then, let $G'$ be $H$, as desired.

For the case of reverse transition, it can be proven similarly, we omit it.
\end{proof}

\begin{theorem}[Unique solution of equations for FR weak pomset bisimulation]
Let the guarded and sequential expressions $\widetilde{E}$ contain free variables $\subseteq \widetilde{X}$, then,

If $\widetilde{P}\approx_p^{fr} \widetilde{E}\{\widetilde{P}/\widetilde{X}\}$ and $\widetilde{Q}\approx_p^{fr} \widetilde{E}\{\widetilde{Q}/\widetilde{X}\}$, then 
$\widetilde{P}\approx_p^{fr} \widetilde{Q}$.
\end{theorem}

\begin{proof}
We only prove the case of forward transition. 

Like the corresponding theorem in CCS, without loss of generality, we only consider a single equation $X=E$. So we assume $P\approx_p^{fr} E(P)$, $Q\approx_p^{fr} E(Q)$, then $P\approx_p^{fr} Q$.

We will prove $\{(H(P),H(Q)): H\}$ sequential, if $\langle H(P),s\rangle\xrightarrow{\{\alpha_1,\cdots,\alpha_n\}}\langle P',s'\rangle$, then, for some $Q'$,
$\langle H(Q),s\rangle\xRightarrow{\{\alpha_1.\cdots,\alpha_n\}}\langle Q',s'\rangle$ and $P'\approx_p^{fr} Q'$.

Let $\langle H(P),s\rangle\xrightarrow{\{\alpha_1,\cdot,\alpha_n\}}\langle P',s'\rangle$, then $\langle H(E(P)),s\rangle\xRightarrow{\{\alpha_1,\cdots,\alpha_n\}}\langle P'',s''\rangle$
and $P'\approx_p^{fr} P''$.

By Lemma \ref{LUSWW06}, we know there is a sequential $H'$ such that $\langle H(E(P)),s\rangle\xRightarrow{\{\alpha_1,\cdots,\alpha_n\}}\langle H'(P),s'\rangle\Rightarrow P''\approx_p^{fr} P'$.

And, $\langle H(E(Q)),s\rangle\xRightarrow{\{\alpha_1,\cdots,\alpha_n\}}\langle H'(Q),s'\rangle\Rightarrow Q''$ and $P''\approx_p^{fr} Q''$. And $\langle H(Q),s\rangle\xrightarrow{\{\alpha_1,\cdots,\alpha_n\}}\langle Q',s'\rangle\approx_p^{fr} \Rightarrow Q'\approx_p^{fr} Q''$.
Hence, $P'\approx_p^{fr} Q'$, as desired.

For the case of reverse transition, it can be proven similarly, we omit it.
\end{proof}

\begin{theorem}[Unique solution of equations for FR weak step bisimulation]
Let the guarded and sequential expressions $\widetilde{E}$ contain free variables $\subseteq \widetilde{X}$, then,

If $\widetilde{P}\approx_s^{fr} \widetilde{E}\{\widetilde{P}/\widetilde{X}\}$ and $\widetilde{Q}\approx_s^{fr} \widetilde{E}\{\widetilde{Q}/\widetilde{X}\}$, then
$\widetilde{P}\approx_s^{fr} \widetilde{Q}$.
\end{theorem}

\begin{proof}
We only prove the case of forward transition.

Like the corresponding theorem in CCS, without loss of generality, we only consider a single equation $X=E$. So we assume $P\approx_s^{fr} E(P)$, $Q\approx_s^{fr} E(Q)$, then $P\approx_s^{fr} Q$.

We will prove $\{(H(P),H(Q)): H\}$ sequential, if $\langle H(P),s\rangle\xrightarrow{\{\alpha_1,\cdots,\alpha_n\}}\langle P',s'\rangle$, then, for some $Q'$,
$\langle H(Q),s\rangle\xRightarrow{\{\alpha_1.\cdots,\alpha_n\}}\langle Q',s'\rangle$ and $P'\approx_s^{fr} Q'$.

Let $\langle H(P),s\rangle\xrightarrow{\{\alpha_1,\cdot,\alpha_n\}}\langle P',s'\rangle$, then $\langle H(E(P)),s\rangle\xRightarrow{\{\alpha_1,\cdots,\alpha_n\}}\langle P'',s''\rangle$
and $P'\approx_s^{fr} P''$.

By Lemma \ref{LUSWW06}, we know there is a sequential $H'$ such that $\langle H(E(P)),s\rangle\xRightarrow{\{\alpha_1,\cdots,\alpha_n\}}\langle H'(P),s'\rangle\Rightarrow P''\approx_s^{fr} P'$.

And, $\langle H(E(Q)),s\rangle\xRightarrow{\{\alpha_1,\cdots,\alpha_n\}}\langle H'(Q),s'\rangle\Rightarrow Q''$ and $P''\approx_s^{fr} Q''$. And $\langle H(Q),s\rangle\xrightarrow{\{\alpha_1,\cdots,\alpha_n\}}\langle Q',s'\rangle\approx_s^{fr} \Rightarrow Q'\approx_s^{fr} Q''$.
Hence, $P'\approx_s^{fr} Q'$, as desired.

For the case of reverse transition, it can be proven similarly, we omit it.
\end{proof}

\begin{theorem}[Unique solution of equations for FR weak hp-bisimulation]
Let the guarded and sequential expressions $\widetilde{E}$ contain free variables $\subseteq \widetilde{X}$, then,

If $\widetilde{P}\approx_{hp}^{fr} \widetilde{E}\{\widetilde{P}/\widetilde{X}\}$ and $\widetilde{Q}\approx_{hp}^{fr} \widetilde{E}\{\widetilde{Q}/\widetilde{X}\}$, then
$\widetilde{P}\approx_{hp}^{fr} \widetilde{Q}$.
\end{theorem}

\begin{proof}
We only prove the case of forward transition.

Like the corresponding theorem in CCS, without loss of generality, we only consider a single equation $X=E$. So we assume $P\approx_{hp}^{fr} E(P)$, $Q\approx_{hp}^{fr} E(Q)$, then $P\approx_{hp}^{fr} Q$.

We will prove $\{(H(P),H(Q)): H\}$ sequential, if $\langle H(P),s\rangle\xrightarrow{\{\alpha_1,\cdots,\alpha_n\}}\langle P',s'\rangle$, then, for some $Q'$,
$\langle H(Q),s\rangle\xRightarrow{\{\alpha_1.\cdots,\alpha_n\}}\langle Q',s'\rangle$ and $P'\approx_{hp}^{fr} Q'$.

Let $\langle H(P),s\rangle\xrightarrow{\{\alpha_1,\cdot,\alpha_n\}}\langle P',s'\rangle$, then $\langle H(E(P)),s\rangle\xRightarrow{\{\alpha_1,\cdots,\alpha_n\}}\langle P'',s''\rangle$
and $P'\approx_{hp}^{fr} P''$.

By Lemma \ref{LUSWW06}, we know there is a sequential $H'$ such that $\langle H(E(P)),s\rangle\xRightarrow{\{\alpha_1,\cdots,\alpha_n\}}\langle H'(P),s'\rangle\Rightarrow P''\approx_{hp}^{fr} P'$.

And, $\langle H(E(Q)),s\rangle\xRightarrow{\{\alpha_1,\cdots,\alpha_n\}}\langle H'(Q),s'\rangle\Rightarrow Q''$ and $P''\approx_{hp}^{fr} Q''$. And $\langle H(Q),s\rangle\xrightarrow{\{\alpha_1,\cdots,\alpha_n\}}\langle Q',s'\rangle\approx_{hp}^{fr} \Rightarrow Q'\approx_{hp}^{fr} Q''$.
Hence, $P'\approx_{hp}^{fr} Q'$, as desired.

For the case of reverse transition, it can be proven similarly, we omit it.
\end{proof}

\begin{theorem}[Unique solution of equations for FR weak hhp-bisimulation]
Let the guarded and sequential expressions $\widetilde{E}$ contain free variables $\subseteq \widetilde{X}$, then,

If $\widetilde{P}\approx_{hhp}^{fr} \widetilde{E}\{\widetilde{P}/\widetilde{X}\}$ and $\widetilde{Q}\approx_{hhp}^{fr} \widetilde{E}\{\widetilde{Q}/\widetilde{X}\}$, then
$\widetilde{P}\approx_{hhp}^{fr} \widetilde{Q}$.
\end{theorem}

\begin{proof}
We only prove the case of forward transition.

Like the corresponding theorem in CCS, without loss of generality, we only consider a single equation $X=E$. So we assume $P\approx_{hhp}^{fr} E(P)$, $Q\approx_{hhp}^{fr} E(Q)$, then $P\approx_{hhp}^{fr} Q$.

We will prove $\{(H(P),H(Q)): H\}$ sequential, if $\langle H(P),s\rangle\xrightarrow{\{\alpha_1,\cdots,\alpha_n\}}\langle P',s'\rangle$, then, for some $Q'$,
$\langle H(Q),s\rangle\xRightarrow{\{\alpha_1.\cdots,\alpha_n\}}\langle Q',s'\rangle$ and $P'\approx_{hhp}^{fr} Q'$.

Let $\langle H(P),s\rangle\xrightarrow{\{\alpha_1,\cdot,\alpha_n\}}\langle P',s'\rangle$, then $\langle H(E(P)),s\rangle\xRightarrow{\{\alpha_1,\cdots,\alpha_n\}}\langle P'',s''\rangle$
and $P'\approx_{hhp}^{fr} P''$.

By Lemma \ref{LUSWW06}, we know there is a sequential $H'$ such that $\langle H(E(P)),s\rangle\xRightarrow{\{\alpha_1,\cdots,\alpha_n\}}\langle H'(P),s'\rangle\Rightarrow P''\approx_{hhp}^{fr} P'$.

And, $\langle H(E(Q)),s\rangle\xRightarrow{\{\alpha_1,\cdots,\alpha_n\}}\langle H'(Q),s'\rangle\Rightarrow Q''$ and $P''\approx_{hhp}^{fr} Q''$. And $\langle H(Q),s\rangle\xrightarrow{\{\alpha_1,\cdots,\alpha_n\}}\langle Q',s'\rangle\approx_{hhp}^{fr} \Rightarrow Q'\approx_{hhp}^{fr} Q''$.
Hence, $P'\approx_{hhp}^{fr} Q'$, as desired.

For the case of reverse transition, it can be proven similarly, we omit it.
\end{proof}

\newpage\section{Putting All the Things into a Whole}\label{pa}

In this chapter, we design the calculus CTC with reversibility, probabilism and guards. This chapter is organized as follows. We introduce the operational semantics in section \ref{ospa}, its syntax and operational
semantics in section \ref{sospa}, and its properties for strong bisimulations in section \ref{sftcbpa}, its properties for weak bisimulations in section \ref{wftcbpa}.

\subsection{Operational Semantics}\label{ospa}

\begin{definition}[Probabilistic transitions]
Let $\mathcal{E}$ be a PES and let $C\in\mathcal{C}(\mathcal{E})$, the transition $\langle C,s\rangle\xrsquigarrow{\pi} \langle C^{\pi},s\rangle$ is called a probabilistic
transition
from $\langle C,s\rangle$ to $\langle C^{\pi},s\rangle$.
\end{definition}

\begin{definition}[FR probabilistic pomset, step bisimulation]\label{PSBG}
Let $\mathcal{E}_1$, $\mathcal{E}_2$ be PESs. A FR probabilistic pomset bisimulation is a relation $R\subseteq\langle\mathcal{C}(\mathcal{E}_1),S\rangle\times\langle\mathcal{C}(\mathcal{E}_2),S\rangle$,
such that (1) if $(\langle C_1,s\rangle,\langle C_2,s\rangle)\in R$, and $\langle C_1,s\rangle\xrightarrow{X_1}\langle C_1',s'\rangle$ then
$\langle C_2,s\rangle\xrightarrow{X_2}\langle C_2',s'\rangle$, with $X_1\subseteq \mathbb{E}_1$, $X_2\subseteq \mathbb{E}_2$, $X_1\sim X_2$ and
$(\langle C_1',s'\rangle,\langle C_2',s'\rangle)\in R$ for all $s,s'\in S$, and vice-versa;
(2) if $(\langle C_1,s\rangle,\langle C_2,s\rangle)\in R$, and $\langle C_1,s\rangle\xtworightarrow{X_1[\mathcal{K}_1]}\langle C_1',s'\rangle$ then
$\langle C_2,s\rangle\xrightarrow{X_2[\mathcal{K}_2]}\langle C_2',s'\rangle$, with $X_1\subseteq \mathbb{E}_1$, $X_2\subseteq \mathbb{E}_2$, $X_1\sim X_2$ and
$(\langle C_1',s'\rangle,\langle C_2',s'\rangle)\in R$ for all $s,s'\in S$, and vice-versa;
(3) if $(\langle C_1,s\rangle,\langle C_2,s\rangle)\in R$, and $\langle C_1,s\rangle\xrsquigarrow{\pi}\langle C_1^{\pi},s\rangle$
then $\langle C_2,s\rangle\xrsquigarrow{\pi}\langle C_2^{\pi},s\rangle$ and $(\langle C_1^{\pi},s\rangle,\langle C_2^{\pi},s\rangle)\in R$, and vice-versa; (4) if $(\langle C_1,s\rangle,\langle C_2,s\rangle)\in R$,
then $\mu(C_1,C)=\mu(C_2,C)$ for each $C\in\mathcal{C}(\mathcal{E})/R$; (5) $[\surd]_R=\{\surd\}$. We say that $\mathcal{E}_1$, $\mathcal{E}_2$ are FR probabilistic pomset bisimilar, written
$\mathcal{E}_1\sim_{pp}^{fr}\mathcal{E}_2$, if there exists a probabilistic pomset bisimulation $R$, such that $(\langle\emptyset,\emptyset\rangle,\langle\emptyset,\emptyset\rangle)\in R$.
By replacing FR probabilistic pomset transitions with FR probabilistic steps, we can get the definition of FR probabilistic step bisimulation. When PESs $\mathcal{E}_1$ and $\mathcal{E}_2$ are FR
probabilistic step bisimilar, we write $\mathcal{E}_1\sim_{ps}^{fr}\mathcal{E}_2$.
\end{definition}

\begin{definition}[FR weakly probabilistic pomset, step bisimulation]
Let $\mathcal{E}_1$, $\mathcal{E}_2$ be PESs. A FR weakly probabilistic pomset bisimulation is a relation $R\subseteq\langle\mathcal{C}(\mathcal{E}_1),S\rangle\times\langle\mathcal{C}(\mathcal{E}_2),S\rangle$,
such that (1) if $(\langle C_1,s\rangle,\langle C_2,s\rangle)\in R$, and $\langle C_1,s\rangle\xRightarrow{X_1}\langle C_1',s'\rangle$ then
$\langle C_2,s\rangle\xRightarrow{X_2}\langle C_2',s'\rangle$, with $X_1\subseteq \hat{\mathbb{E}_1}$, $X_2\subseteq \hat{\mathbb{E}_2}$, $X_1\sim X_2$ and
$(\langle C_1',s'\rangle,\langle C_2',s'\rangle)\in R$ for all $s,s'\in S$, and vice-versa;
(2) if $(\langle C_1,s\rangle,\langle C_2,s\rangle)\in R$, and $\langle C_1,s\rangle\xTworightarrow{X_1[\mathcal{K}_1]}\langle C_1',s'\rangle$ then
$\langle C_2,s\rangle\xTworightarrow{X_2[\mathcal{K}_2]}\langle C_2',s'\rangle$, with $X_1\subseteq \hat{\mathbb{E}_1}$, $X_2\subseteq \hat{\mathbb{E}_2}$, $X_1\sim X_2$ and
$(\langle C_1',s'\rangle,\langle C_2',s'\rangle)\in R$ for all $s,s'\in S$, and vice-versa;
(3) if $(\langle C_1,s\rangle,\langle C_2,s\rangle)\in R$, and $\langle C_1,s\rangle\xrsquigarrow{\pi}\langle C_1^{\pi},s\rangle$
then $\langle C_2,s\rangle\xrsquigarrow{\pi}\langle C_2^{\pi},s\rangle$ and $(\langle C_1^{\pi},s\rangle,\langle C_2^{\pi},s\rangle)\in R$, and vice-versa; (4) if $(\langle C_1,s\rangle,\langle C_2,s\rangle)\in R$,
then $\mu(C_1,C)=\mu(C_2,C)$ for each $C\in\mathcal{C}(\mathcal{E})/R$; (5) $[\surd]_R=\{\surd\}$. We say that $\mathcal{E}_1$, $\mathcal{E}_2$ are FR weakly probabilistic pomset bisimilar,
written $\mathcal{E}_1\approx_{pp}^{fr}\mathcal{E}_2$, if there exists a FR weakly probabilistic pomset bisimulation $R$, such that
$(\langle\emptyset,\emptyset\rangle,\langle\emptyset,\emptyset\rangle)\in R$. By replacing FR weakly probabilistic pomset transitions with FR weakly probabilistic steps, we can get the
definition of FR weakly probabilistic step bisimulation. When PESs $\mathcal{E}_1$ and $\mathcal{E}_2$ are FR weakly probabilistic step bisimilar, we write
$\mathcal{E}_1\approx_{ps}^{FR}\mathcal{E}_2$.
\end{definition}

\begin{definition}[Posetal product]
Given two PESs $\mathcal{E}_1$, $\mathcal{E}_2$, the posetal product of their configurations, denoted
$\langle\mathcal{C}(\mathcal{E}_1),S\rangle\overline{\times}\langle\mathcal{C}(\mathcal{E}_2),S\rangle$, is defined as

$$\{(\langle C_1,s\rangle,f,\langle C_2,s\rangle)|C_1\in\mathcal{C}(\mathcal{E}_1),C_2\in\mathcal{C}(\mathcal{E}_2),f:C_1\rightarrow C_2 \textrm{ isomorphism}\}.$$

A subset $R\subseteq\langle\mathcal{C}(\mathcal{E}_1),S\rangle\overline{\times}\langle\mathcal{C}(\mathcal{E}_2),S\rangle$ is called a posetal relation. We say that $R$ is downward
closed when for any $(\langle C_1,s\rangle,f,\langle C_2,s\rangle),(\langle C_1',s'\rangle,f',\langle C_2',s'\rangle)\in \langle\mathcal{C}(\mathcal{E}_1),S\rangle\overline{\times}\langle\mathcal{C}(\mathcal{E}_2),S\rangle$,
if $(\langle C_1,s\rangle,f,\langle C_2,s\rangle)\subseteq (\langle C_1',s'\rangle,f',\langle C_2',s'\rangle)$ pointwise and
$(\langle C_1',s'\rangle,f',\langle C_2',s'\rangle)\in R$, then $(\langle C_1,s\rangle,f,\langle C_2,s\rangle)\in R$.

For $f:X_1\rightarrow X_2$, we define $f[x_1\mapsto x_2]:X_1\cup\{x_1\}\rightarrow X_2\cup\{x_2\}$, $z\in X_1\cup\{x_1\}$,(1)$f[x_1\mapsto x_2](z)=
x_2$,if $z=x_1$;(2)$f[x_1\mapsto x_2](z)=f(z)$, otherwise. Where $X_1\subseteq \mathbb{E}_1$, $X_2\subseteq \mathbb{E}_2$, $x_1\in \mathbb{E}_1$, $x_2\in \mathbb{E}_2$.
\end{definition}

\begin{definition}[Weakly posetal product]
Given two PESs $\mathcal{E}_1$, $\mathcal{E}_2$, the weakly posetal product of their configurations, denoted
$\langle\mathcal{C}(\mathcal{E}_1),S\rangle\overline{\times}\langle\mathcal{C}(\mathcal{E}_2),S\rangle$, is defined as

$$\{(\langle C_1,s\rangle,f,\langle C_2,s\rangle)|C_1\in\mathcal{C}(\mathcal{E}_1),C_2\in\mathcal{C}(\mathcal{E}_2),f:\hat{C_1}\rightarrow \hat{C_2} \textrm{ isomorphism}\}.$$

A subset $R\subseteq\langle\mathcal{C}(\mathcal{E}_1),S\rangle\overline{\times}\langle\mathcal{C}(\mathcal{E}_2),S\rangle$ is called a weakly posetal relation. We say that $R$ is
downward closed when for any $(\langle C_1,s\rangle,f,\langle C_2,s\rangle),(\langle C_1',s'\rangle,f,\langle C_2',s'\rangle)\in \langle\mathcal{C}(\mathcal{E}_1),S\rangle\overline{\times}\langle\mathcal{C}(\mathcal{E}_2),S\rangle$,
if $(\langle C_1,s\rangle,f,\langle C_2,s\rangle)\subseteq (\langle C_1',s'\rangle,f',\langle C_2',s'\rangle)$ pointwise and
$(\langle C_1',s'\rangle,f',\langle C_2',s'\rangle)\in R$, then $(\langle C_1,s\rangle,f,\langle C_2,s\rangle)\in R$.

For $f:X_1\rightarrow X_2$, we define $f[x_1\mapsto x_2]:X_1\cup\{x_1\}\rightarrow X_2\cup\{x_2\}$, $z\in X_1\cup\{x_1\}$,(1)$f[x_1\mapsto x_2](z)=
x_2$,if $z=x_1$;(2)$f[x_1\mapsto x_2](z)=f(z)$, otherwise. Where $X_1\subseteq \hat{\mathbb{E}_1}$, $X_2\subseteq \hat{\mathbb{E}_2}$, $x_1\in \hat{\mathbb{E}}_1$,
$x_2\in \hat{\mathbb{E}}_2$. Also, we define $f(\tau^*)=f(\tau^*)$.
\end{definition}

\begin{definition}[FR probabilistic (hereditary) history-preserving bisimulation]
A FR probabilistic history-preserving (hp-) bisimulation is a posetal relation
$R\subseteq\langle\mathcal{C}(\mathcal{E}_1),S\rangle\overline{\times}\langle\mathcal{C}(\mathcal{E}_2),S\rangle$ such that (1) if $(\langle C_1,s\rangle,f,\langle C_2,s\rangle)\in R$,
and $\langle C_1,s\rangle\xrightarrow{e_1} \langle C_1',s'\rangle$, then $\langle C_2,s\rangle\xrightarrow{e_2} \langle C_2',s'\rangle$, with
$(\langle C_1',s'\rangle,f[e_1\mapsto e_2],\langle C_2',s'\rangle)\in R$ for all $s,s'\in S$, and vice-versa;
(2) if $(\langle C_1,s\rangle,f,\langle C_2,s\rangle)\in R$,
and $\langle C_1,s\rangle\xtworightarrow{e_1[m]} \langle C_1',s'\rangle$, then $\langle C_2,s\rangle\xtworightarrow{e_2[n]} \langle C_2',s'\rangle$, with
$(\langle C_1',s'\rangle,f[e_1[m]\mapsto e_2[n]],\langle C_2',s'\rangle)\in R$ for all $s,s'\in S$, and vice-versa;
(3) if $(\langle C_1,s\rangle,f,\langle C_2,s\rangle)\in R$, and
$\langle C_1,s\rangle\xrsquigarrow{\pi}\langle C_1^{\pi},s\rangle$ then $\langle C_2,s\rangle\xrsquigarrow{\pi}\langle C_2^{\pi},s\rangle$ and $(\langle C_1^{\pi},s\rangle,f,\langle C_2^{\pi},s\rangle)\in R$,
and vice-versa; (4) if $(C_1,f,C_2)\in R$, then $\mu(C_1,C)=\mu(C_2,C)$ for each $C\in\mathcal{C}(\mathcal{E})/R$; (5) $[\surd]_R=\{\surd\}$. $\mathcal{E}_1,\mathcal{E}_2$ are
probabilistic history-preserving (hp-)bisimilar and are written $\mathcal{E}_1\sim_{php}\mathcal{E}_2$ if there exists a probabilistic hp-bisimulation $R$ such that
$(\langle\emptyset,\emptyset\rangle,\emptyset,\langle\emptyset,\emptyset\rangle)\in R$.

A FR probabilistic hereditary history-preserving (hhp-)bisimulation is a downward closed FR probabilistic hp-bisimulation. $\mathcal{E}_1,\mathcal{E}_2$ are FR probabilistic hereditary
history-preserving (hhp-)bisimilar and are written $\mathcal{E}_1\sim_{phhp}^{fr}\mathcal{E}_2$.
\end{definition}

\begin{definition}[FR weakly probabilistic (hereditary) history-preserving bisimulation]
A FR weakly probabilistic history-preserving (hp-) bisimulation is a weakly posetal relation\\
$R\subseteq\langle\mathcal{C}(\mathcal{E}_1),S\rangle\overline{\times}\langle\mathcal{C}(\mathcal{E}_2),S\rangle$ such that (1) if $(\langle C_1,s\rangle,f,\langle C_2,s\rangle)\in R$,
and $\langle C_1,s\rangle\xRightarrow{e_1} \langle C_1',s'\rangle$, then $\langle C_2,s\rangle\xRightarrow{e_2} \langle C_2',s'\rangle$, with
$(\langle C_1',s'\rangle,f[e_1\mapsto e_2],\langle C_2',s'\rangle)\in R$ for all $s,s'\in S$, and vice-versa;
(2) if $(\langle C_1,s\rangle,f,\langle C_2,s\rangle)\in R$,
and $\langle C_1,s\rangle\xTworightarrow{e_1[m]} \langle C_1',s'\rangle$, then $\langle C_2,s\rangle\xTworightarrow{e_2[n]} \langle C_2',s'\rangle$, with
$(\langle C_1',s'\rangle,f[e_1[m]\mapsto e_2[n]],\langle C_2',s'\rangle)\in R$ for all $s,s'\in S$, and vice-versa;
(3) if $(\langle C_1,s\rangle,f,\langle C_2,s\rangle)\in R$, and
$\langle C_1,s\rangle\xrsquigarrow{\pi}\langle C_1^{\pi},s\rangle$ then $\langle C_2,s\rangle\xrsquigarrow{\pi}\langle C_2^{\pi},s\rangle$ and
$(\langle C_1^{\pi},s\rangle,f,\langle C_2^{\pi},s\rangle)\in R$, and vice-versa; (4) if $(C_1,f,C_2)\in R$, then $\mu(C_1,C)=\mu(C_2,C)$ for each $C\in\mathcal{C}(\mathcal{E})/R$;
(5) $[\surd]_R=\{\surd\}$. $\mathcal{E}_1,\mathcal{E}_2$ are FR weakly probabilistic history-preserving (hp-)bisimilar and are written $\mathcal{E}_1\approx_{php}^{fr}\mathcal{E}_2$ if there
exists a FR weakly probabilistic hp-bisimulation $R$ such that $(\langle\emptyset,\emptyset\rangle,\emptyset,\langle\emptyset,\emptyset\rangle)\in R$.

A FR weakly probabilistic hereditary history-preserving (hhp-)bisimulation is a downward closed FR weakly probabilistic hp-bisimulation. $\mathcal{E}_1,\mathcal{E}_2$ are FR weakly
probabilistic hereditary history-preserving (hhp-)bisimilar and are written $\mathcal{E}_1\approx_{phhp}^{fr}\mathcal{E}_2$.
\end{definition}

\subsection{Syntax and Operational Semantics}\label{sospa}

We assume an infinite set $\mathcal{N}$ of (action or event) names, and use $a,b,c,\cdots$ to range over $\mathcal{N}$. We denote by $\overline{\mathcal{N}}$ the set of co-names and
let $\overline{a},\overline{b},\overline{c},\cdots$ range over $\overline{\mathcal{N}}$. Then we set $\mathcal{L}=\mathcal{N}\cup\overline{\mathcal{N}}$ as the set of labels, and use
$l,\overline{l}$ to range over $\mathcal{L}$. We extend complementation to $\mathcal{L}$ such that $\overline{\overline{a}}=a$. Let $\tau$ denote the silent step (internal action or
event) and define $Act=\mathcal{L}\cup\{\tau\}\cup\mathcal{L}[\mathcal{K}]$ to be the set of actions, $\alpha,\beta$ range over $Act$. And $K,L$ are used to stand for subsets of
$\mathcal{L}$ and $\overline{L}$ is used for the set of complements of labels in $L$. A relabelling function $f$ is a function from $\mathcal{L}$ to $\mathcal{L}$ such that
$f(\overline{l})=\overline{f(l)}$. By defining $f(\tau)=\tau$, we extend $f$ to $Act$. We write $\mathcal{P}$ for the set of processes. Sometimes, we use $I,J$ to stand for an indexing
set, and we write $E_i:i\in I$ for a family of expressions indexed by $I$. $Id_D$ is the identity function or relation over set $D$.

For each process constant schema $A$, a defining equation of the form

$$A\overset{\text{def}}{=}P$$

is assumed, where $P$ is a process.

Let $G_{at}$ be the set of atomic guards, $\delta$ be the deadlock constant, and $\epsilon$ be the empty action, and extend $Act$ to $Act\cup\{\epsilon\}\cup\{\delta\}$. We extend
$G_{at}$ to the set of basic guards $G$ with element $\phi,\psi,\cdots$, which is generated by the following formation rules:

$$\phi::=\delta|\epsilon|\neg\phi|\psi\in G_{at}|\phi+\psi|\phi\cdot\psi$$

The predicate $test(\phi,s)$ represents that $\phi$ holds in the state $s$, and $test(\epsilon,s)$ holds and $test(\delta,s)$ does not hold. $effect(e,s)\in S$ denotes $s'$ in
$s\xrightarrow{e}s'$. The predicate weakest precondition $wp(e,\phi)$ denotes that $\forall s,s'\in S, test(\phi,effect(e,s))$ holds.

\subsubsection{Syntax}

We use the Prefix $.$ to model the causality relation $\leq$ in true concurrency, the Summation $+$ to model the conflict relation $\sharp$ in true concurrency, and the Composition
$\parallel$ to explicitly model concurrent relation in true concurrency. And we follow the conventions of process algebra.

\begin{definition}[Syntax]\label{syntax07}
CTC with reversibility, probabilism and guards are defined inductively by the following formation rules:

\begin{enumerate}
  \item $A\in\mathcal{P}$;
  \item $\phi\in\mathcal{P}$;
  \item $\textbf{nil}\in\mathcal{P}$;
  \item if $P\in\mathcal{P}$, then the Prefix $\alpha.P\in\mathcal{P}$ and $P.\alpha[m]\in\mathcal{P}$, for $\alpha\in Act$ and $m\in\mathcal{K}$;
  \item if $P\in\mathcal{P}$, then the Prefix $\phi.P\in\mathcal{P}$, for $\phi\in G_{at}$;
  \item if $P,Q\in\mathcal{P}$, then the Summation $P+Q\in\mathcal{P}$;
  \item if $P,Q\in\mathcal{P}$, then the Box-Summation $P\boxplus_{\pi}Q\in\mathcal{P}$;
  \item if $P,Q\in\mathcal{P}$, then the Composition $P\parallel Q\in\mathcal{P}$;
  \item if $P\in\mathcal{P}$, then the Prefix $(\alpha_1\parallel\cdots\parallel\alpha_n).P\in\mathcal{P}\quad(n\in I)$ and $P.(\alpha_1[m]\parallel\cdots\parallel\alpha_n[m])\in\mathcal{P}\quad(n\in I)$, for $\alpha_,\cdots,\alpha_n\in Act$ and $m\in\mathcal{K}$;
  \item if $P\in\mathcal{P}$, then the Restriction $P\setminus L\in\mathcal{P}$ with $L\in\mathcal{L}$;
  \item if $P\in\mathcal{P}$, then the Relabelling $P[f]\in\mathcal{P}$.
\end{enumerate}

The standard BNF grammar of syntax of CTC with reversibility, probabilism and guards can be summarized as follows:

$P::=A|\textbf{nil}|\alpha.P| P.\alpha[m]|\phi.P| P+P |P\boxplus_{\pi} P| P\parallel P | (\alpha_1\parallel\cdots\parallel\alpha_n).P|  P.(\alpha_1[m]\parallel\cdots\parallel\alpha_n[m])  | P\setminus L | P[f].$
\end{definition}

\subsubsection{Operational Semantics}

The operational semantics is defined by LTSs (labelled transition systems), and it is detailed by the following definition.

\begin{definition}[Semantics]\label{semantics07}
The operational semantics of CTC with reversibility, probabilism and guards corresponding to the syntax in Definition \ref{syntax07} is defined by a series of transition rules, they are shown in Table \ref{FTRForPS07},
\ref{RTRForPS07}, \ref{FTRForCom07}, \ref{RTRForCom07}, \ref{FTRForRRC07} and \ref{RTRForRRC07}. And the predicate
$\xrightarrow{\alpha}\alpha[m]$ represents successful forward termination after execution of the action $\alpha$, the predicate $\xtworightarrow{\alpha[m]}\alpha$ represents successful
reverse termination after execution of the event $\alpha[m]$, the the predicate $\textrm{Std(P)}$ represents that $p$ is a standard process containing no past events, the the predicate
$\textrm{NStd(P)}$ represents that $P$ is a process full of past events.
\end{definition}

\begin{center}
    \begin{table}
        $$\frac{}{\alpha.P\rightsquigarrow\breve{\alpha}.P}$$
        $$\frac{P\rightsquigarrow P'\quad Q\rightsquigarrow Q'}{P+Q\rightsquigarrow P'+Q'}$$
        $$\frac{P\rightsquigarrow P'}{P\boxplus_{\pi}Q\rightsquigarrow P'}\quad \frac{Q\rightsquigarrow Q'}{P\boxplus_{\pi}Q\rightsquigarrow Q'}$$
        $$\frac{P\rightsquigarrow P'\quad Q\rightsquigarrow Q'}{P\parallel Q\rightsquigarrow P'+Q'}$$
        $$\frac{}{(\alpha_1\parallel\cdots\parallel\alpha_n).P\rightsquigarrow(\breve{\alpha_1}\parallel\cdots\parallel\breve{\alpha_n}).P}$$
        $$\frac{P\rightsquigarrow P'}{P\setminus L\rightsquigarrow P'\setminus L}$$
        $$\frac{P\rightsquigarrow P'}{P[f]\rightsquigarrow P'[f]}$$
        $$\frac{P\rightsquigarrow P'}{A\rightsquigarrow P'}\quad (A\overset{\text{def}}{=}P)$$
        \caption{Probabilistic transition rules}
        \label{PTRForCTC07}
    \end{table}
\end{center}

The forward transition rules for Prefix and Summation are shown in Table \ref{FTRForPS07}.

\begin{center}
    \begin{table}
        $$\frac{}{\langle\breve{\alpha},s\rangle\xrightarrow{\alpha}\langle\alpha[m],s'\rangle}$$
        $$\frac{}{\langle\breve{\epsilon},s\rangle\rightarrow\langle\surd,s\rangle}$$
        $$\frac{}{\langle\phi,s\rangle\rightarrow\langle\surd,s\rangle}\textrm{ if }test(\phi,s)$$

        $$\frac{\langle P,s\rangle\xrightarrow{\alpha}\langle\alpha[m],s'\rangle}{\langle P+Q,s\rangle\xrightarrow{\alpha}\langle\alpha[m],s'\rangle}
        \quad\frac{\langle P,s\rangle\xrightarrow{\alpha}\langle P',s'\rangle}{\langle P+Q,s\rangle\xrightarrow{\alpha}\langle P',s'\rangle}$$
        $$\frac{\langle Q,s\rangle\xrightarrow{\alpha}\langle \alpha[m],s'\rangle}{\langle P+Q,s\rangle\xrightarrow{\alpha}\langle\alpha[m],s'\rangle}
        \quad\frac{\langle Q,s\rangle\xrightarrow{\alpha}\langle Q',s'\rangle}{\langle P+Q,s\rangle\xrightarrow{\alpha}\langle Q',s'\rangle}$$


        $$\frac{\langle P,s\rangle\xrightarrow{\alpha}\langle \alpha[m],s'\rangle\quad\textrm{Std}(Q)}{\langle P. Q,s\rangle\xrightarrow{\alpha} \langle\alpha[m]. Q,s'\rangle}
        \quad\frac{\langle P,s\rangle\xrightarrow{\alpha}\langle P',s'\rangle \quad \textrm{Std}(Q)}{\langle P. Q,s\rangle\xrightarrow{\alpha}\langle P'. Q,s'\rangle}$$
        $$\frac{\langle Q,s\rangle\xrightarrow{\beta}\langle\beta[n],s'\rangle\quad \textrm{NStd}(P)}{\langle P. Q,s\rangle\xrightarrow{\beta}\langle P. \beta[n],s'\rangle}
        \quad\frac{\langle Q,s\rangle\xrightarrow{\beta}\langle Q',s'\rangle\quad \textrm{NStd}(P)}{\langle P. Q,s\rangle\xrightarrow{\beta}\langle P. Q',s'\rangle}$$
        \caption{Forward transition rules of Prefix and Summation}
        \label{FTRForPS07}
    \end{table}
\end{center}

The reverse transition rules for Prefix and Summation are shown in Table \ref{RTRForPS07}.

\begin{center}
    \begin{table}
        $$\frac{}{\langle \breve{\alpha[m]},s\rangle\xtworightarrow{\alpha[m]}\langle\alpha,s'\rangle}$$
        $$\frac{}{\langle\breve{\epsilon},s\rangle\xtworightarrow{ }\langle\surd,s\rangle}$$
        $$\frac{}{\langle\phi,s\rangle\xtworightarrow{ }\langle\surd,s\rangle}\textrm{ if }test(\phi,s)$$

        $$\frac{\langle P,s\rangle\xtworightarrow{\alpha[m]}\langle \alpha,s'\rangle}{\langle P+Q,s\rangle\xtworightarrow{\alpha[m]}\langle\alpha,s'\rangle}
        \quad\frac{\langle P,s\rangle\xtworightarrow{\alpha[m]}\langle P',s'\rangle}{\langle P+Q,s\rangle\xtworightarrow{\alpha[m]}\langle P',s'\rangle}$$
        $$\frac{\langle Q,s\rangle\xtworightarrow{\alpha[m]}\langle\alpha,s'\rangle}{\langle P+Q,s\rangle\xtworightarrow{\alpha[m]}\langle\alpha,s'\rangle}
        \quad\frac{\langle Q,s\rangle\xtworightarrow{\alpha[m]}\langle Q',s'\rangle}{\langle P+Q,s\rangle\xtworightarrow{\alpha[m]}\langle Q',s'\rangle}$$


        $$\frac{\langle P,s\rangle\xtworightarrow{\alpha[m]}\langle\alpha,s'\rangle \quad \textrm{Std}(Q)}{\langle P. Q,s\rangle\xtworightarrow{\alpha[m]} \langle\alpha. Q,s'\rangle}
        \quad\frac{\langle P,s\rangle\xtworightarrow{\alpha[m]}\langle P',s'\rangle\quad \textrm{Std}(Q)}{\langle P. Q,s\rangle\xtworightarrow{\alpha[m]}\langle P'. Q,s'\rangle}$$
        $$\frac{\langle Q,s\rangle\xtworightarrow{\beta[n]}\langle\beta,s'\rangle \quad \textrm{NStd}(P)}{\langle P. Q,s\rangle\xtworightarrow{\beta[n]}\langle P. \beta,s'\rangle}\quad
        \frac{\langle Q,s\rangle\xtworightarrow{\beta[n]}\langle Q',s'\rangle \quad \textrm{NStd}(P)}{\langle P. Q,s\rangle\xtworightarrow{\beta[n]}\langle P. Q',s'\rangle}$$
        \caption{Reverse transition rules of Prefix and Summation}
        \label{RTRForPS07}
    \end{table}
\end{center}

The forward transition rules for Composition are shown in Table \ref{FTRForCom07}.

\begin{center}
    \begin{table}
        $$\frac{\langle P,s\rangle\xrightarrow{\alpha}\langle P',s'\rangle\quad \langle Q,s\rangle\nrightarrow}{\langle P\parallel Q,s\rangle\xrightarrow{\alpha}\langle P'\parallel Q,s'\rangle}$$
        $$\frac{\langle Q,s\rangle\xrightarrow{\alpha}\langle Q',s'\rangle\quad \langle P,s\rangle\nrightarrow}{\langle P\parallel Q,s\rangle\xrightarrow{\alpha}\langle P\parallel Q',s'\rangle}$$
        $$\frac{\langle P,s\rangle\xrightarrow{\alpha}\langle P',s'\rangle\quad \langle Q,s\rangle\xrightarrow{\beta}\langle Q',s''\rangle}{\langle P\parallel Q,s\rangle\xrightarrow{\{\alpha,\beta\}}\langle P'\parallel Q',s'\cup s''\rangle}\quad (\beta\neq\overline{\alpha})$$
        $$\frac{\langle P,s\rangle\xrightarrow{l}\langle P',s'\rangle\quad \langle Q,s\rangle\xrightarrow{\overline{l}}\langle Q',s''\rangle}{\langle P\parallel Q,s\rangle\xrightarrow{\tau}\langle P'\parallel Q',s'\cup s''\rangle}$$
        \caption{Forward transition rules of Composition}
        \label{FTRForCom07}
    \end{table}
\end{center}

The reverse transition rules for Composition are shown in Table \ref{RTRForCom07}.

\begin{center}
    \begin{table}
        $$\frac{\langle P,s\rangle\xtworightarrow{\alpha[m]}\langle P',s'\rangle\quad \langle Q,s\rangle\xntworightarrow{}}{\langle P\parallel Q,s\rangle\xtworightarrow{\alpha[m]}\langle P'\parallel Q,s'\rangle}$$
        $$\frac{\langle Q,s\rangle\xtworightarrow{\alpha[m]}\langle Q',s'\rangle\quad \langle P,s\rangle\xntworightarrow{}}{\langle P\parallel Q,s\rangle\xtworightarrow{\alpha[m]}\langle P\parallel Q',s'\rangle}$$
        $$\frac{\langle P,s\rangle\xtworightarrow{\alpha[m]}\langle P',s'\rangle\quad \langle Q,s\rangle\xtworightarrow{\beta[m]}\langle Q',s''\rangle}{\langle P\parallel Q,s\rangle\xtworightarrow{\{\alpha[m],\beta[m]\}}\langle P'\parallel Q',s'\cup s''\rangle}\quad (\beta\neq\overline{\alpha})$$
        $$\frac{\langle P,s\rangle\xtworightarrow{l[m]}\langle P',s'\rangle\quad \langle Q,s\rangle\xtworightarrow{\overline{l}[m]}\langle Q',s''\rangle}{\langle P\parallel Q,s\rangle\xtworightarrow{\tau}\langle P'\parallel Q',s'\cup s''\rangle}$$
        \caption{Reverse transition rules of Composition}
        \label{RTRForCom07}
    \end{table}
\end{center}

The forward transition rules for Restriction, Relabelling and Constants are shown in Table \ref{FTRForRRC07}.

\begin{center}
    \begin{table}
        $$\frac{\langle P,s\rangle\xrightarrow{\alpha}\langle P',s'\rangle}{\langle P\setminus L,s\rangle\xrightarrow{\alpha}\langle P'\setminus L,s'\rangle}\quad (\alpha,\overline{\alpha}\notin L)$$
        $$\frac{\langle P,s\rangle\xrightarrow{\{\alpha_1,\cdots,\alpha_n\}}\langle P',s'\rangle}{\langle P\setminus L,s\rangle\xrightarrow{\{\alpha_1,\cdots,\alpha_n\}}\langle P'\setminus L,s'\rangle}\quad (\alpha_1,\overline{\alpha_1},\cdots,\alpha_n,\overline{\alpha_n}\notin L)$$
        $$\frac{\langle P,s\rangle\xrightarrow{\alpha}\langle P',s'\rangle}{\langle P[f],s\rangle\xrightarrow{f(\alpha)}\langle P'[f],s'\rangle}$$
        $$\frac{\langle P,s\rangle\xrightarrow{\{\alpha_1,\cdots,\alpha_n\}}\langle P',s'\rangle}{\langle P[f],s\rangle\xrightarrow{\{f(\alpha_1),\cdots,f(\alpha_n)\}}\langle P'[f],s'\rangle}$$
        $$\frac{\langle P,s\rangle\xrightarrow{\alpha}\langle P',s'\rangle}{\langle A,s\rangle\xrightarrow{\alpha}\langle P',s'\rangle}\quad (A\overset{\text{def}}{=}P)$$
        $$\frac{\langle P,s\rangle\xrightarrow{\{\alpha_1,\cdots,\alpha_n\}}\langle P',s'\rangle}{\langle A,s\rangle\xrightarrow{\{\alpha_1,\cdots,\alpha_n\}}\langle P',s'\rangle}\quad (A\overset{\text{def}}{=}P)$$
        \caption{Forward transition rules of Restriction, Relabelling and Constants}
        \label{FTRForRRC07}
    \end{table}
\end{center}

The reverse transition rules for Restriction, Relabelling and Constants are shown in Table \ref{RTRForRRC07}.

\begin{center}
    \begin{table}
        $$\frac{\langle P,s\rangle\xtworightarrow{\alpha[m]}\langle P',s'\rangle}{\langle P\setminus L,s\rangle\xtworightarrow{\alpha[m]}\langle P'\setminus L,s'\rangle}\quad (\alpha,\overline{\alpha}\notin L)$$
        $$\frac{\langle P,s\rangle\xtworightarrow{\{\alpha_1[m],\cdots,\alpha_n[m]\}}\langle P',s'\rangle}{\langle P\setminus L,s\rangle\xtworightarrow{\{\alpha_1[m],\cdots,\alpha_n[m]\}}\langle P'\setminus L,s'\rangle}\quad (\alpha_1,\overline{\alpha_1},\cdots,\alpha_n,\overline{\alpha_n}\notin L)$$
        $$\frac{\langle P,s\rangle\xtworightarrow{\alpha[m]}\langle P',s'\rangle}{\langle P[f],s\rangle\xtworightarrow{f(\alpha[m])}\langle P'[f],s'\rangle}$$
        $$\frac{\langle P,s\rangle\xtworightarrow{\{\alpha_1[m],\cdots,\alpha_n[m]\}}\langle P',s'\rangle}{\langle P[f],s\rangle\xtworightarrow{\{f(\alpha_1)[m],\cdots,f(\alpha_n)[m]\}}\langle P'[f],s'\rangle}$$
        $$\frac{\langle P,s\rangle\xtworightarrow{\alpha[m]}\langle P',s'\rangle}{\langle A,s\rangle\xtworightarrow{\alpha[m]}\langle P',s'\rangle}\quad (A\overset{\text{def}}{=}P)$$
        $$\frac{\langle P,s\rangle\xtworightarrow{\{\alpha_1[m],\cdots,\alpha_n[m]\}}\langle P',s'\rangle}{\langle A,s\rangle\xtworightarrow{\{\alpha_1[m],\cdots,\alpha_n[m]\}}\langle P',s'\rangle}\quad (A\overset{\text{def}}{=}P)$$
        \caption{Reverse transition rules of Restriction, Relabelling and Constants}
        \label{RTRForRRC07}
    \end{table}
\end{center}

\subsubsection{Properties of Transitions}

\begin{definition}[Sorts]\label{sorts07}
Given the sorts $\mathcal{L}(A)$ and $\mathcal{L}(X)$ of constants and variables, we define $\mathcal{L}(P)$ inductively as follows.

\begin{enumerate}
  \item $\mathcal{L}(l.P)=\{l\}\cup\mathcal{L}(P)$;
  \item $\mathcal{L}(P.l[m])=\{l\}\cup\mathcal{L}(P)$;
  \item $\mathcal{L}((l_1\parallel \cdots\parallel l_n).P)=\{l_1,\cdots,l_n\}\cup\mathcal{L}(P)$;
  \item $\mathcal{L}(P.(l_1[m]\parallel \cdots\parallel l_n[m]))=\{l_1,\cdots,l_n\}\cup\mathcal{L}(P)$;
  \item $\mathcal{L}(\tau.P)=\mathcal{L}(P)$;
  \item $\mathcal{L}(\epsilon.P)=\mathcal{L}(P)$;
  \item $\mathcal{L}(\phi.P)=\mathcal{L}(P)$;
  \item $\mathcal{L}(P+Q)=\mathcal{L}(P)\cup\mathcal{L}(Q)$;
  \item $\mathcal{L}(P\boxplus_{\pi}Q)=\mathcal{L}(P)\cup\mathcal{L}(Q)$;
  \item $\mathcal{L}(P\parallel Q)=\mathcal{L}(P)\cup\mathcal{L}(Q)$;
  \item $\mathcal{L}(P\setminus L)=\mathcal{L}(P)-(L\cup\overline{L})$;
  \item $\mathcal{L}(P[f])=\{f(l):l\in\mathcal{L}(P)\}$;
  \item for $A\overset{\text{def}}{=}P$, $\mathcal{L}(P)\subseteq\mathcal{L}(A)$.
\end{enumerate}
\end{definition}

Now, we present some properties of the transition rules defined in Definition \ref{semantics07}.

\begin{proposition}
If $P\xrightarrow{\alpha}P'$, then
\begin{enumerate}
  \item $\alpha\in\mathcal{L}(P)\cup\{\tau\}\cup\{\epsilon\}$;
  \item $\mathcal{L}(P')\subseteq\mathcal{L}(P)$.
\end{enumerate}

If $P\xrightarrow{\{\alpha_1,\cdots,\alpha_n\}}P'$, then
\begin{enumerate}
  \item $\alpha_1,\cdots,\alpha_n\in\mathcal{L}(P)\cup\{\tau\}\cup\{\epsilon\}$;
  \item $\mathcal{L}(P')\subseteq\mathcal{L}(P)$.
\end{enumerate}
\end{proposition}

\begin{proof}
By induction on the inference of $P\xrightarrow{\alpha}P'$ and $P\xrightarrow{\{\alpha_1,\cdots,\alpha_n\}}P'$, there are several cases corresponding to the forward transition rules in
Definition \ref{semantics07}, we omit them.
\end{proof}

\begin{proposition}
If $P\xtworightarrow{\alpha[m]}P'$, then
\begin{enumerate}
  \item $\alpha\in\mathcal{L}(P)\cup\{\tau\}\cup\{\epsilon\}$;
  \item $\mathcal{L}(P')\subseteq\mathcal{L}(P)$.
\end{enumerate}

If $P\xtworightarrow{\{\alpha_1[m],\cdots,\alpha_n[m]\}}P'$, then
\begin{enumerate}
  \item $\alpha_1,\cdots,\alpha_n\in\mathcal{L}(P)\cup\{\tau\}\cup\{\epsilon\}$;
  \item $\mathcal{L}(P')\subseteq\mathcal{L}(P)$.
\end{enumerate}
\end{proposition}

\begin{proof}
By induction on the inference of $P\xtworightarrow{\alpha}P'$ and $P\xtworightarrow{\{\alpha_1,\cdots,\alpha_n\}}P'$, there are several cases corresponding to the forward transition
rules in Definition \ref{semantics07}, we omit them.
\end{proof}

\subsection{Strong Bisimulations}\label{sftcbpa}

\subsubsection{Laws and Congruence}

Based on the concepts of strongly FR truly concurrent bisimulation equivalences, we get the following laws.

\begin{proposition}[Monoid laws for FR strongly probabilistic pomset bisimulation] The monoid laws for FR strongly probabilistic pomset bisimulation are as follows.

\begin{enumerate}
  \item $P+Q\sim_{pp}^{fr} Q+P$;
  \item $P+(Q+R)\sim_{pp}^{fr} (P+Q)+R$;
  \item $P+P\sim_{pp}^{fr} P$;
  \item $P+\textbf{nil}\sim_{pp}^{fr} P$.
\end{enumerate}

\end{proposition}

\begin{proof}
\begin{enumerate}
  \item $P+Q\sim_{pp}^{fr} Q+P$. It is sufficient to prove the relation $R=\{(P+Q, Q+P)\}\cup \textbf{Id}$ is a FR strongly probabilistic pomset bisimulation, we omit it;
  \item $P+(Q+R)\sim_{pp}^{fr} (P+Q)+R$. It is sufficient to prove the relation $R=\{(P+(Q+R), (P+Q)+R)\}\cup \textbf{Id}$ is a FR strongly probabilistic pomset bisimulation, we omit it;
  \item $P+P\sim_{pp}^{fr} P$. It is sufficient to prove the relation $R=\{(P+P, P)\}\cup \textbf{Id}$ is a FR strongly probabilistic pomset bisimulation, we omit it;
  \item $P+\textbf{nil}\sim_{pp}^{fr} P$. It is sufficient to prove the relation $R=\{(P+\textbf{nil}, P)\}\cup \textbf{Id}$ is a FR strongly probabilistic pomset bisimulation, we omit it.
\end{enumerate}
\end{proof}

\begin{proposition}[Monoid laws for FR strongly probabilistic step bisimulation] The monoid laws for FR strongly probabilistic step bisimulation are as follows.
\begin{enumerate}
  \item $P+Q\sim_{ps}^{fr} Q+P$;
  \item $P+(Q+R)\sim_{ps}^{fr} (P+Q)+R$;
  \item $P+P\sim_{ps}^{fr} P$;
  \item $P+\textbf{nil}\sim_{ps}^{fr} P$.
\end{enumerate}
\end{proposition}

\begin{proof}
\begin{enumerate}
  \item $P+Q\sim_{ps}^{fr} Q+P$. It is sufficient to prove the relation $R=\{(P+Q, Q+P)\}\cup \textbf{Id}$ is a FR strongly probabilistic step bisimulation, we omit it;
  \item $P+(Q+R)\sim_{ps}^{fr} (P+Q)+R$. It is sufficient to prove the relation $R=\{(P+(Q+R), (P+Q)+R)\}\cup \textbf{Id}$ is a FR strongly probabilistic step bisimulation, we omit it;
  \item $P+P\sim_{ps}^{fr} P$. It is sufficient to prove the relation $R=\{(P+P, P)\}\cup \textbf{Id}$ is a FR strongly probabilistic step bisimulation, we omit it;
  \item $P+\textbf{nil}\sim_{ps}^{fr} P$. It is sufficient to prove the relation $R=\{(P+\textbf{nil}, P)\}\cup \textbf{Id}$ is a FR strongly probabilistic step bisimulation, we omit it.
\end{enumerate}
\end{proof}

\begin{proposition}[Monoid laws for FR strongly probabilistic hp-bisimulation] The monoid laws for FR strongly probabilistic hp-bisimulation are as follows.
\begin{enumerate}
  \item $P+Q\sim_{php}^{fr} Q+P$;
  \item $P+(Q+R)\sim_{php}^{fr} (P+Q)+R$;
  \item $P+P\sim_{php}^{fr} P$;
  \item $P+\textbf{nil}\sim_{php}^{fr} P$.
\end{enumerate}
\end{proposition}

\begin{proof}
\begin{enumerate}
  \item $P+Q\sim_{php}^{fr} Q+P$. It is sufficient to prove the relation $R=\{(P+Q, Q+P)\}\cup \textbf{Id}$ is a FR strongly probabilistic hp-bisimulation, we omit it;
  \item $P+(Q+R)\sim_{php}^{fr} (P+Q)+R$. It is sufficient to prove the relation $R=\{(P+(Q+R), (P+Q)+R)\}\cup \textbf{Id}$ is a FR strongly probabilistic hp-bisimulation, we omit it;
  \item $P+P\sim_{php}^{fr} P$. It is sufficient to prove the relation $R=\{(P+P, P)\}\cup \textbf{Id}$ is a FR strongly probabilistic hp-bisimulation, we omit it;
  \item $P+\textbf{nil}\sim_{php}^{fr} P$. It is sufficient to prove the relation $R=\{(P+\textbf{nil}, P)\}\cup \textbf{Id}$ is a FR strongly probabilistic hp-bisimulation, we omit it.
\end{enumerate}
\end{proof}

\begin{proposition}[Monoid laws for FR strongly probabilistic hhp-bisimulation] The monoid laws for FR strongly probabilistic hhp-bisimulation are as follows.
\begin{enumerate}
  \item $P+Q\sim_{phhp}^{fr} Q+P$;
  \item $P+(Q+R)\sim_{phhp}^{fr} (P+Q)+R$;
  \item $P+P\sim_{phhp}^{fr} P$;
  \item $P+\textbf{nil}\sim_{phhp}^{fr} P$.
\end{enumerate}
\end{proposition}

\begin{proof}
\begin{enumerate}
  \item $P+Q\sim_{phhp}^{fr} Q+P$. It is sufficient to prove the relation $R=\{(P+Q, Q+P)\}\cup \textbf{Id}$ is a FR strongly probabilistic hhp-bisimulation, we omit it;
  \item $P+(Q+R)\sim_{phhp}^{fr} (P+Q)+R$. It is sufficient to prove the relation $R=\{(P+(Q+R), (P+Q)+R)\}\cup \textbf{Id}$ is a FR strongly probabilistic hhp-bisimulation, we omit it;
  \item $P+P\sim_{phhp}^{fr} P$. It is sufficient to prove the relation $R=\{(P+P, P)\}\cup \textbf{Id}$ is a FR strongly probabilistic hhp-bisimulation, we omit it;
  \item $P+\textbf{nil}\sim_{phhp}^{fr} P$. It is sufficient to prove the relation $R=\{(P+\textbf{nil}, P)\}\cup \textbf{Id}$ is a FR strongly probabilistic hhp-bisimulation, we omit it.
\end{enumerate}
\end{proof}

\begin{proposition}[Monoid laws 2 for FR strongly probabilistic pomset bisimulation]
The monoid laws 2 for FR strongly probabilistic pomset bisimulation are as follows.

\begin{enumerate}
  \item $P\boxplus_{\pi} Q\sim_{pp}^{fr} Q\boxplus_{1-\pi} P$;
  \item $P\boxplus_{\pi}(Q\boxplus_{\rho} R)\sim_{pp}^{fr} (P\boxplus_{\frac{\pi}{\pi+\rho-\pi\rho}}Q)\boxplus_{\pi+\rho-\pi\rho} R$;
  \item $P\boxplus_{\pi}P\sim_{pp}^{fr} P$;
  \item $P\boxplus_{\pi}\textbf{nil}\sim_{pp}^{fr} P$.
\end{enumerate}
\end{proposition}

\begin{proof}
\begin{enumerate}
  \item $P\boxplus_{\pi} Q\sim_{pp}^{fr} Q\boxplus_{1-\pi} P$. It is sufficient to prove the relation $R=\{(P\boxplus_{\pi} Q, Q\boxplus_{1-\pi} P)\}\cup \textbf{Id}$ is a FR strongly probabilistic pomset bisimulation, we omit it;
  \item $P\boxplus_{\pi}(Q\boxplus_{\rho} R)\sim_{pp}^{fr} (P\boxplus_{\frac{\pi}{\pi+\rho-\pi\rho}}Q)\boxplus_{\pi+\rho-\pi\rho} R$. It is sufficient to prove the relation $R=\{(P\boxplus_{\pi}(Q\boxplus_{\rho} R), (P\boxplus_{\frac{\pi}{\pi+\rho-\pi\rho}}Q)\boxplus_{\pi+\rho-\pi\rho} R)\}\cup \textbf{Id}$ is a FR strongly probabilistic pomset bisimulation, we omit it;
  \item $P\boxplus_{\pi}P\sim_{pp}^{fr} P$. It is sufficient to prove the relation $R=\{(P\boxplus_{\pi}P, P)\}\cup \textbf{Id}$ is a FR strongly probabilistic pomset bisimulation, we omit it;
  \item $P\boxplus_{\pi}\textbf{nil}\sim_{pp}^{fr} P$. It is sufficient to prove the relation $R=\{(P\boxplus_{\pi}\textbf{nil}, P)\}\cup \textbf{Id}$ is a FR strongly probabilistic pomset bisimulation, we omit it.
\end{enumerate}
\end{proof}

\begin{proposition}[Monoid laws 2 for FR strongly probabilistic step bisimulation]
The monoid laws 2 for FR strongly probabilistic step bisimulation are as follows.

\begin{enumerate}
  \item $P\boxplus_{\pi} Q\sim_{ps}^{fr} Q\boxplus_{1-\pi} P$;
  \item $P\boxplus_{\pi}(Q\boxplus_{\rho} R)\sim_{ps}^{fr} (P\boxplus_{\frac{\pi}{\pi+\rho-\pi\rho}}Q)\boxplus_{\pi+\rho-\pi\rho} R$;
  \item $P\boxplus_{\pi}P\sim_{ps}^{fr} P$;
  \item $P\boxplus_{\pi}\textbf{nil}\sim_{ps}^{fr} P$.
\end{enumerate}
\end{proposition}

\begin{proof}
\begin{enumerate}
  \item $P\boxplus_{\pi} Q\sim_{ps}^{fr} Q\boxplus_{1-\pi} P$. It is sufficient to prove the relation $R=\{(P\boxplus_{\pi} Q, Q\boxplus_{1-\pi} P)\}\cup \textbf{Id}$ is a FR strongly probabilistic step bisimulation, we omit it;
  \item $P\boxplus_{\pi}(Q\boxplus_{\rho} R)\sim_{ps}^{fr} (P\boxplus_{\frac{\pi}{\pi+\rho-\pi\rho}}Q)\boxplus_{\pi+\rho-\pi\rho} R$. It is sufficient to prove the relation $R=\{(P\boxplus_{\pi}(Q\boxplus_{\rho} R), (P\boxplus_{\frac{\pi}{\pi+\rho-\pi\rho}}Q)\boxplus_{\pi+\rho-\pi\rho} R)\}\cup \textbf{Id}$ is a FR strongly probabilistic step bisimulation, we omit it;
  \item $P\boxplus_{\pi}P\sim_{ps}^{fr} P$. It is sufficient to prove the relation $R=\{(P\boxplus_{\pi}P, P)\}\cup \textbf{Id}$ is a FR strongly probabilistic step bisimulation, we omit it;
  \item $P\boxplus_{\pi}\textbf{nil}\sim_{ps}^{fr} P$. It is sufficient to prove the relation $R=\{(P\boxplus_{\pi}\textbf{nil}, P)\}\cup \textbf{Id}$ is a FR strongly probabilistic step bisimulation, we omit it.
\end{enumerate}
\end{proof}

\begin{proposition}[Monoid laws 2 for FR strongly probabilistic hp-bisimulation]
The monoid laws 2 for FR strongly probabilistic hp-bisimulation are as follows.

\begin{enumerate}
  \item $P\boxplus_{\pi} Q\sim_{php}^{fr} Q\boxplus_{1-\pi} P$;
  \item $P\boxplus_{\pi}(Q\boxplus_{\rho} R)\sim_{php}^{fr} (P\boxplus_{\frac{\pi}{\pi+\rho-\pi\rho}}Q)\boxplus_{\pi+\rho-\pi\rho} R$;
  \item $P\boxplus_{\pi}P\sim_{php}^{fr} P$;
  \item $P\boxplus_{\pi}\textbf{nil}\sim_{php}^{fr} P$.
\end{enumerate}
\end{proposition}

\begin{proof}
\begin{enumerate}
  \item $P\boxplus_{\pi} Q\sim_{php}^{fr} Q\boxplus_{1-\pi} P$. It is sufficient to prove the relation $R=\{(P\boxplus_{\pi} Q, Q\boxplus_{1-\pi} P)\}\cup \textbf{Id}$ is a FR strongly probabilistic hp-bisimulation, we omit it;
  \item $P\boxplus_{\pi}(Q\boxplus_{\rho} R)\sim_{php}^{fr} (P\boxplus_{\frac{\pi}{\pi+\rho-\pi\rho}}Q)\boxplus_{\pi+\rho-\pi\rho} R$. It is sufficient to prove the relation $R=\{(P\boxplus_{\pi}(Q\boxplus_{\rho} R), (P\boxplus_{\frac{\pi}{\pi+\rho-\pi\rho}}Q)\boxplus_{\pi+\rho-\pi\rho} R)\}\cup \textbf{Id}$ is a FR strongly probabilistic hp-bisimulation, we omit it;
  \item $P\boxplus_{\pi}P\sim_{php}^{fr} P$. It is sufficient to prove the relation $R=\{(P\boxplus_{\pi}P, P)\}\cup \textbf{Id}$ is a FR strongly probabilistic hp-bisimulation, we omit it;
  \item $P\boxplus_{\pi}\textbf{nil}\sim_{php}^{fr} P$. It is sufficient to prove the relation $R=\{(P\boxplus_{\pi}\textbf{nil}, P)\}\cup \textbf{Id}$ is a FR strongly probabilistic hp-bisimulation, we omit it.
\end{enumerate}
\end{proof}

\begin{proposition}[Monoid laws 2 for FR strongly probabilistic hhp-bisimulation]
The monoid laws 2 for FR strongly probabilistic hhp-bisimulation are as follows.

\begin{enumerate}
  \item $P\boxplus_{\pi} Q\sim_{phhp}^{fr} Q\boxplus_{1-\pi} P$;
  \item $P\boxplus_{\pi}(Q\boxplus_{\rho} R)\sim_{phhp}^{fr} (P\boxplus_{\frac{\pi}{\pi+\rho-\pi\rho}}Q)\boxplus_{\pi+\rho-\pi\rho} R$;
  \item $P\boxplus_{\pi}P\sim_{phhp}^{fr} P$;
  \item $P\boxplus_{\pi}\textbf{nil}\sim_{phhp}^{fr} P$.
\end{enumerate}
\end{proposition}

\begin{proof}
\begin{enumerate}
  \item $P\boxplus_{\pi} Q\sim_{phhp}^{fr} Q\boxplus_{1-\pi} P$. It is sufficient to prove the relation $R=\{(P\boxplus_{\pi} Q, Q\boxplus_{1-\pi} P)\}\cup \textbf{Id}$ is a FR strongly probabilistic hhp-bisimulation, we omit it;
  \item $P\boxplus_{\pi}(Q\boxplus_{\rho} R)\sim_{phhp}^{fr} (P\boxplus_{\frac{\pi}{\pi+\rho-\pi\rho}}Q)\boxplus_{\pi+\rho-\pi\rho} R$. It is sufficient to prove the relation $R=\{(P\boxplus_{\pi}(Q\boxplus_{\rho} R), (P\boxplus_{\frac{\pi}{\pi+\rho-\pi\rho}}Q)\boxplus_{\pi+\rho-\pi\rho} R)\}\cup \textbf{Id}$ is a FR strongly probabilistic hhp-bisimulation, we omit it;
  \item $P\boxplus_{\pi}P\sim_{phhp}^{fr} P$. It is sufficient to prove the relation $R=\{(P\boxplus_{\pi}P, P)\}\cup \textbf{Id}$ is a FR strongly probabilistic hhp-bisimulation, we omit it;
  \item $P\boxplus_{\pi}\textbf{nil}\sim_{phhp}^{fr} P$. It is sufficient to prove the relation $R=\{(P\boxplus_{\pi}\textbf{nil}, P)\}\cup \textbf{Id}$ is a FR strongly probabilistic hhp-bisimulation, we omit it.
\end{enumerate}
\end{proof}

\begin{proposition}[Static laws for FR strongly probabilistic pomset bisimulation]
The static laws for FR strongly probabilistic pomset bisimulation are as follows.
\begin{enumerate}
  \item $P\parallel Q\sim_{pp}^{fr} Q\parallel P$;
  \item $P\parallel(Q\parallel R)\sim_{pp}^{fr} (P\parallel Q)\parallel R$;
  \item $P\parallel \textbf{nil}\sim_{pp}^{fr} P$;
  \item $P\setminus L\sim_{pp}^{fr} P$, if $\mathcal{L}(P)\cap(L\cup\overline{L})=\emptyset$;
  \item $P\setminus K\setminus L\sim_{pp}^{fr} P\setminus(K\cup L)$;
  \item $P[f]\setminus L\sim_{pp}^{fr} P\setminus f^{-1}(L)[f]$;
  \item $(P\parallel Q)\setminus L\sim_{pp}^{fr} P\setminus L\parallel Q\setminus L$, if $\mathcal{L}(P)\cap\overline{\mathcal{L}(Q)}\cap(L\cup\overline{L})=\emptyset$;
  \item $P[Id]\sim_{pp}^{fr} P$;
  \item $P[f]\sim_{pp}^{fr} P[f']$, if $f\upharpoonright\mathcal{L}(P)=f'\upharpoonright\mathcal{L}(P)$;
  \item $P[f][f']\sim_{pp}^{fr} P[f'\circ f]$;
  \item $(P\parallel Q)[f]\sim_{pp}^{fr} P[f]\parallel Q[f]$, if $f\upharpoonright(L\cup\overline{L})$ is one-to-one, where $L=\mathcal{L}(P)\cup\mathcal{L}(Q)$.
\end{enumerate}
\end{proposition}

\begin{proof}
\begin{enumerate}
  \item $P\parallel Q\sim_{pp}^{fr} Q\parallel P$. It is sufficient to prove the relation $R=\{(P\parallel Q, Q\parallel P)\}\cup \textbf{Id}$ is a FR strongly probabilistic pomset bisimulation, we omit it;
  \item $P\parallel(Q\parallel R)\sim_{pp}^{fr} (P\parallel Q)\parallel R$. It is sufficient to prove the relation $R=\{(P\parallel(Q\parallel R), (P\parallel Q)\parallel R)\}\cup \textbf{Id}$ is a FR strongly probabilistic pomset bisimulation, we omit it;
  \item $P\parallel \textbf{nil}\sim_{pp}^{fr} P$. It is sufficient to prove the relation $R=\{(P\parallel \textbf{nil}, P)\}\cup \textbf{Id}$ is a FR strongly probabilistic pomset bisimulation, we omit it;
  \item $P\setminus L\sim_{pp}^{fr} P$, if $\mathcal{L}(P)\cap(L\cup\overline{L})=\emptyset$. It is sufficient to prove the relation $R=\{(P\setminus L, P)\}\cup \textbf{Id}$, if $\mathcal{L}(P)\cap(L\cup\overline{L})=\emptyset$, is a FR strongly probabilistic pomset bisimulation, we omit it;
  \item $P\setminus K\setminus L\sim_{pp}^{fr} P\setminus(K\cup L)$. It is sufficient to prove the relation $R=\{(P\setminus K\setminus L, P\setminus(K\cup L))\}\cup \textbf{Id}$ is a FR strongly probabilistic pomset bisimulation, we omit it;
  \item $P[f]\setminus L\sim_{pp}^{fr} P\setminus f^{-1}(L)[f]$. It is sufficient to prove the relation $R=\{(P[f]\setminus L, P\setminus f^{-1}(L)[f])\}\cup \textbf{Id}$ is a FR strongly probabilistic pomset bisimulation, we omit it;
  \item $(P\parallel Q)\setminus L\sim_{pp}^{fr} P\setminus L\parallel Q\setminus L$, if $\mathcal{L}(P)\cap\overline{\mathcal{L}(Q)}\cap(L\cup\overline{L})=\emptyset$. It is sufficient to prove the relation
  $R=\{((P\parallel Q)\setminus L, P\setminus L\parallel Q\setminus L)\}\cup \textbf{Id}$, if $\mathcal{L}(P)\cap\overline{\mathcal{L}(Q)}\cap(L\cup\overline{L})=\emptyset$, is a FR strongly probabilistic pomset bisimulation, we omit it;
  \item $P[Id]\sim_{pp}^{fr} P$. It is sufficient to prove the relation $R=\{(P[Id], P)\}\cup \textbf{Id}$ is a FR strongly probabilistic pomset bisimulation, we omit it;
  \item $P[f]\sim_{pp}^{fr} P[f']$, if $f\upharpoonright\mathcal{L}(P)=f'\upharpoonright\mathcal{L}(P)$. It is sufficient to prove the relation $R=\{(P[f], P[f'])\}\cup \textbf{Id}$, if $f\upharpoonright\mathcal{L}(P)=f'\upharpoonright\mathcal{L}(P)$, is a FR strongly probabilistic pomset bisimulation, we omit it;
  \item $P[f][f']\sim_{pp}^{fr} P[f'\circ f]$. It is sufficient to prove the relation $R=\{(P[f][f'], P[f'\circ f])\}\cup \textbf{Id}$ is a FR strongly probabilistic pomset bisimulation, we omit it;
  \item $(P\parallel Q)[f]\sim_{pp}^{fr} P[f]\parallel Q[f]$, if $f\upharpoonright(L\cup\overline{L})$ is one-to-one, where $L=\mathcal{L}(P)\cup\mathcal{L}(Q)$. It is sufficient to prove the
  relation $R=\{((P\parallel Q)[f], P[f]\parallel Q[f])\}\cup \textbf{Id}$, if $f\upharpoonright(L\cup\overline{L})$ is one-to-one, where $L=\mathcal{L}(P)\cup\mathcal{L}(Q)$, is a FR strongly probabilistic pomset bisimulation, we omit it.
\end{enumerate}
\end{proof}

\begin{proposition}[Static laws for FR strongly probabilistic step bisimulation]
The static laws for FR strongly probabilistic step bisimulation are as follows.
\begin{enumerate}
  \item $P\parallel Q\sim_{ps}^{fr} Q\parallel P$;
  \item $P\parallel(Q\parallel R)\sim_{ps}^{fr} (P\parallel Q)\parallel R$;
  \item $P\parallel \textbf{nil}\sim_{ps}^{fr} P$;
  \item $P\setminus L\sim_{ps}^{fr} P$, if $\mathcal{L}(P)\cap(L\cup\overline{L})=\emptyset$;
  \item $P\setminus K\setminus L\sim_{ps}^{fr} P\setminus(K\cup L)$;
  \item $P[f]\setminus L\sim_{ps}^{fr} P\setminus f^{-1}(L)[f]$;
  \item $(P\parallel Q)\setminus L\sim_{ps}^{fr} P\setminus L\parallel Q\setminus L$, if $\mathcal{L}(P)\cap\overline{\mathcal{L}(Q)}\cap(L\cup\overline{L})=\emptyset$;
  \item $P[Id]\sim_{ps}^{fr} P$;
  \item $P[f]\sim_{ps}^{fr} P[f']$, if $f\upharpoonright\mathcal{L}(P)=f'\upharpoonright\mathcal{L}(P)$;
  \item $P[f][f']\sim_{ps}^{fr} P[f'\circ f]$;
  \item $(P\parallel Q)[f]\sim_{ps}^{fr} P[f]\parallel Q[f]$, if $f\upharpoonright(L\cup\overline{L})$ is one-to-one, where $L=\mathcal{L}(P)\cup\mathcal{L}(Q)$.
\end{enumerate}
\end{proposition}

\begin{proof}
\begin{enumerate}
  \item $P\parallel Q\sim_{ps}^{fr} Q\parallel P$. It is sufficient to prove the relation $R=\{(P\parallel Q, Q\parallel P)\}\cup \textbf{Id}$ is a FR strongly probabilistic step bisimulation, we omit it;
  \item $P\parallel(Q\parallel R)\sim_{ps}^{fr} (P\parallel Q)\parallel R$. It is sufficient to prove the relation $R=\{(P\parallel(Q\parallel R), (P\parallel Q)\parallel R)\}\cup \textbf{Id}$ is a FR strongly probabilistic step bisimulation, we omit it;
  \item $P\parallel \textbf{nil}\sim_{ps}^{fr} P$. It is sufficient to prove the relation $R=\{(P\parallel \textbf{nil}, P)\}\cup \textbf{Id}$ is a FR strongly probabilistic step bisimulation, we omit it;
  \item $P\setminus L\sim_{ps}^{fr} P$, if $\mathcal{L}(P)\cap(L\cup\overline{L})=\emptyset$. It is sufficient to prove the relation $R=\{(P\setminus L, P)\}\cup \textbf{Id}$, if $\mathcal{L}(P)\cap(L\cup\overline{L})=\emptyset$, is a FR strongly probabilistic step bisimulation, we omit it;
  \item $P\setminus K\setminus L\sim_{ps}^{fr} P\setminus(K\cup L)$. It is sufficient to prove the relation $R=\{(P\setminus K\setminus L, P\setminus(K\cup L))\}\cup \textbf{Id}$ is a FR strongly probabilistic step bisimulation, we omit it;
  \item $P[f]\setminus L\sim_{ps}^{fr} P\setminus f^{-1}(L)[f]$. It is sufficient to prove the relation $R=\{(P[f]\setminus L, P\setminus f^{-1}(L)[f])\}\cup \textbf{Id}$ is a FR strongly probabilistic step bisimulation, we omit it;
  \item $(P\parallel Q)\setminus L\sim_{ps}^{fr} P\setminus L\parallel Q\setminus L$, if $\mathcal{L}(P)\cap\overline{\mathcal{L}(Q)}\cap(L\cup\overline{L})=\emptyset$. It is sufficient to prove the relation
  $R=\{((P\parallel Q)\setminus L, P\setminus L\parallel Q\setminus L)\}\cup \textbf{Id}$, if $\mathcal{L}(P)\cap\overline{\mathcal{L}(Q)}\cap(L\cup\overline{L})=\emptyset$, is a FR strongly probabilistic step bisimulation, we omit it;
  \item $P[Id]\sim_{ps}^{fr} P$. It is sufficient to prove the relation $R=\{(P[Id], P)\}\cup \textbf{Id}$ is a FR strongly probabilistic step bisimulation, we omit it;
  \item $P[f]\sim_{ps}^{fr} P[f']$, if $f\upharpoonright\mathcal{L}(P)=f'\upharpoonright\mathcal{L}(P)$. It is sufficient to prove the relation $R=\{(P[f], P[f'])\}\cup \textbf{Id}$, if $f\upharpoonright\mathcal{L}(P)=f'\upharpoonright\mathcal{L}(P)$, is a FR strongly probabilistic step bisimulation, we omit it;
  \item $P[f][f']\sim_{ps}^{fr} P[f'\circ f]$. It is sufficient to prove the relation $R=\{(P[f][f'], P[f'\circ f])\}\cup \textbf{Id}$ is a FR strongly probabilistic step bisimulation, we omit it;
  \item $(P\parallel Q)[f]\sim_{ps}^{fr} P[f]\parallel Q[f]$, if $f\upharpoonright(L\cup\overline{L})$ is one-to-one, where $L=\mathcal{L}(P)\cup\mathcal{L}(Q)$. It is sufficient to prove the
  relation $R=\{((P\parallel Q)[f], P[f]\parallel Q[f])\}\cup \textbf{Id}$, if $f\upharpoonright(L\cup\overline{L})$ is one-to-one, where $L=\mathcal{L}(P)\cup\mathcal{L}(Q)$, is a FR strongly probabilistic step bisimulation, we omit it.
\end{enumerate}
\end{proof}

\begin{proposition}[Static laws for FR strongly probabilistic hp-bisimulation]
The static laws for FR strongly probabilistic hp-bisimulation are as follows.
\begin{enumerate}
  \item $P\parallel Q\sim_{php}^{fr} Q\parallel P$;
  \item $P\parallel(Q\parallel R)\sim_{php}^{fr} (P\parallel Q)\parallel R$;
  \item $P\parallel \textbf{nil}\sim_{php}^{fr} P$;
  \item $P\setminus L\sim_{php}^{fr} P$, if $\mathcal{L}(P)\cap(L\cup\overline{L})=\emptyset$;
  \item $P\setminus K\setminus L\sim_{php}^{fr} P\setminus(K\cup L)$;
  \item $P[f]\setminus L\sim_{php}^{fr} P\setminus f^{-1}(L)[f]$;
  \item $(P\parallel Q)\setminus L\sim_{php}^{fr} P\setminus L\parallel Q\setminus L$, if $\mathcal{L}(P)\cap\overline{\mathcal{L}(Q)}\cap(L\cup\overline{L})=\emptyset$;
  \item $P[Id]\sim_{php}^{fr} P$;
  \item $P[f]\sim_{php}^{fr} P[f']$, if $f\upharpoonright\mathcal{L}(P)=f'\upharpoonright\mathcal{L}(P)$;
  \item $P[f][f']\sim_{php}^{fr} P[f'\circ f]$;
  \item $(P\parallel Q)[f]\sim_{php}^{fr} P[f]\parallel Q[f]$, if $f\upharpoonright(L\cup\overline{L})$ is one-to-one, where $L=\mathcal{L}(P)\cup\mathcal{L}(Q)$.
\end{enumerate}
\end{proposition}

\begin{proof}
\begin{enumerate}
  \item $P\parallel Q\sim_{php}^{fr} Q\parallel P$. It is sufficient to prove the relation $R=\{(P\parallel Q, Q\parallel P)\}\cup \textbf{Id}$ is a FR strongly probabilistic hp-bisimulation, we omit it;
  \item $P\parallel(Q\parallel R)\sim_{php}^{fr} (P\parallel Q)\parallel R$. It is sufficient to prove the relation $R=\{(P\parallel(Q\parallel R), (P\parallel Q)\parallel R)\}\cup \textbf{Id}$ is a FR strongly probabilistic hp-bisimulation, we omit it;
  \item $P\parallel \textbf{nil}\sim_{php}^{fr} P$. It is sufficient to prove the relation $R=\{(P\parallel \textbf{nil}, P)\}\cup \textbf{Id}$ is a FR strongly probabilistic hp-bisimulation, we omit it;
  \item $P\setminus L\sim_{php}^{fr} P$, if $\mathcal{L}(P)\cap(L\cup\overline{L})=\emptyset$. It is sufficient to prove the relation $R=\{(P\setminus L, P)\}\cup \textbf{Id}$, if $\mathcal{L}(P)\cap(L\cup\overline{L})=\emptyset$, is a FR strongly probabilistic hp-bisimulation, we omit it;
  \item $P\setminus K\setminus L\sim_{php}^{fr} P\setminus(K\cup L)$. It is sufficient to prove the relation $R=\{(P\setminus K\setminus L, P\setminus(K\cup L))\}\cup \textbf{Id}$ is a FR strongly probabilistic hp-bisimulation, we omit it;
  \item $P[f]\setminus L\sim_{php}^{fr} P\setminus f^{-1}(L)[f]$. It is sufficient to prove the relation $R=\{(P[f]\setminus L, P\setminus f^{-1}(L)[f])\}\cup \textbf{Id}$ is a FR strongly probabilistic hp-bisimulation, we omit it;
  \item $(P\parallel Q)\setminus L\sim_{php}^{fr} P\setminus L\parallel Q\setminus L$, if $\mathcal{L}(P)\cap\overline{\mathcal{L}(Q)}\cap(L\cup\overline{L})=\emptyset$. It is sufficient to prove the relation
  $R=\{((P\parallel Q)\setminus L, P\setminus L\parallel Q\setminus L)\}\cup \textbf{Id}$, if $\mathcal{L}(P)\cap\overline{\mathcal{L}(Q)}\cap(L\cup\overline{L})=\emptyset$, is a FR strongly probabilistic hp-bisimulation, we omit it;
  \item $P[Id]\sim_{php}^{fr} P$. It is sufficient to prove the relation $R=\{(P[Id], P)\}\cup \textbf{Id}$ is a FR strongly probabilistic hp-bisimulation, we omit it;
  \item $P[f]\sim_{php}^{fr} P[f']$, if $f\upharpoonright\mathcal{L}(P)=f'\upharpoonright\mathcal{L}(P)$. It is sufficient to prove the relation $R=\{(P[f], P[f'])\}\cup \textbf{Id}$, if $f\upharpoonright\mathcal{L}(P)=f'\upharpoonright\mathcal{L}(P)$, is a FR strongly probabilistic hp-bisimulation, we omit it;
  \item $P[f][f']\sim_{php}^{fr} P[f'\circ f]$. It is sufficient to prove the relation $R=\{(P[f][f'], P[f'\circ f])\}\cup \textbf{Id}$ is a FR strongly probabilistic hp-bisimulation, we omit it;
  \item $(P\parallel Q)[f]\sim_{php}^{fr} P[f]\parallel Q[f]$, if $f\upharpoonright(L\cup\overline{L})$ is one-to-one, where $L=\mathcal{L}(P)\cup\mathcal{L}(Q)$. It is sufficient to prove the
  relation $R=\{((P\parallel Q)[f], P[f]\parallel Q[f])\}\cup \textbf{Id}$, if $f\upharpoonright(L\cup\overline{L})$ is one-to-one, where $L=\mathcal{L}(P)\cup\mathcal{L}(Q)$, is a FR strongly probabilistic hp-bisimulation, we omit it.
\end{enumerate}
\end{proof}

\begin{proposition}[Static laws for FR strongly probabilistic hhp-bisimulation]
The static laws for FR strongly probabilistic hhp-bisimulation are as follows.
\begin{enumerate}
  \item $P\parallel Q\sim_{phhp}^{fr} Q\parallel P$;
  \item $P\parallel(Q\parallel R)\sim_{phhp}^{fr} (P\parallel Q)\parallel R$;
  \item $P\parallel \textbf{nil}\sim_{phhp}^{fr} P$;
  \item $P\setminus L\sim_{phhp}^{fr} P$, if $\mathcal{L}(P)\cap(L\cup\overline{L})=\emptyset$;
  \item $P\setminus K\setminus L\sim_{phhp}^{fr} P\setminus(K\cup L)$;
  \item $P[f]\setminus L\sim_{phhp}^{fr} P\setminus f^{-1}(L)[f]$;
  \item $(P\parallel Q)\setminus L\sim_{phhp}^{fr} P\setminus L\parallel Q\setminus L$, if $\mathcal{L}(P)\cap\overline{\mathcal{L}(Q)}\cap(L\cup\overline{L})=\emptyset$;
  \item $P[Id]\sim_{phhp}^{fr} P$;
  \item $P[f]\sim_{phhp}^{fr} P[f']$, if $f\upharpoonright\mathcal{L}(P)=f'\upharpoonright\mathcal{L}(P)$;
  \item $P[f][f']\sim_{phhp}^{fr} P[f'\circ f]$;
  \item $(P\parallel Q)[f]\sim_{phhp}^{fr} P[f]\parallel Q[f]$, if $f\upharpoonright(L\cup\overline{L})$ is one-to-one, where $L=\mathcal{L}(P)\cup\mathcal{L}(Q)$.
\end{enumerate}
\end{proposition}

\begin{proof}
\begin{enumerate}
  \item $P\parallel Q\sim_{phhp}^{fr} Q\parallel P$. It is sufficient to prove the relation $R=\{(P\parallel Q, Q\parallel P)\}\cup \textbf{Id}$ is a FR strongly probabilistic hhp-bisimulation, we omit it;
  \item $P\parallel(Q\parallel R)\sim_{phhp}^{fr} (P\parallel Q)\parallel R$. It is sufficient to prove the relation $R=\{(P\parallel(Q\parallel R), (P\parallel Q)\parallel R)\}\cup \textbf{Id}$ is a FR strongly probabilistic hhp-bisimulation, we omit it;
  \item $P\parallel \textbf{nil}\sim_{phhp}^{fr} P$. It is sufficient to prove the relation $R=\{(P\parallel \textbf{nil}, P)\}\cup \textbf{Id}$ is a FR strongly probabilistic hhp-bisimulation, we omit it;
  \item $P\setminus L\sim_{phhp}^{fr} P$, if $\mathcal{L}(P)\cap(L\cup\overline{L})=\emptyset$. It is sufficient to prove the relation $R=\{(P\setminus L, P)\}\cup \textbf{Id}$, if $\mathcal{L}(P)\cap(L\cup\overline{L})=\emptyset$, is a FR strongly probabilistic hhp-bisimulation, we omit it;
  \item $P\setminus K\setminus L\sim_{phhp}^{fr} P\setminus(K\cup L)$. It is sufficient to prove the relation $R=\{(P\setminus K\setminus L, P\setminus(K\cup L))\}\cup \textbf{Id}$ is a FR strongly probabilistic hhp-bisimulation, we omit it;
  \item $P[f]\setminus L\sim_{phhp}^{fr} P\setminus f^{-1}(L)[f]$. It is sufficient to prove the relation $R=\{(P[f]\setminus L, P\setminus f^{-1}(L)[f])\}\cup \textbf{Id}$ is a FR strongly probabilistic hhp-bisimulation, we omit it;
  \item $(P\parallel Q)\setminus L\sim_{phhp}^{fr} P\setminus L\parallel Q\setminus L$, if $\mathcal{L}(P)\cap\overline{\mathcal{L}(Q)}\cap(L\cup\overline{L})=\emptyset$. It is sufficient to prove the relation
  $R=\{((P\parallel Q)\setminus L, P\setminus L\parallel Q\setminus L)\}\cup \textbf{Id}$, if $\mathcal{L}(P)\cap\overline{\mathcal{L}(Q)}\cap(L\cup\overline{L})=\emptyset$, is a FR strongly probabilistic hhp-bisimulation, we omit it;
  \item $P[Id]\sim_{phhp}^{fr} P$. It is sufficient to prove the relation $R=\{(P[Id], P)\}\cup \textbf{Id}$ is a FR strongly probabilistic hhp-bisimulation, we omit it;
  \item $P[f]\sim_{phhp}^{fr} P[f']$, if $f\upharpoonright\mathcal{L}(P)=f'\upharpoonright\mathcal{L}(P)$. It is sufficient to prove the relation $R=\{(P[f], P[f'])\}\cup \textbf{Id}$, if $f\upharpoonright\mathcal{L}(P)=f'\upharpoonright\mathcal{L}(P)$, is a FR strongly probabilistic hhp-bisimulation, we omit it;
  \item $P[f][f']\sim_{phhp}^{fr} P[f'\circ f]$. It is sufficient to prove the relation $R=\{(P[f][f'], P[f'\circ f])\}\cup \textbf{Id}$ is a FR strongly probabilistic hhp-bisimulation, we omit it;
  \item $(P\parallel Q)[f]\sim_{phhp}^{fr} P[f]\parallel Q[f]$, if $f\upharpoonright(L\cup\overline{L})$ is one-to-one, where $L=\mathcal{L}(P)\cup\mathcal{L}(Q)$. It is sufficient to prove the
  relation $R=\{((P\parallel Q)[f], P[f]\parallel Q[f])\}\cup \textbf{Id}$, if $f\upharpoonright(L\cup\overline{L})$ is one-to-one, where $L=\mathcal{L}(P)\cup\mathcal{L}(Q)$, is a FR strongly probabilistic hhp-bisimulation, we omit it.
\end{enumerate}
\end{proof}

\begin{proposition}[Guards laws for FR strongly probabilistic pomset bisimulation] The guards laws for FR strongly probabilistic pomset bisimulation are as follows.

\begin{enumerate}
  \item $P+\delta \sim_{pp}^{fr} P$;
  \item $\delta.P \sim_{pp}^{fr} \delta$;
  \item $\epsilon.P \sim_{pp}^{fr} P$;
  \item $P.\epsilon \sim_{pp}^{fr} P$;
  \item $\phi.\neg\phi \sim_{pp}^{fr} \delta$;
  \item $\phi+\neg\phi \sim_{pp}^{fr} \epsilon$;
  \item $\phi\boxplus_{\pi}\neg\phi \sim_{pp}^{fr} \epsilon$;
  \item $\phi.\delta \sim_{pp}^{fr} \delta$;
  \item $\phi.(P+Q)\sim_{pp}^{fr}\phi.P+\phi.Q\quad (Std(P),Std(Q))$;
  \item $(P+Q).\phi\sim_{pp}^{fr} P.\phi+ Q.\phi\quad(NStd(P),NStd(Q))$;
  \item $\phi.(P\boxplus_{\pi}Q)\sim_{pp}^{fr}\phi.P\boxplus_{\pi}\phi.Q\quad (Std(P),Std(Q))$;
  \item $(P\boxplus_{\pi}Q).\phi\sim_{pp}^{fr} P.\phi\boxplus_{\pi} Q.\phi\quad(NStd(P),NStd(Q))$;
  \item $\phi.(P.Q)\sim_{pp}^{fr} \phi.P.Q\quad (Std(P),Std(Q))$;
  \item $(P.Q).\phi\sim_{pp}^{fr} P.Q.\phi\quad(NStd(P),NStd(Q))$;
  \item $(\phi+\psi).P \sim_{pp}^{fr} \phi.P + \psi.P\quad (Std(P))$;
  \item $P.(\phi+\psi) \sim_{pp}^{fr} P.\phi + P.\psi\quad(NStd(P))$;
  \item $(\phi\boxplus_{\pi}\psi).P \sim_{pp}^{fr} \phi.P \boxplus_{\pi} \psi.P\quad (Std(P))$;
  \item $P.(\phi\boxplus_{\pi}\psi) \sim_{pp}^{fr} P.\phi \boxplus_{\pi} P.\psi\quad(NStd(P))$;
  \item $(\phi.\psi).P \sim_{pp}^{fr} \phi.(\psi.P)\quad(Std(P))$;
  \item $P.(\phi.\psi) \sim_{pp}^{fr}(P.\phi).\psi\quad(NStd(P))$;
  \item $\phi\sim_{pp}^{fr}\epsilon$ if $\forall s\in S.test(\phi,s)$;
  \item $\phi_0\cdot\cdots\cdot\phi_n \sim_{pp}^{fr} \delta$ if $\forall s\in S,\exists i\leq n.test(\neg\phi_i,s)$;
  \item $wp(\alpha,\phi).\alpha.\phi\sim_{pp}^{fr} wp(\alpha,\phi).\alpha$;
  \item $\phi. \alpha[m]. wp(\alpha[m],\phi)\sim_{pp}^{fr}\alpha[m].wp(\alpha[m],\phi)$;
  \item $\neg wp(\alpha,\phi).\alpha.\neg\phi\sim_{pp}^{fr}\neg wp(\alpha,\phi).\alpha$;
  \item $\neg\phi .\alpha[m] .\neg wp(\alpha[m],\phi)\sim_{pp}^{fr} \alpha[m]. \neg wp(\alpha[m],\phi)$;
  \item $\delta\parallel P \sim_{pp}^{fr} \delta$;
  \item $P\parallel \delta \sim_{pp}^{fr} \delta$;
  \item $\epsilon\parallel P \sim_{pp}^{fr} P$;
  \item $P\parallel \epsilon \sim_{pp}^{fr} P$;
  \item $\phi.(P\parallel Q) \sim_{pp}^{fr}\phi.P\parallel \phi.Q$;
  \item $\phi\parallel \delta \sim_{pp}^{fr} \delta$;
  \item $\delta\parallel \phi \sim_{pp}^{fr} \delta$;
  \item $\phi\parallel \epsilon \sim_{pp}^{fr} \phi$;
  \item $\epsilon\parallel \phi \sim_{pp}^{fr} \phi$;
  \item $\phi\parallel\neg\phi \sim_{pp}^{fr} \delta$;
  \item $\phi_0\parallel\cdots\parallel\phi_n \sim_{pp}^{fr} \delta$ if $\forall s_0,\cdots,s_n\in S,\exists i\leq n.test(\neg\phi_i,s_0\cup\cdots\cup s_n)$.
\end{enumerate}
\end{proposition}

\begin{proof}
\begin{enumerate}
  \item $P+\delta \sim_{pp}^{fr} P$. It is sufficient to prove the relation $R=\{(P+\delta, P)\}\cup \textbf{Id}$ is a FR strongly probabilistic pomset bisimulation, and we omit it;
  \item $\delta.P \sim_{pp}^{fr} \delta$. It is sufficient to prove the relation $R=\{(\delta.P, \delta)\}\cup \textbf{Id}$ is a FR strongly probabilistic pomset bisimulation, and we omit it;
  \item $\epsilon.P \sim_{pp}^{fr} P$. It is sufficient to prove the relation $R=\{(\epsilon.P, P)\}\cup \textbf{Id}$ is a FR strongly probabilistic pomset bisimulation, and we omit it;
  \item $P.\epsilon \sim_{pp}^{fr} P$. It is sufficient to prove the relation $R=\{(P.\epsilon, P)\}\cup \textbf{Id}$ is a FR strongly probabilistic pomset bisimulation, and we omit it;
  \item $\phi.\neg\phi \sim_{pp}^{fr} \delta$. It is sufficient to prove the relation $R=\{(\phi.\neg\phi, \delta)\}\cup \textbf{Id}$ is a FR strongly probabilistic pomset bisimulation, and we omit it;
  \item $\phi+\neg\phi \sim_{pp}^{fr} \epsilon$. It is sufficient to prove the relation $R=\{(\phi+\neg\phi, \epsilon)\}\cup \textbf{Id}$ is a FR strongly probabilistic pomset bisimulation, and we omit it;
  \item $\phi\boxplus_{\pi}\neg\phi \sim_{pp}^{fr} \epsilon$. It is sufficient to prove the relation $R=\{(\phi\boxplus_{\pi}\neg\phi, \epsilon)\}\cup \textbf{Id}$ is a FR strongly probabilistic pomset bisimulation, and we omit it;
  \item $\phi.\delta \sim_{pp}^{fr} \delta$. It is sufficient to prove the relation $R=\{(\phi.\delta, \delta)\}\cup \textbf{Id}$ is a FR strongly probabilistic pomset bisimulation, and we omit it;
  \item $\phi.(P+Q)\sim_{pp}^{fr}\phi.P+\phi.Q\quad (Std(P),Std(Q))$. It is sufficient to prove the relation $R=\{(\phi.(P+Q), \phi.P+\phi.Q)\}\cup \textbf{Id}\quad (Std(P),Std(Q))$ is a FR strongly probabilistic pomset bisimulation, and we omit it;
  \item $(P+Q).\phi\sim_{pp}^{fr} P.\phi+ Q.\phi\quad(NStd(P),NStd(Q))$. It is sufficient to prove the relation $R=\{((P+Q).\phi, P.\phi+ Q.\phi)\}\cup \textbf{Id}\quad(NStd(P),NStd(Q))$ is a FR strongly probabilistic pomset bisimulation, and we omit it;
  \item $\phi.(P\boxplus_{\pi}Q)\sim_{pp}^{fr}\phi.P\boxplus_{\pi}\phi.Q\quad (Std(P),Std(Q))$. It is sufficient to prove the relation $R=\{(\phi.(P\boxplus_{\pi}Q), \phi.P\boxplus_{\pi}\phi.Q)\}\cup \textbf{Id}\quad (Std(P),Std(Q))$ is a FR strongly probabilistic pomset bisimulation, and we omit it;
  \item $(P\boxplus_{\pi}Q).\phi\sim_{pp}^{fr} P.\phi\boxplus_{\pi} Q.\phi\quad(NStd(P),NStd(Q))$. It is sufficient to prove the relation $R=\{((P\boxplus_{\pi}Q).\phi, P.\phi\boxplus_{\pi} Q.\phi)\}\cup \textbf{Id}\quad(NStd(P),NStd(Q))$ is a FR strongly probabilistic pomset bisimulation, and we omit it;
  \item $\phi.(P.Q)\sim_{pp}^{fr} \phi.P.Q\quad (Std(P),Std(Q))$. It is sufficient to prove the relation $R=\{(\phi.(P.Q), \phi.P.Q)\}\cup \textbf{Id}\quad (Std(P),Std(Q))$ is a FR strongly probabilistic pomset bisimulation, and we omit it;
  \item $(P.Q).\phi\sim_{pp}^{fr} P.Q.\phi\quad(NStd(P),NStd(Q))$. It is sufficient to prove the relation $R=\{((P.Q).\phi, P.Q.\phi)\}\cup \textbf{Id}\quad(NStd(P),NStd(Q))$ is a FR strongly probabilistic pomset bisimulation, and we omit it;
  \item $(\phi+\psi).P \sim_{pp}^{fr} \phi.P + \psi.P\quad (Std(P))$. It is sufficient to prove the relation $R=\{((\phi+\psi).P, \phi.P + \psi.P)\}\cup \textbf{Id}\quad (Std(P))$, is a FR strongly probabilistic pomset bisimulation, and we omit it;
  \item $P.(\phi+\psi) \sim_{pp}^{fr} P.\phi + P.\psi\quad(NStd(P))$. It is sufficient to prove the relation $R=\{(P.(\phi+\psi), P.\phi + P.\psi)\}\cup \textbf{Id}\quad(NStd(P))$, is a FR strongly probabilistic pomset bisimulation, and we omit it;
  \item $(\phi\boxplus_{\pi}\psi).P \sim_{pp}^{fr} \phi.P \boxplus_{\pi} \psi.P\quad (Std(P))$. It is sufficient to prove the relation $R=\{((\phi\boxplus_{\pi}\psi).P, \phi.P \boxplus_{\pi} \psi.P)\}\cup \textbf{Id}\quad (Std(P))$, is a FR strongly probabilistic pomset bisimulation, and we omit it;
  \item $P.(\phi\boxplus_{\pi}\psi) \sim_{pp}^{fr} P.\phi \boxplus_{\pi} P.\psi\quad(NStd(P))$. It is sufficient to prove the relation $R=\{(P.(\phi\boxplus_{\pi}\psi), P.\phi \boxplus_{\pi} P.\psi)\}\cup \textbf{Id}\quad(NStd(P))$, is a FR strongly probabilistic pomset bisimulation, and we omit it;
  \item $(\phi.\psi).P \sim_{pp}^{fr} \phi.(\psi.P)\quad(Std(P))$. It is sufficient to prove the relation $R=\{((\phi.\psi).P, \phi.(\psi.P))\}\cup \textbf{Id}\quad(Std(P))$ is a FR strongly probabilistic pomset bisimulation, and we omit it;
  \item $P.(\phi.\psi) \sim_{pp}^{fr}(P.\phi).\psi\quad(NStd(P))$. It is sufficient to prove the relation $R=\{(P.(\phi.\psi), (P.\phi).\psi)\}\cup \textbf{Id}\quad(NStd(P))$ is a FR strongly probabilistic pomset bisimulation, and we omit it;
  \item $\phi\sim_{pp}^{fr}\epsilon$ if $\forall s\in S.test(\phi,s)$. It is sufficient to prove the relation $R=\{(\phi, \epsilon)\}\cup \textbf{Id}$, if $\forall s\in S.test(\phi,s)$, is a FR strongly probabilistic pomset bisimulation, and we omit it;
  \item $\phi_0\cdot\cdots\cdot\phi_n \sim_{pp}^{fr} \delta$ if $\forall s\in S,\exists i\leq n.test(\neg\phi_i,s)$. It is sufficient to prove the relation $R=\{(\phi_0\cdot\cdots\cdot\phi_n, \delta)\}\cup \textbf{Id}$, if $\forall s\in S,\exists i\leq n.test(\neg\phi_i,s)$, is a FR strongly probabilistic pomset bisimulation, and we omit it;
  \item $wp(\alpha,\phi).\alpha.\phi\sim_{pp}^{fr} wp(\alpha,\phi).\alpha$. It is sufficient to prove the relation $R=\{(wp(\alpha,\phi).\alpha.\phi, wp(\alpha,\phi).\alpha)\}\cup \textbf{Id}$ is a FR strongly probabilistic pomset bisimulation, and we omit it;
  \item $\phi. \alpha[m]. wp(\alpha[m],\phi)\sim_{pp}^{fr}\alpha[m].wp(\alpha[m],\phi)$. It is sufficient to prove the relation $R=\{(\phi. \alpha[m]. wp(\alpha[m],\phi), \alpha[m].wp(\alpha[m],\phi))\}\cup \textbf{Id}$ is a FR strongly probabilistic pomset bisimulation, and we omit it;
  \item $\neg wp(\alpha,\phi).\alpha.\neg\phi\sim_{pp}^{fr}\neg wp(\alpha,\phi).\alpha$. It is sufficient to prove the relation \\$R=\{(\neg wp(\alpha,\phi).\alpha.\neg\phi, \neg wp(\alpha,\phi).\alpha)\}\cup \textbf{Id}$, is a FR strongly probabilistic pomset bisimulation, and we omit it;
  \item $\neg\phi .\alpha[m] .\neg wp(\alpha[m],\phi)\sim_{pp}^{fr} \alpha[m]. \neg wp(\alpha[m],\phi)$. It is sufficient to prove the relation $R=\{(\neg\phi .\alpha[m] .\neg wp(\alpha[m],\phi), \alpha[m]. \neg wp(\alpha[m],\phi))\}\cup \textbf{Id}$, is a FR strongly probabilistic pomset bisimulation, and we omit it;
  \item $\delta\parallel P \sim_{pp}^{fr} \delta$. It is sufficient to prove the relation $R=\{(\delta\parallel P, \delta)\}\cup \textbf{Id}$ is a FR strongly probabilistic pomset bisimulation, and we omit it;
  \item $P\parallel \delta \sim_{pp}^{fr} \delta$. It is sufficient to prove the relation $R=\{(P\parallel \delta, \delta)\}\cup \textbf{Id}$ is a FR strongly probabilistic pomset bisimulation, and we omit it;
  \item $\epsilon\parallel P \sim_{pp}^{fr} P$. It is sufficient to prove the relation $R=\{(\epsilon\parallel P, P)\}\cup \textbf{Id}$ is a FR strongly probabilistic pomset bisimulation, and we omit it;
  \item $P\parallel \epsilon \sim_{pp}^{fr} P$. It is sufficient to prove the relation $R=\{(P\parallel \epsilon, P)\}\cup \textbf{Id}$ is a FR strongly probabilistic pomset bisimulation, and we omit it;
  \item $\phi.(P\parallel Q) \sim_{pp}^{fr}\phi.P\parallel \phi.Q$. It is sufficient to prove the relation $R=\{(\phi.(P\parallel Q), \phi.P\parallel \phi.Q)\}\cup \textbf{Id}$ is a FR strongly probabilistic pomset bisimulation, and we omit it;
  \item $\phi\parallel \delta \sim_{pp}^{fr} \delta$. It is sufficient to prove the relation $R=\{(\phi\parallel \delta, \delta)\}\cup \textbf{Id}$ is a FR strongly probabilistic pomset bisimulation, and we omit it;
  \item $\delta\parallel \phi \sim_{pp}^{fr} \delta$. It is sufficient to prove the relation $R=\{(\delta\parallel \phi, \delta)\}\cup \textbf{Id}$ is a FR strongly probabilistic pomset bisimulation, and we omit it;
  \item $\phi\parallel \epsilon \sim_{pp}^{fr} \phi$. It is sufficient to prove the relation $R=\{(\phi\parallel \epsilon, \phi)\}\cup \textbf{Id}$ is a FR strongly probabilistic pomset bisimulation, and we omit it;
  \item $\epsilon\parallel \phi \sim_{pp}^{fr} \phi$. It is sufficient to prove the relation $R=\{(\epsilon\parallel \phi, \phi)\}\cup \textbf{Id}$ is a FR strongly probabilistic pomset bisimulation, and we omit it;
  \item $\phi\parallel\neg\phi \sim_{pp}^{fr} \delta$. It is sufficient to prove the relation $R=\{(\phi\parallel\neg\phi, \delta)\}\cup \textbf{Id}$ is a FR strongly probabilistic pomset bisimulation, and we omit it;
  \item $\phi_0\parallel\cdots\parallel\phi_n \sim_{pp}^{fr} \delta$ if $\forall s_0,\cdots,s_n\in S,\exists i\leq n.test(\neg\phi_i,s_0\cup\cdots\cup s_n)$. It is sufficient to prove the relation $R=\{(\phi_0\parallel\cdots\parallel\phi_n, \delta)\}\cup \textbf{Id}$, if $\forall s_0,\cdots,s_n\in S,\exists i\leq n.test(\neg\phi_i,s_0\cup\cdots\cup s_n)$, is a FR strongly probabilistic pomset bisimulation, and we omit it.
\end{enumerate}
\end{proof}

\begin{proposition}[Guards laws for FR strongly probabilistic step bisimulation] The guards laws for FR strongly probabilistic step bisimulation are as follows.

\begin{enumerate}
  \item $P+\delta \sim_{ps}^{fr} P$;
  \item $\delta.P \sim_{ps}^{fr} \delta$;
  \item $\epsilon.P \sim_{ps}^{fr} P$;
  \item $P.\epsilon \sim_{ps}^{fr} P$;
  \item $\phi.\neg\phi \sim_{ps}^{fr} \delta$;
  \item $\phi+\neg\phi \sim_{ps}^{fr} \epsilon$;
  \item $\phi\boxplus_{\pi}\neg\phi \sim_{ps}^{fr} \epsilon$;
  \item $\phi.\delta \sim_{ps}^{fr} \delta$;
  \item $\phi.(P+Q)\sim_{ps}^{fr}\phi.P+\phi.Q\quad (Std(P),Std(Q))$;
  \item $(P+Q).\phi\sim_{ps}^{fr} P.\phi+ Q.\phi\quad(NStd(P),NStd(Q))$;
  \item $\phi.(P\boxplus_{\pi}Q)\sim_{ps}^{fr}\phi.P\boxplus_{\pi}\phi.Q\quad (Std(P),Std(Q))$;
  \item $(P\boxplus_{\pi}Q).\phi\sim_{ps}^{fr} P.\phi\boxplus_{\pi} Q.\phi\quad(NStd(P),NStd(Q))$;
  \item $\phi.(P.Q)\sim_{ps}^{fr} \phi.P.Q\quad (Std(P),Std(Q))$;
  \item $(P.Q).\phi\sim_{ps}^{fr} P.Q.\phi\quad(NStd(P),NStd(Q))$;
  \item $(\phi+\psi).P \sim_{ps}^{fr} \phi.P + \psi.P\quad (Std(P))$;
  \item $P.(\phi+\psi) \sim_{ps}^{fr} P.\phi + P.\psi\quad(NStd(P))$;
  \item $(\phi\boxplus_{\pi}\psi).P \sim_{ps}^{fr} \phi.P \boxplus_{\pi} \psi.P\quad (Std(P))$;
  \item $P.(\phi\boxplus_{\pi}\psi) \sim_{ps}^{fr} P.\phi \boxplus_{\pi} P.\psi\quad(NStd(P))$;
  \item $(\phi.\psi).P \sim_{ps}^{fr} \phi.(\psi.P)\quad(Std(P))$;
  \item $P.(\phi.\psi) \sim_{ps}^{fr}(P.\phi).\psi\quad(NStd(P))$;
  \item $\phi\sim_{ps}^{fr}\epsilon$ if $\forall s\in S.test(\phi,s)$;
  \item $\phi_0\cdot\cdots\cdot\phi_n \sim_{ps}^{fr} \delta$ if $\forall s\in S,\exists i\leq n.test(\neg\phi_i,s)$;
  \item $wp(\alpha,\phi).\alpha.\phi\sim_{ps}^{fr} wp(\alpha,\phi).\alpha$;
  \item $\phi. \alpha[m]. wp(\alpha[m],\phi)\sim_{ps}^{fr}\alpha[m].wp(\alpha[m],\phi)$;
  \item $\neg wp(\alpha,\phi).\alpha.\neg\phi\sim_{ps}^{fr}\neg wp(\alpha,\phi).\alpha$;
  \item $\neg\phi .\alpha[m] .\neg wp(\alpha[m],\phi)\sim_{ps}^{fr} \alpha[m]. \neg wp(\alpha[m],\phi)$;
  \item $\delta\parallel P \sim_{ps}^{fr} \delta$;
  \item $P\parallel \delta \sim_{ps}^{fr} \delta$;
  \item $\epsilon\parallel P \sim_{ps}^{fr} P$;
  \item $P\parallel \epsilon \sim_{ps}^{fr} P$;
  \item $\phi.(P\parallel Q) \sim_{ps}^{fr}\phi.P\parallel \phi.Q$;
  \item $\phi\parallel \delta \sim_{ps}^{fr} \delta$;
  \item $\delta\parallel \phi \sim_{ps}^{fr} \delta$;
  \item $\phi\parallel \epsilon \sim_{ps}^{fr} \phi$;
  \item $\epsilon\parallel \phi \sim_{ps}^{fr} \phi$;
  \item $\phi\parallel\neg\phi \sim_{ps}^{fr} \delta$;
  \item $\phi_0\parallel\cdots\parallel\phi_n \sim_{ps}^{fr} \delta$ if $\forall s_0,\cdots,s_n\in S,\exists i\leq n.test(\neg\phi_i,s_0\cup\cdots\cup s_n)$.
\end{enumerate}
\end{proposition}

\begin{proof}
\begin{enumerate}
  \item $P+\delta \sim_{ps}^{fr} P$. It is sufficient to prove the relation $R=\{(P+\delta, P)\}\cup \textbf{Id}$ is a FR strongly probabilistic step bisimulation, and we omit it;
  \item $\delta.P \sim_{ps}^{fr} \delta$. It is sufficient to prove the relation $R=\{(\delta.P, \delta)\}\cup \textbf{Id}$ is a FR strongly probabilistic step bisimulation, and we omit it;
  \item $\epsilon.P \sim_{ps}^{fr} P$. It is sufficient to prove the relation $R=\{(\epsilon.P, P)\}\cup \textbf{Id}$ is a FR strongly probabilistic step bisimulation, and we omit it;
  \item $P.\epsilon \sim_{ps}^{fr} P$. It is sufficient to prove the relation $R=\{(P.\epsilon, P)\}\cup \textbf{Id}$ is a FR strongly probabilistic step bisimulation, and we omit it;
  \item $\phi.\neg\phi \sim_{ps}^{fr} \delta$. It is sufficient to prove the relation $R=\{(\phi.\neg\phi, \delta)\}\cup \textbf{Id}$ is a FR strongly probabilistic step bisimulation, and we omit it;
  \item $\phi+\neg\phi \sim_{ps}^{fr} \epsilon$. It is sufficient to prove the relation $R=\{(\phi+\neg\phi, \epsilon)\}\cup \textbf{Id}$ is a FR strongly probabilistic step bisimulation, and we omit it;
  \item $\phi\boxplus_{\pi}\neg\phi \sim_{ps}^{fr} \epsilon$. It is sufficient to prove the relation $R=\{(\phi\boxplus_{\pi}\neg\phi, \epsilon)\}\cup \textbf{Id}$ is a FR strongly probabilistic step bisimulation, and we omit it;
  \item $\phi.\delta \sim_{ps}^{fr} \delta$. It is sufficient to prove the relation $R=\{(\phi.\delta, \delta)\}\cup \textbf{Id}$ is a FR strongly probabilistic step bisimulation, and we omit it;
  \item $\phi.(P+Q)\sim_{ps}^{fr}\phi.P+\phi.Q\quad (Std(P),Std(Q))$. It is sufficient to prove the relation $R=\{(\phi.(P+Q), \phi.P+\phi.Q)\}\cup \textbf{Id}\quad (Std(P),Std(Q))$ is a FR strongly probabilistic step bisimulation, and we omit it;
  \item $(P+Q).\phi\sim_{ps}^{fr} P.\phi+ Q.\phi\quad(NStd(P),NStd(Q))$. It is sufficient to prove the relation $R=\{((P+Q).\phi, P.\phi+ Q.\phi)\}\cup \textbf{Id}\quad(NStd(P),NStd(Q))$ is a FR strongly probabilistic step bisimulation, and we omit it;
  \item $\phi.(P\boxplus_{\pi}Q)\sim_{ps}^{fr}\phi.P\boxplus_{\pi}\phi.Q\quad (Std(P),Std(Q))$. It is sufficient to prove the relation $R=\{(\phi.(P\boxplus_{\pi}Q), \phi.P\boxplus_{\pi}\phi.Q)\}\cup \textbf{Id}\quad (Std(P),Std(Q))$ is a FR strongly probabilistic step bisimulation, and we omit it;
  \item $(P\boxplus_{\pi}Q).\phi\sim_{ps}^{fr} P.\phi\boxplus_{\pi} Q.\phi\quad(NStd(P),NStd(Q))$. It is sufficient to prove the relation $R=\{((P\boxplus_{\pi}Q).\phi, P.\phi\boxplus_{\pi} Q.\phi)\}\cup \textbf{Id}\quad(NStd(P),NStd(Q))$ is a FR strongly probabilistic step bisimulation, and we omit it;
  \item $\phi.(P.Q)\sim_{ps}^{fr} \phi.P.Q\quad (Std(P),Std(Q))$. It is sufficient to prove the relation $R=\{(\phi.(P.Q), \phi.P.Q)\}\cup \textbf{Id}\quad (Std(P),Std(Q))$ is a FR strongly probabilistic step bisimulation, and we omit it;
  \item $(P.Q).\phi\sim_{ps}^{fr} P.Q.\phi\quad(NStd(P),NStd(Q))$. It is sufficient to prove the relation $R=\{((P.Q).\phi, P.Q.\phi)\}\cup \textbf{Id}\quad(NStd(P),NStd(Q))$ is a FR strongly probabilistic step bisimulation, and we omit it;
  \item $(\phi+\psi).P \sim_{ps}^{fr} \phi.P + \psi.P\quad (Std(P))$. It is sufficient to prove the relation $R=\{((\phi+\psi).P, \phi.P + \psi.P)\}\cup \textbf{Id}\quad (Std(P))$, is a FR strongly probabilistic step bisimulation, and we omit it;
  \item $P.(\phi+\psi) \sim_{ps}^{fr} P.\phi + P.\psi\quad(NStd(P))$. It is sufficient to prove the relation $R=\{(P.(\phi+\psi), P.\phi + P.\psi)\}\cup \textbf{Id}\quad(NStd(P))$, is a FR strongly probabilistic step bisimulation, and we omit it;
  \item $(\phi\boxplus_{\pi}\psi).P \sim_{ps}^{fr} \phi.P \boxplus_{\pi} \psi.P\quad (Std(P))$. It is sufficient to prove the relation $R=\{((\phi\boxplus_{\pi}\psi).P, \phi.P \boxplus_{\pi} \psi.P)\}\cup \textbf{Id}\quad (Std(P))$, is a FR strongly probabilistic step bisimulation, and we omit it;
  \item $P.(\phi\boxplus_{\pi}\psi) \sim_{ps}^{fr} P.\phi \boxplus_{\pi} P.\psi\quad(NStd(P))$. It is sufficient to prove the relation $R=\{(P.(\phi\boxplus_{\pi}\psi), P.\phi \boxplus_{\pi} P.\psi)\}\cup \textbf{Id}\quad(NStd(P))$, is a FR strongly probabilistic step bisimulation, and we omit it;
  \item $(\phi.\psi).P \sim_{ps}^{fr} \phi.(\psi.P)\quad(Std(P))$. It is sufficient to prove the relation $R=\{((\phi.\psi).P, \phi.(\psi.P))\}\cup \textbf{Id}\quad(Std(P))$ is a FR strongly probabilistic step bisimulation, and we omit it;
  \item $P.(\phi.\psi) \sim_{ps}^{fr}(P.\phi).\psi\quad(NStd(P))$. It is sufficient to prove the relation $R=\{(P.(\phi.\psi), (P.\phi).\psi)\}\cup \textbf{Id}\quad(NStd(P))$ is a FR strongly probabilistic step bisimulation, and we omit it;
  \item $\phi\sim_{ps}^{fr}\epsilon$ if $\forall s\in S.test(\phi,s)$. It is sufficient to prove the relation $R=\{(\phi, \epsilon)\}\cup \textbf{Id}$, if $\forall s\in S.test(\phi,s)$, is a FR strongly probabilistic step bisimulation, and we omit it;
  \item $\phi_0\cdot\cdots\cdot\phi_n \sim_{ps}^{fr} \delta$ if $\forall s\in S,\exists i\leq n.test(\neg\phi_i,s)$. It is sufficient to prove the relation $R=\{(\phi_0\cdot\cdots\cdot\phi_n, \delta)\}\cup \textbf{Id}$, if $\forall s\in S,\exists i\leq n.test(\neg\phi_i,s)$, is a FR strongly probabilistic step bisimulation, and we omit it;
  \item $wp(\alpha,\phi).\alpha.\phi\sim_{ps}^{fr} wp(\alpha,\phi).\alpha$. It is sufficient to prove the relation $R=\{(wp(\alpha,\phi).\alpha.\phi, wp(\alpha,\phi).\alpha)\}\cup \textbf{Id}$ is a FR strongly probabilistic step bisimulation, and we omit it;
  \item $\phi. \alpha[m]. wp(\alpha[m],\phi)\sim_{ps}^{fr}\alpha[m].wp(\alpha[m],\phi)$. It is sufficient to prove the relation $R=\{(\phi. \alpha[m]. wp(\alpha[m],\phi), \alpha[m].wp(\alpha[m],\phi))\}\cup \textbf{Id}$ is a FR strongly probabilistic step bisimulation, and we omit it;
  \item $\neg wp(\alpha,\phi).\alpha.\neg\phi\sim_{ps}^{fr}\neg wp(\alpha,\phi).\alpha$. It is sufficient to prove the relation \\$R=\{(\neg wp(\alpha,\phi).\alpha.\neg\phi, \neg wp(\alpha,\phi).\alpha)\}\cup \textbf{Id}$, is a FR strongly probabilistic step bisimulation, and we omit it;
  \item $\neg\phi .\alpha[m] .\neg wp(\alpha[m],\phi)\sim_{ps}^{fr} \alpha[m]. \neg wp(\alpha[m],\phi)$. It is sufficient to prove the relation $R=\{(\neg\phi .\alpha[m] .\neg wp(\alpha[m],\phi), \alpha[m]. \neg wp(\alpha[m],\phi))\}\cup \textbf{Id}$, is a FR strongly probabilistic step bisimulation, and we omit it;
  \item $\delta\parallel P \sim_{ps}^{fr} \delta$. It is sufficient to prove the relation $R=\{(\delta\parallel P, \delta)\}\cup \textbf{Id}$ is a FR strongly probabilistic step bisimulation, and we omit it;
  \item $P\parallel \delta \sim_{ps}^{fr} \delta$. It is sufficient to prove the relation $R=\{(P\parallel \delta, \delta)\}\cup \textbf{Id}$ is a FR strongly probabilistic step bisimulation, and we omit it;
  \item $\epsilon\parallel P \sim_{ps}^{fr} P$. It is sufficient to prove the relation $R=\{(\epsilon\parallel P, P)\}\cup \textbf{Id}$ is a FR strongly probabilistic step bisimulation, and we omit it;
  \item $P\parallel \epsilon \sim_{ps}^{fr} P$. It is sufficient to prove the relation $R=\{(P\parallel \epsilon, P)\}\cup \textbf{Id}$ is a FR strongly probabilistic step bisimulation, and we omit it;
  \item $\phi.(P\parallel Q) \sim_{ps}^{fr}\phi.P\parallel \phi.Q$. It is sufficient to prove the relation $R=\{(\phi.(P\parallel Q), \phi.P\parallel \phi.Q)\}\cup \textbf{Id}$ is a FR strongly probabilistic step bisimulation, and we omit it;
  \item $\phi\parallel \delta \sim_{ps}^{fr} \delta$. It is sufficient to prove the relation $R=\{(\phi\parallel \delta, \delta)\}\cup \textbf{Id}$ is a FR strongly probabilistic step bisimulation, and we omit it;
  \item $\delta\parallel \phi \sim_{ps}^{fr} \delta$. It is sufficient to prove the relation $R=\{(\delta\parallel \phi, \delta)\}\cup \textbf{Id}$ is a FR strongly probabilistic step bisimulation, and we omit it;
  \item $\phi\parallel \epsilon \sim_{ps}^{fr} \phi$. It is sufficient to prove the relation $R=\{(\phi\parallel \epsilon, \phi)\}\cup \textbf{Id}$ is a FR strongly probabilistic step bisimulation, and we omit it;
  \item $\epsilon\parallel \phi \sim_{ps}^{fr} \phi$. It is sufficient to prove the relation $R=\{(\epsilon\parallel \phi, \phi)\}\cup \textbf{Id}$ is a FR strongly probabilistic step bisimulation, and we omit it;
  \item $\phi\parallel\neg\phi \sim_{ps}^{fr} \delta$. It is sufficient to prove the relation $R=\{(\phi\parallel\neg\phi, \delta)\}\cup \textbf{Id}$ is a FR strongly probabilistic step bisimulation, and we omit it;
  \item $\phi_0\parallel\cdots\parallel\phi_n \sim_{ps}^{fr} \delta$ if $\forall s_0,\cdots,s_n\in S,\exists i\leq n.test(\neg\phi_i,s_0\cup\cdots\cup s_n)$. It is sufficient to prove the relation $R=\{(\phi_0\parallel\cdots\parallel\phi_n, \delta)\}\cup \textbf{Id}$, if $\forall s_0,\cdots,s_n\in S,\exists i\leq n.test(\neg\phi_i,s_0\cup\cdots\cup s_n)$, is a FR strongly probabilistic step bisimulation, and we omit it.
\end{enumerate}
\end{proof}

\begin{proposition}[Guards laws for FR strongly probabilistic hp-bisimulation] The guards laws for FR strongly probabilistic hp-bisimulation are as follows.

\begin{enumerate}
  \item $P+\delta \sim_{php}^{fr} P$;
  \item $\delta.P \sim_{php}^{fr} \delta$;
  \item $\epsilon.P \sim_{php}^{fr} P$;
  \item $P.\epsilon \sim_{php}^{fr} P$;
  \item $\phi.\neg\phi \sim_{php}^{fr} \delta$;
  \item $\phi+\neg\phi \sim_{php}^{fr} \epsilon$;
  \item $\phi\boxplus_{\pi}\neg\phi \sim_{php}^{fr} \epsilon$;
  \item $\phi.\delta \sim_{php}^{fr} \delta$;
  \item $\phi.(P+Q)\sim_{php}^{fr}\phi.P+\phi.Q\quad (Std(P),Std(Q))$;
  \item $(P+Q).\phi\sim_{php}^{fr} P.\phi+ Q.\phi\quad(NStd(P),NStd(Q))$;
  \item $\phi.(P\boxplus_{\pi}Q)\sim_{php}^{fr}\phi.P\boxplus_{\pi}\phi.Q\quad (Std(P),Std(Q))$;
  \item $(P\boxplus_{\pi}Q).\phi\sim_{php}^{fr} P.\phi\boxplus_{\pi} Q.\phi\quad(NStd(P),NStd(Q))$;
  \item $\phi.(P.Q)\sim_{php}^{fr} \phi.P.Q\quad (Std(P),Std(Q))$;
  \item $(P.Q).\phi\sim_{php}^{fr} P.Q.\phi\quad(NStd(P),NStd(Q))$;
  \item $(\phi+\psi).P \sim_{php}^{fr} \phi.P + \psi.P\quad (Std(P))$;
  \item $P.(\phi+\psi) \sim_{php}^{fr} P.\phi + P.\psi\quad(NStd(P))$;
  \item $(\phi\boxplus_{\pi}\psi).P \sim_{php}^{fr} \phi.P \boxplus_{\pi} \psi.P\quad (Std(P))$;
  \item $P.(\phi\boxplus_{\pi}\psi) \sim_{php}^{fr} P.\phi \boxplus_{\pi} P.\psi\quad(NStd(P))$;
  \item $(\phi.\psi).P \sim_{php}^{fr} \phi.(\psi.P)\quad(Std(P))$;
  \item $P.(\phi.\psi) \sim_{php}^{fr}(P.\phi).\psi\quad(NStd(P))$;
  \item $\phi\sim_{php}^{fr}\epsilon$ if $\forall s\in S.test(\phi,s)$;
  \item $\phi_0\cdot\cdots\cdot\phi_n \sim_{php}^{fr} \delta$ if $\forall s\in S,\exists i\leq n.test(\neg\phi_i,s)$;
  \item $wp(\alpha,\phi).\alpha.\phi\sim_{php}^{fr} wp(\alpha,\phi).\alpha$;
  \item $\phi. \alpha[m]. wp(\alpha[m],\phi)\sim_{php}^{fr}\alpha[m].wp(\alpha[m],\phi)$;
  \item $\neg wp(\alpha,\phi).\alpha.\neg\phi\sim_{php}^{fr}\neg wp(\alpha,\phi).\alpha$;
  \item $\neg\phi .\alpha[m] .\neg wp(\alpha[m],\phi)\sim_{php}^{fr} \alpha[m]. \neg wp(\alpha[m],\phi)$;
  \item $\delta\parallel P \sim_{php}^{fr} \delta$;
  \item $P\parallel \delta \sim_{php}^{fr} \delta$;
  \item $\epsilon\parallel P \sim_{php}^{fr} P$;
  \item $P\parallel \epsilon \sim_{php}^{fr} P$;
  \item $\phi.(P\parallel Q) \sim_{php}^{fr}\phi.P\parallel \phi.Q$;
  \item $\phi\parallel \delta \sim_{php}^{fr} \delta$;
  \item $\delta\parallel \phi \sim_{php}^{fr} \delta$;
  \item $\phi\parallel \epsilon \sim_{php}^{fr} \phi$;
  \item $\epsilon\parallel \phi \sim_{php}^{fr} \phi$;
  \item $\phi\parallel\neg\phi \sim_{php}^{fr} \delta$;
  \item $\phi_0\parallel\cdots\parallel\phi_n \sim_{php}^{fr} \delta$ if $\forall s_0,\cdots,s_n\in S,\exists i\leq n.test(\neg\phi_i,s_0\cup\cdots\cup s_n)$.
\end{enumerate}
\end{proposition}

\begin{proof}
\begin{enumerate}
  \item $P+\delta \sim_{php}^{fr} P$. It is sufficient to prove the relation $R=\{(P+\delta, P)\}\cup \textbf{Id}$ is a FR strongly probabilistic hp-bisimulation, and we omit it;
  \item $\delta.P \sim_{php}^{fr} \delta$. It is sufficient to prove the relation $R=\{(\delta.P, \delta)\}\cup \textbf{Id}$ is a FR strongly probabilistic hp-bisimulation, and we omit it;
  \item $\epsilon.P \sim_{php}^{fr} P$. It is sufficient to prove the relation $R=\{(\epsilon.P, P)\}\cup \textbf{Id}$ is a FR strongly probabilistic hp-bisimulation, and we omit it;
  \item $P.\epsilon \sim_{php}^{fr} P$. It is sufficient to prove the relation $R=\{(P.\epsilon, P)\}\cup \textbf{Id}$ is a FR strongly probabilistic hp-bisimulation, and we omit it;
  \item $\phi.\neg\phi \sim_{php}^{fr} \delta$. It is sufficient to prove the relation $R=\{(\phi.\neg\phi, \delta)\}\cup \textbf{Id}$ is a FR strongly probabilistic hp-bisimulation, and we omit it;
  \item $\phi+\neg\phi \sim_{php}^{fr} \epsilon$. It is sufficient to prove the relation $R=\{(\phi+\neg\phi, \epsilon)\}\cup \textbf{Id}$ is a FR strongly probabilistic hp-bisimulation, and we omit it;
  \item $\phi\boxplus_{\pi}\neg\phi \sim_{php}^{fr} \epsilon$. It is sufficient to prove the relation $R=\{(\phi\boxplus_{\pi}\neg\phi, \epsilon)\}\cup \textbf{Id}$ is a FR strongly probabilistic hp-bisimulation, and we omit it;
  \item $\phi.\delta \sim_{php}^{fr} \delta$. It is sufficient to prove the relation $R=\{(\phi.\delta, \delta)\}\cup \textbf{Id}$ is a FR strongly probabilistic hp-bisimulation, and we omit it;
  \item $\phi.(P+Q)\sim_{php}^{fr}\phi.P+\phi.Q\quad (Std(P),Std(Q))$. It is sufficient to prove the relation $R=\{(\phi.(P+Q), \phi.P+\phi.Q)\}\cup \textbf{Id}\quad (Std(P),Std(Q))$ is a FR strongly probabilistic hp-bisimulation, and we omit it;
  \item $(P+Q).\phi\sim_{php}^{fr} P.\phi+ Q.\phi\quad(NStd(P),NStd(Q))$. It is sufficient to prove the relation $R=\{((P+Q).\phi, P.\phi+ Q.\phi)\}\cup \textbf{Id}\quad(NStd(P),NStd(Q))$ is a FR strongly probabilistic hp-bisimulation, and we omit it;
  \item $\phi.(P\boxplus_{\pi}Q)\sim_{php}^{fr}\phi.P\boxplus_{\pi}\phi.Q\quad (Std(P),Std(Q))$. It is sufficient to prove the relation $R=\{(\phi.(P\boxplus_{\pi}Q), \phi.P\boxplus_{\pi}\phi.Q)\}\cup \textbf{Id}\quad (Std(P),Std(Q))$ is a FR strongly probabilistic hp-bisimulation, and we omit it;
  \item $(P\boxplus_{\pi}Q).\phi\sim_{php}^{fr} P.\phi\boxplus_{\pi} Q.\phi\quad(NStd(P),NStd(Q))$. It is sufficient to prove the relation $R=\{((P\boxplus_{\pi}Q).\phi, P.\phi\boxplus_{\pi} Q.\phi)\}\cup \textbf{Id}\quad(NStd(P),NStd(Q))$ is a FR strongly probabilistic hp-bisimulation, and we omit it;
  \item $\phi.(P.Q)\sim_{php}^{fr} \phi.P.Q\quad (Std(P),Std(Q))$. It is sufficient to prove the relation $R=\{(\phi.(P.Q), \phi.P.Q)\}\cup \textbf{Id}\quad (Std(P),Std(Q))$ is a FR strongly probabilistic hp-bisimulation, and we omit it;
  \item $(P.Q).\phi\sim_{php}^{fr} P.Q.\phi\quad(NStd(P),NStd(Q))$. It is sufficient to prove the relation $R=\{((P.Q).\phi, P.Q.\phi)\}\cup \textbf{Id}\quad(NStd(P),NStd(Q))$ is a FR strongly probabilistic hp-bisimulation, and we omit it;
  \item $(\phi+\psi).P \sim_{php}^{fr} \phi.P + \psi.P\quad (Std(P))$. It is sufficient to prove the relation $R=\{((\phi+\psi).P, \phi.P + \psi.P)\}\cup \textbf{Id}\quad (Std(P))$, is a FR strongly probabilistic hp-bisimulation, and we omit it;
  \item $P.(\phi+\psi) \sim_{php}^{fr} P.\phi + P.\psi\quad(NStd(P))$. It is sufficient to prove the relation $R=\{(P.(\phi+\psi), P.\phi + P.\psi)\}\cup \textbf{Id}\quad(NStd(P))$, is a FR strongly probabilistic hp-bisimulation, and we omit it;
  \item $(\phi\boxplus_{\pi}\psi).P \sim_{php}^{fr} \phi.P \boxplus_{\pi} \psi.P\quad (Std(P))$. It is sufficient to prove the relation $R=\{((\phi\boxplus_{\pi}\psi).P, \phi.P \boxplus_{\pi} \psi.P)\}\cup \textbf{Id}\quad (Std(P))$, is a FR strongly probabilistic hp-bisimulation, and we omit it;
  \item $P.(\phi\boxplus_{\pi}\psi) \sim_{php}^{fr} P.\phi \boxplus_{\pi} P.\psi\quad(NStd(P))$. It is sufficient to prove the relation $R=\{(P.(\phi\boxplus_{\pi}\psi), P.\phi \boxplus_{\pi} P.\psi)\}\cup \textbf{Id}\quad(NStd(P))$, is a FR strongly probabilistic hp-bisimulation, and we omit it;
  \item $(\phi.\psi).P \sim_{php}^{fr} \phi.(\psi.P)\quad(Std(P))$. It is sufficient to prove the relation $R=\{((\phi.\psi).P, \phi.(\psi.P))\}\cup \textbf{Id}\quad(Std(P))$ is a FR strongly probabilistic hp-bisimulation, and we omit it;
  \item $P.(\phi.\psi) \sim_{php}^{fr}(P.\phi).\psi\quad(NStd(P))$. It is sufficient to prove the relation $R=\{(P.(\phi.\psi), (P.\phi).\psi)\}\cup \textbf{Id}\quad(NStd(P))$ is a FR strongly probabilistic hp-bisimulation, and we omit it;
  \item $\phi\sim_{php}^{fr}\epsilon$ if $\forall s\in S.test(\phi,s)$. It is sufficient to prove the relation $R=\{(\phi, \epsilon)\}\cup \textbf{Id}$, if $\forall s\in S.test(\phi,s)$, is a FR strongly probabilistic hp-bisimulation, and we omit it;
  \item $\phi_0\cdot\cdots\cdot\phi_n \sim_{php}^{fr} \delta$ if $\forall s\in S,\exists i\leq n.test(\neg\phi_i,s)$. It is sufficient to prove the relation $R=\{(\phi_0\cdot\cdots\cdot\phi_n, \delta)\}\cup \textbf{Id}$, if $\forall s\in S,\exists i\leq n.test(\neg\phi_i,s)$, is a FR strongly probabilistic hp-bisimulation, and we omit it;
  \item $wp(\alpha,\phi).\alpha.\phi\sim_{php}^{fr} wp(\alpha,\phi).\alpha$. It is sufficient to prove the relation $R=\{(wp(\alpha,\phi).\alpha.\phi, wp(\alpha,\phi).\alpha)\}\cup \textbf{Id}$ is a FR strongly probabilistic hp-bisimulation, and we omit it;
  \item $\phi. \alpha[m]. wp(\alpha[m],\phi)\sim_{php}^{fr}\alpha[m].wp(\alpha[m],\phi)$. It is sufficient to prove the relation $R=\{(\phi. \alpha[m]. wp(\alpha[m],\phi), \alpha[m].wp(\alpha[m],\phi))\}\cup \textbf{Id}$ is a FR strongly probabilistic hp-bisimulation, and we omit it;
  \item $\neg wp(\alpha,\phi).\alpha.\neg\phi\sim_{php}^{fr}\neg wp(\alpha,\phi).\alpha$. It is sufficient to prove the relation \\$R=\{(\neg wp(\alpha,\phi).\alpha.\neg\phi, \neg wp(\alpha,\phi).\alpha)\}\cup \textbf{Id}$, is a FR strongly probabilistic hp-bisimulation, and we omit it;
  \item $\neg\phi .\alpha[m] .\neg wp(\alpha[m],\phi)\sim_{php}^{fr} \alpha[m]. \neg wp(\alpha[m],\phi)$. It is sufficient to prove the relation $R=\{(\neg\phi .\alpha[m] .\neg wp(\alpha[m],\phi), \alpha[m]. \neg wp(\alpha[m],\phi))\}\cup \textbf{Id}$, is a FR strongly probabilistic hp-bisimulation, and we omit it;
  \item $\delta\parallel P \sim_{php}^{fr} \delta$. It is sufficient to prove the relation $R=\{(\delta\parallel P, \delta)\}\cup \textbf{Id}$ is a FR strongly probabilistic hp-bisimulation, and we omit it;
  \item $P\parallel \delta \sim_{php}^{fr} \delta$. It is sufficient to prove the relation $R=\{(P\parallel \delta, \delta)\}\cup \textbf{Id}$ is a FR strongly probabilistic hp-bisimulation, and we omit it;
  \item $\epsilon\parallel P \sim_{php}^{fr} P$. It is sufficient to prove the relation $R=\{(\epsilon\parallel P, P)\}\cup \textbf{Id}$ is a FR strongly probabilistic hp-bisimulation, and we omit it;
  \item $P\parallel \epsilon \sim_{php}^{fr} P$. It is sufficient to prove the relation $R=\{(P\parallel \epsilon, P)\}\cup \textbf{Id}$ is a FR strongly probabilistic hp-bisimulation, and we omit it;
  \item $\phi.(P\parallel Q) \sim_{php}^{fr}\phi.P\parallel \phi.Q$. It is sufficient to prove the relation $R=\{(\phi.(P\parallel Q), \phi.P\parallel \phi.Q)\}\cup \textbf{Id}$ is a FR strongly probabilistic hp-bisimulation, and we omit it;
  \item $\phi\parallel \delta \sim_{php}^{fr} \delta$. It is sufficient to prove the relation $R=\{(\phi\parallel \delta, \delta)\}\cup \textbf{Id}$ is a FR strongly probabilistic hp-bisimulation, and we omit it;
  \item $\delta\parallel \phi \sim_{php}^{fr} \delta$. It is sufficient to prove the relation $R=\{(\delta\parallel \phi, \delta)\}\cup \textbf{Id}$ is a FR strongly probabilistic hp-bisimulation, and we omit it;
  \item $\phi\parallel \epsilon \sim_{php}^{fr} \phi$. It is sufficient to prove the relation $R=\{(\phi\parallel \epsilon, \phi)\}\cup \textbf{Id}$ is a FR strongly probabilistic hp-bisimulation, and we omit it;
  \item $\epsilon\parallel \phi \sim_{php}^{fr} \phi$. It is sufficient to prove the relation $R=\{(\epsilon\parallel \phi, \phi)\}\cup \textbf{Id}$ is a FR strongly probabilistic hp-bisimulation, and we omit it;
  \item $\phi\parallel\neg\phi \sim_{php}^{fr} \delta$. It is sufficient to prove the relation $R=\{(\phi\parallel\neg\phi, \delta)\}\cup \textbf{Id}$ is a FR strongly probabilistic hp-bisimulation, and we omit it;
  \item $\phi_0\parallel\cdots\parallel\phi_n \sim_{php}^{fr} \delta$ if $\forall s_0,\cdots,s_n\in S,\exists i\leq n.test(\neg\phi_i,s_0\cup\cdots\cup s_n)$. It is sufficient to prove the relation $R=\{(\phi_0\parallel\cdots\parallel\phi_n, \delta)\}\cup \textbf{Id}$, if $\forall s_0,\cdots,s_n\in S,\exists i\leq n.test(\neg\phi_i,s_0\cup\cdots\cup s_n)$, is a FR strongly probabilistic hp-bisimulation, and we omit it.
\end{enumerate}
\end{proof}

\begin{proposition}[Guards laws for FR strongly probabilistic hhp-bisimulation] The guards laws for FR strongly probabilistic hhp-bisimulation are as follows.

\begin{enumerate}
  \item $P+\delta \sim_{phhp}^{fr} P$;
  \item $\delta.P \sim_{phhp}^{fr} \delta$;
  \item $\epsilon.P \sim_{phhp}^{fr} P$;
  \item $P.\epsilon \sim_{phhp}^{fr} P$;
  \item $\phi.\neg\phi \sim_{phhp}^{fr} \delta$;
  \item $\phi+\neg\phi \sim_{phhp}^{fr} \epsilon$;
  \item $\phi\boxplus_{\pi}\neg\phi \sim_{phhp}^{fr} \epsilon$;
  \item $\phi.\delta \sim_{phhp}^{fr} \delta$;
  \item $\phi.(P+Q)\sim_{phhp}^{fr}\phi.P+\phi.Q\quad (Std(P),Std(Q))$;
  \item $(P+Q).\phi\sim_{phhp}^{fr} P.\phi+ Q.\phi\quad(NStd(P),NStd(Q))$;
  \item $\phi.(P\boxplus_{\pi}Q)\sim_{phhp}^{fr}\phi.P\boxplus_{\pi}\phi.Q\quad (Std(P),Std(Q))$;
  \item $(P\boxplus_{\pi}Q).\phi\sim_{phhp}^{fr} P.\phi\boxplus_{\pi} Q.\phi\quad(NStd(P),NStd(Q))$;
  \item $\phi.(P.Q)\sim_{phhp}^{fr} \phi.P.Q\quad (Std(P),Std(Q))$;
  \item $(P.Q).\phi\sim_{phhp}^{fr} P.Q.\phi\quad(NStd(P),NStd(Q))$;
  \item $(\phi+\psi).P \sim_{phhp}^{fr} \phi.P + \psi.P\quad (Std(P))$;
  \item $P.(\phi+\psi) \sim_{phhp}^{fr} P.\phi + P.\psi\quad(NStd(P))$;
  \item $(\phi\boxplus_{\pi}\psi).P \sim_{phhp}^{fr} \phi.P \boxplus_{\pi} \psi.P\quad (Std(P))$;
  \item $P.(\phi\boxplus_{\pi}\psi) \sim_{phhp}^{fr} P.\phi \boxplus_{\pi} P.\psi\quad(NStd(P))$;
  \item $(\phi.\psi).P \sim_{phhp}^{fr} \phi.(\psi.P)\quad(Std(P))$;
  \item $P.(\phi.\psi) \sim_{phhp}^{fr}(P.\phi).\psi\quad(NStd(P))$;
  \item $\phi\sim_{phhp}^{fr}\epsilon$ if $\forall s\in S.test(\phi,s)$;
  \item $\phi_0\cdot\cdots\cdot\phi_n \sim_{phhp}^{fr} \delta$ if $\forall s\in S,\exists i\leq n.test(\neg\phi_i,s)$;
  \item $wp(\alpha,\phi).\alpha.\phi\sim_{phhp}^{fr} wp(\alpha,\phi).\alpha$;
  \item $\phi. \alpha[m]. wp(\alpha[m],\phi)\sim_{phhp}^{fr}\alpha[m].wp(\alpha[m],\phi)$;
  \item $\neg wp(\alpha,\phi).\alpha.\neg\phi\sim_{phhp}^{fr}\neg wp(\alpha,\phi).\alpha$;
  \item $\neg\phi .\alpha[m] .\neg wp(\alpha[m],\phi)\sim_{phhp}^{fr} \alpha[m]. \neg wp(\alpha[m],\phi)$;
  \item $\delta\parallel P \sim_{phhp}^{fr} \delta$;
  \item $P\parallel \delta \sim_{phhp}^{fr} \delta$;
  \item $\epsilon\parallel P \sim_{phhp}^{fr} P$;
  \item $P\parallel \epsilon \sim_{phhp}^{fr} P$;
  \item $\phi.(P\parallel Q) \sim_{phhp}^{fr}\phi.P\parallel \phi.Q$;
  \item $\phi\parallel \delta \sim_{phhp}^{fr} \delta$;
  \item $\delta\parallel \phi \sim_{phhp}^{fr} \delta$;
  \item $\phi\parallel \epsilon \sim_{phhp}^{fr} \phi$;
  \item $\epsilon\parallel \phi \sim_{phhp}^{fr} \phi$;
  \item $\phi\parallel\neg\phi \sim_{phhp}^{fr} \delta$;
  \item $\phi_0\parallel\cdots\parallel\phi_n \sim_{phhp}^{fr} \delta$ if $\forall s_0,\cdots,s_n\in S,\exists i\leq n.test(\neg\phi_i,s_0\cup\cdots\cup s_n)$.
\end{enumerate}
\end{proposition}

\begin{proof}
\begin{enumerate}
  \item $P+\delta \sim_{phhp}^{fr} P$. It is sufficient to prove the relation $R=\{(P+\delta, P)\}\cup \textbf{Id}$ is a FR strongly probabilistic hhp-bisimulation, and we omit it;
  \item $\delta.P \sim_{phhp}^{fr} \delta$. It is sufficient to prove the relation $R=\{(\delta.P, \delta)\}\cup \textbf{Id}$ is a FR strongly probabilistic hhp-bisimulation, and we omit it;
  \item $\epsilon.P \sim_{phhp}^{fr} P$. It is sufficient to prove the relation $R=\{(\epsilon.P, P)\}\cup \textbf{Id}$ is a FR strongly probabilistic hhp-bisimulation, and we omit it;
  \item $P.\epsilon \sim_{phhp}^{fr} P$. It is sufficient to prove the relation $R=\{(P.\epsilon, P)\}\cup \textbf{Id}$ is a FR strongly probabilistic hhp-bisimulation, and we omit it;
  \item $\phi.\neg\phi \sim_{phhp}^{fr} \delta$. It is sufficient to prove the relation $R=\{(\phi.\neg\phi, \delta)\}\cup \textbf{Id}$ is a FR strongly probabilistic hhp-bisimulation, and we omit it;
  \item $\phi+\neg\phi \sim_{phhp}^{fr} \epsilon$. It is sufficient to prove the relation $R=\{(\phi+\neg\phi, \epsilon)\}\cup \textbf{Id}$ is a FR strongly probabilistic hhp-bisimulation, and we omit it;
  \item $\phi\boxplus_{\pi}\neg\phi \sim_{phhp}^{fr} \epsilon$. It is sufficient to prove the relation $R=\{(\phi\boxplus_{\pi}\neg\phi, \epsilon)\}\cup \textbf{Id}$ is a FR strongly probabilistic hhp-bisimulation, and we omit it;
  \item $\phi.\delta \sim_{phhp}^{fr} \delta$. It is sufficient to prove the relation $R=\{(\phi.\delta, \delta)\}\cup \textbf{Id}$ is a FR strongly probabilistic hhp-bisimulation, and we omit it;
  \item $\phi.(P+Q)\sim_{phhp}^{fr}\phi.P+\phi.Q\quad (Std(P),Std(Q))$. It is sufficient to prove the relation $R=\{(\phi.(P+Q), \phi.P+\phi.Q)\}\cup \textbf{Id}\quad (Std(P),Std(Q))$ is a FR strongly probabilistic hhp-bisimulation, and we omit it;
  \item $(P+Q).\phi\sim_{phhp}^{fr} P.\phi+ Q.\phi\quad(NStd(P),NStd(Q))$. It is sufficient to prove the relation $R=\{((P+Q).\phi, P.\phi+ Q.\phi)\}\cup \textbf{Id}\quad(NStd(P),NStd(Q))$ is a FR strongly probabilistic hhp-bisimulation, and we omit it;
  \item $\phi.(P\boxplus_{\pi}Q)\sim_{phhp}^{fr}\phi.P\boxplus_{\pi}\phi.Q\quad (Std(P),Std(Q))$. It is sufficient to prove the relation $R=\{(\phi.(P\boxplus_{\pi}Q), \phi.P\boxplus_{\pi}\phi.Q)\}\cup \textbf{Id}\quad (Std(P),Std(Q))$ is a FR strongly probabilistic hhp-bisimulation, and we omit it;
  \item $(P\boxplus_{\pi}Q).\phi\sim_{phhp}^{fr} P.\phi\boxplus_{\pi} Q.\phi\quad(NStd(P),NStd(Q))$. It is sufficient to prove the relation $R=\{((P\boxplus_{\pi}Q).\phi, P.\phi\boxplus_{\pi} Q.\phi)\}\cup \textbf{Id}\quad(NStd(P),NStd(Q))$ is a FR strongly probabilistic hhp-bisimulation, and we omit it;
  \item $\phi.(P.Q)\sim_{phhp}^{fr} \phi.P.Q\quad (Std(P),Std(Q))$. It is sufficient to prove the relation $R=\{(\phi.(P.Q), \phi.P.Q)\}\cup \textbf{Id}\quad (Std(P),Std(Q))$ is a FR strongly probabilistic hhp-bisimulation, and we omit it;
  \item $(P.Q).\phi\sim_{phhp}^{fr} P.Q.\phi\quad(NStd(P),NStd(Q))$. It is sufficient to prove the relation $R=\{((P.Q).\phi, P.Q.\phi)\}\cup \textbf{Id}\quad(NStd(P),NStd(Q))$ is a FR strongly probabilistic hhp-bisimulation, and we omit it;
  \item $(\phi+\psi).P \sim_{phhp}^{fr} \phi.P + \psi.P\quad (Std(P))$. It is sufficient to prove the relation $R=\{((\phi+\psi).P, \phi.P + \psi.P)\}\cup \textbf{Id}\quad (Std(P))$, is a FR strongly probabilistic hhp-bisimulation, and we omit it;
  \item $P.(\phi+\psi) \sim_{phhp}^{fr} P.\phi + P.\psi\quad(NStd(P))$. It is sufficient to prove the relation $R=\{(P.(\phi+\psi), P.\phi + P.\psi)\}\cup \textbf{Id}\quad(NStd(P))$, is a FR strongly probabilistic hhp-bisimulation, and we omit it;
  \item $(\phi\boxplus_{\pi}\psi).P \sim_{phhp}^{fr} \phi.P \boxplus_{\pi} \psi.P\quad (Std(P))$. It is sufficient to prove the relation $R=\{((\phi\boxplus_{\pi}\psi).P, \phi.P \boxplus_{\pi} \psi.P)\}\cup \textbf{Id}\quad (Std(P))$, is a FR strongly probabilistic hhp-bisimulation, and we omit it;
  \item $P.(\phi\boxplus_{\pi}\psi) \sim_{phhp}^{fr} P.\phi \boxplus_{\pi} P.\psi\quad(NStd(P))$. It is sufficient to prove the relation $R=\{(P.(\phi\boxplus_{\pi}\psi), P.\phi \boxplus_{\pi} P.\psi)\}\cup \textbf{Id}\quad(NStd(P))$, is a FR strongly probabilistic hhp-bisimulation, and we omit it;
  \item $(\phi.\psi).P \sim_{phhp}^{fr} \phi.(\psi.P)\quad(Std(P))$. It is sufficient to prove the relation $R=\{((\phi.\psi).P, \phi.(\psi.P))\}\cup \textbf{Id}\quad(Std(P))$ is a FR strongly probabilistic hhp-bisimulation, and we omit it;
  \item $P.(\phi.\psi) \sim_{phhp}^{fr}(P.\phi).\psi\quad(NStd(P))$. It is sufficient to prove the relation $R=\{(P.(\phi.\psi), (P.\phi).\psi)\}\cup \textbf{Id}\quad(NStd(P))$ is a FR strongly probabilistic hhp-bisimulation, and we omit it;
  \item $\phi\sim_{phhp}^{fr}\epsilon$ if $\forall s\in S.test(\phi,s)$. It is sufficient to prove the relation $R=\{(\phi, \epsilon)\}\cup \textbf{Id}$, if $\forall s\in S.test(\phi,s)$, is a FR strongly probabilistic hhp-bisimulation, and we omit it;
  \item $\phi_0\cdot\cdots\cdot\phi_n \sim_{phhp}^{fr} \delta$ if $\forall s\in S,\exists i\leq n.test(\neg\phi_i,s)$. It is sufficient to prove the relation $R=\{(\phi_0\cdot\cdots\cdot\phi_n, \delta)\}\cup \textbf{Id}$, if $\forall s\in S,\exists i\leq n.test(\neg\phi_i,s)$, is a FR strongly probabilistic hhp-bisimulation, and we omit it;
  \item $wp(\alpha,\phi).\alpha.\phi\sim_{phhp}^{fr} wp(\alpha,\phi).\alpha$. It is sufficient to prove the relation $R=\{(wp(\alpha,\phi).\alpha.\phi, wp(\alpha,\phi).\alpha)\}\cup \textbf{Id}$ is a FR strongly probabilistic hhp-bisimulation, and we omit it;
  \item $\phi. \alpha[m]. wp(\alpha[m],\phi)\sim_{phhp}^{fr}\alpha[m].wp(\alpha[m],\phi)$. It is sufficient to prove the relation $R=\{(\phi. \alpha[m]. wp(\alpha[m],\phi), \alpha[m].wp(\alpha[m],\phi))\}\cup \textbf{Id}$ is a FR strongly probabilistic hhp-bisimulation, and we omit it;
  \item $\neg wp(\alpha,\phi).\alpha.\neg\phi\sim_{phhp}^{fr}\neg wp(\alpha,\phi).\alpha$. It is sufficient to prove the relation \\$R=\{(\neg wp(\alpha,\phi).\alpha.\neg\phi, \neg wp(\alpha,\phi).\alpha)\}\cup \textbf{Id}$, is a FR strongly probabilistic hhp-bisimulation, and we omit it;
  \item $\neg\phi .\alpha[m] .\neg wp(\alpha[m],\phi)\sim_{phhp}^{fr} \alpha[m]. \neg wp(\alpha[m],\phi)$. It is sufficient to prove the relation $R=\{(\neg\phi .\alpha[m] .\neg wp(\alpha[m],\phi), \alpha[m]. \neg wp(\alpha[m],\phi))\}\cup \textbf{Id}$, is a FR strongly probabilistic hhp-bisimulation, and we omit it;
  \item $\delta\parallel P \sim_{phhp}^{fr} \delta$. It is sufficient to prove the relation $R=\{(\delta\parallel P, \delta)\}\cup \textbf{Id}$ is a FR strongly probabilistic hhp-bisimulation, and we omit it;
  \item $P\parallel \delta \sim_{phhp}^{fr} \delta$. It is sufficient to prove the relation $R=\{(P\parallel \delta, \delta)\}\cup \textbf{Id}$ is a FR strongly probabilistic hhp-bisimulation, and we omit it;
  \item $\epsilon\parallel P \sim_{phhp}^{fr} P$. It is sufficient to prove the relation $R=\{(\epsilon\parallel P, P)\}\cup \textbf{Id}$ is a FR strongly probabilistic hhp-bisimulation, and we omit it;
  \item $P\parallel \epsilon \sim_{phhp}^{fr} P$. It is sufficient to prove the relation $R=\{(P\parallel \epsilon, P)\}\cup \textbf{Id}$ is a FR strongly probabilistic hhp-bisimulation, and we omit it;
  \item $\phi.(P\parallel Q) \sim_{phhp}^{fr}\phi.P\parallel \phi.Q$. It is sufficient to prove the relation $R=\{(\phi.(P\parallel Q), \phi.P\parallel \phi.Q)\}\cup \textbf{Id}$ is a FR strongly probabilistic hhp-bisimulation, and we omit it;
  \item $\phi\parallel \delta \sim_{phhp}^{fr} \delta$. It is sufficient to prove the relation $R=\{(\phi\parallel \delta, \delta)\}\cup \textbf{Id}$ is a FR strongly probabilistic hhp-bisimulation, and we omit it;
  \item $\delta\parallel \phi \sim_{phhp}^{fr} \delta$. It is sufficient to prove the relation $R=\{(\delta\parallel \phi, \delta)\}\cup \textbf{Id}$ is a FR strongly probabilistic hhp-bisimulation, and we omit it;
  \item $\phi\parallel \epsilon \sim_{phhp}^{fr} \phi$. It is sufficient to prove the relation $R=\{(\phi\parallel \epsilon, \phi)\}\cup \textbf{Id}$ is a FR strongly probabilistic hhp-bisimulation, and we omit it;
  \item $\epsilon\parallel \phi \sim_{phhp}^{fr} \phi$. It is sufficient to prove the relation $R=\{(\epsilon\parallel \phi, \phi)\}\cup \textbf{Id}$ is a FR strongly probabilistic hhp-bisimulation, and we omit it;
  \item $\phi\parallel\neg\phi \sim_{phhp}^{fr} \delta$. It is sufficient to prove the relation $R=\{(\phi\parallel\neg\phi, \delta)\}\cup \textbf{Id}$ is a FR strongly probabilistic hhp-bisimulation, and we omit it;
  \item $\phi_0\parallel\cdots\parallel\phi_n \sim_{phhp}^{fr} \delta$ if $\forall s_0,\cdots,s_n\in S,\exists i\leq n.test(\neg\phi_i,s_0\cup\cdots\cup s_n)$. It is sufficient to prove the relation $R=\{(\phi_0\parallel\cdots\parallel\phi_n, \delta)\}\cup \textbf{Id}$, if $\forall s_0,\cdots,s_n\in S,\exists i\leq n.test(\neg\phi_i,s_0\cup\cdots\cup s_n)$, is a FR strongly probabilistic hhp-bisimulation, and we omit it.
\end{enumerate}
\end{proof}

\begin{proposition}[Expansion law for FR strongly probabilistic pomset bisimulation]
Let $P\equiv (P_1[f_1]\parallel\cdots\parallel P_n[f_n])\setminus L$, with $n\geq 1$. Then

\begin{eqnarray}
P\sim_{pp}^{f} \{(f_1(\alpha_1)\parallel\cdots\parallel f_n(\alpha_n)).(P_1'[f_1]\parallel\cdots\parallel P_n'[f_n])\setminus L: \nonumber\\
\langle P_i,s_i\rangle\rightsquigarrow\xrightarrow{\alpha_i}\langle P_i',s_i'\rangle,i\in\{1,\cdots,n\},f_i(\alpha_i)\notin L\cup\overline{L}\} \nonumber\\
+\sum\{\tau.(P_1[f_1]\parallel\cdots\parallel P_i'[f_i]\parallel\cdots\parallel P_j'[f_j]\parallel\cdots\parallel P_n[f_n])\setminus L: \nonumber\\
\langle P_i,s_i\rangle\rightsquigarrow\xrightarrow{l_1}\langle P_i',s_i'\rangle,\langle P_j,s_j\rangle\rightsquigarrow\xrightarrow{l_2}\langle P_j',s_j'\rangle,f_i(l_1)=\overline{f_j(l_2)},i<j\}\nonumber
\end{eqnarray}
\begin{eqnarray}
P\sim_{pp}^{r} \{(P_1'[f_1]\parallel\cdots\parallel P_n'[f_n]).(f_1(\alpha_1[m])\parallel\cdots\parallel f_n(\alpha_n)[m])\setminus L: \nonumber\\
\langle P_i,s_i\rangle\rightsquigarrow\xtworightarrow{\alpha_i[m]}\langle P_i',s_i'\rangle,i\in\{1,\cdots,n\},f_i(\alpha_i)\notin L\cup\overline{L}\} \nonumber\\
+\sum\{(P_1[f_1]\parallel\cdots\parallel P_i'[f_i]\parallel\cdots\parallel P_j'[f_j]\parallel\cdots\parallel P_n[f_n]).\tau\setminus L: \nonumber\\
\langle P_i,s_i\rangle\rightsquigarrow\xtworightarrow{l_1[m]}\langle P_i',s_i'\rangle,\langle P_j,s_j\rangle\rightsquigarrow\xtworightarrow{l_2[m]}\langle P_j',s_j'\rangle,f_i(l_1)=\overline{f_j(l_2)},i<j\}\nonumber
\end{eqnarray}
\end{proposition}

\begin{proof}
(1) The case of forward strongly probabilistic pomset bisimulation.

Firstly, we consider the case without Restriction and Relabeling. That is, we suffice to prove the following case by induction on the size $n$.

For $P\equiv P_1\parallel\cdots\parallel P_n$, with $n\geq 1$, we need to prove

\begin{eqnarray}
P\sim_{pp} \{(\alpha_1\parallel\cdots\parallel \alpha_n).(P_1'\parallel\cdots\parallel P_n'): \langle P_i,s_i\rangle\rightsquigarrow\xrightarrow{\alpha_i}\langle P_i',s_i'\rangle,i\in\{1,\cdots,n\}\nonumber\\
+\sum\{\tau.(P_1\parallel\cdots\parallel P_i'\parallel\cdots\parallel P_j'\parallel\cdots\parallel P_n): \langle P_i,s_i\rangle\rightsquigarrow\xrightarrow{l}\langle P_i',s_i'\rangle,\langle P_j,s_j\rangle\rightsquigarrow\xrightarrow{\overline{l}}\langle P_j',s_j'\rangle,i<j\} \nonumber
\end{eqnarray}

For $n=1$, $P_1\sim_{pp}^{f} \alpha_1.P_1':\langle P_1,s_1\rangle\rightsquigarrow\xrightarrow{\alpha_1}\langle P_1',s_1'\rangle$ is obvious. Then with a hypothesis $n$, we consider
$R\equiv P\parallel P_{n+1}$. By the forward transition rules of Composition, we can get

\begin{eqnarray}
R\sim_{pp}^{f} \{(p\parallel \alpha_{n+1}).(P'\parallel P_{n+1}'): \langle P,s\rangle\rightsquigarrow\xrightarrow{p}\langle P',s'\rangle,\langle P_{n+1},s_{n+1}\rangle\rightsquigarrow\xrightarrow{\alpha_{n+1}}\langle P_{n+1}',s_{n+1}'\rangle,p\subseteq P\}\nonumber\\
+\sum\{\tau.(P'\parallel P_{n+1}'): \langle P,s\rangle\rightsquigarrow\xrightarrow{l}\langle P',s'\rangle,\langle P_{n+1},s_{n+1}\rangle\rightsquigarrow\xrightarrow{\overline{l}}\langle P_{n+1}',s_{n+1}'\rangle\} \nonumber
\end{eqnarray}

Now with the induction assumption $P\equiv P_1\parallel\cdots\parallel P_n$, the right-hand side can be reformulated as follows.

\begin{eqnarray}
\{(\alpha_1\parallel\cdots\parallel \alpha_n\parallel \alpha_{n+1}).(P_1'\parallel\cdots\parallel P_n'\parallel P_{n+1}'): \nonumber\\
\langle P_i,s_i\rangle\rightsquigarrow\xrightarrow{\alpha_i}\langle P_i',s_i'\rangle,i\in\{1,\cdots,n+1\}\nonumber\\
+\sum\{\tau.(P_1\parallel\cdots\parallel P_i'\parallel\cdots\parallel P_j'\parallel\cdots\parallel P_n\parallel P_{n+1}): \nonumber\\
\langle P_i,s_i\rangle\rightsquigarrow\xrightarrow{l}\langle P_i',s_i'\rangle,\langle P_j,s_j\rangle\rightsquigarrow\xrightarrow{\overline{l}}\langle P_j',s_j'\rangle,i<j\} \nonumber\\
+\sum\{\tau.(P_1\parallel\cdots\parallel P_i'\parallel\cdots\parallel P_j\parallel\cdots\parallel P_n\parallel P_{n+1}'): \nonumber\\
\langle P_i,s_i\rangle\rightsquigarrow\xrightarrow{l}\langle P_i',s_i'\rangle,\langle P_{n+1},s_{n+1}\rangle\rightsquigarrow\xrightarrow{\overline{l}}\langle P_{n+1}',s_{n+1}'\rangle,i\in\{1,\cdots, n\}\} \nonumber
\end{eqnarray}

So,

\begin{eqnarray}
R\sim_{pp}^{f} \{(\alpha_1\parallel\cdots\parallel \alpha_n\parallel \alpha_{n+1}).(P_1'\parallel\cdots\parallel P_n'\parallel P_{n+1}'): \nonumber\\
\langle P_i,s_i\rangle\rightsquigarrow\xrightarrow{\alpha_i}\langle P_i',s_i'\rangle,i\in\{1,\cdots,n+1\}\nonumber\\
+\sum\{\tau.(P_1\parallel\cdots\parallel P_i'\parallel\cdots\parallel P_j'\parallel\cdots\parallel P_n): \nonumber\\
\langle P_i,s_i\rangle\rightsquigarrow\xrightarrow{l}\langle P_i',s_i'\rangle,\langle P_j,s_j\rangle\rightsquigarrow\xrightarrow{\overline{l}}\langle P_j',s_j'\rangle,1 \leq i<j\geq n+1\} \nonumber
\end{eqnarray}

Then, we can easily add the full conditions with Restriction and Relabeling.

(2) The case of reverse strongly probabilistic pomset bisimulation.

Firstly, we consider the case without Restriction and Relabeling. That is, we suffice to prove the following case by induction on the size $n$.

For $P\equiv P_1\parallel\cdots\parallel P_n$, with $n\geq 1$, we need to prove

\begin{eqnarray}
P\sim_{pp}^{r} \{(P_1'\parallel\cdots\parallel P_n').(\alpha_1[m]\parallel\cdots\parallel \alpha_n[m]): \langle P_i,s_i\rangle\rightsquigarrow\xtworightarrow{\alpha_i[m]}\langle P_i',s_i'\rangle,i\in\{1,\cdots,n\}\nonumber\\
+\sum\{(P_1\parallel\cdots\parallel P_i'\parallel\cdots\parallel P_j'\parallel\cdots\parallel P_n).\tau: \langle P_i,s_i\rangle\rightsquigarrow\xtworightarrow{l[m]}\langle P_i',s_i'\rangle,\langle P_j,s_j\rangle\rightsquigarrow\xtworightarrow{\overline{l}[m]}\langle P_j',s_j'\rangle,i<j\} \nonumber
\end{eqnarray}

For $n=1$, $P_1\sim_{pp}^{r} P_1'.\alpha_1[m]:\langle P_1,s_1\rangle\rightsquigarrow\xtworightarrow{\alpha_1[m]}\langle P_1',s_1'\rangle$ is obvious. Then with a hypothesis $n$, we consider
$R\equiv P\parallel P_{n+1}$. By the reverse transition rules of Composition, we can get

\begin{eqnarray}
R\sim_{pp}^{r} \{(P'\parallel P_{n+1}').(p[m]\parallel \alpha_{n+1}[m]): \langle P,s\rangle\rightsquigarrow\xtworightarrow{p[m]}\langle P',s'\rangle,\langle P_{n+1},s_{n+1}\rangle\rightsquigarrow\xtworightarrow{\alpha_{n+1}[m]}\langle P_{n+1}',s_{n+1}'\rangle,p\subseteq P\}\nonumber\\
+\sum\{(P'\parallel P_{n+1}').\tau: \langle P,s\rangle\rightsquigarrow\xtworightarrow{l[m]}\langle P',s'\rangle,\langle P_{n+1},s_{n+1}\rangle\rightsquigarrow\xtworightarrow{\overline{l}[m]}\langle P_{n+1}',s_{n+1}'\rangle\} \nonumber
\end{eqnarray}

Now with the induction assumption $P\equiv P_1\parallel\cdots\parallel P_n$, the right-hand side can be reformulated as follows.

\begin{eqnarray}
\{(P_1'\parallel\cdots\parallel P_n'\parallel P_{n+1}).(\alpha_1[m]\parallel\cdots\parallel \alpha_n[m]\parallel \alpha_{n+1}[m]): \nonumber\\
\langle P_i,s_i\rangle\rightsquigarrow\xtworightarrow{\alpha_i[m]}\langle P_i',s_i'\rangle,i\in\{1,\cdots,n+1\}\nonumber\\
+\sum\{(P_1\parallel\cdots\parallel P_i'\parallel\cdots\parallel P_j'\parallel\cdots\parallel P_n\parallel P_{n+1}).\tau: \nonumber\\
\langle P_i,s_i\rangle\rightsquigarrow\xtworightarrow{l[m]}\langle P_i',s_i'\rangle,\langle P_j,s_j\rangle\rightsquigarrow\xtworightarrow{\overline{l}[m]}\langle P_j',s_j'\rangle,i<j\} \nonumber\\
+\sum\{(P_1\parallel\cdots\parallel P_i'\parallel\cdots\parallel P_j'\parallel\cdots\parallel P_n\parallel P_{n+1}).\tau: \nonumber\\
\langle P_i,s_i\rangle\rightsquigarrow\xtworightarrow{l[m]}\langle P_i',s_i'\rangle,\langle P_{n+1},s_{n+1}\rangle\rightsquigarrow\xtworightarrow{\overline{l}[m]}\langle P_{n+1}',s_{n+1}'\rangle,i\in\{1,\cdots, n\}\} \nonumber
\end{eqnarray}

So,

\begin{eqnarray}
R\sim_{pp}^{r} \{(P_1'\parallel\cdots\parallel P_n'\parallel P_{n+1}').(\alpha_1[m]\parallel\cdots\parallel \alpha_n[m]\parallel \alpha_{n+1}[m]): \nonumber\\
\langle P_i,s_i\rangle\rightsquigarrow\xtworightarrow{\alpha_i[m]}\langle P_i',s_i'\rangle,i\in\{1,\cdots,n+1\}\nonumber\\
+\sum\{(P_1\parallel\cdots\parallel P_i'\parallel\cdots\parallel P_j'\parallel\cdots\parallel P_n).\tau: \nonumber\\
\langle P_i,s_i\rangle\rightsquigarrow\xtworightarrow{l[m]}\langle P_i',s_i'\rangle,\langle P_j,s_j\rangle\rightsquigarrow\xtworightarrow{\overline{l}[m]}\langle P_j',s_j'\rangle,1 \leq i<j\geq n+1\} \nonumber
\end{eqnarray}

Then, we can easily add the full conditions with Restriction and Relabeling.
\end{proof}

\begin{proposition}[Expansion law for FR strongly probabilistic step bisimulation]
Let $P\equiv (P_1[f_1]\parallel\cdots\parallel P_n[f_n])\setminus L$, with $n\geq 1$. Then

\begin{eqnarray}
P\sim_{ps}^{f} \{(f_1(\alpha_1)\parallel\cdots\parallel f_n(\alpha_n)).(P_1'[f_1]\parallel\cdots\parallel P_n'[f_n])\setminus L: \nonumber\\
\langle P_i,s_i\rangle\rightsquigarrow\xrightarrow{\alpha_i}\langle P_i',s_i'\rangle,i\in\{1,\cdots,n\},f_i(\alpha_i)\notin L\cup\overline{L}\} \nonumber\\
+\sum\{\tau.(P_1[f_1]\parallel\cdots\parallel P_i'[f_i]\parallel\cdots\parallel P_j'[f_j]\parallel\cdots\parallel P_n[f_n])\setminus L: \nonumber\\
\langle P_i,s_i\rangle\rightsquigarrow\xrightarrow{l_1}\langle P_i',s_i'\rangle,\langle P_j,s_j\rangle\rightsquigarrow\xrightarrow{l_2}\langle P_j',s_j'\rangle,f_i(l_1)=\overline{f_j(l_2)},i<j\}\nonumber
\end{eqnarray}
\begin{eqnarray}
P\sim_{ps}^{r} \{(P_1'[f_1]\parallel\cdots\parallel P_n'[f_n]).(f_1(\alpha_1[m])\parallel\cdots\parallel f_n(\alpha_n)[m])\setminus L: \nonumber\\
\langle P_i,s_i\rangle\rightsquigarrow\xtworightarrow{\alpha_i[m]}\langle P_i',s_i'\rangle,i\in\{1,\cdots,n\},f_i(\alpha_i)\notin L\cup\overline{L}\} \nonumber\\
+\sum\{(P_1[f_1]\parallel\cdots\parallel P_i'[f_i]\parallel\cdots\parallel P_j'[f_j]\parallel\cdots\parallel P_n[f_n]).\tau\setminus L: \nonumber\\
\langle P_i,s_i\rangle\rightsquigarrow\xtworightarrow{l_1[m]}\langle P_i',s_i'\rangle,\langle P_j,s_j\rangle\rightsquigarrow\xtworightarrow{l_2[m]}\langle P_j',s_j'\rangle,f_i(l_1)=\overline{f_j(l_2)},i<j\}\nonumber
\end{eqnarray}
\end{proposition}

\begin{proof}
(1) The case of forward strongly probabilistic step bisimulation.

Firstly, we consider the case without Restriction and Relabeling. That is, we suffice to prove the following case by induction on the size $n$.

For $P\equiv P_1\parallel\cdots\parallel P_n$, with $n\geq 1$, we need to prove

\begin{eqnarray}
P\sim_{ps} \{(\alpha_1\parallel\cdots\parallel \alpha_n).(P_1'\parallel\cdots\parallel P_n'): \langle P_i,s_i\rangle\rightsquigarrow\xrightarrow{\alpha_i}\langle P_i',s_i'\rangle,i\in\{1,\cdots,n\}\nonumber\\
+\sum\{\tau.(P_1\parallel\cdots\parallel P_i'\parallel\cdots\parallel P_j'\parallel\cdots\parallel P_n): \langle P_i,s_i\rangle\rightsquigarrow\xrightarrow{l}\langle P_i',s_i'\rangle,\langle P_j,s_j\rangle\rightsquigarrow\xrightarrow{\overline{l}}\langle P_j',s_j'\rangle,i<j\} \nonumber
\end{eqnarray}

For $n=1$, $P_1\sim_{ps}^{f} \alpha_1.P_1':\langle P_1,s_1\rangle\rightsquigarrow\xrightarrow{\alpha_1}\langle P_1',s_1'\rangle$ is obvious. Then with a hypothesis $n$, we consider
$R\equiv P\parallel P_{n+1}$. By the forward transition rules of Composition, we can get

\begin{eqnarray}
R\sim_{ps}^{f} \{(p\parallel \alpha_{n+1}).(P'\parallel P_{n+1}'): \langle P,s\rangle\rightsquigarrow\xrightarrow{p}\langle P',s'\rangle,\langle P_{n+1},s_{n+1}\rangle\rightsquigarrow\xrightarrow{\alpha_{n+1}}\langle P_{n+1}',s_{n+1}'\rangle,p\subseteq P\}\nonumber\\
+\sum\{\tau.(P'\parallel P_{n+1}'): \langle P,s\rangle\rightsquigarrow\xrightarrow{l}\langle P',s'\rangle,\langle P_{n+1},s_{n+1}\rangle\rightsquigarrow\xrightarrow{\overline{l}}\langle P_{n+1}',s_{n+1}'\rangle\} \nonumber
\end{eqnarray}

Now with the induction assumption $P\equiv P_1\parallel\cdots\parallel P_n$, the right-hand side can be reformulated as follows.

\begin{eqnarray}
\{(\alpha_1\parallel\cdots\parallel \alpha_n\parallel \alpha_{n+1}).(P_1'\parallel\cdots\parallel P_n'\parallel P_{n+1}'): \nonumber\\
\langle P_i,s_i\rangle\rightsquigarrow\xrightarrow{\alpha_i}\langle P_i',s_i'\rangle,i\in\{1,\cdots,n+1\}\nonumber\\
+\sum\{\tau.(P_1\parallel\cdots\parallel P_i'\parallel\cdots\parallel P_j'\parallel\cdots\parallel P_n\parallel P_{n+1}): \nonumber\\
\langle P_i,s_i\rangle\rightsquigarrow\xrightarrow{l}\langle P_i',s_i'\rangle,\langle P_j,s_j\rangle\rightsquigarrow\xrightarrow{\overline{l}}\langle P_j',s_j'\rangle,i<j\} \nonumber\\
+\sum\{\tau.(P_1\parallel\cdots\parallel P_i'\parallel\cdots\parallel P_j\parallel\cdots\parallel P_n\parallel P_{n+1}'): \nonumber\\
\langle P_i,s_i\rangle\rightsquigarrow\xrightarrow{l}\langle P_i',s_i'\rangle,\langle P_{n+1},s_{n+1}\rangle\rightsquigarrow\xrightarrow{\overline{l}}\langle P_{n+1}',s_{n+1}'\rangle,i\in\{1,\cdots, n\}\} \nonumber
\end{eqnarray}

So,

\begin{eqnarray}
R\sim_{ps}^{f} \{(\alpha_1\parallel\cdots\parallel \alpha_n\parallel \alpha_{n+1}).(P_1'\parallel\cdots\parallel P_n'\parallel P_{n+1}'): \nonumber\\
\langle P_i,s_i\rangle\rightsquigarrow\xrightarrow{\alpha_i}\langle P_i',s_i'\rangle,i\in\{1,\cdots,n+1\}\nonumber\\
+\sum\{\tau.(P_1\parallel\cdots\parallel P_i'\parallel\cdots\parallel P_j'\parallel\cdots\parallel P_n): \nonumber\\
\langle P_i,s_i\rangle\rightsquigarrow\xrightarrow{l}\langle P_i',s_i'\rangle,\langle P_j,s_j\rangle\rightsquigarrow\xrightarrow{\overline{l}}\langle P_j',s_j'\rangle,1 \leq i<j\geq n+1\} \nonumber
\end{eqnarray}

Then, we can easily add the full conditions with Restriction and Relabeling.

(2) The case of reverse strongly probabilistic step bisimulation.

Firstly, we consider the case without Restriction and Relabeling. That is, we suffice to prove the following case by induction on the size $n$.

For $P\equiv P_1\parallel\cdots\parallel P_n$, with $n\geq 1$, we need to prove

\begin{eqnarray}
P\sim_{ps}^{r} \{(P_1'\parallel\cdots\parallel P_n').(\alpha_1[m]\parallel\cdots\parallel \alpha_n[m]): \langle P_i,s_i\rangle\rightsquigarrow\xtworightarrow{\alpha_i[m]}\langle P_i',s_i'\rangle,i\in\{1,\cdots,n\}\nonumber\\
+\sum\{(P_1\parallel\cdots\parallel P_i'\parallel\cdots\parallel P_j'\parallel\cdots\parallel P_n).\tau: \langle P_i,s_i\rangle\rightsquigarrow\xtworightarrow{l[m]}\langle P_i',s_i'\rangle,\langle P_j,s_j\rangle\rightsquigarrow\xtworightarrow{\overline{l}[m]}\langle P_j',s_j'\rangle,i<j\} \nonumber
\end{eqnarray}

For $n=1$, $P_1\sim_{ps}^{r} P_1'.\alpha_1[m]:\langle P_1,s_1\rangle\rightsquigarrow\xtworightarrow{\alpha_1[m]}\langle P_1',s_1'\rangle$ is obvious. Then with a hypothesis $n$, we consider
$R\equiv P\parallel P_{n+1}$. By the reverse transition rules of Composition, we can get

\begin{eqnarray}
R\sim_{ps}^{r} \{(P'\parallel P_{n+1}').(p[m]\parallel \alpha_{n+1}[m]): \langle P,s\rangle\rightsquigarrow\xtworightarrow{p[m]}\langle P',s'\rangle,\langle P_{n+1},s_{n+1}\rangle\rightsquigarrow\xtworightarrow{\alpha_{n+1}[m]}\langle P_{n+1}',s_{n+1}'\rangle,p\subseteq P\}\nonumber\\
+\sum\{(P'\parallel P_{n+1}').\tau: \langle P,s\rangle\rightsquigarrow\xtworightarrow{l[m]}\langle P',s'\rangle,\langle P_{n+1},s_{n+1}\rangle\rightsquigarrow\xtworightarrow{\overline{l}[m]}\langle P_{n+1}',s_{n+1}'\rangle\} \nonumber
\end{eqnarray}

Now with the induction assumption $P\equiv P_1\parallel\cdots\parallel P_n$, the right-hand side can be reformulated as follows.

\begin{eqnarray}
\{(P_1'\parallel\cdots\parallel P_n'\parallel P_{n+1}).(\alpha_1[m]\parallel\cdots\parallel \alpha_n[m]\parallel \alpha_{n+1}[m]): \nonumber\\
\langle P_i,s_i\rangle\rightsquigarrow\xtworightarrow{\alpha_i[m]}\langle P_i',s_i'\rangle,i\in\{1,\cdots,n+1\}\nonumber\\
+\sum\{(P_1\parallel\cdots\parallel P_i'\parallel\cdots\parallel P_j'\parallel\cdots\parallel P_n\parallel P_{n+1}).\tau: \nonumber\\
\langle P_i,s_i\rangle\rightsquigarrow\xtworightarrow{l[m]}\langle P_i',s_i'\rangle,\langle P_j,s_j\rangle\rightsquigarrow\xtworightarrow{\overline{l}[m]}\langle P_j',s_j'\rangle,i<j\} \nonumber\\
+\sum\{(P_1\parallel\cdots\parallel P_i'\parallel\cdots\parallel P_j'\parallel\cdots\parallel P_n\parallel P_{n+1}).\tau: \nonumber\\
\langle P_i,s_i\rangle\rightsquigarrow\xtworightarrow{l[m]}\langle P_i',s_i'\rangle,\langle P_{n+1},s_{n+1}\rangle\rightsquigarrow\xtworightarrow{\overline{l}[m]}\langle P_{n+1}',s_{n+1}'\rangle,i\in\{1,\cdots, n\}\} \nonumber
\end{eqnarray}

So,

\begin{eqnarray}
R\sim_{ps}^{r} \{(P_1'\parallel\cdots\parallel P_n'\parallel P_{n+1}').(\alpha_1[m]\parallel\cdots\parallel \alpha_n[m]\parallel \alpha_{n+1}[m]): \nonumber\\
\langle P_i,s_i\rangle\rightsquigarrow\xtworightarrow{\alpha_i[m]}\langle P_i',s_i'\rangle,i\in\{1,\cdots,n+1\}\nonumber\\
+\sum\{(P_1\parallel\cdots\parallel P_i'\parallel\cdots\parallel P_j'\parallel\cdots\parallel P_n).\tau: \nonumber\\
\langle P_i,s_i\rangle\rightsquigarrow\xtworightarrow{l[m]}\langle P_i',s_i'\rangle,\langle P_j,s_j\rangle\rightsquigarrow\xtworightarrow{\overline{l}[m]}\langle P_j',s_j'\rangle,1 \leq i<j\geq n+1\} \nonumber
\end{eqnarray}

Then, we can easily add the full conditions with Restriction and Relabeling.
\end{proof}

\begin{proposition}[Expansion law for FR strongly probabilistic hp-bisimulation]
Let $P\equiv (P_1[f_1]\parallel\cdots\parallel P_n[f_n])\setminus L$, with $n\geq 1$. Then

\begin{eqnarray}
P\sim_{php}^{f} \{(f_1(\alpha_1)\parallel\cdots\parallel f_n(\alpha_n)).(P_1'[f_1]\parallel\cdots\parallel P_n'[f_n])\setminus L: \nonumber\\
\langle P_i,s_i\rangle\rightsquigarrow\xrightarrow{\alpha_i}\langle P_i',s_i'\rangle,i\in\{1,\cdots,n\},f_i(\alpha_i)\notin L\cup\overline{L}\} \nonumber\\
+\sum\{\tau.(P_1[f_1]\parallel\cdots\parallel P_i'[f_i]\parallel\cdots\parallel P_j'[f_j]\parallel\cdots\parallel P_n[f_n])\setminus L: \nonumber\\
\langle P_i,s_i\rangle\rightsquigarrow\xrightarrow{l_1}\langle P_i',s_i'\rangle,\langle P_j,s_j\rangle\rightsquigarrow\xrightarrow{l_2}\langle P_j',s_j'\rangle,f_i(l_1)=\overline{f_j(l_2)},i<j\}\nonumber
\end{eqnarray}
\begin{eqnarray}
P\sim_{php}^{r} \{(P_1'[f_1]\parallel\cdots\parallel P_n'[f_n]).(f_1(\alpha_1[m])\parallel\cdots\parallel f_n(\alpha_n)[m])\setminus L: \nonumber\\
\langle P_i,s_i\rangle\rightsquigarrow\xtworightarrow{\alpha_i[m]}\langle P_i',s_i'\rangle,i\in\{1,\cdots,n\},f_i(\alpha_i)\notin L\cup\overline{L}\} \nonumber\\
+\sum\{(P_1[f_1]\parallel\cdots\parallel P_i'[f_i]\parallel\cdots\parallel P_j'[f_j]\parallel\cdots\parallel P_n[f_n]).\tau\setminus L: \nonumber\\
\langle P_i,s_i\rangle\rightsquigarrow\xtworightarrow{l_1[m]}\langle P_i',s_i'\rangle,\langle P_j,s_j\rangle\rightsquigarrow\xtworightarrow{l_2[m]}\langle P_j',s_j'\rangle,f_i(l_1)=\overline{f_j(l_2)},i<j\}\nonumber
\end{eqnarray}
\end{proposition}

\begin{proof}
(1) The case of forward strongly probabilistic hp-bisimulation.

Firstly, we consider the case without Restriction and Relabeling. That is, we suffice to prove the following case by induction on the size $n$.

For $P\equiv P_1\parallel\cdots\parallel P_n$, with $n\geq 1$, we need to prove

\begin{eqnarray}
P\sim_{php} \{(\alpha_1\parallel\cdots\parallel \alpha_n).(P_1'\parallel\cdots\parallel P_n'): \langle P_i,s_i\rangle\rightsquigarrow\xrightarrow{\alpha_i}\langle P_i',s_i'\rangle,i\in\{1,\cdots,n\}\nonumber\\
+\sum\{\tau.(P_1\parallel\cdots\parallel P_i'\parallel\cdots\parallel P_j'\parallel\cdots\parallel P_n): \langle P_i,s_i\rangle\rightsquigarrow\xrightarrow{l}\langle P_i',s_i'\rangle,\langle P_j,s_j\rangle\rightsquigarrow\xrightarrow{\overline{l}}\langle P_j',s_j'\rangle,i<j\} \nonumber
\end{eqnarray}

For $n=1$, $P_1\sim_{php}^{f} \alpha_1.P_1':\langle P_1,s_1\rangle\rightsquigarrow\xrightarrow{\alpha_1}\langle P_1',s_1'\rangle$ is obvious. Then with a hypothesis $n$, we consider
$R\equiv P\parallel P_{n+1}$. By the forward transition rules of Composition, we can get

\begin{eqnarray}
R\sim_{php}^{f} \{(p\parallel \alpha_{n+1}).(P'\parallel P_{n+1}'): \langle P,s\rangle\rightsquigarrow\xrightarrow{p}\langle P',s'\rangle,\langle P_{n+1},s_{n+1}\rangle\rightsquigarrow\xrightarrow{\alpha_{n+1}}\langle P_{n+1}',s_{n+1}'\rangle,p\subseteq P\}\nonumber\\
+\sum\{\tau.(P'\parallel P_{n+1}'): \langle P,s\rangle\rightsquigarrow\xrightarrow{l}\langle P',s'\rangle,\langle P_{n+1},s_{n+1}\rangle\rightsquigarrow\xrightarrow{\overline{l}}\langle P_{n+1}',s_{n+1}'\rangle\} \nonumber
\end{eqnarray}

Now with the induction assumption $P\equiv P_1\parallel\cdots\parallel P_n$, the right-hand side can be reformulated as follows.

\begin{eqnarray}
\{(\alpha_1\parallel\cdots\parallel \alpha_n\parallel \alpha_{n+1}).(P_1'\parallel\cdots\parallel P_n'\parallel P_{n+1}'): \nonumber\\
\langle P_i,s_i\rangle\rightsquigarrow\xrightarrow{\alpha_i}\langle P_i',s_i'\rangle,i\in\{1,\cdots,n+1\}\nonumber\\
+\sum\{\tau.(P_1\parallel\cdots\parallel P_i'\parallel\cdots\parallel P_j'\parallel\cdots\parallel P_n\parallel P_{n+1}): \nonumber\\
\langle P_i,s_i\rangle\rightsquigarrow\xrightarrow{l}\langle P_i',s_i'\rangle,\langle P_j,s_j\rangle\rightsquigarrow\xrightarrow{\overline{l}}\langle P_j',s_j'\rangle,i<j\} \nonumber\\
+\sum\{\tau.(P_1\parallel\cdots\parallel P_i'\parallel\cdots\parallel P_j\parallel\cdots\parallel P_n\parallel P_{n+1}'): \nonumber\\
\langle P_i,s_i\rangle\rightsquigarrow\xrightarrow{l}\langle P_i',s_i'\rangle,\langle P_{n+1},s_{n+1}\rangle\rightsquigarrow\xrightarrow{\overline{l}}\langle P_{n+1}',s_{n+1}'\rangle,i\in\{1,\cdots, n\}\} \nonumber
\end{eqnarray}

So,

\begin{eqnarray}
R\sim_{php}^{f} \{(\alpha_1\parallel\cdots\parallel \alpha_n\parallel \alpha_{n+1}).(P_1'\parallel\cdots\parallel P_n'\parallel P_{n+1}'): \nonumber\\
\langle P_i,s_i\rangle\rightsquigarrow\xrightarrow{\alpha_i}\langle P_i',s_i'\rangle,i\in\{1,\cdots,n+1\}\nonumber\\
+\sum\{\tau.(P_1\parallel\cdots\parallel P_i'\parallel\cdots\parallel P_j'\parallel\cdots\parallel P_n): \nonumber\\
\langle P_i,s_i\rangle\rightsquigarrow\xrightarrow{l}\langle P_i',s_i'\rangle,\langle P_j,s_j\rangle\rightsquigarrow\xrightarrow{\overline{l}}\langle P_j',s_j'\rangle,1 \leq i<j\geq n+1\} \nonumber
\end{eqnarray}

Then, we can easily add the full conditions with Restriction and Relabeling.

(2) The case of reverse strongly probabilistic hp-bisimulation.

Firstly, we consider the case without Restriction and Relabeling. That is, we suffice to prove the following case by induction on the size $n$.

For $P\equiv P_1\parallel\cdots\parallel P_n$, with $n\geq 1$, we need to prove

\begin{eqnarray}
P\sim_{php}^{r} \{(P_1'\parallel\cdots\parallel P_n').(\alpha_1[m]\parallel\cdots\parallel \alpha_n[m]): \langle P_i,s_i\rangle\rightsquigarrow\xtworightarrow{\alpha_i[m]}\langle P_i',s_i'\rangle,i\in\{1,\cdots,n\}\nonumber\\
+\sum\{(P_1\parallel\cdots\parallel P_i'\parallel\cdots\parallel P_j'\parallel\cdots\parallel P_n).\tau: \langle P_i,s_i\rangle\rightsquigarrow\xtworightarrow{l[m]}\langle P_i',s_i'\rangle,\langle P_j,s_j\rangle\rightsquigarrow\xtworightarrow{\overline{l}[m]}\langle P_j',s_j'\rangle,i<j\} \nonumber
\end{eqnarray}

For $n=1$, $P_1\sim_{php}^{r} P_1'.\alpha_1[m]:\langle P_1,s_1\rangle\rightsquigarrow\xtworightarrow{\alpha_1[m]}\langle P_1',s_1'\rangle$ is obvious. Then with a hypothesis $n$, we consider
$R\equiv P\parallel P_{n+1}$. By the reverse transition rules of Composition, we can get

\begin{eqnarray}
R\sim_{php}^{r} \{(P'\parallel P_{n+1}').(p[m]\parallel \alpha_{n+1}[m]): \langle P,s\rangle\rightsquigarrow\xtworightarrow{p[m]}\langle P',s'\rangle,\langle P_{n+1},s_{n+1}\rangle\rightsquigarrow\xtworightarrow{\alpha_{n+1}[m]}\langle P_{n+1}',s_{n+1}'\rangle,p\subseteq P\}\nonumber\\
+\sum\{(P'\parallel P_{n+1}').\tau: \langle P,s\rangle\rightsquigarrow\xtworightarrow{l[m]}\langle P',s'\rangle,\langle P_{n+1},s_{n+1}\rangle\rightsquigarrow\xtworightarrow{\overline{l}[m]}\langle P_{n+1}',s_{n+1}'\rangle\} \nonumber
\end{eqnarray}

Now with the induction assumption $P\equiv P_1\parallel\cdots\parallel P_n$, the right-hand side can be reformulated as follows.

\begin{eqnarray}
\{(P_1'\parallel\cdots\parallel P_n'\parallel P_{n+1}).(\alpha_1[m]\parallel\cdots\parallel \alpha_n[m]\parallel \alpha_{n+1}[m]): \nonumber\\
\langle P_i,s_i\rangle\rightsquigarrow\xtworightarrow{\alpha_i[m]}\langle P_i',s_i'\rangle,i\in\{1,\cdots,n+1\}\nonumber\\
+\sum\{(P_1\parallel\cdots\parallel P_i'\parallel\cdots\parallel P_j'\parallel\cdots\parallel P_n\parallel P_{n+1}).\tau: \nonumber\\
\langle P_i,s_i\rangle\rightsquigarrow\xtworightarrow{l[m]}\langle P_i',s_i'\rangle,\langle P_j,s_j\rangle\rightsquigarrow\xtworightarrow{\overline{l}[m]}\langle P_j',s_j'\rangle,i<j\} \nonumber\\
+\sum\{(P_1\parallel\cdots\parallel P_i'\parallel\cdots\parallel P_j'\parallel\cdots\parallel P_n\parallel P_{n+1}).\tau: \nonumber\\
\langle P_i,s_i\rangle\rightsquigarrow\xtworightarrow{l[m]}\langle P_i',s_i'\rangle,\langle P_{n+1},s_{n+1}\rangle\rightsquigarrow\xtworightarrow{\overline{l}[m]}\langle P_{n+1}',s_{n+1}'\rangle,i\in\{1,\cdots, n\}\} \nonumber
\end{eqnarray}

So,

\begin{eqnarray}
R\sim_{php}^{r} \{(P_1'\parallel\cdots\parallel P_n'\parallel P_{n+1}').(\alpha_1[m]\parallel\cdots\parallel \alpha_n[m]\parallel \alpha_{n+1}[m]): \nonumber\\
\langle P_i,s_i\rangle\rightsquigarrow\xtworightarrow{\alpha_i[m]}\langle P_i',s_i'\rangle,i\in\{1,\cdots,n+1\}\nonumber\\
+\sum\{(P_1\parallel\cdots\parallel P_i'\parallel\cdots\parallel P_j'\parallel\cdots\parallel P_n).\tau: \nonumber\\
\langle P_i,s_i\rangle\rightsquigarrow\xtworightarrow{l[m]}\langle P_i',s_i'\rangle,\langle P_j,s_j\rangle\rightsquigarrow\xtworightarrow{\overline{l}[m]}\langle P_j',s_j'\rangle,1 \leq i<j\geq n+1\} \nonumber
\end{eqnarray}

Then, we can easily add the full conditions with Restriction and Relabeling.
\end{proof}

\begin{proposition}[Expansion law for FR strongly probabilistic hhp-bisimulation]
Let $P\equiv (P_1[f_1]\parallel\cdots\parallel P_n[f_n])\setminus L$, with $n\geq 1$. Then

\begin{eqnarray}
P\sim_{phhp}^{f} \{(f_1(\alpha_1)\parallel\cdots\parallel f_n(\alpha_n)).(P_1'[f_1]\parallel\cdots\parallel P_n'[f_n])\setminus L: \nonumber\\
\langle P_i,s_i\rangle\rightsquigarrow\xrightarrow{\alpha_i}\langle P_i',s_i'\rangle,i\in\{1,\cdots,n\},f_i(\alpha_i)\notin L\cup\overline{L}\} \nonumber\\
+\sum\{\tau.(P_1[f_1]\parallel\cdots\parallel P_i'[f_i]\parallel\cdots\parallel P_j'[f_j]\parallel\cdots\parallel P_n[f_n])\setminus L: \nonumber\\
\langle P_i,s_i\rangle\rightsquigarrow\xrightarrow{l_1}\langle P_i',s_i'\rangle,\langle P_j,s_j\rangle\rightsquigarrow\xrightarrow{l_2}\langle P_j',s_j'\rangle,f_i(l_1)=\overline{f_j(l_2)},i<j\}\nonumber
\end{eqnarray}
\begin{eqnarray}
P\sim_{phhp}^{r} \{(P_1'[f_1]\parallel\cdots\parallel P_n'[f_n]).(f_1(\alpha_1[m])\parallel\cdots\parallel f_n(\alpha_n)[m])\setminus L: \nonumber\\
\langle P_i,s_i\rangle\rightsquigarrow\xtworightarrow{\alpha_i[m]}\langle P_i',s_i'\rangle,i\in\{1,\cdots,n\},f_i(\alpha_i)\notin L\cup\overline{L}\} \nonumber\\
+\sum\{(P_1[f_1]\parallel\cdots\parallel P_i'[f_i]\parallel\cdots\parallel P_j'[f_j]\parallel\cdots\parallel P_n[f_n]).\tau\setminus L: \nonumber\\
\langle P_i,s_i\rangle\rightsquigarrow\xtworightarrow{l_1[m]}\langle P_i',s_i'\rangle,\langle P_j,s_j\rangle\rightsquigarrow\xtworightarrow{l_2[m]}\langle P_j',s_j'\rangle,f_i(l_1)=\overline{f_j(l_2)},i<j\}\nonumber
\end{eqnarray}
\end{proposition}

\begin{proof}
(1) The case of forward strongly probabilistic hhp-bisimulation.

Firstly, we consider the case without Restriction and Relabeling. That is, we suffice to prove the following case by induction on the size $n$.

For $P\equiv P_1\parallel\cdots\parallel P_n$, with $n\geq 1$, we need to prove

\begin{eqnarray}
P\sim_{phhp} \{(\alpha_1\parallel\cdots\parallel \alpha_n).(P_1'\parallel\cdots\parallel P_n'): \langle P_i,s_i\rangle\rightsquigarrow\xrightarrow{\alpha_i}\langle P_i',s_i'\rangle,i\in\{1,\cdots,n\}\nonumber\\
+\sum\{\tau.(P_1\parallel\cdots\parallel P_i'\parallel\cdots\parallel P_j'\parallel\cdots\parallel P_n): \langle P_i,s_i\rangle\rightsquigarrow\xrightarrow{l}\langle P_i',s_i'\rangle,\langle P_j,s_j\rangle\rightsquigarrow\xrightarrow{\overline{l}}\langle P_j',s_j'\rangle,i<j\} \nonumber
\end{eqnarray}

For $n=1$, $P_1\sim_{phhp}^{f} \alpha_1.P_1':\langle P_1,s_1\rangle\rightsquigarrow\xrightarrow{\alpha_1}\langle P_1',s_1'\rangle$ is obvious. Then with a hypothesis $n$, we consider
$R\equiv P\parallel P_{n+1}$. By the forward transition rules of Composition, we can get

\begin{eqnarray}
R\sim_{phhp}^{f} \{(p\parallel \alpha_{n+1}).(P'\parallel P_{n+1}'): \langle P,s\rangle\rightsquigarrow\xrightarrow{p}\langle P',s'\rangle,\langle P_{n+1},s_{n+1}\rangle\rightsquigarrow\xrightarrow{\alpha_{n+1}}\langle P_{n+1}',s_{n+1}'\rangle,p\subseteq P\}\nonumber\\
+\sum\{\tau.(P'\parallel P_{n+1}'): \langle P,s\rangle\rightsquigarrow\xrightarrow{l}\langle P',s'\rangle,\langle P_{n+1},s_{n+1}\rangle\rightsquigarrow\xrightarrow{\overline{l}}\langle P_{n+1}',s_{n+1}'\rangle\} \nonumber
\end{eqnarray}

Now with the induction assumption $P\equiv P_1\parallel\cdots\parallel P_n$, the right-hand side can be reformulated as follows.

\begin{eqnarray}
\{(\alpha_1\parallel\cdots\parallel \alpha_n\parallel \alpha_{n+1}).(P_1'\parallel\cdots\parallel P_n'\parallel P_{n+1}'): \nonumber\\
\langle P_i,s_i\rangle\rightsquigarrow\xrightarrow{\alpha_i}\langle P_i',s_i'\rangle,i\in\{1,\cdots,n+1\}\nonumber\\
+\sum\{\tau.(P_1\parallel\cdots\parallel P_i'\parallel\cdots\parallel P_j'\parallel\cdots\parallel P_n\parallel P_{n+1}): \nonumber\\
\langle P_i,s_i\rangle\rightsquigarrow\xrightarrow{l}\langle P_i',s_i'\rangle,\langle P_j,s_j\rangle\rightsquigarrow\xrightarrow{\overline{l}}\langle P_j',s_j'\rangle,i<j\} \nonumber\\
+\sum\{\tau.(P_1\parallel\cdots\parallel P_i'\parallel\cdots\parallel P_j\parallel\cdots\parallel P_n\parallel P_{n+1}'): \nonumber\\
\langle P_i,s_i\rangle\rightsquigarrow\xrightarrow{l}\langle P_i',s_i'\rangle,\langle P_{n+1},s_{n+1}\rangle\rightsquigarrow\xrightarrow{\overline{l}}\langle P_{n+1}',s_{n+1}'\rangle,i\in\{1,\cdots, n\}\} \nonumber
\end{eqnarray}

So,

\begin{eqnarray}
R\sim_{phhp}^{f} \{(\alpha_1\parallel\cdots\parallel \alpha_n\parallel \alpha_{n+1}).(P_1'\parallel\cdots\parallel P_n'\parallel P_{n+1}'): \nonumber\\
\langle P_i,s_i\rangle\rightsquigarrow\xrightarrow{\alpha_i}\langle P_i',s_i'\rangle,i\in\{1,\cdots,n+1\}\nonumber\\
+\sum\{\tau.(P_1\parallel\cdots\parallel P_i'\parallel\cdots\parallel P_j'\parallel\cdots\parallel P_n): \nonumber\\
\langle P_i,s_i\rangle\rightsquigarrow\xrightarrow{l}\langle P_i',s_i'\rangle,\langle P_j,s_j\rangle\rightsquigarrow\xrightarrow{\overline{l}}\langle P_j',s_j'\rangle,1 \leq i<j\geq n+1\} \nonumber
\end{eqnarray}

Then, we can easily add the full conditions with Restriction and Relabeling.

(2) The case of reverse strongly probabilistic hhp-bisimulation.

Firstly, we consider the case without Restriction and Relabeling. That is, we suffice to prove the following case by induction on the size $n$.

For $P\equiv P_1\parallel\cdots\parallel P_n$, with $n\geq 1$, we need to prove

\begin{eqnarray}
P\sim_{phhp}^{r} \{(P_1'\parallel\cdots\parallel P_n').(\alpha_1[m]\parallel\cdots\parallel \alpha_n[m]): \langle P_i,s_i\rangle\rightsquigarrow\xtworightarrow{\alpha_i[m]}\langle P_i',s_i'\rangle,i\in\{1,\cdots,n\}\nonumber\\
+\sum\{(P_1\parallel\cdots\parallel P_i'\parallel\cdots\parallel P_j'\parallel\cdots\parallel P_n).\tau: \langle P_i,s_i\rangle\rightsquigarrow\xtworightarrow{l[m]}\langle P_i',s_i'\rangle,\langle P_j,s_j\rangle\rightsquigarrow\xtworightarrow{\overline{l}[m]}\langle P_j',s_j'\rangle,i<j\} \nonumber
\end{eqnarray}

For $n=1$, $P_1\sim_{phhp}^{r} P_1'.\alpha_1[m]:\langle P_1,s_1\rangle\rightsquigarrow\xtworightarrow{\alpha_1[m]}\langle P_1',s_1'\rangle$ is obvious. Then with a hypothesis $n$, we consider
$R\equiv P\parallel P_{n+1}$. By the reverse transition rules of Composition, we can get

\begin{eqnarray}
R\sim_{phhp}^{r} \{(P'\parallel P_{n+1}').(p[m]\parallel \alpha_{n+1}[m]): \langle P,s\rangle\rightsquigarrow\xtworightarrow{p[m]}\langle P',s'\rangle,\langle P_{n+1},s_{n+1}\rangle\rightsquigarrow\xtworightarrow{\alpha_{n+1}[m]}\langle P_{n+1}',s_{n+1}'\rangle,p\subseteq P\}\nonumber\\
+\sum\{(P'\parallel P_{n+1}').\tau: \langle P,s\rangle\rightsquigarrow\xtworightarrow{l[m]}\langle P',s'\rangle,\langle P_{n+1},s_{n+1}\rangle\rightsquigarrow\xtworightarrow{\overline{l}[m]}\langle P_{n+1}',s_{n+1}'\rangle\} \nonumber
\end{eqnarray}

Now with the induction assumption $P\equiv P_1\parallel\cdots\parallel P_n$, the right-hand side can be reformulated as follows.

\begin{eqnarray}
\{(P_1'\parallel\cdots\parallel P_n'\parallel P_{n+1}).(\alpha_1[m]\parallel\cdots\parallel \alpha_n[m]\parallel \alpha_{n+1}[m]): \nonumber\\
\langle P_i,s_i\rangle\rightsquigarrow\xtworightarrow{\alpha_i[m]}\langle P_i',s_i'\rangle,i\in\{1,\cdots,n+1\}\nonumber\\
+\sum\{(P_1\parallel\cdots\parallel P_i'\parallel\cdots\parallel P_j'\parallel\cdots\parallel P_n\parallel P_{n+1}).\tau: \nonumber\\
\langle P_i,s_i\rangle\rightsquigarrow\xtworightarrow{l[m]}\langle P_i',s_i'\rangle,\langle P_j,s_j\rangle\rightsquigarrow\xtworightarrow{\overline{l}[m]}\langle P_j',s_j'\rangle,i<j\} \nonumber\\
+\sum\{(P_1\parallel\cdots\parallel P_i'\parallel\cdots\parallel P_j'\parallel\cdots\parallel P_n\parallel P_{n+1}).\tau: \nonumber\\
\langle P_i,s_i\rangle\rightsquigarrow\xtworightarrow{l[m]}\langle P_i',s_i'\rangle,\langle P_{n+1},s_{n+1}\rangle\rightsquigarrow\xtworightarrow{\overline{l}[m]}\langle P_{n+1}',s_{n+1}'\rangle,i\in\{1,\cdots, n\}\} \nonumber
\end{eqnarray}

So,

\begin{eqnarray}
R\sim_{phhp}^{r} \{(P_1'\parallel\cdots\parallel P_n'\parallel P_{n+1}').(\alpha_1[m]\parallel\cdots\parallel \alpha_n[m]\parallel \alpha_{n+1}[m]): \nonumber\\
\langle P_i,s_i\rangle\rightsquigarrow\xtworightarrow{\alpha_i[m]}\langle P_i',s_i'\rangle,i\in\{1,\cdots,n+1\}\nonumber\\
+\sum\{(P_1\parallel\cdots\parallel P_i'\parallel\cdots\parallel P_j'\parallel\cdots\parallel P_n).\tau: \nonumber\\
\langle P_i,s_i\rangle\rightsquigarrow\xtworightarrow{l[m]}\langle P_i',s_i'\rangle,\langle P_j,s_j\rangle\rightsquigarrow\xtworightarrow{\overline{l}[m]}\langle P_j',s_j'\rangle,1 \leq i<j\geq n+1\} \nonumber
\end{eqnarray}

Then, we can easily add the full conditions with Restriction and Relabeling.
\end{proof}

\begin{theorem}[Congruence for FR strongly probabilistic pomset bisimulation] \label{CSSB05}
We can enjoy the congruence for FR strongly probabilistic pomset bisimulation as follows.
\begin{enumerate}
  \item If $A\overset{\text{def}}{=}P$, then $A\sim_{pp}^{fr} P$;
  \item Let $P_1\sim_{pp}^{fr} P_2$. Then
        \begin{enumerate}
           \item $\alpha.P_1\sim_{pp}^f \alpha.P_2$;
           \item $\phi.P_1\sim_{pp}^f \phi.P_2$;
           \item $(\alpha_1\parallel\cdots\parallel\alpha_n).P_1\sim_{pp}^f (\alpha_1\parallel\cdots\parallel\alpha_n).P_2$;
           \item $P_1.\alpha[m]\sim_{pp}^r P_2.\alpha[m]$;
           \item $P_1.\phi\sim_{pp}^r P_2.\phi$;
           \item $P_1.(\alpha_1[m]\parallel\cdots\parallel\alpha_n[m])\sim_{pp}^r P_2.(\alpha_1[m]\parallel\cdots\parallel\alpha_n[m])$;
           \item $P_1+Q\sim_{pp}^{fr} P_2 +Q$;
           \item $P_1\boxplus_{\pi}Q\sim_{pp}^{fr} P_2 \boxplus_{\pi}Q$;
           \item $P_1\parallel Q\sim_{pp}^{fr} P_2\parallel Q$;
           \item $P_1\setminus L\sim_{pp}^{fr} P_2\setminus L$;
           \item $P_1[f]\sim_{pp}^{fr} P_2[f]$.
         \end{enumerate}
\end{enumerate}
\end{theorem}

\begin{proof}
\begin{enumerate}
  \item If $A\overset{\text{def}}{=}P$, then $A\sim_{pp}^{fr} P$. It is obvious.
  \item Let $P_1\sim_{pp}^{fr} P_2$. Then
        \begin{enumerate}
           \item $\alpha.P_1\sim_{pp}^f \alpha.P_2$. It is sufficient to prove the relation $R=\{(\alpha.P_1, \alpha.P_2)\}\cup \textbf{Id}$ is a F strongly probabilistic pomset bisimulation, we omit it;
           \item $\phi.P_1\sim_{pp}^f \phi.P_2$. It is sufficient to prove the relation $R=\{(\phi.P_1, \phi.P_2)\}\cup \textbf{Id}$ is a F strongly probabilistic pomset bisimulation, we omit it;
           \item $(\alpha_1\parallel\cdots\parallel\alpha_n).P_1\sim_{pp}^f (\alpha_1\parallel\cdots\parallel\alpha_n).P_2$. It is sufficient to prove the relation $R=\{((\alpha_1\parallel\cdots\parallel\alpha_n).P_1, (\alpha_1\parallel\cdots\parallel\alpha_n).P_2)\}\cup \textbf{Id}$ is a F strongly probabilistic pomset bisimulation, we omit it;
           \item $P_1.\alpha[m]\sim_{pp}^r P_2.\alpha[m]$. It is sufficient to prove the relation $R=\{(P_1.\alpha[m], P_2.\alpha[m])\}\cup \textbf{Id}$ is a R strongly probabilistic pomset bisimulation, we omit it;
           \item $P_1.\phi\sim_{pp}^r P_2.\phi$. It is sufficient to prove the relation $R=\{(P_1.\phi, P_2.\phi)\}\cup \textbf{Id}$ is a R strongly probabilistic pomset bisimulation, we omit it;
           \item $P_1.(\alpha_1[m]\parallel\cdots\parallel\alpha_n[m])\sim_{pp}^r P_2.(\alpha_1[m]\parallel\cdots\parallel\alpha_n[m])$. It is sufficient to prove the relation $R=\{(P_1.(\alpha_1[m]\parallel\cdots\parallel\alpha_n[m]), P_2.(\alpha_1[m]\parallel\cdots\parallel\alpha_n[m]))\}\cup \textbf{Id}$ is a R strongly probabilistic pomset bisimulation, we omit it;
           \item $P_1+Q\sim_{pp}^{fr} P_2 +Q$. It is sufficient to prove the relation $R=\{(P_1+Q, P_2+Q)\}\cup \textbf{Id}$ is a FR strongly probabilistic pomset bisimulation, we omit it;
           \item $P_1\boxplus_{\pi}Q\sim_{pp}^{fr} P_2 \boxplus_{\pi}Q$. It is sufficient to prove the relation $R=\{(P_1\boxplus_{\pi}Q, P_2\boxplus_{\pi}Q)\}\cup \textbf{Id}$ is a FR strongly probabilistic pomset bisimulation, we omit it;
           \item $P_1\parallel Q\sim_{pp}^{fr} P_2\parallel Q$. It is sufficient to prove the relation $R=\{(P_1\parallel Q, P_2\parallel Q)\}\cup \textbf{Id}$ is a FR strongly probabilistic pomset bisimulation, we omit it;
           \item $P_1\setminus L\sim_{pp}^{fr} P_2\setminus L$. It is sufficient to prove the relation $R=\{(P_1\setminus L, P_2\setminus L)\}\cup \textbf{Id}$ is a FR strongly probabilistic pomset bisimulation, we omit it;
           \item $P_1[f]\sim_{pp}^{fr} P_2[f]$. It is sufficient to prove the relation $R=\{(P_1[f], P_2[f])\}\cup \textbf{Id}$ is a FR strongly probabilistic pomset bisimulation, we omit it.
         \end{enumerate}
\end{enumerate}
\end{proof}

\begin{theorem}[Congruence for FR strongly probabilistic step bisimulation] \label{CSSB05}
We can enjoy the congruence for FR strongly probabilistic step bisimulation as follows.
\begin{enumerate}
  \item If $A\overset{\text{def}}{=}P$, then $A\sim_{ps}^{fr} P$;
  \item Let $P_1\sim_{ps}^{fr} P_2$. Then
        \begin{enumerate}
           \item $\alpha.P_1\sim_{ps}^f \alpha.P_2$;
           \item $\phi.P_1\sim_{ps}^f \phi.P_2$;
           \item $(\alpha_1\parallel\cdots\parallel\alpha_n).P_1\sim_{ps}^f (\alpha_1\parallel\cdots\parallel\alpha_n).P_2$;
           \item $P_1.\alpha[m]\sim_{ps}^r P_2.\alpha[m]$;
           \item $P_1.\phi\sim_{ps}^r P_2.\phi$;
           \item $P_1.(\alpha_1[m]\parallel\cdots\parallel\alpha_n[m])\sim_{ps}^r P_2.(\alpha_1[m]\parallel\cdots\parallel\alpha_n[m])$;
           \item $P_1+Q\sim_{ps}^{fr} P_2 +Q$;
           \item $P_1\boxplus_{\pi}Q\sim_{ps}^{fr} P_2 \boxplus_{\pi}Q$;
           \item $P_1\parallel Q\sim_{ps}^{fr} P_2\parallel Q$;
           \item $P_1\setminus L\sim_{ps}^{fr} P_2\setminus L$;
           \item $P_1[f]\sim_{ps}^{fr} P_2[f]$.
         \end{enumerate}
\end{enumerate}
\end{theorem}

\begin{proof}
\begin{enumerate}
  \item If $A\overset{\text{def}}{=}P$, then $A\sim_{ps}^{fr} P$. It is obvious.
  \item Let $P_1\sim_{ps}^{fr} P_2$. Then
        \begin{enumerate}
           \item $\alpha.P_1\sim_{ps}^f \alpha.P_2$. It is sufficient to prove the relation $R=\{(\alpha.P_1, \alpha.P_2)\}\cup \textbf{Id}$ is a F strongly probabilistic step bisimulation, we omit it;
           \item $\phi.P_1\sim_{ps}^f \phi.P_2$. It is sufficient to prove the relation $R=\{(\phi.P_1, \phi.P_2)\}\cup \textbf{Id}$ is a F strongly probabilistic step bisimulation, we omit it;
           \item $(\alpha_1\parallel\cdots\parallel\alpha_n).P_1\sim_{ps}^f (\alpha_1\parallel\cdots\parallel\alpha_n).P_2$. It is sufficient to prove the relation $R=\{((\alpha_1\parallel\cdots\parallel\alpha_n).P_1, (\alpha_1\parallel\cdots\parallel\alpha_n).P_2)\}\cup \textbf{Id}$ is a F strongly probabilistic step bisimulation, we omit it;
           \item $P_1.\alpha[m]\sim_{ps}^r P_2.\alpha[m]$. It is sufficient to prove the relation $R=\{(P_1.\alpha[m], P_2.\alpha[m])\}\cup \textbf{Id}$ is a R strongly probabilistic step bisimulation, we omit it;
           \item $P_1.\phi\sim_{ps}^r P_2.\phi$. It is sufficient to prove the relation $R=\{(P_1.\phi, P_2.\phi)\}\cup \textbf{Id}$ is a R strongly probabilistic step bisimulation, we omit it;
           \item $P_1.(\alpha_1[m]\parallel\cdots\parallel\alpha_n[m])\sim_{ps}^r P_2.(\alpha_1[m]\parallel\cdots\parallel\alpha_n[m])$. It is sufficient to prove the relation $R=\{(P_1.(\alpha_1[m]\parallel\cdots\parallel\alpha_n[m]), P_2.(\alpha_1[m]\parallel\cdots\parallel\alpha_n[m]))\}\cup \textbf{Id}$ is a R strongly probabilistic step bisimulation, we omit it;
           \item $P_1+Q\sim_{ps}^{fr} P_2 +Q$. It is sufficient to prove the relation $R=\{(P_1+Q, P_2+Q)\}\cup \textbf{Id}$ is a FR strongly probabilistic step bisimulation, we omit it;
           \item $P_1\boxplus_{\pi}Q\sim_{ps}^{fr} P_2 \boxplus_{\pi}Q$. It is sufficient to prove the relation $R=\{(P_1\boxplus_{\pi}Q, P_2\boxplus_{\pi}Q)\}\cup \textbf{Id}$ is a FR strongly probabilistic step bisimulation, we omit it;
           \item $P_1\parallel Q\sim_{ps}^{fr} P_2\parallel Q$. It is sufficient to prove the relation $R=\{(P_1\parallel Q, P_2\parallel Q)\}\cup \textbf{Id}$ is a FR strongly probabilistic step bisimulation, we omit it;
           \item $P_1\setminus L\sim_{ps}^{fr} P_2\setminus L$. It is sufficient to prove the relation $R=\{(P_1\setminus L, P_2\setminus L)\}\cup \textbf{Id}$ is a FR strongly probabilistic step bisimulation, we omit it;
           \item $P_1[f]\sim_{ps}^{fr} P_2[f]$. It is sufficient to prove the relation $R=\{(P_1[f], P_2[f])\}\cup \textbf{Id}$ is a FR strongly probabilistic step bisimulation, we omit it.
         \end{enumerate}
\end{enumerate}
\end{proof}

\begin{theorem}[Congruence for FR strongly probabilistic hp-bisimulation] \label{CSSB05}
We can enjoy the congruence for FR strongly probabilistic hp-bisimulation as follows.
\begin{enumerate}
  \item If $A\overset{\text{def}}{=}P$, then $A\sim_{php}^{fr} P$;
  \item Let $P_1\sim_{php}^{fr} P_2$. Then
        \begin{enumerate}
           \item $\alpha.P_1\sim_{php}^f \alpha.P_2$;
           \item $\phi.P_1\sim_{php}^f \phi.P_2$;
           \item $(\alpha_1\parallel\cdots\parallel\alpha_n).P_1\sim_{php}^f (\alpha_1\parallel\cdots\parallel\alpha_n).P_2$;
           \item $P_1.\alpha[m]\sim_{php}^r P_2.\alpha[m]$;
           \item $P_1.\phi\sim_{php}^r P_2.\phi$;
           \item $P_1.(\alpha_1[m]\parallel\cdots\parallel\alpha_n[m])\sim_{php}^r P_2.(\alpha_1[m]\parallel\cdots\parallel\alpha_n[m])$;
           \item $P_1+Q\sim_{php}^{fr} P_2 +Q$;
           \item $P_1\boxplus_{\pi}Q\sim_{php}^{fr} P_2 \boxplus_{\pi}Q$;
           \item $P_1\parallel Q\sim_{php}^{fr} P_2\parallel Q$;
           \item $P_1\setminus L\sim_{php}^{fr} P_2\setminus L$;
           \item $P_1[f]\sim_{php}^{fr} P_2[f]$.
         \end{enumerate}
\end{enumerate}
\end{theorem}

\begin{proof}
\begin{enumerate}
  \item If $A\overset{\text{def}}{=}P$, then $A\sim_{php}^{fr} P$. It is obvious.
  \item Let $P_1\sim_{php}^{fr} P_2$. Then
        \begin{enumerate}
           \item $\alpha.P_1\sim_{php}^f \alpha.P_2$. It is sufficient to prove the relation $R=\{(\alpha.P_1, \alpha.P_2)\}\cup \textbf{Id}$ is a F strongly probabilistic hp-bisimulation, we omit it;
           \item $\phi.P_1\sim_{php}^f \phi.P_2$. It is sufficient to prove the relation $R=\{(\phi.P_1, \phi.P_2)\}\cup \textbf{Id}$ is a F strongly probabilistic hp-bisimulation, we omit it;
           \item $(\alpha_1\parallel\cdots\parallel\alpha_n).P_1\sim_{php}^f (\alpha_1\parallel\cdots\parallel\alpha_n).P_2$. It is sufficient to prove the relation $R=\{((\alpha_1\parallel\cdots\parallel\alpha_n).P_1, (\alpha_1\parallel\cdots\parallel\alpha_n).P_2)\}\cup \textbf{Id}$ is a F strongly probabilistic hp-bisimulation, we omit it;
           \item $P_1.\alpha[m]\sim_{php}^r P_2.\alpha[m]$. It is sufficient to prove the relation $R=\{(P_1.\alpha[m], P_2.\alpha[m])\}\cup \textbf{Id}$ is a R strongly probabilistic hp-bisimulation, we omit it;
           \item $P_1.\phi\sim_{php}^r P_2.\phi$. It is sufficient to prove the relation $R=\{(P_1.\phi, P_2.\phi)\}\cup \textbf{Id}$ is a R strongly probabilistic hp-bisimulation, we omit it;
           \item $P_1.(\alpha_1[m]\parallel\cdots\parallel\alpha_n[m])\sim_{php}^r P_2.(\alpha_1[m]\parallel\cdots\parallel\alpha_n[m])$. It is sufficient to prove the relation $R=\{(P_1.(\alpha_1[m]\parallel\cdots\parallel\alpha_n[m]), P_2.(\alpha_1[m]\parallel\cdots\parallel\alpha_n[m]))\}\cup \textbf{Id}$ is a R strongly probabilistic hp-bisimulation, we omit it;
           \item $P_1+Q\sim_{php}^{fr} P_2 +Q$. It is sufficient to prove the relation $R=\{(P_1+Q, P_2+Q)\}\cup \textbf{Id}$ is a FR strongly probabilistic hp-bisimulation, we omit it;
           \item $P_1\boxplus_{\pi}Q\sim_{php}^{fr} P_2 \boxplus_{\pi}Q$. It is sufficient to prove the relation $R=\{(P_1\boxplus_{\pi}Q, P_2\boxplus_{\pi}Q)\}\cup \textbf{Id}$ is a FR strongly probabilistic hp-bisimulation, we omit it;
           \item $P_1\parallel Q\sim_{php}^{fr} P_2\parallel Q$. It is sufficient to prove the relation $R=\{(P_1\parallel Q, P_2\parallel Q)\}\cup \textbf{Id}$ is a FR strongly probabilistic hp-bisimulation, we omit it;
           \item $P_1\setminus L\sim_{php}^{fr} P_2\setminus L$. It is sufficient to prove the relation $R=\{(P_1\setminus L, P_2\setminus L)\}\cup \textbf{Id}$ is a FR strongly probabilistic hp-bisimulation, we omit it;
           \item $P_1[f]\sim_{php}^{fr} P_2[f]$. It is sufficient to prove the relation $R=\{(P_1[f], P_2[f])\}\cup \textbf{Id}$ is a FR strongly probabilistic hp-bisimulation, we omit it.
         \end{enumerate}
\end{enumerate}
\end{proof}

\begin{theorem}[Congruence for FR strongly probabilistic hhp-bisimulation] \label{CSSB05}
We can enjoy the congruence for FR strongly probabilistic hhp-bisimulation as follows.
\begin{enumerate}
  \item If $A\overset{\text{def}}{=}P$, then $A\sim_{phhp}^{fr} P$;
  \item Let $P_1\sim_{phhp}^{fr} P_2$. Then
        \begin{enumerate}
           \item $\alpha.P_1\sim_{phhp}^f \alpha.P_2$;
           \item $\phi.P_1\sim_{phhp}^f \phi.P_2$;
           \item $(\alpha_1\parallel\cdots\parallel\alpha_n).P_1\sim_{phhp}^f (\alpha_1\parallel\cdots\parallel\alpha_n).P_2$;
           \item $P_1.\alpha[m]\sim_{phhp}^r P_2.\alpha[m]$;
           \item $P_1.\phi\sim_{phhp}^r P_2.\phi$;
           \item $P_1.(\alpha_1[m]\parallel\cdots\parallel\alpha_n[m])\sim_{phhp}^r P_2.(\alpha_1[m]\parallel\cdots\parallel\alpha_n[m])$;
           \item $P_1+Q\sim_{phhp}^{fr} P_2 +Q$;
           \item $P_1\boxplus_{\pi}Q\sim_{phhp}^{fr} P_2 \boxplus_{\pi}Q$;
           \item $P_1\parallel Q\sim_{phhp}^{fr} P_2\parallel Q$;
           \item $P_1\setminus L\sim_{phhp}^{fr} P_2\setminus L$;
           \item $P_1[f]\sim_{phhp}^{fr} P_2[f]$.
         \end{enumerate}
\end{enumerate}
\end{theorem}

\begin{proof}
\begin{enumerate}
  \item If $A\overset{\text{def}}{=}P$, then $A\sim_{phhp}^{fr} P$. It is obvious.
  \item Let $P_1\sim_{phhp}^{fr} P_2$. Then
        \begin{enumerate}
           \item $\alpha.P_1\sim_{phhp}^f \alpha.P_2$. It is sufficient to prove the relation $R=\{(\alpha.P_1, \alpha.P_2)\}\cup \textbf{Id}$ is a F strongly probabilistic hhp-bisimulation, we omit it;
           \item $\phi.P_1\sim_{phhp}^f \phi.P_2$. It is sufficient to prove the relation $R=\{(\phi.P_1, \phi.P_2)\}\cup \textbf{Id}$ is a F strongly probabilistic hhp-bisimulation, we omit it;
           \item $(\alpha_1\parallel\cdots\parallel\alpha_n).P_1\sim_{phhp}^f (\alpha_1\parallel\cdots\parallel\alpha_n).P_2$. It is sufficient to prove the relation $R=\{((\alpha_1\parallel\cdots\parallel\alpha_n).P_1, (\alpha_1\parallel\cdots\parallel\alpha_n).P_2)\}\cup \textbf{Id}$ is a F strongly probabilistic hhp-bisimulation, we omit it;
           \item $P_1.\alpha[m]\sim_{phhp}^r P_2.\alpha[m]$. It is sufficient to prove the relation $R=\{(P_1.\alpha[m], P_2.\alpha[m])\}\cup \textbf{Id}$ is a R strongly probabilistic hhp-bisimulation, we omit it;
           \item $P_1.\phi\sim_{phhp}^r P_2.\phi$. It is sufficient to prove the relation $R=\{(P_1.\phi, P_2.\phi)\}\cup \textbf{Id}$ is a R strongly probabilistic hhp-bisimulation, we omit it;
           \item $P_1.(\alpha_1[m]\parallel\cdots\parallel\alpha_n[m])\sim_{phhp}^r P_2.(\alpha_1[m]\parallel\cdots\parallel\alpha_n[m])$. It is sufficient to prove the relation $R=\{(P_1.(\alpha_1[m]\parallel\cdots\parallel\alpha_n[m]), P_2.(\alpha_1[m]\parallel\cdots\parallel\alpha_n[m]))\}\cup \textbf{Id}$ is a R strongly probabilistic hhp-bisimulation, we omit it;
           \item $P_1+Q\sim_{phhp}^{fr} P_2 +Q$. It is sufficient to prove the relation $R=\{(P_1+Q, P_2+Q)\}\cup \textbf{Id}$ is a FR strongly probabilistic hhp-bisimulation, we omit it;
           \item $P_1\boxplus_{\pi}Q\sim_{phhp}^{fr} P_2 \boxplus_{\pi}Q$. It is sufficient to prove the relation $R=\{(P_1\boxplus_{\pi}Q, P_2\boxplus_{\pi}Q)\}\cup \textbf{Id}$ is a FR strongly probabilistic hhp-bisimulation, we omit it;
           \item $P_1\parallel Q\sim_{phhp}^{fr} P_2\parallel Q$. It is sufficient to prove the relation $R=\{(P_1\parallel Q, P_2\parallel Q)\}\cup \textbf{Id}$ is a FR strongly probabilistic hhp-bisimulation, we omit it;
           \item $P_1\setminus L\sim_{phhp}^{fr} P_2\setminus L$. It is sufficient to prove the relation $R=\{(P_1\setminus L, P_2\setminus L)\}\cup \textbf{Id}$ is a FR strongly probabilistic hhp-bisimulation, we omit it;
           \item $P_1[f]\sim_{phhp}^{fr} P_2[f]$. It is sufficient to prove the relation $R=\{(P_1[f], P_2[f])\}\cup \textbf{Id}$ is a FR strongly probabilistic hhp-bisimulation, we omit it.
         \end{enumerate}
\end{enumerate}
\end{proof}

\subsubsection{Recursion}

\begin{definition}[Weakly guarded recursive expression]
$X$ is weakly guarded in $E$ if each occurrence of $X$ is with some subexpression $\alpha.F$ or $(\alpha_1\parallel\cdots\parallel\alpha_n).F$ or $F.\alpha[m]$ or
$F.(\alpha_1[m]\parallel\cdots\parallel\alpha_n[m])$ of $E$.
\end{definition}

\begin{lemma}\label{LUS06}
If the variables $\widetilde{X}$ are weakly guarded in $E$, and $\langle E\{\widetilde{P}/\widetilde{X}\},s\rangle\rightsquigarrow\xrightarrow{\{\alpha_1,\cdots,\alpha_n\}}\langle P',s'\rangle$ or
$\langle E\{\widetilde{P}/\widetilde{X}\},s\rangle\rightsquigarrow\xtworightarrow{\{\alpha_1[m],\cdots,\alpha_n[m]\}}\langle P',s'\rangle$, then $P'$ takes the form
$E'\{\widetilde{P}/\widetilde{X}\}$ for some expression $E'$, and moreover, for any $\widetilde{Q}$,
$\langle E\{\widetilde{Q}/\widetilde{X}\},s\rangle\rightsquigarrow\xrightarrow{\{\alpha_1,\cdots,\alpha_n\}}\langle E'\{\widetilde{Q}/\widetilde{X}\},s'\rangle$ or
$\langle E\{\widetilde{Q}/\widetilde{X}\},s\rangle\rightsquigarrow\xtworightarrow{\{\alpha_1[m],\cdots,\alpha_n[m]\}}\langle E'\{\widetilde{Q}/\widetilde{X}\},s'\rangle$.
\end{lemma}

\begin{proof}
We only prove the case of forward transition.

It needs to induct on the depth of the inference of $\langle E\{\widetilde{P}/\widetilde{X}\},s\rangle\rightsquigarrow\xrightarrow{\{\alpha_1,\cdots,\alpha_n\}}\langle P',s'\rangle$.

\begin{enumerate}
  \item Case $E\equiv Y$, a variable. Then $Y\notin \widetilde{X}$. Since $\widetilde{X}$ are weakly guarded, $\langle Y\{\widetilde{P}/\widetilde{X}\equiv Y\},s\rangle\nrightarrow$, this case is
  impossible.
  \item Case $E\equiv\beta.F$. Then we must have $\alpha=\beta$, and $P'\equiv F\{\widetilde{P}/\widetilde{X}\}$, and
  $\langle E\{\widetilde{Q}/\widetilde{X}\},s\rangle\equiv \langle \beta.F\{\widetilde{Q}/\widetilde{X}\},s\rangle \rightsquigarrow\xrightarrow{\beta}\langle F\{\widetilde{Q}/\widetilde{X}\},s'\rangle$,
  then, let $E'$ be $F$, as desired.
  \item Case $E\equiv(\beta_1\parallel\cdots\parallel\beta_n).F$. Then we must have $\alpha_i=\beta_i$ for $1\leq i\leq n$, and $P'\equiv F\{\widetilde{P}/\widetilde{X}\}$, and
  $\langle E\{\widetilde{Q}/\widetilde{X}\},s\rangle\equiv \langle(\beta_1\parallel\cdots\parallel\beta_n).F\{\widetilde{Q}/\widetilde{X}\},s\rangle \rightsquigarrow\xrightarrow{\{\beta_1,\cdots,\beta_n\}}\langle F\{\widetilde{Q}/\widetilde{X}\},s'\rangle$,
  then, let $E'$ be $F$, as desired.
  \item Case $E\equiv E_1+E_2$. Then either $\langle E_1\{\widetilde{P}/\widetilde{X}\},s\rangle \rightsquigarrow\xrightarrow{\{\alpha_1,\cdots,\alpha_n\}}\langle P',s'\rangle$ or
  $\langle E_2\{\widetilde{P}/\widetilde{X}\},s\rangle \rightsquigarrow\xrightarrow{\{\alpha_1,\cdots,\alpha_n\}}\langle P',s'\rangle$, then, we can apply this lemma in either case, as desired.
  \item Case $E\equiv E_1\parallel E_2$. There are four possibilities.
  \begin{enumerate}
    \item We may have $\langle E_1\{\widetilde{P}/\widetilde{X}\},s\rangle \rightsquigarrow\xrightarrow{\alpha}\langle P_1',s'\rangle$ and $\langle E_2\{\widetilde{P}/\widetilde{X}\},s\rangle\nrightarrow$
    with $P'\equiv P_1'\parallel (E_2\{\widetilde{P}/\widetilde{X}\})$, then by applying this lemma, $P_1'$ is of the form $E_1'\{\widetilde{P}/\widetilde{X}\}$, and for any $Q$,
    $\langle E_1\{\widetilde{Q}/\widetilde{X}\},s\rangle\rightsquigarrow\xrightarrow{\alpha} \langle E_1'\{\widetilde{Q}/\widetilde{X}\},s'\rangle$. So, $P'$ is of the form
    $E_1'\parallel E_2\{\widetilde{P}/\widetilde{X}\}$, and for any $Q$,
    $\langle E\{\widetilde{Q}/\widetilde{X}\}\equiv E_1\{\widetilde{Q}/\widetilde{X}\}\parallel E_2\{\widetilde{Q}/\widetilde{X}\},s\rangle\rightsquigarrow\xrightarrow{\alpha} \langle(E_1'\parallel E_2)\{\widetilde{Q}/\widetilde{X}\},s'\rangle$,
    then, let $E'$ be $E_1'\parallel E_2$, as desired.
    \item We may have $\langle E_2\{\widetilde{P}/\widetilde{X}\},s\rangle \rightsquigarrow\xrightarrow{\alpha}\langle P_2',s'\rangle$ and $\langle E_1\{\widetilde{P}/\widetilde{X}\},s\rangle\nrightarrow$
    with $P'\equiv P_2'\parallel (E_1\{\widetilde{P}/\widetilde{X}\})$, this case can be prove similarly to the above subcase, as desired.
    \item We may have $\langle E_1\{\widetilde{P}/\widetilde{X}\},s\rangle \rightsquigarrow\xrightarrow{\alpha}\langle P_1',s'\rangle$ and
    $\langle E_2\{\widetilde{P}/\widetilde{X}\},s\rangle\rightsquigarrow\xrightarrow{\beta}\langle P_2',s''\rangle$ with $\alpha\neq\overline{\beta}$ and $P'\equiv P_1'\parallel P_2'$, then by
    applying this lemma, $P_1'$ is of the form $E_1'\{\widetilde{P}/\widetilde{X}\}$, and for any $Q$,
    $\langle E_1\{\widetilde{Q}/\widetilde{X}\},s\rangle\rightsquigarrow\xrightarrow{\alpha} \langle E_1'\{\widetilde{Q}/\widetilde{X}\},s'\rangle$; $P_2'$ is of the form
    $E_2'\{\widetilde{P}/\widetilde{X}\}$, and for any $Q$, $\langle E_2\{\widetilde{Q}/\widetilde{X}\},s\rangle\rightsquigarrow\xrightarrow{\alpha} \langle E_2'\{\widetilde{Q}/\widetilde{X}\},s''\rangle$.
    So, $P'$ is of the form $E_1'\parallel E_2'\{\widetilde{P}/\widetilde{X}\}$, and for any $Q$,
    $\langle E\{\widetilde{Q}/\widetilde{X}\}\equiv E_1\{\widetilde{Q}/\widetilde{X}\}\parallel E_2\{\widetilde{Q}/\widetilde{X}\},s\rangle\rightsquigarrow\xrightarrow{\{\alpha,\beta\}}
    \langle (E_1'\parallel E_2')\{\widetilde{Q}/\widetilde{X}\},s'\cup s''\rangle$, then, let $E'$ be $E_1'\parallel E_2'$, as desired.
    \item We may have $\langle E_1\{\widetilde{P}/\widetilde{X}\},s\rangle \rightsquigarrow\xrightarrow{l}\langle P_1',s'\rangle$ and
    $\langle E_2\{\widetilde{P}/\widetilde{X}\},s\rangle\rightsquigarrow\xrightarrow{\overline{l}}\langle P_2',s''\rangle$ with $P'\equiv P_1'\parallel P_2'$, then by applying this lemma,
    $P_1'$ is of the form $E_1'\{\widetilde{P}/\widetilde{X}\}$, and for any $Q$, $\langle E_1\{\widetilde{Q}/\widetilde{X}\},s\rangle\rightsquigarrow\xrightarrow{l} \langle E_1'\{\widetilde{Q}/\widetilde{X}\},s'\rangle$;
    $P_2'$ is of the form $E_2'\{\widetilde{P}/\widetilde{X}\}$, and for any $Q$, $\langle E_2\{\widetilde{Q}/\widetilde{X}\},s\rangle\rightsquigarrow\xrightarrow{\overline{l}}\langle E_2'\{\widetilde{Q}/\widetilde{X}\},s''\rangle$.
    So, $P'$ is of the form $E_1'\parallel E_2'\{\widetilde{P}/\widetilde{X}\}$, and for any $Q$, $\langle E\{\widetilde{Q}/\widetilde{X}\}\equiv E_1\{\widetilde{Q}/\widetilde{X}\}\parallel E_2\{\widetilde{Q}/\widetilde{X}\},s\rangle
    \rightsquigarrow\xrightarrow{\tau} \langle (E_1'\parallel E_2')\{\widetilde{Q}/\widetilde{X}\},s'\cup s''\rangle$, then, let $E'$ be $E_1'\parallel E_2'$, as desired.
  \end{enumerate}
  \item Case $E\equiv F[R]$ and $E\equiv F\setminus L$. These cases can be prove similarly to the above case.
  \item Case $E\equiv C$, an agent constant defined by $C\overset{\text{def}}{=}R$. Then there is no $X\in\widetilde{X}$ occurring in $E$, so
  $\langle C,s\rangle\rightsquigarrow\xrightarrow{\{\alpha_1,\cdots,\alpha_n\}}\langle P',s'\rangle$, let $E'$ be $P'$, as desired.
\end{enumerate}

For the case of reverse transition, it can be proven similarly, we omit it.
\end{proof}

\begin{theorem}[Unique solution of equations for FR strongly probabilistic pomset bisimulation]
Let the recursive expressions $E_i(i\in I)$ contain at most the variables $X_i(i\in I)$, and let each $X_j(j\in I)$ be weakly guarded in each $E_i$. Then,

If $\widetilde{P}\sim_{pp}^{fr} \widetilde{E}\{\widetilde{P}/\widetilde{X}\}$ and $\widetilde{Q}\sim_{pp}^{fr} \widetilde{E}\{\widetilde{Q}/\widetilde{X}\}$, then
$\widetilde{P}\sim_{pp}^{fr} \widetilde{Q}$.
\end{theorem}

\begin{proof}
We only prove the case of forward transition.

It is sufficient to induct on the depth of the inference of $\langle E\{\widetilde{P}/\widetilde{X}\},s\rangle\rightsquigarrow\xrightarrow{\{\alpha_1,\cdots,\alpha_n\}}\langle P',s'\rangle$.

\begin{enumerate}
  \item Case $E\equiv X_i$. Then we have $\langle E\{\widetilde{P}/\widetilde{X}\},s\rangle\equiv \langle P_i,s\rangle\rightsquigarrow\xrightarrow{\{\alpha_1,\cdots,\alpha_n\}}\langle P',s'\rangle$,
  since $P_i\sim_{pp}^{fr} E_i\{\widetilde{P}/\widetilde{X}\}$, we have $\langle E_i\{\widetilde{P}/\widetilde{X}\},s\rangle\rightsquigarrow\xrightarrow{\{\alpha_1,\cdots,\alpha_n\}}\langle P'',s'\rangle\sim_{pp}^{fr} \langle P',s'\rangle$.
  Since $\widetilde{X}$ are weakly guarded in $E_i$, by Lemma \ref{LUS06}, $P''\equiv E'\{\widetilde{P}/\widetilde{X}\}$ and $\langle E_i\{\widetilde{P}/\widetilde{X}\},s\rangle
  \rightsquigarrow\xrightarrow{\{\alpha_1,\cdots,\alpha_n\}} \langle E'\{\widetilde{P}/\widetilde{X}\},s'\rangle$. Since
  $E\{\widetilde{Q}/\widetilde{X}\}\equiv X_i\{\widetilde{Q}/\widetilde{X}\} \equiv Q_i\sim_{pp}^{fr} E_i\{\widetilde{Q}/\widetilde{X}\}$, $\langle E\{\widetilde{Q}/\widetilde{X}\},s\rangle\rightsquigarrow\xrightarrow{\{\alpha_1,\cdots,\alpha_n\}}\langle Q',s'\rangle\sim_{pp}^{fr} \langle E'\{\widetilde{Q}/\widetilde{X}\},s'\rangle$.
  So, $P'\sim_{pp}^{fr} Q'$, as desired.
  \item Case $E\equiv\alpha.F$. This case can be proven similarly.
  \item Case $E\equiv(\alpha_1\parallel\cdots\parallel\alpha_n).F$. This case can be proven similarly.
  \item Case $E\equiv E_1+E_2$. We have $\langle E_i\{\widetilde{P}/\widetilde{X}\},s\rangle \rightsquigarrow\xrightarrow{\{\alpha_1,\cdots,\alpha_n\}}\langle P',s'\rangle$,
  $\langle E_i\{\widetilde{Q}/\widetilde{X}\},s\rangle \rightsquigarrow\xrightarrow{\{\alpha_1,\cdots,\alpha_n\}}\langle Q',s'\rangle$, then, $P'\sim_{pp}^{fr} Q'$, as desired.
  \item Case $E\equiv E_1\parallel E_2$, $E\equiv F[R]$ and $E\equiv F\setminus L$, $E\equiv C$. These cases can be prove similarly to the above case.
\end{enumerate}

For the case of reverse transition, it can be proven similarly, we omit it.
\end{proof}

\begin{theorem}[Unique solution of equations for FR strongly probabilistic step bisimulation]
Let the recursive expressions $E_i(i\in I)$ contain at most the variables $X_i(i\in I)$, and let each $X_j(j\in I)$ be weakly guarded in each $E_i$. Then,

If $\widetilde{P}\sim_{ps}^{fr} \widetilde{E}\{\widetilde{P}/\widetilde{X}\}$ and $\widetilde{Q}\sim_{ps}^{fr} \widetilde{E}\{\widetilde{Q}/\widetilde{X}\}$, then
$\widetilde{P}\sim_{ps}^{fr} \widetilde{Q}$.
\end{theorem}

\begin{proof}
We only prove the case of forward transition.

It is sufficient to induct on the depth of the inference of $\langle E\{\widetilde{P}/\widetilde{X}\},s\rangle\rightsquigarrow\xrightarrow{\{\alpha_1,\cdots,\alpha_n\}}\langle P',s'\rangle$.

\begin{enumerate}
  \item Case $E\equiv X_i$. Then we have $\langle E\{\widetilde{P}/\widetilde{X}\},s\rangle\equiv \langle P_i,s\rangle\rightsquigarrow\xrightarrow{\{\alpha_1,\cdots,\alpha_n\}}\langle P',s'\rangle$,
  since $P_i\sim_{ps}^{fr} E_i\{\widetilde{P}/\widetilde{X}\}$, we have $\langle E_i\{\widetilde{P}/\widetilde{X}\},s\rangle\rightsquigarrow\xrightarrow{\{\alpha_1,\cdots,\alpha_n\}}\langle P'',s'\rangle\sim_{ps}^{fr} \langle P',s'\rangle$.
  Since $\widetilde{X}$ are weakly guarded in $E_i$, by Lemma \ref{LUS06}, $P''\equiv E'\{\widetilde{P}/\widetilde{X}\}$ and $\langle E_i\{\widetilde{P}/\widetilde{X}\},s\rangle
  \rightsquigarrow\xrightarrow{\{\alpha_1,\cdots,\alpha_n\}} \langle E'\{\widetilde{P}/\widetilde{X}\},s'\rangle$. Since
  $E\{\widetilde{Q}/\widetilde{X}\}\equiv X_i\{\widetilde{Q}/\widetilde{X}\} \equiv Q_i\sim_{ps}^{fr} E_i\{\widetilde{Q}/\widetilde{X}\}$, $\langle E\{\widetilde{Q}/\widetilde{X}\},s\rangle\rightsquigarrow\xrightarrow{\{\alpha_1,\cdots,\alpha_n\}}\langle Q',s'\rangle\sim_{ps}^{fr} \langle E'\{\widetilde{Q}/\widetilde{X}\},s'\rangle$.
  So, $P'\sim_{ps}^{fr} Q'$, as desired.
  \item Case $E\equiv\alpha.F$. This case can be proven similarly.
  \item Case $E\equiv(\alpha_1\parallel\cdots\parallel\alpha_n).F$. This case can be proven similarly.
  \item Case $E\equiv E_1+E_2$. We have $\langle E_i\{\widetilde{P}/\widetilde{X}\},s\rangle \rightsquigarrow\xrightarrow{\{\alpha_1,\cdots,\alpha_n\}}\langle P',s'\rangle$,
  $\langle E_i\{\widetilde{Q}/\widetilde{X}\},s\rangle \rightsquigarrow\xrightarrow{\{\alpha_1,\cdots,\alpha_n\}}\langle Q',s'\rangle$, then, $P'\sim_{ps}^{fr} Q'$, as desired.
  \item Case $E\equiv E_1\parallel E_2$, $E\equiv F[R]$ and $E\equiv F\setminus L$, $E\equiv C$. These cases can be prove similarly to the above case.
\end{enumerate}

For the case of reverse transition, it can be proven similarly, we omit it.
\end{proof}

\begin{theorem}[Unique solution of equations for FR strongly probabilistic hp-bisimulation]
Let the recursive expressions $E_i(i\in I)$ contain at most the variables $X_i(i\in I)$, and let each $X_j(j\in I)$ be weakly guarded in each $E_i$. Then,

If $\widetilde{P}\sim_{php}^{fr} \widetilde{E}\{\widetilde{P}/\widetilde{X}\}$ and $\widetilde{Q}\sim_{php}^{fr} \widetilde{E}\{\widetilde{Q}/\widetilde{X}\}$, then
$\widetilde{P}\sim_{php}^{fr} \widetilde{Q}$.
\end{theorem}

\begin{proof}
We only prove the case of forward transition.

It is sufficient to induct on the depth of the inference of $\langle E\{\widetilde{P}/\widetilde{X}\},s\rangle\rightsquigarrow\xrightarrow{\{\alpha_1,\cdots,\alpha_n\}}\langle P',s'\rangle$.

\begin{enumerate}
  \item Case $E\equiv X_i$. Then we have $\langle E\{\widetilde{P}/\widetilde{X}\},s\rangle\equiv \langle P_i,s\rangle\rightsquigarrow\xrightarrow{\{\alpha_1,\cdots,\alpha_n\}}\langle P',s'\rangle$,
  since $P_i\sim_{php}^{fr} E_i\{\widetilde{P}/\widetilde{X}\}$, we have $\langle E_i\{\widetilde{P}/\widetilde{X}\},s\rangle\rightsquigarrow\xrightarrow{\{\alpha_1,\cdots,\alpha_n\}}\langle P'',s'\rangle\sim_{php}^{fr} \langle P',s'\rangle$.
  Since $\widetilde{X}$ are weakly guarded in $E_i$, by Lemma \ref{LUS06}, $P''\equiv E'\{\widetilde{P}/\widetilde{X}\}$ and $\langle E_i\{\widetilde{P}/\widetilde{X}\},s\rangle
  \rightsquigarrow\xrightarrow{\{\alpha_1,\cdots,\alpha_n\}} \langle E'\{\widetilde{P}/\widetilde{X}\},s'\rangle$. Since
  $E\{\widetilde{Q}/\widetilde{X}\}\equiv X_i\{\widetilde{Q}/\widetilde{X}\} \equiv Q_i\sim_{php}^{fr} E_i\{\widetilde{Q}/\widetilde{X}\}$, $\langle E\{\widetilde{Q}/\widetilde{X}\},s\rangle\rightsquigarrow\xrightarrow{\{\alpha_1,\cdots,\alpha_n\}}\langle Q',s'\rangle\sim_{php}^{fr} \langle E'\{\widetilde{Q}/\widetilde{X}\},s'\rangle$.
  So, $P'\sim_{php}^{fr} Q'$, as desired.
  \item Case $E\equiv\alpha.F$. This case can be proven similarly.
  \item Case $E\equiv(\alpha_1\parallel\cdots\parallel\alpha_n).F$. This case can be proven similarly.
  \item Case $E\equiv E_1+E_2$. We have $\langle E_i\{\widetilde{P}/\widetilde{X}\},s\rangle \rightsquigarrow\xrightarrow{\{\alpha_1,\cdots,\alpha_n\}}\langle P',s'\rangle$,
  $\langle E_i\{\widetilde{Q}/\widetilde{X}\},s\rangle \rightsquigarrow\xrightarrow{\{\alpha_1,\cdots,\alpha_n\}}\langle Q',s'\rangle$, then, $P'\sim_{php}^{fr} Q'$, as desired.
  \item Case $E\equiv E_1\parallel E_2$, $E\equiv F[R]$ and $E\equiv F\setminus L$, $E\equiv C$. These cases can be prove similarly to the above case.
\end{enumerate}

For the case of reverse transition, it can be proven similarly, we omit it.
\end{proof}

\begin{theorem}[Unique solution of equations for FR strongly probabilistic hhp-bisimulation]
Let the recursive expressions $E_i(i\in I)$ contain at most the variables $X_i(i\in I)$, and let each $X_j(j\in I)$ be weakly guarded in each $E_i$. Then,

If $\widetilde{P}\sim_{phhp}^{fr} \widetilde{E}\{\widetilde{P}/\widetilde{X}\}$ and $\widetilde{Q}\sim_{phhp}^{fr} \widetilde{E}\{\widetilde{Q}/\widetilde{X}\}$, then
$\widetilde{P}\sim_{phhp}^{fr} \widetilde{Q}$.
\end{theorem}

\begin{proof}
We only prove the case of forward transition.

It is sufficient to induct on the depth of the inference of $\langle E\{\widetilde{P}/\widetilde{X}\},s\rangle\rightsquigarrow\xrightarrow{\{\alpha_1,\cdots,\alpha_n\}}\langle P',s'\rangle$.

\begin{enumerate}
  \item Case $E\equiv X_i$. Then we have $\langle E\{\widetilde{P}/\widetilde{X}\},s\rangle\equiv \langle P_i,s\rangle\rightsquigarrow\xrightarrow{\{\alpha_1,\cdots,\alpha_n\}}\langle P',s'\rangle$,
  since $P_i\sim_{phhp}^{fr} E_i\{\widetilde{P}/\widetilde{X}\}$, we have $\langle E_i\{\widetilde{P}/\widetilde{X}\},s\rangle\rightsquigarrow\xrightarrow{\{\alpha_1,\cdots,\alpha_n\}}\langle P'',s'\rangle\sim_{phhp}^{fr} \langle P',s'\rangle$.
  Since $\widetilde{X}$ are weakly guarded in $E_i$, by Lemma \ref{LUS06}, $P''\equiv E'\{\widetilde{P}/\widetilde{X}\}$ and $\langle E_i\{\widetilde{P}/\widetilde{X}\},s\rangle
  \rightsquigarrow\xrightarrow{\{\alpha_1,\cdots,\alpha_n\}} \langle E'\{\widetilde{P}/\widetilde{X}\},s'\rangle$. Since
  $E\{\widetilde{Q}/\widetilde{X}\}\equiv X_i\{\widetilde{Q}/\widetilde{X}\} \equiv Q_i\sim_{phhp}^{fr} E_i\{\widetilde{Q}/\widetilde{X}\}$, $\langle E\{\widetilde{Q}/\widetilde{X}\},s\rangle\rightsquigarrow\xrightarrow{\{\alpha_1,\cdots,\alpha_n\}}\langle Q',s'\rangle\sim_{phhp}^{fr} \langle E'\{\widetilde{Q}/\widetilde{X}\},s'\rangle$.
  So, $P'\sim_{phhp}^{fr} Q'$, as desired.
  \item Case $E\equiv\alpha.F$. This case can be proven similarly.
  \item Case $E\equiv(\alpha_1\parallel\cdots\parallel\alpha_n).F$. This case can be proven similarly.
  \item Case $E\equiv E_1+E_2$. We have $\langle E_i\{\widetilde{P}/\widetilde{X}\},s\rangle \rightsquigarrow\xrightarrow{\{\alpha_1,\cdots,\alpha_n\}}\langle P',s'\rangle$,
  $\langle E_i\{\widetilde{Q}/\widetilde{X}\},s\rangle \rightsquigarrow\xrightarrow{\{\alpha_1,\cdots,\alpha_n\}}\langle Q',s'\rangle$, then, $P'\sim_{phhp}^{fr} Q'$, as desired.
  \item Case $E\equiv E_1\parallel E_2$, $E\equiv F[R]$ and $E\equiv F\setminus L$, $E\equiv C$. These cases can be prove similarly to the above case.
\end{enumerate}

For the case of reverse transition, it can be proven similarly, we omit it.
\end{proof}

\subsection{Weak Bisimulations}\label{wftcbpa}

\subsubsection{Laws}

Remembering that $\tau$ can neither be restricted nor relabeled, we know that the monoid laws, the static laws, the guards laws, and the new expansion law
still hold with respect to the corresponding FR weakly probabilistic truly concurrent bisimulations. And also, we can enjoy the congruence of Prefix, Summation, Composition, Restriction, Relabelling
and Constants with respect to corresponding FR weakly probabilistic truly concurrent bisimulations. We will not retype these laws, and just give the $\tau$-specific laws. The forward and reverse
transition rules of $\tau$ are shown in Table \ref{TRForTAU07}, where $\rightsquigarrow\xrightarrow{\tau}\surd$ is a predicate which represents a successful termination after execution of the silent
step $\tau$.

\begin{center}
    \begin{table}
        $$\frac{}{\langle \tau,s\rangle\rightsquigarrow\xrightarrow{\tau}\langle\surd,\tau(s)\rangle}$$
        $$\frac{}{\langle \tau,s\rangle\rightsquigarrow\xtworightarrow{\tau}\langle\surd,\tau(s)\rangle}$$
        \caption{Forward and reverse transition rules of $\tau$}
        \label{TRForTAU07}
    \end{table}
\end{center}

\begin{proposition}[$\tau$ laws for FR weakly probabilistic pomset bisimulation]
The $\tau$ laws for FR weakly probabilistic pomset bisimulation is as follows.
\begin{enumerate}
  \item $P\approx_{pp}^f \tau.P$;
  \item $P\approx_{pp}^r P.\tau$;
  \item $\alpha.\tau.P\approx_{pp}^f \alpha.P$;
  \item $P.\tau.\alpha[m]\approx_{pp}^r P.\alpha[m]$;
  \item $(\alpha_1\parallel\cdots\parallel\alpha_n).\tau.P\approx_{pp}^f (\alpha_1\parallel\cdots\parallel\alpha_n).P$;
  \item $P.\tau.(\alpha_1[m]\parallel\cdots\parallel\alpha_n[m])\approx_{pp}^r P.(\alpha_1[m]\parallel\cdots\parallel\alpha_n[m])$;
  \item $P+\tau.P\approx_{pp}^f \tau.P$;
  \item $P+P.\tau\approx_{pp}^r P.\tau$;
  \item $P\cdot((Q+\tau\cdot(Q+R))\boxplus_{\pi}S)\approx_{pp}^{f}P\cdot((Q+R)\boxplus_{\pi}S)$;
  \item $((Q+(Q+R)\cdot\tau)\boxplus_{\pi}S)\cdot P\approx_{pp}^{r}((Q+R)\boxplus_{\pi}S)\cdot P$;
  \item $P\approx_{pp}^{fr} \tau\parallel P$.
\end{enumerate}
\end{proposition}

\begin{proof}
\begin{enumerate}
  \item $P\approx_{pp}^f \tau.P$. It is sufficient to prove the relation $R=\{(P, \tau.P)\}\cup \textbf{Id}$ is a F weakly probabilistic pomset bisimulation, we omit it;
  \item $P\approx_{pp}^r P.\tau$. It is sufficient to prove the relation $R=\{(P, P.\tau)\}\cup \textbf{Id}$ is a R weakly probabilistic pomset bisimulation, we omit it;
  \item $\alpha.\tau.P\approx_{pp}^f \alpha.P$. It is sufficient to prove the relation $R=\{(\alpha.\tau.P, \alpha.P)\}\cup \textbf{Id}$ is a F weakly probabilistic pomset bisimulation, we omit it;
  \item $P.\tau.\alpha[m]\approx_{pp}^r P.\alpha[m]$. It is sufficient to prove the relation $R=\{(P.\tau.\alpha[m], P.\alpha[m])\}\cup \textbf{Id}$ is a R weakly probabilistic pomset bisimulation, we omit it;
  \item $(\alpha_1\parallel\cdots\parallel\alpha_n).\tau.P\approx_{pp}^f (\alpha_1\parallel\cdots\parallel\alpha_n).P$. It is sufficient to prove the relation $R=\{((\alpha_1\parallel\cdots\parallel\alpha_n).\tau.P, (\alpha_1\parallel\cdots\parallel\alpha_n).P)\}\cup \textbf{Id}$ is a F weakly probabilistic pomset bisimulation, we omit it;
  \item $P.\tau.(\alpha_1[m]\parallel\cdots\parallel\alpha_n[m])\approx_{pp}^r P.(\alpha_1[m]\parallel\cdots\parallel\alpha_n[m])$. It is sufficient to prove the relation $R=\{(P.\tau.(\alpha_1[m]\parallel\cdots\parallel\alpha_n[m]), P.(\alpha_1[m]\parallel\cdots\parallel\alpha_n[m]))\}\cup \textbf{Id}$ is a R weakly probabilistic pomset bisimulation, we omit it;
  \item $P+\tau.P\approx_{pp}^f \tau.P$. It is sufficient to prove the relation $R=\{(P+\tau.P, \tau.P)\}\cup \textbf{Id}$ is a F weakly probabilistic pomset bisimulation, we omit it;
  \item $P+P.\tau\approx_{pp}^r P.\tau$. It is sufficient to prove the relation $R=\{(P+P.\tau, P.\tau)\}\cup \textbf{Id}$ is a R weakly probabilistic pomset bisimulation, we omit it;
  \item $P\cdot((Q+\tau\cdot(Q+R))\boxplus_{\pi}S)\approx_{pp}^{f}P\cdot((Q+R)\boxplus_{\pi}S)$. It is sufficient to prove the relation $R=\{(P\cdot((Q+\tau\cdot(Q+R))\boxplus_{\pi}S), P\cdot((Q+R)\boxplus_{\pi}S))\}\cup \textbf{Id}$ is a F weakly probabilistic pomset bisimulation, we omit it;
  \item $((Q+(Q+R)\cdot\tau)\boxplus_{\pi}S)\cdot P\approx_{pp}^{r}((Q+R)\boxplus_{\pi}S)\cdot P$. It is sufficient to prove the relation $R=\{(((Q+(Q+R)\cdot\tau)\boxplus_{\pi}S)\cdot P, ((Q+R)\boxplus_{\pi}S)\cdot P)\}\cup \textbf{Id}$ is a R weakly probabilistic pomset bisimulation, we omit it;  
  \item $P\approx_{pp}^{fr} \tau\parallel P$. It is sufficient to prove the relation $R=\{(P, \tau\parallel P)\}\cup \textbf{Id}$ is a FR weakly probabilistic pomset bisimulation, we omit it.
\end{enumerate}
\end{proof}

\begin{proposition}[$\tau$ laws for FR weakly probabilistic step bisimulation]
The $\tau$ laws for FR weakly probabilistic step bisimulation is as follows.
\begin{enumerate}
  \item $P\approx_{ps}^f \tau.P$;
  \item $P\approx_{ps}^r P.\tau$;
  \item $\alpha.\tau.P\approx_{ps}^f \alpha.P$;
  \item $P.\tau.\alpha[m]\approx_{ps}^r P.\alpha[m]$;
  \item $(\alpha_1\parallel\cdots\parallel\alpha_n).\tau.P\approx_{ps}^f (\alpha_1\parallel\cdots\parallel\alpha_n).P$;
  \item $P.\tau.(\alpha_1[m]\parallel\cdots\parallel\alpha_n[m])\approx_{ps}^r P.(\alpha_1[m]\parallel\cdots\parallel\alpha_n[m])$;
  \item $P+\tau.P\approx_{ps}^f \tau.P$;
  \item $P+P.\tau\approx_{ps}^r P.\tau$;
  \item $P\cdot((Q+\tau\cdot(Q+R))\boxplus_{\pi}S)\approx_{ps}^{f}P\cdot((Q+R)\boxplus_{\pi}S)$;
  \item $((Q+(Q+R)\cdot\tau)\boxplus_{\pi}S)\cdot P\approx_{ps}^{r}((Q+R)\boxplus_{\pi}S)\cdot P$;
  \item $P\approx_{ps}^{fr} \tau\parallel P$.
\end{enumerate}
\end{proposition}

\begin{proof}
\begin{enumerate}
  \item $P\approx_{ps}^f \tau.P$. It is sufficient to prove the relation $R=\{(P, \tau.P)\}\cup \textbf{Id}$ is a F weakly probabilistic step bisimulation, we omit it;
  \item $P\approx_{ps}^r P.\tau$. It is sufficient to prove the relation $R=\{(P, P.\tau)\}\cup \textbf{Id}$ is a R weakly probabilistic step bisimulation, we omit it;
  \item $\alpha.\tau.P\approx_{ps}^f \alpha.P$. It is sufficient to prove the relation $R=\{(\alpha.\tau.P, \alpha.P)\}\cup \textbf{Id}$ is a F weakly probabilistic step bisimulation, we omit it;
  \item $P.\tau.\alpha[m]\approx_{ps}^r P.\alpha[m]$. It is sufficient to prove the relation $R=\{(P.\tau.\alpha[m], P.\alpha[m])\}\cup \textbf{Id}$ is a R weakly probabilistic step bisimulation, we omit it;
  \item $(\alpha_1\parallel\cdots\parallel\alpha_n).\tau.P\approx_{ps}^f (\alpha_1\parallel\cdots\parallel\alpha_n).P$. It is sufficient to prove the relation $R=\{((\alpha_1\parallel\cdots\parallel\alpha_n).\tau.P, (\alpha_1\parallel\cdots\parallel\alpha_n).P)\}\cup \textbf{Id}$ is a F weakly probabilistic step bisimulation, we omit it;
  \item $P.\tau.(\alpha_1[m]\parallel\cdots\parallel\alpha_n[m])\approx_{ps}^r P.(\alpha_1[m]\parallel\cdots\parallel\alpha_n[m])$. It is sufficient to prove the relation $R=\{(P.\tau.(\alpha_1[m]\parallel\cdots\parallel\alpha_n[m]), P.(\alpha_1[m]\parallel\cdots\parallel\alpha_n[m]))\}\cup \textbf{Id}$ is a R weakly probabilistic step bisimulation, we omit it;
  \item $P+\tau.P\approx_{ps}^f \tau.P$. It is sufficient to prove the relation $R=\{(P+\tau.P, \tau.P)\}\cup \textbf{Id}$ is a F weakly probabilistic step bisimulation, we omit it;
  \item $P+P.\tau\approx_{ps}^r P.\tau$. It is sufficient to prove the relation $R=\{(P+P.\tau, P.\tau)\}\cup \textbf{Id}$ is a R weakly probabilistic step bisimulation, we omit it;
  \item $P\cdot((Q+\tau\cdot(Q+R))\boxplus_{\pi}S)\approx_{ps}^{f}P\cdot((Q+R)\boxplus_{\pi}S)$. It is sufficient to prove the relation $R=\{(P\cdot((Q+\tau\cdot(Q+R))\boxplus_{\pi}S), P\cdot((Q+R)\boxplus_{\pi}S))\}\cup \textbf{Id}$ is a F weakly probabilistic step bisimulation, we omit it;
  \item $((Q+(Q+R)\cdot\tau)\boxplus_{\pi}S)\cdot P\approx_{ps}^{r}((Q+R)\boxplus_{\pi}S)\cdot P$. It is sufficient to prove the relation $R=\{(((Q+(Q+R)\cdot\tau)\boxplus_{\pi}S)\cdot P, ((Q+R)\boxplus_{\pi}S)\cdot P)\}\cup \textbf{Id}$ is a R weakly probabilistic step bisimulation, we omit it;
  \item $P\approx_{ps}^{fr} \tau\parallel P$. It is sufficient to prove the relation $R=\{(P, \tau\parallel P)\}\cup \textbf{Id}$ is a FR weakly probabilistic step bisimulation, we omit it.
\end{enumerate}
\end{proof}

\begin{proposition}[$\tau$ laws for FR weakly probabilistic hp-bisimulation]
The $\tau$ laws for FR weakly probabilistic hp-bisimulation is as follows.
\begin{enumerate}
  \item $P\approx_{php}^f \tau.P$;
  \item $P\approx_{php}^r P.\tau$;
  \item $\alpha.\tau.P\approx_{php}^f \alpha.P$;
  \item $P.\tau.\alpha[m]\approx_{php}^r P.\alpha[m]$;
  \item $(\alpha_1\parallel\cdots\parallel\alpha_n).\tau.P\approx_{php}^f (\alpha_1\parallel\cdots\parallel\alpha_n).P$;
  \item $P.\tau.(\alpha_1[m]\parallel\cdots\parallel\alpha_n[m])\approx_{php}^r P.(\alpha_1[m]\parallel\cdots\parallel\alpha_n[m])$;
  \item $P+\tau.P\approx_{php}^f \tau.P$;
  \item $P+P.\tau\approx_{php}^r P.\tau$;
  \item $P\cdot((Q+\tau\cdot(Q+R))\boxplus_{\pi}S)\approx_{php}^{f}P\cdot((Q+R)\boxplus_{\pi}S)$;
  \item $((Q+(Q+R)\cdot\tau)\boxplus_{\pi}S)\cdot P\approx_{php}^{r}((Q+R)\boxplus_{\pi}S)\cdot P$;
  \item $P\approx_{php}^{fr} \tau\parallel P$.
\end{enumerate}
\end{proposition}

\begin{proof}
\begin{enumerate}
  \item $P\approx_{php}^f \tau.P$. It is sufficient to prove the relation $R=\{(P, \tau.P)\}\cup \textbf{Id}$ is a F weakly probabilistic hp-bisimulation, we omit it;
  \item $P\approx_{php}^r P.\tau$. It is sufficient to prove the relation $R=\{(P, P.\tau)\}\cup \textbf{Id}$ is a R weakly probabilistic hp-bisimulation, we omit it;
  \item $\alpha.\tau.P\approx_{php}^f \alpha.P$. It is sufficient to prove the relation $R=\{(\alpha.\tau.P, \alpha.P)\}\cup \textbf{Id}$ is a F weakly probabilistic hp-bisimulation, we omit it;
  \item $P.\tau.\alpha[m]\approx_{php}^r P.\alpha[m]$. It is sufficient to prove the relation $R=\{(P.\tau.\alpha[m], P.\alpha[m])\}\cup \textbf{Id}$ is a R weakly probabilistic hp-bisimulation, we omit it;
  \item $(\alpha_1\parallel\cdots\parallel\alpha_n).\tau.P\approx_{php}^f (\alpha_1\parallel\cdots\parallel\alpha_n).P$. It is sufficient to prove the relation $R=\{((\alpha_1\parallel\cdots\parallel\alpha_n).\tau.P, (\alpha_1\parallel\cdots\parallel\alpha_n).P)\}\cup \textbf{Id}$ is a F weakly probabilistic hp-bisimulation, we omit it;
  \item $P.\tau.(\alpha_1[m]\parallel\cdots\parallel\alpha_n[m])\approx_{php}^r P.(\alpha_1[m]\parallel\cdots\parallel\alpha_n[m])$. It is sufficient to prove the relation $R=\{(P.\tau.(\alpha_1[m]\parallel\cdots\parallel\alpha_n[m]), P.(\alpha_1[m]\parallel\cdots\parallel\alpha_n[m]))\}\cup \textbf{Id}$ is a R weakly probabilistic hp-bisimulation, we omit it;
  \item $P+\tau.P\approx_{php}^f \tau.P$. It is sufficient to prove the relation $R=\{(P+\tau.P, \tau.P)\}\cup \textbf{Id}$ is a F weakly probabilistic hp-bisimulation, we omit it;
  \item $P+P.\tau\approx_{php}^r P.\tau$. It is sufficient to prove the relation $R=\{(P+P.\tau, P.\tau)\}\cup \textbf{Id}$ is a R weakly probabilistic hp-bisimulation, we omit it;
  \item $P\cdot((Q+\tau\cdot(Q+R))\boxplus_{\pi}S)\approx_{php}^{f}P\cdot((Q+R)\boxplus_{\pi}S)$. It is sufficient to prove the relation $R=\{(P\cdot((Q+\tau\cdot(Q+R))\boxplus_{\pi}S), P\cdot((Q+R)\boxplus_{\pi}S))\}\cup \textbf{Id}$ is a F weakly probabilistic hp-bisimulation, we omit it;
  \item $((Q+(Q+R)\cdot\tau)\boxplus_{\pi}S)\cdot P\approx_{php}^{r}((Q+R)\boxplus_{\pi}S)\cdot P$. It is sufficient to prove the relation $R=\{(((Q+(Q+R)\cdot\tau)\boxplus_{\pi}S)\cdot P, ((Q+R)\boxplus_{\pi}S)\cdot P)\}\cup \textbf{Id}$ is a R weakly probabilistic hp-bisimulation, we omit it;
  \item $P\approx_{php}^{fr} \tau\parallel P$. It is sufficient to prove the relation $R=\{(P, \tau\parallel P)\}\cup \textbf{Id}$ is a FR weakly probabilistic hp-bisimulation, we omit it.
\end{enumerate}
\end{proof}

\begin{proposition}[$\tau$ laws for FR weakly probabilistic hhp-bisimulation]
The $\tau$ laws for FR weakly probabilistic hhp-bisimulation is as follows.
\begin{enumerate}
  \item $P\approx_{phhp}^f \tau.P$;
  \item $P\approx_{phhp}^r P.\tau$;
  \item $\alpha.\tau.P\approx_{phhp}^f \alpha.P$;
  \item $P.\tau.\alpha[m]\approx_{phhp}^r P.\alpha[m]$;
  \item $(\alpha_1\parallel\cdots\parallel\alpha_n).\tau.P\approx_{phhp}^f (\alpha_1\parallel\cdots\parallel\alpha_n).P$;
  \item $P.\tau.(\alpha_1[m]\parallel\cdots\parallel\alpha_n[m])\approx_{phhp}^r P.(\alpha_1[m]\parallel\cdots\parallel\alpha_n[m])$;
  \item $P+\tau.P\approx_{phhp}^f \tau.P$;
  \item $P+P.\tau\approx_{phhp}^r P.\tau$;
  \item $P\cdot((Q+\tau\cdot(Q+R))\boxplus_{\pi}S)\approx_{phhp}^{f}P\cdot((Q+R)\boxplus_{\pi}S)$;
  \item $((Q+(Q+R)\cdot\tau)\boxplus_{\pi}S)\cdot P\approx_{phhp}^{r}((Q+R)\boxplus_{\pi}S)\cdot P$;
  \item $P\approx_{phhp}^{fr} \tau\parallel P$.
\end{enumerate}
\end{proposition}

\begin{proof}
\begin{enumerate}
  \item $P\approx_{phhp}^f \tau.P$. It is sufficient to prove the relation $R=\{(P, \tau.P)\}\cup \textbf{Id}$ is a F weakly probabilistic hhp-bisimulation, we omit it;
  \item $P\approx_{phhp}^r P.\tau$. It is sufficient to prove the relation $R=\{(P, P.\tau)\}\cup \textbf{Id}$ is a R weakly probabilistic hhp-bisimulation, we omit it;
  \item $\alpha.\tau.P\approx_{phhp}^f \alpha.P$. It is sufficient to prove the relation $R=\{(\alpha.\tau.P, \alpha.P)\}\cup \textbf{Id}$ is a F weakly probabilistic hhp-bisimulation, we omit it;
  \item $P.\tau.\alpha[m]\approx_{phhp}^r P.\alpha[m]$. It is sufficient to prove the relation $R=\{(P.\tau.\alpha[m], P.\alpha[m])\}\cup \textbf{Id}$ is a R weakly probabilistic hhp-bisimulation, we omit it;
  \item $(\alpha_1\parallel\cdots\parallel\alpha_n).\tau.P\approx_{phhp}^f (\alpha_1\parallel\cdots\parallel\alpha_n).P$. It is sufficient to prove the relation $R=\{((\alpha_1\parallel\cdots\parallel\alpha_n).\tau.P, (\alpha_1\parallel\cdots\parallel\alpha_n).P)\}\cup \textbf{Id}$ is a F weakly probabilistic hhp-bisimulation, we omit it;
  \item $P.\tau.(\alpha_1[m]\parallel\cdots\parallel\alpha_n[m])\approx_{phhp}^r P.(\alpha_1[m]\parallel\cdots\parallel\alpha_n[m])$. It is sufficient to prove the relation $R=\{(P.\tau.(\alpha_1[m]\parallel\cdots\parallel\alpha_n[m]), P.(\alpha_1[m]\parallel\cdots\parallel\alpha_n[m]))\}\cup \textbf{Id}$ is a R weakly probabilistic hhp-bisimulation, we omit it;
  \item $P+\tau.P\approx_{phhp}^f \tau.P$. It is sufficient to prove the relation $R=\{(P+\tau.P, \tau.P)\}\cup \textbf{Id}$ is a F weakly probabilistic hhp-bisimulation, we omit it;
  \item $P+P.\tau\approx_{phhp}^r P.\tau$. It is sufficient to prove the relation $R=\{(P+P.\tau, P.\tau)\}\cup \textbf{Id}$ is a R weakly probabilistic hhp-bisimulation, we omit it;
  \item $P\cdot((Q+\tau\cdot(Q+R))\boxplus_{\pi}S)\approx_{phhp}^{f}P\cdot((Q+R)\boxplus_{\pi}S)$. It is sufficient to prove the relation $R=\{(P\cdot((Q+\tau\cdot(Q+R))\boxplus_{\pi}S), P\cdot((Q+R)\boxplus_{\pi}S))\}\cup \textbf{Id}$ is a F weakly probabilistic hhp-bisimulation, we omit it;
  \item $((Q+(Q+R)\cdot\tau)\boxplus_{\pi}S)\cdot P\approx_{phhp}^{r}((Q+R)\boxplus_{\pi}S)\cdot P$. It is sufficient to prove the relation $R=\{(((Q+(Q+R)\cdot\tau)\boxplus_{\pi}S)\cdot P, ((Q+R)\boxplus_{\pi}S)\cdot P)\}\cup \textbf{Id}$ is a R weakly probabilistic hhp-bisimulation, we omit it;
  \item $P\approx_{phhp}^{fr} \tau\parallel P$. It is sufficient to prove the relation $R=\{(P, \tau\parallel P)\}\cup \textbf{Id}$ is a FR weakly probabilistic hhp-bisimulation, we omit it.
\end{enumerate}
\end{proof}

\subsubsection{Recursion}

\begin{definition}[Sequential]
$X$ is sequential in $E$ if every subexpression of $E$ which contains $X$, apart from $X$ itself, is of the form $\alpha.F$ or $F.\alpha[m]$, or
$(\alpha_1\parallel\cdots\parallel\alpha_n).F$ or $F.(\alpha_1[m]\parallel\cdots\parallel\alpha_n[m])$, or $\sum\widetilde{F}$.
\end{definition}

\begin{definition}[Guarded recursive expression]
$X$ is guarded in $E$ if each occurrence of $X$ is with some subexpression $l.F$ or $F.l[m]$, or $(l_1\parallel\cdots\parallel l_n).F$ or $F.(l_1[m]\parallel\cdots\parallel l_n[m])$ of
$E$.
\end{definition}

\begin{lemma}\label{LUSWW07}
Let $G$ be guarded and sequential, $Vars(G)\subseteq\widetilde{X}$, and let $\langle G\{\widetilde{P}/\widetilde{X}\},s\rangle\rightsquigarrow\xrightarrow{\{\alpha_1,\cdots,\alpha_n\}}\langle P',s'\rangle$
or $\langle G\{\widetilde{P}/\widetilde{X}\},s\rangle\rightsquigarrow\xtworightarrow{\{\alpha_1[m],\cdots,\alpha_n[m]\}}\langle P',s'\rangle$. Then there is an expression $H$ such that
$\langle G,s\rangle\rightsquigarrow\xrightarrow{\{\alpha_1,\cdots,\alpha_n\}}\langle H,s'\rangle$ or $\langle G,s\rangle\rightsquigarrow\xtworightarrow{\{\alpha_1[m],\cdots,\alpha_n[m]\}}\langle H,s'\rangle$,
$P'\equiv H\{\widetilde{P}/\widetilde{X}\}$, and for any $\widetilde{Q}$, $\langle G\{\widetilde{Q}/\widetilde{X}\},s\rangle\rightsquigarrow\xrightarrow{\{\alpha_1,\cdots,\alpha_n\}} \langle H\{\widetilde{Q}/\widetilde{X}\},s'\rangle$
or $\langle G\{\widetilde{Q}/\widetilde{X}\},s\rangle\rightsquigarrow\xtworightarrow{\{\alpha_1[m],\cdots,\alpha_n[m]\}} \langle H\{\widetilde{Q}/\widetilde{X}\},s'\rangle$. Moreover $H$ is sequential,
$Vars(H)\subseteq\widetilde{X}$, and if $\alpha_1=\cdots=\alpha_n=\alpha_1[m]=\cdots=\alpha_n[m]=\tau$, then $H$ is also guarded.
\end{lemma}

\begin{proof}
We only prove the case of forward transition.

We need to induct on the structure of $G$.

If $G$ is a Constant, a Composition, a Restriction or a Relabeling then it contains no variables, since $G$ is sequential and guarded, then
$\langle G,s\rangle\rightsquigarrow\xrightarrow{\{\alpha_1,\cdots,\alpha_n\}}\langle P',s'\rangle$, then let $H\equiv P'$, as desired.

$G$ cannot be a variable, since it is guarded.

If $G\equiv G_1+G_2$. Then either $\langle G_1\{\widetilde{P}/\widetilde{X}\},s\rangle \rightsquigarrow\xrightarrow{\{\alpha_1,\cdots,\alpha_n\}}\langle P',s'\rangle$ or
$\langle G_2\{\widetilde{P}/\widetilde{X}\},s\rangle \rightsquigarrow\xrightarrow{\{\alpha_1,\cdots,\alpha_n\}}\langle P',s'\rangle$, then, we can apply this lemma in either case, as desired.

If $G\equiv\beta.H$. Then we must have $\alpha=\beta$, and $P'\equiv H\{\widetilde{P}/\widetilde{X}\}$, and
$\langle G\{\widetilde{Q}/\widetilde{X}\},s\rangle\equiv \langle\beta.H\{\widetilde{Q}/\widetilde{X}\},s\rangle \rightsquigarrow\xrightarrow{\beta}\langle H\{\widetilde{Q}/\widetilde{X}\},s'\rangle$,
then, let $G'$ be $H$, as desired.

If $G\equiv(\beta_1\parallel\cdots\parallel\beta_n).H$. Then we must have $\alpha_i=\beta_i$ for $1\leq i\leq n$, and $P'\equiv H\{\widetilde{P}/\widetilde{X}\}$, and
$\langle G\{\widetilde{Q}/\widetilde{X}\},s\rangle\equiv \langle(\beta_1\parallel\cdots\parallel\beta_n).H\{\widetilde{Q}/\widetilde{X}\},s\rangle \rightsquigarrow\xrightarrow{\{\beta_1,\cdots,\beta_n\}}\langle H\{\widetilde{Q}/\widetilde{X}\},s'\rangle$,
then, let $G'$ be $H$, as desired.

If $G\equiv\tau.H$. Then we must have $\tau=\tau$, and $P'\equiv H\{\widetilde{P}/\widetilde{X}\}$, and
$\langle G\{\widetilde{Q}/\widetilde{X}\},s\rangle\equiv \langle\tau.H\{\widetilde{Q}/\widetilde{X}\},s\rangle \rightsquigarrow\xrightarrow{\tau}\langle H\{\widetilde{Q}/\widetilde{X}\},s'\rangle$,
then, let $G'$ be $H$, as desired.

For the case of reverse transition, it can be proven similarly, we omit it.
\end{proof}

\begin{theorem}[Unique solution of equations for FR weakly probabilistic pomset bisimulation]
Let the guarded and sequential expressions $\widetilde{E}$ contain free variables $\subseteq \widetilde{X}$, then,

If $\widetilde{P}\approx_{pp}^{fr} \widetilde{E}\{\widetilde{P}/\widetilde{X}\}$ and $\widetilde{Q}\approx_{pp}^{fr} \widetilde{E}\{\widetilde{Q}/\widetilde{X}\}$, then
$\widetilde{P}\approx_{pp}^{fr} \widetilde{Q}$.
\end{theorem}

\begin{proof}
We only prove the case of forward transition.

Like the corresponding theorem in CCS, without loss of generality, we only consider a single equation $X=E$. So we assume $P\approx_{pp}^{fr} E(P)$, $Q\approx_{pp}^{fr} E(Q)$, then $P\approx_{pp}^{fr} Q$.

We will prove $\{(H(P),H(Q)): H\}$ sequential, if $\langle H(P),s\rangle\rightsquigarrow\xrightarrow{\{\alpha_1,\cdots,\alpha_n\}}\langle P',s'\rangle$, then, for some $Q'$,
$\langle H(Q),s\rangle\rightsquigarrow\xRightarrow{\{\alpha_1.\cdots,\alpha_n\}}\langle Q',s'\rangle$ and $P'\approx_{pp}^{fr} Q'$.

Let $\langle H(P),s\rangle\rightsquigarrow\xrightarrow{\{\alpha_1,\cdot,\alpha_n\}}\langle P',s'\rangle$, then $\langle H(E(P)),s\rangle\rightsquigarrow\xRightarrow{\{\alpha_1,\cdots,\alpha_n\}}\langle P'',s''\rangle$
and $P'\approx_{pp}^{fr} P''$.

By Lemma \ref{LUSWW07}, we know there is a sequential $H'$ such that $\langle H(E(P)),s\rangle\rightsquigarrow\xRightarrow{\{\alpha_1,\cdots,\alpha_n\}}\langle H'(P),s'\rangle\Rightarrow P''\approx_{pp}^{fr} P'$.

And, $\langle H(E(Q)),s\rangle\rightsquigarrow\xRightarrow{\{\alpha_1,\cdots,\alpha_n\}}\langle H'(Q),s'\rangle\Rightarrow Q''$ and $P''\approx_{pp}^{fr} Q''$. And $\langle H(Q),s\rangle\rightsquigarrow\xrightarrow{\{\alpha_1,\cdots,\alpha_n\}}\langle Q',s'\rangle\approx_{pp}^{fr} \Rightarrow Q'\approx_{pp}^{fr} Q''$.
Hence, $P'\approx_{pp}^{fr} Q'$, as desired.

For the case of reverse transition, it can be proven similarly, we omit it.
\end{proof}

\begin{theorem}[Unique solution of equations for FR weakly probabilistic step bisimulation]
Let the guarded and sequential expressions $\widetilde{E}$ contain free variables $\subseteq \widetilde{X}$, then,

If $\widetilde{P}\approx_{ps}^{fr} \widetilde{E}\{\widetilde{P}/\widetilde{X}\}$ and $\widetilde{Q}\approx_{ps}^{fr} \widetilde{E}\{\widetilde{Q}/\widetilde{X}\}$, then
$\widetilde{P}\approx_{ps}^{fr} \widetilde{Q}$.
\end{theorem}

\begin{proof}
We only prove the case of forward transition.

Like the corresponding theorem in CCS, without loss of generality, we only consider a single equation $X=E$. So we assume $P\approx_{ps}^{fr} E(P)$, $Q\approx_{ps}^{fr} E(Q)$, then $P\approx_{ps}^{fr} Q$.

We will prove $\{(H(P),H(Q)): H\}$ sequential, if $\langle H(P),s\rangle\rightsquigarrow\xrightarrow{\{\alpha_1,\cdots,\alpha_n\}}\langle P',s'\rangle$, then, for some $Q'$,
$\langle H(Q),s\rangle\rightsquigarrow\xRightarrow{\{\alpha_1.\cdots,\alpha_n\}}\langle Q',s'\rangle$ and $P'\approx_{ps}^{fr} Q'$.

Let $\langle H(P),s\rangle\rightsquigarrow\xrightarrow{\{\alpha_1,\cdot,\alpha_n\}}\langle P',s'\rangle$, then $\langle H(E(P)),s\rangle\rightsquigarrow\xRightarrow{\{\alpha_1,\cdots,\alpha_n\}}\langle P'',s''\rangle$
and $P'\approx_{ps}^{fr} P''$.

By Lemma \ref{LUSWW07}, we know there is a sequential $H'$ such that $\langle H(E(P)),s\rangle\rightsquigarrow\xRightarrow{\{\alpha_1,\cdots,\alpha_n\}}\langle H'(P),s'\rangle\Rightarrow P''\approx_{ps}^{fr} P'$.

And, $\langle H(E(Q)),s\rangle\rightsquigarrow\xRightarrow{\{\alpha_1,\cdots,\alpha_n\}}\langle H'(Q),s'\rangle\Rightarrow Q''$ and $P''\approx_{ps}^{fr} Q''$. And $\langle H(Q),s\rangle\rightsquigarrow\xrightarrow{\{\alpha_1,\cdots,\alpha_n\}}\langle Q',s'\rangle\approx_{ps}^{fr} \Rightarrow Q'\approx_{ps}^{fr} Q''$.
Hence, $P'\approx_{ps}^{fr} Q'$, as desired.

For the case of reverse transition, it can be proven similarly, we omit it.
\end{proof}

\begin{theorem}[Unique solution of equations for FR weakly probabilistic hp-bisimulation]
Let the guarded and sequential expressions $\widetilde{E}$ contain free variables $\subseteq \widetilde{X}$, then,

If $\widetilde{P}\approx_{php}^{fr} \widetilde{E}\{\widetilde{P}/\widetilde{X}\}$ and $\widetilde{Q}\approx_{php}^{fr} \widetilde{E}\{\widetilde{Q}/\widetilde{X}\}$, then
$\widetilde{P}\approx_{php}^{fr} \widetilde{Q}$.
\end{theorem}

\begin{proof}
We only prove the case of forward transition.

Like the corresponding theorem in CCS, without loss of generality, we only consider a single equation $X=E$. So we assume $P\approx_{php}^{fr} E(P)$, $Q\approx_{php}^{fr} E(Q)$, then $P\approx_{php}^{fr} Q$.

We will prove $\{(H(P),H(Q)): H\}$ sequential, if $\langle H(P),s\rangle\rightsquigarrow\xrightarrow{\{\alpha_1,\cdots,\alpha_n\}}\langle P',s'\rangle$, then, for some $Q'$,
$\langle H(Q),s\rangle\rightsquigarrow\xRightarrow{\{\alpha_1.\cdots,\alpha_n\}}\langle Q',s'\rangle$ and $P'\approx_{php}^{fr} Q'$.

Let $\langle H(P),s\rangle\rightsquigarrow\xrightarrow{\{\alpha_1,\cdot,\alpha_n\}}\langle P',s'\rangle$, then $\langle H(E(P)),s\rangle\rightsquigarrow\xRightarrow{\{\alpha_1,\cdots,\alpha_n\}}\langle P'',s''\rangle$
and $P'\approx_{php}^{fr} P''$.

By Lemma \ref{LUSWW07}, we know there is a sequential $H'$ such that $\langle H(E(P)),s\rangle\rightsquigarrow\xRightarrow{\{\alpha_1,\cdots,\alpha_n\}}\langle H'(P),s'\rangle\Rightarrow P''\approx_{php}^{fr} P'$.

And, $\langle H(E(Q)),s\rangle\rightsquigarrow\xRightarrow{\{\alpha_1,\cdots,\alpha_n\}}\langle H'(Q),s'\rangle\Rightarrow Q''$ and $P''\approx_{php}^{fr} Q''$. And $\langle H(Q),s\rangle\rightsquigarrow\xrightarrow{\{\alpha_1,\cdots,\alpha_n\}}\langle Q',s'\rangle\approx_{php}^{fr} \Rightarrow Q'\approx_{php}^{fr} Q''$.
Hence, $P'\approx_{php}^{fr} Q'$, as desired.

For the case of reverse transition, it can be proven similarly, we omit it.
\end{proof}

\begin{theorem}[Unique solution of equations for FR weakly probabilistic hhp-bisimulation]
Let the guarded and sequential expressions $\widetilde{E}$ contain free variables $\subseteq \widetilde{X}$, then,

If $\widetilde{P}\approx_{phhp}^{fr} \widetilde{E}\{\widetilde{P}/\widetilde{X}\}$ and $\widetilde{Q}\approx_{phhp}^{fr} \widetilde{E}\{\widetilde{Q}/\widetilde{X}\}$, then
$\widetilde{P}\approx_{phhp}^{fr} \widetilde{Q}$.
\end{theorem}

\begin{proof}
We only prove the case of forward transition.

Like the corresponding theorem in CCS, without loss of generality, we only consider a single equation $X=E$. So we assume $P\approx_{phhp}^{fr} E(P)$, $Q\approx_{phhp}^{fr} E(Q)$, then $P\approx_{phhp}^{fr} Q$.

We will prove $\{(H(P),H(Q)): H\}$ sequential, if $\langle H(P),s\rangle\rightsquigarrow\xrightarrow{\{\alpha_1,\cdots,\alpha_n\}}\langle P',s'\rangle$, then, for some $Q'$,
$\langle H(Q),s\rangle\rightsquigarrow\xRightarrow{\{\alpha_1.\cdots,\alpha_n\}}\langle Q',s'\rangle$ and $P'\approx_{phhp}^{fr} Q'$.

Let $\langle H(P),s\rangle\rightsquigarrow\xrightarrow{\{\alpha_1,\cdot,\alpha_n\}}\langle P',s'\rangle$, then $\langle H(E(P)),s\rangle\rightsquigarrow\xRightarrow{\{\alpha_1,\cdots,\alpha_n\}}\langle P'',s''\rangle$
and $P'\approx_{phhp}^{fr} P''$.

By Lemma \ref{LUSWW07}, we know there is a sequential $H'$ such that $\langle H(E(P)),s\rangle\rightsquigarrow\xRightarrow{\{\alpha_1,\cdots,\alpha_n\}}\langle H'(P),s'\rangle\Rightarrow P''\approx_{phhp}^{fr} P'$.

And, $\langle H(E(Q)),s\rangle\rightsquigarrow\xRightarrow{\{\alpha_1,\cdots,\alpha_n\}}\langle H'(Q),s'\rangle\Rightarrow Q''$ and $P''\approx_{phhp}^{fr} Q''$. And $\langle H(Q),s\rangle\rightsquigarrow\xrightarrow{\{\alpha_1,\cdots,\alpha_n\}}\langle Q',s'\rangle\approx_{phhp}^{fr} \Rightarrow Q'\approx_{phhp}^{fr} Q''$.
Hence, $P'\approx_{phhp}^{fr} Q'$, as desired.

For the case of reverse transition, it can be proven similarly, we omit it.
\end{proof}


%




\newpage\begin{thebibliography}{Lam94}
\bibitem{CC}R. Milner. (1989). Communication and concurrency. Printice Hall.

\bibitem{CCS}R. Milner. (1980). A calculus of communicating systems. LNCS 92, Springer.

\bibitem{ACP}W. Fokkink. (2007). Introduction to process algebra 2nd ed. Springer-Verlag.


\bibitem{PI1}R. Milner, J. Parrow, and D. Walker. (1992). A Calculus of Mobile Processes, Part I. Information and Computation, 100(1):1-40.

\bibitem{PI2}R. Milner, J. Parrow, and D. Walker. (1992). A calculus of mobile processes, Part II. Information and Computation, 100(1):41-77.

\bibitem{CTC}Y. Wang. (2017). A calculus for true concurrency. Manuscript, arxiv: 1703.00159.

\bibitem{ATC}Y. Wang. (2016). Algebraic laws for true concurrency. Manuscript, arXiv: 1611.09035.

\bibitem{PITC}Y. Wang. (2017). A calculus of truly concurrent mobile processes. Manuscript, arXiv: 1704.07774.

\bibitem{APRTC}Y. Wang. (2018). Truly Concurrent Process Algebra Is Reversible. Manuscript, arXiv: 1810.00868.

\bibitem{APPTC}Y. Wang. (2021). Probabilistic Process Algebra for True Concurrency. Manuscript, arXiv: 2107.08453.

\bibitem{PPA}S. Andova. (2002). Probabilistic process algebra. Annals of Operations Research 128(2002):204-219.

\bibitem{PPA2}S. Andova, J. Baeten, T. Willemse. (2006). A Complete Axiomatisation of Branching Bisimulation for Probabilistic Systems with an Application in Protocol
Verification. International Conference on Concurrency Theory. Springer Berlin Heidelberg.

\bibitem{PPA3}S. Andova, S. Georgievska. (2009). On Compositionality, Efficiency, and Applicability of Abstraction in Probabilistic Systems. Conference on Current Trends in Theory and
Practice of Computer Science. Springer-Verlag.

\bibitem{HLPA}J. F. Groote, A. Ponse. (1994). Process algebra with guards: combining hoare logic with process algebra. Formal Aspects of Computing, 6(2): 115-164.
\end{thebibliography}
\end{document}